\documentclass[a4paper,12pt,times,numbered,print,index]{Classes/PhDThesisPSnPDF}

\input{Preamble/preamble}

\crest{\includegraphics[width=\textwidth]{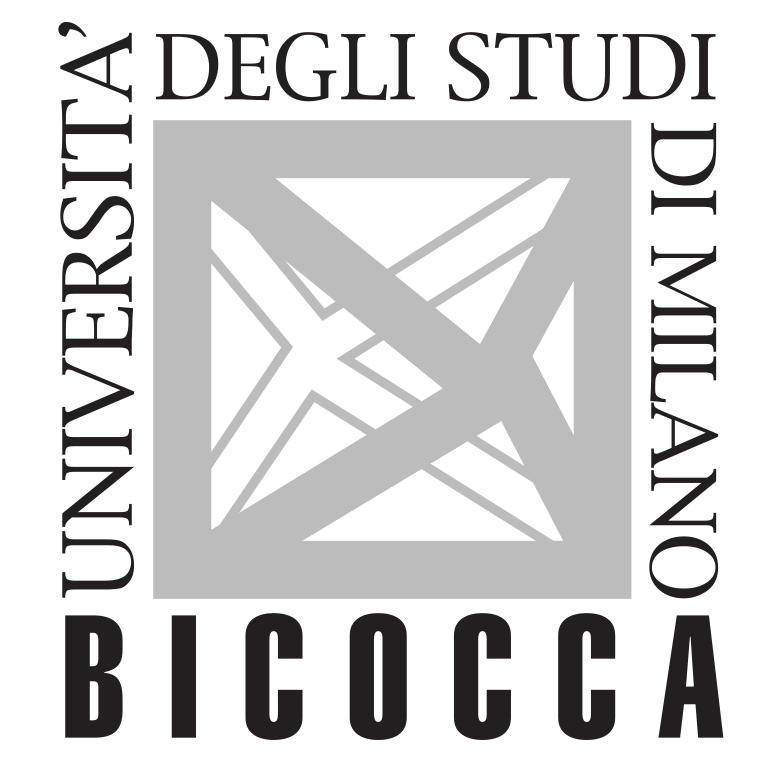}}

\university{\textsc{University of Milano-Bicocca}}

\dept{Department of Informatics, Systems and Communication \\
Ph.D. Program in Computer Science -- Cycle XXXI}

\title{Computer-Assisted Analysis of Biomedical Images}

\author{Leonardo Rundo}

\supervisorrole{Supervisor: }
\supervisor{Prof. Giancarlo Mauri}
\tutor{Prof. Giuseppe Vizzari}
\coordinator{Prof. Stefania Bandini}

\degreedate{\textsc{Academic Year 2017-2018}}

\subject{Computer-Assisted Analysis of Biomedical Images} 

\ifdefineAbstract
\fi

\ifdefineChapter
\fi

\begin{document}

\frontmatter

\maketitle


\begin{dedication} 

\begin{flushright}
\textit{To my parents and Miriana,}\\
\textit{for your loving support.}
\end{flushright}

\end{dedication}

\begin{acknowledgements}

\scriptsize
First of all, I want to thank my parents to instill in me the strong values about the scientific work and its social impact.
I am essentially the result of your infinite love and perseverance.
Obviously, the motivations for my Ph.D. have been strengthened by the presence of a unique person.
Miriana, you always supported my choices, even the most difficult to accept, with love and passion.
Many thanks also to every single component of your family, especially your parents Carmela and Francesco for treating me like a son during my short weekends.
A loving thought goes to my little nieces (Maria Francesca, Noemi, Maria), nephews (Lorenzo, Angelo, Francesco), and cousins (Giuseppe and Antonino) who managed to gladden me even during the most difficult moments in these intense years.
I am grateful to my entire family, from the South to the North of Italy, for cheering me throughout this fantastic journey.
My uncles Benedetto e Marianna have been like parents and my cousins Benedetta and Adriano, Cristina and Salvatore like real brothers and sisters.
In addition, my cousins Anna Maria and Salvatore hosted me when my Ph.D. admission seemed to be a crazy idea, while Mariella and Nino were fundamental in my settling-in period in Milan.

I would like to sincerely thank all the people who discovered potential in me, even when I was a shy Master's student.
Prof. Roberto Pirrone, Prof. Orazio Gambino, and Dr. Vincenzo Cannella presented the fascinating medical imaging context for my Master's Thesis activities at the University of Palermo. 
I thank Prof. Maria Carla Gilardi, Dr. Carmelo Militello, Dr. Giorgio Russo, Dr. Alessandro Stefano, for providing valuable opportunities for my Research Fellowship at the Institute of Molecular Bioimaging and Physiology -- Italian National Research Council.
In particular, Carmelo has been my affectionate ``elder brother'' in the last five years for my scientific advances.
I am also grateful to Prof. Salvatore Vitabile (University of Palermo) and Prof. Vincenzo Conti (University of Enna KORE) for trustfully involving me in challenging projects.

My Ph.D. program has been an amazing experience mainly thanks to Prof. Giancarlo Mauri.
He is not just my Supervisor, but he has been a perfect mentor who always demonstrated to believe in me.
In addition to a significant impact on my education, his kindness forged my approach to research and daily life.
I wish to thank the people working at DISCo for their friendship and for welcoming me from the first moment.
I absolutely learned a lot from all of you.
Especially, the research group composed of Prof. Daniela Besozzi, Prof. Paolo Cazzaniga, Dr. Marco Nobile, Andrea Tangherloni, and Simone Spolaor.
In addition to introducing me to the Computational Systems Biology field, I have been growing up as a scientist thanks to your methodological rigor and dedication.
I am very lucky for really enjoying these years with you and your students (especially Simone and Riccardo).
Andrea, recalling our memories at the Vanderbilt University and at the University of Cambridge, we are going to spend a lot of new memorable experiences together in Cambridge.
A great thank to Prof. Giuseppe Vizzari for his tutoring and precious support in every moment, in particular at the very beginning of my Ph.D. activities.

Special thanks are directed to: Prof. Carlos Lopez, along with Prof. Vito Quaranta and Prof. Darren Tyson, for inviting me at the Vanderbilt University, Nashville, TN, USA; Prof. Hideki Nakayama (The University of Tokyo, Tokyo, Japan) together with Changhee Han (a real friend who literally hosted me in his own home, so spending wonderful Japanese moments with the other guys in the Lab) for our international collaboration project; last but not least, Prof. Pietro Liò (University of Cambridge, Cambridge, UK) who practically transmits the love for science in everyday life.
The experience as a Visiting Scholar in your laboratories and the fruitful collaborations with your group members positively changed my vision about scientific research.

In this regard, I am grateful to Prof. Evis Sala who is investing in me for my next exciting adventure at the University of Cambridge.
Tons of thanks go to my new colleagues at the Department of Radiology, since you have made me feel at ease from the first moment.

\end{acknowledgements}

\begin{abstract}
\small
Nowadays, the amount of heterogeneous biomedical data is increasing more and more thanks to novel sensing techniques and high-throughput technologies.
In reference to biomedical image analysis, the advances in image acquisition modalities and high-throughput imaging experiments are creating new challenges.
This huge information ensemble could overwhelm the analytic capabilities needed by physicians in their daily decision-making tasks as well as by biologists investigating complex biochemical systems.
In particular, quantitative imaging methods convey scientifically and clinically relevant information in prediction, prognosis or treatment response assessment, by also considering radiomics approaches.
Therefore, the computational analysis of medical and biological images plays a key role in radiology and laboratory applications.
In this regard, frameworks based on advanced Machine Learning and Computational Intelligence can significantly improve traditional Image Processing and Pattern Recognition approaches.
However, conventional Artificial Intelligence techniques must be tailored to address the unique challenges concerning biomedical imaging data.

This thesis aims at proposing novel and advanced computer-assisted methods for biomedical image analysis, also as an instrument in the development of Clinical Decision Support Systems, by always keeping in mind the clinical feasibility of the developed solutions.
The devised classical Image Processing algorithms, with particular interest to region-based and morphological approaches in biomedical image segmentation, are first described.
Afterwards, Pattern Recognition techniques are introduced, applying unsupervised fuzzy clustering and graph-based models (i.e., Random Walker and Cellular Automata) to multispectral and multimodal medical imaging data processing.
Taking into account Computational Intelligence, an evolutionary framework based on Genetic Algorithms for medical image enhancement and segmentation is presented.
Moreover, multimodal image co-registration using Particle Swarm Optimization is discussed.
Finally, Deep Neural Networks are investigated: (\textit{i}) the generalization abilities of Convolutional Neural Networks in medical image segmentation for multi-institutional datasets are addressed by conceiving an architecture that integrates adaptive feature recalibration blocks, and (\textit{ii}) the generation of realistic medical images based on Generative Adversarial Networks is applied to data augmentation purposes.
In conclusion, the ultimate goal of these research studies is to gain clinically and biologically useful insights that can guide differential diagnosis and therapies, leading towards biomedical data integration for personalized medicine.
As a matter of fact, the proposed computer-assisted bioimage analysis methods can be beneficial for the definition of imaging biomarkers, as well as for quantitative medicine and biology.

\end{abstract}


\tableofcontents

\listoffigures

\listoftables


\printnomenclature

\mainmatter

\renewcommand*{\thefootnote}{\fnsymbol{footnote}}

\chapter{Introduction}  
\label{chap1}
\graphicspath{{Chapter1/Figs/}}

\section{Aims and motivations}
\label{sec:AimsMotiv}


Nowadays, the amount of heterogeneous biomedical data is increasing more and more thanks to novel sensing techniques (i.e., wearable devices and ubiquitous computing paradigms) \cite{hung2015} and high-throughput technologies (i.e., high-content imaging and multi-omics studies) \cite{hayer2010,libbrecht2015}, generating data typically characterized by an underlying complex structure \cite{gomes2012}.
In addition, electronic health (e-health) \cite{gray2011} and mobile health (m-health) \cite{alepis2013} can be properly integrated to develop automated mining engines \cite{alaa2016}, as well as support real-time personalized diagnosis \cite{pasero2017} and continuous monitoring (also by means of wearable biophysical and electrochemical sensors) \cite{randazzo2018}.
These emerging trends bring life to the so-called medical Internet of Things (mIoTs), which provides network-based platforms for the ubiquitous connectivity of patients and medical doctors in future medical applications \cite{metcalf2016}.
Therefore, specific network security mechanisms are required by such sensor-based medical devices \cite{aram2016}.
Cutting-edge Information and Communication Technology (ICT) can enable the shift from an organization-centric to a patient-centric model, leading to collaborative multi-centric healthcare service delivery processes \cite{serbanati2011}.

With particular interest to the biomedical image analysis field, the advances in image acquisition modalities \cite{peng2008} and high-throughput imaging experiments \cite{shamir2010} are creating new challenges.
This huge and valuable information ensemble, deriving from heterogeneous large-scale datasets, could overwhelm the analytic capabilities needed by physicians in their decision-making tasks as well as by biologists investigating complex biological systems \cite{meijering2016}.
In this regard, medical imaging comprises minimally invasive techniques for acquiring images that provide detailed information about the anatomy and physiology of the imaged organs \cite{duncan2000}, while live cell microscopy imaging allows for the visualization and analysis of the dynamic processes of living specimens \cite{meijering2012} that underwent treatments (e.g., radiation therapy or chemotherapy \cite{militelloCBM2017}).

Quantitative imaging methods convey scientifically and clinically relevant information in prediction, prognosis or treatment response assessment \cite{yankeelov2016}, by also considering radiomics approaches \cite{aerts2014}.
In such a context, Computational Intelligence and Machine Learning can significantly improve traditional Image Processing and Pattern Recognition techniques \cite{wang2012}, playing a key role in radiology \cite{wernick2010} and laboratory applications \cite{peng2012}. 
However, conventional Machine Learning and Computational Intelligence techniques must be tailored to address the unique challenges pertaining to biomedical images \cite{kraus2017,shen2017}.
Machine Learning studies computer algorithms that can learn automatically complex relationships or patterns from empirical data (i.e., features) and are able to make accurate decisions \cite{bishop2006}.
In addition, Computational Intelligence comprises a set of nature-inspired computational techniques to address complex real-world problems that cannot be tackled by means of exact algorithmic approaches.
As a matter of fact, these tasks---which could be difficult to model with constrained-based approaches relying on dynamic or  multi-parametric programming \cite{faisca2008,faisca2009}---might be stochastic in nature and can be suitably solved by exploiting global search metaheuristics \cite{siddique2013}.
Relevant models, features, and characteristics can be indeed automatically learned from images by relying on these computational frameworks \cite{rueckert2016}.

However, interpretable computational models \cite{castelvecchi2016}, allowing for the understandability of the results, must be taken into account in life sciences.
The issues related to the interpretability of Machine Learning and Computational Intelligence methods in medicine are compelling for the adoption and the feasibility of a novel Clinical Decision Support System (CDSS).
As a matter of fact, over-reliance on CDSS platforms could also cause a subtle loss of self-confidence and affect the willingness of a physician to provide a definitive interpretation or diagnosis \cite{cabitza2017}.
Therefore, considering the unique challenges encountered in clinical scenarios, interactive segmentation
algorithms \cite{boykov2000} often represent a more feasible and safe solution for
physician in clinical practice with respect to fully automatic approaches \cite{hamamci2012}.

It is worth noting that in this Ph.D. thesis, the design choice among the devised computational methods has been always motivated by the availability of biomedical data (considering also their labeling).
In addition, my research has been always tightly linked to the feasibility of the proposed clinical or laboratory applications.

\section{Computer science contributions to biomedical imaging}
\label{sec:CScontribut}

Considering these compelling issues related to biomedical big data \cite{regge2017}, proposing novel automated computational approaches can involve very diverse computer science areas and domains:
\begin{itemize}
    \item Information Retrieval \cite{drias2016} and Data Mining \cite{meystre2008} to extract knowledge from Electronic Health Records (EHRs)---for epidemiological research studies \cite{powell2005}---relying on structured and unstructured textual data;
    \item Computational Intelligence \cite{allmendinger2018} and Machine Learning \cite{wang2012,wernick2010} to enhance conventional digital Image Processing techniques \cite{bankman2009} on pictorial data.
\end{itemize}

\begin{figure}[!t]
	\centering
	\includegraphics[width=\textwidth]{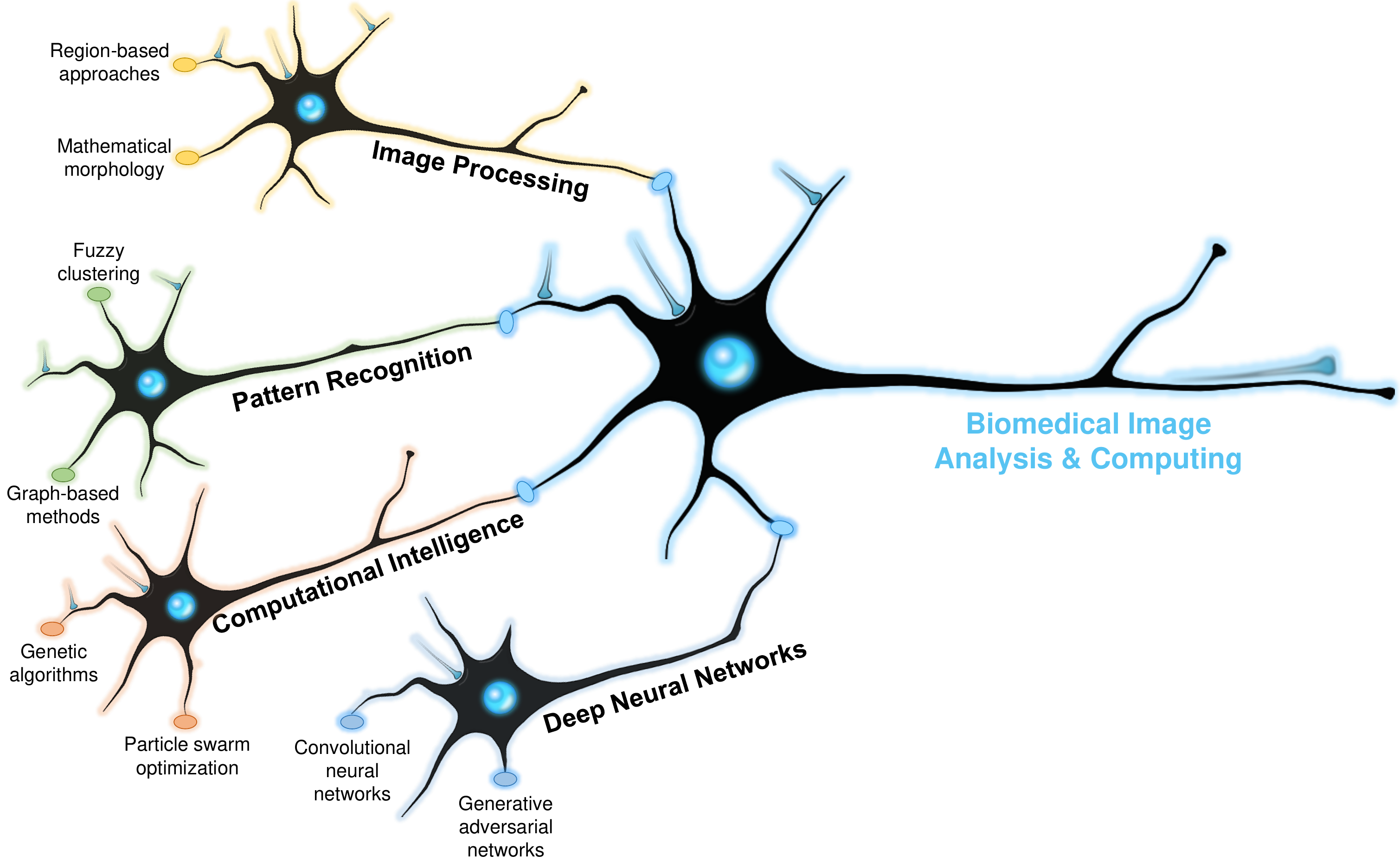}
	\caption[Conceptual scheme of the main topics treated in the thesis]{Conceptual scheme of the main topics treated in this thesis. This ``neural-based scheme'' aims at showing the interconnections (represented by synapses) among the different computational methods and biomedical image analysis and computing. The terminal synapses report the approaches specifically encountered during my research activities, reflecting the structure of the thesis. The colored halos of the four left-most neurons denote the four computer science areas related to biomedical imaging: Image Processing (Chapter 3), Pattern Recognition (Chapter 4), Computational Intelligence (Chapter 5), Deep Neural Networks (Chapter 6). Under a biological inspiration, the axons of these four neurons send out their output signal to the dendrites of the central light-blue neuron, by activating its own synapses. The effective combination emerging from the different computational approaches involves the excitation of the neuron representing biomedical image analysis and computing, which suddenly generates an action potential originating at the soma and propagates along the axon.}
	\label{fig:graphAbstract}
\end{figure}

The conceptual ``neural-based scheme'' of the main topics treated in this thesis is shown in Fig. \ref{fig:graphAbstract}, aiming at representing the interconnections among the devised computational methods and biomedical image analysis and computing.
Therefore, the principal contribution of computer science research in biomedicine concerns the proper fusion of diverse datasets \cite{lahat2015,serra2018} (i.e., imaging modalities, high-throughput technologies, EHRs) to provide a complete clinical knowledge for precision medicine \cite{brady2016}.

\section{Scientific production and thesis overview}
\label{sec:SciProd}

This Ph.D. thesis aims at proposing novel computer-assisted methods for biomedical image analysis, also as an instrument in the development of CDSSs, by always keeping in mind the clinical feasibility of the developed solutions.
Therefore, my Ph.D. program has been directed towards a two-fold objective: (\textit{i}) computer-assisted segmentation methods in cancer imaging, and (\textit{ii}) automatic cell image analysis.

This thesis is structured as follows.
The basic concepts related to medical imaging and bioimage informatics, as well as the investigated clinical and laboratory scenarios are recalled in \textbf{Chapter \ref{chap2}}.
The devised classical Image Processing algorithms, especially region-based and morphological approaches in biomedical image segmentation, are described in \textbf{Chapter \ref{chap3}}.
\textbf{Chapter \ref{chap4}} introduces Pattern Recognition techniques, with particular interest to unsupervised fuzzy clustering and graph-based models (i.e., Cellular Automata and Random Walker) to multispectral and multimodal medical imaging data processing.
Computational Intelligence methods are presented in \textbf{Chapter \ref{chap5}}, wherein a novel evolutionary framework based on Genetic Algorithms for medical image enhancement and segmentation is presented.
Moreover, multimodal image co-registration using Particle Swarm Optimization is discussed.
Finally, Deep Neural Networks are investigated in \textbf{Chapter \ref{chap6}}: (\textit{i}) the generalization abilities of Convolutional Neural Networks in medical image segmentation of multi-institutional datasets are addressed by conceiving an architecture that integrates adaptive feature recalibration, and (\textit{ii}) the generation of realistic medical images based on Generative Adversarial Networks is applied to data augmentation purposes.
\textbf{Chapter \ref{chap7}} concludes with final remarks and future developments.
\textbf{Appendix \ref{appendixA}} contains the evaluation metrics and procedures used for the validation of the proposed computational approaches.

\clearpage

\paragraph{Journal articles}

\begin{itemize}

\item \textbf{Rundo L.}$^*$, Tangherloni A.$^*$, Nobile M.S., Militello C., Mauri G., Besozzi D., Cazzaniga P. (2019) MedGA: A novel evolutionary method for image enhancement in medical imaging systems. \textit{Expert Systems with Applications}, 119, 387--399.\\ \texttt{DOI: 10.1016/j.eswa.2018.11.013}.

\item Gambino O.\footnote[1]{Equal contribution}, \textbf{Rundo L.}$^*$, Cannella V., Vitabile S., Pirrone R. (2018) A framework for data-driven adaptive GUI generation based on DICOM. \textit{Journal of Biomedical Informatics}, 88, 37--52.\\ \texttt{DOI: 10.1016/j.jbi.2018.10.009}

\item \textbf{Rundo L.}, Militello C., Russo G., Vitabile S., Gilardi M.C., Mauri G. (2018) \textsc{GTVcut} for neuro-radiosurgery treatment planning: an MRI brain cancer seeded image segmentation method based on a Cellular Automata model. \textit{Natural Computing}, 17(3), 521--536. Special Issue on ``Cellular Automata for Research and Industry (ACRI 2016)''.\\ \texttt{DOI: 10.1007/s11047-017-9636-z}

\item \textbf{Rundo L.}, Militello C., Tangherloni A., Russo G., Vitabile S., Gilardi M.C., Mauri G. (2018) NeXt for neuro-radiosurgery: a fully automatic approach for necrosis extraction in brain tumor MRI using an unsupervised Machine Learning technique. \textit{International Journal of Imaging Systems and Technology}, 28(1), 21--37.\\ \texttt{DOI: 10.1002/ima.22253}

\item Militello C., \textbf{Rundo L.}, Conti V., Minafra L., Cammarata F.P., Vitabile S., Mauri G., Gilardi M.C., Porcino N. (2017) Area-based cell colony surviving fraction evaluation: a novel fully automatic approach using general-purpose acquisition hardware. \textit{Computers in Biology and Medicine}, 89, 454--465.\\ \texttt{DOI: 10.1016/j.compbiomed.2017.08.005}

\item \textbf{Rundo L.}, Militello C., Russo G., Garufi A., Vitabile S., Mauri G., Gilardi M.C. (2017) Automated Prostate gland segmentation based on an unsupervised fuzzy C-means clustering technique using multispectral T1w and T2w MR imaging. \textit{Information}, 8(2), 49, pp. 1--28, Special Issue on ``Fuzzy Logic for Image Processing''.\\ \texttt{DOI: 10.3390/info8020049}

\item \textbf{Rundo L.}, Stefano A., Militello C., Russo, G. Sabini M.G., D’Arrigo C., Marletta F., Ippolito M., Mauri G., Vitabile S., Gilardi M.C. (2017) A fully automatic approach for multimodal PET and MR image segmentation in Gamma Knife treatment planning. \textit{Computer Methods and Programs in Biomedicine}, 144, 77--96.\\ \texttt{DOI: 10.1016/j.cmpb.2017.03.011}

\end{itemize}

\paragraph{Conference Proceedings and Book Chapters}

\begin{itemize}

\item \textbf{Rundo L.}, Han C., Zhang J., Hataya R., Nagano Y., Militello C., Ferretti C., Nobile M.S., Tangherloni A., Gilardi M.C., Vitabile S., Nakayama H., Mauri G. (2018) CNN-based prostate zonal segmentation on T2-weighted MR images: a cross-dataset study. Accepted by the: \textit{28th Italian Workshop on Neural Networks (WIRN 2018)}, Vietri sul Mare, Salerno, Italy, June 13--15, 2018.

\item Han C., \textbf{Rundo L.}, Araki R., Furukawa Y., Mauri G., Nakayama H., Hayashi H. (2018) Infinite brain MR images: PGGAN-based data augmentation for tumor detection. Accepted by the: \textit{28th Italian Workshop on Neural Networks (WIRN 2018)}, Vietri sul Mare, Salerno, Italy, June 13--15, 2018.

\item Han C., Hayashi H., \textbf{Rundo L.}, Araki R., Shimoda W., Muramatsu S., Yujiro F., Mauri G., Nakayama H. (2018) GAN-Based synthetic brain MR image generation. Proceedings of the \textit{2018 IEEE 15th International Symposium on Biomedical Imaging (ISBI 2018)}, pp. 734--738, Washington, DC, USA, April 4--7, 2018.\\ \texttt{DOI: 10.1109/ISBI.2018.8363678}

\item \textbf{Rundo L.}, Militello C., Tangherloni A., Russo G., Lagalla R., Mauri G., Gilardi M.C., Vitabile S. (2019) Computer-assisted approaches for uterine fibroid segmentation in MRgFUS treatments: quantitative evaluation and clinical feasibility analysis. In: Esposito, A., Faundez-Zanuy, M., Morabito, F.C., Pasero, E. (Editors), Quantifying and Processing Biomedical and Behavioral Signals, \textit{Smart Innovation, Systems and Technologies}, Vol. 103, Chap. 22, pp. 229--241, Springer International Publishing, 2018. Proceedings of the 27th Italian Workshop on Neural Networks (WIRN 2017), Vietri sul Mare, Salerno, Italy, June 14--16, 2017.\\ \texttt{DOI: 10.1007/978-3-319-95095-2\_22}

\item \textbf{Rundo L.}, Tangherloni A., Militello C., Gilardi M.C., Mauri G. (2017) Multimodal medical image registration using particle swarm optimization: a review. Proceedings of the \textit{2016 IEEE Symposium Series on Computational Intelligence (IEEE SSCI 2016)} -- Swarm Intelligence Symposium (SIS), pp. 1--8, Athens, Greece, December 6--9, 2016.\\ \texttt{DOI: 10.1109/SSCI.2016.7850261}

\item \textbf{Rundo L.}, Militello C., Russo G., D’Urso D., Valastro L.M., Garufi A., Mauri G., Vitabile S., Gilardi M.C. (2017) Fully automatic multispectral MR image segmentation of prostate gland based on the fuzzy C-means clustering algorithm. In: Esposito, A., Faundez-Zanuy, M., Morabito, F.C. (Editors), Multidisciplinary Approaches to Neural Computing, \textit{Smart Innovation, Systems and Technologies}, Vol. 69, Chap. 3, pp. 23--37, Springer International Publishing, 2017. Proceedings of the 26th Italian Workshop on Neural Networks (WIRN 2016), Vietri sul Mare, Salerno, Italy, May 18-20, 2016.\\ \texttt{DOI: 10.1007/978-3-319-56904-8\_3}

\end{itemize}

\paragraph{Abstracts}
\begin{itemize}
\item \textbf{Rundo L.}, Militello C., Vitabile S., Russo G., Pisciotta P., Sabini M.G., Marletta F., Ippolito M., D’Arrigo C., Midiri M., Gilardi M.C. (2016) A.197 -- Clinical support in radiation therapy scenarios: MR brain tumor segmentation using an unsupervised fuzzy C-means clustering technique. In: 9th National Congress of AIFM 2016. \textit{Physica Medica, European Journal of Medical Physics}, Vol. 32, Suppl. 1, Feb. 2016, p. 58. \\ \texttt{DOI: 10.1016/j.ejmp.2016.01.201}

\item Pisciotta P., Militello C., \textbf{Rundo L.}, Stefano A., Russo G., Vitabile S., Sabini M.G., D’Arrigo C., Marletta F., D’Urso D., Ippolito M., Midiri M., Gilardi M.C. (2016) A.179 -- Using anatomic and metabolic imaging in stereotactic radio neuro-surgery treatments. In: 9th National Congress of AIFM 2016. \textit{Physica Medica, European Journal of Medical Physics}, Vol. 32, Suppl. 1, Feb. 2016, p. 53.\\ \texttt{DOI: 10.1016/j.ejmp.2016.01.183}
\end{itemize}

\paragraph{Submitted papers and manuscripts in preparation}

\begin{itemize}

\item \textbf{Rundo L.}$^*$, Tangherloni A.$^*$, Nobile M.S., Cazzaniga P., Besozzi D., Russo G., Vitabile S., Gilardi M.C., Mauri G., Militello C. (2018) Genetic Algorithms improve thresholding-based segmentation of bimodal Magnetic Resonance images. Submitted to: \textit{Computer Methods and Programs in Biomedicine} -- Elsevier. ISSN: 0169-2607.

\item \textbf{Rundo L.}$^*$, Han C.$^*$, Nagano Y., Zhang J., Hataya R., Militello C., Tangherloni A., Nobile M.S., Ferretti C., Besozzi D., Gilardi M.C., Vitabile S., Mauri G., Nakayama H., Cazzaniga P. (2018) USE-Net: incorporating Squeeze-and-Excitation blocks into U-Net for prostate zonal segmentation of multi-centric MRI datasets. Submitted to: \textit{Neurocomputing} -- Elsevier. ISSN: 0925-2312.

\item \textbf{Rundo L.}$^*$, Tangherloni A.$^*$, Tyson D.R.$^*$, Betta R., Nobile M.S., Spolaor S., Militello C., Lubbock A.L.R., Mauri G., Quaranta V., Besozzi D., Lopez C.F., Cazzaniga P. (2019) ACDC: Automated Cell Detection and Counting for time-lapse fluorescence microscopy. \textit{Medical \& Biological Engineering \& Computing} -- Springer [Manuscript in preparation]

\item Militello C., \textbf{Rundo L.}, Conti V., Toia P., La Grutta L., Midiri M., Vitabile S. (2019) A semi-automatic approach for epicardial adipose tissue segmentation and quantification on non-enhanced cardiac CT scans. \textit{Symmetry} -- MDPI [Manuscript in preparation].

\end{itemize}

\section{Other publications}
\label{sec:OtherPub}
Along with my main research on biomedical image analysis, to achieve a broader perspective on personalized medicine, I have been involved in activities regarding Systems Biology and Genome Analysis.
In particular, some challenging computational issues were addressed exploiting High-Performance Computing (HPC).
These efforts resulted in some publications.
\paragraph{Journal articles}
\begin{itemize}
    \item Tangherloni A., Spolaor S., \textbf{Rundo L.}, Nobile M.S., Cazzaniga P., Mauri G., Liò P., Merelli I., Besozzi D. (2018) GenHap: a novel computational method based on Genetic Algorithms for haplotype assembly. Accepted by: BMC Bioinformatics – BioMed Central; Special Issue of the 17th Workshop on Network Tools and Applications for Biology (NETTAB 2017). ISSN: 1471-2105. [In press] \\arXiv preprint \texttt{arXiv:1812.07689}.
    
    \item Tangherloni A., Spolaor S., Cazzaniga P., Besozzi D., \textbf{Rundo L.}, Mauri G., Nobile M.S. (2018) Benchmark functions do not capture the complexity of biochemical parameter estimation. Submitted to: \textit{Applied Soft Computing} (Virtual Special Issue on ``Benchmarking of Computational Intelligence Algorithms'') -- Elsevier. ISSN: 1568-4946.
    
    \item Tangherloni A., Nobile M.S., Cazzaniga P., Capitoli G., Spolaor S., \textbf{Rundo L.}, Mauri G., Besozzi D. (2018) FiCoS: a fine- and coarse-grained GPU-powered deterministic simulator for biochemical networks. \textit{Scientific Reports} -- Nature Publishing Group. [Manuscript in preparation]
\end{itemize}

\paragraph{Conference proceedings}
\begin{itemize}

    \item Spolaor S., Tangherloni A., \textbf{Rundo L.}, Nobile M.S., Cazzaniga P. (2019) Estimation of kinetic reaction constants: exploiting reboot strategies to improve PSO's performance. Proceedings of the \textit{14th International Meeting on Computational Intelligence Methods for Bioinformatics and Biostatistics (CIBB 2017)}, Cagliari, Italy, September 7--9, 2017.
    
    \item Tangherloni A., \textbf{Rundo L.}, Spolaor S., Nobile M.S., Merelli I., Besozzi D., Mauri G., Cazzaniga P., Liò P. (2018) High performance computing for haplotyping: models and platforms.  In: Mencagli, G., \textit{et al.} (Editors),  Euro-Par 2018 Workshops, Lecture Notes in Computer Science, Vol. 11339, pp. 650--661, Springer International Publishing, 2019.  \textit{24th International European Conference on Parallel and Distributed Computing (Euro-Par 2018)}, Workshop on Advances in High-Performance Bioinformatics, Systems Biology (Med-HPC 2018), Turin, Italy, August 27--28, 2018.\\ \texttt{DOI:10.1007/978-3-030-10549-5\_51}
    
    \item Nobile M.S., Tangherloni A., \textbf{Rundo L.}, Spolaor S., Besozzi D., Mauri G., Cazzaniga P. (2018) Computational Intelligence for parameter estimation of biochemical systems. Proceedings of the \textit{IEEE World Congress on Computational Intelligence (IEEE WCCI 2018) -- 2018 IEEE Congress on Evolutionary Computation (IEEE CEC 2018)}, pp. 1--8, Rio de Janeiro, Brazil, July 8--13, 2018.\\ \texttt{DOI: 10.1109/CEC.2018.8477873}
    
    \item Tangherloni A., \textbf{Rundo L.}, Spolaor S., Cazzaniga P., Nobile M.S. (2018) GPU-powered multi-swarm parameter estimation of biological systems: a master-slave approach. Proceedings of the \textit{2018 Parallel, Distributed, and Network-Based Processing (PDP 2018)}, Special Session on: Parallel and distributed high-performance computing solutions in Systems Biology, pp. 698--705, Cambridge, UK, March 21--23, 2018.\\ \texttt{DOI: 10.1109/PDP2018.2018.00115}
    
    \item Spolaor S., Tangherloni A., \textbf{Rundo L.}, Nobile M.S., Cazzaniga P. (2017) Reboot strategies in Particle Swarm Optimization and their Impact on parameter estimation of biochemical systems. Proceedings of the \textit{IEEE International Conference on Computational Intelligence in Bioinformatics and Computational Biology (IEEE CIBCB 2017)} -- Special Session on Parallel and Distributed High Performance Computing Solutions for Computational Intelligence Methods, pp. 1--8, Manchester, United Kingdom, August 23--25, 2017.\\ \texttt{DOI: 10.1109/CIBCB.2017.8058550}

    \item Tangherloni A., \textbf{Rundo L.}, Nobile M.S. (2017) Proactive particles in swarm optimization: a settings-free algorithm for real-parameter single objective optimization problems. Proceedings of the \textit{2017 IEEE Congress on Evolutionary Computation (IEEE CEC 2017)} -- Special Session \& Competitions on Real-Parameter Single Objective Optimization, pp. 1940--1947, Donostia – San Sebastián, Spain, June 5--8, 2017.\\ \texttt{DOI: 10.1109/CEC.2017.7969538}
    
\end{itemize}

In addition, the following works concern applications of Field Programmable Gate Arrays (FPGAs) in biometry and digital signal processing:
\begin{itemize}

    \item Conti V., \textbf{Rundo L.}, Billeci G.D., Militello C., Vitabile S. (2018) Energy efficiency evaluation of dynamic partial reconfiguration in Field Programmable Gate Arrays: an experimental case study. \textit{Energies}, 11(4), 739.\\ \texttt{DOI:10.3390/en11040739}
    
    \item Conti V., \textbf{Rundo L.}, Militello C., Mauri G., Vitabile S. (2017) Resource-efficient hardware implementation of a neural-based node for automatic fingerprint classification. \textit{Journal of Wireless Mobile Networks, Ubiquitous Computing, and Dependable Applications}, 8(4), 19--36.\\ \texttt{DOI: 10.22670/JOWUA.2017.12.31.019}
\end{itemize}


\chapter{Biomedical image analysis and computing}
\label{chap2}
\graphicspath{{Chapter2/Figs/}}

\section{Background}
\label{sec:background}

This section first introduces the basic concepts in medical imaging and bioimage informatics.
Afterwards, the practical scenarios under investigation are described.

\subsection{Medical imaging}
\label{sec:medImaging}

Nowadays, medical imaging plays a crucial role in the clinical workflow, thanks to its capability of representing anatomical and physiological features that are otherwise inaccessible for inspection, thus producing accurate imaging biomarkers and clinically useful information \cite{lambin2017,rueckert2016}.
Medical images are considerably different from the pictures usually analyzed in Pattern Recognition and Computer Vision, regarding the appearance of the depicted objects as well as the information conveyed by the pixels \cite{gillies2015}.
Indeed, medical imaging techniques exploit several different principles to measure spatial distributions of physical attributes of the human body, allowing us to better understand complex or rare diseases \cite{toennies2012}.
The effectiveness of such techniques can be reduced by a plethora of phenomena, such as noise and Partial Volume Effect (PVE) \cite{soret2007,toennies2012}, which might affect the measurement processes involved in imaging modalities and data acquisition devices.
In addition, computer-aided medical image acquisition procedures generally include reconstruction methods (producing two, three, or even four dimensional imaging data), which could cause the appearance of artifacts.
Image contrast and details might also be impaired by the procedures used in medical imaging, as well as by the physiological nature of the body part under investigation.

The most common imaging modalities as well as diagnostic workstation software issues are briefly outlined in what follows.

\subsubsection{X-ray radiography and Computed Tomography}
X-ray (plain) radiography is a projectional $2$D image acquisition technique that utilizes ionizing radiation.
The X-rays are directed towards the body and then detected, in order to yield an image representing the amount of rays that reached the detector without being absorbed or scattered by the patient.

In the early 1970s, X-ray Computed Tomography (CT) gave rise to a significant impact in technological development in medicine as well as in industry \cite{hsieh2009}.
CT is a diagnostic imaging modality that combines a finite number of conventional X-ray projections with computing algorithms.
CT scans are acquired by rotating an X-ray source around the patient.
The detectors are located on the opposite side with respect to the X-ray tube.
In more recent CT scanners, the patient's bed is continuously moved to obtain a spiral acquisition mode \cite{kalender1994}.
X-ray beam intensity exponentially decreases according to the mass attenuation coefficient: the beam is more attenuated by tissues with a high atomic number.
On the contrary, whether the beam crosses a low-density tissue, the resulting attenuation is lower.
By so doing, high-density tissues appear as hyper-intense regions (i.e., higher attenuation) and, conversely, low-density tissue are hypo-intense (lower attenuation).
After the tomographic reconstruction (usually by means of the filtered back projection algorithm \cite{geyer2015}), the CT images measure the attenuation of each voxel according to the Hounsfield scale, where the air has a radio-density of $-1000$ Hounsfield Units (HU) and distilled water has a radio-density of $0$ HU.
CT provides high resolution anatomical images of the body, with an excellent contrast between air, adipose tissue, soft-tissue, and bone.
However, the soft-tissue contrast is poor compared to other imaging modalities, such as Magnetic Resonance Imaging (MRI).
Contrast agents may additionally be used to highlight some regions, such as the gastrointestinal tract or the blood vessels.

\subsubsection{Magnetic Resonance Imaging}

MRI relies on the different relaxation times of tissues after to a high and stimulated by means of a radiofrequency impulse \cite{brown2014,mcrobbie2017}.
MRI basically measures a proton density map of the tissues \cite{evans2008}.
This technique is a multiparametric modality that can acquire both anatomical and functional imaging data \cite{brindle2008}.
According to the different relaxation times, different sequences can be obtained:
\begin{itemize}
    \item T1-weighted (T1w) images: the spin-lattice time (i.e., longitudinal) relaxation time is a measurement of the time needed by the protons to return to the initial equilibrium conditions, through the transfer of energy to the surrounding microenvironment (lattice), in order to obtain a T1w SE sequence, using a short relaxation time (TR) associated with a short echo time (TE). In T1w images, the cerebrospinal fluid is dark while the fat is bright;
    \item T2-weighted (T2w) images: the spin-spin (i.e., transverse) relaxation time is a measurement of the time taken by the spin of protons to get out of synchronization. This progressive desynchronization strongly reduces the transverse magnetization. A T2w sequence generally has a long TR associated with a long TE. Liquids or very hydrated tissues appear hyper-intense on T2w images.
\end{itemize}

The relaxation time of a tissue depends on its water content.
MRI provides excellent soft-tissue contrast.
For this reason, MRI is widely applied in the diagnosis and treatment of neurological, cardiovascular, musculoskeletal, liver, and gastrointestinal diseases.
Moreover, image contrast can be further enhanced through the injection of a CE agent.

Advanced MRI techniques can be used to convey information about physiological tissue characteristics \cite{mcrobbie2017}:
\begin{itemize}
    \item Dynamic Contrast Enhanced (DCE) imaging allows for inferring the perfusion and vascularity in the tumor micro-environment by means of the shortening of the T1 time induced by a Gadolinium-based contrast bolus \cite{tofts1999};
    \item Diffusion Weighted Imaging (DWI) estimates the random Brownian motion of water molecules within tissue voxels \cite{hagmann2006}.
    The resistance of water molecule diffusion is quantitatively measured using the Apparent Diffusion Coefficient (ADC) value \cite{leBihan2013}, calculating clinically useful ADC maps pertaining to the imaged body part;
    \item Magnetic Resonance Spectroscopic Imaging (MRSI) quantifies further information about metabolic activity, by providing spectroscopic information regarding metabolites in addition to anatomical MR images \cite{posse2013};
    \item functional MRI (fMRI) is valuable in neuroimaging \cite{logothetis2001} to measure the brain activity by detecting the changes associated with blood flow and oxygenation patterns, thus acquiring functional activation maps \cite{frahm1993}.
\end{itemize}

\subsubsection{Medical ultrasonography}

Medical ultrasonography is well-established in clinical practice since it is relatively portable, cost-effective, safe (in terms of radiation risk), and allows for real-time capabilities in image-guided interventional procedures \cite{hoskins2010}.
In particular, this modality is useful for soft-tissue imaging.
Ultrasounds (US), representing mechanical pressure waves with frequencies higher than 20 kHz, can be both emitted and received by a hand-held probe (i.e., a piezoelectric transducer).
These transducers are composed of multiple piezoelectric crystals that can convert electrical signals to mechanical waves as well as mechanical pressure to electrical signals.
In addition, US scans are tomographic, i.e., offering cross-sections in the imaging of anatomical structures.

US waves propagating through tissues are partially transmitted to deeper body parts, reflected back to the transducer as echoes, scattered, and also converted to heat.
For imaging purposes, the echoes reflected back to the transducer are the most valuable to investigate the tissue properties \cite{chan2011}.
The measurement of the echo resulting from the interaction with a tissue interface is performed by directly employing the acoustic impedance, which is an intrinsic physical property \cite{buddemeyer1975}.

The echo principle defines the most commonly used diagnostic US-based techniques:
\begin{itemize}
	\item Amplitude modulation (A-mode) refers to a one-dimensional signal acquisition technique based on the echoes received are displayed as vertical deflections;
	\item Brightness modulation (B-mode) is a technique wherein the echo amplitude is depicted on a gray-scale image. It is mostly used as a $2$D B-scan by combining multiple ultrasound beams. This technique allows for real-time acquisitions thanks to its high frame rate;
	\item Motion modulation (M-mode) relies on a stationary ultrasound field. The images are continuously recorded over time. M-mode has its main application in echocardiography imaging;
	\item Doppler techniques exploit the Doppler effect as a further source of information. As a matter of fact, the US waves reflected by a moving object reveal a difference, in terms of frequency with respect to the emitted frequencies, which is proportional to the speed of the moving reflector. Color-Doppler techniques integrate Doppler information on the standard B-mode scans by means of pseudo-colored images \cite{mitchell1990}.
\end{itemize}

\subsubsection{Elastography and Tactile Imaging}
Mechanical Imaging (MI) is a relatively recent medical diagnostics modality based on the reconstruction of tissue structures and viscoelastic properties exploiting biomechanical sensors.
This inverse problem relies on the acquisition of data regarding stress patterns on the surface of tissue compressed by a pressure sensor array.
Thus, the imaged tissue structures are presented in terms of their viscoelastic properties, so allowing for the characterization of the tissue, differentiating normal and diseased conditions and detecting tumors and other lesions \cite{sarvazyan1998}.
Unlike the existing medical imaging methods that use sophisticated hardware (e.g., superconductive magnets, expensive CT equipment, and complex ultrasonic phased arrays), MI hardware consists of inexpensive mechanical sensors and a positioning system connected to a general-purpose computer.
Therefore, the structures are visualized by sensing the pattern of mechanical stresses on the surface of an organ.

Elastography maps the elastic properties and stiffness of tissues, specifically enabling the visualization and assessment of mechanical properties of soft-tissues \cite{sarvazyan2011}.
Mechanical properties of tissues---i.e., elastic modulus and viscosity---are highly sensitive to tissue structural changes related to several physiological and pathological processes.

Tactile Imaging (TI) yields a tissue elasticity map, by computationally emulating the manual palpation process by means of a pressure sensor array mounted onto a probe that compresses soft-tissues and detects the changes in the pressure patterns \cite{egorov2010}.

\subsubsection{Nuclear Medicine}

Nuclear Medicine (NM) includes functional imaging techniques that require the delivery of a gamma-emitting radioisotope (i.e., a radionuclide) into the patient, generally by means of an injection \cite{mettler2012}.

Single-Photon Emission Computed Tomography (SPECT) is an NM tomographic imaging technique using gamma rays \cite{patton2008}.
Extending the conventional NM planar imaging based on a gamma camera (i.e., scintigraphy), SPECT can acquire 3D information through cross-sectional slices.

Positron Emission Tomography (PET) is a non-invasive nuclear medical imaging technique for studying functional characteristics of the human body \cite{wernick2004}.
PET scans reveal functional processes that show complementary information with respect to anatomical imaging \cite{evanko2008}, by providing an \textit{in vivo} measurement of the tumor biological
processes \cite{gambhir2001}.

Common clinical applications include brain imaging, neurology, oncology, and recently also in cardiology.
The patient is injected with a dose of radioactive tracer isotope that selectively concentrates in tissues of interest in the body.
Typically, the most active cells in the tissue reveal a higher metabolism, thus uptaking more tracer isotope than cells which are less active.
The isotope undergoes radioactive decay involving the emission of a positron.
As soon as the emitted positron interacts with an electron, this annihilation produces a pair of $511$ keV gamma photons at almost $180$ degrees to each other and is detected by the PET scanner.
Gathering all these pairs allow for the PET measurements from which the distribution of the relevant readiotracer is reconstructed by means of direct or iterative approaches \cite{burger2014}.
Therefore, PET is a type of emissive imaging differently to CT, which represents a transmissive technique.

Among the different PET radiotracers labeled with radioactive or stable isotopes, $^{18}$F-fludeoxyglucose (FDG) is a glucose analog widely used in the evaluation of several neoplastic pathologies as well as in radiotherapy planning.
The FDG uptake is increased in tissues with a high metabolic rate, such as tumor or inflammation regions.
These areas appear as hyper-intense regions on PET images.
Accordingly, PET studies are frequently used in oncological imaging, offering an adequate staging and patient follow-up tool to several cancer types.
In addition, metabolic changes are often faster and more indicative of the effects of the therapy with respect to morphological changes \cite{wahl2009}.
Unfortunately, due to the poor resolution in PET images, an accurate quantification is strongly affected by the PVE \cite{soret2007}.
Several literature methods dealt with the PVE correction \cite{gallivanone2011,stefano2014}.
Modern scanners can estimate attenuation maps using integrated X-ray CT equipment \cite{zaidi2003}, allowing for attenuation correction relying on CT scans.

Hybrid PET/MRI scanners can acquire both PET and MR images simultaneously, so conveying new insights in molucular imaging \cite{judenhofer2008,pichler2008}.
These MRI-compatible PET systems were realized thanks to technological developments in the detectors of PET scanners, i.e., replacing  scintillators coupled to photomultiplier tubes with magnetic field-insensitive avalanche photodiodes \cite{catana2008}.

\subsubsection{Medical photography and video acquisition}
High-resolution digital cameras have been revolutionizing medical applications concerning photography and video acquisition.

Considering m-health applications, modern smartphone cameras can be considered as medical devices for disease diagnosis or medical condition monitoring \cite{agu2013}, with potential applications in telemedicine for chronic diseases \cite{vanNetten2017}.
In this context, smartphones equipped with electrochemical sensors can be suitably used for real-time health monitoring \cite{guo2018}

In the context of video acquisition, endoscopy has become fundamental in many surgical settings.
The endoscopy procedure uses an endoscope to examine the interior of an organ or cavity inside the body.
Unlike other medical imaging techniques, endoscopes are introduced directly into the patient's body.
During surgery, the patient's anatomy can be displayed as a $2$D video (captured by a camera mounted onto the endoscopy probe) and specialized surgical instruments are introduced into the body accordingly.
More recently, Video Capsule Endoscopy (VCE) has revolutionized the diagnostic workflow in the field of small bowel diseases \cite{iakovidis2015}.
Although endoscopy is widely used to guide surgical treatments, it allows the surgeon to investigate only the anatomical surface of the surgical site with a limited FOV.
The $2$D conventional endoscopy does not provide any depth information of the surgical scene.
Therefore, Augmented Reality (AR) techniques have been employed to address these limitations \cite{wang2018AR}. 
Moreover, CT scans can be used to non-invasively reconstruct the corresponding virtual colonoscopy, such as in the case of polyp detection \cite{summers2005}.

\subsubsection{Diagnostic medical imaging software}

Computer applications for diagnostic medical imaging generally provide a wide range of tools to support physicians in their daily diagnosis activities.
Unfortunately, some functionalities are specialized for specific diseases or imaging modalities, while other ones are useless for the images under investigation \cite{cannella2009}.
Nevertheless, the corresponding Graphical User Interface (GUI) widgets are still present on the screen reducing the image visualization area.
As a consequence, the physician may be affected by cognitive overload and visual stress, causing a degradation of performances mainly due to unuseful widgets.
In clinical environments, a GUI must represent a sequence of steps for image investigation following a well-defined workflow.

These issues were addressed in a software framework proposed in \cite{gambino2018}.
Specifically, we designed a Digital Imaging and COmmunications in Medicine (DICOM) standard \cite{bidgood1992, PS3.1} based mechanism of data-driven GUI generation, referring to the examined body part and imaging modality as well as to the medical image analysis task to perform.
In this way, the self-configuring GUI is generated on-the-fly, so that just specific functionalities are active according to the current clinical scenario.
Such a solution provides also a tight integration with the DICOM standard, which considers various aspects of the technology in medicine but does not address GUI specification issues.

The proposed workflow is designed for diagnostic workstations with a local file system on an interchange media inside or outside the hospital ward, not for Picture Archiving and Communication System (PACS) platforms---connected by means of the Radiology Information System (RIS)---which have functionalities mainly oriented to fundamental imaging tools and network-based services.
The framework is designed to manage high-level multiple functionalities, such as image segmentation, image fusion, and advanced measurements.
As a matter of fact, a diagnostic workstation is not only installed in a hospital ward, but it may also belong to an external facility for a second opinion given by highly specialized physicians \cite{zan2010} or multi-disciplinary boards \cite{garcia2018} who have to examine the data.
This is a typical scenario of a second opinion, where an external physician who does not belong to the hospital institution is consulted to perform an alternative reporting/diagnosis.

We implemented a proof-of-concept as a suitable plug-in of the OsiriX imaging software (Pixmeo SARL, Bernex, Geneva, Switzerland) \cite{osirix,rosset2004}, by exploiting the DICOM standard as an enabling technology for an auto-consistent solution in medical diagnostic applications.
Accordingly, the DICOMDIR conceptual data model, defined by the hierarchical structure of the Basic Directory  Information Object Definition (IOD) Information Model \cite{PS3.3}, is exploited and extended to include the GUI information thanks to a new Information Object Module (IOM), which reuses the DICOM information model.

\subsection{Bioimage informatics}
\label{sec:bioimageInform}
Modern biological research increasingly relies on bioimages as a primary source of information in unraveling the complex mechanisms and functioning of live cells \cite{peng2008}.
The quantity and complexity of the data generated by the most recent microscopes preclude visual or manual analyses and require advanced computational methods to fully explore this wealth of information \cite{meijering2016}.
As a matter of fact, such a kind of peculiar challenges have given rise to a separate research area that focuses on the development of novel image processing, data mining, and visualization techniques to extract and integrate the biological knowledge in these data-intensive problems.
Nowadays, this emerging scientific field involving bioinformatics and bioimage computing can be called ``bioimage informatics'' \cite{peng2012}.

\section{Investigated contexts and applications}
This section provides the background about the clinical and biological scenarios considered in this thesis.

\subsection{Medical image analysis}

\subsubsection{MRgFUS treatments for uterine fibroids}
Magnetic Resonance guided Focused Ultrasound Surgery (MRgFUS) represents a non-invasive surgical technique that exploits thermal ablation principles to treat several oncological and neurological pathologies, carefully preserving neighboring healthy tissues (Organs at Risk, OARs) \cite{jolesz2009}.
High Intensity Focused Ultrasound (HIFU) waves are used to rapidly increase the temperature (reaching temperatures above $50^\circ$C) inside the target solid tumors \cite{ismail2013} leading the neoplastic tissue to either apoptosis or coagulative necrosis \cite{wu2002}.
Theoretically, the HIFU technology can be applied in both small and large irregularly-shaped tumors in any anatomical district where the path to the focus is free of bones and air interfaces \cite{cline1992}.
Image guidance is needed to treat these deep-lying tumors, because accurate HIFU beam positioning and reliable acoustic power delivery are mandatory.
Especially, MRI allows for excellent soft-tissue contrast as well as real-time visualization of the heat distribution for thermal dose calculation on target areas, thanks to the Proton Resonance Frequency (PRF) shift thermometry \cite{agnello2015}.

Uterine leiomyomas, more commonly called fibroids or myomas, are benign clonal tumors growing from the smooth-muscle tissue of the uterus.
They are clinically apparent in about $25\%$ of women during middle and late reproductive years \cite{buttram1981,ryan2005}, and with recent \textit{in vivo} acquisition modalities for soft-tissue imaging, especially US and MRI, the actual clinical prevalence may be higher, so representing a relevant public health problem \cite{cramer1990}.
Most fibroids are asymptomatic, but many women have significant symptoms that could negatively impact on their life quality, so requiring effective therapies.
Generally, these symptoms include abnormal uterine bleeding, pelvic pressure or pain, and reproductive dysfunction \cite{stewart2001}.
Depending on the severity of the symptoms, different treatment options for uterine fibroids are available: traditional surgery (i.e., abdominal/laparoscopic hysterectomy or myomectomy), mini-invasive treatments (i.e., hysteroscopic myomectomy, uterine artery embolization), and non-invasive approaches (i.e., MRgFUS, pharmacological therapy) \cite{machtinger2012}.

In MRgFUS uterine fibroid therapy \cite{chapman2007,roberts2008}, MR image analysis is required in all clinical phases: (\textit{i}) imaging-aided diagnosis; (\textit{ii}) treatment planning and real-time temperature monitoring, by detecting the fibroids to be treated---i.e., Region of Treatment (ROT)---as well as the OARs near the uterus---i.e., Region of Interest (ROI)---to be preserved from the HIFU beam; (\textit{iii}) patient's follow-up for measuring the Non-Perfused Volume (NPV), which is the fibroid area actually ablated with HIFU therapy \cite{militelloTCR2014}.

Nowadays, the treatment outcome is evaluated on complete infarction of the ablated fibroids and normal perfusion of the surrounding myometrium \cite{masciocchi2017}.
This estimation is accomplished on post-treatment T1w contrast-enhanced (CE) MRI, wherein the ablated ROT does not uptake the administered Gadolinium-based contrast medium and are imaged as unenhanced hypo-intense with respect to the uterus.
In clinical practice, the NPV is measured by means of manual contouring by an experienced radiologist.
The current operator-dependent methodology could affect the subsequent follow-up phases, due to the lack of results repeatability.
In addition, this fully manual procedure is definitely time-consuming, considerably increasing treatment times.
These critical issues can be addressed only by accurate and efficient computer-assisted MR image segmentation approaches based on Pattern Recognition techniques.

\paragraph{Related work}
Guyon \textit{et al.} \cite{guyon2003} developed a tool called Volume Estimation and Tracking over Time (VETOT) to track the volume of fibroids in patient’s follow-up.
VETOT provides two different types of Level Set Functions (LSFs) for fibroid segmentation: (\textit{i}) fast marching level sets, and (\textit{ii}) geodesic active contours.
In \cite{benZadok2009}, a two-step semi-automatic method was described: an initial automatic segmentation, based on LSFs, is followed by an interactive manual refinement (with user feedback).
This method was applied to uterine fibroid MR scans acquired prior to MRgFUS treatment execution.
The authors of \cite{yao2006} proposed a semi-automatic approach based on LSFs, by combining the fast marching level set and the Laplacian level set methods.
However, this approach was not tailored to HIFU treatments.
In \cite{khotanlou2014}, a two-step method for segmentation of uterine fibroids was developed: (\textit{i}) a coarse semi-automatic segmentation is performed using the Chan-Vese level set method; (\textit{ii}) segmentation refinements based on the prior-shape model are applied, by exploiting training data regarding an ellipse model based on Principal Component Analysis (PCA).
Fallahi \textit{et al.} \cite{fallahi2010} proposed an automated approach for fibroid volume evaluation on MR images.
By exploiting the method described in \cite{fallahi2011} the uterus is firstly segmented using the Fuzzy C-Means (FCM) algorithm from the T1w CE-MRI series.
Some refinement operations are applied, and redundant parts are removed by masking the co-registered T1w MRI series.
The fibroids are then segmented by applying first the Modified Possibilistic Fuzzy C-Means (MPFCM) algorithm on registered T2w MR images and finally by performing some post-processing operations.
Image registration steps are mandatory and are performed using the external software tool Medical Image Processing, Analysis, and Visualization (MIPAV) \cite{mcauliffe2001}.
In \cite{antila2014}, an automatic method for treated fibroid segmentation is proposed.
Both the post-operative CE-MR image stack (interpolated to a $3$D volume) and the treatment plan (i.e., file that stores the sonication cells used during the MRgFUS treatment) are required.
In fact, the coordinates and the size of each treatment cell are exploited as \textit{a priori} information for the fibroid localization.
In the first step, a surface model is created on the sonication cell data and it is deformed with an Active Shape Model (ASM) \cite{cootes1995} by moving the mesh nodes towards the strongest image edge gradients. Finally, the voxels are classified in the treated NPV and the untreated (Perfused Volume, PV) tissues, by an Expectation-Maximization (EM) \cite{moon1996} segmentation algorithm.

In recent years, our research group has already proposed computer-assisted segmentation approaches for uterine fibroids in MRgFUS treatments \cite{militelloCBM2015,militello2013,rundoMBEC2016}.
The clinical feasibility of these methodologies was evaluated and critically discussed in \cite{rundo2019SIST}.

\subsubsection{Neuro-radiosurgery for brain tumors}
Stereotactic radiosurgery allows for an accurate external irradiation (with a single, high dose and steep dose gradient) to minimize the doses delivered to adjacent critical brain structures.
Stereotactic neuro-radiosurgery is a well-established mini-invasive surgical therapy for the treatment of intracranial diseases \cite{luxton1993}, especially brain metastases and highly invasive cancers \cite{nieder2014,soliman2016}.
Early clinical trials showed that tumor control rates are superior to Whole Brain Radiotherapy (WBRT) alone \cite{andrews2004,kondziolka1999}.
As a result, WBRT combined with stereotactic radiosurgery has been widely adopted for patients with a limited number of brain metastases (i.e., $1$-$4$ metastases).
The main difference with conventional radiotherapy is that the treatment is executed in one session alone using a single and high dose by means of a conformal technique.

In particular, Leksell Gamma Knife\textsuperscript{\textregistered} (Elekta, Stockholm, Sweden) is a stereotactic radiosurgical device for the treatment of different brain disorders that are often inaccessible for conventional surgery, such as benign or malignant tumors, arteriovenous malformations and trigeminal neuralgia \cite{leksell1949,luxton1993}.
The gamma rays (generated by Cobalt-60 [$^{60}$Co] radioactive sources) are focused on the target through a metal helmet.
Multiple separate radiation rays converge onto the target tumor, delivering precise high dose radiation to the tumor, but a negligible dose to surrounding health tissues \cite{gerber2008}.
Unlike surgical tumor resection, which eliminates the tumor at the time of the operation, stereotactic radiosurgery causes the tumor to undergo necrosis or growth arrest over time \cite{morantz1995,peterson1999}.
The Leksell stereotactic frame is applied onto the patient (under mild sedation and a local anesthesia) before MRI scanning and treatment, to accurately locate the target areas and to prevent head movement during imaging and treatment.
A personalized treatment plan is implemented using the Leksell GammaPlan\textsuperscript{\textregistered} Treatment Planning System (TPS).
Currently, the Gamma Plan software does not support DICOM-RT objects (generated for the radiation therapy workflow \cite{law2009}) and does not permit to import/export any external ROIs (neither manually nor automatically calculated).
In addition, the Gamma Plan software does not provide automatic image processing methods.

CyberKnife\textsuperscript{\textregistered} (Accuray Inc., Sunnyvale, CA, USA) is an alternative neuro-radiosurgery system that employs different technologies and treats several anatomical districts \cite{adler1997,chang2017}.
As a matter of fact, the main difference between Gamma Knife and CyberKnife is represented by the design concept of the two devices.
Gamma Knife uses [$^{60}$Co] Cobalt-60 radioactive sources, placed inside the machine in well-defined positions, and the patient is moved during the treatment.
In Leksell Gamma Knife\textsuperscript{\textregistered} Model C, the patient is fixed onto the Automatic Positioning System (APS) that is part of the patient bed.
In the most recent Gamma Knife\textsuperscript{\textregistered} Icon\texttrademark, an X-ray system is employed to acquire a cone beam CT.
CyberKnife is a frameless system that utilizes a linear accelerator mounted onto a robotic arm, which allows for dose delivery with several degrees of freedom while the patient stays hold on the treatment table.
The CyberKnife\textsuperscript{\textregistered} uses just a thermoplastic mask, since the position of the skull is checked online by means of two X-ray tubes. 

MRI provides high-quality detailed images, especially concerning the extent of the oncological disease, providing a support in the radiotherapy planning of solid brain tumors \cite{beavis1998}.
MRI is a prominent modality in radiation therapy planning and patient follow-up, complementing the use of CT in target delineation.
MRI is considered to be superior to CT in determining the extent of tumor infiltration, although the literature shows that some histologic evidence of malignancy may extend beyond the margin of enhancement \cite{earnest1988,greene1989,joe1999,johnson1989}.
In particular, MRI is currently the most used imaging modality in radiation therapy for soft-tissue anatomical districts, such as brain and prostate, by providing several advantages over CT.
Furthermore, the excellent soft-tissue contrast of MRI has led to an increasing role of this technique in target volume delineation for therapy applications, such as MRI-guided surgery and radiotherapy treatment, and for patient follow-up (staging and assessing tumor response) \cite{khoo2006}.
Indeed, MRI is considered the best imaging modality, with significant improvement, in terms of contrast, among lesions and normal tissue \cite{metcalfe2013}.
In clinical practice, typical MRI protocols, used for stereotactic radiosurgery treatment planning, include pre- and post-contrast (Gadolinium-enhanced) T1w MR images of the head volume.
Because of the infiltrative nature of glial and metastatic tumors, accurate delineation is difficult and the possibility for a precise determination of the appropriate target volume for treatment is limited.
Moreover, in patients who underwent surgery, CT and MR imaging may not effectively define tumor recurrence.
Furthermore, CT attenuation maps, used in radiotherapy for dose planning, are not available in MRI and geometric distortion could be manifested due to the static magnetic field non-uniformities \cite{evans2008}.
Nevertheless, stereotactic dose calculation algorithms utilize a simplified isocentric technique with a known beam profile and a constant linear attenuation through tissue \cite{moskvin2002}, not performing dose painting for non-uniform radiation dose distribution.
The Gamma Plan system also assumes that the brain is composed entirely of unit density material \cite{wu1992} and dose delivery is based on the unit “shot” (i.e. a dose distribution approximately with spherical shape).
Multiple shots are used in a single Gamma Knife treatment to deliver a conformal dose to an irregular radiosurgical target \cite{wu1999}.

Gross Tumor Volume (GTV) segmentation is a great challenge in stereotactic neuro-radiosurgery therapy.
In treatment planning phase, this delineation task is generally carried out by expert neurosurgeons or radiation oncologists through a fully manual procedure, since the current TPS platforms do not implement automatic or semi-automatic segmentation tools.
This methodology is time-consuming and strongly dependent on the operator: target contouring is affected by high variability in radiation therapy treatment plans \cite{altman2015}.
Only automatic or semi-automatic methods, which assist the clinicians in the segmentation task, can enable the repeatability of the delineated lesion boundaries \cite{joe1999}, representing an operator-independent solution for the assessment of the treatment response during patient's follow-up and cancer staging.
Nowadays, the GTV is usually delineated on anatomical MRI alone, acquired a few hours before treatment, by means of a fully manual process.
Manual segmentation has two main drawbacks: (\textit{i}) time-expensiveness, because dozens of slices have to be manually segmented in a short time; (\textit{ii}) operator-dependence in target volume definition.
The repeatability of the tumor volume delineation can be ensured only by using effective computer-assisted methods, to support the treatment planning phase.
Automated or semi-automated approaches may be of great help, providing higher intra- and inter-operator reliability compared to conventional manual tracing \cite{meier2016}.
Consequently, stereotactic neuro-radiosurgery treatment effectiveness can be optimized using automatic methods to support clinicians in the planning phase and to improve treatment response assessment.

In addition, detecting and analyzing necrotic material within the whole tumor region is an important problem in cancer progression assessment \cite{hamamci2012}.
These pathological necrotic regions in heterogeneous tumors are generally composed of cells adapted to survive within a hypoxic microenvironment.
Hypoxia is implicated in several aspects of tumor development, angiogenesis, and growth in many different human cancers.
The characteristics of highly aggressive brain tumors, particularly the glioblastoma multiforme (GBM) and metastases with necrotic tissue, are supposed to be affected similarly by hypoxia.
This pathological condition is hypothesized to be the mechanism by which an aggressive tumor phenotype is developed through increased invasion, metastatic potential, loss of apoptosis, as well as chemoresistance, resistance to anti-angiogenic therapy, and radiation resistance.
In this context, evaluating hypoxic regions could predict the likelihood of metastases and tumor recurrence \cite{jensen2009}.
Sub-volumes reflecting hypoxic cell populations can be identified by considering functional imaging techniques, such as CE-MRI \cite{cao2006}.
Also in in treatment response assessment, quantifying the extent of necrotic cores within a tumor could be relevant: delayed radiation necrosis, which typically occurs three months or more after the treatment, is indeed the primary risk associated with stereotactic radiosurgery \cite{chin2001,shah2012}.
Necrotic material could develop after aggressive radiation therapy options as well as by the tumor progression itself, such as in high grade gliomas and metastases \cite{lerhun2016}.

The integration of metabolic PET imaging may add another layer of sophistication to the use of radiosurgery in the treatment of gliomas and metastases.
In addition, metabolic changes are often faster and more indicative of the effects caused by the therapy with respect to anatomical imaging \cite{wahl2009}.
Levivier \textit{et al.} \cite{levivier2002} used PET functional imaging in stereotactic conditions for the management of brain tumors, and found that it conveys independent metabolic information that is complementary to the anatomical information derived from CT or MR imaging.

In this regard, [$^{11}$C]-Methionine (MET) seems to have a potential role in conveying additional information, although MRI remains the gold standard for diagnosis, treatment planning, and follow-up in radiation therapy \cite{grosu2005a,miwa2012}.
Methionine is a natural amino acid avidly taken up by brain cancer cells, whereas its uptake by normal cells is low. As a matter of fact, MET uptake is mainly from the activation of the L-mediated and A-mediated amino acid transport at the level of the Blood-Brain Barrier (BBB) \cite{heiss1999}.
In this way, MET-PET discerns malign and benign tissues in brain tumors with great sensitivity and specificity, by localizing selectively in cancer regions of the brain.
Numerous studies have shown that the specificity of the MET-PET for marking tumor delineation and for the differentiation relapse versus radiation necrosis is higher compared with MRI.
In \cite{grosu2005b}, metabolic imaging was used for biological target delineation in $36$ patients that showed a significantly longer median survival compared with the group of patients in which target volume was merely defined by MRI.
In contrast, FDG-PET for metabolic target delineation, which was systematically analyzed by Gross \textit{et al.} \cite{gross1998}, showed limited clinical value when comparing brain tumor volume defined by PET with the corresponding volume defined by PET/MRI fusion images.
Only in few patients, additional information was derived from FDG-PET for radiation treatment planning because of the low contrast between viable tumor and normal brain tissue, although FDG uptake is regionally related to anaplastic areas.
The FDG distribution is thus not limited to malignant tissue because FDG is uptaken according to glucose transport mechanism \cite{delbeke1999}.

The higher diagnostic accuracy of MET-PET is the rationale for using this diagnostic technique in target volume delineation of brain tumors: T1w MRI alone cannot differentiate between treatment-related changes and residual tumor after neurosurgery, chemotherapy or radiation therapy \cite{grosu2010}.
Although the high contrast between tumor and normal tissue on PET images can reduce the inter- and intra-observer variability in tumor localization, the variability in tumor delineation with the qualitative use of PET is still high and often inconsistent with anatomically defined GTV \cite{grosu2010}.
Due to the nature of PET images (low spatial resolution, high noise and weak boundary), the Biological Target Volume (BTV) varies substantially depending on the algorithm used to segment functional lesions: the choice of a standard method for PET volume contouring is a very challenging yet unresolved step \cite{stefano2015}.
Anatomical GTV often does not match with metabolic BTV at all.
For this reason, MET-PET metabolic imaging is valuable and it could be used to provide additional information useful for treatment planning and enhanced tumor characterization \cite{georgiadis2008}: the BTV may be used to modify the GTV in order to treat the actual cancer region more precisely.
To improve the determination of the lesion margin, it is necessary to combine the complementary information of tissues from both anatomical and functional imaging.
Therefore, a reproducible multimodal PET/MRI segmentation method, which contextually segments tumors in each image domain, is mandatory.
This task, named joint segmentation or co-segmentation, is a challenging problem due to: (\textit{i}) unique demands and peculiarities brought by each imaging modality, and (\textit{ii}) a lack of one-to-one region and boundary correspondences of lesions in different imaging modalities \cite{xu2015}.
The implementation of molecular imaging, such as PET, into Gamma Knife treatments allows us to better understand the biological activity about the cancer imaged on MRI.
Especially for the brain tumors gaining deep insights about the cancer region is not always easy investigating MRI alone and it is needed to compare and, often, combine the information acquired by different imaging modalities.
Applying contextually PET and MRI in Gamma Knife clinical scenarios showed that sometimes the BTV may contain metabolically active regions outside the GTV.
In this case, combining both PET and MR imaging modalities is very important to improve the clinical outcome.
Moreover, PET imaging requires computer-assisted segmentation methods to obtain an accurate BTV, minimizing operator-dependence and increasing result repeatability.
As a matter of fact, this data integration approach is prominent for effective brain lesion therapy.

\paragraph{Related work}
\subparagraph{Brain tumor segmentation}
Computer-assisted brain tumor segmentation on structural medical images, i.e., CT and MRI, represents a widely-investigated research field \cite{wu2014}.

Current literature does not present works closely related to the segmentation of target volumes for Gamma Knife treatment support.
Consequently, a broader bibliographical research about the state-of-the-art in lesion segmentation on brain MRI was carried out.
This section focuses particularly on studies dealing with tumor identification on brain MR images, and especially with target segmentation in radiotherapy.

Hamamci \textit{et al.} \cite{hamamci2012,hamamci2010} presented a segmentation tool for radiosurgery clinical applications, especially in CyberKnife\textsuperscript{\textregistered} treatments.
The proposed semi-automatic method for brain tumors segmentation uses a seeded Cellular Automata (CA) \cite{kari2005} approach on input contrast enhanced T1w MR images.
First, the iterative CA framework solves the shortest path problem.
Then, an implicit Level Set surface is evolved on a tumor probability map constructed from CA states to handle the heterogeneous tumor regions.
Because there are heterogeneous intensity distributions in many application fields using MR images, such as cancer imaging, Aslian \textit{et al.} \cite{aslian2013} argued that it is better to use local image statistics instead of a global one.
A localized region-based Active Contour semi-automatic segmentation method (by means of LSFs) was applied on brain MR images of patients to obtain the target volumes.
A similar method was reported in \cite{xie2005}, in which Hybrid Level Sets (HLSs) are exploited.
The segmentation is driven by region and boundary information simultaneously.
In \cite{thapaliya2014}, owing to the non-overlapping property of Zernike polynomials, MR images are analyzed and the different shapes of the tumors present in the brain were retrieved.
Local statistics values obtained from the low- and high-order Zernike moments are thus used to calculate the appropriate threshold value for efficient tumor region extraction.

It was also found that segmentation approaches borrowed from Artificial Intelligence techniques are frequently used.
Mazzara \textit{et al.} \cite{mazzara2004} presented and evaluated two fully automated brain MRI tumor segmentation approaches: supervised $k$-Nearest Neighbors (kNN) and automatic Knowledge-Guided (KG) methods.
Analogously, an approach based on Support Vector Machine (SVM) discriminative classification is proposed in \cite{bauer2011}.
However, for these supervised methods, a training step is always required.
Hall \textit{et al.} \cite{hall1992} extensively compared literal and approximate FCM (i.e., FCM \cite{bezdek1984} and AFCM \cite{cannon1986}, respectively) unsupervised clustering algorithms, and a supervised computational Neural Network, in brain MRI segmentation \cite{bezdek1993}.
The comparison study suggested comparable experimental results (qualitative visual evaluation made by expert radiologists) between supervised and unsupervised learning.
Another system for unsupervised feature extraction for brain MRIs was presented in \cite{zhang2014}.
In particular, T1w brain MR images are partitioned into three main tissue types: white matter (WM), gray matter (GM), and cerebrospinal fluid (CSF).
First, a multidimensional feature vector is constructed based on intensity, low-frequency sub-band of Dual-Tree Complex Wavelet Transform (DT-CWT), and spatial position image information.
Then, a Spatial Constrained Self-Organizing Tree Map (SCSOTM) is used as the actual segmentation system.
SCSOTM applies a dual-thresholding method for automatic growing of the tree map, which uses the information from the high-frequency sub-bands of DT-CWT.
Several Soft Computing techniques have been thus proposed for brain tumor region identification, and a complete automatic segmentation method using optimization and clustering techniques was proposed in \cite{govindaraj2014}.
The resulting hybrid FCM algorithm, combined with Particle Swarm Optimization (PSO) \cite{kennedy1995} during the centroid vector initialization step, segments the tissues and identifies the edema and tumor affected regions in the brain.
On the other hand, Verma \textit{et al.} \cite{verma2014} presented an Improved Fuzzy Entropy Clustering (IFEC) algorithm to segment brain MR images, characterized by noisy data.
A new fuzzy factor, which incorporates both local spatial and gray-level information, was introduced.
The developed IFEC method handles noisy data, preserves image details during clustering, and is setting-free.

\subparagraph{Brain tumor tissue classification}
Several approaches aimed to process multispectral MRI data, by generally employing supervised Machine Learning techniques even though large-scale labeled input datasets are required to learn highly discriminative features.
However, brain tumor segmentation and tissue classification is still a difficult problem, due to the large number of cancer types that differ greatly in size, shape, location, tissue composition, and tissue homogeneity \cite{warfield2000}.
Bauer \textit{et al.} \cite{bauer2011} proposed a supervised fully automatic approach for brain tumor tissue classification using an SVM model for multispectral MR image intensities/textures classification.
Since this SVM assumes voxels to be independent from their neighbors, a subsequent hierarchical regularization based on Conditional Random Fields (CRFs) was integrated to introduce spatial constraints and neighboring context information that encounter the dependence of voxel labels on their neighbors.
The whole process results in a two-step classification: (\textit{i}) healthy and tumor tissue distinction; (\textit{ii}) sub-classification in CSF, WM, GM and necrotic, active, edema regions, respectively.
The approach was tested on a dataset composed of $10$ patients affected by gliomas, including T1w, T1w-CE, T2w, and Fluid Attenuation Inversion Recovery (FLAIR) MR images.
Similarly, in Wu \textit{et al.} \cite{wu2014} multispectral MRI data (T1w, T1w-CE, T2w, and FLAIR images) are first segmented into superpixels to alleviate the sampling issue  and to improve the sample representativeness.
Afterwards, features were extracted from the superpixels using multi-level Gabor wavelet filters.
According to the features, an SVM classifier and an affinity metric model for tumors were trained to overcome the limitations of previous generative models.
Based on the output of the SVM and spatial affinity models, CRFs were applied to segment the tumor using a maximum \textit{a posteriori} (MAP) estimation, given the smoothness prior defined by the affinity model.
The same classification procedure is repeated for edema segmentation.
Finally, a structural denoising process on the labels is performed by relying on symmetrical and continuous characteristics of the tumor in spatial domain.
Li \textit{et al.} \cite{li2016AIM} presented a model that combines sparse representation and Markov Random Fields (MRFs) to solve the spatial and structural variability problem.
The tumor segmentation problem was formulated as a multi-classification task wherein each voxel is labeled estimating the MAP probability, by introducing the sparse representation into a likelihood probability and an MRF into the prior probability.
Since the MAP is an NP-hard problem, this estimation is converted in a minimum energy optimization problem and the Graph cuts algorithm is exploited to find the solution to the MAP estimation.
The Brain Tumor Segmentation (BRATS) 2015 database was used for the evaluation \cite{menze2015}.
Zikic \textit{et al.} \cite{zikic2012} presented a method for automatic segmentation of high-grade gliomas and the following differentiation between active cells, necrotic core, and edema.
The authors integrated a discriminative model, based on decision Random Forests (RFs) using context-aware spatial features, with a generative model of tissue appearance probabilities, obtained by tissue-specific Gaussian Mixture Models (GMMs), as additional input for the RF.
The approach was tested on a set of multi-channel 3D MRI data for $40$ patients suffering from high-grade gliomas.

In the latest year, Deep Neural Networks (DNNs) have been used to learn a hierarchy of increasingly complex features from the processed data, enabling multiple levels of abstraction \cite{lecun2015}.
Havaei \textit{et al.} \cite{havaei2017} adapted Convolutional Neural Network (CNN) architectures specifically to the task of brain tumor segmentation on multispectral MRI data.
The prediction of the label of each pixel is influenced by local features (i.e., details of the region around that pixel) as well as more global contextual features (i.e., considering a larger “context” about the position of the patch is in the brain), simultaneously.
Various architectures are compared by concatenating feature maps from different layers.
In such cascaded CNN architectures, a two-phase training procedure was employed to deal with imbalanced label distribution.
An alternative deep architecture, consisting in an $11$-layers three-dimensional CNN, is proposed in Kamnitsas \textit{et al.} \cite{kamnitas2017}.
A dense training scheme was used, by processing adjacent 3D image patches (i.e., set of connected voxels) into one pass adapting to the inherent class imbalance.
A dual pathway architecture for local and larger contextual information simultaneous processing.
Finally, a CRF is used as a post-processing step for regularization, by removing false positives.
The proposed pipeline was evaluated on three challenging tasks of lesion segmentation in multispectral MRI patient datasets concerning traumatic brain injuries, brain tumors, and ischemic strokes.
Top ranking performance was achieved on the public benchmarks BRATS 2015 \cite{menze2015} and Ischemic Stroke Lesion Segmentation (ISLES) 2015 \cite{maier2017} challenges.

Most of the methods presented above used supervised Machine Learning techniques, thus requiring a training phase.
Nowadays, the availability of medical imaging data for both training and validation phases represents still a serious issue, especially when a computational approach is tailored for a specific clinical application.
Indeed, unsupervised techniques do not need input labeled images for the training and could be clinically feasible for supporting neurosurgeons and radiation oncologists in their decision-making tasks.
Unsupervised Machine Learning techniques have gained ground in biomedical applications, particularly when dealing with an amount of input data that does not allow for a significant training sample.
CA models have been also recently used for MRI brain tumor segmentation.
Hamamci \textit{et al.} \cite{hamamci2012} presented a semi-automatic segmentation tool, called “Tumor-Cut”, for brain tumors segmentation in Cyberknife\textsuperscript{\textregistered} radiosurgery treatments. A method based on CA is used on input contrast enhanced T1w MR images.
First, the iterative CA framework solves the shortest path problem.
Secondly, an implicit Level Set surface is evolved on a tumor probability map, which combines the tumor and background strengths obtained separately by two independent CA algorithm executions.
User interaction consists in drawing the maximum diameter of the tumor on the corresponding axial slice, according to the Response Evaluation Criteria In Solid Tumors (RECIST) rules \cite{eisenhauer2009}.
Subsequently, a bounding volume is built, by enlarging the traced line, and the seeds are computed.
The method was validated on three different T1w CE-MRI datasets: $5$ synthetically generated brain tumors from the University of Utah, $10$ real brain tumors from the Harvard Brain Tumor Repository, $19$ tumors from patients who underwent CyberKnife radiosurgery treatment.
A further processing phase based on CA was proposed to distinguish necrotic and enhancing regions within the previously segmented tumor.
The initial seed selection procedure exploits the threshold computed by the Otsu’s adaptive thresholding method \cite{otsu1975}, by employing the $25\%$ of the most hypo-intense necrotic and the $25\%$ of the most hyper-intense enhanced volumes as necrosis (i.e., foreground) and enhancement (i.e., background) seeds, respectively.
This approach for enhancement and necrotic region differentiation was evaluated on $6$ tumors with necrosis taken from the clinical dataset of patients who underwent neuro-radiosurgery.
Sompong and Wongthanavasu \cite{sompong2017} presented the Improved Tumor-Cut (ITC) to cope with the robustness of seed growing in the standard Tumor-Cut algorithm.
The multimodal MRI brain tumor dataset BRATS 2013, including T1w, T2w, T1w-CE, and FLAIR images, was processed.
The authors focused on the segmentation of edema tissues, which generally exhibit ambiguous tumor boundaries and are imaged with a brighter hyper-intense signal than normal brain on T2w MRI.
Firstly, MR images are enhanced using a novel Gray Level Co-occurrence Matrix based CA (GLCM-based CA) \cite{haralick1979,haralick1973}, by transforming the original MRI in a new GLCM feature space.
Secondly, the proposed CA algorithm maps these features to the target image, by employing a patch-weighted feature in the similarity function and consequently defining a patch-weighted distance.
Cluster analysis methods have been widely used also due to the computational efficiency since it does not require training data, so increasing operator-independency and responding to potential inconsistencies of supervised classification techniques \cite{fletcher2001}.
A superpixel-based graph for GBM from multimodal MR images was used in \cite{su2013}.
First, a local $K$-means clustering with weighted distances is employed to segment the multispectral MR images, including T1w, T1w-CE, T2w and FLAIR, into superpixels.
Then, the spectral clustering algorithm is utilized to extract the enhanced tumor, necrosis and non-enhancing T2w hyper-intense regions by considering the superpixel map as a graph.
Another critical issue is represented by the border between pathological and normal tissues that is often not well imaged in heterogeneous tumors.
Moreover, PVEs (i.e., one pixel may be actually composed of multiple tissue types) and the noise due to the MRI acquisition process can affect segmentation accuracy.
Fuzzy Logic techniques can cope with these difficulties, by modeling multiple membership degrees for each voxel and fusing multispectral MRI information \cite{dou2007}.
Therefore, introducing Fuzzy Logic in clustering-based classification has represented an active research area in recent years.
Prakash and Kumari \cite{meena2016} proposed an automated segmentation method of MR brain images into three tissue types: WM, GM, and CSF based on FCM clustering, by integrating spatial information and contrast enhancement.
Similarly, Agnello \textit{et al.} \cite{agnello2016} proposed an unsupervised tissue classification method of brain for Voxel‐Based Morphometry (VBM) analysis on T1w MRI data alone.
The FCM algorithm for defining the training set, composed of tuple wherein each intensity value of the MRI data corresponds to a specific cluster, in order to train a Fully Connected Cascade Neural Network (FCCNN) to learn the brain tissue classification in WM, GM, and CSF.
Govindaraj and Murugan \cite{govindaraj2014} presented a hybrid FCM algorithm, combined with the PSO technique in the centroid vector initialization step, to segment brain tumor tissues and identify edema on multispectral MRI.
Hemanth \textit{et al.} \cite{hemanth2012} introduced an additional data reduction pre-processing step in the conventional FCM algorithm to minimize the computational complexity and the convergence rate.
The resulting Fast and Accurate FCM (FAFCM) algorithm consists of two stages: (\textit{i}) binary representation and pre-clustering for intensity value reduction; (\textit{ii}) FCM clustering with the reduced MRI sequence.
Kaya \textit{et al.} \cite{kaya2017} focused on reduction techniques, based on different PCA versions, as a data pre-processing step to allow for an appropriate low-dimensional MRI data representation for the subsequent clustering process using the \textit{K}-means and FCM algorithms. 

\subparagraph{Multimodal PET/MRI segmentation}
A huge number of monomodal segmentation methods on single image modalities (i.e. CT, MRI, or PET) are present in the literature.
Only few studies address the segmentation of multimodal imaging data.
In particular, this section reviews the approaches that combine the complementary functional information conveyed by PET with the corresponding anatomical image scans (CT or MRI).
In fact, in clinical use, it is highly desirable to acquire both functional and structural quantifiable information by means of a single scan so the disease can be both identified and localized, potentially resulting in an earlier diagnosis and more effective treatment plan.
In parallel to the developments of multimodal scanners (PET/CT and PET/MRI), there have been recent literature works trying to bring the usefulness of integrating anatomical and functional information into a common place for separating tumor tissues from normal structures.
A concise but complete review of the state-of-the-art regarding multimodal co-segmentation approaches is reported in the following, wherein each research paper is briefly explained.

Several studies deal with PET/CT tumor identification and characterization in radiation therapy scenarios.
In \cite{yu2009}, co-registered FDG-PET/CT images were used for the textural characterization of head and neck cancer (HNC) in radiotherapy treatment planning.
After a manual segmentation on co-registered PET/CT images (performed by an experienced radiation oncologist), useful textural features were selected for distinguishing tumor from normal tissue in HNC subjects.
Both $k$-NNs and decision tree (DT)-based $k$-NN classifiers were employed to discriminate images of cancerous and healthy tissues.
Han \textit{et al.} \cite{han2011} presented an MRF-based co-segmentation of the PET/CT image pair with a regularized term that penalizes the segmentation difference between PET and CT.
This graph-based method utilizes the strength of PET and CT modalities for target delineation in a group of $16$ patients with HNC.
Background and foreground seed voxels must be always manually identified by the user. A similar approach is reported in \cite{song2013}, where the segmentation is seen as a minimization problem of an MRF model, which encodes the information from both modalities.
This optimization is solved using a Graph cuts based method \cite{boykov2001b,boykov2001a}, by constructing two sub-graphs for PET and CT segmentation, respectively, and adding inter-subgraph arcs.
The overall approach is semi-automatic, because initial seed points are required for the graph-based co-segmentation: one center point and two radii are given by the user.
The algorithm was validated in robust delineation of lung tumors on $23$ PET/CT datasets and two HNC subjects.
A further MRF-based systematic solution for the automated co-segmentation of brain PET/CT images into GM, WM, and CSF regions is exposed in \cite{xia2008}.
A PET/CT image pair and its segmentation result are modeled as an MRF triplet, and segmentation is eventually achieved by solving a maximum \textit{a posteriori} (MAP) problem using the EM \cite{dempster1977} algorithm with simulated annealing.
The overall MRF-MAP model was tested both on simulated and real patient PET/CT data. The same authors presented in \cite{xia2012} a more advanced brain PET/CT segmentation model.
This dual-modality image segmentation approach converts PET/CT image segmentation into an optimization process controlled simultaneously by PET and CT voxel values and spatial constraints.
A modality discriminatory power (MDP) coefficient is applied as a weighing scheme to adaptively combine functional and anatomical information to separate voxels from different tissue types.
The authors of \cite{potesil2007} proposed a method for automated delineation of tumor boundaries in whole-body PET/CT by jointly using information from both PET and diagnostic CT images.
After an initial robust hot-spot detection and segmentation performed in PET, a model for tumor appearance and shape in corresponding CT structures is learned by the weighted non-parametric density estimate.
This voxel-based CT classification is then probabilistically integrated with the PET classification using the joint likelihood ratio test technique to derive the final segmentation.
The algorithm was tested on patient studies with lung and liver tumors identifiable in both the PET and CT images acquired by the same scanner.
In this context, a recent study \cite{vallieres2015} aims at developing a radiomics model from joint FDG-PET and MRI texture analysis for the early evaluation of lung metastasis risk in soft-tissue sarcomas (STSs).
The creation of new composite textures from the combination of FDG-PET and MR imaging information, to better identify aggressive tumors, was investigated.
The results showed that FDG-PET and MRI texture features could act as strong prognostic factors of STSs.
Yezzi \textit{et al.} \cite{yezzi2001} introduced a geometric variational framework that uses active contours to simultaneously segment and register features from multiple images.
The key aspect of this approach is that multiple images may be segmented by evolving a single contour as well as the mappings of that contour into each image during feature-based realignment steps.
The results of three experiments on MRI/CT images of the head and the spine were reported.
In \cite{wojak2010}, a variational segmentation method, based on the minimization of the total variation semi-norm and a convex formulation, was used for segmenting lung tumors and lymph node disease on thoracic PET/CT image pairs, in the context of radiotherapy planning.
Also the authors of \cite{elNaqa2007} developed a method based on multi-valued level set deformable models for simultaneous $2$D or $3$D segmentation of multimodality images consisting of combinations of co-registered PET, CT, or MRI datasets.
In particular, only three patients are considered: a non-small cell lung cancer case with PET/CT, a cervix cancer case with PET/CT, and a prostate patient case with CT/MRI.
In addition, CT, PET, and MRI phantom data were used for quantitative validation of the proposed multimodality segmentation approach.
However, here we focus on PET/MRI joint segmentation approaches reported in the literature.
An automatic algorithm for the co-segmentation of HNC based on PET/MRI data was proposed in \cite{leibfarth2015}, in order to standardize tumor volume delineation.
For both imaging modalities tumor probability maps were derived, assigning each voxel a probability of being cancerous according to its signal intensity.
A combination of these maps was subsequently segmented using a threshold level set algorithm. The algorithm processes both the anatomical T2w MRI and FDG-PET data concerning $10$ HNC patient datasets acquired by a combined PET/MRI system.
The group led by Bagci and Mollura developed some co-segmentation approaches in multimodal medical imaging, which mostly used the Random Walker (RW) algorithm \cite{grady2006}.
In \cite{bagci2013a} a computer-assisted volume quantification method for PET/MRI dual modality images using automated PET-guided RW MR image co-segmentation was presented.
A small-animal breast cancer model was built by using $30$ female nude mice.
A more comprehensive and general approach was proposed in \cite{bagci2013MedIA}.
The RW algorithm is extended for jointly delineating multiple objects from multimodal images by unifying graph representation of each image modality in a single product lattice.
The overall method results in an automatic and unified framework, providing an automated object detection \textit{via} an Interesting Uptake Region (IUR) algorithm to avoid users having to provide foreground and background seeds.
Afterwards, prior to the initiation of the segmentation process, these identified seeds are propagated to the corresponding anatomical images.
Although no significant anatomical and functional changes between the scans have to be assumed, the study used PET, PET/CT, MRI/PET, and fused MRI/PET/CT scans ($77$ studies in total) from $56$ patients who had various lesions in different body regions.
Finally, a $3$D computer-aided co-segmentation tool for lesion detection and quantification from hybrid PET/MRI and PET/CT images was described in \cite{xu2015}.
This method used a modality-specific visibility weighting scheme built upon a Fuzzy Connectedness (FC) image segmentation algorithm.

\subsubsection{Prostate MRI analysis}

According to the American Cancer Society, in 2018 the Prostate Cancer (PCa) is the most common malignant tumor for American males~\cite{siegel2018}.
The only well-established risk factors for prostate cancer are increasing age, a family history of the disease, and genetic conditions.
Prostate image analysis is still a critical and challenging issue in diagnosis, therapy, and staging of PCa.

Currently, high-resolution multiparametric MRI has been shown to have higher accuracy than Transrectal Ultrasound (TRUS) when used to ascertain the presence of prostate cancer \cite{liu2009}.
In \cite{roethke2012}, MRI-guided biopsy showed a high detection rate of clinically significant PCa in patients with persisting cancer suspicion, also after a previous negative systematic TRUS-guided biopsy, which represents the gold-standard for diagnosis of prostate cancer \cite{heidenreich2008}.
In addition, MRI scanning before a biopsy can also serve as a triage test in men with raised serum Prostate-Specific Antigen (PSA) levels, in order to select those for biopsy with significant cancer that requires treatment \cite{ahmed2009}.
For instance, PCa treatment by radiotherapy requires an accurate localization of the prostate. In addition, neighboring tissues and organs at risk, i.e., rectum and bladder, must be carefully preserved from radiation, whereas the tumor should receive a prescribed dose \cite{klein2008}.
CT has been traditionally used for radiotherapy treatment planning, but MRI is rapidly gaining relevance because of its superior soft-tissue contrast, especially in conformal and intensity-modulated radiation therapy \cite{villeirs2007}.
Manual detection and segmentation of both prostate gland and prostate carcinoma on multispectral MRI is a time-consuming task, which has to be performed by experienced radiologists \cite{hambrock2013}.
In fact, in addition to conventional structural T1w and T2w MRI protocols, complementary and powerful functional information about the tumor can be extracted \cite{chilali2014,ghose2012,lemaitre2015} from: DCE \cite{fabijanska2016}, DWI \cite{haider2007}, and MRSI \cite{futterer2006,matulewicz2014}.
A standardized interpretation of multiparametric MRI is very difficult and significant inter-observer variability has been reported in the literature \cite{rosenkrantz2013}.
The increasing number and complexity of multispectral MRI sequences could overwhelm analytic capabilities of radiologists in their decision-making processes.
Therefore, automated and computer-aided segmentation methods are needed to ensure results repeatability.
Accurate prostate segmentation on different imaging modalities is a mandatory step in clinical activities. For instance, prostate boundaries are used in several treatments of prostate diseases: radiation therapy, brachytherapy, HIFU ablation, cryotherapy, and transurethral microwave therapy \cite{ghose2012}.
Prostate volume, which can be directly calculated from the prostate ROI segmentation, aids in diagnosis of benign prostate hyperplasia and prostate bulging.
Precise prostate volume estimation is useful for calculating PSA density and for evaluating post-treatment response.
However, these tasks require several degrees of manual intervention, and may not always yield accurate estimates \cite{toth2011}.

Given a clinical context, several imaging modalities can be used for PCa diagnosis, such as TRUS, CT, and MRI.
For an in-depth investigation, structural T1w and T2w MRI sequences can be combined with the functional information from  DCE-MRI, DWI, and MRSI~\cite{lemaitre2015}.
Despite the recent technological advances in MRI scanners and coils, prostate images are prone to artifacts related to magnetic susceptibility.
Although the shift from $1.5$ T to $3$ T magnetic field strength systems theoretically results in a doubled Signal-to-Noise-Ratio (SNR), it also involves different T1 (spin-lattice) and T2 (spin-spin) relaxation times and greater magnetic field inhomogeneity in the organs and tissues of interest \cite{caivano2012,rouviere2006}.
On the other hand, $3$ T MRI scanners allow for using pelvic coils instead of endorectal coils, obtaining good quality images and avoiding invasiveness as well as prostate gland compression and deformation \cite{heijmink2007,kim2008}.
So, challenges for automatic segmentation of the prostate in MR images include the presence of imaging artifacts due to air in the rectum (i.e., susceptibility artifacts) and large intensity inhomogeneities of the magnetic field (i.e., streaking or shadowing artifacts), the large anatomical variability between subjects, the differences in rectum and bladder filling, and the lack of a normalized “HU” for MRI, like in CT scans \cite{klein2008}.
Therefore, MRI plays a decisive role in PCa diagnosis and disease monitoring (even in an advanced status~\cite{padhani2017}), revealing the internal prostatic anatomy, prostatic margins, and PCa extent~\cite{villeirs2007}.
According to the zonal compartment system proposed by McNeal, the prostate Whole Gland (WG) can be partitioned into the Central Gland (CG) and Peripheral Zone (PZ) \cite{selman2011}.
In prostate imaging, T2w MRI serves as the principal sequence~\cite{scheenen2015}, thanks to its high resolution that enables to differentiate the hyper-intense PZ and hypo-intense CG in young male subjects~\cite{hoeks2011}.

Besides manual detection/delineation of the WG and PCa on MR images, distinguishing between the CG and PZ is clinically essential, since the frequency and severity of tumors differ in these regions~\cite{choi2007,niaf2012}.
As a matter of fact, the PZ harbors $70$-$80\%$ of PCa and represents a target for prostate biopsy~\cite{haffner2009}. Furthermore, the PZ volume ratio (i.e., the PZ volume divided by the WG volume) can be considered for PCa diagnostic refinement~\cite{chang2017}, while the CG volume ratio may help monitoring prostate hyperplasia~\cite{kirby2002}. 
Therefore, according to the Prostate Imaging-Reporting and Data System version 2 (PI-RADS\textsuperscript{TM} v2)~\cite{weinreb2016}, radiologists must perform a zonal partitioning before assessing the suspicion of PCa on multi-parametric MRI.

\paragraph{Related work}
\subparagraph{Whole prostate gland segmentation}
From a computer science perspective, especially in automatic MR image analysis for PCa segmentation and characterization, the computed prostate ROI is very useful for the subsequent more advanced processing steps \cite{lemaitre2015}.
Klein \textit{et al.} \cite{klein2008} presented an automatic method, based on non-rigid registration of a set of pre-labeled atlas images, for delineating the prostate in $3$D MR scans.
Each atlas image is non-rigidly registered, employing Localized Mutual Information (LMI) as similarity measure \cite{studholme2006}, with the target patient image.
The registration is performed in two steps: (\textit{i}) coarse alignment of the two images by a rigid registration; (\textit{ii}) fine non-rigid registration, using a coordinate transformation that is parameterized by cubic B-splines \cite{rueckert1999}.
The registration stage yields a set of transformations that can be applied to the atlas label images, resulting in a set of deformed label images that must be combined into a single segmentation of the patient image, considering majority voting (VOTE) and Simultaneous Truth And Performance Level Estimation (STAPLE) \cite{warfield2004}.
The method was evaluated on $50$ clinical scans (probably T2w images, but the acquisition protocol is not explicitly reported in the paper), which had been manually segmented by three experts.
The Dice Similarity Coefficient (\emph{DSC}) \cite{dice1945}, to quantify the overlap between the automatic and manual segmented regions, and the shortest Euclidean distance, between the manual and automatic segmentation boundaries, were computed.
The differences among the used four fusion methods are small and mostly not statistically significant.
For most test images, the accuracy of the automatic segmentation method was, on a large part of the prostate surface, close to the level of inter-observer variability.
With the best parameter setting, a median \emph{DSC} of around $0.85$ was achieved.
The segmentation quality is especially good at the prostate-rectum interface, whereas most serious errors occurred around the tips of the seminal vesicles and at the anterior side of the prostate.
Vincent \textit{et al.} \cite{vincent2012} proposed a fully automatic model-based system for segmenting the prostate in MR images.
The segmentation method is based on Active Appearance Models (AAMs) \cite{heimann2009} built from $50$ manually segmented examples provided by the Medical Image Computing and Computer-Assisted Intervention (MICCAI) 2012 Prostate MR Image Segmentation 2012 (PROMISE12) team \cite{litjens2014}.
High-quality correspondences for the model are generated using a variant of the Minimum Description Length approach to Groupwise Image Registration (MDL-GIR) \cite{cootes2005}, which finds the set of deformations for registering all the images together as efficiently as possible.
A multi-start optimization scheme is used to robustly match the model to new images.
To test the performance of the algorithm, \emph{DSC}, mean absolute distance and the $95\%$-th Hausdorff distance (\emph{HD}) \cite{cardenes2009} were calculated.
The model was evaluated using a Leave-One-Out Cross Validation (LOOCV) on the training data obtaining a good degree of accuracy and successfully segmented all the test data.
The system was used also to segment $30$ test cases (without reference segmentations), considering the results very similar to the LOOCV experiment by a visual assessment. 
In \cite{gao2010}, a unified shape-based framework to extract the prostate from MR images was proposed.
First, the authors address the registration problem by representing the shapes of a training set as point clouds.
In this way, they are able to exploit the more global aspects of registration \textit{via} a particle filtering based scheme \cite{sandhu2008}.
In addition, once the shapes have been registered, a cost functional is designed to incorporate both the local image statistics and the learned shape prior (used as constrain to perform the segmentation task).
Satisfying experimental results were obtained on several challenging clinical datasets, considering \emph{DSC} and the $95\%$-th \emph{HD}, also compared to \cite{tsai2003}, which employs the Chan-Vese Level Set model \cite{chan2001} assuming a bimodal image.
Martin \textit{et al.} \cite{martin2008} presented the preliminary results of a semi-automatic approach for prostate segmentation of MR images that aims to be integrated into a navigation system for prostate brachytherapy.
The method is based on the registration of an anatomical atlas computed from a population of $18$ MRI studies along with each input patient image.
A hybrid registration framework, which combines an intensity-based registration \cite{hellier2003} and a Robust Point-Matching (RPM) algorithm introducing a geometric constraint (i.e., the distance between the two point sets) \cite{rangarajan1997}, is used for both atlas building and atlas registration.
This approach incorporates statistical shape information to improve and regularize the quality of the resulting segmentations, using Active Shape Models (ASMs) \cite{cootes1994}.
The method was validated on the same dataset used for atlas construction, using the LOOCV.
Better segmentation accuracy, in terms of volume-based and distance-based metrics, was obtained in the apex zone and in the central zone with respect to the base of the prostate.
The same authors proposed a fully automatic algorithm for the segmentation of the prostate in $3$D MR images \cite{martin2010}.
The approach requires the use of a probabilistic anatomical atlas that is built by computing transformation fields (mapping a set of manually segmented images to a common reference), which are then applied to the manually segmented structures of the training set in order to get a probabilistic map on the atlas.
The segmentation procedure is composed of two phases: (\textit{i}) the processed image is registered to the probabilistic atlas and then a probabilistic segmentation (allowing the introduction of a spatial constraint) is performed by aligning the probabilistic map of the atlas against the patient’s anatomy; (\textit{ii}) a deformable surface evolves towards the prostate boundaries by merging information coming from the probabilistic segmentation, an image feature model, and a statistical shape model.
During the evolution of the surface, the probabilistic segmentation allows for the introduction of a spatial constraint that prevents the deformable surface from leaking in an unlikely configuration.
This method was evaluated on $36$ MRI exams.
The results showed that introducing a spatial constraint increases the segmentation robustness of the deformable model compared to another one that is only driven by an image appearance model.
Toth \textit{et al.} \cite{toth2011} used a Multifeature Active Shape (MFA) model \cite{cootes1995}, which extends ASMs by training the multiple features for each voxel using a GMM based on the log-likelihood maximization scheme, for estimating prostate volume from T2w MRI.
Using a set of training images, the MFA learns the most discriminating statistical texture descriptors of the prostate boundary \textit{via} a forward feature selection algorithm.
After the identification of the optimal image features, the MFA is deformed to accurately fit the prostate border.
The MFA-based volume estimation scheme was evaluated on a total $45$ \textit{in vivo} T2w MRI studies, corresponding to both $1.5$ T and $3.0$ T field strengths.
A correlation with the ground truth volume estimations showed that the MFA obtained higher Pearson's correlation coefficient values with respect to the current clinical volume estimation schemes, such as ellipsoid, Myschetzky, and prolate spheroid methodologies.
In \cite{toth2011}, the same research team proposed a scheme to automatically initialize an ASM for prostate segmentation on endorectal \textit{in vivo} multiprotocol MRI, exploiting MRSI data and identifying the MRI spectra that lie within the prostate.
As a matter of fact, MRSI is able to measure relative metabolic concentrations and the metavoxels near the prostate have distinct spectral signals from metavoxels outside the prostate.
A replicated clustering scheme is employed to distinguish prostatic from extra-prostatic MR spectra in the midgland.
The detected spatial locations of the prostate are then used to initialize a multi-feature ASM, which employs also statistical texture features in addition to intensity information.
This scheme was quantitatively compared with another ASM initialization method for TRUS imaging by Cosìo \cite{cosio2008}, showing superior average segmentation performance on a total of $388$ $2$D MRI sections obtained from $32$ $3$D endorectal \textit{in vivo} patient studies.
Other segmentation approaches are also able to segment the zonal anatomy of the prostate \cite{choi2007}, differentiating the CG from the PZ \cite{hricak1987}.
The method in \cite{makni2011} is based on a modified version of the Evidential C-Means (ECM) clustering algorithm \cite{masson2008}, introducing spatial neighborhood information.
\textit{A priori} knowledge of the prostate zonal morphology was modeled as a geometric criterion and used as an additional data source to enhance the differentiation of the two zones.
$31$ clinical MRI series were used to validate the accuracy of the method, taking into account inter-observer variability.
The mean \emph{DSC} was $89\%$ for the CG and $80\%$ for the PZ, calculated against an expert radiologist segmentation.

\subparagraph{Prostate zonal segmentation}
Due to the crucial role of MR image analysis~\cite{lemaitre2015} in PCa diagnosis and staging, researchers have paid specific attention to automatic WG detection/segmentation.
Classic methods mainly leveraged atlases~\cite{klein2008,martin2008} or statistical shape priors~\cite{martin2010}:
atlas-based approaches realized accurate segmentation when new prostate instances resemble the atlas, relying on a non-rigid registration algorithm~\cite{martin2010,toth2013}.
Unsupervised clustering techniques allowed for segmentation without manual labeling of large-scale MRI datasets~\cite{rundo2018SIST,rundo2017Inf}.
In the latest years, Deep Learning techniques \cite{litjens2017} have achieved accurate prostate segmentation results by using deep feature learning combined with shape models~\cite{guo2017} or location-prior maps~\cite{sun2017}.
Moreover, CNNs were used with patch-based ensemble learning~\cite{jia2018} or dense prediction schemes~\cite{milletari2016}.

Differently to WG segmentation, less attention has been paid to CG and PZ segmentation despite its clinical importance in PCa diagnosis~\cite{niaf2012}.
In this context, classic Computer Vision techniques have been mainly exploited on T2w MRI.
For instance, early studies combined classifiers with statistical shape models~\cite{allen2006} or deformable models~\cite{yin2012}; Toth \textit{et al.}~\cite{toth2013} employed active appearance models with multiple level sets for simultaneous zonal segmentation; Qiu~\textit{et al.}~\cite{qiu2014} used a continuous max-flow model---the dual formulation of convex relaxed optimization with region consistency constraints~\cite{yuan2010}; in contrast, Makni~\textit{et al.}~\cite{makni2011} fused and processed 3D T2w, DWI, and T1w CE-MR images by means of an ECM algorithm~\cite{masson2008}.
Litjens \textit{et al.} \cite{litjens2012} proposed a zonal prostate segmentation approach based on Pattern Recognition techniques.
First of all, they use the multi-atlas segmentation approach in \cite{klein2008} for prostate segmentation, by extending it for using multi-parametric data.
T2w MR images and the quantitative ADC maps were registered simultaneously.
As a matter of fact, the ADC map conveys additional information on the zonal distribution within the prostate in order to differentiate between the CG and PZ.
Then, for the voxel classification, the authors determined a set of features that represent the difference between the two zones, by considering three categories: anatomy (positional), intensity and texture.
This voxel classification approach was validated on $48$ multiparametric MRI studies with manual segmentations of the whole prostate.
\emph{DSC} and Jaccard index (\emph{JI}) \cite{jaccard1901} showed good results in zonal segmentation of the prostate and outperformed the multiparametric multi-atlas based method in \cite{makni2011} in segmenting both CG and PZ.

As the first CNN-based method, Clark \textit{et al.}~\cite{clark2017} detected DWI MR images with prostate relying on Visual Geometry Group (VGG) net~\cite{simonyan2015}, and then sequentially segmented WG and CG using U-Net~\cite{ronneberger2015}.
However, no literature method so far coped with the generalization ability among multi-institutional MRI datasets, making their clinical applicability difficult~\cite{albadawy2018}.
In a preliminary study~\cite{rundoWIRN2018}, we compared existing CNN-based architectures---namely, SegNet~\cite{badrinarayanan2017}, U-Net~\cite{ronneberger2015}, and pix2pix~\cite{isola2016}---on two multi-institutional MRI datasets.
According to our results, U-Net generally achieves the most accurate performance.

\subsection{Biological image analysis}

\subsubsection{Clonogenic assays}
Clonogenic assay (or colony formation assay) is an \textit{in vitro} cell survival technique based on the ability of a single cell to grow into a colony.
It is widely used to study the number and the size of cell colonies after irradiation or cytotoxic agent administration, serving as a measure for the anti-proliferative effect of these treatments.
After plating, incubation and treatment of cells, the standard procedure consists in colony counting under a stereomicroscope \cite{puck1956}.
Traditionally, clonogenic assay evaluation has been performed by counting the colonies composed of at least $50$ densely-packed cells \cite{puck1956}: to evaluate the effect of the treatment on cell survival, the Plating Efficiency (PE), which is the fraction of colonies obtained from untreated cells, and the Surviving Fraction (SF) of cells after any treatment, are measured \cite{franken2006}.
It is well known that cells after any type of treatment have a reduction in growth rate, and the colonies are clearly smaller than the ones of the untreated control.
The conventional counting process can underestimate the actual effect of the therapy because, as explained in \cite{franken2006}, the treatment could not decrease the colony number while reducing the colony size. This counting procedure is definitely time-consuming and tedious.
Especially, accurate counting with high numbers of Colony Forming Units (CFUs) on a plate is difficult and error-prone, since it has to be carefully carried out by human operators.
Manual counting is strongly subjective and it was shown that the results may also be affected by intra-operator variability \cite{lumley1996}.
From the variation observed in the measurements, the automated counts are generally more consistent with respect to the manual ones, representing the only approach able to ensure result repeatability and operator independence.
The concept of computer-aided colony counting has been already investigated in literature.
As a matter of fact, several methods to segment images of cell colonies were proposed \cite{barber2001,bernard2001,chiang2015,dahle2004,geissmann2013,guzman2014}, but none of them has been widely used in practice.
Their apparent failure to be adopted could be explained by: (\textit{i}) the inability to split adjacent merged colonies, representing a common problem with mammalian cell cultures, and (\textit{ii}) long processing times or, more generally, the need for interactive inputs since fully automatic approaches have not been proposed yet.
Moreover, no related works address all the problems simultaneously.
Some approaches analyze only single-well images, while the ones that use multi-well plates do not perform a fully automatic well detection, by strongly relying on plate measurements and characteristics during well extraction phases \cite{dahle2004,guzman2014}.
In addition, some methods use different image processing techniques depending on the type of the analyzed cells, leaving the choice of the most suitable method case-by-case to the user \cite{barber2001,dahle2004}.

Considering the above mentioned open issues and that the well Area Covered by Colonies (ACC) is directly correlated with the number and size of colonies, we decided to tackle these challenges by measuring the percentage of ACC.
In this way, we overcome the necessity to split the possible merged colonies, by taking into account just the colony size, which has not currently been considered, since it is correlated with the effect of the treatment delivered to the cells \cite{harada2012,maycotte2014}.

\paragraph{Related work}
In the literature, numerous computer-aided approaches were proposed for colony counting in clonogenic assays, even though they strongly depend on the analyzed cell type.
Especially, when cell colonies are small, characterized by a circular shape or with few adjacent merging colonies, such as in isolated and diluted bacteria cultures, the problem is easier to solve and estimating the number of colonies is not particularly difficult \cite{bernard2001,chiang2015,geissmann2013,clarke2010}.
On the contrary, colonies with irregular shapes and greater sizes, as well as colonies deriving from plating a conspicuous number of cells,  often cause colony merging and overlapping, which could represent a critical problem when the aim is determining the number of grown colonies.
Moreover, the conventional counting process can underestimate the effect of therapy, since it is possible that the treatment does not reduce the colony number but shrinks the colony size, as reported in \cite{lumley1996}.
In \cite{dahle2004}, the authors designed a cheap and efficient device, which employs a flatbed scanner to acquire $12$ Petri dishes at the same time, for automated counting of mammalian cell colonies.
A sequence of traditional image processing algorithms, including low-pass and high-pass filtering, was applied to enhance the image for shape analysis of colonies.
Colony detection was performed by placing the ROIs automatically onto a fixed location of the Petri dishes.
Finally, gray level global thresholding was performed by means of user interaction on the histogram of each ROI.
Guzmán \textit{et al.} \cite{guzman2014} developed ColonyArea, a plug-in for the open-source image analysis software ImageJ (National Institutes of Health, Bethesda, MD, USA) \cite{schneider2012}, optimized for an immediate analysis of focus formation assays.
ColonyArea processes image data of multi-well dishes for colony formation quantification, after separating, concentrically cropping and background correcting individual well images.
Instead of counting the number of colonies, ColonyArea determines the percentage of area covered by crystal violet stained cell colonies.
In the well separation and cropping steps, this approach uses a mask obtained considering the size and shape of the used plate.
The characteristics and typical dimensions of $6$- to $24$-well plates, as reported in the datasheets from a specific manufacturer, are stored in the plug-in.
In \cite{chiang2015}, an automated counting system was proposed, by using a hardware apparatus (designed and built ad hoc by the authors themselves) for Petri dish image acquisition.
The processing pipeline is based on several steps: (\textit{i}) the image is converted from color to gray levels using PCA; (\textit{ii}) the gradient magnitude image is calculated and thresholded to detect the imaged dish; (\textit{iii}) the bacterial colonies are extracted using the Otsu’s method \cite{otsu1975}.
Colonies in the central area and in the rim area are analyzed differently, both in the extraction and counting phases.
In the case of overlapping colonies, the distance transform \cite{kimmel1996,maurer2003} and the watershed transform \cite{beucher1992,vincent1991} are used to divide the overlapping colonies.
Clarke \textit{et al.} \cite{clarke2010} described a colony counting tool, named NIST’s Integrated Colony Enumerator (NICE), exploiting either an ordinary camera or a flatbed scanner.
The developed software combines extended minima, employed for colony center localization, and automatic or manual thresholding approaches \cite{otsu1975}.
Image denoising is performed by means of Gaussian smoothing and then extended minima are computed.
The author of \cite{geissmann2013} presented OpenCFU: a method for circular object segmentation on images and video sequences.
The input color image, acquired by means of either a high-definition camera or a webcam (relying on a white trans-illuminator to optimize image contrast in both cases), is pre-processed separately channel-by-channel to correct background intensity inhomogeneities and expand the contrast.
The first processing stage, which is applied on the image resulting from the normalization and fusion of the multiple image channels, yields a score-map based on valid region annotation, by taking advantage of thresholding with multiple values and then evaluating the connected-components by means of particle filtering.
The second stage consists in labeling the connected-components in the thresholded score-map and combining the distance transform and the watershed algorithm for the segmentation \cite{marotz2001,najman2005}.
Sometimes, cell counting approaches require to use specific hardware (designed \textit{ad hoc} or commercially available) for the acquisition of the images to be processed.
Barber \textit{et al.} \cite{barber2001} developed an automated colony counter composed of a hardware device and a software for image processing.
The system was tested with four different cell types and the achieved results were compared against the corresponding manual counting measurements from a staff composed of four experts.
The specific hardware enabled high-quality image acquisition of the culture flasks.
As a matter of fact, a monochrome Charge-Coupled Device (CCD) camera with a wide-angle lens ($f=2.8$ mm) was used, defining a field of view of $75 \times 56$ mm at the flask bottom.
In the acquisition phase, the average of $10$ images was utilized to deal with random noise.
However, a geometric deformation was employed for coping with the barrel distortion due to the wide-angle lens.
Two different approaches based on classic image processing techniques were used to identify cell colonies: the first method is tailored to highly-contrasted cell images; the second one deals with irregular-shaped colonies with fuzzy boundaries.
However, the most suitable between these two image processing methods is selected by the user according to the cell type.
In \cite{bernard2001}, a model-based image segmentation method is introduced, by relying on prior knowledge concerning the colony.
This strategy addressed the issue related to recognizing isolated, touching and overlapping cell colonies.
As first step, a statistical method proposes a set of candidate Gaussian model instances, by estimating the model parameters and the discrepancy between an instance and the current image.
Afterwards, the MDL principle selects the model instances that better match the distinct colonies in the cell image.
This method was tested on images of Chinese hamster lung fibroblast cell lines.
The hardware apparatus included a customized homogeneous diffuse lighting module that provides a Petri dish holder.
A CCD monochrome camera with C-mount lens ($f=50$ mm) was placed over the dish and connected to a Personal Computer.
The acquired $8$-bit images had a size of $512 \times 512$ pixels.
However, the computational method could not consider colonies lying on the Petri dish rim, since this border region is excluded \textit{a priori}.
In this case, a masking operation for each image is sufficient, by exploiting the fixed position and dimension of the imaged Petri dishes.

\subsubsection{Fluorescence microscopy}
Thanks to the advances in microscopy imaging technologies, it is possible to visualize living specimens' dynamic processes by time-lapse microscopy images \cite{kanade2011}.
Typical application examples, such as high-throughput/high-content phenotyping and atlas building for model organisms, reveal the importance of bioimage informatics \cite{peng2008,peng2012}.
However, these compelling challenges require specialized image analysis approaches.
In addition, the imaging data captured during even a single experiment may consist of hundreds of objects over thousands of images, making manual inspection a tedious and rather unfeasible task \cite{kanade2011}.
In practice, usually a single biological image stack has a large size (hundreds of MB or even GB) over multiple channels \cite{georgescu2011}.
The objects of interest in cellular images might be characterized by high variations of morphology and intensity from image to image \cite{peng2012}.
The high dimensionality and heterogeneity of the cell data acquired by means of the modern imaging technologies play a key role in the understanding of complex biological processes.
With more detail, cellular morphology is an important phenotypic feature that is indicative of the physiological state of a cell.
Therefore, the problem of cell segmentation has gained increasing attention during the past years since accurate cell boundaries are often required for subsequent analysis of intra-cellular processes and cell interactions.
These aspects give rise to new challenges with respect to existing image processing techniques, due to the high complexity and information content in bioimages.

Experiments where live cell specimens are imaged over prolonged time periods can potentially help us to understand how cells behave and respond to changes in the local environment.
High-content time-lapse imaging monitors the responses in living cells over a specific period of time exploiting the activity of fluorescently labeled and unlabeled cells.
These technologies have enabled large-scale imaging experiments for studying cell biology and for drug screening: indeed, the modern systems can acquire hundreds of thousands of microscopy images per-day, however their utility is hindered by inefficient image analysis methods.
Modern imaging experiments, with ever increasing scale, point out the limits of manual/visual image processing and analysis \cite{sadanandan2017}, requiring new effective automated methods.
One of the main problems in many automated analysis approaches, especially those following individual cells over time, is the image segmentation.
Unfortunately, automated segmentation of unstained cells imaged by bright-field and fluorescence microscopy is typically very challenging \cite{meijering2012}.
In practice, automated analysis is highly desirable due to manual analysis being subjective, biased and extremely time-consuming for large-scale datasets \cite{grah2017}.

\paragraph{Related work}
This section briefly outlines and reviews the state-of-the-art of cell fluorescence microscopy image segmentation.
As a matter of fact, several tools exist for automated bioimage analysis and processing \cite{eliceiri2012}.

The most common free and open-source software tools for microscopy applications in laboratory routine are: (\textit{i}) ImageJ \cite{schneider2012} or Fiji (Fiji Is Just ImageJ) \cite{schindelin2012} (National Institutes of Health, Bethesda, MD, USA), and (\textit{ii}) CellProfiler \cite{carpenter2006} (Broad Institute of Harvard and Massachusetts Institute of Technology, Cambridge, MA, USA).
Although these tools offer customization capabilities,  they do not provide suitable functionalities for fast and efficient high-throughput cell image analysis.
Moreover, CellAnimation \cite{georgescu2011} is similar to CellProfiler but provides a modular architecture as well as high-throughput options.
CellProfiler Analyst \cite{dao2016} was designed for the  exploration and visualization of image-based data, along with the classification of complex biological phenotypes, by means of an interactive graphical user interface.
This tools implements classic supervised Machine Learning (e.g., RFs, SVMs).

Mathematical morphology has been extensively used in cell imaging \cite{meijering2012}.
W{\"a}hlby \textit{et al.} \cite{wahlby2004} presented a region-based segmentation approach in which both foreground and background seeds are selected by means of morphological filtering---based on $h$-maxima and $h$-minima, respectively---applied on the original image and the gradient magnitude of the image, respectively.
These seeds are then used as starting points for the watershed segmentation of the gradient magnitude image.
Since more than one seed could be assigned to a single object, initial over-segmentation might occur, i.e., a boundary is created also in the case of weak edges.
Finally, a merging procedure is performed on neighboring objects according to gradient magnitude.
In \cite{kaliman2016}, the authors developed an automated approach based on the Voronoi tessellation \cite{honda1978} built from the centers of mass of the cell nuclei, to estimate morphological features of epithelial cells.
The full procedure is optimized for the imaging of cell mono-layers, where nuclei of cells do not overlap.
The authors leveraged the watershed algorithm in order to correctly segment cell nuclei.
In this case, the Voronoi tessellation is well-suited for measuring morphological properties of cells (such as area, perimeter, elongation) in a tissue composed of adjacent cells.

Machine Learning techniques have been applied to bioimage analysis \cite{angermueller2016}.
CellCognition aims at annotating complex cellular dynamics in live-cell microscopic movies \cite{held2010}.
This computational framework combines Machine Learning methods for supervised classification and Hidden Markov Models (HMMs)---exploiting the Viterbi algorithm \cite{viterbi1967} to incorporate the time information---in order to evaluate the progression by means of morphologically distinct biological states.

More recently, it has been shown that methods based on CNNs \cite{ciresan2012} can successfully address segmentation problems that are difficult to solve exploiting traditional image processing methods \cite{sadanandan2017}.
For instance, Deep Learning has been applied in biological image analysis, with particular interest to yeast images \cite{kraus2017} and also digital histopathology \cite{hatipoglu2017}.
Recently, an ensemble of multiple CNNs was used for segmentation of cell images \cite{hiramatsu2018}.
A gating network automatically divides the input image into several sub-problems and assigns them to specialized networks, allowing for a more efficient learning with respect to a single CNN.
This semantic segmentation approach was assessed on the segmentation problem of cell membrane and nucleus.
Abousamra \textit{et al.} \cite{abousamra2018} proposed a CNN-based architecture for localization, classification, and tracking in 4D fluorescence microscopy imaging.
In particular, this CNN-based approach was tested in the case of microtubule fibers' bridge formation during the cell division of zebrafish embryos.

Traditional image segmentation approaches often require experiment-specific parameter tuning, whilst CNNs require a huge amount of training data.
Large amounts of high-quality annotated samples or ground truth is typically required for the CNN training.
The ground truth, representing the extent to which an object is actually present, is usually delineated by a domain expert.
Obviously, visual and manual annotation is tedious, cumbersome and time-consuming.
The annotated samples for one dataset may not be useful for another dataset, so new ground truth generation may be needed for every new dataset, limiting the effectiveness of CNNs.
The authors of \cite{sadanandan2017} proposed an approach for automatically creating high-quality experiment-specific ground truth for segmentation of bright-field images of cultured cells based on end-point fluorescent staining (for nucleic and cytoplasmic regions).
This gold standard was obtained by means of standard automated methods available in CellProfiler \cite{carpenter2006}, exploiting the information conveyed by the fluorescent channels.
Finally, these automatically created segmentation results were exploited as labels along with the corresponding bright-field image data to train a CCN \cite{kraus2016}.
In general, applying CNNs to microscopy images remains still challenging due to the lack of large datasets labeled at the single cell level \cite{kraus2016}.
Thus, unsupervised image analysis techniques that do not require a training phase represent valuable solutions in this context \cite{held2010}.

\chapter{Classical Image Processing algorithms}
\label{chap3}
\graphicspath{{Chapter3/Figs/}}

\section{Region-based approaches}
\label{sec:regionBased}

\subsection{Image thresholding}
\label{sec:imageThresholding}

\subsubsection{Global image thresholding}
The simplest Image Processing technique for automatic image segmentation is global thresholding, which generally consists in classifying the pixels according to fixed criteria, usually specified as ranges of intensities \cite{rogowska2000}.
In particular, binarization is a segmentation approach that partitions the input image into two classes by considering a certain intensity threshold value $\theta$.
Despite its simplicity, this strategy provides an efficient and effective segmentation technique, according to the different intensities in the foreground and background regions of an image.
The threshold value $\theta$ must be carefully chosen, considering the features of the image underlying the pixel intensity values.
Consequently, given an image $\mathcall{I}$ consisting of $M \times N$ pixels, this intensity threshold defines two different classes, by dividing the histogram of the gray levels $\mathcall{H}$ into two parts, namely $\mathcall{H}_1$ and $\mathcall{H}_2$, according to the threshold intensity value $\theta$.
The pixels in the image $\mathcall{I}$ are partitioned into the two sub-regions $\mathcall{R}_1 = \left\{ \mathcall{I}(x,y) : \mathcall{I}(x,y) > \theta \right\}$ and $\mathcall{R}_2 = \left\{ \mathcall{I}(x,y) : \mathcall{I}(x,y) \leq \theta \right\}$, for every $x = 1, \ldots, M$ and $y = 1, \ldots, N$.

Several literature methods have been proposed to implement adaptive thresholding methods, able to automatically select a proper value for each analyzed image.
The most widespread algorithms for dynamic thresholding  are: the Iterative Optimal Threshold Selection (IOTS) \cite{ridler1978}, the method proposed by Otsu \cite{otsu1975}, and the Minimum-Error Thresholding (MET) method proposed by  Kittler and Illingworth \cite{kittler1986} later extended by Ye and Danielsson \cite{ye1988}.
All these approaches are closely related and strongly rely on images characterized by bimodal histograms (see \cite{xue2012}).
In the case of two-class image segmentation, the two sub-histograms (i.e., associated to the foreground and background pixels, respectively), assumed to be nearly Gaussian distributions, should be characterized by approximately equal size and variance \cite{kurita1992}.
When these assumptions are not satisfied---i.e., the pixel intensity distribution is not approximately bimodal---the aforementioned algorithms show some limitations.
As a matter of fact, the optimal threshold $\theta_\text{opt}$---especially in the case of the Otsu's method---either over- or under-estimates the ROI, since the computed threshold tends to split the class with larger size and to bias towards the class with larger variance.
Under these conditions, the IOTS method \cite{ridler1978, trussell1979} could provide better results than the Otsu's method \cite{otsu1975} when the sizes of the two classes are highly different \cite{xu2011}.
In addition, Medina-Carnicer \emph{et al.} in \cite{medina2011} showed that the above mentioned algorithms often perform poorly with unimodal distributions of gray levels.
Moreover, in the case of images affected by intensity overlap, the IOTS algorithm is less likely to either over- or under-estimate the threshold, when compared to other techniques selecting a threshold in the valley between two peaks of the histogram, even if the histogram is not strongly bimodal \cite{sonka2007}, in particular when applied to medical images \cite{kwok2004}.

Algorithm \ref{pc:IOTS} shows the pseudo-code of the IOTS algorithm \cite{ridler1978}, by using an efficient implementation that computes the image histogram only once \cite{trussell1979}.
Therefore, a single histogram (i.e., the initial image histogram) is processed during the iterations until convergence.
This iterative approach aims at dividing the whole histogram $\mathcall{H} = h(r)$ (with $r \in [0, 1, \ldots, L-1]$, where $L$ is the maximum gray level) into two classes $\mathcall{H}_1$ and $\mathcall{H}_2$, separated by an adaptive global threshold $\theta_{\text{opt}}$.

\begin{algorithm}
	\caption{Pseudo-code of the iterative optimal threshold selection algorithm.}
	\label{pc:IOTS}
	\textbf{Input:} Image $\mathcall{I}$\\
	\textbf{Output:} Optimal threshold $\theta_{\text{opt}}$ \\
	\begin{algorithmic}[1]
	    \State  Calculation of the histogram $\mathcall{H}=h(r)$, with $r \in [0, 1, \ldots, L-1]$
	    \LeftComment  Initial estimated value for the threshold $\theta^{(t)}$ (for example, the average intensity of $\mathcall{I}$)
	    \State $\theta^{(0)} \gets \mu_\text{global} = \frac{\sum\limits_{r=0}^{L-1} h(r) \cdot r}{\sum\limits_{r=0}^{L-1} h(r)}$
	    \While{$| \theta^{(t+1)} - \theta^{(t)} | \le  \varepsilon_\text{tol}$} \Comment{Convergence condition achievement}
	    \State The histogram $\mathcall{H}$ is divided into two classes $\mathcall{H}_1$ and $\mathcall{H}_2$ separated by $\theta^{(t)}$
	    \LeftComment Average intensity values $\mu_1$ and $\mu_2$ for the sub-histograms $\mathcall{H}_1$ and $\mathcall{H}_2$, respectively
	    \State $\mu_1^{(t)} = \frac{\sum\limits_{r=0}^{\theta^{(t)}} h(r) \cdot r}{\sum\limits_{r=0}^{\theta^{(t)}} h(r)}$; $\mu_2^{(t)} = \frac{\sum\limits_{r=\theta^{(t)}+1}^{L-1} h(r) \cdot r}{\sum\limits_{r=\theta^{(t)}+1}^{L-1} h(r)}$ \State $\theta^{(t+1)} \gets \frac{\mu_1^{(t)} + \mu_2^{(t)}}{2}$ \Comment{Update the current threshold value}
	    \EndWhile
	    \State $\theta_{\text{opt}} \gets \theta^{(t+1)}$
	    \end{algorithmic}
\end{algorithm}

\paragraph{Uterine fibroid segmentation on MR images}

\begin{figure}
	\centering
	\includegraphics[width=0.3\linewidth]{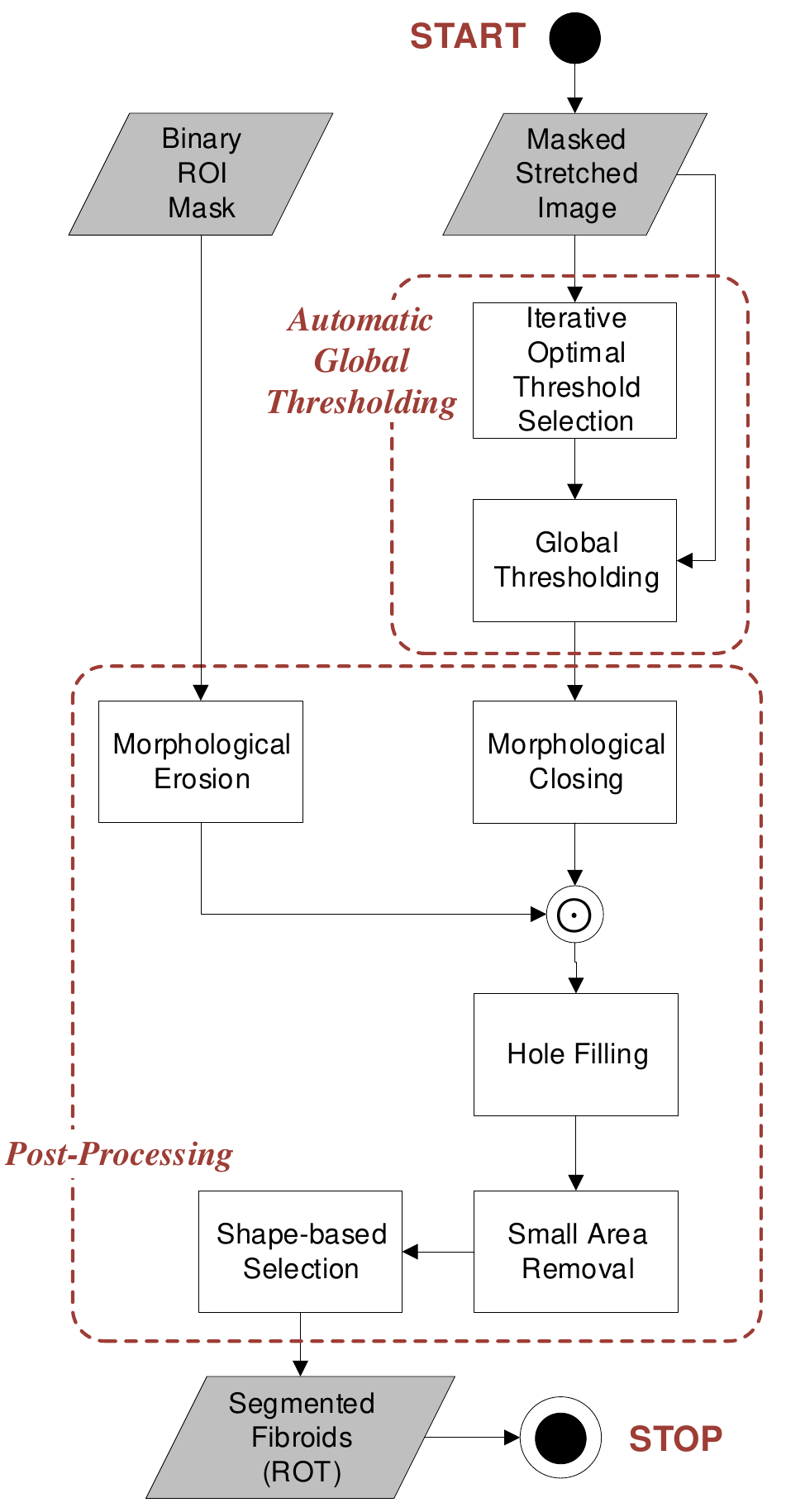}
	\caption[Flow diagram of the IOTS-based fibroid segmentation method]{Flow diagram representing the steps of the fibroid segmentation method based on the IOTS algorithm.}
	\label{fig:AFS-flowDiagram}
\end{figure}

\begin{figure}
	\centering
	\includegraphics[width=0.9\linewidth]{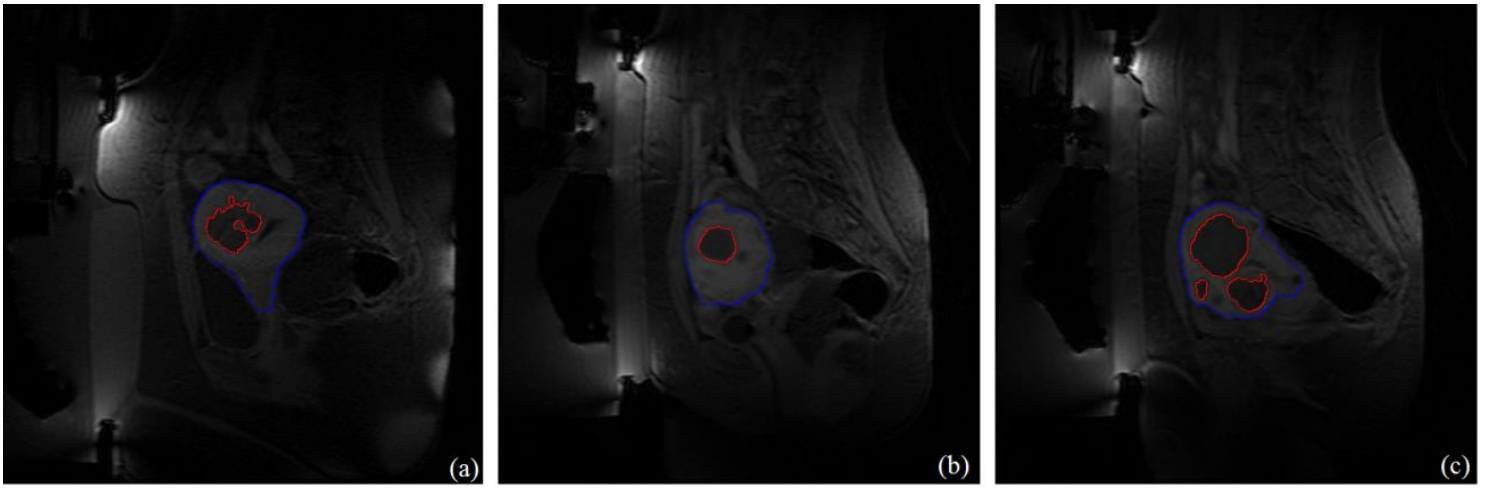}
	\caption[Uterine fibroid segmentation results achieved by the IOTS-based method]{Uterine fibroid segmentation results achieved by the IOTS-based method. The blue contour represents the uterus, while the red contour represents the ROT fibroids.}
	\label{fig:AFS-Results}
\end{figure}

In \cite{militelloCBM2015}, uterine fibroids were segmented using the efficient version of the IOTS algorithm \cite{ridler1978,trussell1979} described in Algorithm \ref{pc:IOTS}.
After the uterus segmentation based on the FCM clustering algorithm presented in \cite{militelloCBM2015}, the particular characteristics of the region were identified using an adaptive thresholding applied on gray levels, because fibroids have intensity values much lower than the untreated uterus pixels.
Although IOTS works generally well on images characterized by a basically bimodal histogram, some post-processing and mathematical morphological refinements are applied on the detected fibroid ROT to improve segmentation reliability.
The overall processing pipeline is presented in Fig. \ref{fig:AFS-flowDiagram}.
Fig. \ref{fig:AFS-Results} shows three segmentation examples of the uterus ROI (blue contour) and of the fibroid ROT (red contour).

The experimental fibroid MRI dataset as well as the achieved segmentation evaluation metrics are reported in Section \ref{sec:RegionGrowing}, wherein a quantitative comparison against the method in \cite{rundoMBEC2016} is shown (according to the evaluation metrics provided in Appendix \ref{sec:segEval}).

\subsubsection{Local image thresholding}

Local adaptive thresholding techniques estimate the threshold locally over sub-regions of the entire image, by considering only a neighborhood (i.e., sliding window) with a specified size and exploiting local image properties to calculate a variable threshold \cite{gonzalez2002}.

These local approaches can be classified into two strategies.
The first one refers to the approach in \cite{chow1972}, in which the entire image is divided into an array of overlapping sub-images: the optimal threshold is searched for each sub-image by investigating its histogram.
The threshold for each single pixel is found by interpolating the results of the sub-images.
The main drawback of this method is the computational cost, making it often not suitable for real-time applications.
The second class, also called local adaptive thresholding, finds the threshold by locally examining the intensity values of the neighborhood of each pixel according to image intensity statistics. The most appropriate first-order statistics depends strongly on the analyzed input image \cite{sauvola2000}.
Mean and median of the local intensity distribution are the most used measures of central tendency.
The size of the examined neighborhood has to be large enough to cover sufficient foreground and background pixels, so defining a significant statistical sample.
On the other hand, choosing too large regions could violate the assumption of approximately uniform illumination.
This method is less computationally expensive than the approach in \cite{chow1972} and yields good segmentation results in several practical applications, such as in document binarization \cite{bataineh2011,sauvola2000}, cell image analysis \cite{huang2008,zhou2016}, and medical image analysis \cite{kom2007}.
The assumption underlying these methods is that smaller image regions are more likely to show approximately uniform illumination, thus being more suitable for thresholding.

In our implementation, to calculate the local threshold $\theta_i$, we used local adaptive thresholding considering the mean value in the neighborhood as first-order statistics.
In particular, let $\mathcall{N}_i$ be the $i$-th neighborhood in which the input $\mathcall{I}$ (with size $M \times N$) was divided, the corresponding local threshold $\theta_i$ is computed for each pixel according to Eq. (\ref{eq:localTh}), where the index $i=1,2, \ldots, M \times N$ scans the vectorization of the image matrix $\mathcall{I}$:

\begin{equation}
    \label{eq:localTh}
    \theta_i = \text{mean}\left\{ \sum\limits_{(x,y) \in \mathcall{N}_i} \mathcall{I}(x,y) \right\}, \text{with } i=1,2, \ldots, M \times N.
\end{equation}

The size of each neighborhood is $\text{size}_\mathcall{N}$, depending only on the size of the well sub-image $\mathcall{I}$, is a pair of positive odd integers according to Eq. (\ref{eq:sizeNeigh}):

\begin{equation}
    \label{eq:sizeNeigh}
    \text{size}_\mathcall{N}=\left( 2 \times \left\lfloor \frac{M}{16} \right\rfloor + 1 \text{, } 2 \times \left\lfloor \frac{N}{16} \right\rfloor + 1 \right).
\end{equation}

\paragraph{Area-based cell colony surviving fraction evaluation}

\begin{figure}
	\centering
	\includegraphics[width=0.6\linewidth]{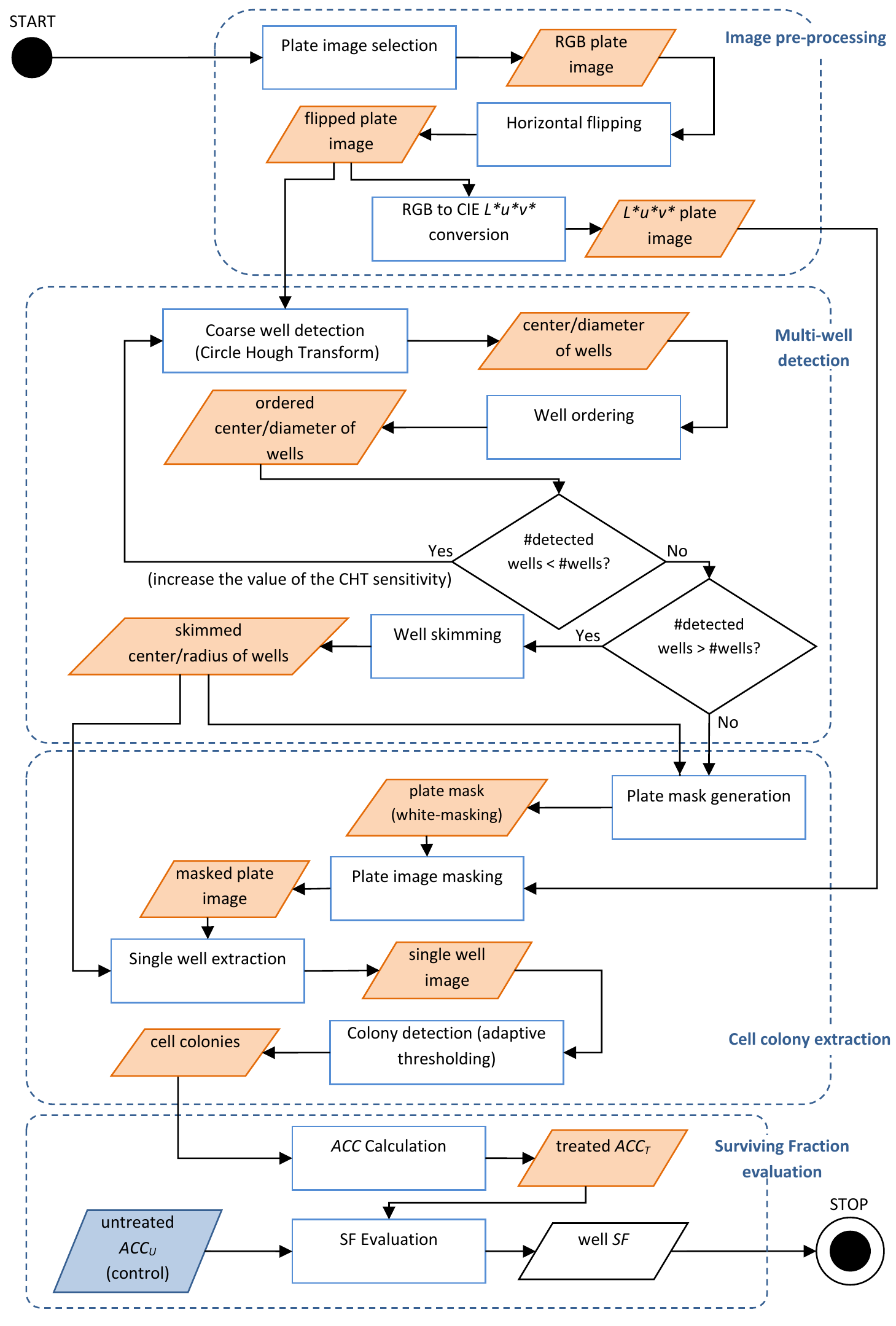}
	\caption[Overall flow diagram of the processing pipeline of the proposed fully automatic approach for ACC]{Overall flow diagram of the processing pipeline of the proposed fully automatic approach for ACC. Four main stages can be identified: (\textit{i}) image-preprocessing, basically dealing with the RGB to CIE \textit{L*u*v*} conversion; (\textit{ii}) multi-well detection, which identifies and selects the wells inside the plate image; (\textit{iii}) colony extraction, which segments cell colonies from the well background; (\textit{iv}) SF evaluation, which estimates the SF of the current well according to the treated ACC\textsubscript{T} against the reference untreated ACC\textsubscript{U} (control).}
	\label{fig:ACC-flowDiagram}
\end{figure}

This section describes a novel computational method for automated SF evaluation, based on ACC alone.
Fig. \ref{fig:ACC-flowDiagram} shows the overall flow diagram with the processing steps performed in the pipeline that, starting from the acquired plate image, automatically detects the wells, extracts the colonies, calculates the ACC, making it possible to measure the SF.
The proposed method was entirely developed using the MatLab\textsuperscript{\textregistered} environment (The MathWorks, Natick, MA, USA).

The proposed approach exploits the Circle Hough Transform (CHT) and local adaptive thresholding algorithms: these two effective and efficient techniques are combined in an original pipeline, not yet considered in the literature, enabling to address several open issues in clonogenic assay SF evaluation.

The proposed fully automatic approach presents several key novelties and contributions with respect to related literature works:
\begin{itemize}
    \item independence on the colony number, which in some cases may lead to inconsistencies between the computed SF values and the real cell growth. As a matter of fact, plates with the same number of colonies but different ACCs, due to different growth rate, are indicative of different treatment responses. The traditional approach, based on the number of colonies alone, is not able to differentiate among these cases;
    \item evaluation of the SF based on the percentage of the ACC, which allows us to establish a correlation between the growth rate and the dose administered during the treatment;
    \item unlike some of the literature works that use specific cameras or equipment (sometimes expensive), our approach uses general-purpose acquisition hardware (a conventional flatbed scanner);
    \item no calibration and parameter-settings are required, differently to \cite{geissmann2013};
    \item only a pre-processing step, performing a Red Green Blue (RGB) to International Commission on Illumination (CIE) \textit{L*u*v*} color space conversion, is applied to reduce luminance inhomogeneities (decoupling the intensity and chromaticity components), thus not affecting the original pictorial data \cite{sangwine1998};
    \item automated wells detection inside the plate image by means of the CHT \cite{duda1972,yuen1990}, unlike several literature approaches \cite{bernard2001,dahle2004,guzman2014} that require the plate dimensions and the spatial arrangement of the wells;
    \item colony segmentation exploiting just a local adaptive thresholding algorithm \cite{gonzalez2002,jain1999}, which deals with both small and large colonies;
    \item no ``aggressive'' post-processing operations are needed (e.g., morphological refinement could have stronger effects when performed on small colonies and, consequently, can lead to significant errors in the ACC calculation).
\end{itemize}

\subparagraph{Materials}
In the following, the characteristics of the analyzed cell cultures as well as the procedure for multi-well plate image acquisition are reported.

This study analyzed MCF7 and MCF10A cell lines: MCF7 is a human breast epithelial carcinoma estrogen receptor-positive cell line, while MCF10A is a human non-tumorigenic breast epithelial cell line.
At the end of a $12$-$15$ day incubation period, the cultures were fixed and stained with crystal violet according to the protocol described in \cite{franken2006} and were analyzed manually by the experienced staff at the Cellular and Genomic Methodologies Laboratory of the Institute of Molecular Bioimaging and Physiology (IBFM-CNR) in Cefalù (PA), Italy.
The obtained SF values, averaged over three counting measurements for each well, performed by a staff composed of three different expert biologists, represent our ground-truth used as a reference for the fully automated approach.

MCF7 and MCF10A cellular lines were placed on $6$-well plates (Corning\textsuperscript{\textregistered} Inc., Corning, NY, USA).
Several culture plates were prepared, considering different configurations of either irradiation or cytotoxic agent.
The proposed approach is not just limited to $6$-well plates, since it is able to automatically detect the wells, regardless of the well spatial arrangement and location in the plate (only the well number and well diameter are needed).
Table \ref{table:ACC-plateCharacteristics} depicts the main characteristics of the different plate types employed in this study.
The plate images were acquired using the Epson Expression 10000XL (Seiko Epson Corporation, Nagano, Japan), which is a conventional flatbed scanner, with a spatial resolution of $800$ dots-per-inch (dpi) and $24$ bits ($8$ bits per-color channel) as bit-depth.
The scanned images were saved in the uncompressed TIFF format.
The used resolution represents a good compromise between image quality (lower resolutions could lead to the elimination of small colonies) and the required processing-time.

\begin{table}
    \centering
	\caption[Characteristics of the analyzed multi-well plates]{Some characteristics of the multi-well plates employed in this study. Differently to literature approaches that use static techniques (i.e., plate measurements and well spatial arrangement) for well detection, the proposed approach, based on the CHT, requires only the well number and diameter.}
	\label{table:ACC-plateCharacteristics}
	\begin{scriptsize}
		\begin{tabular}{lcccc}
			\hline\hline
			Plate type	& Number of wells	& Well diameter [ms]	& Spatial arrangement of wells [pixels]	& Plate size [mm$^2$] \\
			\hline
			Corning\textsuperscript{\textregistered} 6 wells	& $6$		& $34.80$ 		& $2 \times 3$ 	& $127.76 \times 85.47$ \\
			Corning\textsuperscript{\textregistered} 12 wells	& $12$		& $22.11$ 		& $3 \times 4$ 	& $127.89 \times 85.60$ \\
			Corning\textsuperscript{\textregistered} 24 wells	& $24$		& $15.62$ 		& $4 \times 6$ 	& $127.89 \times 85.60$ \\
			Corning\textsuperscript{\textregistered} 48 wells	& $48$		& $11.05$ 		& $6 \times 8$ 	& $127.89 \times 85.60$ \\
			\hline\hline
		\end{tabular}
	\end{scriptsize}
\end{table}

\subparagraph{Image pre-processing}
After the multi-well plate image acquisition phase, some image artifacts and inhomogeneities could affect the accuracy achieved in the subsequent processing steps.
Thus, some pre-processing steps are performed.

Firstly, the image is horizontally flipped to correct the specular reflection that occurs in the acquisition phase, allowing for the correct correspondence between the well number and its real position.
However, due to small parallax errors, a circular dark shadow on the rim of each well was observed.
In order to reduce these shadowing artifacts and make the illumination more uniform, a conversion of the color space from RGB to CIE \textit{L*u*v*} is performed \cite{sangwine1998}.
As a matter of fact, the \textit{L*u*v*} is an excellent intensity (represented by lightness \textit{L*}) and chromaticity (denoted by \textit{u*} and \textit{v*} components) decoupler \cite{gonzalez2002}.
The lightness \textit{L*} component, which conveys greater information related to intensity, is not considered in the subsequent processing steps for colony cell segmentation.
By exploiting this decoupling between intensity and chromatic components, we are able to deal better with possible brightness non-uniformities inside the wells.

\subparagraph{Multi-well detection based on circle Hough transform}

The wells inside the plate are detected automatically applying the HT on the horizontally flipped RGB plate image.
The HT is an efficient feature extraction technique for detecting objects with a specific shape \cite{hough1962}, which brings the problem of curve detection in a simpler search procedure for maxima in an accumulator array \cite{calatroni2017}.
Especially, the CHT was used since it is specialized to find circle-shaped objects \cite{duda1972,yao2016,yuen1990}.
A plane curve $\mathcal{F}(x,y):\mathbb{R}^2 \rightarrow \mathbb{R}$ is defined analytically using a set of parameters.
Eq. (\ref{eq:analCart}) maps the parameters to Cartesian coordinates in implicit form:
\begin{equation}
    \label{eq:analCart}
    \mathcal{F}((x,y); (a_1,a_2, \ldots,a_n))=0,
\end{equation}

where $(x,y)$ is a point of the two-dimensional curve, and $(a_1,a_2, \ldots,a_n)$ is an $n$-ple of values that defines a point in the parameter space (i.e., a point in the space of parameters uniquely identifies an analytical curve).
Each point in the image space corresponds to a hypersurface in the parameter space and $n$ points in the image space, belonging to the same curve, generate $n$ surfaces that intersect at the same point in the parameter space.
An intersection of many surfaces in the parameter space is the clue to the presence of a particular instance of the analytical curve to be found.
In our case, the purpose is to detect circles, which represent the wells in the plate image.
The CHT is able to find circles of known radius as well as circles having any radius.
Since the well diameter $r_w$ is known, we are interested in the first case, in which the parameter space is two-dimensional and the curve generated from each point $(x,y)$ in the image space is a circle itself, according to Eq. (\ref{eq:circ}).
The parameter space is generated, considering each pixel $(x,y)$, by the center coordinates $\mathbf{c} \equiv (x_c,y_c)$:

\begin{equation}
    \label{eq:circ}
    (x-x_c)^2 + (y-y_c)^2 = r_w^2.
\end{equation}

Generally, a number of points at least equal to the number of parameters has to be considered to define a curve (i.e., its features).
Therefore, the CHT allows to convert a searching problem for curves in a simpler surface intersection search by means of a so-called accumulator matrix \cite{illingworth1987}.
The circle candidates are produced by voting in the parameter space center points $\mathbf{c}$, and then the local maxima are selected in the accumulator matrix.
Formally, each point $(x_c,y_c )$ in the parameter space, representing a circle center, must meet the following conditions in polar representation:

\begin{equation}
    \label{eq:polarRep}
    \begin{cases}
       x_c = x - r_w \cdot \cos{\theta} \\
       y_c = y - r_w \cdot \sin{\theta}
     \end{cases},
\end{equation}
where $r_w$ is the radius of the well and $\theta \in [0,2 \pi]$ is the angular coordinate with respect to the positive $x$-axis, by using a complex phase coding scheme \cite{atherton1999}.
The pseudo-code of the implemented robust well detection strategy based on the CHT technique is outlined in Algorithm \ref{pc:wellDetectionCHT}.

\begin{algorithm}
	\caption{Pseudo-code of the well detection strategy based on the CHT technique.}
	\label{pc:wellDetectionCHT}
	\textbf{Input:} plate image $\mathcall{I}_\text{plate}$, radius of each well $r_w$, total number of wells $N_\text{wells}$ \\
	\textbf{Output:} $N_\text{wells}$ circles identified by the center coordinates $\mathbf{c} \equiv (x_c, y_c)$  \\
	\begin{algorithmic}[1]
	    \State $\mathbf{A}(\theta,r_w) \gets \mathbf{0}$ \Comment{Define and initialize the accumulator matrix}
	    \State $N_\text{DetWells} \gets 0$ \Comment{Number of detected wells}
		\ForEach{center-candidate point $\mathbf{c} \equiv (x_c, y_c) \in \mathcall{I}_\text{plate}$}
		    \State Increase all the corresponding points in $\mathbf{A}$ \Comment{Local maxima in the accumulator correspond to circular objects}
		    \If{$N_\text{DetWells} < N_\text{wells}$}
		        \State Increase the CHT sensitivity value
		        \State \textbf{continue}
		    \EndIf
		    \If{$N_\text{DetWells} > N_\text{wells}$} \Comment{Well-skimming procedure (according to CHT metrics)}
		        \State Select the first $N_\text{wells}$ detected circles sorted by circle strength value
		    \EndIf
		\EndFor
	\end{algorithmic}
\end{algorithm}

\begin{figure}
	\centering
	\includegraphics[width=0.9\linewidth]{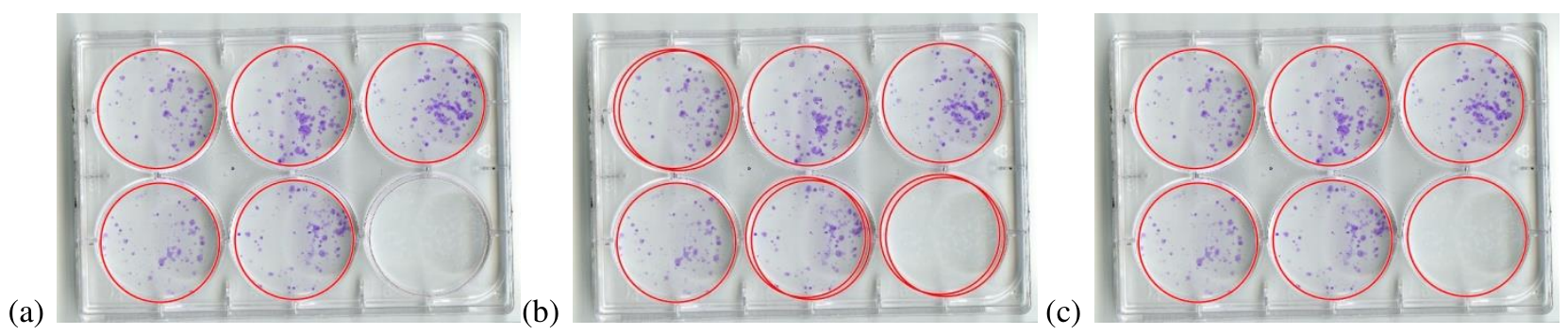}
	\caption[Examples of the control strategy used to enhance the automatic well detection performed using the CHT]{Some examples of the control strategy used to enhance the automatic well detection performed using the CHT in the case of $6$-well plates: (a) the initial CHT detects only $5$ wells; (b) after the sensitivity increasing, the CHT detects $9$ wells; (c) in order to skim the correct wells, only the first $6$ detected wells are selected after well ordering according to the CHT metrics (i.e., circle strength value).}
	\label{fig:ACC-wellSkim}
\end{figure}

In all the analyzed plates, the well detection using the CHT is performed robustly without errors directly in the original RGB color space.
However, in order to make the procedure more robust, in case of incorrect well identification, a more selective control strategy was implemented.
In particular, if the initial CHT identified less than the expected number of wells (i.e., $6$, $12$, $24$ or $48$), then the CHT is repeated increasing the sensitivity factor for the accumulator matrix (from $0.98$ to $0.99$), including also weak and partially occluded or overlapping circles (e.g., possible shadowing artifacts in well edges or rims).
Otherwise, if the identified wells are less than the expected number, then a skimming procedure is performed considering the value of the metrics that the CHT computes for each found circular object.
We noticed remarkable differences between the metrics values of the correct wells and those of incorrectly identified wells.
For this reason, we decided to select exactly the number of expected wells reporting the highest CHT metrics (i.e., the first $6$, $12$, $24$ or $48$ wells in the candidate list sorted by the circle strength value).
Fig. \ref{fig:ACC-wellSkim} shows a particular case, in which the initial CHT detects only $5$ wells (Fig. \ref{fig:ACC-wellSkim}a).
The following CHT iteration with increased sensitivity detects $9$ wells (Fig. \ref{fig:ACC-wellSkim}b).
The final skimming phase extracts the correct wells (Fig. \ref{fig:ACC-wellSkim}c).
Overall, the CHT enhanced with this control strategy was shown to be robust and reliable, allowing for an accurate well localization in $100\%$ of the examined plates.

\begin{figure}
	\centering
	\includegraphics[width=0.6\linewidth]{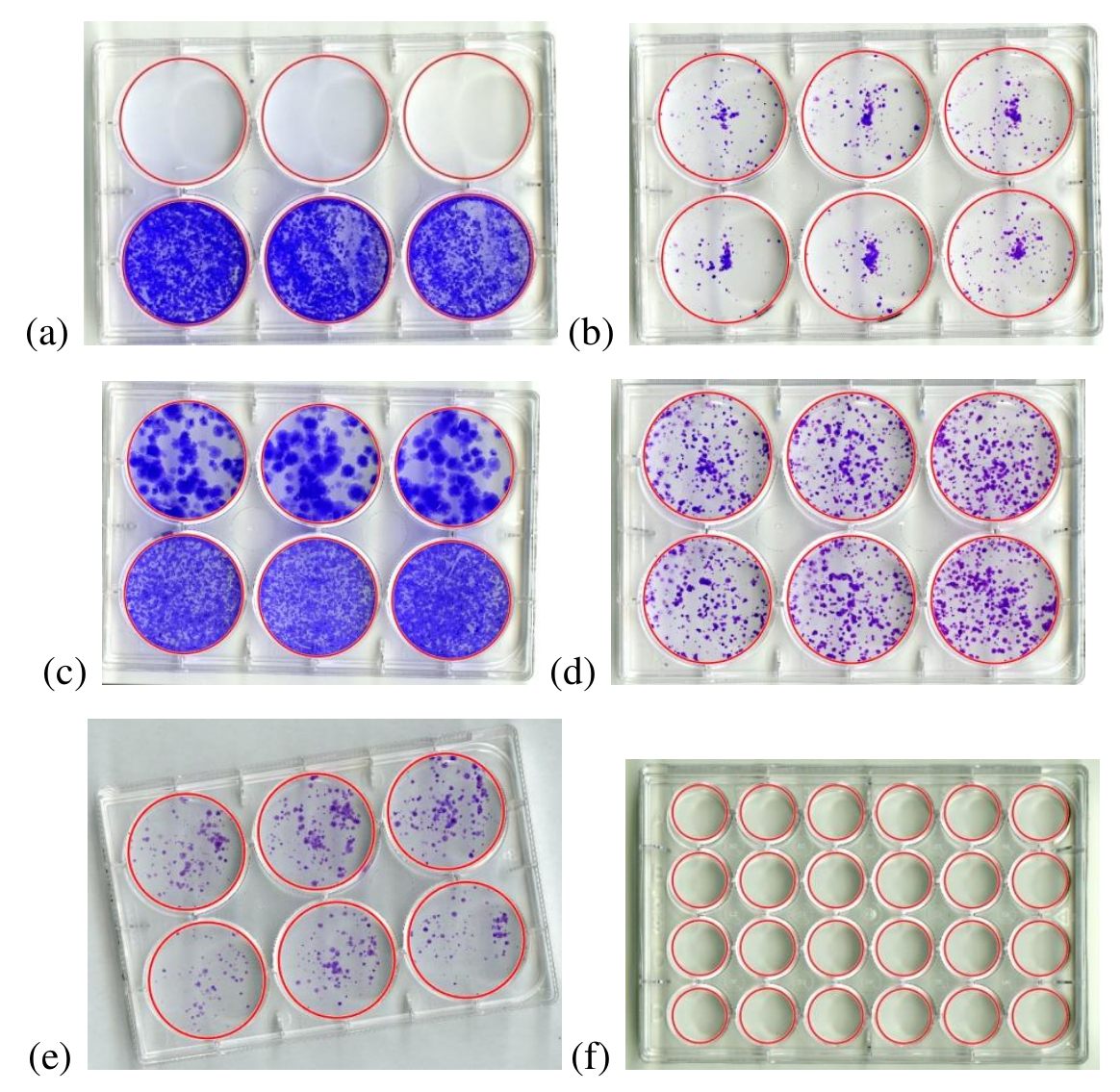}
	\caption[Correct detection of the wells in typical operating conditions]{The proposed approach correctly detects wells in typical operating conditions (a-d), as well as when the acquired plate image is not aligned (e), and (f) with different plate types (e.g., $24$-well plate).}
	\label{fig:ACC-wellDetect}
\end{figure}

Unlike some literature works \cite{dahle2004,guzman2014} that identify the wells statically, by considering either the plate characteristics (e.g., well radius, horizontal and vertical distance between wells, thickness of well wall, number of wells per-row, number of well per column) or by using a fixed position mask \cite{bernard2001}, our method is fully automatic.
An approach based on specific sizes can lead to errors in case of misalignment between the plate and the reference position.
Differently, the well detection implemented in our method requires only the number of wells in the plate and the well radius regardless the type and position of the plate.
The examples reported in Fig. \ref{fig:ACC-wellDetect} show the robustness of the well detection phase to rotation and translation.
In Figs. \ref{fig:ACC-wellDetect}(a, b, c, d), some well detection results in four different typical real conditions are depicted.
Moreover, as depicted in Fig. \ref{fig:ACC-wellDetect}e, it is possible to identify automatically the wells even if the image is not perfectly aligned horizontally (due to arbitrary misalignments in the acquisition phase) and regardless the well spatial arrangement ($2$ rows and $3$ columns in $6$-well plates).
Fig. \ref{fig:ACC-wellDetect}f shows an example of detection on a $24$-well plate.
After the well detection, an ordering step is performed, to establish a unique correspondence between the center coordinates of the found circles and their own position in the plate.

\subparagraph{Cell colony extraction using local adaptive thresholding}
Cell colony segmentation is performed using a local adaptive thresholding algorithm.
Both global and local thresholding techniques aim to separate desirable foreground objects from the background in an image, considering differences in pixel intensities.
Global thresholding determines a single threshold for all pixels and works only if the histogram of the input image contains well-separated peaks corresponding to the desired subject and background.
Consequently, such an approach cannot deal with images containing a strong illumination gradient \cite{jain1999}.
As a matter of fact, robust and accurate image segmentation methods represent an active research topic in image processing and graphics \cite{oh2017}.

Nevertheless, the RGB to CIE \textit{L*u*v*} conversion allows us to reduce most illumination issues, in some areas remain some non-uniformities that adaptive thresholding can easily handle, making it possible to handle changes in lighting conditions in the image.

Several literature approaches \cite{chiang2015,clark2017,dahle2004,geissmann2013,guzman2014} use global thresholding for colony extraction.
Although global adaptive thresholding algorithms, such as \cite{kittler1986,otsu1975,ridler1978}, generally yield excellent segmentation results, they could fail under certain conditions.
For instance, when the colonies cover the whole neighborhood, the adaptive thresholding could not work properly.
As a matter of fact, if the analyzed local neighborhood contains only cells, the local adaptive thresholding algorithm is not able to estimate a valid local threshold $\theta_i$ and, consequently, this area will be discarded.
These conditions occur if a huge number of cells are plated, or if cells have a high growth rate to cover the whole well (or at least the currently examined neighborhood).
To overcome this issue, which may occur mainly in areas near the edge of the plate, instead of a typical black-mask we adopted a white-mask, obtained automatically by considering centers and diameters of the circles identified in the previous well detection phase, to mask the wells in the plate image.
Using this simple solution, the adaptive thresholding works correctly, so considerably enhancing colony extraction.

\begin{figure}
	\centering
	\includegraphics[width=0.6\linewidth]{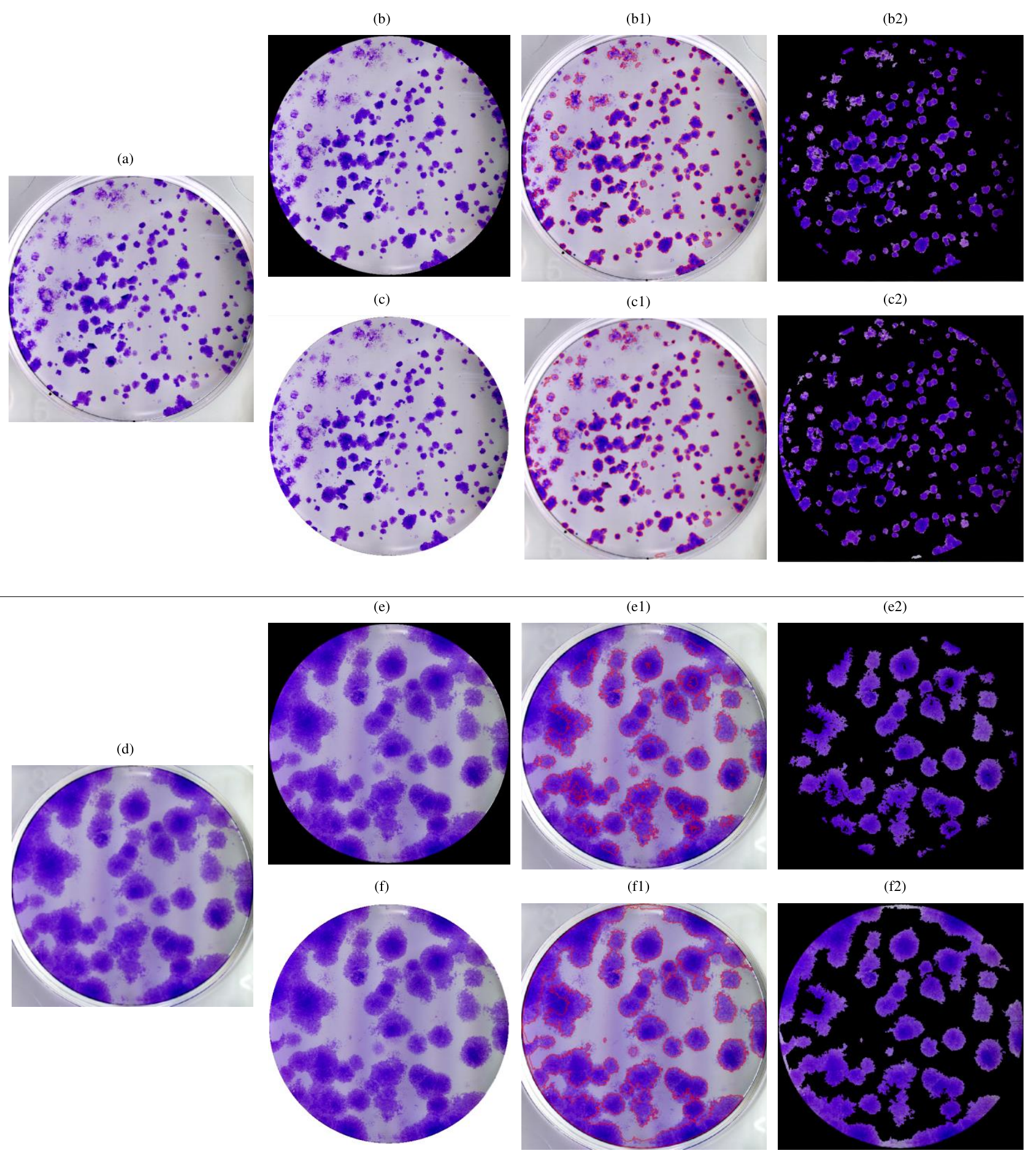}
	\caption[Examples of colony extraction performed using the proposed approach based on local adaptive thresholding]{Examples of colony extraction performed using the proposed approach based on local adaptive thresholding using white-masking compared against the same local adaptive thresholding with previous black-masking.
	Two wells with $500$ MCF7 and $150$ MCF10 plated cells are shown in the first two rows and in the last two rows, respectively: (a, d) input wells before colony detection; (b1, e1) and (b2, e2) colonies detected using local adaptive thresholding and black-masking (image with superimposed red contour and masked with the achieved segmentation, respectively); (c1, f1) and (c2, f2) colonies detected using local adaptive thresholding and white-masking (image with superimposed red contour and masked with the achieved segmentation, respectively).
	The segmentation combining adaptive thresholding and white-masking enhances colony extraction close to the well rim, where cell growth created a uniform area, which classic adaptive thresholding with black-masking is not able to segment properly.}
	\label{fig:ACC-colonyExtr}
\end{figure}

Fig. \ref{fig:ACC-colonyExtr} shows two series of results: in the first two rows a well with $500$ plated MCF7 cells, and in the last two rows a well with $150$ plated MCF10 cells. It is appreciable how the used white-masking operation (Figs. \ref{fig:ACC-colonyExtr}(c1, c2, f1, f2)) enhances the output with respect to adaptive thresholding after typical black-masking (Figs. \ref{fig:ACC-colonyExtr}(b1, b2, e1, e2)).
After colony detection, the only post-processing step is represented by a hole filling algorithm \cite{soille2013}, aimed to include the central region of the colonies that, appearing generally darker, can be excluded from the detected regions.

\subparagraph{Results}

The SF values obtained by considering the percentage of ACC were compared against the conventional measurement based on colony counting alone, performed manually by three different expert biologist operators (see Appendix \ref{sec:validClonAssay} for a detailed explanation).

\begin{figure}
	\centering
	\includegraphics[width=0.9\linewidth]{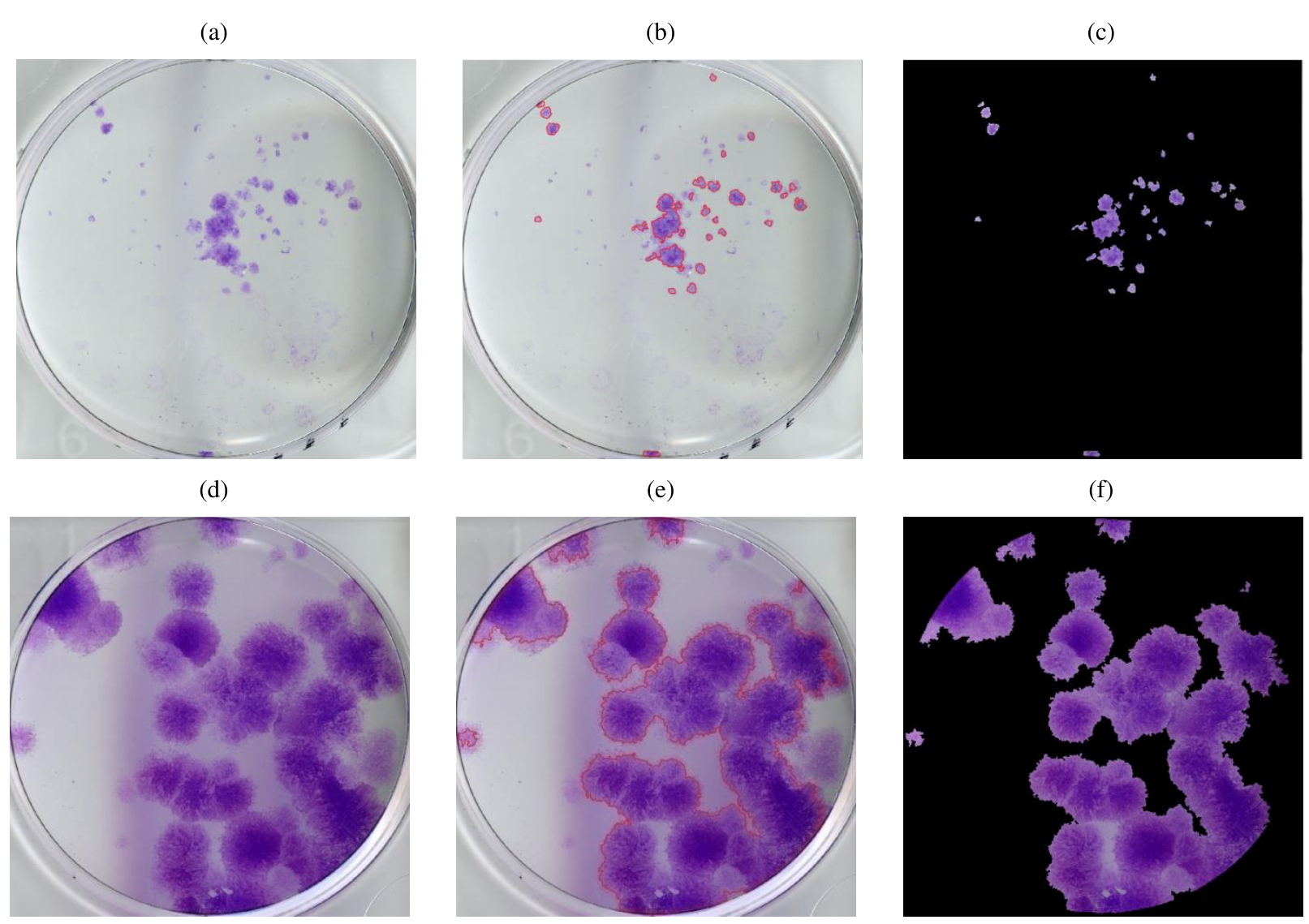}
	\caption[Examples of cell colony detection in different real scenarios]{Examples of cell colony detection in different real scenarios. The first row concerns MCF7 cell colonies with very small size: (a) original well; (b) and (c) show the detected colonies with red contour and background masking, respectively. The second row concerns MCF10A cell colonies characterized by a high growth rate: (d) original well; (e) and (f) show the detected colonies with red contour and background masking, respectively.}
	\label{fig:ACC-colonyDetect}
\end{figure}

The performed experimental tests aim to demonstrate the validity of the proposed approach and the accuracy of the obtained results, in terms of correlation between the SF calculated considering manually counted cell colonies (SF\textsubscript{Manual}) and the SF based on ACC computed using our automated method (SF\textsubscript{\%Area}).
As shown in Fig. \ref{fig:ACC-colonyDetect}, the proposed approach allows for the correct detection of the colonies in a wide variety of scenarios, including cells characterized by a high growth rate, which develop in very large colonies with significantly variable shape (and that merging together they would create problems to conventional approaches based on counting the number of colonies), as well as with cells that form very small colonies.

\begin{figure}
	\centering
	\includegraphics[width=0.5\linewidth]{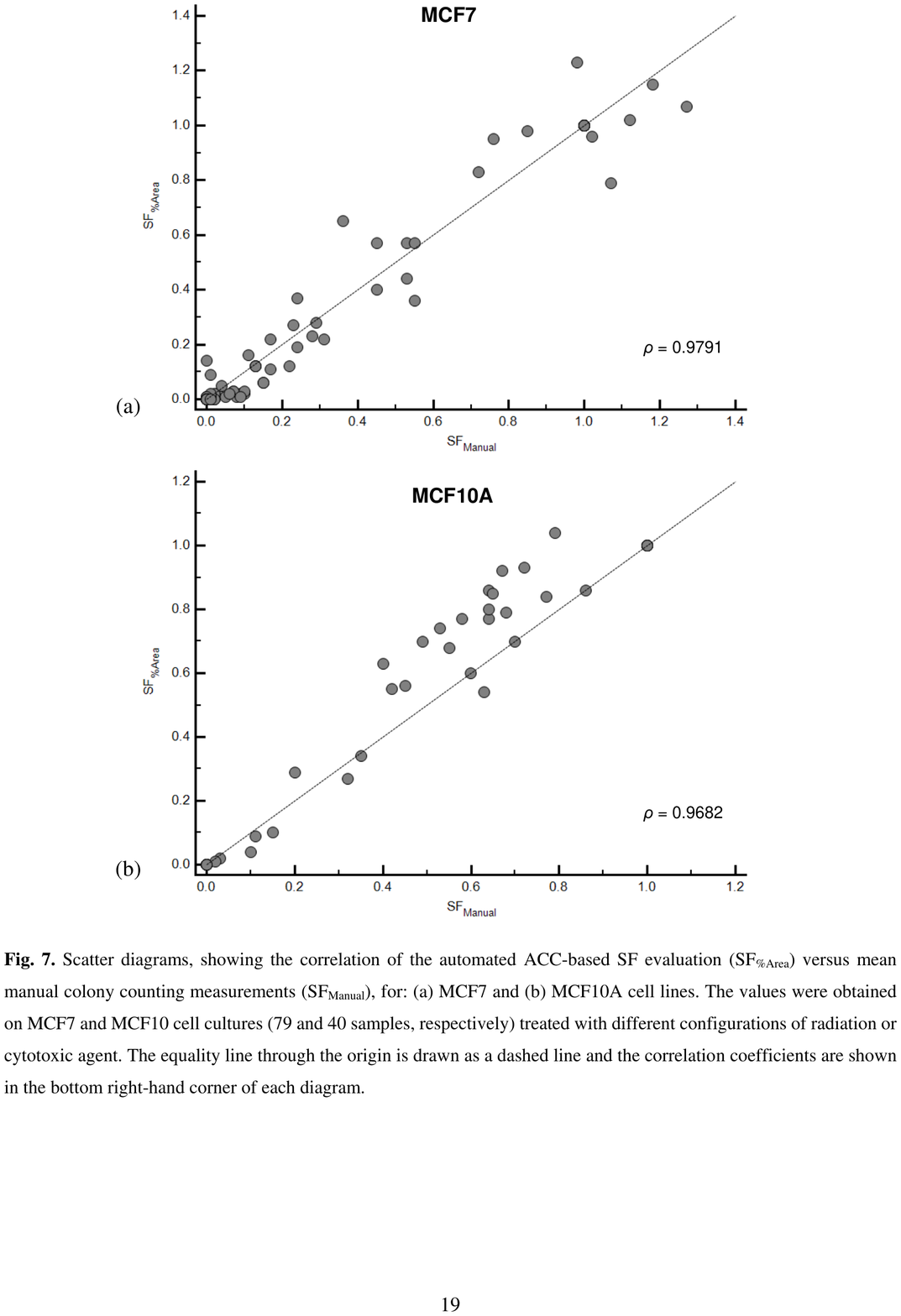}
	\caption[Scatter diagrams, showing the correlation of the automated ACC-based SF evaluation (SF\textsubscript{\%Area}) versus mean manual colony counting measurements (SF\textsubscript{Manual})]{Scatter diagrams, showing the correlation of the automated ACC-based SF evaluation (SF\textsubscript{\%Area}) versus mean manual colony counting measurements (SF\textsubscript{Manual}), for: (a) MCF7 and (b) MCF10A cell lines. The values were obtained on MCF7 and MCF10 cell cultures ($79$ and $40$ samples, respectively) treated with different configurations of radiation or cytotoxic agent. The equality line through the origin is drawn as a dashed line and the correlation coefficients are shown in the bottom right-hand corner of each diagram.}
	\label{fig:ACC-scatterPlot}
\end{figure}

Fig. \ref{fig:ACC-scatterPlot} shows the scatter diagrams concerning the automated SF evaluation based on the percentage of ACC (SF\textsubscript{\%Area}) versus the manual colony counts (SF\textsubscript{Manual}) averaged over three measurements by three different biologists.
The achieved results ($\rho = 0.9791$ on MCF7 cells and $\rho = 0.9682$ on MCF10A cells obtained on $79$ and $40$ samples (i.e., different wells), respectively) showed a high level of correlation between our fully automatic method and those conventionally obtained with manual cell colony counting.
The achieved high Pearson's coefficient $\rho$ values prove the accuracy of the proposed approach, so carrying out a validation against the conventional counting procedure by three experienced biologists.
More details about the Pearson's linear correlation coefficient calculation are depicted in Table \ref{table:ACC-Pearson}.

\begin{table}[!t]
\centering
	\caption[Pearson's linear correlation coefficient calculated between the automatic ACC-based SF values (SF\textsubscript{\%Area}) and the conventional colony counting measurements (SF\textsubscript{Manual})]{Pearson's linear correlation coefficient calculated between the automatic ACC-based SF values (SF\textsubscript{\%Area}) and the conventional colony counting measurements (SF\textsubscript{Manual}) on the analyzed MCF7 and MCF10A cell lines ($79$ and $40$ samples, respectively). The corresponding significance level ($p$-value) and $95\%$ confidence interval are also reported.}
	\label{table:ACC-Pearson}
	\begin{scriptsize}
		\begin{tabular}{ccccc}
			\hline\hline
			Cell culture	& Number of samples	& Pearson's correlation coefficient	& Significance level ($p$-value)	& $95\%$ confidence interval \\
			\hline
			MCF7 &	$79$ &	$\rho = 0.9791$ &	$p < 0.0001$ &	$0.9674$ to $0.9866$ \\
		    MCF10A &	$40$ &	$\rho = 0.9682$ &	$p < 0.0001$ &	$0.9403$ to $0.9832$ \\
			\hline\hline
		\end{tabular}
	\end{scriptsize}
\end{table}

A strong positive correlation between the SF computed automatically by the proposed method and the conventional manual SF measurements was observed for the two cell lines analyzed in this study.
Automated SF measurements are highly consistent with the corresponding manual evaluations for the MCF7 cell samples, by denoting also a unity slope, as shown in Fig. \ref{fig:ACC-scatterPlot}a.
In the case of MCF10A cell line (Fig. \ref{fig:ACC-scatterPlot}b), the actual regression straight-line reveals a positive offset in the automated SF measurement based on ACC alone.
As a matter of fact, the SFs calculated by the proposed method slightly over-estimated the manual SF measurements in approximately $45\%$ of cases.
This experimental finding could depend on the colony morphology and size: the conventional cell counting procedure for the SF evaluation is based on the colony number alone and does not consider the colony area. Especially, correct SF measurements could be affected in cell lines with high growth rate, such as MCF10A.
As an example, two different wells with the same cell colony number but with different area coverage will have the same SF\textsubscript{Manual} value.
On the contrary, the proposed SF evaluation methodology based on the percentage of ACC can appropriately deal with these situations, by discriminating both small and large cell colonies using the local adaptive thresholding algorithm, and yields coherent SF\textsubscript{\%Area} measurements.
In conclusion, according to the findings in \cite{freshney2015}, the conventional counting procedure underestimated the effect of the treatment in highly proliferative cell lines, because the therapy could not reduce the colony number by influencing the colony size alone.

\subparagraph{Discussion}
As reported in the literature, cells after any type of treatment have a reduction in growth rate, and the colonies are generally smaller than the colonies of the untreated control \cite{harada2012,maycotte2014}.
The conventional counting procedure, based on colony number alone, can underestimate the effect of treatment, because sometimes it is possible that the therapy does not reduce the colony number but decreases just the colony size \cite{freshney2015}.
Several authors highlighted the problems associated with detecting colonies near the rim of a culture well/flask.
Possible solutions can be to exclude the rim area from processing or to use different segmentation methods for central and rim zones of the well \cite{chiang2015}.
In addition, when cell colonies present adherence in areas near the rim, discarding the rim area could strongly affect the SF evaluation, so worsening result accuracy.
Many literature approaches \cite{chiang2015,dahle2004,geissmann2013,guzman2014} use algorithms based on global thresholding techniques.
Otsu’s method \cite{otsu1975} alone is not able to process the well/flask edges appropriately nor is suitable for less-discrete colonies. The approaches proposed in \cite{chiang2015,clarke2010,dahle2004} were tested on small cell colonies, such as bacterial ones that are generally isolated with a circular shape and adjacent colonies rarely merging together.
Conversely, human cells colonies often present an irregular shape and a greater size giving rise to large overlapping colonies. These issues have also proved to be difficult to address \cite{thielmann1985}.
More sophisticated approaches (e.g., watershed and distance transforms \cite{roerdink2000} were also employed to overcome the problems regarding less-discrete or merging colonies with fuzzy boundaries by combining multiple counting measured, thus resulting in total colony area based statistical counts \cite{mukherjee1995} with higher computational cost.
Overlapping and heterogeneous colonies should be processed properly and effectively, without requiring statistical correction or estimation.

For these reasons, the proposed approach takes into account the actual size of each colony, by using a method that measures the ACC.
In this way, the effect of the treatment on cell growth rate is also considered.
Although estimating manually the ACC is unfeasible, this measurement conveys valuable information on the effects of the treatment.
By considering the percentage of ACC, the number of colonies is not mandatory to evaluate the SF.
The correct identification of the area wherein the cells were plated (e.g., well, Petri dish, flask) is crucial, because it can affect all the subsequent processing steps. Some literature approaches use binary masks with fixed position \cite{bernard2001,dahle2004,guzman2014} or require manual alignment performed by the operator \cite{guzman2014}.
Often user intervention is needed also in the following steps.
In \cite{dahle2004} the operator interacts for threshold selection, while in \cite{barber2001} two different methods were implemented and the choice is made by the operator according to the cell type (colonies presented a low or a high contrast with respect to the background).
To reduce operator dependence and possible errors in the processing pipeline (e.g., due to a misalignment between the well mask and the plate when fixed-positions/fixed-masks are used), we propose a fully automated approach.
First of all, we employed a robust strategy, based on the CHT, for well detection.
A control system was implemented \textit{ad hoc} to enhance the CHT technique when errors occurs, making it possible to detect all the wells properly in $100\%$ of the cases.
Successively, to detect cell colonies, the extracted wells are processed by using the local adaptive thresholding algorithm.

Finally, the need for dedicated hardware for the images acquisition is not to be underestimated. In \cite{barber2001,bernard2001,chiang2015}, specific hardware is required (i.e., CCD cameras, lenses, light diffuser, or positioning devices) that can be also expensive.
We exploited a conventional flatbed scanner to acquire multi-well plate images.

\subparagraph{Conclusion}

Nowadays, clonogenic cell survival assays are evaluated by means of manual procedures, which could lead to inconsistencies related to operator dependence, so affecting the result repeatability. 
Unlike other literature approaches, the proposed method aims to address the issues that are yet open in the evaluation of the clonogenic assay (i.e., subjectivity in cell colony counting, need for dedicated hardware for acquire the analyzed plate images, manual well delineation in multi-well plates), by proposing a fully automatic approach that attempts to overcome the limitations of the related works.

The proposed approach represents an alternative to the problem of clonogenic assays, presenting some novelties not yet considered in literature:
\begin{itemize}
    \item fully automated well detection based on the CHT (properly enhanced with a control system to handle possible wrong detections) to recognize the wells inside the plate image (unlike several literature approaches that require plate dimensions and the spatial arrangement of the wells);
    \item differently to some related works, which use special cameras or hardware, our approach does not require any specific acquisition device;
    \item only a pre-processing step is applied, performing a conversion in the CIE \textit{L*u*v*} color model to reduce luminance inhomogenities with respect to classical grayscale or RGB color space;
    \item colony segmentation exploiting adaptive thresholding to extract cell colonies from the background, which makes it effective for both small and large colonies;
    \item no post-processing is needed (for instance, morphological refinement \cite{breen2000,soille2013} can heavily affect ACC calculation, especially when performed on small cell colonies);
    \item SF evaluation based on the percentage of ACC instead that on the colony number (not applicable when the high growth rate leads to adjacent colony merging), which allows to establish a correlation between the growth rate and the dose administered during the treatment.
\end{itemize}

Comparisons on the SF calculated using the percentage of ACC against the conventional counting, averaged over three measurements by three expert operators, were performed for each experiment.
The achieved results and the correlation between automatic and manual SF, demonstrate the reliability (in terms of robust identification of the wells in different plate types as well as correct colony extraction in various operative conditions) and accuracy (in terms of high correlation degree between the SFs measured by the human operators and those calculated automatically) of our approach.
The proposed methodology to calculate the SF enables to overcome some of the most challenging open issues in clonogenic assay evaluation.
In our opinion, this approach represents a valuable and effective solution to be integrated in an expert system to support biologists in their own decision making tasks.
Future work is aimed at extending and improving the proposed approach on other cell lines, which specifically show a lower violet color uptake during colony fixation/staining phase (e.g., invasive breast cancer, such as MDA cells).
A more sensitive classification approach could be realized by extracting appropriate textural feature descriptors, such as the Haralick's features \cite{haralick1979,haralick1973}.

\subsection{Region splitting-and-merging}
\label{sec:SplitMerge}
The split-and-merge algorithm \cite{horowitz1976} is a direct region detection approach that works in a hybrid fashion: the initial image can be alternatively subdivided into a set of disjoint regions by splitting (top-down procedure) and/or merging (bottom-up procedure) them by attempting to match the segmentation conditions in terms of the homogeneity criteria expressed by a Boolean predicate $\mathsf{P}$ \cite{faruquzzaman2008}.
Therefore, the split-and-merge algorithm starts with an arbitrary partition $\mathcall{I}$ (i.e., entire image region) and yields an output composed of uniform sub-regions: $\mathcall{R}_i$, with $i=1,2, \ldots, R$.
At the $t$-th step, a generic region $\mathcall{R}_i$ is split if uniformity measures are not matched ($\mathsf{P}(\mathcall{R}_i) = \mathsf{false}$). Therefore, the initial square image region $\mathcall{I}$ (whose dimensions are a power of $2$) can be recursively divided into four smaller quadrants (quad-regions).

In this context, the most critical choice is the logical predicate $\mathsf{P}$, as it is fundamental to get satisfactory outputs.
For instance, homogeneity criteria can defined in terms of statistical measures computed in the ROI (e.g., mean value or standard deviation).
In order to control the fibroid segmentation accuracy, the parameter $\rho_\text{min}$ sets the minimum quad-region size beyond which no further splitting is carried out.
This parameter must be a trade-off between precision and computational effort.In our case, The best results are obtained with $\rho_\text{min} = 4 \times 4$ pixels, because even small regions can be detected.

\subsubsection{Uterine fibroid detection on MR images}

The aim of the overall approach is to automatically segment the ablated fibroid ROT that are within the uterus, in order to optimize the current methodology for MRgFUS treatments.

\begin{figure}
	\centering
	\includegraphics[width=0.4\linewidth]{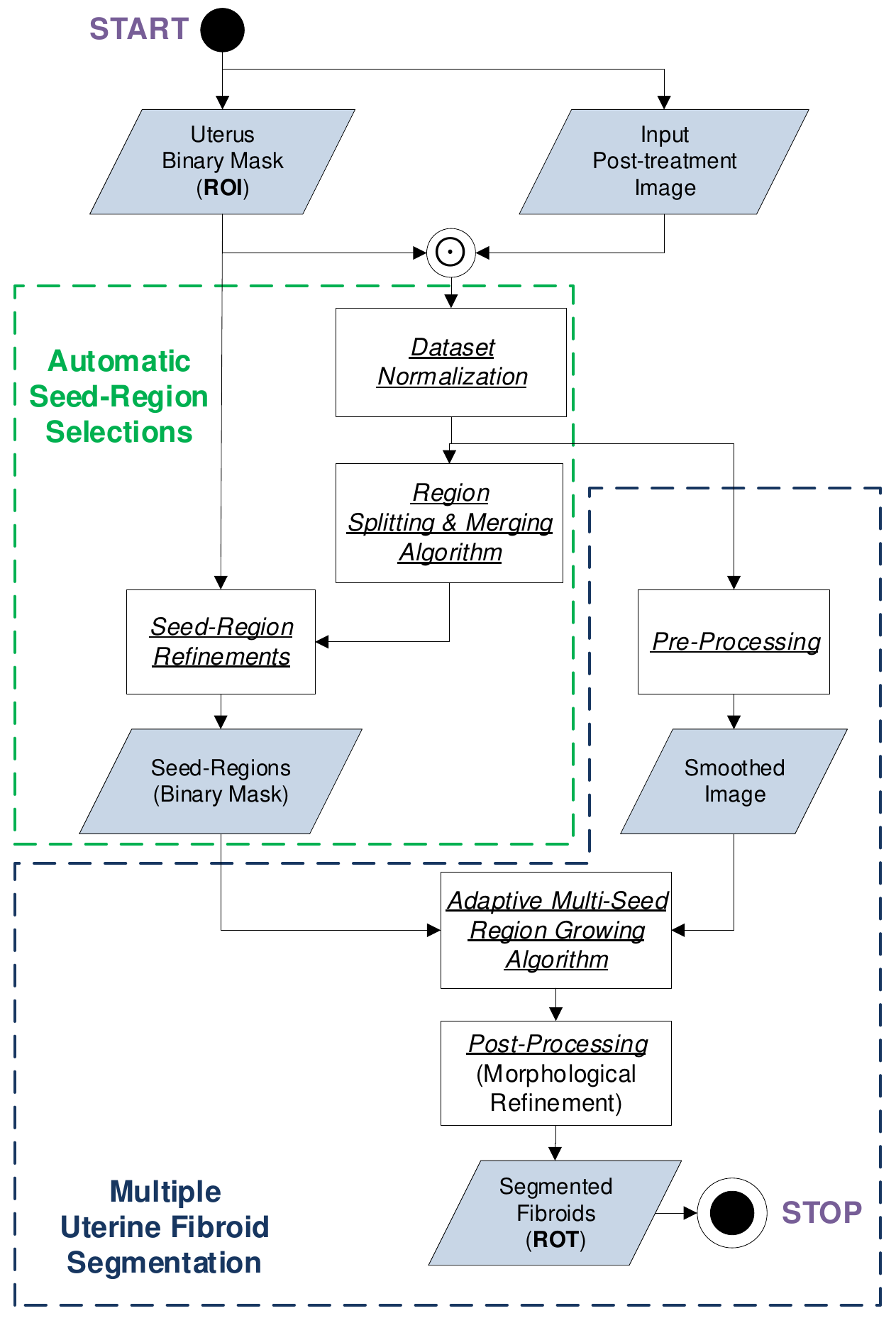}
	\caption[Flow diagram of the SM \& RG method]{Flow diagram of the SM \& RG method, divided into two phases: (\textit{i}) automatic seed-region selections; (\textit{ii}) multiple uterine fibroid segmentation.}
	\label{fig:SMRG-FlowDiagram}
\end{figure}

Our approach combines region splitting-and-merging and multi-seeded region growing that relies on two phases: (\textit{i}) automatic seed-region selections and (\textit{ii}) uterine fibroid segmentation (Fig. \ref{fig:SMRG-FlowDiagram}).
By taking inspiration from \cite{saad2010,saad2012}, where a method for brain lesion segmentation in DWI is described, ROT detection is accomplished through a split-and-merge algorithm \cite{horowitz1976} in order to determine meaningful seeds.
Moreover, the split-and-merge output obtained with our approach, after some refinement operations, is directly exploited as a partial result for the subsequent region growing step by implementing an incremental procedure.
Region growing can segment a single area at a time, since a seed must be chosen for each fibroid, in order to start the inclusion step based on homogeneity measures.
For these reasons, a seed selection method based purely on intensity levels (i.e., histogram inspection) may not be completely reliable and sufficient to detect ROT in a pathological scenario with multiple fibroids, where the number of lesions is not known \textit{a priori}.
Therefore, a region analysis method is required to correctly detect the fibroids within the uterus ROI.
The split-and-merge algorithm represents a weel suited solution \cite{manousakas1998}, because it is able to find homogeneous regions in terms of uniformity criteria.
Seeds are then exactly calculated by excluding connected-components located near the uterus ROI boundaries.
Then, starting from the seed-regions selected by the split-and-merge approach, segmentation is achieved through a region growing procedure guided by appropriate similarity properties that describe ROT intensity features.
Finally, some post-processing operations are used for refining the segmented fibroids.

In order to identify the ROT, homogeneity criteria regarding the growing-region $\mathcall{R}_G$ are defined in terms of its mean value $\mu_{\mathcall{R}_G}$.
By tuning different parameters, we found the following values:

\begin{equation}
    \label{eq:predSM}
    \mathsf{P} := \begin{cases}
       \mathsf{true}, & \text{if } (\mu_{\mathcall{R}_G} > 0.0) \wedge (\mu_{\mathcall{R}_G} < 0.58) \\
       \mathsf{false}, & \text{otherwise}
     \end{cases}.
\end{equation}

However, many splits are inevitably performed at the uterus boundaries due to the strong intensity difference between ROI edge pixels and the background, which involves non-compliance with the homogeneity criteria defined by the predicate  $\mathsf{P}$.
Thus, these connected-components are removed before seed-region selection.
First, loose connections between the splitting-and-merging output and ROI edges are removed by morphological opening (more conservative than erosion) with a diamond-shaped structuring element, to easily disconnect also diagonally connected pixels among $8$-neighborhood, which are very frequent in ROI boundaries.
Diamond-shaped structuring elements yield satisfying results even with vertical/horizontal edges and are preferred to squared or circular filters when object boundaries are roughly in the diagonal direction \cite{soille2001,sun2006}.

The remaining parts at ROI edges must be cut out before seed selections through a connected-component labeling procedure using a flood-fill algorithm \cite{diStefano1999}.
In fact, we calculate the pixel intersection (logical product) between each connected-component and the dilated ROI edge image, using a squared structuring element.
Constraints on the cardinality of this intersection are defined according to the different granularity of the various connected-region areas, in order to consider the occurrence of large fibroids situated at the ROI boundary.
Hence, the current analyzed connected-component is included into the output binary mask only if the control is matched, otherwise it will be removed.
The splitting-and-merging output, besides being used for ROT detection, is thus exploited as an initial result for the subsequent fibroid segmentation through the region growing algorithm, making it possible to considerably optimize execution time.

\subsection{Region growing}
\label{sec:RegionGrowing}
Region growing is a bottom-up segmentation method that produces a homogeneous region by successively merging primitive regions (sub-regions or also single pixels) \cite{gonzalez2002}.
Therefore, the region growing algorithm starts from seed values and attempts to find a local connected-region depending on given homogeneity criteria \cite{saad2012}.
Since the traditional region growing approach is able to perform segmentation of a single area only, when the currently examined patient can be affected by more than one fibroid, it is necessary to implement a multi-seeded region growing segmentation. As mentioned before, in a basic seeded region growing algorithm, each region begins its own growth from a seed.
Therefore, the initial set of seeds $\mathcall{S} = \left\{ s_1, s_2, \ldots, s_n \right\}$  must be a representative and expressive sample of the region to segment.
Since seed choice affects overall segmentation results, this task is fundamental and very critical \cite{adams1994}.
Another issue is the stopping rule formalization.
In fact, the region growing procedure must be interrupted if no more pixels match the membership criterion \cite{gonzalez2002}.
The inclusion in a region can be stated in terms of a segmentation threshold $\Theta$, which specifies the maximum allowed deviation of pixel features within that region.
Therefore, segmentation threshold $\Theta$ is a region-dependent parameter that guides the homogeneity decision test \cite{chang1994}.
These criteria can be local (i.e., pixel intensity \cite{militello2013}) or region descriptor-based (region shape or size, likeness between a candidate pixel and the current grown region in terms of image features \cite{gambino2010}) that are more powerful because they take into account the “history” of region growth when some \textit{a priori} knowledge about the region is available.
As explained in \cite{adams1994}, to prevent a poor starting estimate, it is recommended that seed-regions be used instead of single pixels (seed-points).
In fact, when dealing with noisy image segmentation, each seed-area should be sufficiently large to ensure a stable estimate of its region’s mean. Moreover, these seed-regions are used as input for the region growing incremental process, exploiting the previous split-and-merge segmentation results.

\subsubsection{Uterine fibroid segmentation on MR images}
The algorithm proceeds by considering one seed-region at a time, which identifies a single region growth.
The current region $\mathcall{R}_G$ is iteratively grown by analyzing all unallocated pixels belonging to the boundary list $\mathcall{B}$, obtained by considering $8$-neighborhoods as candidate pixels.
As stated in \cite{militello2013}, region growing is ruled by three conditions concerning the examined pixel $\mathbf{p}_\mathcall{B} \equiv (\bar{x}, \bar{y})$:
\begin{itemize}
    \item $\mathbf{p}_\mathcall{B}$ is inside the image $\mathcall{I}$ (with size $M \times N$ pixels);
    \item $\mathbf{p}_\mathcall{B}$ has not already been included in $\mathcall{R}_G$;
    \item $\mathbf{p}_\mathcall{B}$ value satisfies similarity criteria based on the segmentation threshold $\Theta$ in terms of the region mean intensity, i.e., $\mu_{\mathcall{R}_G} \gets \frac{\sum_{(x,y) \in \mathcall{R}_G} f(x,y)}{\left| \mathcall{R}_G \right|}$.
\end{itemize}
These constraints can be more formally expressed by the following membership predicate:

\begin{equation}
    \label{eq:predRG}
    \mathsf{P} := \begin{cases}
       \mathsf{true}, & \text{if } (0 \leq \bar{x} \leq M-1) \wedge (0 \leq \bar{y} \leq N-1) \wedge \left(\mathbf{p}_\mathcall{B} \notin \mathcall{R}_G) \right) \wedge \left( \left| f(\bar{x}, \bar{y} - \mu_{\mathcall{R}_G}) \right| < \theta \right) \\
       \mathsf{false}, & \text{otherwise}
     \end{cases}.
\end{equation}
 
This similarity criterion is thus defined in terms of absolute distance from the region mean intensity $\mu_{\mathcall{R}_G}$, which is the most representative property in images with speckle noise, such as MRI \cite{ferrari2013,pizurica2006}.
Moreover, the seed-regions $\mathcall{S}$ accurately estimate mean intensity for each growth region $\mathcall{R}_G$.
The value of $\mu_{\mathcall{R}_G}$, the region mean intensity, is also exploited for the automatic and unsupervised estimation of the segmentation threshold $\Theta$, because the manual selection of a single threshold, suitable for all datasets, could be time-consuming and error-prone.
In fact, since normalized images are characterized by a bimodal histogram, an optimal threshold $\theta_\text{opt}$ can be selected through the Otsu's method \cite{otsu1975} that attempts to maximize the inter-class variance between two regions with different intensity values.
The adaptive threshold $\Theta_\text{adapt}^{(t)}$ is defined as:

\begin{equation}
    \label{eq:adaptT}
    \Theta_\text{adapt}^{(t)} = \theta_\text{opt} - \mu_{\mathcall{R}_G}^{(t)},
\end{equation}

\noindent and it is updated during each iteration $t$, by estimating the corresponding average value $\mu_{\mathcall{R}_G}^{(t)}$ while the region grows.
In this way, a different stopping rule is defined for each growth region $\mathcall{R}_G$ by calculating an adaptive segmentation threshold $\Theta_\text{adapt}$ for each fibroid in multi-fibroid pathological contexts.

\begin{figure}
	\centering
	\includegraphics[width=0.5\linewidth]{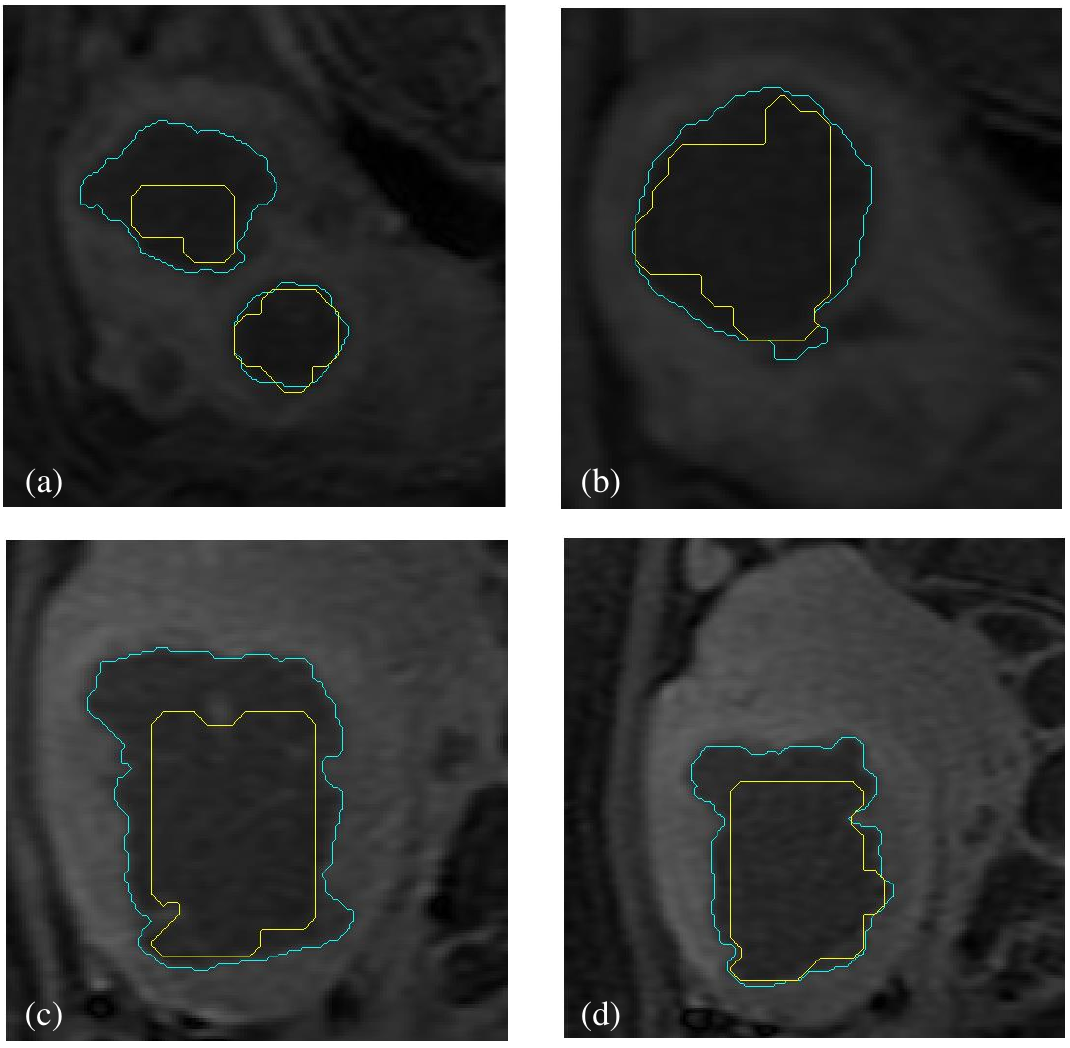}
	\caption[Uterine fibroid segmentation examples achieved by the SM \& RG incremental procedure]{Uterine fibroid segmentation examples achieved by the SM \& RG incremental procedure: (a) patient with multiple fibroid; (b, c, d) patients with large fibroids. Yellow contours represent Phase (\textit{i}) outputs, while cyan contours are the final region growing results (Phase (\textit{ii})).}
	\label{fig:SMRG-incrRes}
\end{figure}

In the initialization step, morphological erosion on the initial seed binary masks $\mathcall{B}$ is necessary to avoid that boundary points of seed-regions are eventually outside the ROT, resulting in leaking during the region growing process.
Furthermore, seed-points $\delta_j$ are determined by sampling $1$ every $5$ edge pixels from the array $\mathcal{D}$ to prevent the overlap between the $8$-neighborhoods of adjacent edge pixels.
Fig. \ref{fig:SMRG-incrRes} shows some examples of the incremental process results.
It is visible as most of the ROT has already been detected and successfully segmented by the split-and-merge algorithm.
The use of seeded region growing, described in \cite{chang1994}, requires seed-points for both fibroids and the rest of the uterus.
In fact, the entire image is divided into a set of disjoined homogeneous regions, achieving a tessellation similar to the watershed algorithm \cite{sijbers1997}.
This mechanism is more computationally expensive in terms of homogeneity tests (uterus region growth in addition to fibroid segmentation) and the need of a selection policy when a given pixel would be assigned simultaneously to several regions during the growth process.
Moreover, small dark areas (due to acquisition artifacts) could be present in post-treatment images, invalidating the overall segmentation process.

\paragraph{Results}

To quantify the efficiency of the proposed method in terms of execution time, the (sequential) implementation of the proposed incremental region growing approach is compared with a classic multi-seed region growing implementation (where a single seed-point is defined for each fibroid by calculating centroids of detected regions in Phase (\textit{i})).
Our method was developed and tested under MatLab, on a PC with a 2.8 GHz Intel\textsuperscript{\textregistered} Core\textsuperscript{TM} i5 processor and 8 GB RAM.
In particular, this incremental procedure reduces the segmentation time with a $1.5 \times$ speed-up factor, when compared to the traditional multi-seeded region growing approach (developed with a single seed-point selection for each fibroid).

\paragraph{MRI dataset composition}

\begin{figure}
	\centering
	\includegraphics[width=0.8\linewidth]{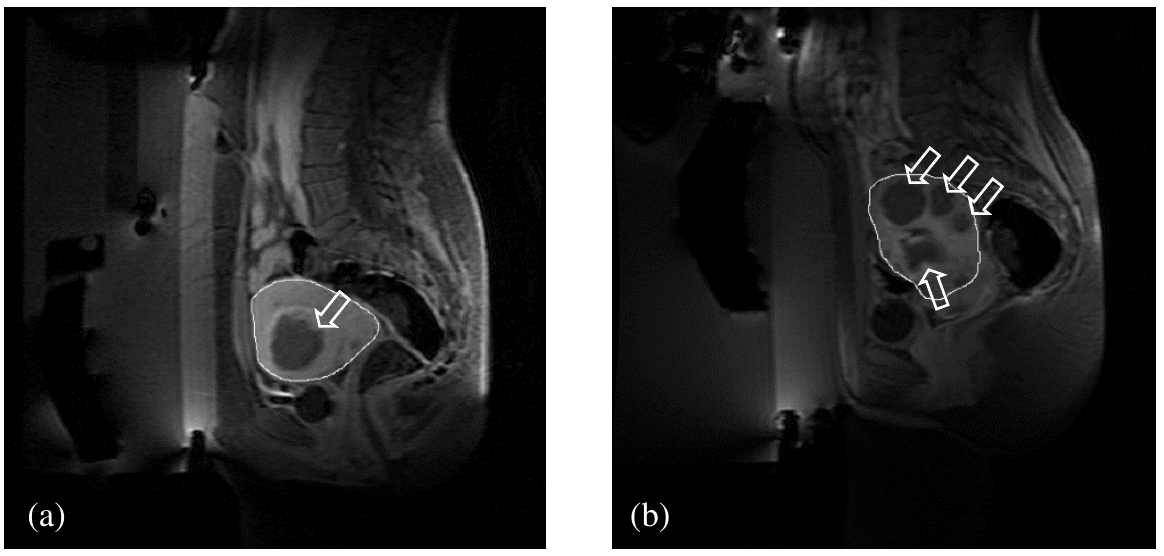}
	\caption[Examples of input uterine fibroid T1w MR sagittal slices]{Examples of input uterine fibroid T1w “FSPGR+FS+C” MR sagittal slices: (a) large fibroid; (b) multi-fibroid pathological scenario. The uterus ROI is delineated (white contour) and uterine fibroid ROTs are indicated with white arrows.}
	\label{fig:SMRG-inputImages}
\end{figure}

The study was carried out on an MRI dataset composed of 18 patients (average age: $42.22 \pm 6.29$ years) affected by symptomatic uterine fibroids underwent MRgFUS therapy.
The total number of the examined fibroids was $27$, since some patients presented a pathological scenario with multiple fibroids.
The analyzed images were acquired using a Signa HDxt MRI $1.5$T scanner (General Electric Medical Systems, Milwaukee, WI, USA) at two different institutions.
Segmentation tests were performed on MR images acquired after the MRgFUS treatment, executed with the ExAblate 2100 (Insightec Ltd., Carmel, Israel) HIFU equipment.
The considered MR slices were scanned using the T1w “Fast Spoiled Gradient Echo + Fat Suppression + Contrast mean” (“FSPGR+FS+C”) protocol.
This MRI protocol is usually employed for NPV assessment \cite{masciocchi2017}, since ablated fibroids appear as hypo-intense areas due to low perfusion of the contrast mean.
Sagittal MRI sections were processed, in compliance with current clinical practice for therapy response assessment. MRI data are encoded in the $16$-bit DICOM format.
Table \ref{table:UF-MRIcharacteristics} depicts the relevant MRI acquisition parameters, by reporting the range when different values were used.
Two instances of input uterine MR images are shown in Fig. \ref{fig:SMRG-inputImages}, where the white contour and the white arrows indicate the uterus ROI and the fibroid ROTs, respectively.

\begin{table}[t]
\centering
	\caption[Acquisition parameters of the MRI uterine fibroid dataset]{Acquisition parameters of the MRI uterine fibroid dataset.}
	\label{table:UF-MRIcharacteristics}
	\begin{tiny}
		\begin{tabular}{cccccccc}
			\hline\hline
			MRI sequence	& TR [ms]	& TE [ms]	& Matrix size [pixels]	& Slice spacing	[mm]	& Slice thickness [mm]	& Pixel spacing [mm] & Slices' number \\
			\hline
			T1w FSPGR+FS+C	& $150$-$260$	& $1.392$-$1.544$		& $512 \times 512$ & $0.6641$-$0.7031	$	& $5.0$ 	& $6.0$ &  $16$-$29$ \\
			\hline\hline
		\end{tabular}
	\end{tiny}
\end{table}

\begin{figure}[t]
	\centering
	\includegraphics[width=0.6\linewidth]{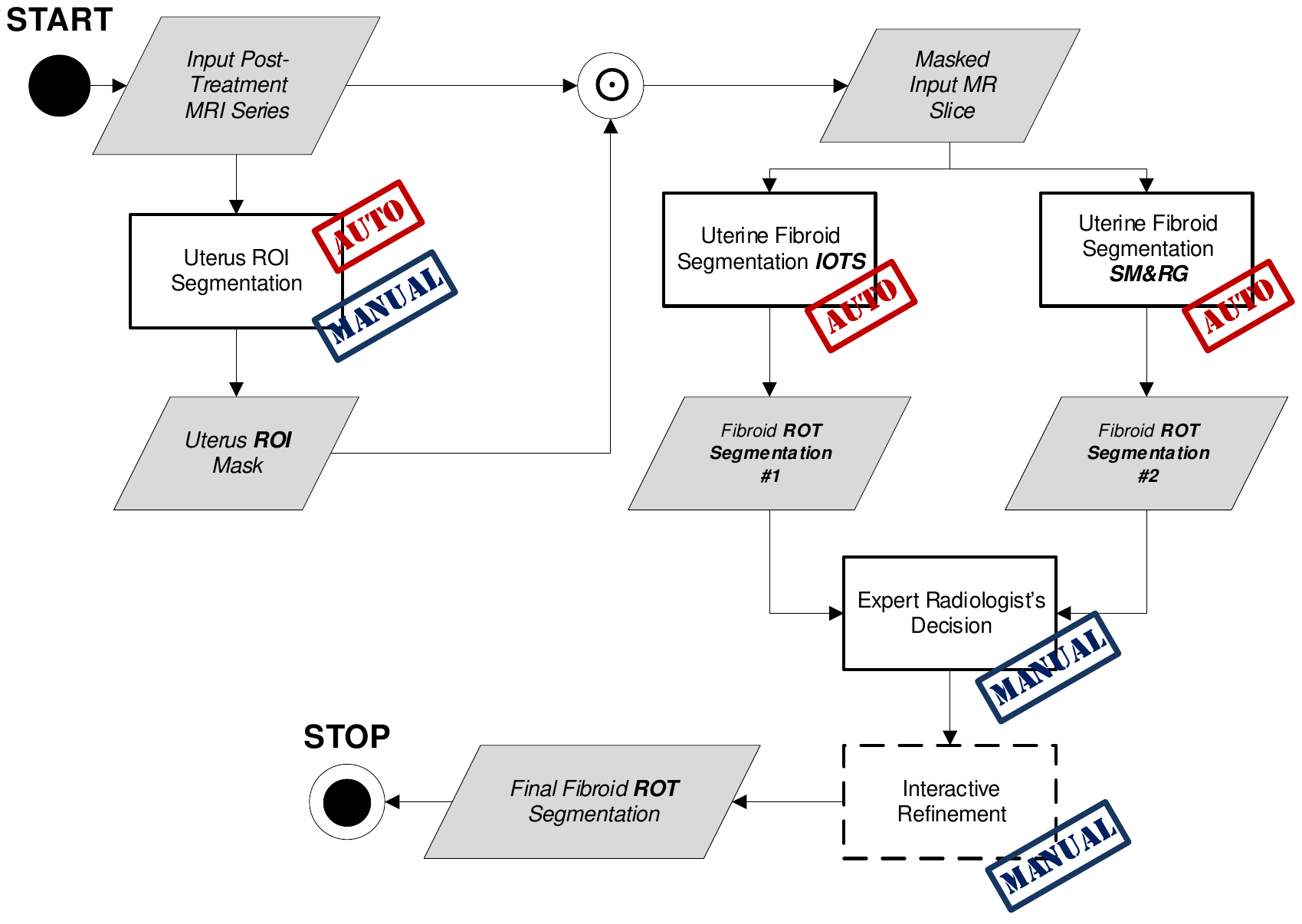}
	\caption[Workflow of uterine fibroid MRgFUS treatment response evaluation]{Workflow of uterine fibroid MRgFUS treatment response evaluation. The whole process can be divided into two main stages: (\textit{i}) uterus ROI segmentation, performed either by an automated approach (for instance, the FCM-based approach in \cite{militelloCBM2015}) or manually by an expert physician; (\textit{ii}) uterine fibroid ROT segmentation, using the proposed approaches based either on the Iterative Optimal Threshold Selection (IOTS) algorithm \cite{militelloCBM2015} or on combined Split-and-Merge and Region Growing (SM \& RG) \cite{rundoMBEC2016}. All the steps involving the radiologist's intervention are also reported, by labeling automatic or manual operations. Solid line boxes are mandatory processes, while dashed line blocks represent interactive steps that are not always required.}
	\label{fig:UF-workflow}
\end{figure}

\paragraph{Quantitative evaluation and clinical feasibility analysis}
We evaluated and compared two validated computer-assisted segmentation methods, which we have already presented in \cite{militelloCBM2017} and \cite{rundoMBEC2016}, for uterine fibroid segmentation in MRgFUS treatments.
A quantitative comparison on segmentation accuracy and reliability, in terms of spatial overlap-based and distance-based metrics, was performed.
The clinical feasibility of these approaches was assessed from physicians' perspective, proposing an integrated solution that considers also the workflow in real environments.
Fig. \ref{fig:UF-workflow} sketches the overall workflow for uterine fibroid MRgFUS treatment response assessment on post-treatments MRI series.

\begin{figure}
	\centering
	\includegraphics[width=0.9\linewidth]{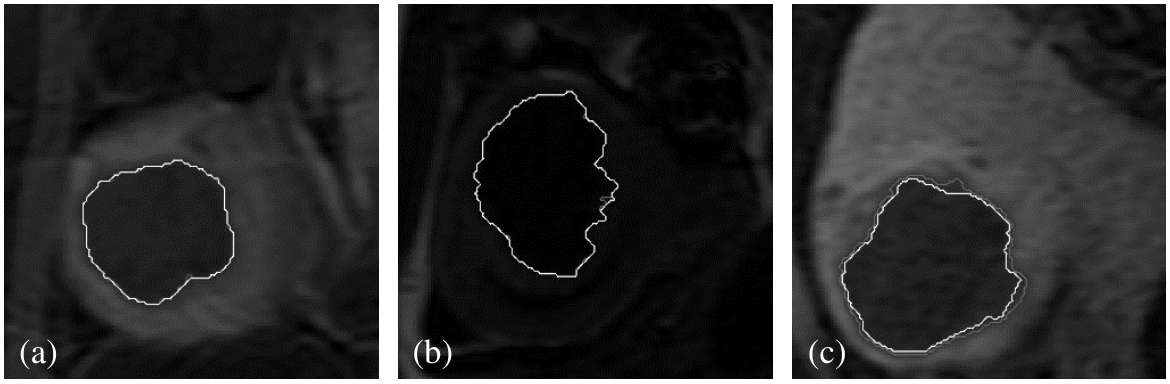}
	\caption[Uterine fibroid ROT segmentation results: IOTS versus SM \& RG]{Uterine fibroid ROT segmentation results: segmentation based on IOTS (gray contour) and segmentation using SM \& RG (white contour). ($8\times$ zoom factor).}
	\label{fig:UF-Results}
\end{figure}

The two methods presented in \cite{militelloCBM2015} and \cite{rundoCMPB2017} may be interchangeably used in the whole workflow defined in Fig. \ref{fig:UF-workflow}, representing two different options to effectively support radiologists in accurate NPV evaluation.
In both cases, the overall medical image analysis pipeline is unsupervised in all its components, enabling its integration in a comprehensive decision support system that provides clinically useful and quantifiable information (i.e., NPV measurement, ROT shape and location) for personalized medicine.
Fig. \ref{fig:UF-Results} shows the uterine fibroid ROT delineations, achieved by the IOTS and SM \& RG segmentation methods (over-imposed gray and white contours, respectively).

\begin{figure}
	\centering
	\includegraphics[width=\linewidth]{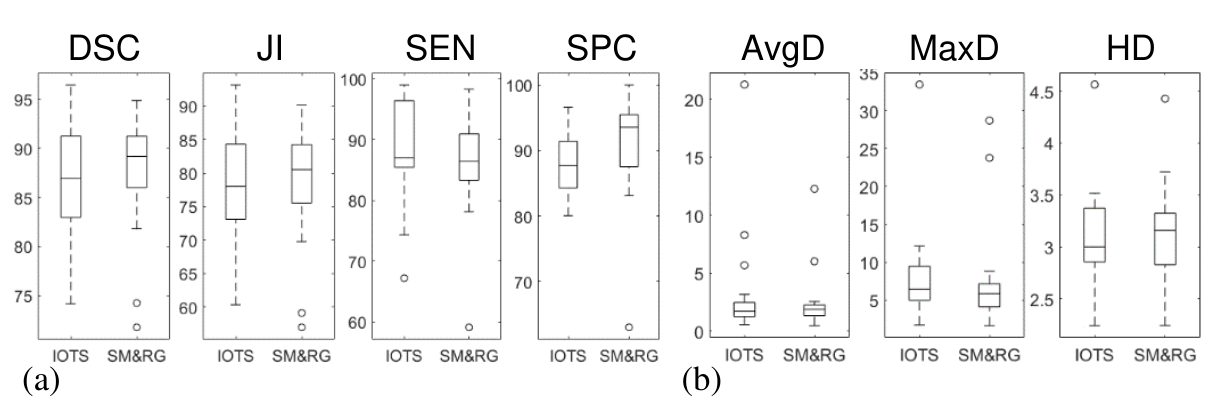}
	\caption[Boxplots of the uterine fibroid ROT segmentation results]{Boxplots of the uterine fibroid ROT segmentation results: (a) overlap-based metrics; (b) distance-based metrics.}
	\label{fig:UF-boxplots}
\end{figure}

The values of both overlap-based and distance-based metrics, achieved by the IOTS \cite{militelloCBM2015} and SM \& RG \cite{rundoCMPB2017} methods on the experimental dataset analyzed in this study, are shown in Table \ref{table:UF-Results}.
To provide a comprehensive graphical statistical summary of the different evaluation metrics, boxplots are illustrated in Fig. \ref{fig:UF-boxplots}.

\begin{table}[!t]
\centering
	\caption[Overlap-based and distance-based metrics achieved by the IOTS and SM \& RG methods on the complete uterine fibroid MRI dataset]{Values of overlap-based and distance-based metrics achieved by the IOTS and SM \& RG methods on the complete MRI dataset composed of $18$ patients undergone MRgFUS therapy for uterine fibroid ROT ablation. The results are expressed as average value $\pm$ standard deviation.}
	\label{table:UF-Results}
	\begin{scriptsize}
		\begin{tabular}{lccccccc}
			\hline\hline
			Method	& DSC	& JI	& SEN	& SPC	& AvgD	& MaxD & HD \\
			\hline
			IOTS &	$87.25 \pm 5.86$ & $78.75 \pm 8.74$ &	$88.16 \pm 8.48$ &	$87.88 \pm 4.77$ &	$3.308 \pm 4.863$	& $7.997 \pm 6.952$	& $3.100 \pm 0.502$ \\
			SM \& RG &	$87.47 \pm 6.29$ &	$78.57 \pm 9.30$ &	$86.07 \pm 8.45$ &	$90.86 \pm 8.23$ &	$2.498 \pm 2.702$	& $7.890 \pm 6.988$ &	$3.123 \pm 0.494$ \\
			\hline\hline
		\end{tabular}
	\end{scriptsize}
\end{table}

\emph{DSC} and \emph{JI} mean values are approximately equal, proving high segmentation accuracy in both cases.
Although SM \& RG results are characterized by slightly higher standard deviation, the interquartile range in boxplots is more concentrated and two outliers are observed because leaking (i.e., over-estimation of fibroid size) in the case of poor contrast images could occur.
\textit{SEN} values show that the IOTS-based global thresholding solution is often more sensitive than the SM \& RG approach, especially in little fibroid detected in initial or final ROT slices.
Indeed, the region splitting-and-merging procedure could not detect small regions with fuzzy boundaries.
However, as visible from the obtained \emph{SPC} values and distance-based metrics, the region growing algorithm generally delineates the ROT contours more precisely than the IOTS algorithm.

Our approaches have been already compared with similar literature works in the original publications \cite{militelloCBM2015,rundoCMPB2017}.
In both cases, the proposed methods remarkably outperformed the other implemented methods.
In addition, the clinical feasibility of many literature approaches is limited. For instance, LSF-based methods are highly sensitive to initial user input and require a labor-intensive parameter setting procedure by clinicians to obtain satisfactory segmentations. Differently to the method proposed in \cite{fallahi2011,fallahi2010}, where multispectral MRI sequences and external co-registration operations are mandatory, our methods require T1w contrast-enhanced MRI sequence alone and can be easily integrated in a single software tool.

\paragraph{Discussion and conclusions}

The evaluation measures achieved by these two methods, namely IOTS \cite{militelloCBM2015} and SM \& RG \cite{rundoMBEC2016} segmentation pipelines, were comparable in terms of delineation accuracy: the former showed higher sensitivity and the latter was characterized by higher specificity.
Hence, we can argue that the IOTS and SM \& RG uterine fibroid segmentation approaches have almost complementary properties, allowing for an integrated image processing pipeline by adapting on the different analyzed images.
According to Fig. \ref{fig:UF-workflow}, a possible clinically feasible solution could be described by the following use case that is easily implementable in the real workflow for uterine fibroid MRgFUS treatment response evaluation:
\begin{enumerate}
    \item the segmentation results obtained by the two different automatic approaches based on Pattern Recognition algorithms are proposed, by avoiding cognitive overload;
    \item the radiologist chooses the best result between the two proposed segmentations, also combining them appropriately in a late integration phase;
    \item the user can interactively refine the achieved segmentation by selecting only the correct ROTs (i.e., removing false positives and adding false negatives), as well as by moving the control points that define the computed ROT boundaries.
\end{enumerate}

A volumetric reconstruction of uterus ROI and ablated fibroid ROT could be also useful for radiologists during their own decision-making tasks.
Especially, in pathological scenarios with multiple fibroids.
However, tridimensional uterine fibroid MR image segmentation approaches are not easy to realize, since fibroids, particularly if they are pedunculated, can be displaced by presenting gaps also in adjacent slices \cite{yao2006}.
In such cases, leaking is likely to occur, especially in $3$D region growing implementations \cite{militello2015IJAIS}.

In conclusion, the resulting segmentation pipeline is an integrated solution that attempts to solve the clinical problems regarding MRgFUS treatment response evaluation for uterine fibroid, by improving significantly the current operative methodology in terms of segmentation accuracy, result repeatability, and execution time.
Generally, reliable segmentation results were observed, even when dealing with fibroid ROTs characterized by irregular or inhomogeneous necrotic material, so that operator de-pendency is reduced.
Future developments aimed at the realization of advanced image enhancement methods to improve the segmentation process for medical images characterized by bimodal histograms, such as the aproach presented in Section \ref{sec:MedGAenhancement}.

\section{Morphological approaches}
\label{sec:morphApproaches}

\subsection{Watershed transform}
\label{sec:watershedTransf}
The watershed transform \cite{soille1990} is one of the most used approaches in cell image segmentation \cite{meijering2012}, even though it was originally proposed in the field of mathematical morphology \cite{soille2013}.

The intuitive description of this transform is straightforward: assuming an image as a topographic relief, where the height of each point is directly related to its gray level, and considering rain gradually falling on the terrain, then the watersheds are the lines that separate the resulting catchment basins \cite{vincent1991}.
This technique is valuable because the watershed lines generally correspond to the most significant edges among the markers \cite{beucher1992} and is useful to separate overlapping objects.
Even when no strong edges between the markers exist, the watershed method is able to detect a contour in the area.
This contour is detected on the pixels with higher contrast \cite{soille1990}.
As a matter of fact, edge-based segmentation techniques---which strongly rely on local discontinuities in gray levels---often do not yield unbroken edge curves, thus heavily limiting their performance in cell segmentation \cite{wahlby2004}.
Unfortunately, it is also well-known that the watershed transform may be affected by over-segmentation issues, so requiring further processing \cite{grau2004}.

From a computational perspective, the watershed algorithm analyzes a gray-scale image by means of a flooding-based procedure.
Since the flooding process is performed either on a gradient image or edge map, the basins should emerge along the edges.
As a matter of fact, during the watershed process, the edge information allows for a better discrimination of the boundary pixels with respect to the original image.
Finally, only the markers of the resulting foreground cells are selected.

Efficient versions of the watershed algorithm exploit a priority queue to store the pixels according to the pixel value (i.e., the height in the gray-scale image landscape) and the entry order into the queue (giving precedence to the closest marker).
More specifically, during the flooding procedure, this process sorts the pixels in increasing order of their intensity value by relying on a breadth-first scan of the plateaus based on a first-in-first-out data structure \cite{vincent1991}.

\subsubsection{Automatic Cell Detection and Counting}

In \cite{rundo2018ACDC}, we presented Automated Cell Detection and Counting (ACDC), a novel pipeline designed for time-lapse microscopy activity detection of fluorescently labeled cell nuclei that is capable of overcoming the practical limitations of the literature methods, mainly related to the training phases or time constraints, to be considered as a feasible solution in laboratory practice.
The proposed fully automatic method is based on the watershed transform \cite{soille1990,vincent1991} and morphological filtering \cite{beucher1992,soille2013}.
Bilateral filtering \cite{tomasi1998} on the original image is formerly applied to smooth the input cell images while preserving edge sharpness.
ACDC efficiently yields reliable results just using few parameters, which may conveniently be altered by the biologist.
Thus, the resulting interpretable segmentation model makes the user sufficiently aware of the underlying automated analysis, unlike sophisticated solutions that do not provide any interactive control suitable for the end users.
The proposed computational approach was validated on two different cell imaging datasets in order to show its reliability in different experimental conditions.
ACDC can also leverage multi-core architectures and computer clusters to reduce the running time required to analyze a large single image stack.
\clearpage

\paragraph{Fluorescence microscopy imaging data}
\label{acdc:Materials}

\begin{figure}
\centering
	\subfloat[][]
    {\includegraphics[width=0.4\textwidth]{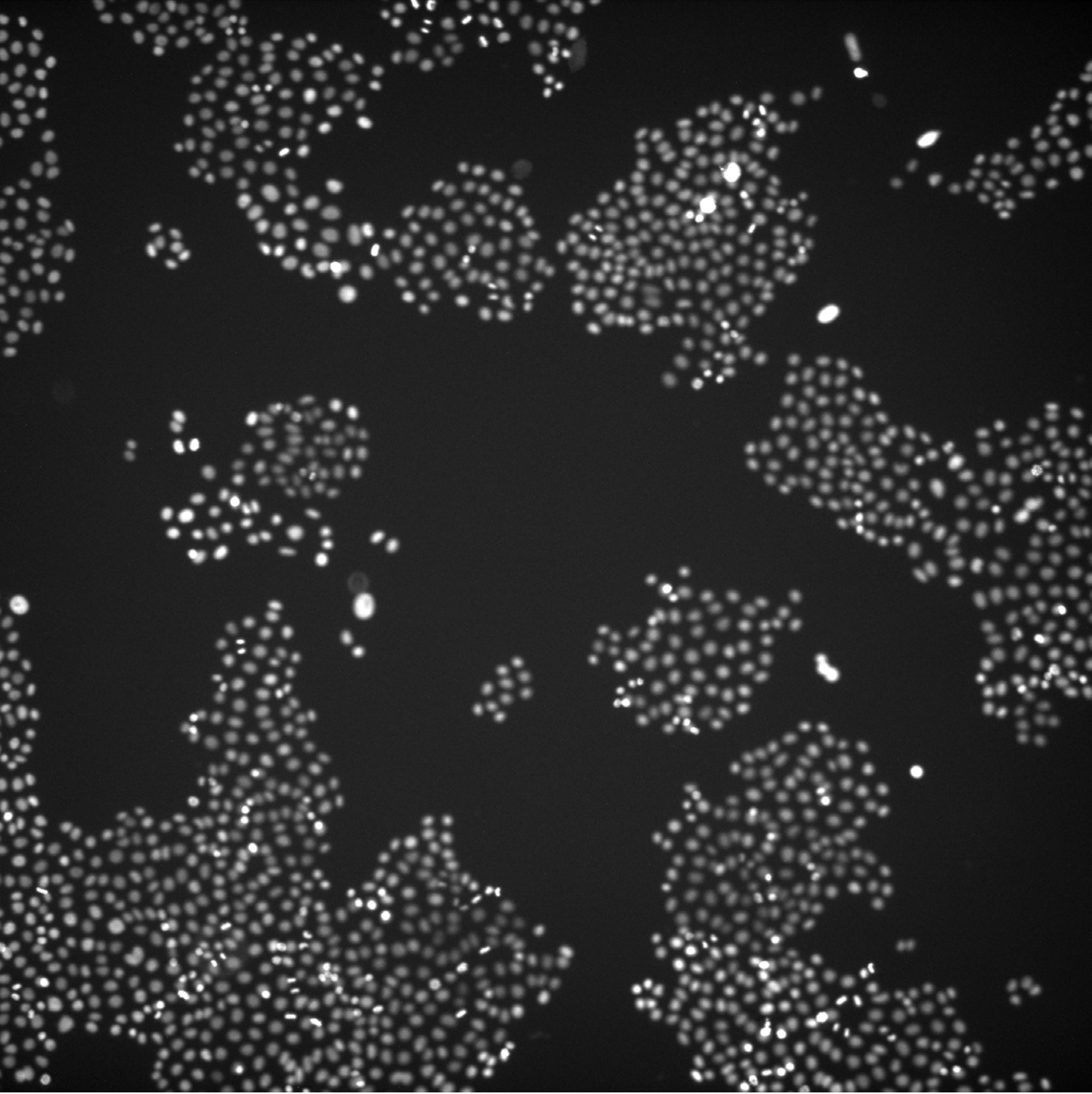}\label{sfig:inputVU1}} \quad
    \subfloat[][]
    {\includegraphics[width=0.4\textwidth]{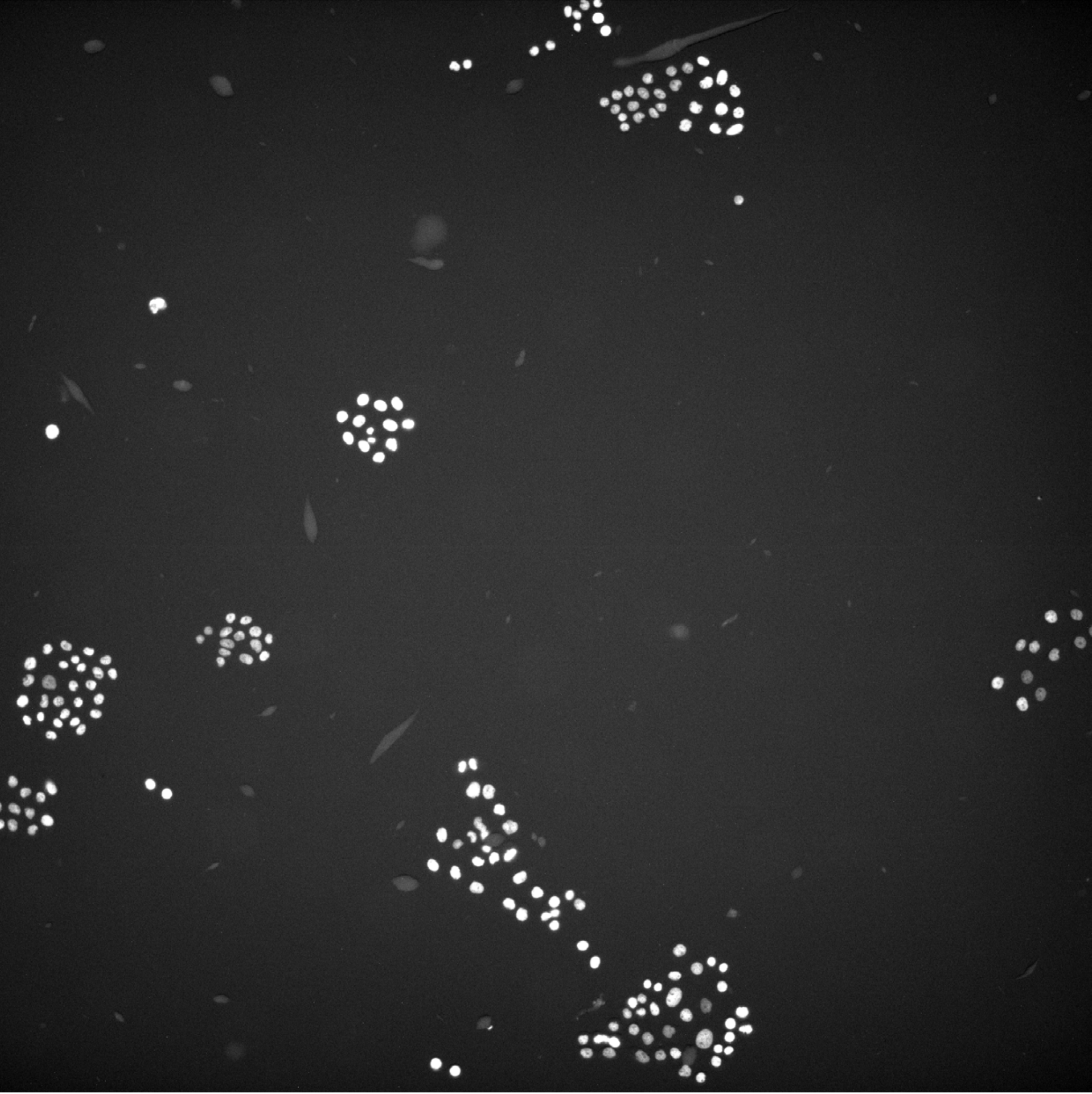}\label{sfig:inputVU2}} \\ 
  \caption[Examples of the analyzed microscopy fluorescence images provided by the Department of Biochemistry of the Vanderbilt University]{Examples of the analyzed microscopy fluorescence images provided by the Department of Biochemistry of the VU. The images were displayed by automatically adjusting the brightness and contrast according to an histogram-based procedure.}
  \label{fig:InputVanderbilt}
\end{figure}

\subparagraph{Vanderbilt University dataset}
\label{sec:datasetVU}
This dataset collects time-lapse microscopy images from two experiments performed at the Department of Biochemistry of the Vanderbilt University (VU) School of Medicine, Nashville, TN, USA.
All the images were acquired using a Molecular Devices (San Jose, CA, USA) from ImageXpress Micro XL with a $10\times$ objective in the red channel (Cy3); some images were acquired in the green channel (FITC) for the same well and location (overlapping information), using a PCO.SDK Camera (Kelheim, Germany).
The imaged region is $2160 \times 2160$ pixels and the pixel-depth is $16$ bits.
In this study, we consider the PC-9 human lung adenocarinoma cell line \cite{jones2018,tyson2012}.
The total number of the analyzed images with the corresponding gold-standard was $46$, related to two different experiments.
Two examples of input images are shown in Fig. \ref{fig:InputVanderbilt}.

\subparagraph{2018 Data Science Bowl}
\label{sec:datasetDSB}
In order to validate ACDC on imaging data coming from different sources, we considered a selection of the training set from the 2018 Data Science Bowl (DSB) \cite{DSB2018}, which is a competition organized by Kaggle (San Francisco, CA, USA).
We had to use the training set alone, since the labels for the test set are not publicly provided by the organizers.
The goal of the 2018 DSB regards the detection of cells' nuclei to identify each individual cell in a sample.
This operation is mandatory to understand the underlying biological processes.

\begin{figure}
\centering
	\subfloat[][]
    {\includegraphics[width=0.3\textwidth]{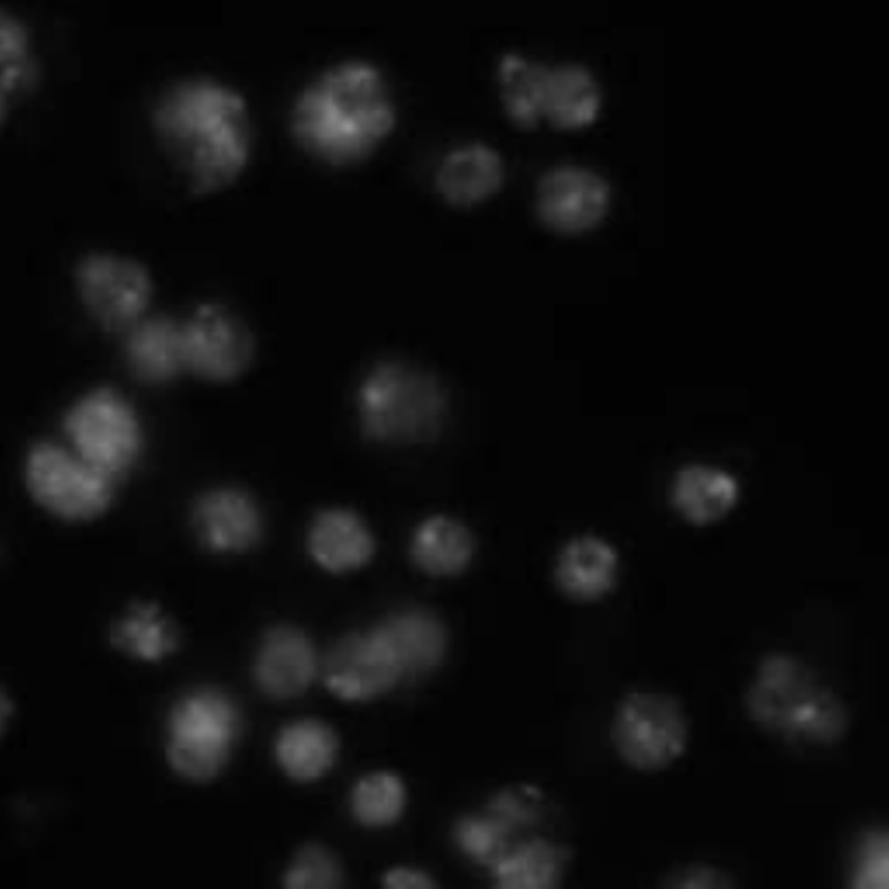}\label{sfig:inputDSB1}} \quad
    \subfloat[][]{\includegraphics[width=0.3\textwidth]{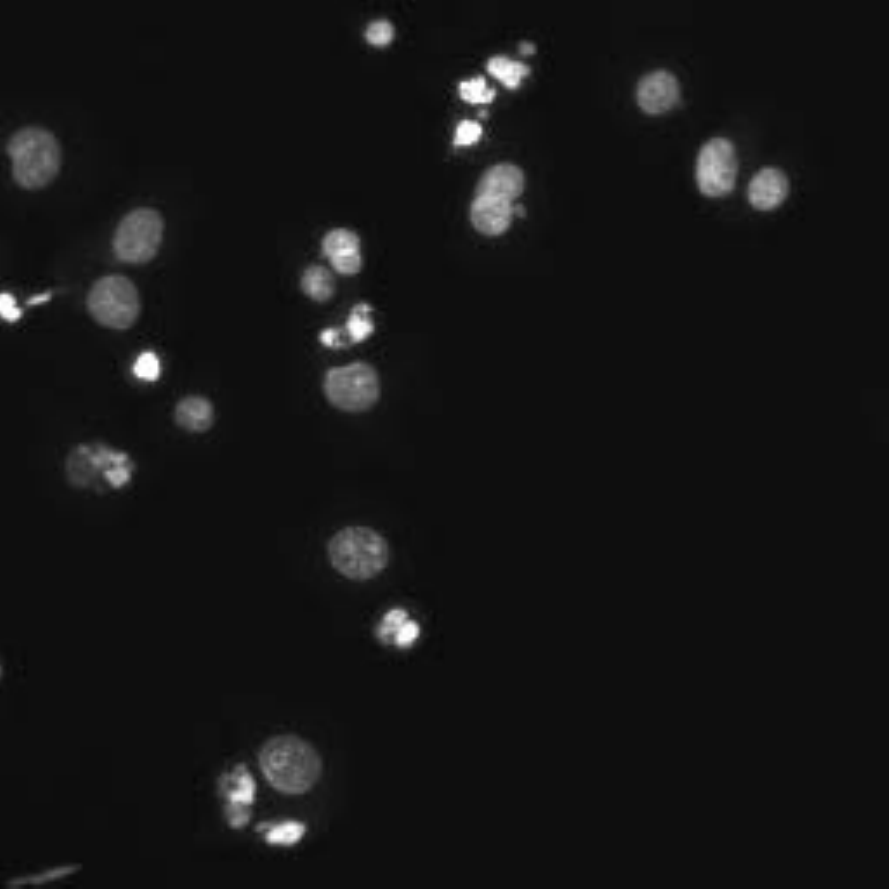}\label{sfig:inputDSB2}} \quad
    \subfloat[][]
    {\includegraphics[width=0.3\textwidth]{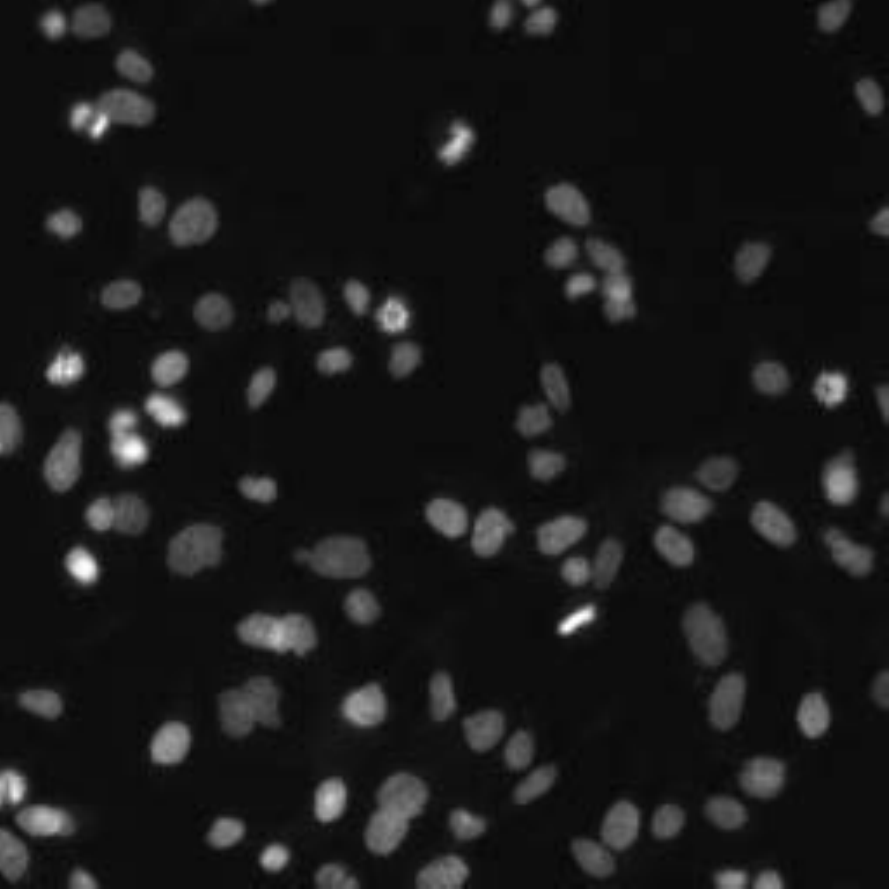}\label{sfig:inputDSB3}} \\ 
  \caption[Examples of the analyzed microscopy fluorescence from the 2018 Data Science Bowl dataset]{Examples of the analyzed microscopy fluorescence from the 2018 DSB dataset.}
  \label{fig:InputDSB}
\end{figure}

This dataset includes a large number of labelled nuclei images acquired under a variety of conditions, magnification, and imaging modality (i.e., bright-field and fluorescence microscopy).
Image size varies in $256 \times 256$, $256 \times 320$, $260 \times 347$, $360 \times 360$, $512 \times 640$, $520 \times 696$, $603 \times 1272$, and $1040 \times 1388$ pixels.
The dataset aims at evaluating the generalization abilities of computational methods when dealing with significantly different data.
We first extracted only the fluorescence microscopy images from the training set.
According to the type of images used during the design phase of ACDC, the selection criterion of the fluorescence microscopy images was based on the maximum size $\max\{\mathrm{size}(\mathcall{R}_\mathrm{cell})\}$ of the region cells $\mathcall{R}_\mathrm{cell}$ segmented in the ground truth.
Therefore, we selected only the images where $\max\{\mathrm{size}(\mathcall{R}_\mathrm{cell})\} \leq 1000$ pixels.
The number of the resulting images is $303$.
Three examples of input images characterized by highly different characteristics are shown in Fig. \ref{fig:InputDSB}.

\paragraph{ACDC: the proposed efficient automatic cell nuclei segmentation method}
\label{sec:Method}

Time-lapse microscopy imaging allows for the visualization and analysis of living specimens' dynamic processes.
A single experiment may consist of hundreds of objects depicted on thousands of images, leading to the need for computational methods that can enable the quantitative analysis of system-level biology automatically.

Advances in optics and imaging systems have enabled biologists to visualize live cell dynamic processes by time-lapse microscopy images.
However, the imaging data recorded during even a single experiment may consist of hundreds of objects over thousands of images, which makes manual inspection a tedious, time-consuming and inaccurate option.
Traditional segmentation techniques proposed in the literature generally exhibit low performance on live unstained cell images.
These limitations are mainly due to low contrast, intensity-variant, and non-uniform illuminated images \cite{kanade2011}.
Therefore, novel automated computational tools for quantitative system-level biology are required.

\begin{figure}[!t]
  \includegraphics[width=0.85\textwidth]{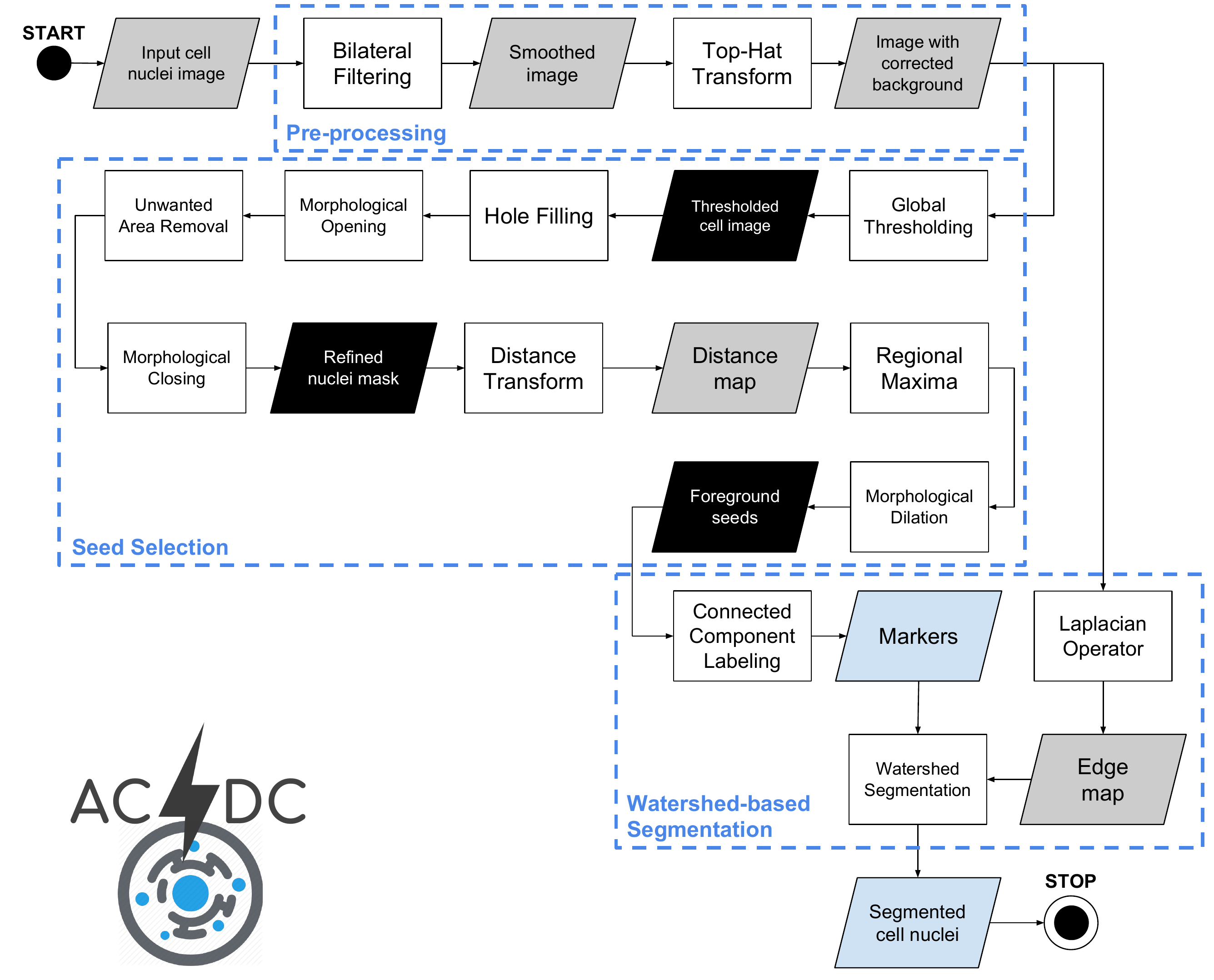}
  \centering
  \caption[Flow diagram of the proposed ACDC pipeline for cell nuclei segmentation]{Flow diagram of the proposed ACDC pipeline for cell nuclei segmentation. The gray, black and light-blue data blocks denote gray-scale images, binary masks and information extracted from the images, respectively.}
  \label{fig:ACDCworkFlow}
\end{figure}

The proposed method is designed for time-lapse microscopy and aims at efficiently overcoming the limitations of the literature methods \cite{carpenter2006,georgescu2011,held2010}, especially in terms of efficiency and execution time.
As a matter of fact, the resulting processing pipeline exploits classical image processing techniques in a smart fashion, allowing for a feasible solution in real laboratory environments.
Fig. \ref{fig:ACDCworkFlow} outlines the overall flow diagram of the ACDC segmentation pipeline.
Each single step is described hereafter.
Our approach is fully automatic, so allowing for reproducible measurements \cite{sansone2012}.
It is worth noting that the number of fixed parameters that need to be set in the underlying image processing operations involves only the kernel size of spatial filters and the structuring element size in morphological operations, allowing for a robust yet simple solution.
Therefore, these few parameters allow the user to have control over the achieved cell segmentation results.
Unlike DCNN-based black or opaque boxes, ACDC offers an interpretable model for the biologists that may adjust conveniently adjust some parameters according to the cell lines currently analyzed.
Differently to supervised Machine Learning approaches \cite{dao2016,sadanandan2017}, ACDC does not require any training phase, thus representing a reliable and practical solution even without labeled datasets.

\subparagraph{Pre-processing}
The input microscopy image is pre-processed to yield a more convenient input to the downstream watershed-based segmentation relying on the following steps:
\begin{enumerate}
\item Bilateral filtering that allows for denoising the image $\mathcall{I}$ while preserving the edges by means of a non-linear combination of nearby image values \cite{tomasi1998}.
This noise-reducing smoothing filter combines gray levels (colors) according to both a geometric closeness function $c$ and a radiometric (photometric) similarity function $s$.
This combination is used to strengthen near values with respect to distant values in both spatial and intensity domains.
More specifically, the intensity of each pixel is replaced with a weighted average of intensity values from neighboring pixels.
The weights depend not only on the Euclidean distance between pixels, but also on the radiometric differences (i.e., range differences).
This dependence on pixel intensity helps to preserve the edge sharpness.
The image $\mathcall{I}_\mathrm{bf}$ smoothed through the bilateral filter can be denoted by the following Equation:
\begin{equation}
    \label{eq:bilFilt}
	\mathcall{I}_\mathrm{bf}(\mathbf{p}) =  \frac{1}{\eta} \sum \limits_{\mathbf{q} \in \Omega} \mathcall{I}(\mathbf{q}) \cdot c(\mathbf{p}, \mathbf{q}) \cdot s(\mathcall{I}(\mathbf{p}), \mathcall{I}(\mathbf{q})),
\end{equation}
where the normalization term $\eta = \sum_{\mathbf{q} \in \Omega} c(\mathbf{p}, \mathbf{q}) \cdot s(\mathcall{I}(\mathbf{p}), \mathcall{I}(\mathbf{q}))$ constrains the sum of weights to one and preserves the local mean \cite{chaudhury2011}.
The window $\Omega$, centered on the current pixel $\mathbf{p}$ for filtering, is a square whose side is calculated as: $\mathrm{size}(\Omega) = \max{\{5, 2 \cdot \lceil 3 \cdot \sigma_s \rceil + 1\}} = 7$ pixels.
This simple yet effective strategy allows for contrast enhancement \cite{schettini2010}.
Bilateral filter has been shown to work properly in fluorescence imaging even preserving the directional information, such as in the case of the F-actin filaments \cite{venkatesh2015}.
This denoising technique was effectively applied to biological electron microscopy \cite{jiang2003} as well as to cell detection \cite{li2008BF}, revealing better performance, with respect to low-pass filtering, in noise reduction without removing the structural features conveyed by strong edges.
The most commonly used version of bilateral filtering is the shift-invariant Gaussian filtering, in which both the closeness function $c$ and the similarity function $s$ are Gaussian functions of the Euclidean distance between their arguments \cite{tomasi1998}.
More specifically, $c$ is radially symmetric:
$c(\mathbf{p}, \mathbf{q}) = e^{-\frac{1}{2}\left(\frac{||\mathbf{p}-\mathbf{q}||}{\sigma_s}\right)^2}$.
Analogously, the similarity function $s$ can be defined as:
$s(\mathbf{p}, \mathbf{q}) =e^{-\frac{1}{2}\left(\frac{||\mathcall{I}(\mathbf{p})-\mathcall{I}(\mathbf{q}||}{\sigma_r})\right)^2}$.
We set $\sigma_c=1$ and $\sigma_s = \sigma_\mathrm{global}$ (where $\sigma_\mathrm{global}$ is the the standard deviation of the input image $I$) for the standard deviation of the Gaussian functions $c$ and $s$, respectively.
This smart denoising approach allows us to keep the edge sharpeness while reducing the noise of the processed image, so avoiding cell region under-estimation.
\item Top-hat transform for background correction with a binary circular structuring element (radius: $21$ pixels) is applied on the smoothed image.
This operation accounts for non-uniform illumination artifacts, by extracting the nuclei from the background.
The white top-hat transform is the difference between the input image $\mathcall{I}$ and the opening of $I$ with a gray-scale structuring element $b$: $\mathcall{T}_\mathrm{w} = \mathcall{I} - \mathcall{I} \circ b$ \cite{gonzalez2002}.
\end{enumerate}

\begin{figure}
\centering
	\subfloat[][]
    {\includegraphics[width=0.35\textwidth]{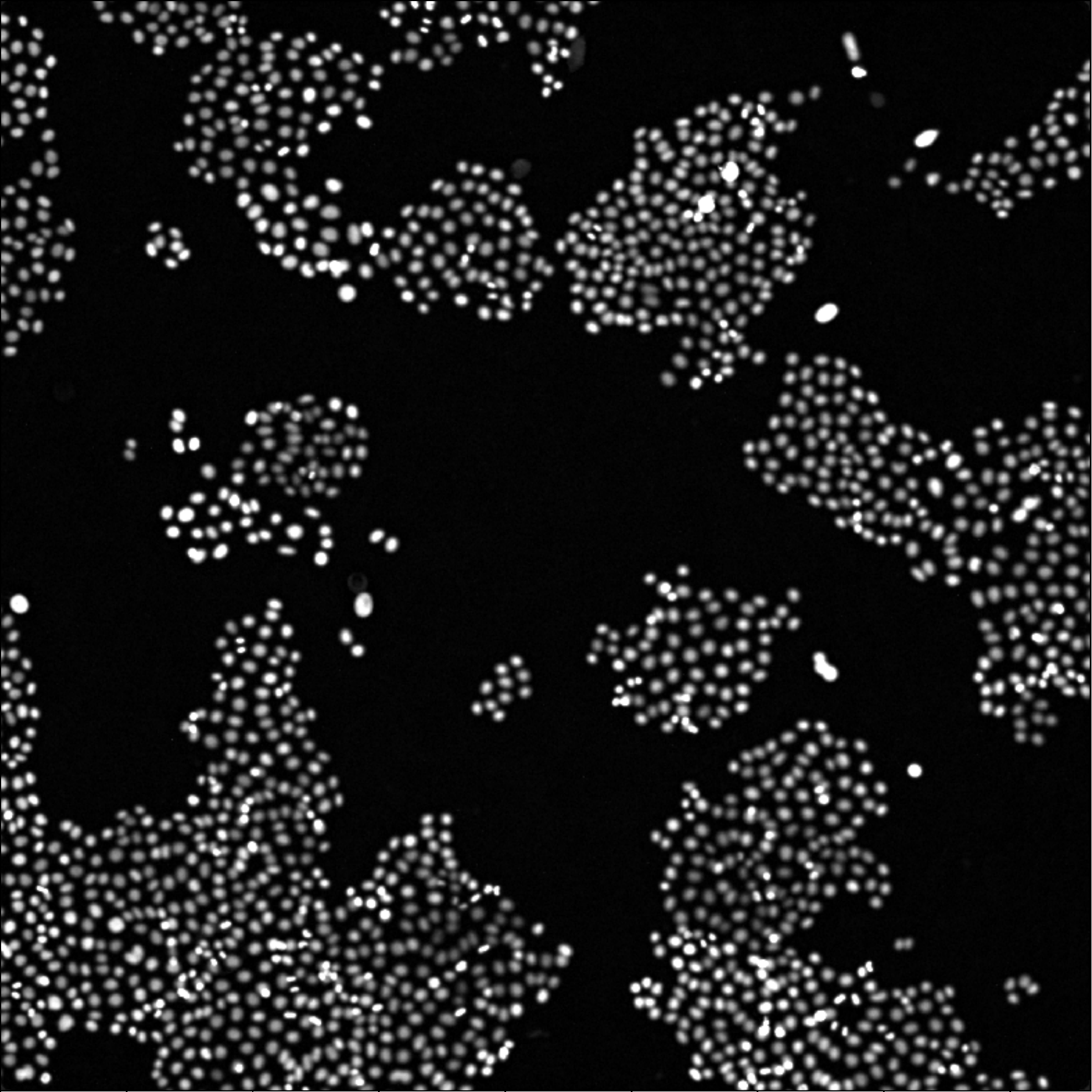}\label{sfig:preProcVU1}} \quad
    \subfloat[][]{\includegraphics[width=0.35\textwidth]{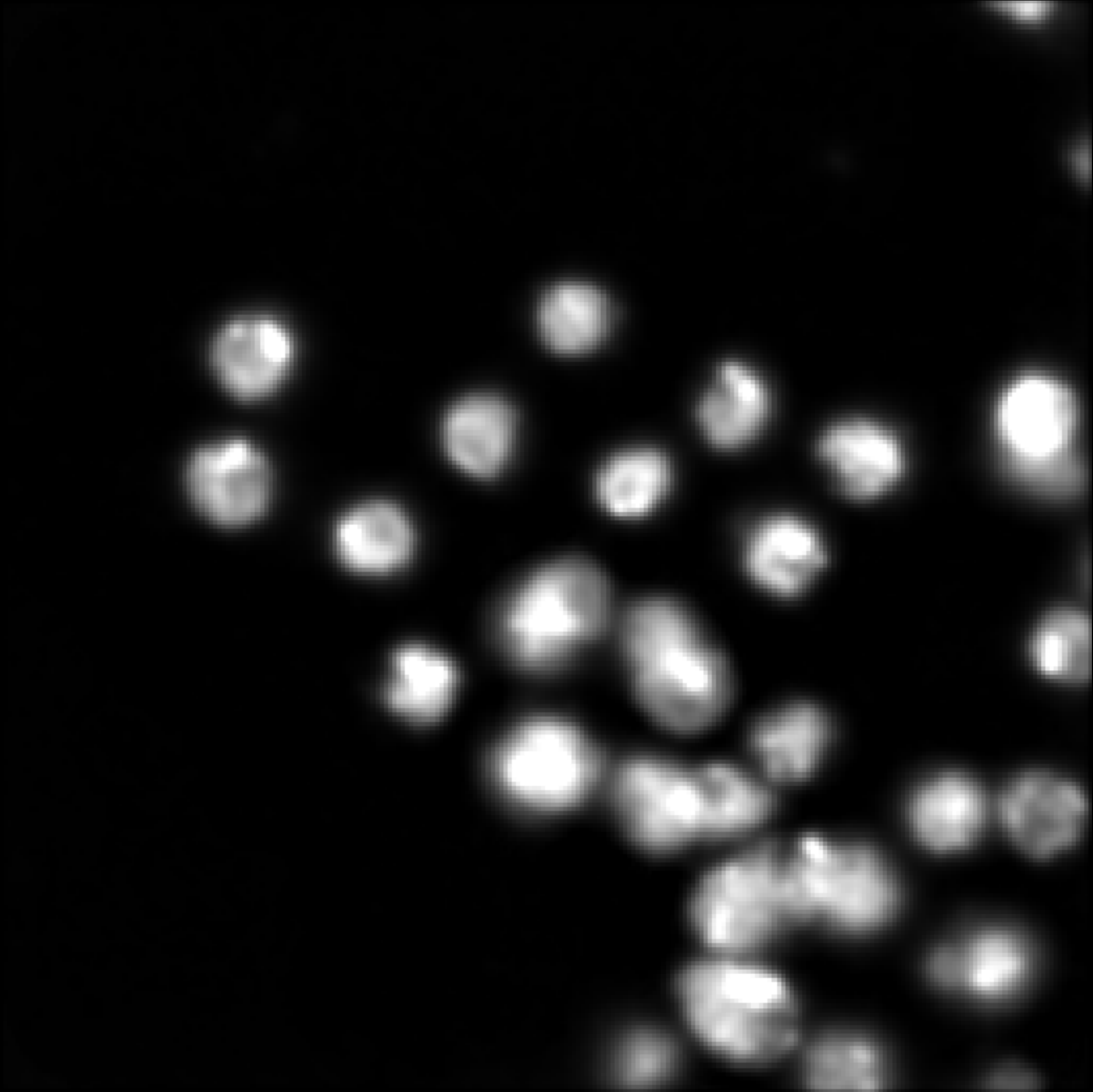}\label{sfig:preProcDSB1}}\\ 
  \caption[Pre-processing results in the ACDC pipeline]{Pre-processed images in the ACDC pipeline: (a) result on the image shown in Fig. \ref{sfig:inputVU1}; (b) result on the image shown in Fig. \ref{sfig:inputDSB1}.}
  \label{fig:preProcImages}
\end{figure}

The results of the pre-processing images applied to Figs. \ref{sfig:inputVU1} and \ref{sfig:inputDSB1} are shown in Figs. \ref{sfig:preProcVU1} and \ref{sfig:preProcDSB1}, respectively.

\subparagraph{Nucleus seed selection}
The cell nuclei have to be accurately extracted from the pre-processed image.
The following steps are applied to obtain a reliable seed selection:
\begin{enumerate}
\item a thresholding technique has to be first applied to detect cell regions.
Both global and local thresholding techniques aim at separating desirable foreground objects from the background in an image, considering differences in pixel intensities \cite{jain1999}.
Global thresholding determines a single threshold for all pixels and works well if the histogram of the input image contains well-separated peaks corresponding to the desired foreground objects and background \cite{otsu1975}.
Local adaptive thresholding techniques estimate the threshold locally over sub-regions of the entire image, by considering only a neighborhood (i.e., a user-defined window) with a specified size and exploiting local image properties to calculate a variable threshold \cite{gonzalez2002,jain1999}.
These algorithms find the threshold by locally examining the intensity values of the neighborhood of each pixel according to image intensity statistics.
In order to avoid unwanted pixels in the thresholded image, mainly due to small noisy hyper-intense regions caused by non-uniform illumination, we applied the Otsu's global thresholding method \cite{otsu1975} instead of local adaptive thresholding based on the mean value in a neighborhood \cite{militelloCBM2017}.
Moreover, global adaptive threshold techniques are significantly faster than local adaptive strategies;

\item hole filling is applied to remove possible holes in the detected nuclei due to small hypo-intense regions included in the nuclei regions;

\item morphological opening (using a $1$-pixel disk as structuring element) is used to remove loosely connected components, such as in the case of almost overlapping cells;

\item unwanted area are removed based on the connected-components size.
Especially, the detected candidate regions with areas smaller than $40$ pixels are removed to refine the achieved segmentation results by robustly avoiding false positives;

\item morphological closing (using a $2$-pixel radius circular structuring element) is applied to smooth the boundaries of the detected nuclei and avoid the under-estimation of the detected nuclei regions;

\item the approximate Euclidean distance transform from the binary mask, achieved by applying the Otsu's algorithm and refined by using the previous four steps, is used to obtain the matrix of distances of each pixel to the background by exploiting the $\ell_2$ Euclidean distance \cite{borgefors1986} (with a $5 \times 5$ pixel mask for a more accurate distance estimation).
This algorithm calculates the distance to the closest background pixel for each pixel of the source image.
Let $\mathcall{G}$ be a regular grid and $f: \mathcall{G} \rightarrow \mathbb{R}$ an arbitrary function on the grid, called a sampled function \cite{felzenszwalb2012}.
We define the distance transform $\mathcall{D}_f :\mathcall{G} \rightarrow \mathbb{R}$ of $f$ as:
\begin{equation}
    \label{eq:distTransf}
	\mathcall{D}_f(\mathbf{p}) = \min\limits_{\mathbf{q} \in \mathcall{G}}{\left( d(\mathbf{p},\mathbf{q}) + f(\mathbf{q}) \right)},
\end{equation}
where $d(\mathbf{p},\mathbf{q})$ is a measure of the distance between the pixels $\mathbf{p}$ and $\mathbf{q}$.
Owing to the fact that cells have a pseudo-circular shape, we used the Euclidean distance, so achieving the Euclidean distance transform (EDT) of $f$.
In the case of binary images, with a set of points $\mathcall{P} \subseteq \mathcall{G}$, the distance transform $\mathcall{D}_\mathcall{P}$ is a real-valued image of the same size:
\begin{equation}
    \label{eq:distTransfBin}
	\mathcall{D}_\mathcall{P} = \min\limits_{\mathbf{q} \in \mathcall{P}}{\left( d(\mathbf{p},\mathbf{q}) + 1(\mathbf{q}) \right)},
\end{equation}
where:
\begin{equation*}
1(\mathbf{q}) =
    \begin{cases}
        0, & \text{if } \mathbf{q} \in \mathcall{P} \\
        \infty, & \text{otherwise}
    \end{cases} 
\end{equation*}
is an indicator function for the membership in $\mathcall{P}$ \cite{felzenszwalb2012}.
The computed distance map is normalized by applying contrast linear stretching to the full $8$-bit dynamic range;

\item regional maxima computation allows for estimating foreground peaks on the normalized distance map.
Regional maxima are connected-components of pixels with a constant intensity value, and whose external boundary pixels have all a lower value \cite{soille2013}.
The resulting binary mask contains pixels that are set to $1$ for identifying regional maxima; all other pixels are set to $0$.
A $5 \times 5$ pixel square was employed as structuring element;

\item morphological dilation (using a $3$-pixel radius circular structuring element) is applied to the foreground peaks previously detected for better defining the foreground regions and merging neighboring local minima into a single seed point.

\begin{figure}
\centering
	\subfloat[][]
    {\includegraphics[width=0.363\textwidth]{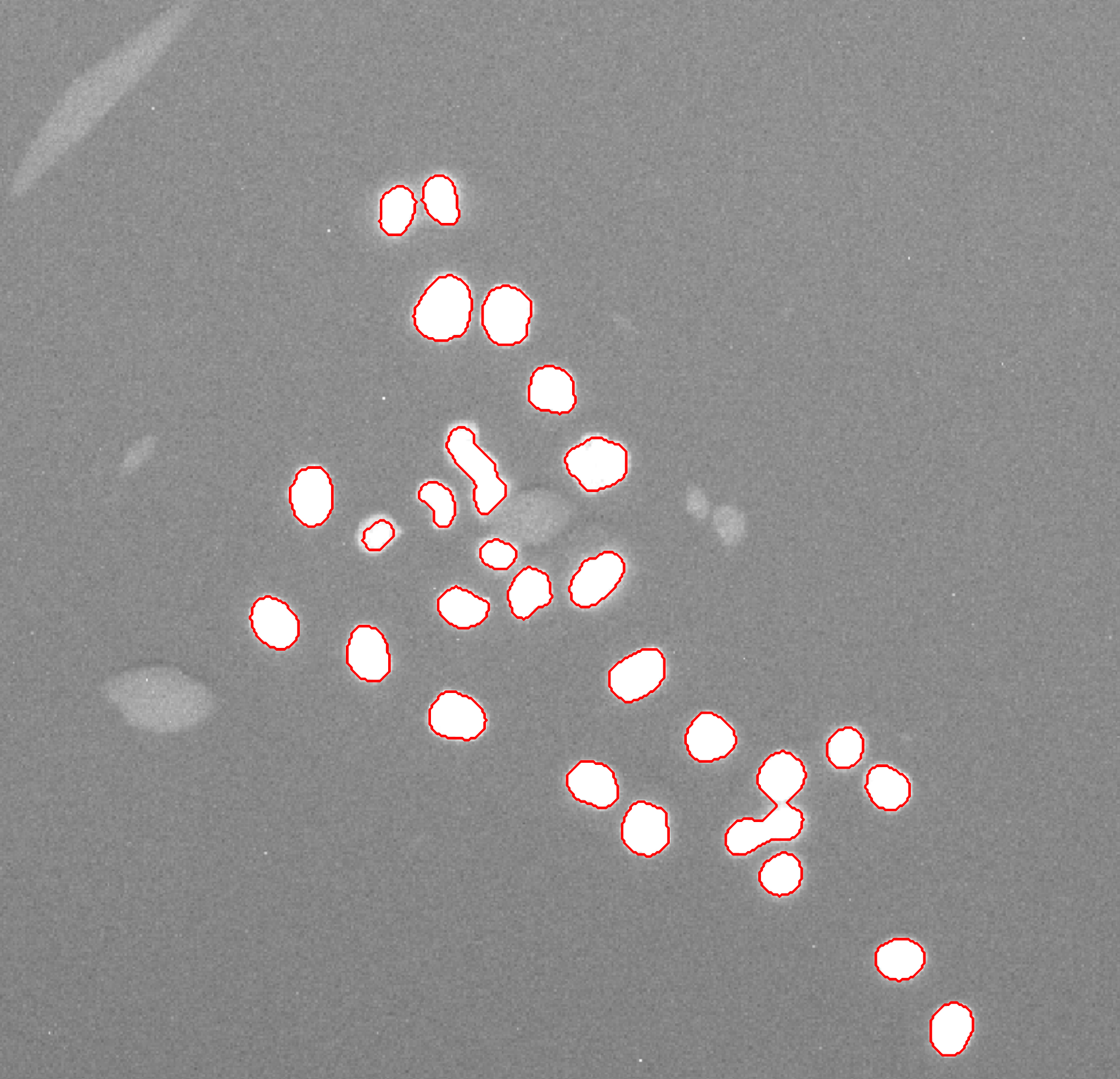}\label{sfig:segVU1}} \quad
    \subfloat[][]{\includegraphics[width=0.35\textwidth]{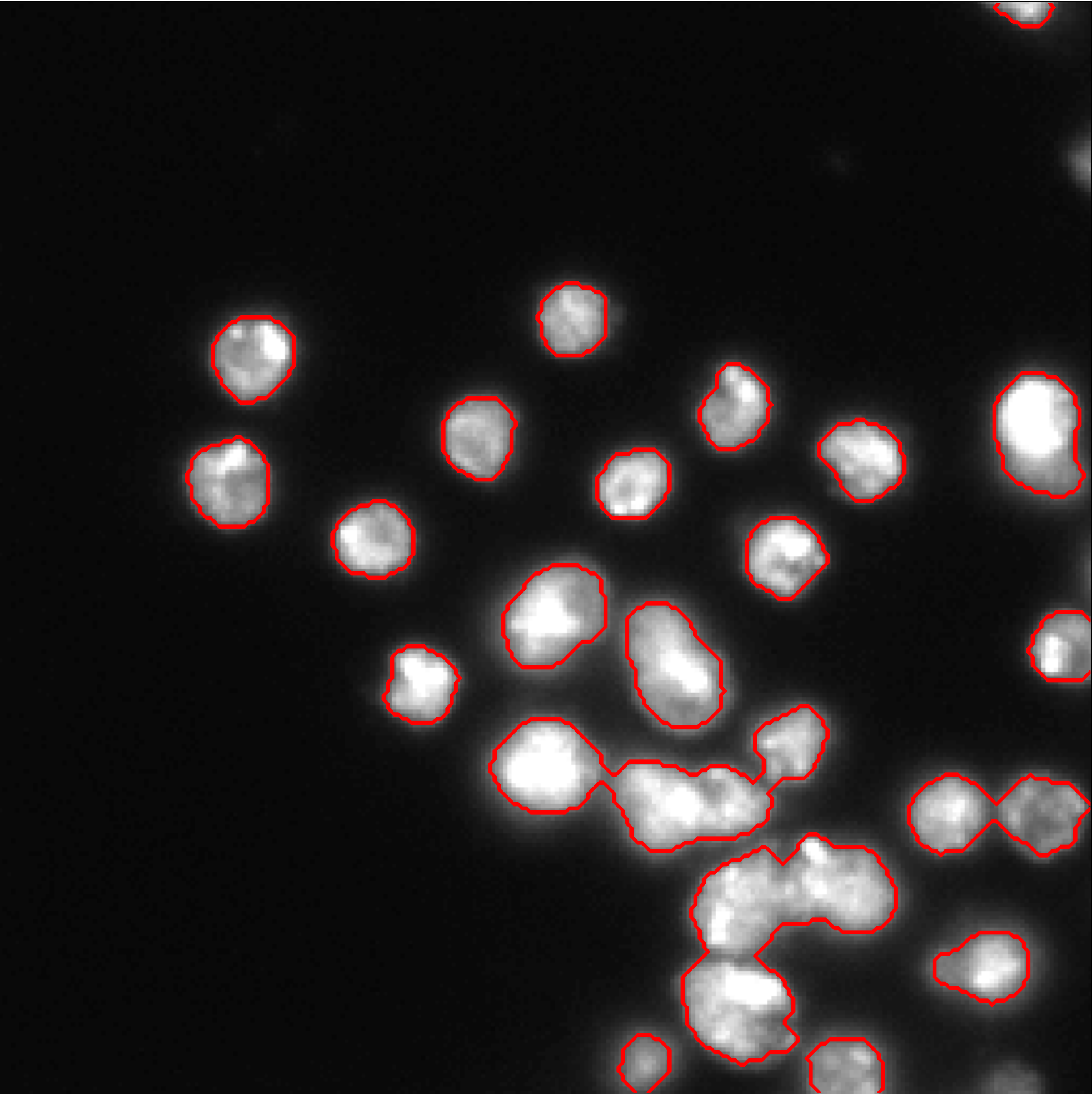}\label{sfig:segDSB1}}\\ 
  \caption[Segmented images after the refinement steps]{Segmented images after the refinement steps, detail of: the (a) result on the image shown in Fig. \ref{sfig:preProcVU1}; the (b) result on the image shown in Fig. \ref{sfig:preProcDSB1}.}
  \label{fig:resImages}
\end{figure}

The segmentation results on the Figs. \ref{sfig:preProcVU1} and \ref{sfig:preProcDSB1} are shown in Figs. \ref{sfig:segVU1} and \ref{sfig:segDSB1}, respectively.
The detail in Fig. \ref{sfig:preProcVU1} shows that ACDC is highly specific to cell nuclei detection, so discarding non-cell regions related to acquisition artifacts.

\end{enumerate}

\subparagraph{Cell nuclei segmentation using the watershed transform}

Although the watershed transform can detect also weak edges, it often may not accurately detect the edge of interest in the case of blurred boundaries \cite{grau2004}.
This sensitivity to noise could be worsened by the use of high pass filters to estimate the gradient and the edges, which amplify the noise.
We address this issue by formerly applying the bilateral filter that reduces the halo effects \cite{tomasi1998}.
Accordingly, we developed the following steps:

\begin{enumerate}

\item Connected-component labeling \cite{suzuki2003cc} of the foreground region binary mask for encoding the markers employed in the following watershed algorithm;

\item Laplacian operator to produce the edge image \cite{gonzalez2002} to feed the edge map as input to the watershed transform;

\item Watershed segmentation on the edge image according to the previously defined markers \cite{meyer1994,najman2005}.

\end{enumerate}

\subparagraph{Implementation details}
The sequential version of ACDC has been entirely developed using the Python programming language (version 2.7.12), exploiting the following libraries and packages: NumPy, SciPy, OpenCV, scikit-image \cite{scikit-image}, and Mahotas \cite{coelho2012}.
The resulting processing pipeline exploits classical image processing techniques in a smart fashion \cite{rundoMBEC2016}, so enabling an efficient and feasible solution in time-lapse microscopy environments.

For laboratory feasibility purposes, an asynchronous job queue, based on a distributed message passing paradigm---namely Advanced Message Queuing Protocol (AMQP)---was developed using Celery \cite{Celery} (implementing workers that execute tasks in parallel) and RabbitMQ \cite{RabbitMQ} (exploited as a message broker to handle communications among workers) for leveraging modern multi-core processors and computer clusters.

\begin{figure}[!t]
	\centering
	\subfloat[]{\includegraphics[width=0.3\textwidth]{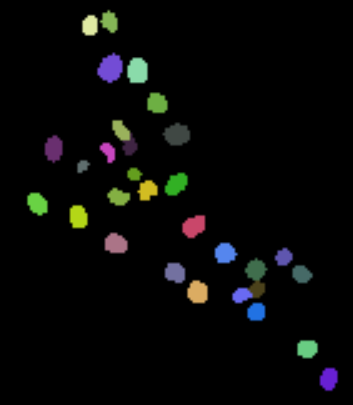}\label{fig:ResVU1}}\quad
	\subfloat[]{\includegraphics[width=0.4\textwidth]{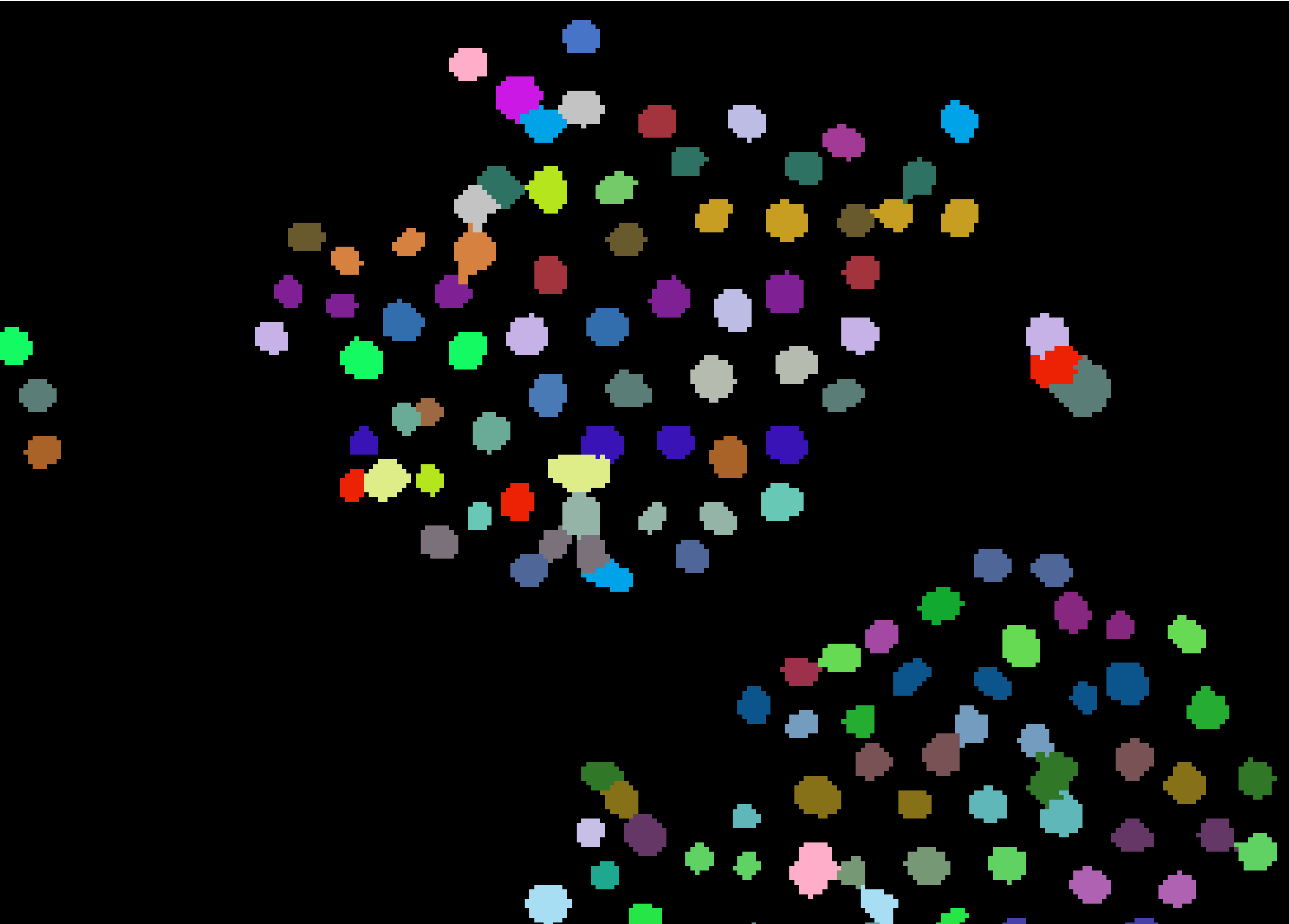}\label{fig:ResVU2}}\\
    \subfloat[]{\includegraphics[width=0.4\textwidth]{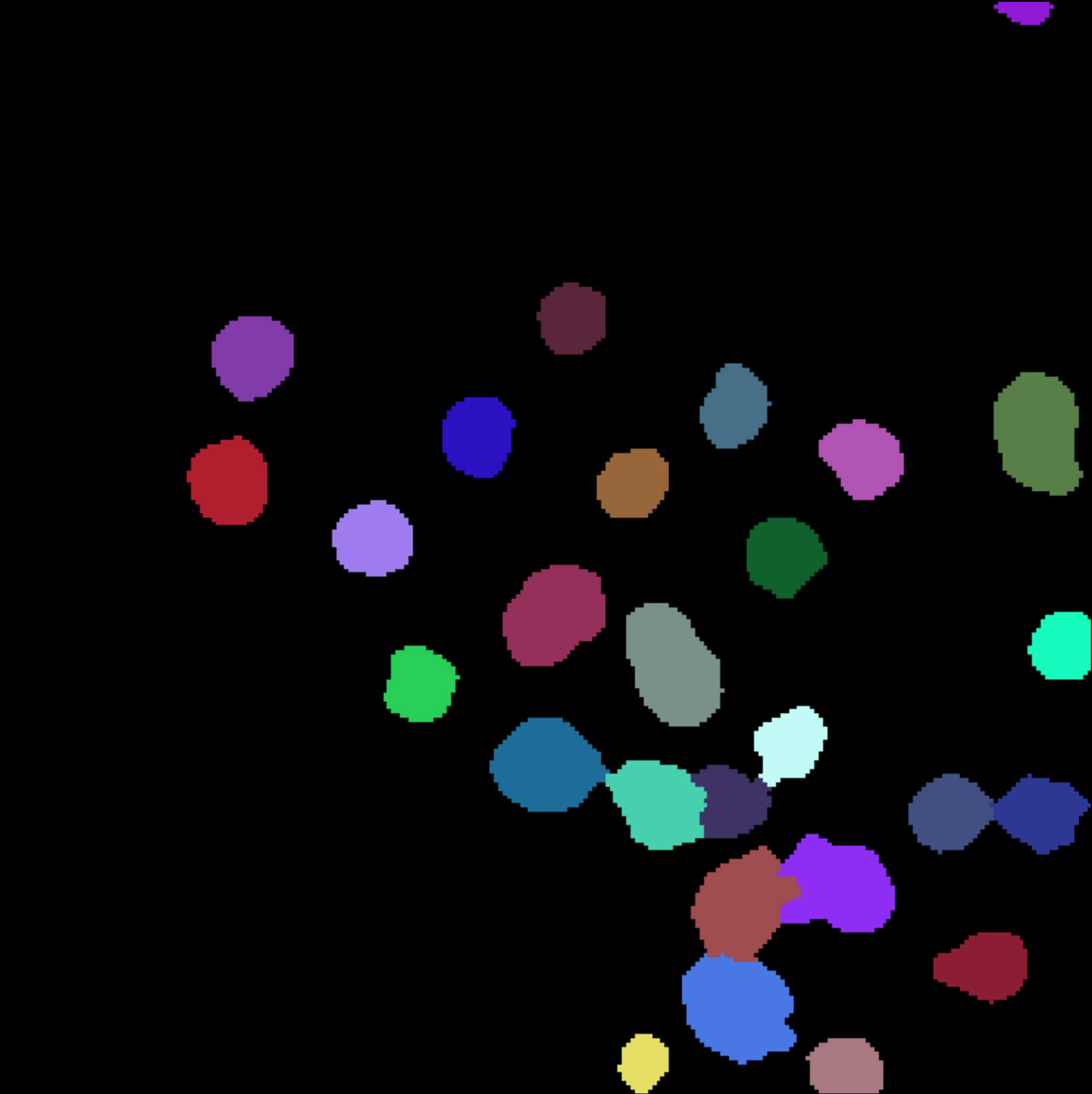}\label{fig:ResDSB3}}\quad
	\subfloat[]{\includegraphics[width=0.4\textwidth]{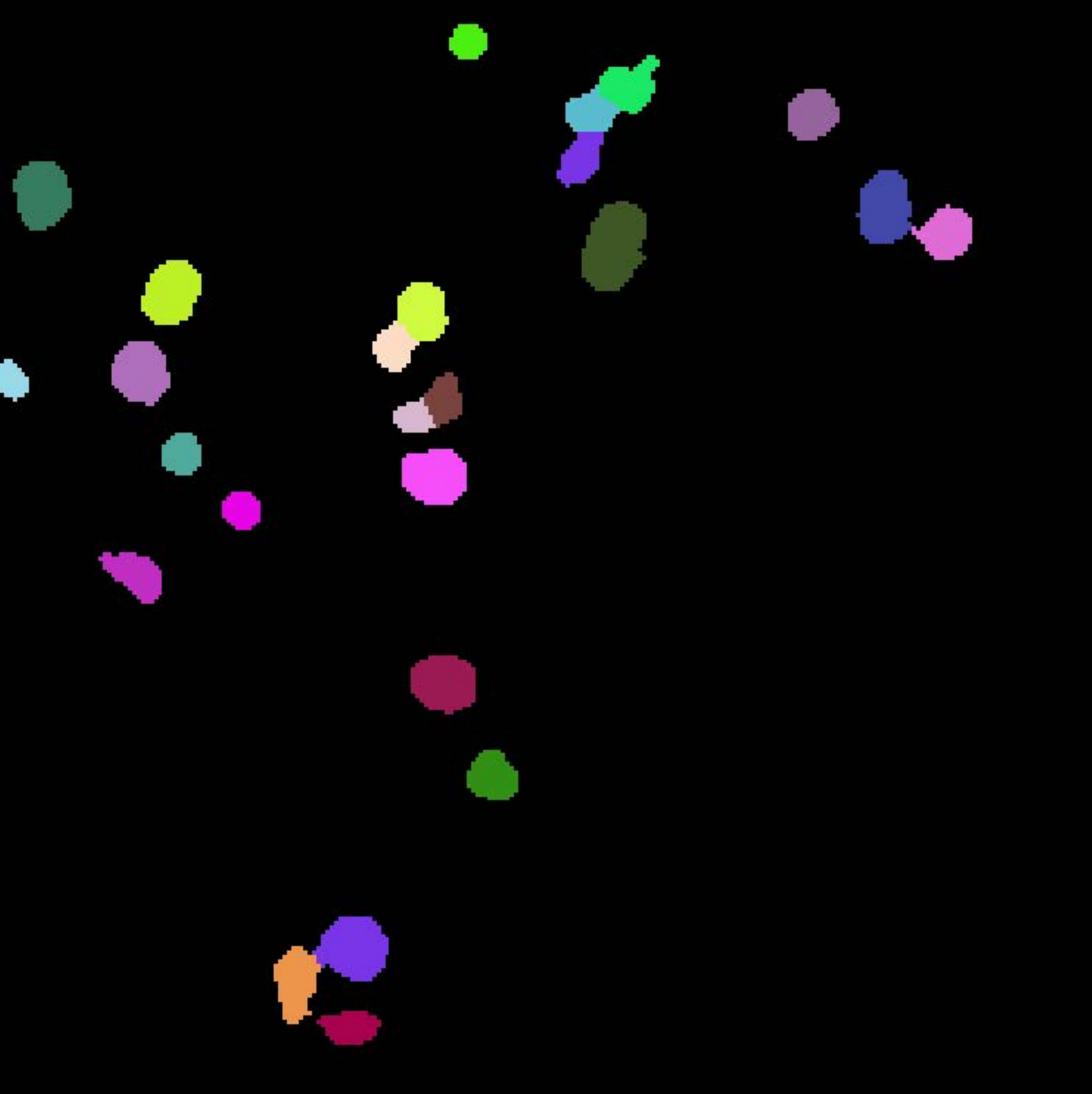}\label{fig:ResDSB4}}\\
	\caption[Examples of cell nuclei segmented by ACDC]{Examples of cell nuclei segmented by ACDC: (a) and (b) show a detail of the results achieved on the images in Figs. \ref{sfig:inputVU1} and \ref{sfig:inputVU2}, respectively; (c) and (d) illustrate the results achieved on the images in \ref{sfig:inputDSB1} and \ref{sfig:inputDSB2}, respectively.}
	\label{fig:segCellImages}	
\end{figure}

\paragraph{Results}
\label{sec:Results}
Fig. \ref{fig:segCellImages} shows two pairs of cell detection results on the VU dataset and the 2018 DSB training dataset, respectively.
The detected cells are displayed with different colors to highlight the separation among overlapping and merging cells.
The analyzed images are characterized by a considerable variability in terms of low or high cell density.

\begin{figure}[!t]
  \includegraphics[width=0.95\textwidth]{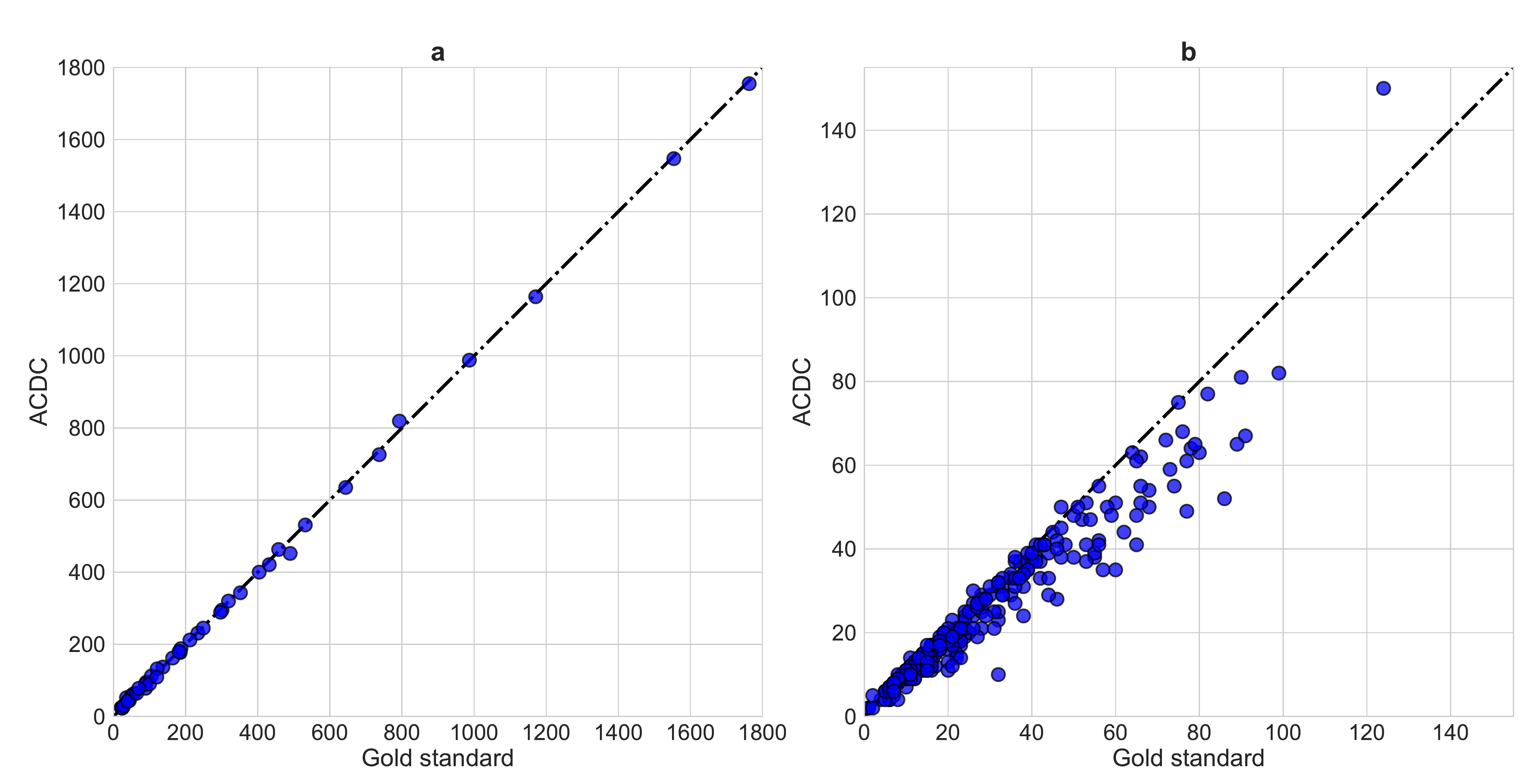}
  \caption[Scatter plots depicting the ACDC results compared to the gold standard in terms of cell nuclei counting]{Scatter plots depicting the ACDC results compared to the gold standard in terms of cell nuclei counting in the case of: (a) time-lapse fluorescence images from the VU; (b) selection of fluorescence images from 2018 DSB.
  The equality line through the origin is drawn as a dashed line.}
  \label{fig:scatterPlot}
\end{figure}

The assessment on cell count is quantified by means of the Pearson's coefficient to measure the linear correlation between the automated and manual cell counts.
Table \ref{tab:ACDCresults} reports the results for both the analyzed datasets.
These results are supported by the scatter plots in Fig. \ref{fig:scatterPlot}.
The accuracy of the achieved segmentation results were quantitatively evaluated with respect to the real measurement by using the Intersection over Union (\emph{IoU}) metrics, also known as \emph{JI} (see Appendix \ref{sec:segEval}).

\begin{table}[ht]
\centering
\caption[Evaluation metrics on cell counting and segmentation achieved by ACDC on the analyzed time-lapse microscopy datasets]{Evaluation metrics on cell counting and segmentation achieved by ACDC on the analyzed time-lapse microscopy datasets.
The results for the \emph{IoU} metrics and the execution time measurements are expressed as mean value $\pm$ standard deviation.}
\label{tab:ACDCresults}
\begin{tabular}{lccc}
\hline\hline
Dataset & Pearson's coeff. & IoU (\%) & Exec. time [s] \\
\hline
Vanderbilt University  &          $\rho = 0.99974$           &$62.84 \pm 8.58$   &$7.49 \pm 0.30$  \\
2018 Data Science Bowl &          $\rho = 0.96121$           &$80.37\pm 10.58$   &$0.12 \pm 0.09$  \\
\hline\hline
\end{tabular}
\end{table}

The achieved high $\rho$ coefficient values confirm the effectiveness of the proposed approach, according to a validation performed against the expert's gold standard.
A strong positive correlation between the cell counts computed automatically by ACDC and the manual measurements was observed for both datasets analyzed in this study.
In particular, in the case of the VU dataset, the automated cell counting is strongly consistent with the corresponding manual measurements, by denoting also a unity slope, as shown in Fig. \ref{fig:scatterPlot}a.
In the case of the 2018 DSB dataset (Fig. \ref{fig:scatterPlot}b), the regression straight-line reveals a negative offset in the ACDC measurements.
This finding means that the cell counts achieved by ACDC slightly under-estimated the gold standard in approximately $55\%$ of cases.
In addition to the high variability of the fluorescence images included in the 2018 DSB dataset, this systematic error often depends on the gold standard binary masks that consider also partial connected-components of cell nuclei smaller than $40$ pixels located at the image borders.
Notice that ACDC's settings were not completely modified, so some very small partial cell components were removed according to the morphological refinements based on the connected-component size.

The \emph{IoU} mean values calculated on the VU images reveals a good agreement between the segmentation results achieved by ACDC and the gold standard.
The standard deviation values confirm the high variability encountered in the input datasets.
In particular, in the case of the VU images, this evidence strongly depends on the density of the cells represented in the input image.
As a matter of fact, the \emph{IoU} metrics is highly affected by the size of the foreground regions with respect to the background.
This behavior is confirmed by the 2018 DSB results, where the \emph{IoU} values are considerably higher than those achieved on the VU images, even though the Pearson's coefficient is slightly lower in this case.
Accordingly, the high standard deviation for the DSB dataset is due to the intrinsic variety of the images---in terms of image size, zoom factor, and specimen type---included in this dataset (see Section \ref{sec:datasetDSB}).

The mean execution times concerning the segmentation tests are shown in Table \ref{tab:ACDCresults}.
These experiments were run on a personal computer equipped with a quad-core Intel\textsuperscript{\textregistered} Core\textsuperscript{TM} 7700HQ (clock frequency $3.80$ GHz), $16$ GB RAM, and Ubuntu 16.04LTS operating system.
The computational efficiency of ACDC is confirmed in both cases, completing the cell detection and counting tasks by coping with the time constraints imposed by high-throughput laboratory routine. 
As expected, the execution times are dependent on the image size.
This trend is mainly due to the bilateral filtering operation.

To conclude, considering the two experimental datasets, ACDC achieved accurate results in terms of cell counting as well as segmentation accuracy.
Therefore, ACDC showed to be a reliable and efficient solution even when dealing with multiple datasets characterized by significant variations, in terms of acquisition devices, specimens, experimental conditions, and imaging characteristics.

\paragraph{Discussion and conclusion}
\label{acdc:Conclusion}

A novel method for cell detection and counting, called ACDC, was proposed, by exploiting a fully automatic pipeline based on the watershed transform \cite{soille1990,vincent1991} and morphological filtering operations \cite{beucher1992,soille2013}.
We benefited also from the edge-preserving smoothing achieved by the bilateral filtering \cite{tomasi1998}.
We tested our approach on two different cell imaging datasets characterized by significantly different acquisition and experimental conditions.
ACDC was shown to be accurate and reliable, thus representing a laboratory feasible solution also thanks to its computational efficiency.
The current ACDC implementation can also distribute the computation on multi-core architectures and computer clusters to further reduce the running time required to analyze large single image stacks.
To this end, an asynchronous job queue, based on a distributed message passing paradigm was developed exploiting Celery and RabbitMQ.

As future developments, we aim at exploiting the most recent Machine Learning techniques \cite{kraus2017} to accurately refine the segmentation results.
The improvements may be achieved by classifying the geometrical and textural features extracted from the detected cells \cite{win2018}.
This advanced computational analysis can allow us to gain biological insights in complex cellular processes \cite{angermueller2016}.
As a biological application in the near future, we plan to focus on the Fluorescent, Ubiquitination-based Cell-Cycle Indicator (FUCCI) reporter system for investigating cell-cycle states \cite{sakaue2008}, by combining accurate nuclei segmentation results with live cell imaging of cell cycle and division.

\chapter{Pattern Recognition techniques}

\graphicspath{{Chapter4/Figs/}}
\label{chap4}

\section{Unsupervised fuzzy clustering}
\label{sec:fuzzyClustering}

The FCM algorithm \cite{bezdek1981,bezdek1984} is a partitional clustering technique, i.e., an input data set is optimally classified in sub-groups \cite{bishop2006}.
Intuitively, a cluster collects a group of data points whose pairwise distances are lower with respect to the distances from points outside the cluster.

Thus, the FCM algorithm operates by minimizing the intra-cluster variance as well as by maximizing the inter-cluster variance, in terms of a distance metrics between the feature vectors \cite{li2010}.
These characteristics make this unsupervised clustering method suitable for pattern recognition and image segmentation, by representing each image region with a cluster.
In addition, FCM is highly robust to initial conditions when well-separable data are analyzed \cite{davenport1988}.
The FCM technique search for the optimal partition of a data set $\mathcall{X} = \left\{ \mathbf{x}_1, \mathbf{x}_2, \ldots, \mathbf{x}_N \right\}$ composed of $N$ feature vectors, which denote data samples $\mathbf{x}_k \in \mathbb{R}^D$  belonging to a $D$-dimensional Euclidean space, into exactly $C$ clusters (i.e., non-empty partitions of the input dataset), as introduced in \cite{bezdek1981, bezdek1984}.
In Pattern Recognition, features are defined as measurable quantities that could be used to classify different regions \cite{lu2015}.
An input dataset is partitioned into groups (regions), and each of them is identified by a centroid.
Distance metrics are used to group data into clusters of similar types and the number of clusters is assumed to be known \textit{a priori} \cite{ghose2012,li2010}.
These feature vectors can be conveniently represented by a data matrix $\mathbf{X} \in \mathbb{R}^{N \times D}$ with $N$ rows containing the input $D$-dimensional feature vectors.
Formally, a partition $\mathcall{P}$  is defined as a set family $\mathcall{P}=\left\{ \mathcall{Y}_1, \mathcall{Y}_2, \ldots, \mathcall{Y}_C \right\}$.
The resulting clusters are always represented by connected-regions with unbroken edges.
The goal of unsupervised clustering methods is to determine an intrinsic partitioning in a set of unlabeled data, which are associated with a feature vector \cite{bishop2006}.

The crisp $K$-means algorithm \cite{lloyd1982} imposes that the clusters must be a proper subset of $\mathcall{X}$ ($\varnothing \subset \mathcall{Y}_i \subseteq \mathcall{X}, \forall i$) and their set union must reconstruct the whole dataset ($\bigcup\limits_{i=1}^C \mathcall{Y}_i = \mathcall{X}$). Moreover, the various clusters are mutually exclusive ($\mathcall{Y}_i \cap \mathcall{Y}_j = \varnothing, \forall i \neq j$), i.e., each feature vector may belong to only one group.
Differently from the $K$-means algorithm \cite{kanungo2002} that yields a hard clustering output, in which each feature vector is strictly assigned to one group, the FCM algorithm defines a fuzzy partition $\mathcall{P}=\left\{ \mathcall{Y}_1, \mathcall{Y}_2, \ldots, \mathcall{Y}_C \right\}$ using a fuzzy set family \cite{zadeh1965}.
Compared to the iterative EM algorithm \cite{dempster1977,moon1996}, which estimates the parameters of a Gaussian mixture model relying on a probabilistic framework  \cite{nguyen2013}, the $K$-Means splits the data into Voronoi cells of the cluster centroids.

The FCM scheme, which employs Fuzzy Logic \cite{zadeh1996}, results in a soft computing solution allowing for partial membership to multiple classes \cite{udupa1996} with varying degree \cite{zhang2004}.
Accordingly, the matrix $\mathbf{U} \in \mathbb{R}^{C \times N}$ denotes a fuzzy $C$-partition of the set $\mathcall{X}$ using $C$ membership functions $\mu_i : \mathcall{X} \rightarrow [0,1]$, whose values $u_{ik} := \mu_i(\mathbf{x}_k)\in[0,1]$ represent the similarity membership of each element $\mathbf{x}_k$ to the $i$-th fuzzy set (i.e., the cluster $\mathcall{Y}_i$) and have to meet the following conditions:

\begin{equation}
    \label{eq:FCMconds}
     \begin{cases}
       0 \leq u_{ik} \leq 1 \\
       \sum\limits_{i=1}^C u_{ik} = 1, \forall k \in \{1, \ldots, N\} \\
       0 < \sum\limits_{k=1}^N u_{ik} < N, \forall i \in \{1, \ldots, C\}
     \end{cases}.
\end{equation}

To clarify, the sets of all hard and fuzzy $C$-partitions of the input dataset $\mathcall{X}$ are defined by \ref{eq:Mhard} and \ref{eq:Mfuzzy}, respectively:

\begin{equation}
    \label{eq:Mhard}
     \mathcall{M}_{\text{hard}} = \left\{ \mathbf{U} \in \mathbb{R}^{C \times N}: u_{ik} \in \{0,1\} \right\},
\end{equation}

\begin{equation}
    \label{eq:Mfuzzy}
     \mathcall{M}_{\text{fuzzy}} = \left\{ \mathbf{U} \in \mathbb{R}^{C \times N}: u_{ik} \in [0,1] \right\},
\end{equation}

\noindent where $\mathcall{M}_{\text{fuzzy}}$ improves the information conveyed by $\mathcall{M}_{\text{hard}}$, allowing for more flexibility than the crisp $K$-Means approach \cite{huang2012}.
As a matter of fact, introducing the fuzzy sets for the partial membership of feature vectors keeps more information from the original image compared to the hard segmentation methods \cite{pham1999}.

The key concept underlying the FCM clustering can be formalized by introducing a set of $D$-dimensional prototype vectors $\mathcall{V} = \left\{ \mathbf{v}_1, \mathbf{v}_2, \ldots, \mathbf{v}_C \right\}$, called centroids, which are associated with the $C$ clusters.
The membership function values are assigned to each feature vector in the data matrix $\mathcall{X}$ according to the relative distance of each feature vector $\mathbf{x}_k$ from the prototype points in $\mathcall{V}$.
The centroids are not forced to coincide with input data points.
So, this clustering algorithm aims to optimize the objective function (also called distortion measure) in Eq. (\ref{eq:LSE_FCM}), representing a generalized least-squares error function:

\begin{equation}
    \label{eq:LSE_FCM}
     \mathcal{J}_m(\mathbf{U}, \mathcall{V}; \mathbf{X}) = \sum\limits_{i=1}^C \sum\limits_{k=1}^N (u_{ik})^m \cdot d_{ik},
\end{equation}

\noindent where:
\begin{itemize}
    \item $m \in [1, \infty)$ is the fuzzification constant that weighs the fuzziness of the classification process. If $m=1$, the FCM algorithm approximates the crisp $K$-means version.
    The role of the weighting exponent $m$ in the FCM model was analyzed in \cite{pal1995} by evaluating the quality of inferences about the validity of FCM $(\hat{\mathbf{U}}, \hat{\mathcall{V}})$ pairs.
    The authors suggested that the best choice for $m$ is in the interval $[1.5, 2.5]$, whose mean and midpoint $m = 2$ is the most used value; 
    \item $d_{ik} = \norm{\mathbf{x}_k - \mathbf{v}_i}^2$ is the squared distance between the elements $\mathbf{x}_k$ and $\mathbf{v}_i$, computed by means of an induced norm $\norm{\cdot}$ on $\mathbb{R}^D$  (usually the Euclidean $\ell_2$ norm).
\end{itemize}

Therefore, the data matrix $\mathbf{X}$ is partitioned by iteratively searching for the optimal fuzzy partition $\mathcall{P}$, denoted by the pair  $(\hat{\mathbf{U}}, \hat{\mathcall{V}})$, which minimizes the objective function  $\mathcal{J}_m$ by means of a local optimization technique.
During the $t$-th iteration, the prototype set $\hat{\mathcall{V}}^{(t)}$, storing the centroid values $\hat{\mathbf{v}}_i^{(t)}$, as well as the elements of the matrix $\hat{\mathbf{U}}^{(t)}$ are estimated and updated according to Eqs. (\ref{eq:v_update}) and (\ref{eq:v_update}), respectively:

\begin{equation}
    \label{eq:v_update}
     \hat{\mathbf{v}}_i^{(t)} = \frac{\sum\limits_{j=1}^N \left( \hat{u}_{ij}^{(t)} \right)^m \mathbf{x}_j }{\sum\limits_{j=1}^N \left( \hat{u}_{ij}^{(t)} \right)^m}
\end{equation}

\begin{equation}
    \label{eq:u_update}
     \hat{u}_{ik}^{(t)} = \left[ \sum\limits_{j=1}^C \left( \frac{\norm{\mathbf{x}_k - \hat{\mathbf{v}}_i^{(t)}}}{\norm{\mathbf{x}_k - \hat{\mathbf{v}}_j^{(t)}}} \right)^{\frac{2}{m-1}}
     \right]^{-1}, \text{with } m>1 \text{ and } \mathbf{x}_k \neq \hat{\mathbf{v}}_j^{(t)} \text{ } \forall j,k.
\end{equation}

Afterwards, each object $\mathbf{x}_k$ is compared with the elements of the centroid vector and is then mapped to the nearest cluster.
The iterative procedure ends when the convergence condition (i.e., the minimum improvement in objective function between two consecutive iterations $\hat{\mathbf{U}}^{(t+1)}$ and $\hat{\mathbf{U}}^{(t)}$ is less than a fixed tolerance value $\epsilon_{\text{tol}}$) or the maximum number of allowed iterations $T_{\text{max}}$ is achieved.

\begin{algorithm}
	\caption{Pseudo-code of the FCM clustering algorithm.}
	\label{pc:FCM}
	\textbf{Input:} Data matrix $\mathbf{X} \in \mathbb{R}^{N \times D}$ that contains the $N$ feature vectors\\
	\textbf{Output:} Fuzzy partition $\mathcall{P}$ into $C$ clusters $\mathcall{M}_{\text{fuzzy}} = \left\{ \mathbf{U} \in \mathbb{R}^{C \times N}: u_{ik} \in [0,1] \right\}$, identified by the prototype set $\mathcall{V}$\\
	\begin{algorithmic}[1]
		\LeftComment Initialization using a uniform pseudorandom number generator $\text{Uniform}(0,1)$
		\State $u_{ij} \sim \text{Uniform}(0,1), \forall u_{ij} \in \mathbf{U}^{(0)}$
		\State $\mathbf{v}_{i,k} \sim \text{Uniform}(0,1), \forall \mathbf{v}_{i} \in \mathcall{V}^{(0)}, k \in \{1, 2, \ldots, C\}$
		\LeftComment Iterate until convergence is not achieved or $T_{\text{max}}$ iterations are executed
		\While{$\left( \left|\mathcal{J}_m^{(t)} - \mathcal{J}_m^{(t-1)} \right|  > \epsilon_{\text{tol}} \textbf{ or } t \leq T_{\text{max}} \right)$}
		    \ForEach{$\mathbf{x}_k \in \mathcall{X}$}
		        \State $\Call{UpdateCentroids}{\hat{\mathcall{V}}^{(t)}}$ \Comment Update the centroids $\hat{\mathcall{V}}^{(t)}$ Eq. (\ref{eq:v_update})
		       \State $\Call{UpdateMemberships}{\hat{\mathbf{U}}^{(t)}}$ \Comment Update the membership matrix $\hat{\mathbf{U}}^{(t)}$ Eq. (\ref{eq:u_update})
		    \EndFor
    		\State $\Call{UpdateObjFunc}{\mathcal{J}_m^{(t)}}$ \Comment Update the objective function $\mathcal{J}_m^{(t)}$ Eq. (\ref{eq:LSE_FCM})
    		\State $\hat{\mathcall{V}}^{(t)} \gets \hat{\mathcall{V}}^{(t+1)}$
            \State $\hat{\mathbf{U}}^{(t)} \gets \hat{\mathbf{U}}^{(t+1)}$
            \State $\mathcal{J}_m^{(t)} \gets \mathcal{J}_m^{(t+1)}$
            \State $t \gets t+1$
		\EndWhile
	\end{algorithmic}
\end{algorithm}

\subsection{Brain tumor segmentation in neuro-radiosurgery}
\label{sec:brainTumorSeg}

\begin{figure}
	\centering
	\includegraphics[width=0.6\linewidth]{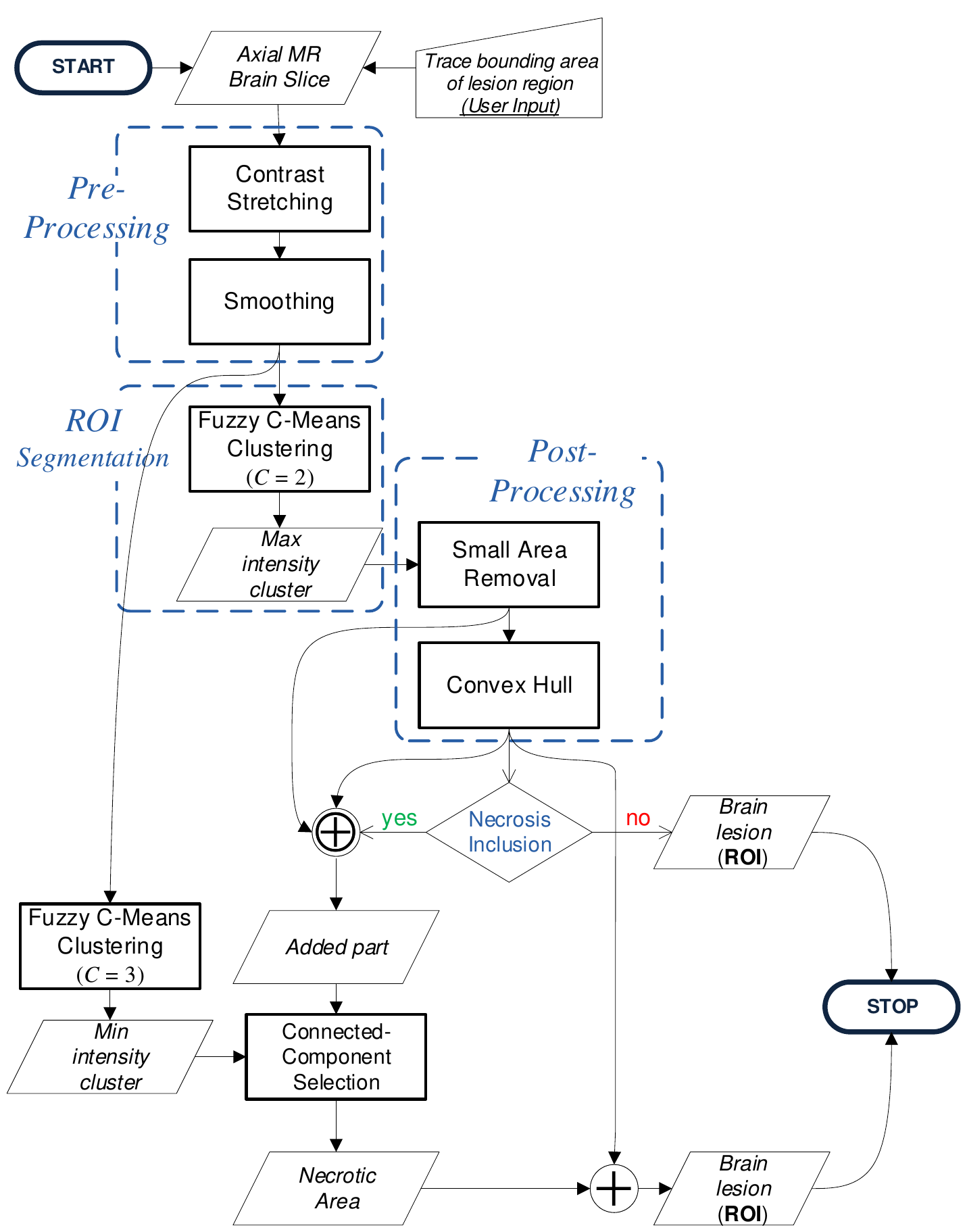}
	\caption[Flow diagram of the FCM-based brain lesion segmentation method]{Flow diagram of the FCM-based brain lesion segmentation method.}
	\label{fig:GK-flowDiagram}
\end{figure}

Hereby, to provide the proper prerequisites on brain lesion segmentation in neuro-radiosurgery, the semi-automatic GTV delineation method published in \cite{militelloIJIST2015,rundo2016WIRN} is recalled.
The operator-dependence is minimized, because user intervention is just limited to select an area containing the tumor and no parameter setting is required.
As illustrated in \cite{joe1999}, semi-automated approaches provide more reproducible measurements compared to conventional manual tracing. This was proved by quantitatively comparing intra- and inter-operator reliability of the two methods.
The flow diagram of the overall segmentation method is shown in Fig. \ref{fig:GK-flowDiagram}.

\subsubsection{Dataset description}
The study was performed on a dataset made up of $15$ patients with brain tumors to be treated with Leksell Gamma Knife\textsuperscript{\textregistered} model C at Cannizzaro Hospital (Catania, Italy).
All the MRI series were acquired on a Gyroscan Intera $1.5$ T MRI scanner (Philips Medical System Eindhoven, the Netherlands), before treatment, for the planning phase.
The total number of processed lesions was $20$ and the average age of the examined patients was $57.80 \pm 10.75$ years.
MR images are T1w Fast Field Echo (“T1w FFE”) CE sequences.
Imaging data were encoded in the $16$-bit DICOM format.
Thanks to the Gadolinium-based contrast agent, brain lesions appeared as hyper-intense zones.
Sometimes a dark area might be present due to necrotic tissue.
The main characteristics of the input data are reported in Table \ref{table:GK-MRIcharacteristics}.
This retrospective research study had no implication on the Leksell Gamma Knife\textsuperscript{\textregistered} treatments involving the enrolled patients.
No information concerning the subjects was accessed; consequently, institutional review board approval was not sought: the proposed image analysis is performed off-line and thus did not change the current treatment protocol.

\begin{table}[!t]
	\caption[MRI acquisition parameters of the brain tumor dataset]{MRI acquisition parameters of the brain tumor dataset.}
	\label{table:GK-MRIcharacteristics}
	\begin{scriptsize}
		\centering
		\begin{tabular}{ccccccc}
			\hline\hline
			MRI sequence	& TR [ms]	& TE [ms]	& Matrix size [pixels]	& Slice spacing	[mm]	& Slice thickness [mm]	& Pixel spacing [mm] \\
			\hline
			T1w FFE			& $25$		& $1.808$-$3.688$ 		& $256 \times 256$ 	& $1.5$ 	& $1.5$ 	& $0.977$-$1.0$ \\
			\hline\hline
		\end{tabular}
	\end{scriptsize}
\end{table}

\begin{figure}
	\centering
	\includegraphics[width=\linewidth]{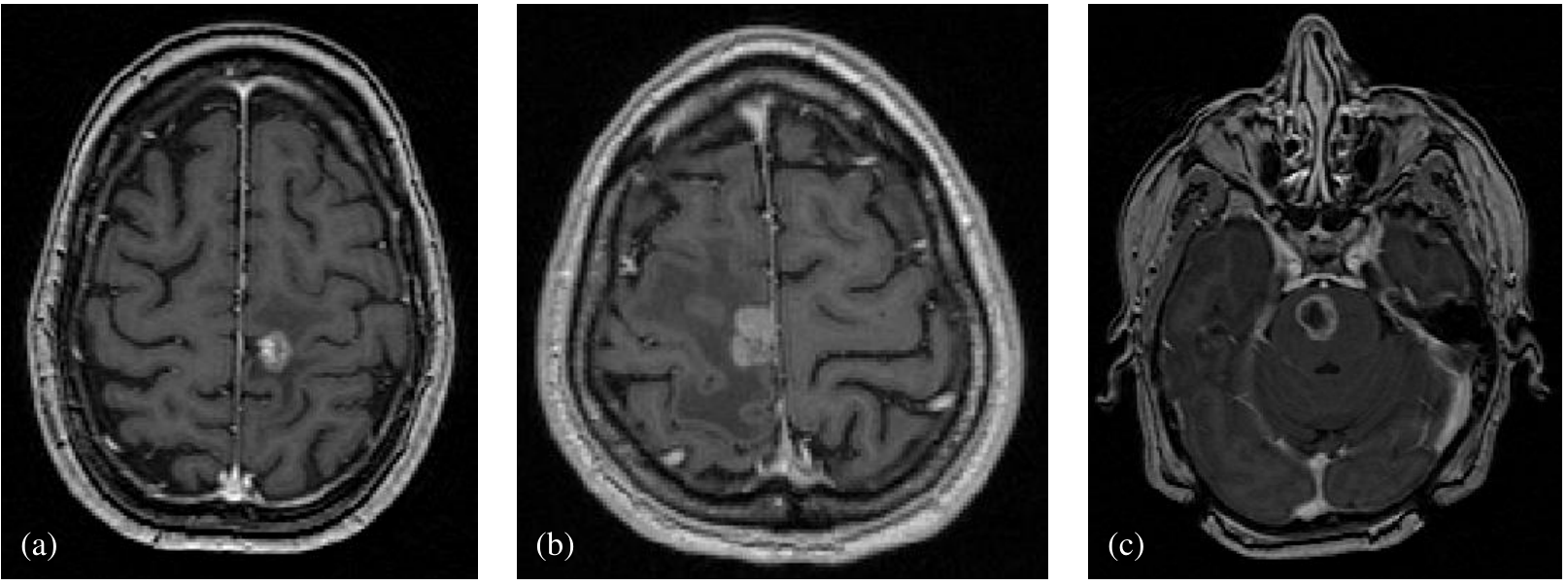}
	\caption[Examples of brain MR axial slices]{Examples of brain MR axial slices: (a) lesion with non-homogeneous enhancement region; (b) irregular tumor shape; (c) tumor with inner necrosis (dark area).}
	\label{fig:GK-brainTumors}	
\end{figure}

These MRI series contain varying levels of segmentation difficulty, with the worst scans consisting of low contrast and large intensity inhomogeneities.
Lesions with a non-homogeneous enhancement region, tumors with irregular shape or with inner necrosis are also present.
Therefore, these case studies represent a meaningful sample of the possible real imaging data.
Representative samples of input brain MR images are shown in Fig. \ref{fig:GK-brainTumors}.

\subsubsection{FCM-based brain tumor segmentation}

\paragraph{ROI selection tool}
The area containing the lesion must be localized within the brain by the operator.
Indeed, the users of medical imaging applications select in the image a specific area, i.e., drawing a boundary around the ROI.

\begin{figure}
	\centering
	\includegraphics[width=0.6\linewidth]{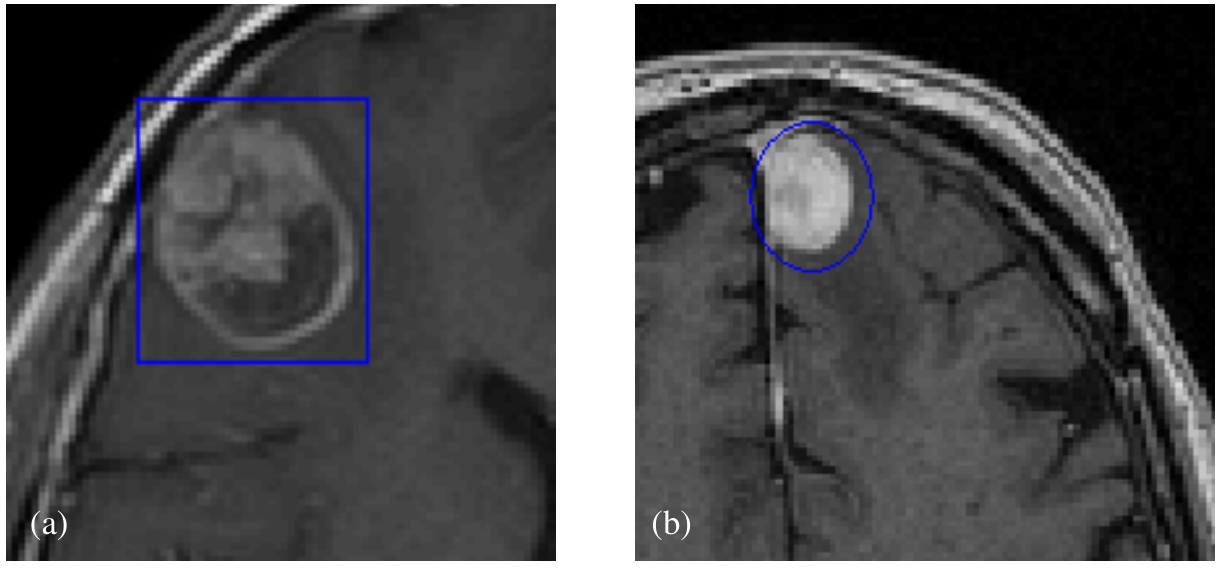}
	\caption[ROI selection tools for brain tumor segmentation]{ROI selection tools for brain tumor segmentation: (a) rectangular versus (b) ellipsoidal regions.}
	\label{fig:GK-selectionTools}	
\end{figure}

Despite the fact that several selection tools are available in current clinical software, often they implement unnecessary or inefficient functions.
On the basis of the evidence reported in \cite{chen2009}, rectangular and free-hand ROI selection tools (bounding region of the lesion) were developed and tested.
When the lesion is located near the cerebral cortex, a rectangular shaped selection tool could include some anatomic parts characterized by similar gray levels (i.e., skull bones or cerebral cortex), as shown in Fig. \ref{fig:GK-selectionTools}a.
Thereby, the clustering algorithm may include, erroneously, these areas into the segmented ROI.
Problems with an ellipsoidal selection tool also occur, if the lesion is near the superior sagittal sinus (both anterior and posterior extremities), as shown in Fig. \ref{fig:GK-selectionTools}b.
Moreover, rectangular o circular settings \cite{rundo2016WIRN} (i.e., resizing and moving) are more complex than free-hand selection during lesion enclosing under critical conditions.
In order to avoid this issue, a free-draw selection enables a more precise contour containing the tumor \cite{militelloIJIST2015}.
In this way, result repeatability is ensured regardless of the form of free-shape bounding region.
As a result, the operator-dependence of the proposed segmentation approach is reduced.
After the freehand drawing, this selected part is then enclosed and zero-padded into the minimum box in order to deal with a rectangular image during the following processing phases.

\paragraph{Pre-processing}
In order to improve segmentation results, some pre-processing operations are applied on masked input images after the ROI selection by the user.
First of all, the range of intensity values of the selected part is expanded to extract the ROI more easily.
Indeed, this operations is performed by means of unsupervised clustering.
Thereby, a linear contrast stretching operation is applied.

\paragraph{FCM-based ROI segmentation}
The FCM algorithm is applied just on pixels included in the drawn bounding region and based upon image intensity levels.
Therefore, the zero-valued background pixels, added with zero-padding, are never considered.
Two clusters are employed ($C=2$) during this step in order to suitably distinguish hyper-intense lesion from the healthy part of the brain.
As a matter of fact, in contrast-enhanced MR brain scans, brain metastases have a brighter core than periphery and distinct borders from the surrounding normal brain tissue \cite{ambrosini2010}.
Finally, during the defuzzification process, the pixels that have the maximum membership with the brightest cluster are selected.

\paragraph{Post-processing}
In some cases, central areas of the lesions could comprise necrotic tissue. During the treatment planning phase, also these necrotic areas (hypo-intense regions) must be included in the target volume.
In order to delete any unwanted connected-components included by the highest intensity cluster, a small area removal operation is employed. These small regions may be due to anatomic ambiguities when the lesion is situated near high valued pixels, i.e. cranial bones or the corpus callosum.
Brain lesions have a nearly spherical or pseudospherical appearance  \cite{ambrosini2010}.
Owing to this fact, a convex hull algorithm \cite{zimmer1997} is used to envelope the segmented lesion into the smallest convex polygon that contains the automatically segmented region.
Furthermore, this operation comprises also hole filling, in order to include any necrosis inside the segmented ROI.
Even when the FCM clustering output is an arch-shaped ROI, perhaps due to poor contrast images, the internal necrosis is properly included by the convex hull execution.

\paragraph{Necrotic area inclusion}
Sometimes, because of images with necrotic tumors and characterized by low contrast, the FCM clustering process might not extract a region such that the following post-processing is not able to include internal or adjacent necrosis.
A particular sequence of steps must be thus defined:
\begin{enumerate}
    \item to detect the necrosis, imaged as a hypo-intense area, the FCM clustering is again applied on the pre-processed image, but this time with $C=3$ clusters. Thereby, the pixels that have the highest membership grade with the minimum intensity cluster are selected;
    \item logical exclusive disjunction (XOR) is calculated on the ROI masks before and after convex hull algorithm execution. The added part by the convex hull operation is thus identified;
    \item the only connected-component, whose intersection with the added part identified by Step 2 has a non-zero value, is selected and represents the desired necrotic area;
    \item lastly, the necrotic area is added to the previous post-processing output (by means of logical OR between the two binary masks), and this represents the resulting binary mask.
\end{enumerate}

\begin{figure}[t]
	\centering
	\includegraphics[width=\linewidth]{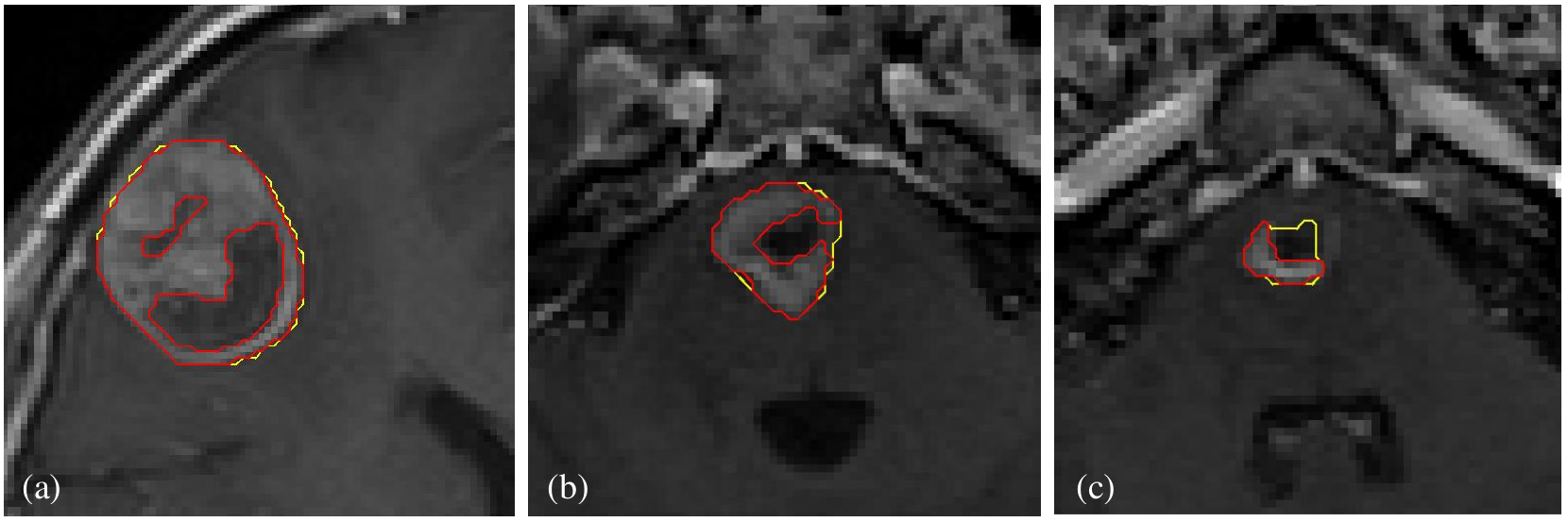}
	\caption[Necrotic area handling examples]{Necrotic area handling examples: (a) ROI segmentation is refined using post-processing (internal necrotic “holes” are filled); (b) arch-shaped lesion is correctly segmented with post-processing steps; (c) necrosis is included using the post-processing steps. Both the boundaries without (red contour) and with the refinements (yellow contour) are superimposed (all images are zoomed by a $4\times$ factor).}
	\label{fig:GK-necrAreaIncl}	
\end{figure}

A very critical case is reported in Fig. \ref{fig:GK-necrAreaIncl}c, where the necrotic hypo-intense area is adjacent to the hyper-intense region of the tumor.
The necrosis is thus correctly included in the resulting output.
As shown in the flow diagram concerning the proposed method (Fig. \ref{fig:GK-flowDiagram}), the physician can select, through a check button, the execution of the necrotic area inclusion steps mentioned above when considered as appropriate.

\subsubsection{Experimental results}
To assess the effectiveness of the proposed approach, several comparative measures between the automatically segmented volume, obtained by the proposed method, and those manually drawn by a neurosurgeon (considered as our gold standard) are provided (for more details see Appendix \ref{sec:segEval}).
Fig. \ref{fig:GK-brainTumRes} shows the segmentation results of the three MR slices reported in Fig. \ref{fig:GK-brainTumors}.

\begin{figure}[t]
	\centering
	\includegraphics[width=\linewidth]{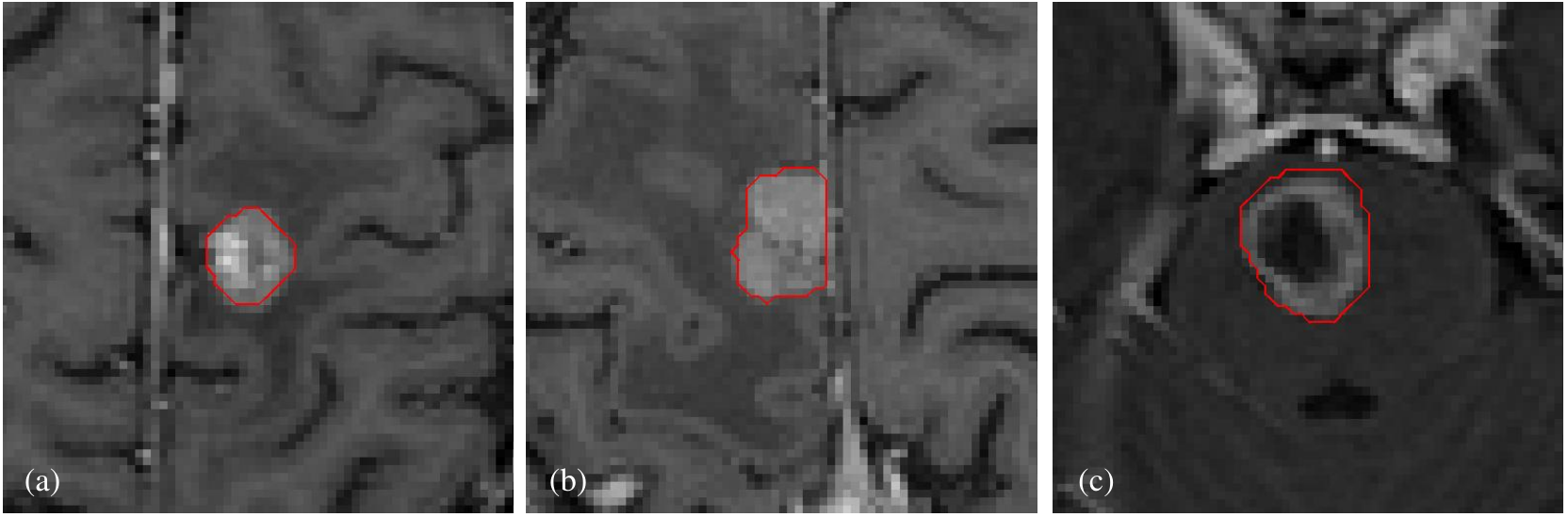}
	\caption[Results achieved by the FCM-based segmentation method]{Results achieved by the FCM-based segmentation method on the input images shown in Fig. \ref{fig:GK-brainTumRes}(a-c), respectively.}
	\label{fig:GK-brainTumRes}	
\end{figure}

\begin{figure}[!t]
\centering
	\subfloat[][]
    {\includegraphics[width=0.5\textwidth]{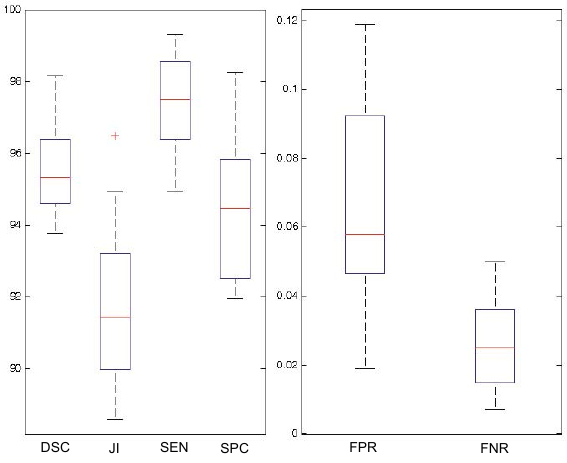}} \\
    \subfloat[][]
    {\includegraphics[width=0.25\textwidth]{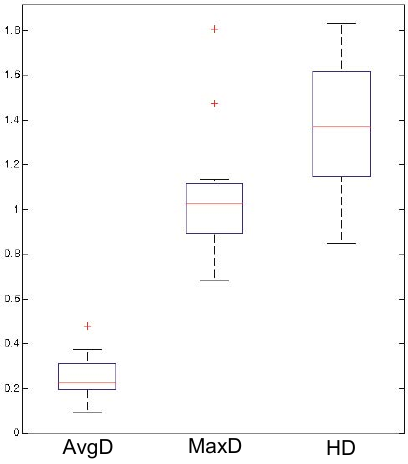}} \\
  \caption[Boxplots of the evaluation metrics results of the FCM-based brain tumor segmentation approach]{Boxplots of the evaluation metrics results of the FCM-based brain tumor segmentation approach: (a) overlap-based metrics; (b) distance-based metrics.}
\label{fig:GK-boxplots}
\end{figure}

The achieved mean values of evaluation metrics (calculated on each dataset) are reported in the boxplots (Fig. \ref{fig:GK-boxplots}).
The short length of the boxplot in Fig. \ref{fig:GK-boxplots}a  means that values are very concentrated.
All index values do not present outliers, thus demonstrating extremely low variability (whisker value is $1$ in all cases).
Fig. \ref{fig:GK-boxplots}b depicts the boxplots of distance-based evaluation metrics.
Even though there are outliers in boxplots (the whisker is set to $1$ pixel, which is the smallest imaging discrete element), they still indicate only a small deviation between the segmentations of the proposed method and those of the radiologist.
Considering the surgery purpose of the application, referring to MRI spatial resolution (see characteristics in Table \ref{table:GK-MRIcharacteristics}), all distance-based metrics mean values are in the range of $1$ mm.
This confirms the great validity and accuracy of the proposed segmentation method.

\begin{figure}[t]
	\centering
	\includegraphics[width=0.7\linewidth]{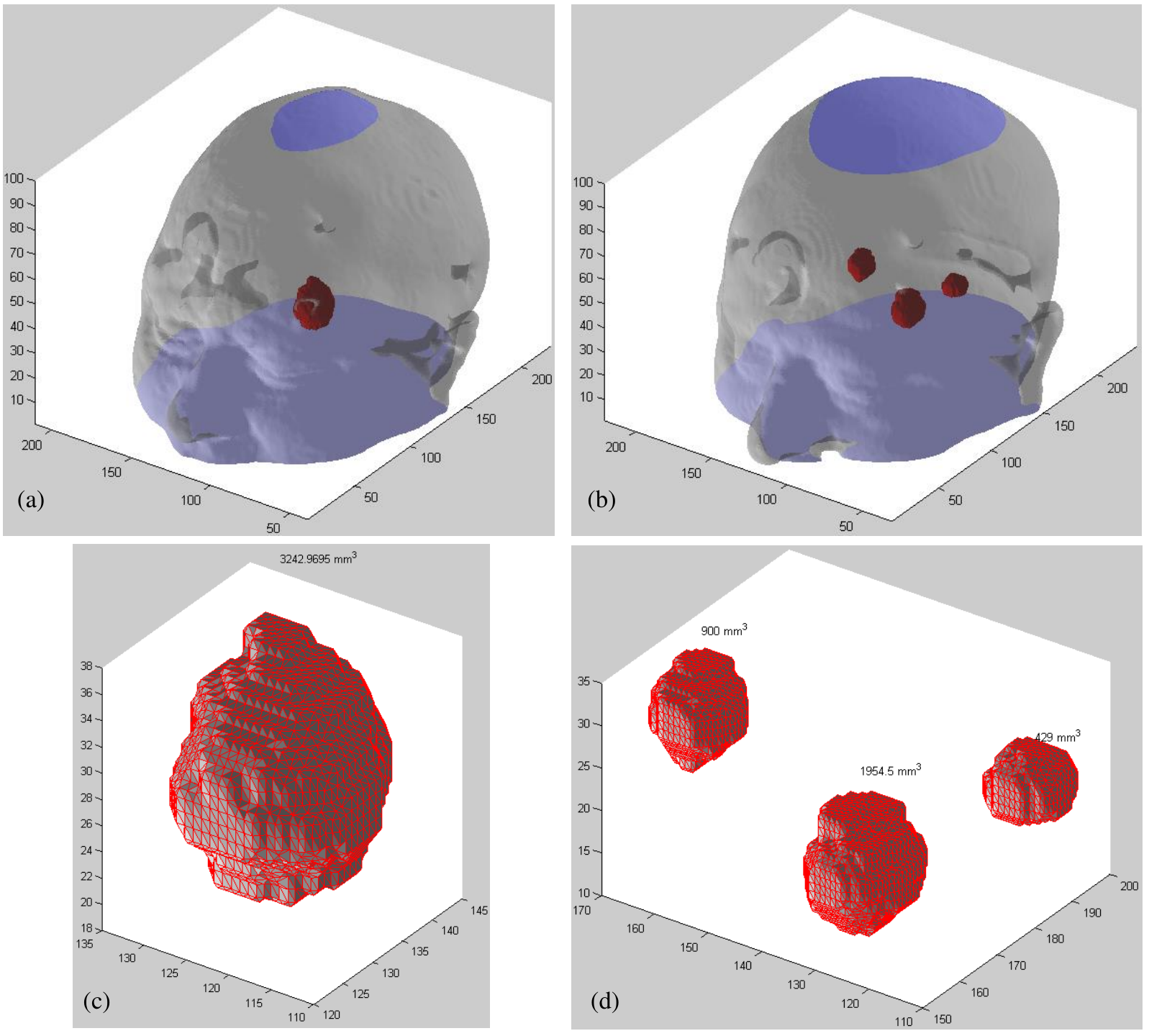}
	\caption[Volume rendering of GTVs]{Volume rendering of GTVs: (a) treated patient with single cerebral lesion, shown in detail in (c); (b) multi-metastasis patient, shown in detail in (d). In both cases, the patient’s head is represented by means of a transparent surface (with  $\alpha = 0.45$), and internal solid tumors (red volumetric models) are visible in their own actual position with respect to the tridimensional reference system. Sub-figures (c) and (d) report also the volume (in mm$^3$) of each segmented lesion.}
	\label{fig:GK-volRendering}	
\end{figure}

Fig. \ref{fig:GK-volRendering} displays two instances of tridimensional reconstruction of the tumors.
In both cases, the patient’s head (extracted using a skull stripping algorithm) is represented by means of a transparent surface (with $\alpha = 0.45$), and internal solid tumors (red volumetric models) are visible in their own actual position with respect to the tridimensional reference system.

For comparison, we developed four different semi-automatic segmentation approaches:
\begin{itemize}
    \item edge detection \cite{canny1986}, after bounding region selection;
    \item adaptive thresholding \cite{otsu1975}, after bounding region selection;
    \item region growing \cite{militello2013}, after seed-point selection;
    \item interactive level set implementing the Distance Regularized Level Set Evolution (DRLSE) method \cite{li2010DRLSE}, after initial region selection.
\end{itemize}

All the implementations were tested on our MRI datasets.
The Canny method fails to successfully segment brain tumors because false edges are wrongly selected when non-homogeneous lesions are processed.
Moreover, sometimes a totally unbroken edge is not returned, due to not connected edge curves.
Lesion segmentation problems might occur with arch-shaped lesions, where an internal necrosis is present.
Necrotic area inclusion is more difficult with respect to region-based segmentation.
Even edge-linking \cite{ghita2002} does not correctly detect the lesion.
Since the FCM algorithm classifies input data, the achieved clusters are represented by connected-regions and the corresponding boundaries are closed even when holes are present. This fact is very important in radiotherapy or radiosurgery treatment planning, because lesion segmentation is a critical step that must be performed accurately and effectively.

For these reasons, we quantitatively compared the segmentation results obtained by our method to the aforementioned segmentation algorithms able to detect a totally unbroken edge (i.e., adaptive thresholding, region growing, interactive level set).
Considering this, the Canny method was not considered in the comparison.
All the developed approaches implement the same pre-processing and post-processing steps provided by the proposed segmentation method. Values of the main overlap-based and distance-based metrics, calculated for each segmentation method, are reported in Table \ref{table:BGK-Results}.

\begin{table}[!t]
\centering
	\caption[Comparison among the developed brain lesion segmentation methods on the same $15$ patients treated with Gamma Knife neuro-radiosurgery]{Comparison among the developed brain lesion segmentation methods on the same $15$ patients treated with Gamma Knife neuro-radiosurgery. The results are expressed as average value $\pm$ standard deviation.}
	\label{table:BGK-Results}
	\begin{tiny}
		\begin{tabular}{lccccccc}
			\hline\hline
			Method	& DSC	& JI	& SEN	& SPC	& AvgD	& MaxD & HD \\
			\hline
			Otsu's method \cite{otsu1975} &	$79.59 \pm 6.62$ &	$68.78 \pm 8.36$ &	$90.49 \pm 8.62$ &	$73.33 \pm 7.65$ &	$1.179 \pm 0.77$ &	$2.534 \pm 1.17$ &	$1.910 \pm 0.37$ \\
			Region growing \cite{militello2013} &	$79.78 \pm 10.93$ &	$70.72 \pm 12.67$ &	$77.64 \pm 11.74$ &	$87.61 \pm 8.54$ &	$1.079 \pm 1.02$ &	$2.425 \pm 2.04$ &	$1.696 \pm 0.28$ \\
			Interactive Level Set \cite{li2010DRLSE} &	$65.16 \pm 18.06$ &	$53.47 \pm 19.05$ &	$76.06 \pm 11.88$ &	$67.97 \pm 23.96$ &	$2.291 \pm 1.91$ &	$4.720 \pm 3.39$ &	$2.021 \pm 0.40$ \\
			FCM \cite{militelloIJIST2015} &	$95.59 \pm 1.28$ &	$91.86 \pm 2.28$ &	$97.39 \pm 1.42$ &	$94.30 \pm 2.04$ &	$0.246 \pm 0.10$ &	$1.050 \pm 0.28$ &	$1.365 \pm 0.31$ \\
			\hline\hline
		\end{tabular}
	\end{tiny}
\end{table}

Several difficulties arose during the trials of the developed segmentation approaches.
Using adaptive thresholding, it is not possible to handle necrotic area inclusion not even with post-processing operations.
In fact, the hypo-intense area cannot be successfully extracted neither through another thresholding operation. On the other hand, leaking occurs with LSFs in low contrast datasets as well as in initial/final slices of the tumor to be treated.
In such cases the actual ROI is over-estimated.
The interactive level set method is highly affected by operator dependence due to the initial region selection.
In addition, a careful parameter tuning must be performed depending on the either image contrast or the tumor region being processed (in particular, possible necroses or arch-shaped lesions).
Weak ROI boundaries may sometimes cause leakage regardless of the parameter settings.
For instance, if a tumor is adjacent to skull bones, the ROI could pass through the actual boundaries.
This is evidenced by the false positive presence, as demonstrated by the higher sensitivity (\emph{SEN}) than specificity (\emph{SPC}).
Region growing also requires a threshold selection according to image features.
This task could be error-prone and tedious.
However segmentation issues occur when there are non-homogeneties or loosely connected enhanced tumor regions (due to internal necroses).
Ambiguities are encountered when the lesion is adjacent to a hyper-intense imaged anatomic area (e.g., sagittal sinus extremities) because the ROI grows towards it.
The achieved distance-based metrics are consistent with overlap-based measures, thus confirming the above discussions
Overall, the proposed segmentation method obtains the best performance with respect to the other implemented approaches.

\subsubsection{Discussion}
A GTV delineation method based on unsupervised FCM clustering was presented.
The developed approach was applied to the planning phase of Gamma Knife treatments for brain lesions.
To evaluate the performance of the proposed approach, segmentation tests on $15$ MRI sequences were performed.
The obtained results, in terms of spatial overlap-based and distance-based metrics, demonstrated the great robustness of the proposed approach even when the MRI series were affected by acquisition noise.
The good segmentation performance was also confirmed with poor contrast datasets or lesions characterized by irregular shapes.
For further discussion, refer to Section \ref{sec:GTVcut}.

\subsection{NeXt: necrosis extraction in neuro-radiosurgery}
\label{sec:necrosisSeg}

This approach distinguishes necrotic and enhancement regions within the complete tumor region, which includes all tumor structures.
Our method is tailored for radiosurgery applications and designed according to the available MRI data used in clinical practice.
As a matter of fact, stereotactic neuro-radiosurgery treatment planning is generally performed on T1w CE-MR images alone by experienced neurosurgeons.
So, the proposed necrosis extraction approach could represent a feasible solution in clinical environments, wherein multispectral MRI data are generally not available.
From a computational perspective, the resulting automated segmentation approach represents an efficient Machine Learning solution.

NeXt should be considered as a second step after the GTV segmentation for a more precise brain cancer characterization.
We exploit an accurate brain tumor segmentation, obtained by our validated GTV segmentation methods for supporting neuro-radiosurgery treatment planning in \cite{militelloIJIST2015,rundo2017NC}.
These computer-assisted segmentation approaches---the former based on the FCM clustering algorithm and the latter on a CA model---have shown to be effective and operator independent solutions.
Especially, both GTV segmentation methods were designed \textit{ad hoc} to robustly deal with edemas and necroses, which are imaged differently to healthy brain tissues in MRI.
As a matter of fact, these hypo-intense areas must be included during the treatment planning phase in the target volume that is going to receive the radiation dose delivered by the neuro-radiosurgery system.
These interactive segmentation algorithms represent a more feasible and safe solution for physicians in clinical practice with respect to fully automatic approaches, as reported in \cite{hamamci2012}.
The overall processing pipeline, built by cascading brain tumor GTV segmentation and necrosis/enhancement region distinction within the yielded GTV, results in an integrated two-stage segmentation scheme to support neuro-radiosurgery therapy.

In practice, necrotic and enhancement region distinction in brain tumors on MR images is clinically valuable in radiation therapy treatment planning as well as in patient's follow-up.
Brain tumor necrosis extraction, yielded by NeXt, provides definitely more insights into the cancer nature and characteristics, gaining clinical significance in both radiation treatment planning and therapy outcome assessment to support personalized medicine \cite{rueckert2016}.
As a matter of fact, necrotic areas are generally characterized by hypoxia, which is strongly involved in several aspects of cancer development.

Since tumors are characterized by heterogeneous and biological properties---such as hypoxia, cell density, proliferation, and irregular vascularization---some regions are more treatment-resistant than others.
Thus, the potential of dose redistribution can gain relevant information by incorporating estimations of oxygen heterogeneity within imaging voxels for optimal dose determination \cite{petit2009}.
In this context, NeXt could be actually exploited for dose escalation, allowing for a more sophisticated strategy to selectively increase radiotherapy dose (i.e., boosting) in hypoxic radioresistant areas with respect to uniformly treat the whole tumor target volume \cite{even2015}.
NeXt is designed to be directly usable in radiation therapy scenarios, especially for neuro-radiosurgery treatments.
In clinical routine, multispectral MRI data are usually not available during neuro-radiosurgery planning phases \cite{hamamci2012}: conventional MRI protocols for radiation treatment planning include T1w and T1w CE-MR imaging of the head volume \cite{quinones2012}.
Accordingly, our approach analyzes CE T1w MR images alone.

\subsubsection{Patient dataset description}
\begin{figure}[t]
	\centering
	\includegraphics[width=\linewidth]{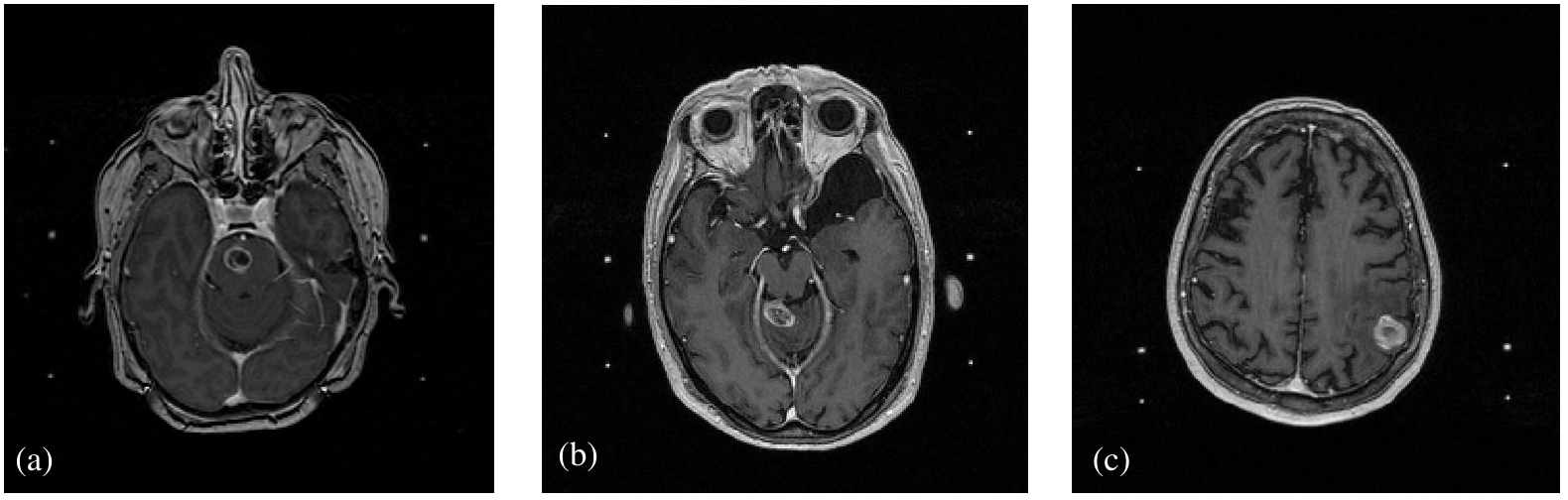}
	\caption[Three instances of input T1w CE-MR axial slices with the rectangular bounding area, selected by the user]{Three instances of input T1w CE-MR axial slices with the rectangular bounding area, selected by the user, containing the tumor: (a) moderately hyper-intense enhancing region including a necrotic core; (b) highly hyper-intense enhancing region with inner necrotic material; (c) heterogenous cancer.}
	\label{fig:NeXt-inputImages}	
\end{figure}

The study was performed on a dataset composed of $32$ brain metastatic cancers that underwent Gamma Knife neuro-radiosurgery.
Among them, $20$ brain tumors presented necrotic areas. These regions have to be carefully considered during treatment planning, since they include hypoxic areas that could lead to recurrent cancers and resistance to oncological therapies \cite{jensen2009}.
The characteristics of the analyzed T1w FFE CE-MR images are reported in Section \ref{sec:brainTumorSeg}.
MRI acquisition parameters are reported in Table \ref{table:GK-MRIcharacteristics}.
Three representative examples of input MR images are shown in Fig. \ref{fig:NeXt-inputImages}.

Typical MRI protocols, employed in stereotactic radiosurgery treatment planning, include T1w CE-MRI of the head volume \cite{quinones2012}.
After the administration of the Gadolinium-based contrast medium to the patient, the enhancing hyper-intense zones in the imaged tumors represent the increased and irregular neovascularity.
On the other hand, hypo-intense areas depict either cystic or necrotic tissues that do not absorb the contrast medium.

\subsubsection{The proposed automatic necrosis extraction approach}
\paragraph{Pre-processing}
Starting from the preliminary complete tumor region segmentation results that include all tumor structures, obtained by using our previous GTV segmentation methods \cite{militelloIJIST2015,rundo2016ACRI,rundo2017NC,rundo2016WIRN}, necrotic parts are distinguished from the tumor enhancing region.
The segmented GTV is firstly eroded by means of a selective strategy to choose a suitable structuring element \cite{breen2000}:
\begin{itemize}
    \item disk with $2$-pixel radius, if the GTV area is greater than $80$ pixels;
    \item disk with $1$-pixel radius, otherwise.
\end{itemize}
Using this robust strategy any hypo-intense region in the segmented GTV boundaries is removed, even when the GTV has been overestimated during the preliminary automatic brain tumor segmentation.
The input MR brain axial slice is then masked with the corresponding eroded GTV binary mask.
This masked image is then linearly stretched in the full dynamic range, exploiting the double precision values in $[0,1]$  and enhancing detail discrimination.
No smoothing operation is required, so avoiding the loss of imaging information about brain necroses.

\paragraph{Necrosis and enhancement region distinction}
In NeXt, the FCM algorithm is applied just on intensity values concerning the pixels included in previously segmented GTV tumor region (i.e.,  $\mathcall{I}_{\text{GTV}}$).
Two clusters ($C=2$) are employed for this segmentation task to distinguish hypo-intense necrotic parts from the enhancement region of the brain tumor.
Finally, to yield a quantifiable and clinically valuable measurement, the pixels that have the highest membership with the hypo-intense cluster are selected during the defuzzification process, so implementing a maximum membership segmentation scheme.

\paragraph{Post-processing}
Some morphological refinements \cite{soille2013} are also employed to enhance necrosis segmentation results and to prevent small hypo-intense region detection.
The post-processing pipeline, which exploits binary mathematical morphology techniques, is the following:
\begin{enumerate}
    \item a small area removal operation to remove small isolated hypo-intense regions and loosely connected pixels using $4$-connectivity (i.e., connected-components with area less than $5$ pixels). Especially, a $4$-connected neighborhood was adopted to consider diagonally connected pixels as different connected-components, so allowing for a more selective labeling strategy;
    \item a hole filling algorithm, based on morphological reconstruction \cite{vincent1993}, to remove possible hyper-intense regions in heterogeneous necrotic cores.
    Holes are determined as disjoint sets of background pixels that cannot be reached by a flood-filling algorithm starting from the background pixels in image borders.
\end{enumerate}

\subsubsection{Results}

The accuracy of the necrosis extraction results yielded by NeXt was quantitatively evaluated by means of the metrics defined in Appendix \ref{sec:segEval}.
Both spatial overlap-based and distance-based metrics were calculated for all tumor slices of each MR image series.
The proposed NeXt method was compared against the only other alternative necrosis/enhancement region segmentation method in literature, presented in Tumor-Cut \cite{hamamci2012}.
For a more complete assessment, a further comparison with the most commonly used computer-assisted image segmentation techniques in medical imaging was carried out.
Therefore, the following image processing methods were implemented and tested:
\begin{itemize}
    \item adaptive thresholding (Otsu's method \cite{otsu1975}), after the GTV segmentation;
    \item seeded region growing \cite{rundoMBEC2016}, after interactive seed-point selection;
    \item interactive LSFs \cite{li2010DRLSE}, after seed-point selection for defining the initial region to segment;
    \item Tumor-Cut based on a CA model \cite{hamamci2012}, after background and foreground seed initialization using Otsu's method \cite{otsu1975} according to the previous GTV segmentation.
\end{itemize}

The necrosis/enhancement region segmentation method presented in Tumor-Cut was implemented by strictly following the description in the original paper \cite{hamamci2012}.
For a fair comparison, the other developed approaches implement exactly the same pre-processing and post-processing pipelines used in NeXt.

\begin{figure}[t]
	\centering
	\includegraphics[width=\linewidth]{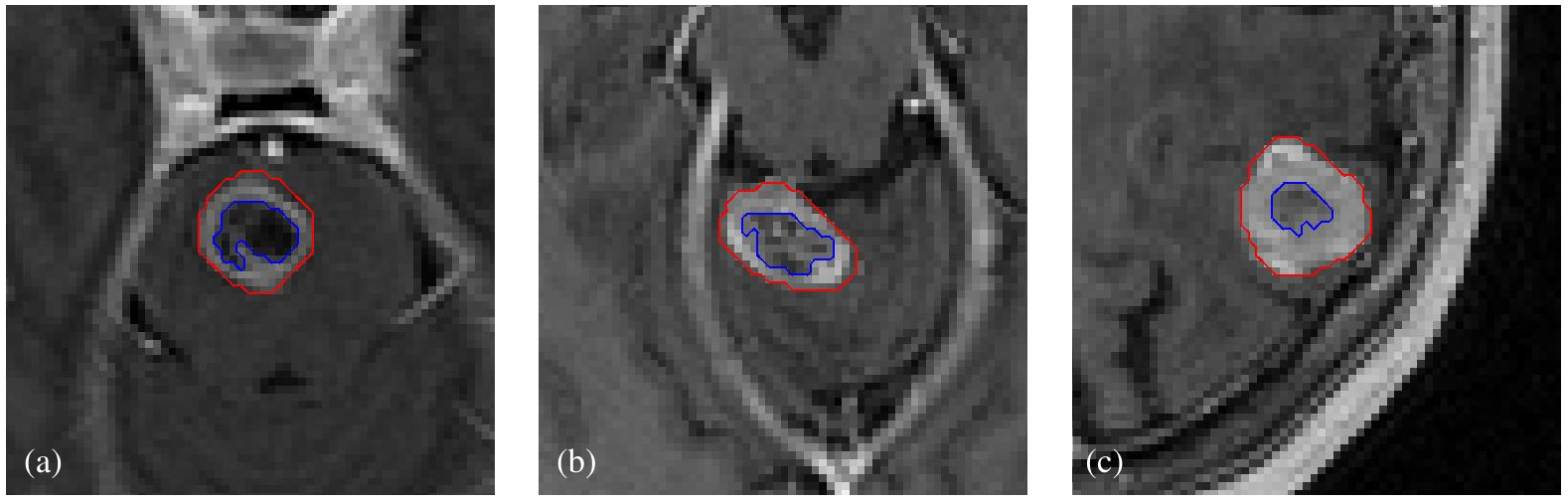}
	\caption[Necroses extracted from the brain tumors using the proposed NeXt approach]{Necroses (blue contour) extracted from the brain tumors (red contour) using the proposed NeXt approach, based on the FCM algorithm, on the images shown in Fig. \ref{fig:NeXt-inputImages}. ($4\times$ zoom factor).}
	\label{fig:NeXt-Results1}	
\end{figure}

\begin{figure}[t]
	\centering
	\includegraphics[width=\linewidth]{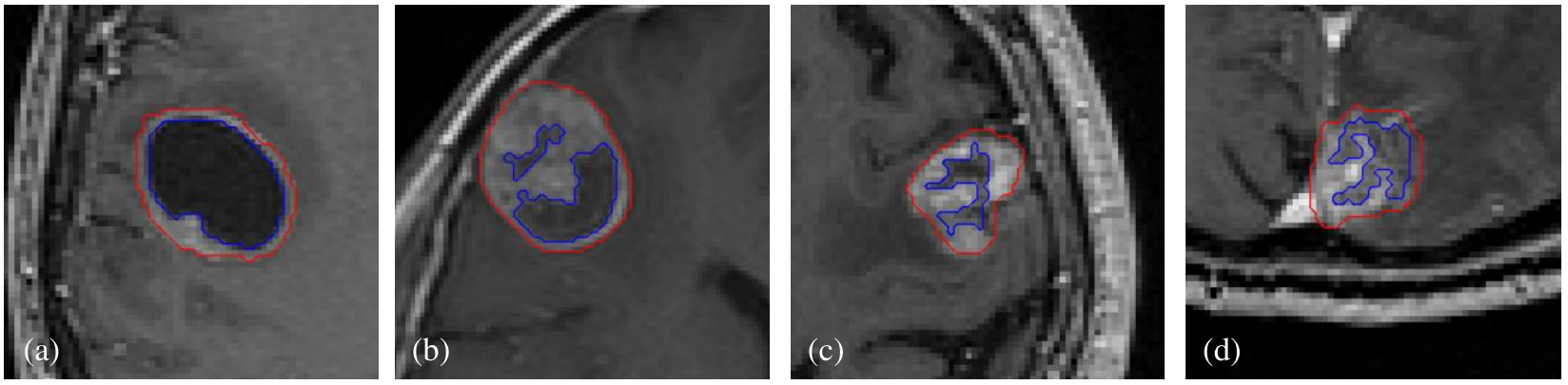}
	\caption[Necrosis segmentation results on representative brain tumors representative necrotic regions]{Necrosis segmentation results on representative brain tumors representative necrotic regions, characterized by very irregular shape or heterogeneous tissue, extracted from the brain tumors using NeXt: (a) homogeneous dark region enclosed by distinct hyper-intense enhancing rims; (b) disjoint necrotic area with diffused necrotic material in the tumor; (c, d) challenging necroses characterized by very irregular shape or heterogeneous tissue. ($4\times$ zoom factor).}
	\label{fig:NeXt-Results2}	
\end{figure}

Fig. \ref{fig:NeXt-Results1} shows the necroses extracted by NeXt from the MRI brain tumors in Fig. \ref{fig:NeXt-inputImages}.
Our approach, based on the FCM clustering algorithm, suitably segments also inhomogeneous necrotic regions characterized by irregular or non-convex shapes.
Additional necrosis segmentation results on representative brain tumors are displayed in Fig. \ref{fig:NeXt-Results2}.
NeXt appropriately detects homogeneous dark regions enclosed by distinct hyper-intense enhancing rims (Fig. \ref{fig:NeXt-Results2}a), as well as disjoint necrotic regions when necrotic material is diffused within the tumor (Fig. \ref{fig:NeXt-Results2}b).
In addition, NeXt correctly delineates challenging necrotic areas, characterized by very irregular shape or heterogeneous tissue (Fig. \ref{fig:NeXt-Results2}(c, d)).

\begin{figure}[t]
	\centering
	\includegraphics[width=0.7\linewidth]{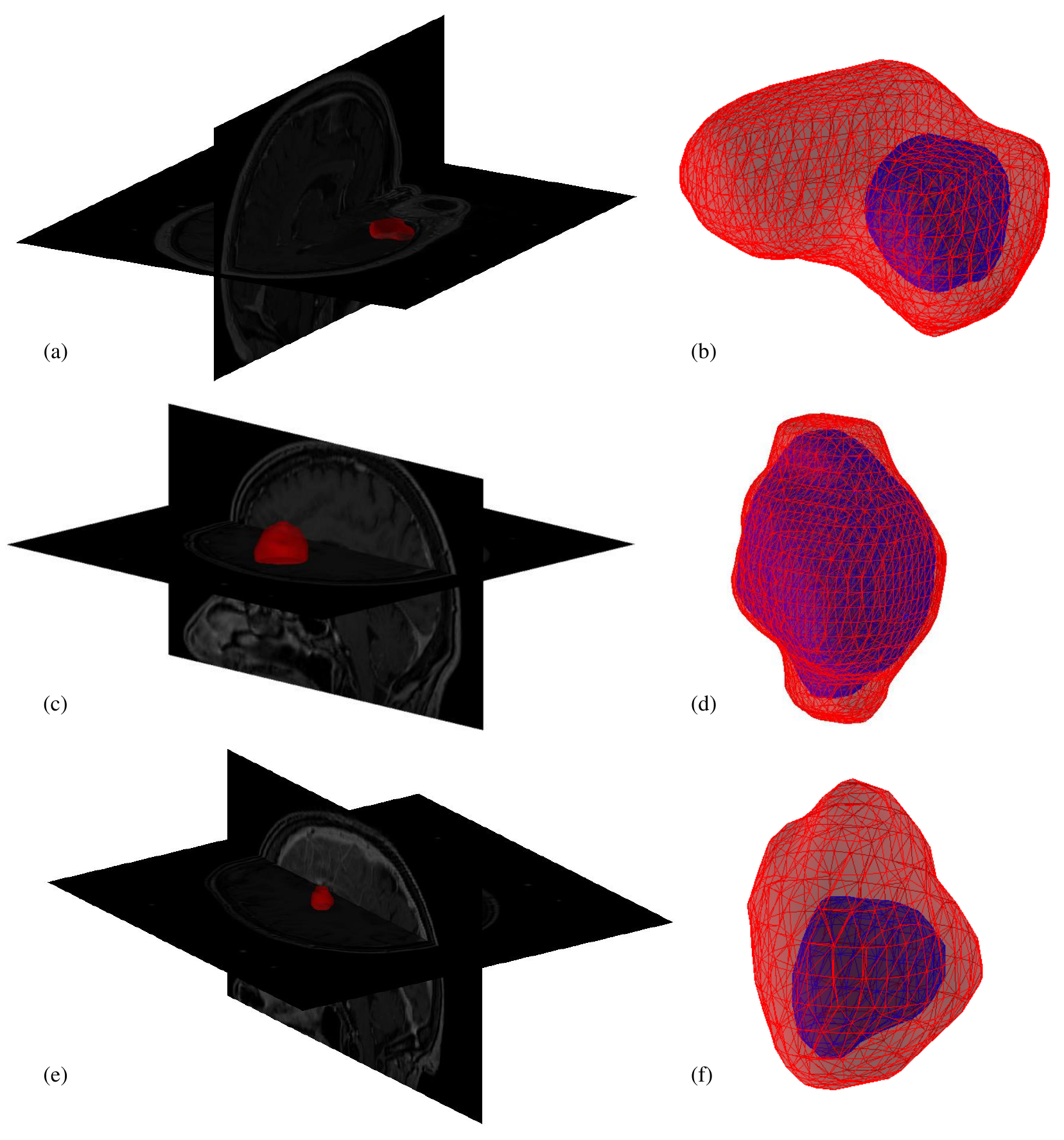}
	\caption[Volumetric reconstruction of brain tumors with necrotic core]{Volumetric reconstruction of brain tumors with necrotic core: (a, c, e) brain tumors (red volumes) represented in a tridimensional space using orthogonal axial and sagittal planes to show their different anatomic positions in the corresponding whole brain MRI series; (b, d, f) the GTV is rendered using a transparent red surface and the inner necrosis (blue volumetric model) is visible in its own actual location. Transparent surfaces are rendered with Alpha blending ($\alpha = 0.40$).}
	\label{fig:NeXt-VolRendering}
\end{figure}

For a more immediate graphical and clinically useful representation \cite{walter2010}, Fig. \ref{fig:NeXt-VolRendering} shows three instances of tridimensional reconstructions of brain tumors with a necrotic core.
The GTV is represented using a transparent red surface, allowing us to visualize the internal necroses (blue volumetric models) in their own actual position with respect to the tridimensional reference system.

\subsubsection{Experimental results}
\begin{figure}
	\centering
	\includegraphics[width=0.8\linewidth]{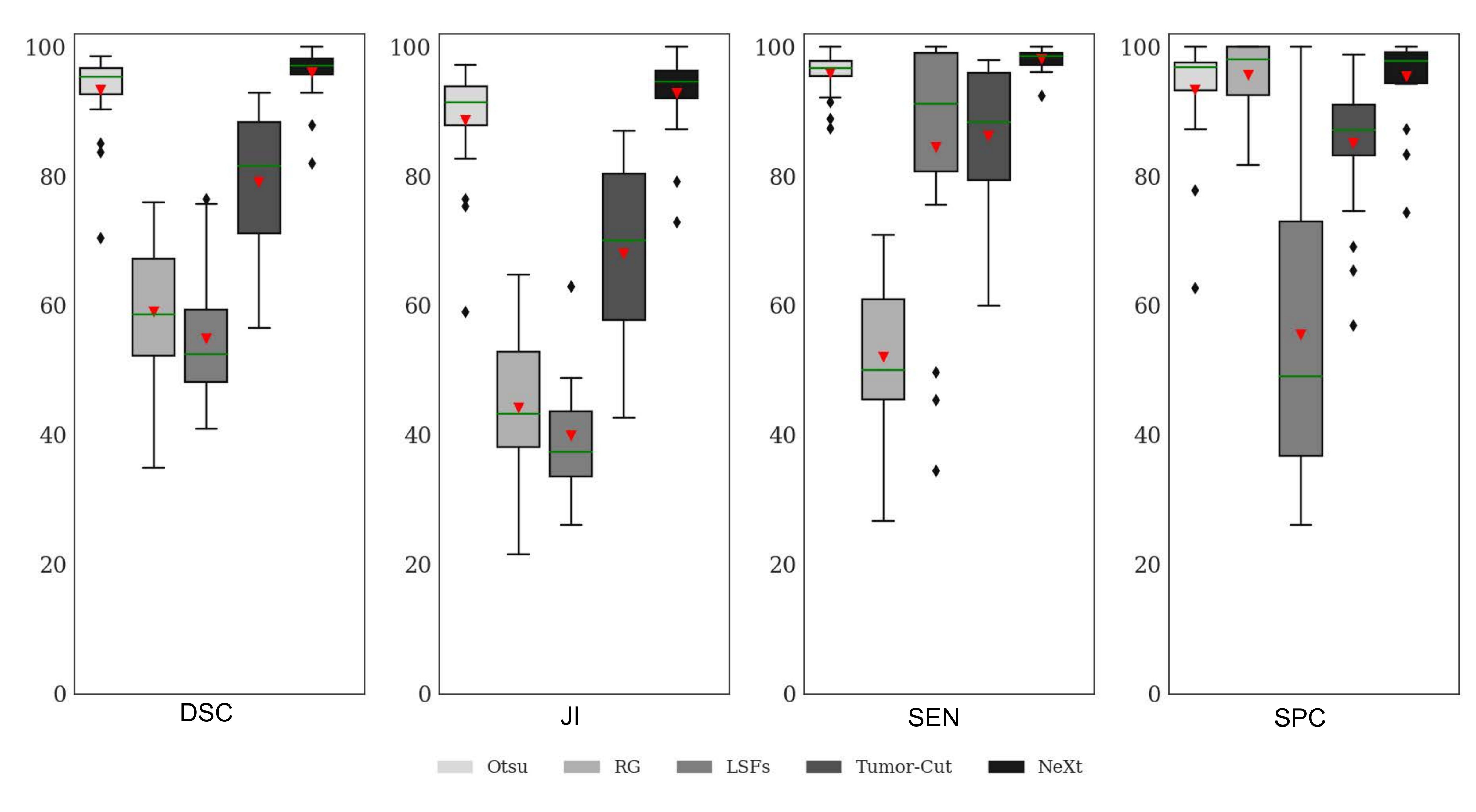}
	\caption[Boxplots of the spatial overlap-based metrics values achieved by the implemented necrosis segmentation approaches]{Boxplots of the spatial overlap-based metrics values achieved by the implemented necrosis segmentation approaches. The lower and the upper bounds of each box represent the first and third quartiles of the statistical distribution, respectively. The median (i.e., the second quartile) and the mean values are represented by a green line and a red triangle, respectively. Whisker value is $1.5$ in all cases and outliers are displayed as black diamonds.}
	\label{fig:NeXt-BoxplotsOM}	
\end{figure}

Table \ref{table:NeXt-ResultsOM} depicts mean and standard deviation values of spatial overlap-based metrics achieved in the experimental tests by the developed necrosis extraction methods.
Fig. \ref{fig:NeXt-BoxplotsOM} shows the boxplots of spatial overlap-based evaluation metrics to give a convenient representation of the statistical distribution of the achieved measurements.
Overall, the achieved segmentation performance shows the great accuracy and reliability of the proposed approach.
This evidence is also confirmed by the boxplots, where distributions for NeXt present approximately only $10\%$ outliers in all the overlap-based metrics, thus evidencing extremely low statistical dispersion.

\begin{figure}[t]
	\centering
	\includegraphics[width=0.8\linewidth]{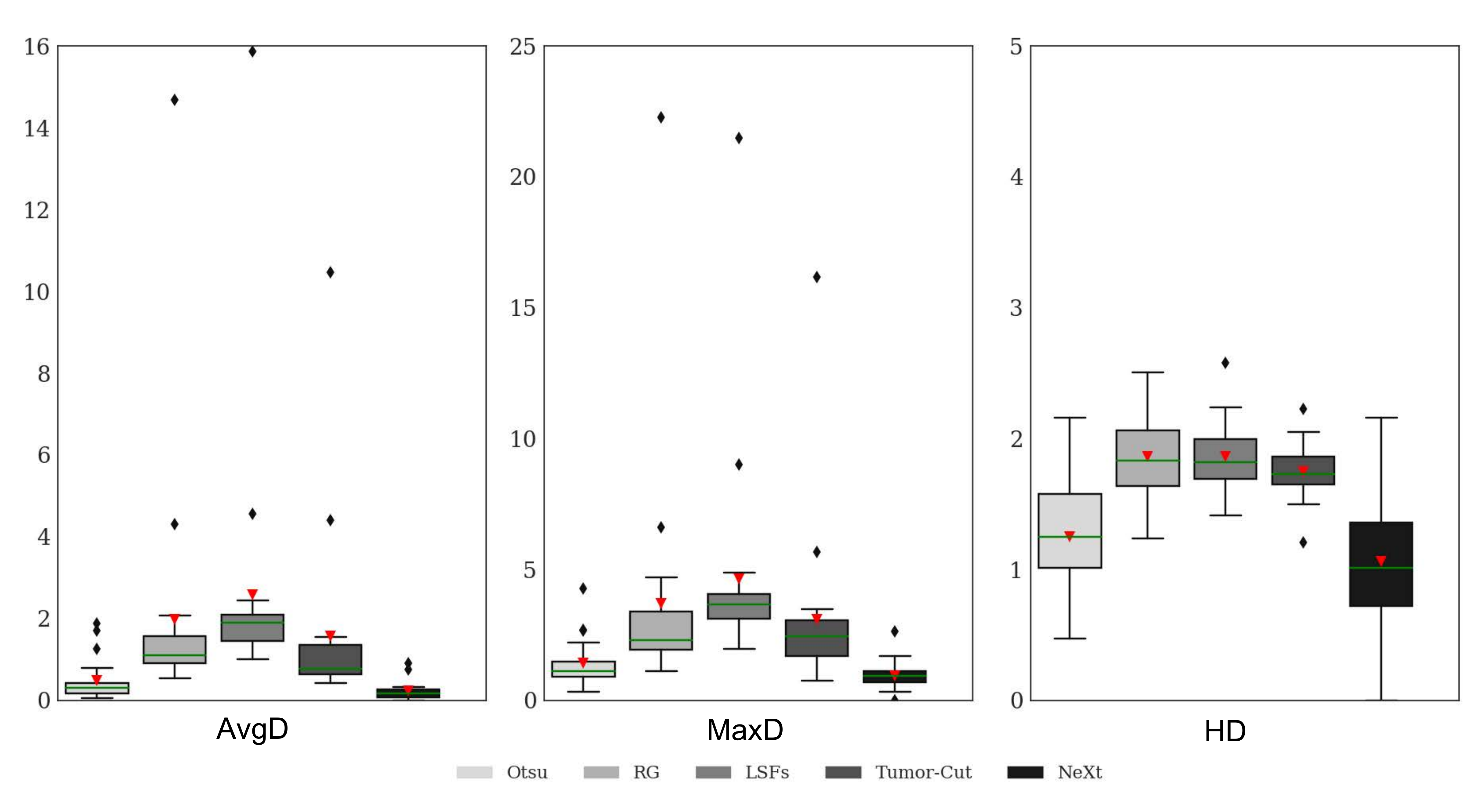}
	\caption[Boxplots of the spatial distance-based metrics achieved by the implemented necrosis segmentation approaches]{Boxplots of the spatial distance-based metrics achieved by the implemented necrosis segmentation approaches. The lower and the upper bounds of each box represent the first and third quartiles of the statistical distribution, respectively. The median (i.e., the second quartile) and the mean values are represented by a green line and a red triangle, respectively. Whisker value is $1.5$ in all cases and outliers are displayed as black diamonds.}
	\label{fig:NeXt-BoxplotsDM}	
\end{figure}

Since it is important to combine the distances from the misclassified voxels with respect to the gold standard for improving the information given by the overlap-based measures, average and standard deviation values of distance-based metrics, achieved on the whole MRI dataset, are reported in Table \ref{table:NeXt-ResultsDM}.
Fig. \ref{fig:NeXt-BoxplotsDM} illustrates the boxplots of the achieved spatial distance-based evaluation metrics.
The achieved spatial distance-based indices are consistent with overlap-based metrics, also observing the corresponding boxplots.
Hence, good performances were obtained also in terms of difference between the automated and the manual boundaries.

\begin{table}[!t]
\centering
	\caption[Values of the spatial overlap-based metrics regarding the necrosis extraction results achieved by the developed segmentation methods]{Values of the spatial overlap-based metrics regarding the necrosis extraction results achieved by the developed segmentation methods. The experimental results are expressed as average value $\pm$ standard deviation.}
	\label{table:NeXt-ResultsOM}
	\begin{scriptsize}
		\begin{tabular}{lcccc}
			\hline\hline
			Method	& DSC	& JI	& SEN	& SPC \\
			\hline
			Adaptive thresholding \cite{otsu1975} &	$93.20 \pm 6.69$ &	$88.65 \pm 9.22$ &	$95.83 \pm 3.38$ &	$93.20 \pm 8.95$ \\
			Seeded Region Growing \cite{rundoMBEC2016} &	$58.98 \pm 10.87$ &	$44.13 \pm 10.94$ &	$51.97 \pm 12.44$ &	$95.53 \pm 5.55$ \\
			Interactive LSFs \cite{li2010DRLSE} &	$54.81 \pm 9.30$ &	$39.89 \pm 9.84$ &	$84.44 \pm 19.67$ &	$55.40 \pm 23.17$ \\
			Tumor-Cut based on CA \cite{hamamci2012} &	$78.98 \pm 10.75$ &	$67.97 \pm 13.34$ &	$86.19 \pm 10.55$ &	$85.09 \pm 10.97$ \\
			NeXt based on FCM &	$95.93 \pm 4.23$ &	$92.81 \pm 6.56$ &	$98.09 \pm 1.75$	& $95.31 \pm 6.56$ \\
			\hline\hline
		\end{tabular}
	\end{scriptsize}
\end{table}

\paragraph{Comparison with other segmentation methods and implementation results}
The proposed method NeXt, based on the FCM clustering algorithm, outperforms all the other implemented necrosis extraction approaches in terms of both spatial overlap-based and distance-based metrics.
Adaptive thresholding using the Otsu's method \cite{otsu1975} achieves good segmentation results, by exploiting the bimodal histogram distribution of the pixels included in the GTV region.
However, the FCM clustering algorithm, thanks to the introduction of Fuzzy Logic enabling partial membership in classes, enables a more flexible classification process with respect to adaptive thresholding \cite{otsu1975}.

\begin{table}[!t]
\centering
	\caption[Values of the spatial distance-based metrics regarding the necrosis extraction results achieved by the developed segmentation methods]{Values of the spatial distance-based metrics regarding the necrosis extraction results achieved by the developed segmentation methods. The experimental results are expressed as average value $\pm$ standard deviation.}
	\label{table:NeXt-ResultsDM}
	\begin{scriptsize}
		\begin{tabular}{lcccc}
			\hline\hline
			Method	& AvgD	& MaxD	& HD \\
			\hline
			Adaptive thresholding \cite{otsu1975} &	$0.481 \pm 0.524$ &	$1.411 \pm 0.919$ &	$1.248 \pm 0.405$ \\
			Seeded Region Growing \cite{rundoMBEC2016} &	$1.977 \pm 3.097$ &	$3.700 \pm 4.546$ &	$1.865 \pm 0.334$ \\
			Interactive LSFs \cite{li2010DRLSE} &	$2.589 \pm 3.211$ &	$4.648 \pm 4.225$ &	$1.864 \pm 0.286$ \\
			Tumor-Cut based on CA \cite{hamamci2012} &	$1.563 \pm 2.266$ &	$3.093 \pm 3.253$	& $1.745 \pm 0.221$ \\
			NeXt based on FCM &	$0.225 \pm 0.229$ &	$0.920 \pm 0.581$ &	$1.060 \pm 0.524$ \\
			\hline\hline
		\end{tabular}
	\end{scriptsize}
\end{table}

The CA-based method in Tumor-Cut \cite{hamamci2012} is characterized by lower sensitivity to small necrotic islands, since it strongly relies on the initial enhancement and necrosis seed selection, based on the Otsu's method \cite{otsu1975}, which considers the $25\%$ of the necrotic and enhanced volumes as necrosis and enhancement seeds, respectively.
So, the authors themselves stated that small islands of necrotic regions with ambiguous low contrast can be missed by the seed selection strategy.
Moreover, this approach could be affected by the introduction of false positive seeds at the brain tumor boundaries, because no refinement steps are described in the original paper by Hamamci \textit{et al.} \cite{hamamci2012}.
The interactive LSF-based approach employed the DRLSE formulation \cite{li2010DRLSE}.
Although this method attempts to address irregularities during the conventional Level Set evolution by means of a distance regularization term that maintains a desired shape of the LSF, leaking (i.e., over-estimation of the segmented necrosis ROI) could occur in the case of non-convex or weak ROI boundaries, see Fig. \ref{fig:NeXt-Results2}(c, d).
This evidence is confirmed by FPs, as shown by the higher average values of SE with respect to SP, involving that the actual necrotic region is generally over-estimated.
In addition, the LSF framework is highly sensitive to parameter setting, by considering image contrast and ROI characteristics (i.e., shape and boundaries).
Region growing also requires a threshold selection according to image features, giving rise to an error-prone procedure.
Segmentation errors could occur when inhomogeneties or loosely connected hypo-intense areas are present: the seeded region growing algorithm can suffer from highly heterogeneous areas, such as in the case of necrotic material.
Especially, increasing the segmentation threshold in the homogeneity criteria could cause leaking, also expanding the segmented ROI to the whole brain region.

To quantify the efficiency of the proposed method in terms of execution times, the (sequential) version of the proposed necrosis extraction approach was compared against the other implemented image segmentation methods.
All the approaches were developed and tested using MatLab\textsuperscript{\textregistered} R2017a $64$-bit, on a computer equipped with an Intel\textsuperscript{\textregistered} Core\textsuperscript{TM} i7-7700HQ quad-core processor (clock frequency $2.80$ GHz), 16 GB RAM and Windows 10 operating system.

\begin{table}[!t]
\centering
	\caption[Comparison of the execution times between NeXt and the implemented approaches]{Comparison of the execution times between NeXt and the implemented approaches.}
	\label{table:NeXt-execTime}
	\begin{scriptsize}
		\begin{tabular}{lcc}
			\hline\hline
			Method	& Average execution time [s] & Standard deviation \\
			\hline
			Adaptive thresholding \cite{otsu1975} &	$2.200 \times 10^{-3}$ &	$2.662 \times 10^{-4}$ \\
			Seeded Region Growing \cite{rundoMBEC2016} &	$7.900 \times 10^{-3}$ &	$6.608 \times 10^{-4}$ \\
			Interactive LSFs \cite{li2010DRLSE} &	$1.683 \times 10^0$ &	$5.190 \times 10^{-1}$ \\
			Tumor-Cut based on CA \cite{hamamci2012} &	$4.350 \times 10^{-2}$ &	$3.200 \times 10^{-3}$ \\
			NeXt based on FCM &	$1.650 \times 10^{-2}$ &	$1.300 \times 10^{-3}$ \\
			\hline\hline
		\end{tabular}
	\end{scriptsize}
\end{table}

Table \ref{table:NeXt-execTime} reports the performances, measured by the MatLab Profiler tool, in terms of average and standard deviation execution times calculated over the complete MRI dataset, by running the algorithms once on each single image.
As expected, adaptive thresholding using Otsu's method \cite{otsu1975}, applied only on the MRI data included in the previously segmented GTV, was the fastest on average and with a low standard deviation.
Since the necrosis ROI area to be segmented is small, seeded region growing achieved segmentation results in the order of milliseconds.
Conversely, LSFs take more time to accomplish the segmentation based on an energy function minimization implementing a variational framework.
However, these two methods require user interaction to select the initial seed-point.
The approach presented in Tumor-Cut \cite{hamamci2012} for the segmentation of necrotic and enhancement regions is fully automatic since foreground and background seeds are selected according to the Otsu's method.
Accordingly, threshold calculation has to be considered in the total execution time.
Our method NeXt, based on FCM clustering, obtains on average a $2.64\times$ speed-up factor with respect to the CA-based approach described in Tumor-Cut \cite{hamamci2012}, showing also a lower standard deviation.

\subsubsection{Conclusion}
We presented a fully automatic approach for necrosis extraction, called NeXt, using an unsupervised FCM clustering technique.
Our method detects and delineates the necrotic regions within the planned GTV to be treated with neuro-radiosurgery.
In compliance with the typical treatment planning practice, in which neurosurgeons generally use neuro-radiosurgery contrast-enhanced T1w MR images alone for GTV segmentation \cite{quinones2012}, NeXt does not need multispectral MRI data.
So, NeXt represents a feasible solution in clinical routine, since the direct transfer of the proposed methodology in clinical environments is strongly reduced.
An MRI dataset including $20$ brain metastases with necrotic material, concerning patients who underwent stereotactic neuro-radiosurgery, was analyzed.
The accuracy of the proposed approach was experimentally evaluated in terms of spatial overlap-based and distance-based metrics.
A quantitative comparison with the most common literature methods for enhancement and necrosis region distinction was also performed.
NeXt achieved the best segmentation accuracy, outperforming all the implemented alternative approaches.

Necrotic and enhancement region distinction in MRI brain tumors by using NeXt has certainly clinical value, in both radiation therapy treatment planning and patient’s follow-up, since necrotic regions present hypoxic areas that could involve recurrent cancers and resistance to cancer therapy damages.
However, conventional MRI could be also not appropriate for assessing the biological or metabolic activity of the tumor, since it does not provide a specific measurement of cancer activity \cite{dhermain2010}.
To overcome these limitations, advanced imaging techniques enable the visualization of the actual biological changes of the cancer, by detecting proliferative activity, hypoxia, apoptosis, necrosis, and tumor vasculature.
Advanced MRI techniques allow for the detection of various metabolic and physiological biomarkers.
As a matter of fact, changes in metabolic tissue profile, tissue blood perfusion, microvessel permeability, and water diffusion can be estimated \cite{cao2006,dhermain2010}.

In a complementary way, PET image analysis quantitatively assesses the metabolic state of cancer.
As shown in \cite{rundoCMPB2017}, it is not always possible to identify the actual tumor extent using the MRI modality alone.
As a matter of fact, MRI and MET-PET convey different but highly correlated imaging information.
Consequently, the BTV, delineated on MET-PET images, is valuable and could be integrated with GTV in neuro-radiosurgery treatment planning.
This applies also in dose painting, where quantitative analysis of MET-PET can provide aid in the differentiation of tumor recurrence from radiation necrosis \cite{terakawa2008}.
In particular, dose painting and dose escalation have been already studied in some anatomic districts, such as non-small cell lung cancer \cite{even2015} or prostate and cervix cancer \cite{hoskin2015}.
Therefore, a comprehensive multimodal PET/MRI segmentation approach, which considers also metabolic and hypoxic sub-volumes, should represent a new way for a more selective and reliable identification of the active cancer regions to be treated with radiosurgery systems that allow for dose painting \cite{luan2009} and personalized radiation therapy plans \cite{ree2015}.

\subsection{Prostate gland segmentation on multispectral MRI}
\label{sec:prostateSeg}

In this section, an automated segmentation approach of the whole prostate gland from axial MRI slices is presented \cite{rundo2018SIST,rundo2017Inf}.
The method is based on an unsupervised Machine Learning technique, resulting in an advanced application of the FCM clustering algorithm on multispectral MR anatomic images.

Nowadays, T2w FSE imaging is the standard protocol for depicting the anatomy of the prostate and for identifying prostate cancer considering reduced T2 signal intensity on MRI \cite{kurhanewicz2008}.
These sequences are sensitive to susceptibility artifacts (i.e., local magnetic field inhomogeneities), for instance due to the presence of air in the rectum, which may affect the correct detection of tissue boundaries \cite{klein2008}.
On the other hand, because the prostate has uniform intermediate signal intensity at T1w MRI, the zonal anatomy cannot be clearly identified on T1w MRI series \cite{choi2007}.
T1w MR images are used mainly to determine the presence of hemorrhage within the prostate and seminal vesicles and to delineate the gland boundaries \cite{barentsz2016}.
Nevertheless, T1w MRI is not highly sensitive to prostate cancers, as well as to inhomogeneous signal in the peripheral zone or adenomatous tissue in the central gland with possible nodules.
We exploit this uniform gray appearance, by combining prostate T2w and T1w MRI, to enhance the clustering-based classification.

Differently to the state-of-the-art approaches, this unsupervised Machine Learning technique does not require training phases, statistical shape priors or atlas pre-labeling.
Therefore, our method could be easily integrated in clinical practice to support radiologists in their daily decision-making tasks.
Another key novelty is the introduction of T1w MRI in the feature vector, composed of the co-registered T1w and T2w MR image series, fed to the proposed processing pipeline.
Although T1w MR images were used in \cite{pasquier2007} for prostate radiotherapy planning, the authors did not combine both multispectral T2w and T1w MR images using a fusion approach.
Their study just highlighted that the Clinical Target Volume (CTV) segmentation results on the T1w and T2w acquisition sequences did not differ significantly in terms of manual CTV.
Our work aims to show that the early integration of T2w and T1w MR image structural information significantly enhances prostate gland segmentation by exploiting the uniform gray appearance of the prostate on T1w MRI \cite{choi2007}.

\subsubsection{Prostate MRI dataset}
\begin{figure}
	\centering
	\includegraphics[width=0.8\linewidth]{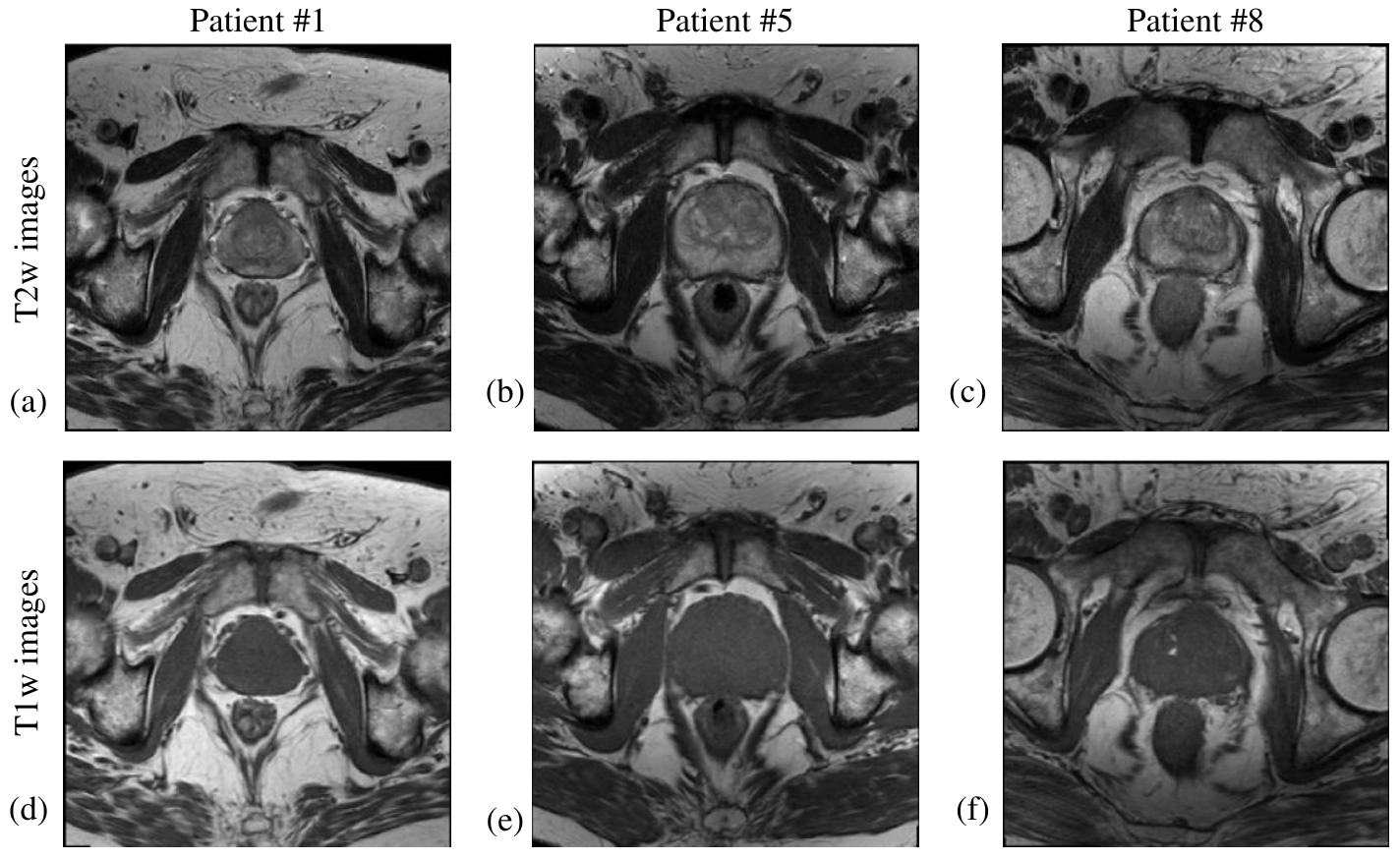}
	\caption[Input prostate T1w and T2w MRI axial slice pairs]{Input prostate MRI axial slice pairs for the Patients $\#1$, $\#5$ and $\#8$: (a, b, c) T2w MR images; (d, e, f) corresponding T1w MR image.}
	\label{fig:PGS-inputImages}
\end{figure}

The study was performed on a clinical dataset composed of $21$ patients with suspicious PCa.
All the analyzed MR images were T1w and T2w Fast Spin Echo (FSE) sequences, acquired with an Achieva $3.0$ T MRI Scanner (Philips Medical Systems, Eindhoven, the Netherlands) using a SENSE XL Torso coil ($16$ elements phased-array pelvic coil), at the Cannizzaro Hospital (Catania, Italy).
The total number of the processed 2D MRI slices, in which prostate gland was imaged, was $185$.
Three pairs of corresponding T1w and T2w MR images are shown in Fig. \ref{fig:PGS-inputImages}.
Imaging data were encoded in the $16$-bit DICOM format.
MRI acquisition parameters are reported in Table \ref{table:PGS-MRIcharacteristics}.

\begin{table}[!t]
\centering
	\caption[MRI acquisition parameters of the prostate dataset]{MRI acquisition parameters of the prostate dataset.}
	\label{table:PGS-MRIcharacteristics}
	\begin{tiny}
		\begin{tabular}{cccccccc}
			\hline\hline
			MRI sequence	& TR [ms]	& TE [ms]	& Matrix size [pixels]	& Slice spacing	[mm]	& Slice thickness [mm]	& Pixel spacing [mm] & Slices' number \\
			\hline
			T1w		& $515.3$		& $10$ 		& $256 \times 256$ 	& $4$ 	& $3$ 	& $0.703$ & $18$ \\
			T2w		& $3035.6$		& $80$ 		& $288 \times 288$ 	& $4$ 	& $3$ 	& $0.625$ & $18$ \\
			\hline\hline
		\end{tabular}
	\end{tiny}
\end{table}

The average age of the examined patients is $65.57 \pm 6.42$ ($53$-$77$) years.
All the enrolled subjects have clinical suspicion about prostate cancer, due to high PSA levels or digital rectal examination (DRE) for prostate screening.
Moreover, the MRI study can be also performed contextually with a previous biopsy, considered as gold standard, or combined with a further targeted biopsy according to the imaging evidences.
In the positive cases, the possible clinical choices are three: prostatectomy, radiotherapy, or chemotherapy.

\subsubsection{The proposed whole prostate gland segmentation method}
Since the organ to be imaged is always positioned approximately near the isocenter of the principal magnetic field to minimize MRI distortions, the whole prostate gland is represented in the imaged Field of View (FOV) center \cite{lagendijk2014}.
In a preliminary work in \cite{rundo2018SIST}, after dividing the entire image in $9$ equal-sized and fixed tiles, we considered the cropped image represented by the central one.
Therefore, only a cropped image, whose size is $1/9$ of the initial input image size, is processed for ROI image segmentation.
Observing the prostate ROIs imaged on morphologic MRI, in terms of the segmented area in each slice, we observed that the prostate gland can be suitably contained in a patch with size $1/9$ of the initial input image size.
Using this strategy, user input is not required and processing time is certainly optimized.

Although the prostate gland is represented in the image FOV center, since the organ to be imaged is always positioned approximately near the isocenter of the principal magnetic field to minimize MRI distortions \cite{lagendijk2014}, sometimes it may appear shifted from the center or the patient could be not correctly positioned.
This fact could affect segmentation accuracy especially in apical and basal prostate MRI slices, when the MR image center is considered for prostate gland segmentation.
To address this issue, a rectangular ROI selection tool is dragged around the prostate by the user \cite{rother2004} for more reliable and precise delineation results \cite{chen2009}.
Interactive segmentation is becoming more and more popular to alleviate the problems regarding fully automatic segmentation approaches, which are not always able to yield accurate results by adapting to all possible clinical scenarios \cite{boykov2001b,boykov2001a}.
Thereby, operator-dependency is reduced but, at the same time, the physician is able to monitor and control the segmentation process.
Accordingly, computerized medical image analysis has given rise to many promising solutions, but, instead of focusing on fully automatic computerized systems, researchers are aimed to propose computational techniques to aid radiologists in their clinical decision-making tasks \cite{lemaitre2015}.

\begin{figure}[!t]
	\centering
	\includegraphics[width=0.6\linewidth]{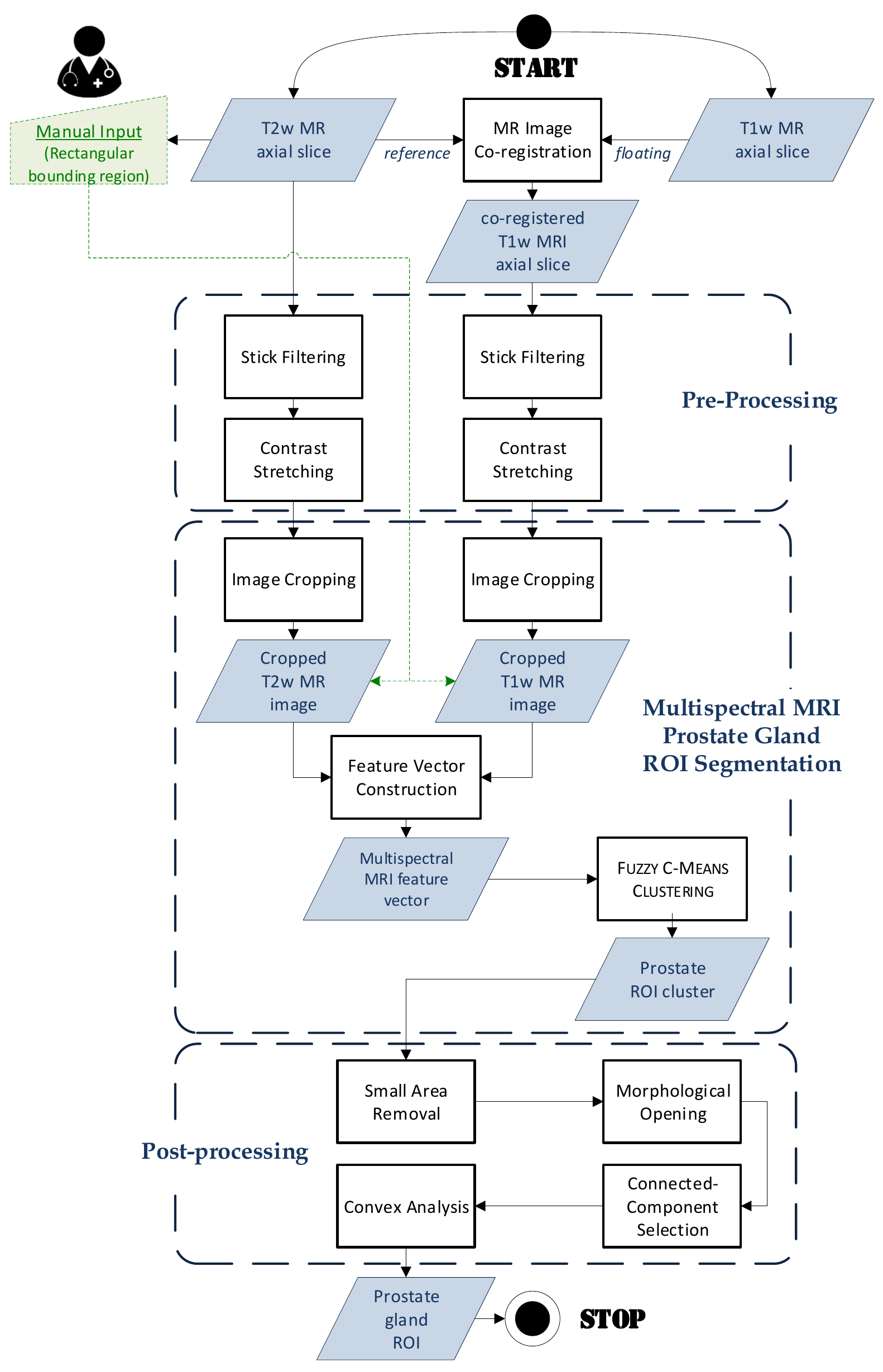}
	\caption[Flow diagram of the proposed prostate gland MRI segmentation method]{Flow diagram of the proposed prostate gland MRI segmentation method. The pipeline can be subdivided into three main stages: (\textit{i}) pre-processing, required to remove speckle noise and enhance the FCM clustering process; (\textit{ii}) FCM clustering on multispectral MR images, to extract the prostate gland ROI; (\textit{iii}) post-processing, well-suited to refine the obtained segmentation results.}
	\label{fig:PGS-FlowDiagram}
\end{figure}

The overall flow diagram of the proposed prostate segmentation method is represented in Fig. \ref{fig:PGS-FlowDiagram}, and a detailed description of each processing phase is provided in the following subsections.

\paragraph{MR image co-registration}
Even though T1w and T2w MRI series are included in the same study, they are not acquired contextually because of the different employed extrinsic parameters that determine T1 and T2 relaxation times.
Thus, an image co-registration step on multispectral MR images is required.
Moreover, T1w and T2w images have different FOVs, in terms of pixel spacing and matrix size (see Table \ref{table:PGS-MRIcharacteristics}).
However, in our case a $2$D image registration method is effective because T1w and T2w MRI sequences are composed of the same number of slices and have the same slice thickness and interslice spacing values.

Image co-registration is accomplished by using an iterative process to align a moving image (T1w) along with a reference image (T2w).
In this way, the two images will be represented in the same reference system, so enabling quantitative analysis and processing on fused imaging data.
T2w MRI was considered as reference image, to use its own reference system in the following advanced image analysis phases for differentiating prostate anatomy and for PCa detection and characterization, by processing also functional data (i.e., DCE, DWI, and MRSI).

\begin{figure}[!t]
	\centering
	\includegraphics[width=0.8\linewidth]{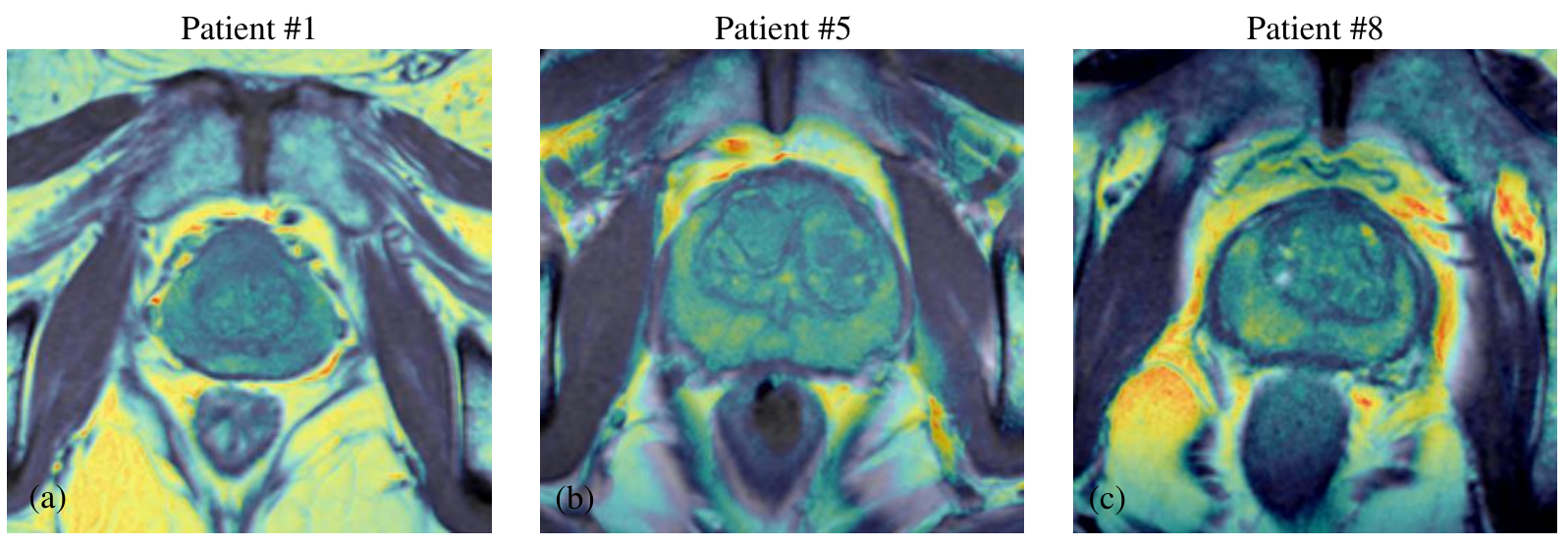}
	\caption[Details of the prostate on a pair of fused T1w and T2w MR images]{Details of the prostate on the fused T1w and T2w MR image pairs in Fig. \ref{fig:PGS-inputImages}, using the gray-scale and the Jet colormap, respectively.}
	\label{fig:PGS-Fusion}
\end{figure}

In intensity-based registration techniques, the accuracy of the registration is evaluated iteratively according to an image similarity metrics (refer to Section \ref{sec:medImageReg} for further information).
We used Mutual Information (MI) that is an information theoretic measure of the correlation degree between two random variables \cite{pluim2003}.
High mutual information means a large reduction in the uncertainty (entropy) between the two probability distributions, revealing that the images are likely better aligned \cite{mattes2001}.
An Evolution Strategy (ES), using a one-plus-one optimization configuration, is exploited to find a set of transformation that yields the optimal registration result. In this optimization technique, both the number of parents and the population size (i.e., number of offspring) are set to one \cite{styner2000}.
Affine transformations, which combine rigid-body transformations (i.e., translations and rotations) with scaling and shearing, are applied to the moving T1w image to be aligned against the reference T2w image \cite{rundo2016SSCI}.
A detail of the prostate on the fused T1w and T2w images after the co-registration step is shown in Fig. \ref{fig:PGS-Fusion}.

\paragraph{Pre-processing}
Pre-processing steps are needed for MR image denoising, while preserving relevant feature details such as organ boundaries, and enhance the FCM clustering process.
These operations are applied on T2w and the corresponding co-registered T1w input MR series in the same study.

\begin{figure}[!t]
	\centering
	\includegraphics[width=0.8\linewidth]{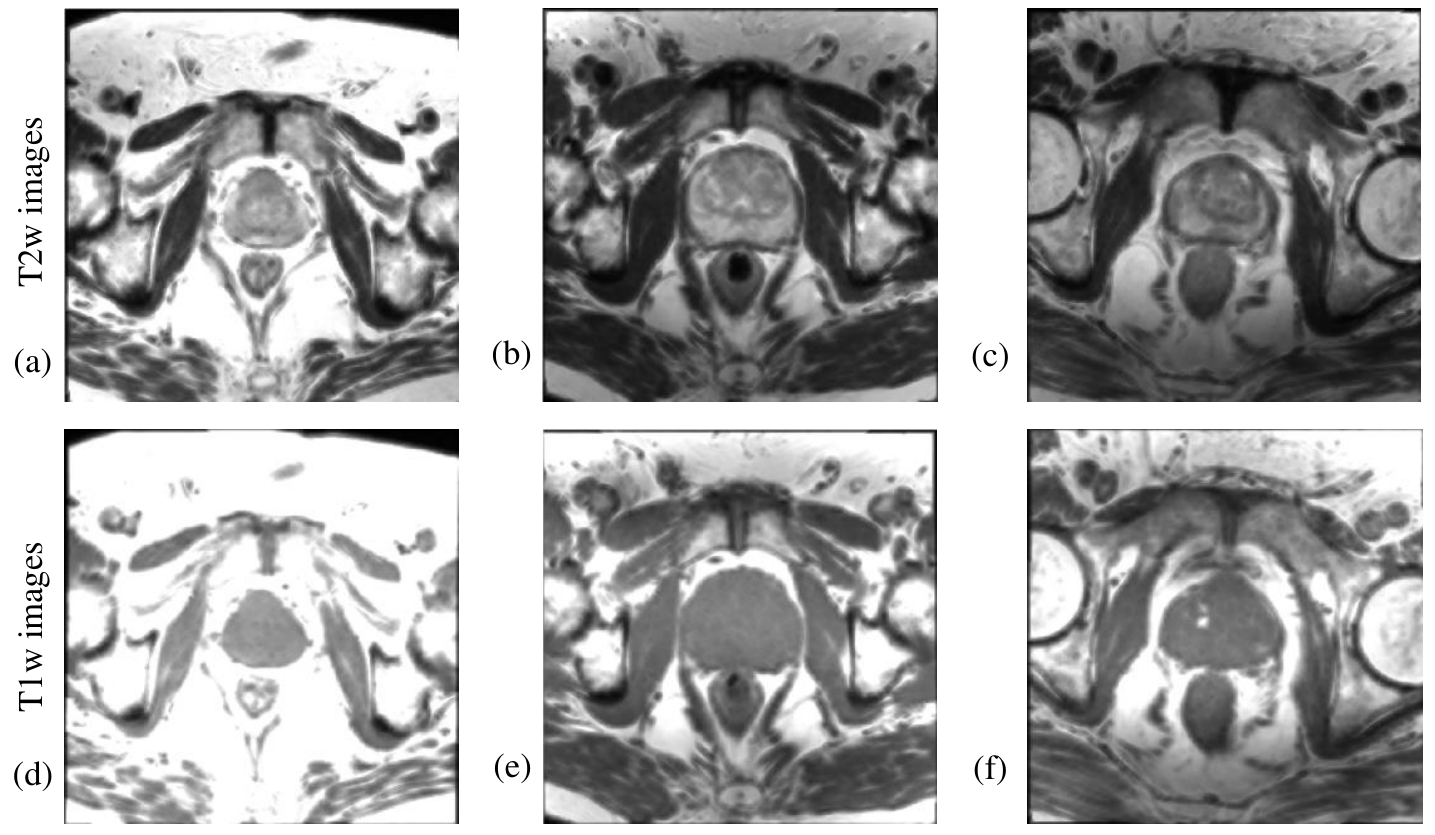}
	\caption[Pre-Processing phase of the prostate gland MRI segmentation]{Pre-Processing phase of the prostate gland MRI segmentation: (a, b, c) T2w images and corresponding (d, e, f) T1w images after pre-processing steps, consisting in stick filtering and linear contrast stretching, on the images in Fig. \ref{fig:PGS-inputImages}.}
	\label{fig:PGS-PreProc}
\end{figure}

Firstly, stick filtering is applied to efficiently remove the speckle noise that affects MR images \cite{czerwinski1998,czerwinski1999}.
The main source of noise in MRI is the thermal noise in the patient \cite{edelstein1986}, influencing both real and imaginary channels with additive white Gaussian noise \cite{macovski1996}.
The MR image is then reconstructed by computing the inverse discrete Fourier transform of the raw data, resulting in complex white Gaussian noise \cite{mcrobbie2017,reyes2013}.
The magnitude of the reconstructed MR image is commonly used for visual inspection and for automatic computer analysis \cite{pizurica2006}.
The magnitude of the MRI signal follows a Rician distribution, since it is calculated as the square root of the sum of the squares of two independent Gaussian variables.
In low-intensity (dark) regions of the magnitude image, the Rician distribution tends to a Rayleigh distribution and in high intensity (bright) regions it tends to a Gaussian distribution \cite{papoulis2001}.
This implies a reduced image contrast, since noise increases the mean value of pixel intensities in dark image regions \cite{pizurica2006}.
By considering a local neighborhood around each pixel, the intensity value of the current pixel is replaced by the maximum of all the stick projections, which are calculated by the convolution of the input image with a long narrow average filter at various orientations \cite{czerwinski1998,czerwinski1999}.
Using a stick filter with odd length $\lambda$ and thickness $\tau < \lambda$, the square $\lambda \times \lambda$  neighborhood can be decomposed into $2 \lambda - 2$  possible stick projections. Thereby, smoothing is performed on homogeneous regions, while preserving resolvable features and enhancing weak edges and linear structures, without producing too many false edges \cite{xiao2005}.
Stick filtering parameters are a good balance: the sticks have to be short enough that they can approximate the edges in images, but long enough that the speckle along the sticks is uncorrelated.
Therefore, the projections along the sticks can smooth out speckle but not damage real edges in the image.
Experimentally, considering the obtained filtering results to be the most suitable input for the FCM clustering algorithm and enhance the achieved classification, the selected stick filter parameters are: stick length $\lambda = 5$ pixels; stick thickness $\tau = 1$ pixel.
All types of low-pass filtering reduce image noise, by removing high-frequency components of the MR signal.
This could cause difficulties in tumor detection.
Even though this smoothing operation could cause difficulties in cancer detection, our method is tailored for prostate gland segmentation alone and other pre-processing strategies could be employed for a better prostate cancer delineation.
Afterwards, a contrast stretching operation is applied by means of a linear intensity transformation that converts the input intensities into the full dynamic range, in order to improve the subsequent segmentation process.
The range of intensity values of the selected part is expanded to enhance the prostate ROI extraction.
Thereby, a contrast stretching operation is applied through a global intensity transformation.
This linear normalization converts input intensity   values into the output full dynamics in $[0,1]$.
This operation implies an expansion of the values, improving detail discrimination.
Since the actual minimum and the maximum intensity values of the MR input were considered, no region can have intensity outside of the cut-off values.
So, no image content or details (e.g., tumors or nodules) might be missed during the following detection phases.
The application of the pre-processing steps is shown in Fig. \ref{fig:PGS-PreProc}.
The image noise is visibly suppressed in order to have a more suitable input for the subsequent clustering process, thus extracting the prostate ROI more easily.

In order to quantitatively justify the selected smoothing operation, we calculated the SNR values on the prostate gland ROI segmentation results for the entire MRI dataset.
Since we masked the different MR images with the corresponding prostate gland ROIs, forcing the background to uniform black, the $SNR$ was computed, in compliance with routine MRI quality assurance testing \cite{firbank1999}, as: $SNR = \mu_\text{ROI}/\sigma_\text{ROI}$, where $\mu_\text{ROI}$ and $\sigma_\text{ROI}$ are the average value and the standard deviation of the signal inside the ROI, respectively.
After the MR image co-registration step, we evaluated and compared SNR by multiplying pixel-by-pixel each ROI binary mask with the corresponding:
\begin{itemize}
    \item original T1w and T2w MR images;
    \item T1w and T2w MR images pre-processed with stick filtering (length $\lambda=5$, thickness $\tau=5$);
    \item T1w and T2w MR images filtered with $3 \times 3$ and $5 \times 5$ Gaussian kernels;
    \item T1w and T2w MR images filtered with $3 \times 3$ and $5 \times 5$ average kernels.
\end{itemize}
For a fair comparison, in all the cases we normalized pixel intensities in $[0,1]$ using a linear stretching transformation, also to be consistent with the proposed pipeline.
The SNR values achieved on the T2w as well as on the co-registered T1w images are reported in (Table \ref{table:PGS-SNR}).
As it can be seen, the images pre-processed with stick filtering achieved the highest SNR in both T1w and T2w MRI series.

\begin{table}[!t]
	\caption[SNR calculated on the entire MRI prostate dataset]{SNR calculated on the entire prostate MRI dataset composed of $21$ co-registered T1w and T2w MRI sequences. The used stick filtering was compared against the most common smoothing strategies.}
	\label{table:PGS-SNR}
	\begin{scriptsize}
		\centering
		\begin{tabular}{ccccccc}
			\hline\hline
			MRI sequence	& Original	& Stick filtering	& $3 \times 3$ Gaussian filter	& $5 \times 5$ Gaussian filter	& $3 \times 3$ Average filter	& $5 \times 5$ Average filter \\
			\hline
			T1w		& $4.2133$ & $5.2598$ & $4.4069$ & $4.4081$ & $4.7523$ & $4.9061$ \\
			T2w		& $3.7856$ & $5.0183$ & $4.0032$ & $4.0044$ & $4.3479$ & $4.8739$ \\
			\hline\hline
		\end{tabular}
	\end{scriptsize}
\end{table}

\paragraph{Prostate gland segmentation using FCM algorithm on multispectral MRI}

\begin{figure}[!t]
	\centering
	\includegraphics[width=0.8\linewidth]{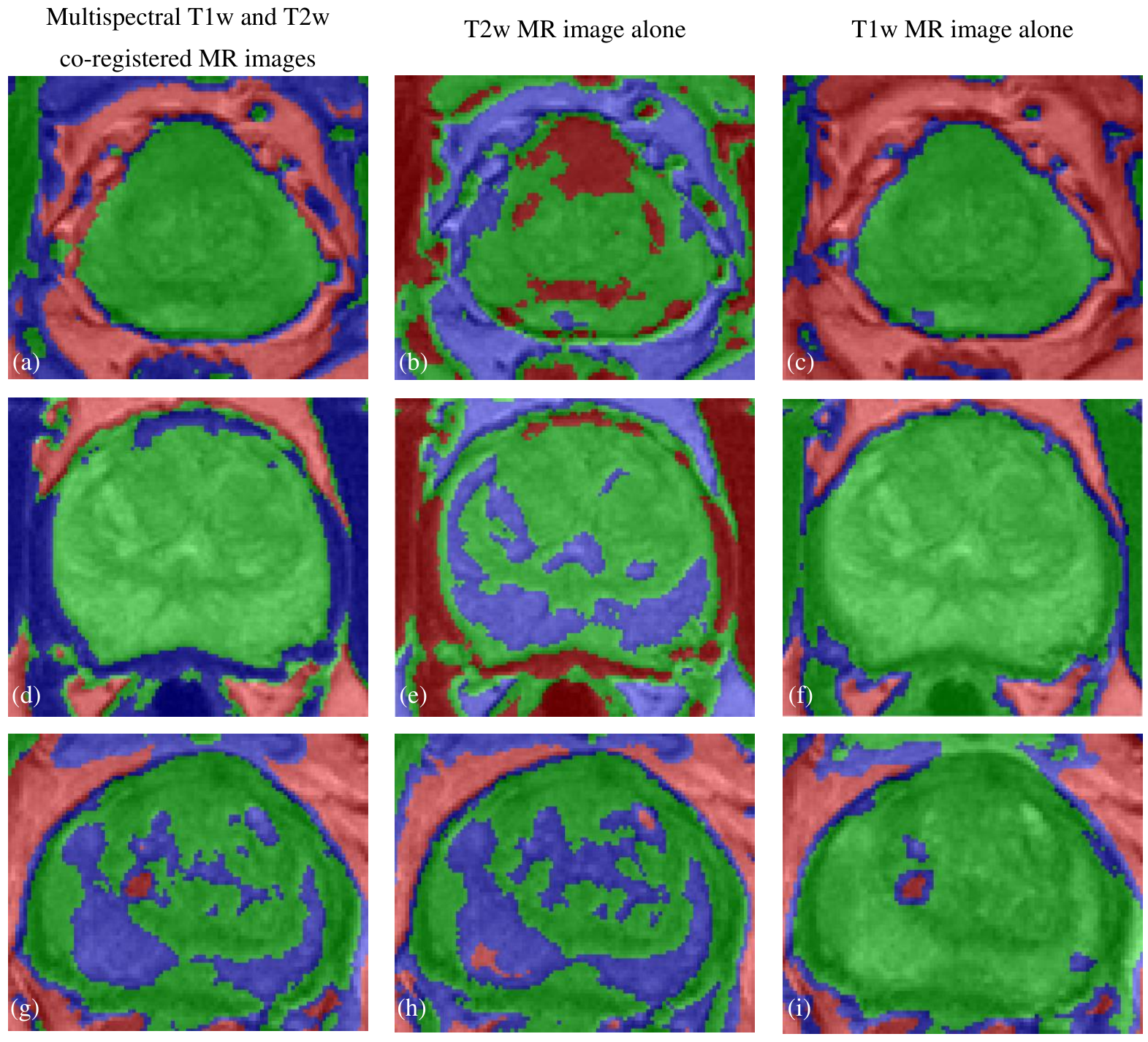}
	\caption[Results of the FCM clustering algorithm with $C=3$ for prostate gland ROI segmentation]{Results of the FCM clustering algorithm with $C=3$ for prostate gland ROI segmentation on the input images in Fig. \ref{fig:PGS-inputImages}, by processing different feature vectors: in the first column (a, d, g) results on the multispectral T1w and T2w co-registered MR images; in the second column (b, e, h) results on T2w MR image alone; in the third column (c, f, i) results on T1w MR image alone. It is possible to appreciate the multispectral MRI contribution with respect to the single series clustering outputs. In all cases, the prostate gland ROI cluster is represented by the green regions. The clusters are superimposed on the original T2w MR image using a transparence with Alpha blending  ($\alpha = 0.40$).}
	\label{fig:PGS-ClusterRes}
\end{figure}

The feature vector, which can include more than one feature, is constructed by concatenating corresponding pixels on T2w and T1w MR images, after the image co-registration step, in an early integration scheme \cite{serra2018}.
The clustering procedure is performed on the feature vector composed of corresponding T2w and T1w pixel values after the co-registration step.
The cluster number is always set to $C=3$, by considering both structural and morphologic properties of the processed prostate MR images.
As a matter of fact, three different classes can essentially be seen by visual inspection according to gray levels.
This choice was also justified and endorsed by experimental trials.
Fig. \ref{fig:PGS-ClusterRes} shows the clustering results achieved by the FCM algorithm with three classes during the prostate gland ROI segmentation stage.
It is appreciable how the FCM clustering output on multispectral MRI data considerably enhances the results achieved by processing either T1w or T2w images alone.
Especially, the introduction of T1w imaging, in the feature vector construction step, yields a more uniform overall clustering output in the prostate ROI cluster.
The resulting feature vector helps decisively to separate the ROI from the other surrounding tissues and organs (i.e., bladder, rectum, seminal vesicles, levator ani muscle).

As visible in Fig. \ref{fig:PGS-ClusterRes}(b, e, h), FCM clustering on T2w images alone often does not detect the whole prostate gland because of the irregularities due to the prostate zonal anatomy or cancer.
However, even though the FCM algorithm on T1w images alone obtains more regular clusters, it can include also external regions that can affect the final ROI segmentation results (see Fig. \ref{fig:PGS-ClusterRes}(f, i)).
Therefore, performing the segmentation on the co-registered T1w and T2w MR image is the best solution for prostate gland segmentation.
After the defuzzification step, which consists in assigning the pixels to the cluster with the greatest membership value, the prostate ROI cluster is identified.

\paragraph{Post-processing}
A sequence of morphological operations \cite{breen2000,soille2013} is applied to the obtained ROI cluster, in order to resolve possible ambiguities resulting from the clustering process and to improve the quality of the segmented prostate gland ROI.
According to Fig. \ref{fig:PGS-FlowDiagram}, the post-processing steps are the following:
\begin{enumerate}
    \item a small area removal operation is employed to delete any unwanted connected-com\-ponents, whose area is less than $500$ pixels, which are included into the ROI cluster. These small regions can have similar gray levels to the prostate ROI and were classified into the same cluster by the FCM clustering algorithm. The areas less than $500$ pixels certainly do not represent the prostate ROI, since the prostate gland in MRI slices is always characterized by a greater area, regardless of the used acquisition protocol. Thus, these small areas can be removed from the prostate ROI cluster to avoid a wrong connected-component selection in the next processing steps;
    \item a morphological opening, using a $5 \times 5$ pixel square structuring element, is a good compromise between precision in the actual detected contours and capability for unlinking poorly connected regions to the prostate ROI;
    \item connected-component selection, using a flood-fill algorithm \cite{wu2009}, in order to determine the connected-component that is the nearest to the cropped image center, since the prostate is always located at the central zone of the cropped image;
    \item convex analysis of the ROI shape since the prostate gland appearance is always convex. A convex hull algorithm \cite{zimmer1997} is suitable to envelope the segmented ROI into the smallest convex polygon that contains it, so considering possible adjacent regions excluded by the FCM clustering output. Finally, a morphological opening operation with a circular structuring element, is performed for smoothing prostate ROI boundaries. The used circular structuring element with $3$-pixel radius allows for smoother and more realistic ROI boundaries without deteriorating significantly the output yielded by the FCM clustering. Accordingly, the value of the radius does not affect the segmentation results and it is not dependent on image resolution.
\end{enumerate}

\subsubsection{Experimental results}
\begin{figure}[!t]
	\centering
	\includegraphics[width=0.8\linewidth]{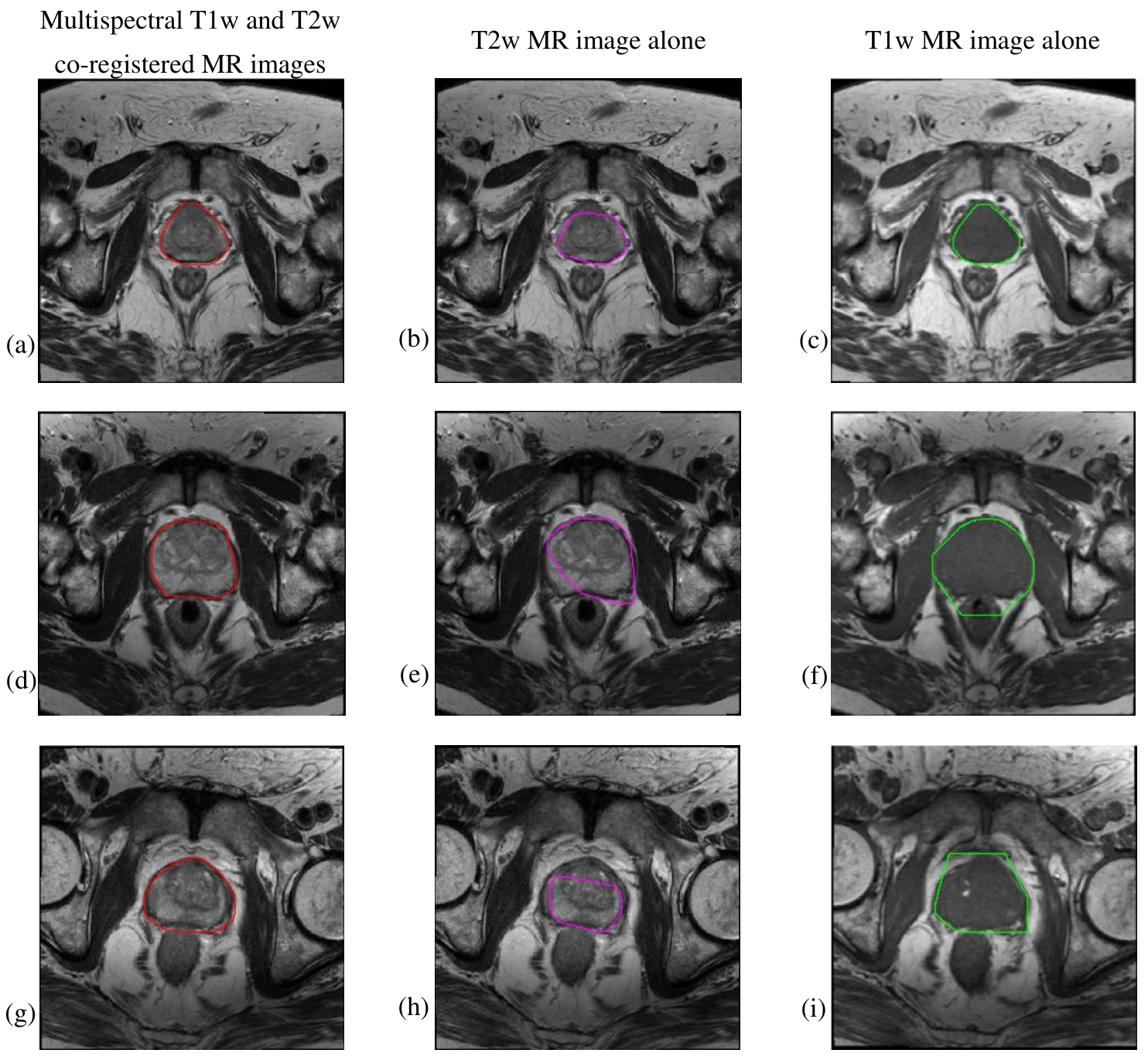}
	\caption[Prostate gland ROI segmentation instances]{Prostate gland ROI segmentation instances on the axial MRI slices shown in Fig. \ref{fig:PGS-inputImages} by processing different input MRI data (i.e., feature vectors) of three different patients (one per-row): first column (a, d, g), results obtained on multispectral T1w and T2w co-registered MR image (red contour superimposed on T2w image); second column (b, e, h), results obtained on T2w MR image alone (magenta contour superimposed on T2w image); third column (c, f, i), results obtained on T1w MR image alone (green contour superimposed on T1w image).}
	\label{fig:PGS-SegRes}
\end{figure}

The results achieved by the proposed segmentation approach on multispectral T1w and T2w co-registered MR images were also compared by applying the same processing pipeline on the corresponding monoparametric T2w and T1w MR images alone.
Fig. \ref{fig:PGS-SegRes} shows three examples of prostate gland ROI obtained by the proposed automated segmentation approach.
It is appreciable how correct segmentation results are achieved even with inhomogeneous and noisy input MRI data.

\begin{figure}[!t]
	\centering
	\includegraphics[width=0.8\linewidth]{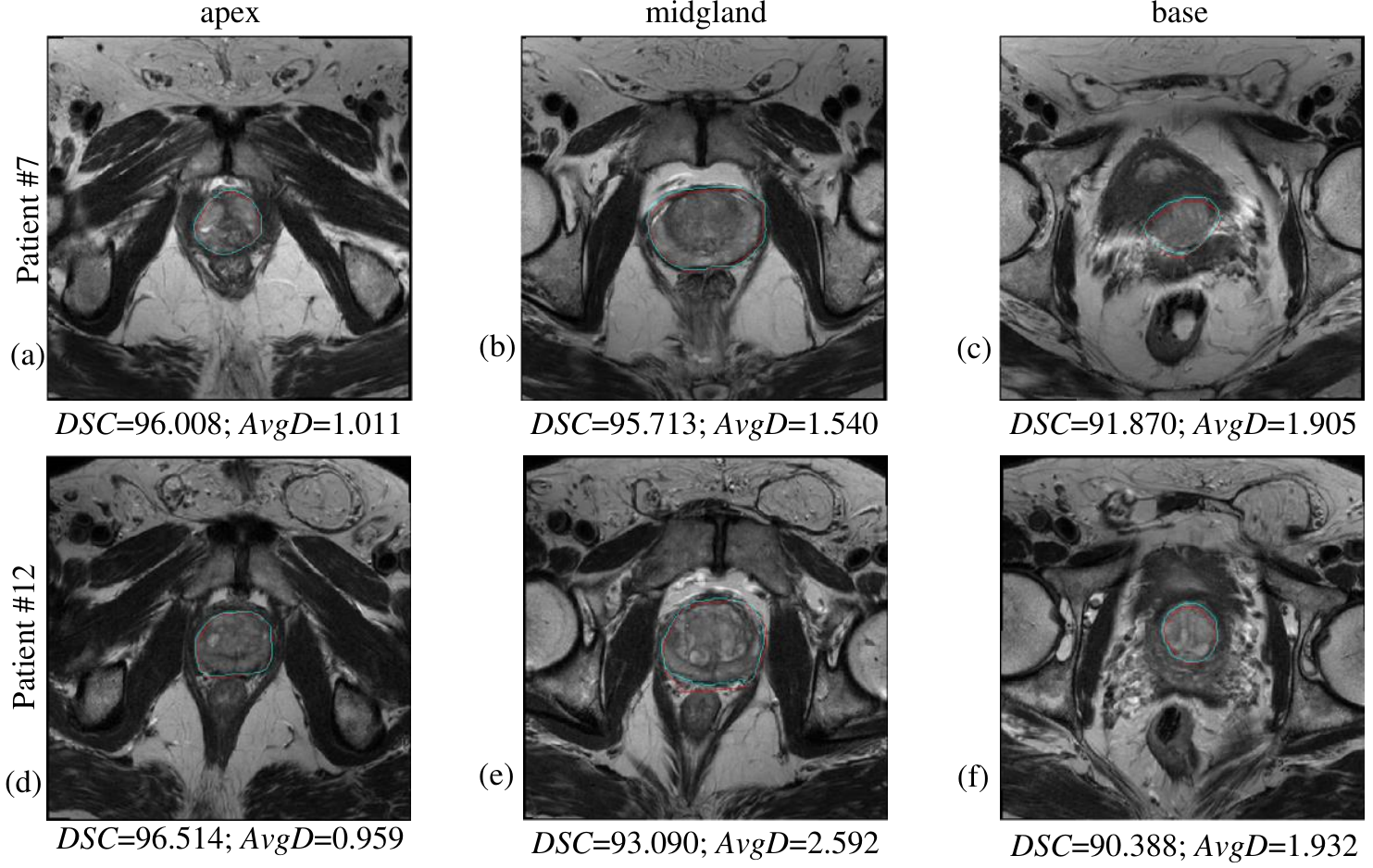}
	\caption[Examples of segmentation results obtained on apical, central, and basal slices]{Examples of segmentation results (red and cyan contours representing the proposed and the manual delineations, respectively, are superimposed on T2w axial prostate MR images) obtained on apical (a, d), central (b, e), and basal (c, f) slices concerning Patients $\#7$ and $\#12$, respectively. At the bottom of each sub-figure, the values for Dice similarity coefficient (\emph{DSC}) and mean average distance (\emph{AvgD}) are reported.}
	\label{fig:PGS-SegResReg}
\end{figure}

According to Fig. \ref{fig:PGS-SegRes}, it is possible to appreciate the differences by processing multispectral anatomic co-registered images, T2w MR image alone, and T1w image alone.
Especially, as shown in Fig. \ref{fig:PGS-SegRes}(b, e, h), FCM clustering on T2w images alone often is not able to detect the whole prostate gland because of the irregularities due to the prostate zonal anatomy or cancer.
On the other hand, even though FCM clustering on T1w images alone obtains an accurate result in Fig. \ref{fig:PGS-SegRes}c, it can segment also wrong anatomic parts, such as bladder or rectum, because they are imaged with similar gray levels to prostatic gland (Fig. \ref{fig:PGS-SegRes}(f, i)).
The proposed method is able to correctly segment the midgland and detects reasonably well the apical to the basal prostate regions.
An example of the achieved segmentations at the base, center (midgland) and apex of the prostate for the Patients $\#7$ and $\#12$ are shown in Fig. \ref{fig:PGS-SegResReg}.
The method is able to separate the prostate gland from seminal vesicles and bladder in basal slices as well as from rectum in central slices.

\begin{figure}[!t]
	\centering
	\includegraphics[width=0.5\linewidth]{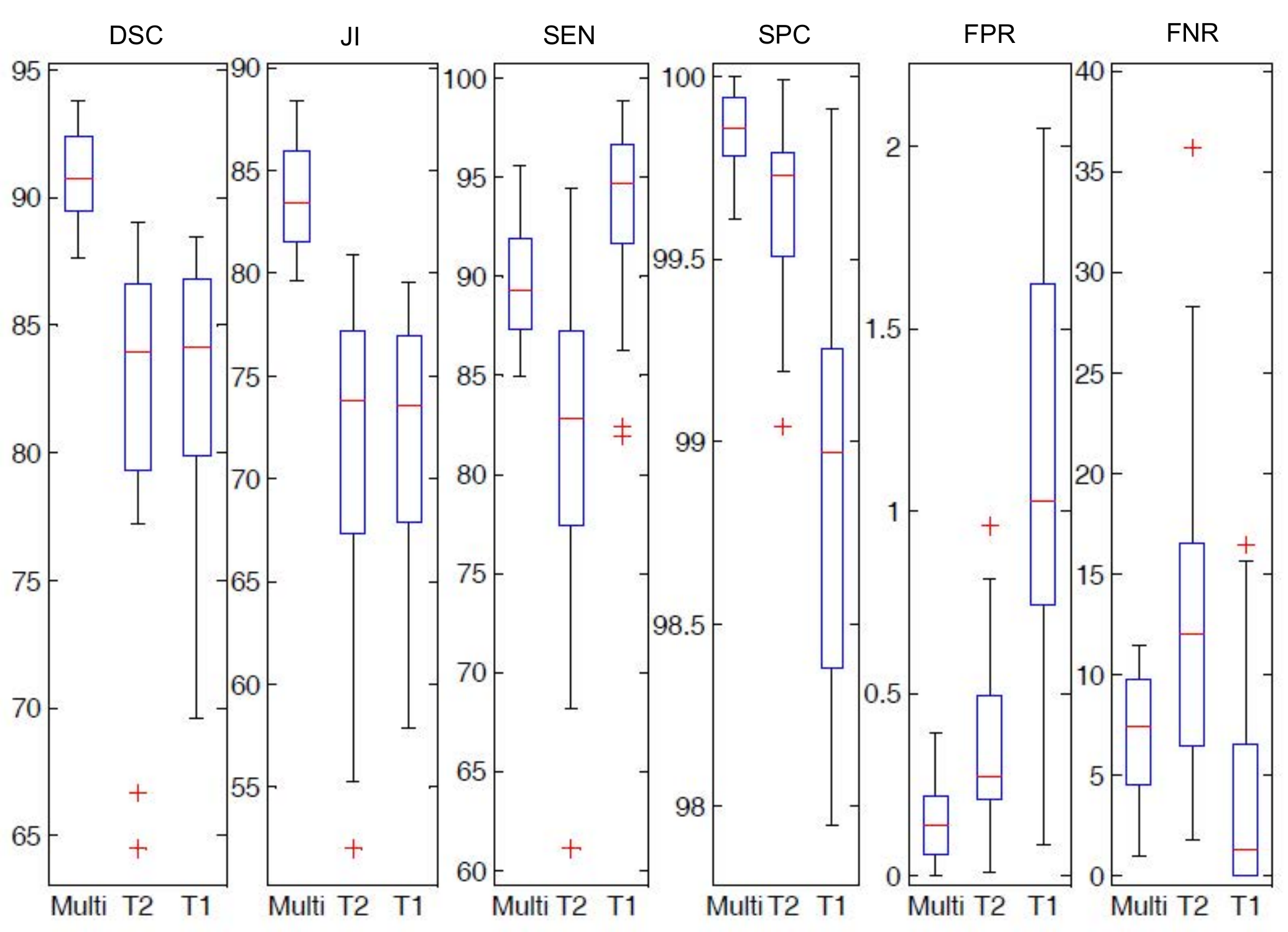}
	\caption[Boxplots of the spatial overlap-based metrics values achieved by the proposed prostate gland ROI segmentation approach based on the FCM algorithm]{Boxplots of the spatial overlap-based metrics values achieved by the proposed prostate gland ROI segmentation approach based on the FCM algorithm. The lower and the upper bounds of each box represent the first and third quartiles of the statistical distribution, respectively. The median value (i.e., the second quartile) is represented by a red line dividing the box. Whisker value is $1.5$ in all cases.}
	\label{fig:PGS-BoxplotsOM}
\end{figure}

Table \ref{table:PGS-ResOM} reports the mean and standard deviation values of the overlap-based metrics obtained in the experimental segmentation tests on $21$ MRI series representing prostate gland executing the proposed segmentation method, based on the FCM clustering algorithm, on multispectral MRI anatomic images, T2w images alone, and T1w images alone, respectively.
High \emph{DSC} and \emph{JI} mean values with very low standard deviation show the segmentation accuracy and reliability.
In addition, \emph{SEN} and \emph{SPC} average values involve the correct detection of the “true” areas and the ability of not detecting “wrong" parts in the segmented prostate gland ROIs, respectively.
This is also shown by the obtained \emph{FPR} and \emph{FNR} values.
The multispectral approach significantly outperforms the monoparametric ones.
The results associated to T1w processing are slightly more sensitive with respect to the proposed multispectral approach, even if they can include also wrong parts as shown by the higher \emph{FPR} values.
However, in all the cases, the variance of the results achieved by the monoparametric approaches is higher compared to the proposed multiparametric method.
To provide a graphical representation of the statistical distribution of the results, the corresponding boxplots of spatial overlap-based evaluation metrics are also reported in Fig. \ref{fig:PGS-BoxplotsOM}.
The short width of the interquartile range (i.e., difference between the third and first quartiles) represented in boxplots implies that the values are considerably concentrated.
All index distributions for the multispectral approach do not present outliers, thus demonstrating extremely low statistical dispersion.

\begin{table}[!t]
\centering
	\caption[Values of the spatial overlap-based metrics regarding the achieved prostate gland ROI segmentation results]{Values of the spatial overlap-based metrics regarding the achieved prostate gland ROI segmentation results. The results are expressed as average value $\pm$ standard deviation.}
	\label{table:PGS-ResOM}
	\begin{scriptsize}
		\begin{tabular}{ccccccc}
			\hline\hline
			MRI data	& DSC	& JI	& SEN	& SPC	& FPR	& FNR \\
			\hline
			Multispectral &	$90.77 \pm 1.75$ &	$83.63 \pm 2.65$ &	$89.56 \pm 3.02$ &	$99.85 \pm 0.11$ &	$0.15 \pm 0.11$ &	$6.89 \pm 3.02$ \\
			T2w alone &	$81.90 \pm 6.49$ &	$71.39 \pm 7.56$ &	$82.21 \pm 8.28$ &	$99.63 \pm 0.25$ &	$0.37 \pm 0.25$ &	$12.58 \pm 8.49$ \\
			T1w alone &	$82.55 \pm 4.93$ &	$71.78 \pm 6.15$ &	$93.27 \pm 4.87$ &	$98.85 \pm 0.58$ &	$1.15 \pm 0.58$	& $3.58 \pm 4.95$ \\
			\hline\hline
		\end{tabular}
	\end{scriptsize}
\end{table}

\begin{table}[!t]
    \centering
	\caption[Values of the spatial distance-based metrics regarding the achieved prostate gland ROI segmentation results]{Values of the spatial distance-based metrics regarding the achieved prostate gland ROI segmentation results. The results are expressed as average value $\pm$ standard deviation.}
	\label{table:PGS-ResDM}
	\begin{scriptsize}
		\begin{tabular}{cccc}
			\hline\hline
			MRI data	& AvgD [pixels]	& MaxD [pixels]	& HD [pixels] \\
			\hline
			Multispectral & $2.676 \pm 0.616$ &	$8.485 \pm 2.091$ &	$4.259 \pm 0.548$ \\
			T2w alone &	$4.941 \pm 1.780$ &	$13.779 \pm 3.430$ &	$4.535 \pm 0.335$ \\
			T1w alone &	$5.566 \pm 1.618$ &	$14.358 \pm 3.722$ &	$5.177 \pm 0.857$ \\
			\hline\hline
		\end{tabular}
	\end{scriptsize}
\end{table}

\begin{figure}[!t]
	\centering
	\includegraphics[width=0.35\linewidth]{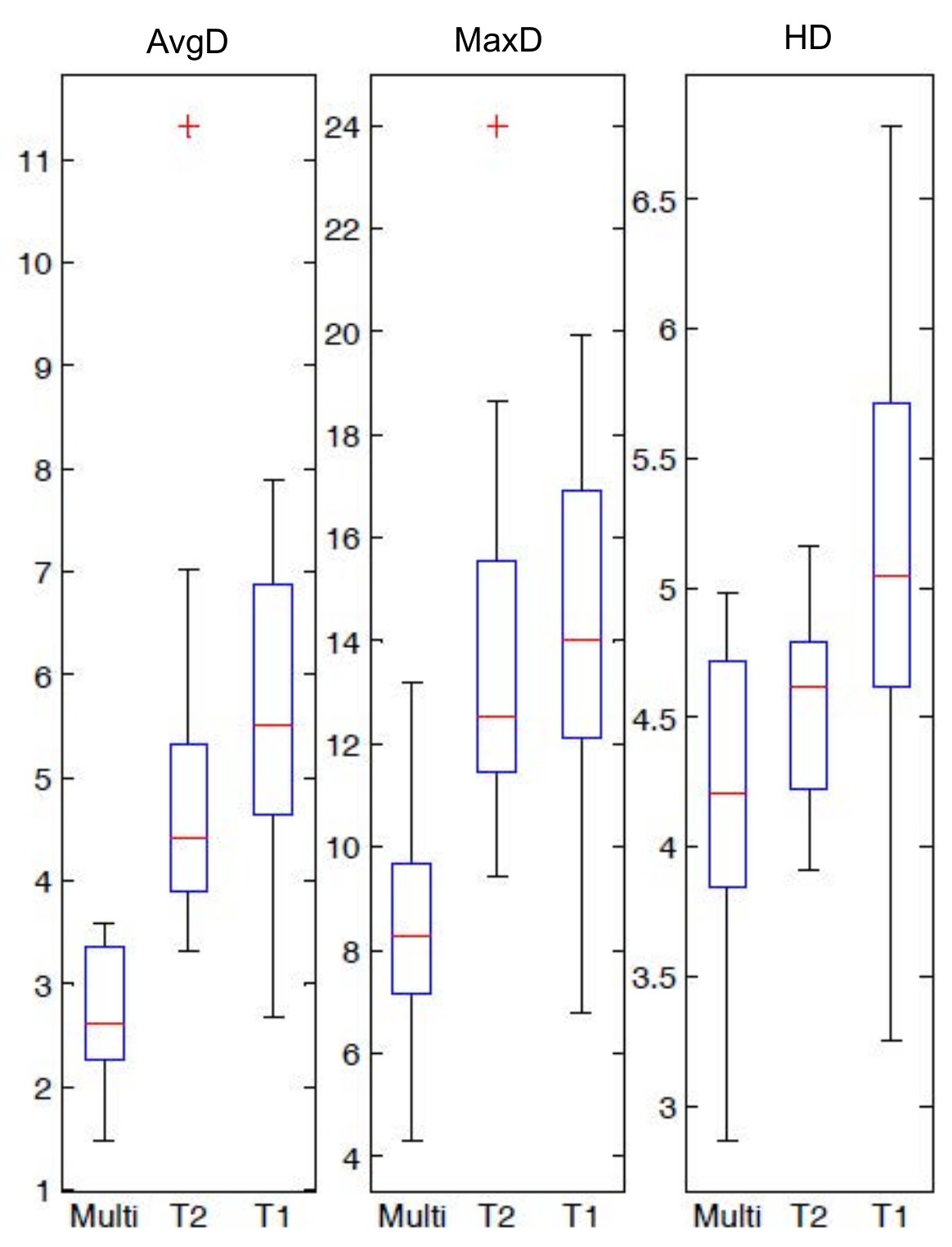}
	\caption[Boxplots of the spatial distance-based metrics values achieved by the proposed prostate gland ROI segmentation approach based on the FCM algorithm]{Boxplots of the spatial distance-based metrics values achieved by the proposed prostate gland ROI segmentation approach based on the FCM algorithm. The lower and the upper bounds of each box represent the first and third quartiles of the statistical distribution, respectively. The median value (i.e., the second quartile) is represented by a red line dividing the box. Whisker value is $1.5$ in all cases.}
	\label{fig:PGS-BoxplotsDM}
\end{figure}
 
Table \ref{table:PGS-ResDM} shows distance-based metrics mean and standard deviation values obtained using the proposed segmentation approach on the dataset composed of $21$ patients, by processing multispectral MRI anatomic images, T2w images alone, and T1w images alone, respectively.
The lower the distance values, the better the segmentation results.
The achieved spatial distance-based indices are consistent with overlap-based metrics.
Hence, good performances were obtained also with heterogeneous prostates. As for the spatial overlap-based metrics, also the results obtained on multiparametric MRI anatomic data outperform the monoparametric ones.
Fig. \ref{fig:PGS-BoxplotsOM} depicts the boxplots of the achieved spatial distance-based evaluation metrics.

Observing the boxplots, just a small deviation between the segmentations of the proposed method and those of the experienced radiologist can be denoted.
Considering the practical purposes of the proposed prostate gland delineation method for volume calculation in clinical applications, referring to MRI spatial resolution (i.e., isotropic pixel resolution of $0.703$ mm in Table \ref{table:PGS-MRIcharacteristics}), the proposed multispectral approach achieved on average $5.965 \pm 1.470$ mm in the worst case (i.e., $MaxD$), so ensuring effective performance.
This confirms the accuracy and reliability of the proposed segmentation method in real clinical contexts.

\begin{figure}[!t]
	\centering
	\includegraphics[width=0.73\linewidth]{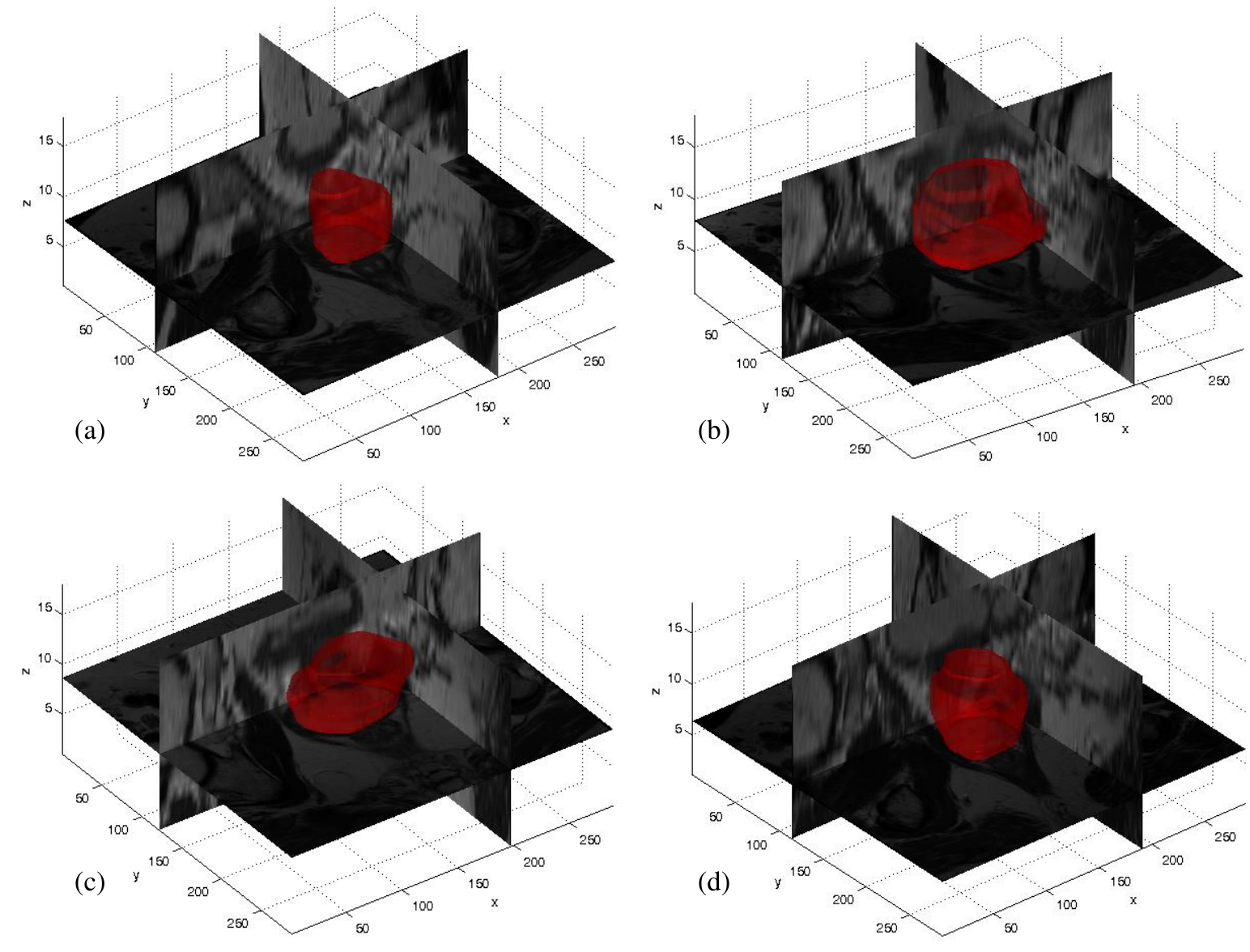}
	\caption[Tridimensional reconstruction of the segmented prostate gland ROI on the $3$D orthogonal tri-planar view]{Tridimensional reconstruction of the segmented prostate gland ROI on the $3$D orthogonal tri-planar view for the Patients: (a) $\#1$; (b) $\#5$; (c) $\#7$; (d) $\#12$. Transparent surfaces of the prostate gland are rendered with Alpha blending ($\alpha = 0.40$).}
	\label{fig:PGS-VolRend}
\end{figure}

For a more complete and intuitive graphical representation, four different instances of volumetric reconstructions of segmented prostate glands are shown in Fig. \ref{fig:PGS-VolRend}.
It is worth to note the variety of the prostate gland dimensions and shapes.

\subsubsection{Discussion}
The proposed approach provided good segmentation results compared to the gold standard boundaries delineated by an experienced radiologist.
Although volume-based metrics do not take into account the intersection, they represent a first and immediate measure of segmentation effectiveness.
Just small differences between the automatic and the manual volume measurements were observed, showing the reliability of the proposed segmentation approach for automatic volume calculation in clinical applications.
The achieved overlap-based indices, which are characterized by high average values and very low standard deviation, reveal the segmentation accuracy, involving the correct detection of the “true” pathological areas as well as the capability of not detecting wrong parts within the segmented prostate.
The achieved spatial distance-based metrics agree with the overlap-based ones, corroborating the aforementioned experimental evidences.

The proposed method is based on an unsupervised Machine Learning technique, without requiring any training phase.
On the contrary, the other literature works use or combine atlases \cite{klein2008,litjens2012,martin2008,martin2010}, AAMs \cite{vincent2012}, or statistical shape priors and ASMs \cite{gao2010,martin2008,martin2010,toth2013,toth2011MedIA}, which require manual labeling of a significant image sample set performed by expert physicians.
In addition, as stated in \cite{martin2010}, atlas-based approaches may be affected by serious errors when the processed prostate instances are dissimilar to the atlas, despite the non-rigid registration.
Literature approaches used T2w MR anatomic images, sometimes combined with ADC maps \cite{litjens2012} or MRSI data \cite{toth2011MedIA}.
Our method integrated T1w and T2w MRI anatomic data to enhance clustering segmentation results.
To the best of our knowledge, we combined T1w and T2w MR image series for the first time in prostate gland segmentation.
As it is stated in \cite{lemaitre2015}, the available public prostate MRI datasets for prostate gland segmentation and prostate cancer detection and delineation, such as PROMISE12 \cite{litjens2014} and the benchmark proposed in \cite{lemaitre2015}, do not provide T1w MR images.
Therefore, our method cannot be applied on these public datasets and compared with the prostate MR image segmentation approaches presented at the MICCAI 2012 PROMISE12 challenge, since our aim is to show that concatenating T2w and T1w pixels during the construction of the feature vector in an early integration phase.
Although a comparison with state-of-the-art methods could be certainly interesting, it is also unfeasible because atlases should be built and supervised methods with \textit{a priori} knowledge (such as Active Appearance Models or statistical shape priors) should be trained on our prostate MRI dataset composed of $21$ patients.
The number of samples is not sufficiently significant to apply suitably supervised Machine Learning techniques or shape-based models without using LOOCV, such as in \cite{vincent2012} and \cite{martin2008}.
As shown in Fig. \ref{fig:PGS-SegResReg}, our approach is also able to correctly segment apical and basal prostate MRI slices, by differentiating also seminal vesicles (see Fig. \ref{fig:PGS-SegResReg}f).
The segmentation quality was also good at the prostate-rectum and prostate-bladder interfaces.
Whereas a segmentation error of a few millimeters is clinically acceptable at boundaries with muscular tissue, the interfaces with rectum and bladder need to be detected and distinguished very accurately \cite{klein2008}.
Moreover, good performances were obtained also with prostate gland instances imaged as heterogeneous regions (i.e., inhomogeneous signal in the peripheral zone or adenomatous central gland).
The principal experimental finding is that the FCM clustering on multispectral MRI anatomic data considerably enhances the achieved prostate ROI segmentations, by taking advantage of the prostate uniform intermediate signal intensity at T1w imaging \cite{choi2007}.
The FCM clustering algorithm on multispectral MR anatomic images allows for more accurate prostate gland ROI segmentation with respect to the clustering results on either T2w or T1w images alone.
However, T1w images are often not able to distinguish among different soft-tissues.
For instance, FCM clustering on T1w images alone is not able to differentiate the prostate-rectum (Fig. \ref{fig:PGS-SegResReg}f) or the prostate-muscle (Figs. \ref{fig:PGS-SegResReg}f and \ref{fig:PGS-SegResReg}i) interfaces.
In conclusion, combining and fusing T2w and T1w MRI data, in the feature vector construction step, allow us to achieve better clustering outputs.
The method was insensitive to variations in patient age, prostate volume, and the presence of tumors (i.e., suspicion of cancer in different prostate regions, inhomogeneous signal in the PZ or adenomatous CG with possible nodules), also considering radiotherapy or chemotherapy treatments, thus increasing its feasibility in clinical practice.

\subsubsection{Conclusions}
NeXt, which is based on an unsupervised Machine Learning technique, was used to effectively support prostate gland delineation, such as in radiation therapy.
The developed approach was tested on a dataset composed of $21$ patients, considering T2w and T1w MRI series.

Spatial overlap-based and distance-based metrics were calculated to evaluate the performance, proving the accuracy of the proposed segmentation approach.
The achieved experimental results demonstrated the great robustness of the proposed approach even when MRI series were affected by acquisition noise or artifacts.
The good segmentation performance was also confirmed with prostate glands characterized by inhomogeneous signal in the peripheral zone or adenomatous central gland with possible nodules.
The developed method could be clinically feasible, by addressing and overcoming typical issues affecting manual segmentation procedures, i.e., time-consuming and operator-dependency.

Future work could be directed towards a more selective segmentation technique, distinguishing between CG and PZ of the prostate anatomy \cite{choi2007,hricak1987}.
The advanced segmentation method can be definitely employed in a two-step prostate cancer delineation approach, in order to focus on pathological regions in the central gland and in the peripheral zone.
The overall prostate cancer segmentation method could be integrated into a Computer-Aided Diagnosis (CAD) systems, which include both Computer-Aided Detection (CADe) and Computer-Aided Diagnosis (CADx), enhancing the diagnosis performance of radiologists \cite{lemaitre2015}.
In this way, a more accurate PCa staging can be performed, and determining the disease burden may also lead to greater benefit for the patients who underwent treatment \cite{ahmed2009}.

\section{Graph-based methods}
\label{sec:graphMethods}

Sinop and Grady \cite{sinop2007} presented a unifying graph theoretical framework for seeded image segmentation depending on the used norm in the objective function optimization.
A weighted graph consists of a pair $\mathcall{G} = (\mathcall{V}, \mathcall{E})$ with vertices $v \in \mathcall{V}$ and edges $e \in \mathcall{E} \subseteq \mathcall{V} \times \mathcall{V}$. Each edge $e_{ij}$, connecting the vertices $v_i$ and $v_j$, is associated to a weight $w_{ij} \in \mathbb{R}^+$.
The graph is assumed to be connected and undirected (i.e., $w_{ij} = w_{ji}$).
In seeded image segmentation, vertices correspond to image pixels while edge weights $w_{ij}$ represent similarity measures between neighboring pixels according to image features.
Moreover, each vertex $v_i$ has an attribute $x_i \in [0, 1]$, which indicates the probability of a label.

Without loss of generality, a two-class image segmentation problem is considered. Starting from initial foreground $\mathcal{F}$ and background $\mathcal{B}$ seeds, the labeling process is solved by the general seeded segmentation optimization in Eq. (\ref{eq:segOpt}):

\begin{equation}
    \label{eq:segOpt}
	x^{\text{opt}} = \arg \min_{x} \left[ \sum_{e_{ij} \in \mathcall{E}} \left( w_{ij} |x_i - x_j| \right)^q\right]^{\frac{1}{q}},
\end{equation}

\noindent with $x(\mathcal{F})=1$, $x(\mathcal{B})=0$, and $q \in \mathbb{R}^+$.
The final segmentation solution $\mathcal{S}$ is defined by:
\begin{equation}
    \label{eq:segSol}
	S_i = 
     \begin{cases}
       1, &\quad\text{if } x_i \ge 0.5\\
       0, &\quad\text{if } x_i < 0.5
     \end{cases}.
\end{equation}

As described by Sinop and Grady \cite{sinop2007}, the solution has been shown to converge to various algorithms by considering different $q$ values and the resulting $\ell_q$ norm:
\begin{itemize}
    \item Graph cuts for $q = 1$ \cite{boykov2001a,greig1989};
    \item Random walker for $q = 2$ \cite{grady2006};
    \item Shortest paths for $q = \infty$ \cite{bellman1956,ford1956}.
\end{itemize}

The main advantage of using CA algorithms against graph-theoretic formulations is the ability to obtain a multi-label solution in a simultaneous iteration.
For instance, the ‘‘Graph cuts’’ algorithm finds a global optimum only for binary labeling; the extensions proposed in literature to achieve approximate solutions on multi-label problems exploit ‘‘Graph cuts’’ iteratively to a sequence of binary problems \cite{boykov2001b,boykov2001a}.
In addition, the local transition rules are simple to interpret, and it is possible to impose prior knowledge, specific to the problem, into the image segmentation approach \cite{hamamci2012}.
Thus, the state of the entire lattice advances in discrete time steps \cite{kauffmann2010}.

\subsection{Shortest Path with Cellular Automata}
\label{sec:shortestPathCA}
Cellular Automata (CA) are one of the oldest models of bio-inspired computing \cite{kari2008} and were introduced by John von Neumann and Stanislaw Ulam in their seminal papers \cite{vonNeumann1966}.
The goal was to design self-replicating artificial systems that make up a discrete universe consisting of a bi-dimensional mesh of finite state machines (cells), interconnected locally to each other, which will produce a complex global emergent behavior \cite{wolfram1984}.
Each cell changes its own state synchronously depending on both: (\textit{i}) its current state, and (\textit{ii}) the states of the neighbor cells at the previous discrete time step, as determined by a local update rule.
All cells use the same update rule so that the system is homogeneous like many physical and biological systems \cite{kari2005}.
The most common CA are discrete, in both space and time, and process a lattice of $d$-dimensional sites $\mathbf{p} \in \mathbb{Z}^d$.
The $1$-D, $2$-D, and $3$-D cases are the most used topologies in practical problems \cite{bhattacharjee2018}.

CA have been also effectively applied in image processing \cite{popovici2002,rosin2010} and Pattern Recognition and classification tasks \cite{raghavan1993,wongthanavasu2016}, also with applications in cancer imaging \cite{mitra2015}.
In image processing, $\mathbf{p}$ represents a pixel ($d = 2$) or a voxel ($d = 3$) of the digital image $\mathcall{I}$.
The image segmentation problem can be solved using a labeling process, where $K$ different types of self-replicating cells compete to occupy the grid space defined by the image $\mathcall{I}$, according to a CA model.
Formally, a deterministic Cellular Automaton is defined by a triplet $\mathcal{A} = (\mathcal{S}, \mathcall{N}, f)$, where: $\mathcal{S}$ is the non-empty state set; $\mathcall{N}$ is the set of the $\nu$ neighbors of a particular cell; $f:\mathcal{S}^\nu \rightarrow \mathcal{S}$ is the local transition rule, which, given the states of the neighborhood cells at the current
time step $t$, maps the cell state at the next $t + 1$ time step.
Let $\mathbf{p} \equiv \left(p_1, p_2, \ldots, p_d\right) \in \mathbb{Z}^d$ be a cell location, the most used neighborhood definitions are:
\begin{itemize}
    \item von Neumann neighborhood $\mathcall{N}(\mathbf{p}) = \left\{\mathbf{q} \in \mathbb{Z}^d: \norm{\mathbf{p} - \mathbf{q}}_1 \leq 1 \right\}$, where $\norm{\mathbf{p} - \mathbf{q}}_1 = \sum_{i=1}^d|p_i - q_i|$ is the $\ell_1$ Manhattan norm;
    \item Moore neighborhood $\mathcall{N}(\mathbf{p}) = \left\{\mathbf{q} \in \mathbb{Z}^d: \norm{\mathbf{p} - \mathbf{q}}_\infty \leq 1 \right\}$, where $\norm{\mathbf{p} - \mathbf{q}}_\infty =$ \sloppy $\smash{\displaystyle\max_{i=1, \ldots, d}}\left\{|p_i - q_i|\right\}$ is the $\ell_\infty$ max-norm.
\end{itemize}

\begin{figure}[!t]
	\centering
	\subfloat[]{\includegraphics[width=0.3\textwidth]{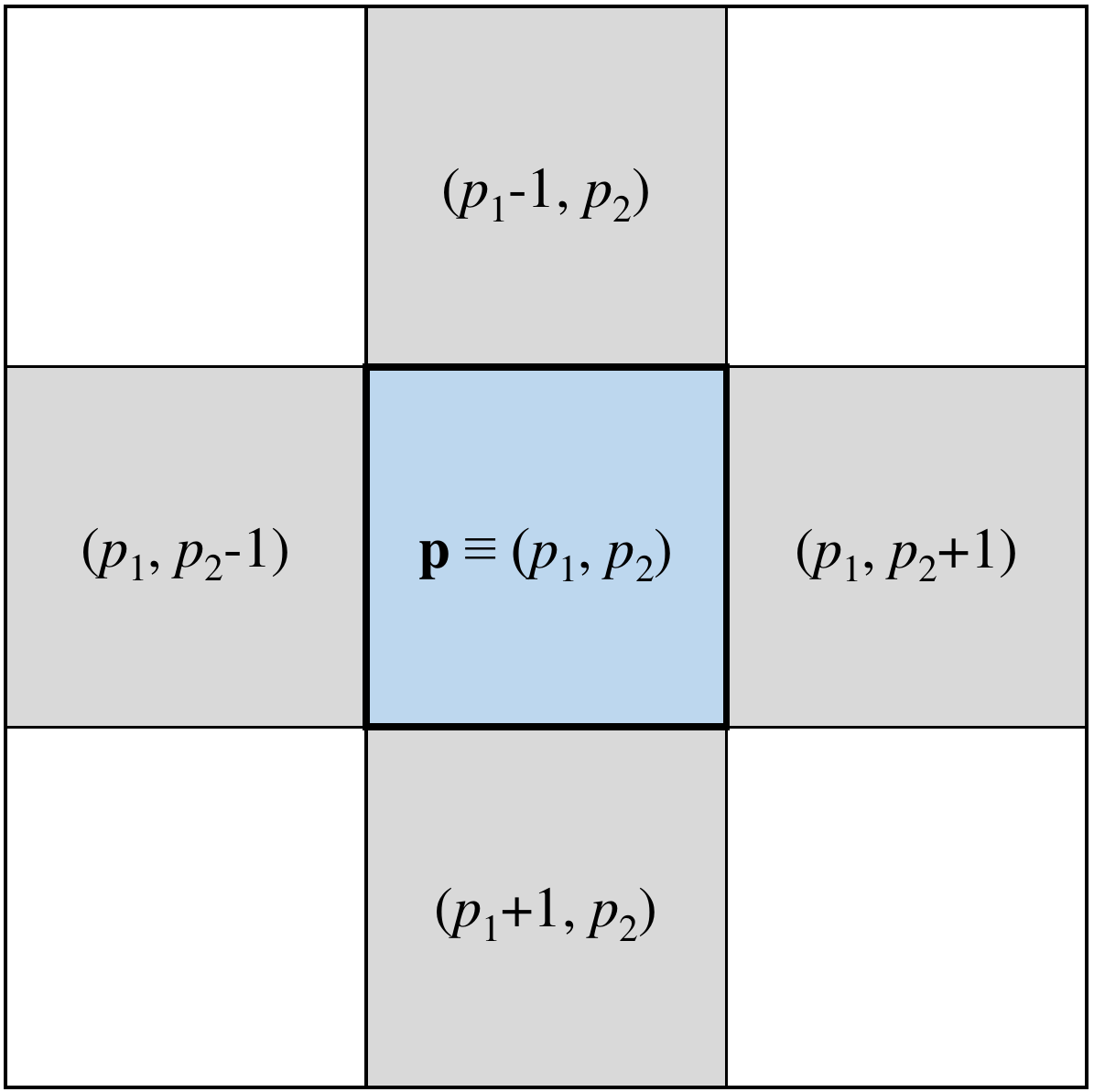}}\qquad
	\subfloat[]{\includegraphics[width=0.3\textwidth]{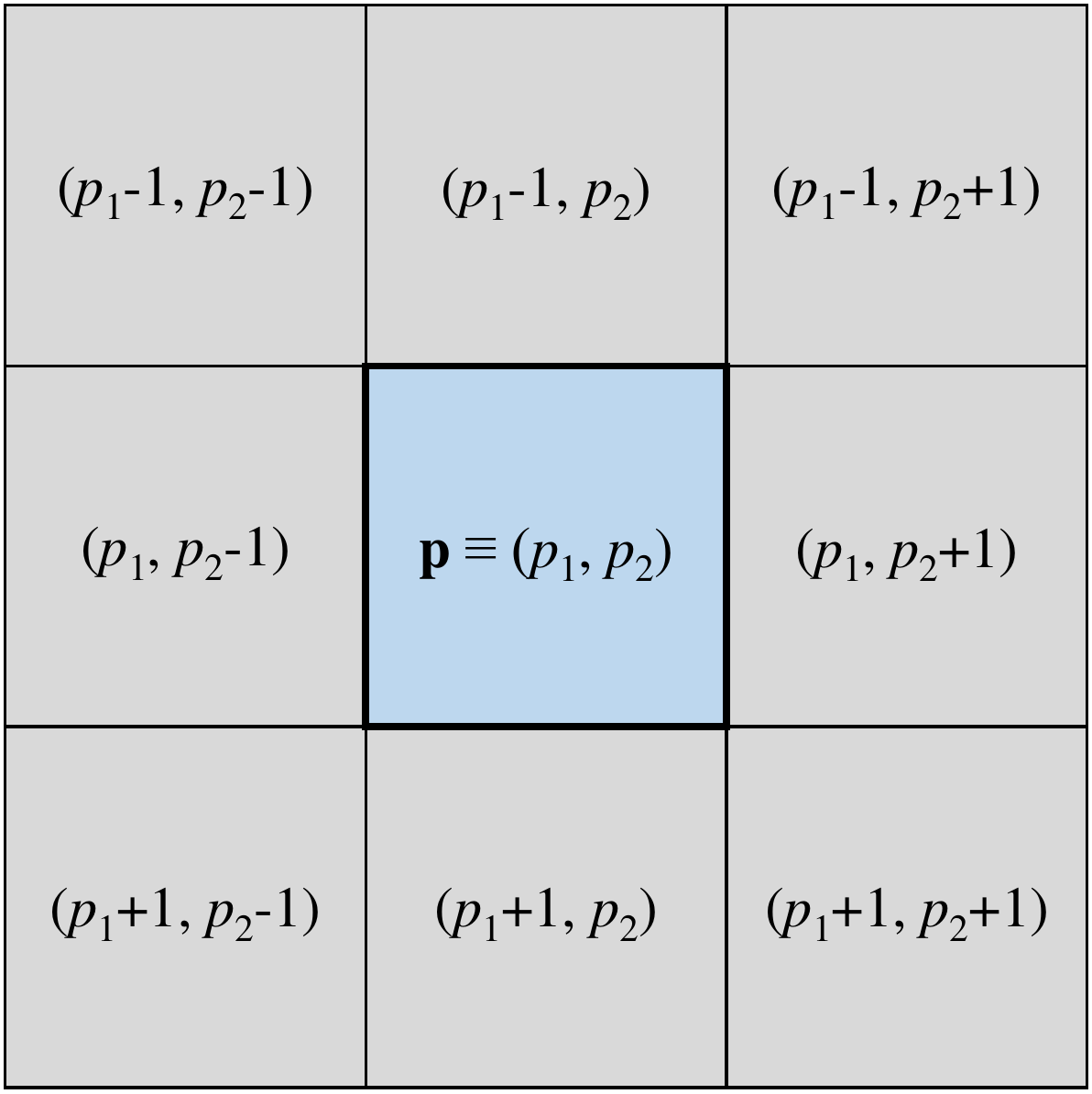}}\\
	\caption[Classical neighborhoods in the $2$-dimensional case: (a) von Neumann neighborhood; (b) Moore neighborhood]{Classical neighborhoods in the $2$-dimensional case: (a) von Neumann neighborhood; (b) Moore neighborhood. Pixel indices are compliant with the conventional image coordinate system for representing digital images as arrays.}
	\label{fig:CAneigh}	
\end{figure}

For greater clarity, the von Neumann and Moore neighborhoods for the $2$-dimensional case are illustrated in Figs. \ref{fig:CAneigh}a and \ref{fig:CAneigh}b, respectively.
See \cite{zaitsev2017} for more details.

In the ‘‘GrowCut’’ CA formulation \cite{vezhnevets2005}, the cell state is a triple $S_\mathbf{p} := \left( l_\mathbf{p}, \theta_\mathbf{p}, \mathbf{c}_\mathbf{p} \right)$, where: $l_\mathbf{p} \in \{1,2, \ldots, K\}$ is the label of the current cell in the $K$-labeled segmentation process; $\theta_\mathbf{p} \in [0,1]$ is the strength value of the current cell; $\mathbf{c}_\mathbf{p} \in \mathbb{R}^n$ is the
feature vector associated to the cell $\mathbf{c}$ according to the image content.
The automaton is initialized with the previously calculated seeds, by labeling foreground and background seeds.
The maximum strength value is assigned to the labeled cells, while unlabeled cells are set $\theta_\mathbf{p}^{(0)}=0$.
Cell states are then updated synchronously at each iteration $t$, until the convergence condition is reached (i.e., any cell does not change its own state).
State transitions occur if the attacker force of a neighbor cell $\mathbf{q} \in \mathcall{N}(\mathbf{p})$ (Moore $8$-neighborood was used), defined by its strength $\theta_\mathbf{q}^{(t)}$ and the distance between $\mathbf{c}_\mathbf{p}$ and $\mathbf{c}_\mathbf{q}$ feature vectors, is greater than the current cell strength $\theta_\mathbf{p}^{(t)}$:
\begin{equation}
    \label{eq:stregths}
	g(\mathbf{c}_\mathbf{p}, \mathbf{c}_\mathbf{q}) \cdot \theta_\mathbf{q}^{(t)} > \theta_\mathbf{p}^{(t)},
\end{equation}
where $g(\mathbf{c}_\mathbf{p}, \mathbf{c}_\mathbf{q})$ is a pixel similarity function, which is monotonically decreasing and bounded.
This property ensures the convergence of the algorithm.

The utilized function in ‘‘GrowCut’’ \cite{vezhnevets2005} is based on the absolute image feature difference:
\begin{equation}
    \label{eq:g_IFD}
	g_\text{IFD}(\mathbf{c}_\mathbf{p}, \mathbf{c}_\mathbf{q}) = 1- \frac{\norm{\mathbf{c}_\mathbf{p} - \mathbf{c}_\mathbf{q}}_2}{\max_{\forall \mathbf{c}_\mathbf{i} \in \mathcall{I}}\norm{\mathbf{c}_\mathbf{i}}_2}.
\end{equation}
The authors of \cite{kauffmann2010} demonstrated that ‘‘GrowCut’’ is equivalent to the Bellman–Ford algorithm.

In \cite{hamamci2012,hamamci2010}, the ‘‘Tumor-Cut’’ CA model implements the gradient magnitude:
\begin{equation}
    \label{eq:g_GM}
	g_\text{GM}(\mathbf{c}_\mathbf{p}, \mathbf{c}_\mathbf{q}) = e^{-\norm{\mathbf{c}_\mathbf{p} - \mathbf{c}_\mathbf{q}}_2}.
\end{equation}
The resulting algorithm is equivalent to the shortest path algorithm, as shown in \cite{sinop2007}.

Recalling the general graph-based theoretical framework: $x_i$ can be seen as the cell strength $\theta_i$ , while the edge weights $w_{ij}$ are related to the values taken by the pixel similarity function $g(\mathbf{c}_i, \mathbf{c}_j)$.

\subsubsection{CA in brain tumor modeling}
CA have been widely employed in dynamic complex system modeling and simulation \cite{bandini2001}.
Thanks to the capability of representing the individual state of each particle, CA are well suited for microscopic modeling and simulation approaches.
Even if each cell takes a local decision, aggregate measures and complex emergent behaviors derive from these decisions.
Thereby, CA models can be applied to diverse real phenomena and research fields.

In particular, CA were also adopted to simulate early tumor growth, by examining the roles of host tissue vascular density and tumor metabolism in the ability of a small number of monoclonal transformed cells to develop into an invasive tumor \cite{patel2001}.
Computationally, the mathematical description of tumor growth can be formulated at different spatial scales: (\textit{i}) simulating the growth at the cellular level (using CA), (\textit{ii}) defining at macroscopic scale how the tumor density will evolve (with Partial Differential Equations (PDEs)).
However, the models attempt to predict the mathematical law of the tumor growth in both cases \cite{angelini2007}.
Kansal \textit{et al.} \cite{kansal2000} developed a three-dimensional CA model of brain tumor growth, showing that macroscopic tumor behavior can be realistically modeled using microscopic parameters.
As a matter of fact, the rules of division and invasion are the key elements of the CA-based approaches.
The hybrid model proposed in \cite{patel2001} incorporates normal cells, tumor cells, necrotic or empty space, and a random network of native microvessels as components of a CA state vector.
More recently, Ibrahim-Hashim \textit{et al.} \cite{ibrahim2017} showed that intra-tumoral subpopulations can be defined using cellular adaptive strategies and intra-tumoral Darwinian interactions rather than molecular properties employed in branching clonal evolution models.
This multi-scale hybrid mathematical model coupled a discrete CA---as an Agent-Based Model (ABM) for describing explicitly the space and state transitions---and a set of reaction-diffusion PDEs---for spatio-temporal interactions.
A reaction-diffusion model describes the change in the distribution and number of tumor cells due to the random movement of tumor cells (diffusion) and proliferation (reaction).

\subsubsection{Brain tumor segmentation based on a CA model}
\label{sec:GTVcut}
This section presents a novel semi-automatic MR brain tumor image segmentation approach (\textsc{GTVcut}) based on a CA model.

\paragraph{Patient dataset description}
The study was performed on a dataset composed of $25$ patients (average age: $59.84 \pm  9.99$ years), affected by metastatic brain cancers, who underwent Gamma Knife neuro-radiosurgery.
Some subjects among the enrolled patients showed multi-metastatic scenarios, so the total number of lesions was $32$.

\begin{figure}[!t]
	\centering
	\includegraphics[width=\linewidth]{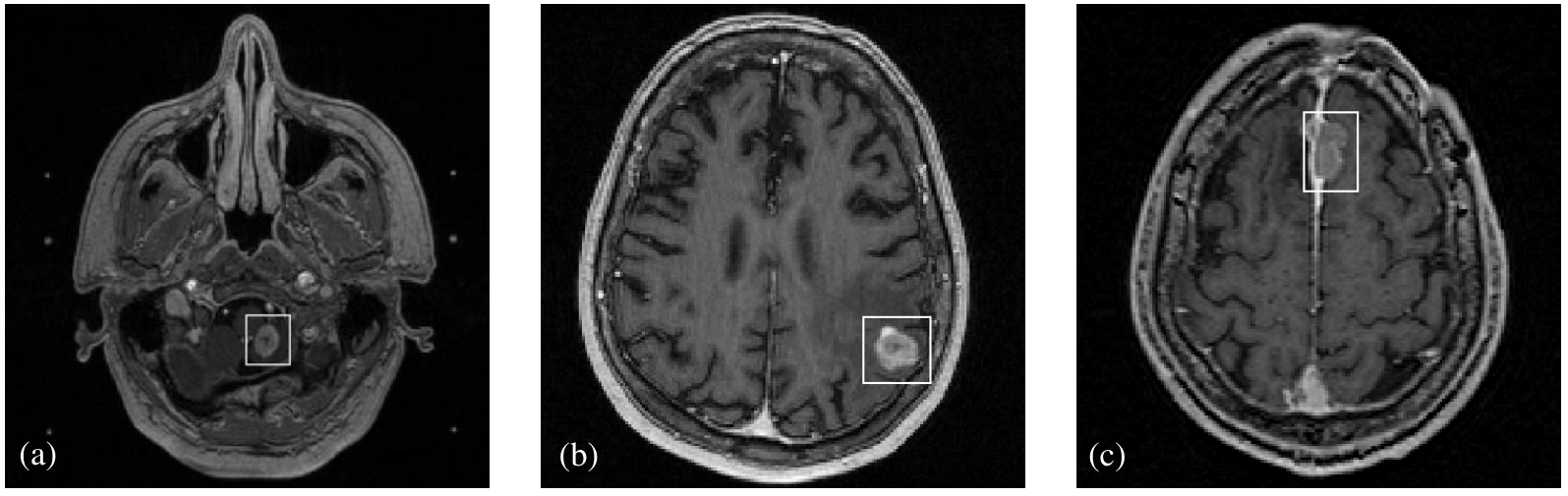}
	\caption[Three instances of input T1w contrast-enhanced MR axial slices with the rectangular bounding area, selected by the user]{Three instances of input T1w contrast-enhanced MR axial slices with the rectangular bounding area, selected by the user, containing the tumor: (a) moderately hyper-intense enhancing region including a necrotic core; (b) highly hyper-intense enhancing region with inner necrotic material; (c) heterogenous cancer.}
	\label{fig:GTVcut-inputImages}
\end{figure}

All available T1w FFE CE-MRI series were acquired on a Philips Gyroscan Intera $1.5$ T MRI scanner, a few hours before the Gamma Knife treatment and then utilized for the planning phase.
Three significant examples of input MR images, representing three brain cancers with different characteristics, are shown in Fig. \ref{fig:GTVcut-inputImages}.
MRI acquisition parameters are reported in Table \ref{table:GK-MRIcharacteristics}.

\paragraph{\textsc{GTVcut}: brain cancer segmentation for neuro-radiosurgery}

\begin{figure}[!t]
	\centering
	\includegraphics[width=0.75\linewidth]{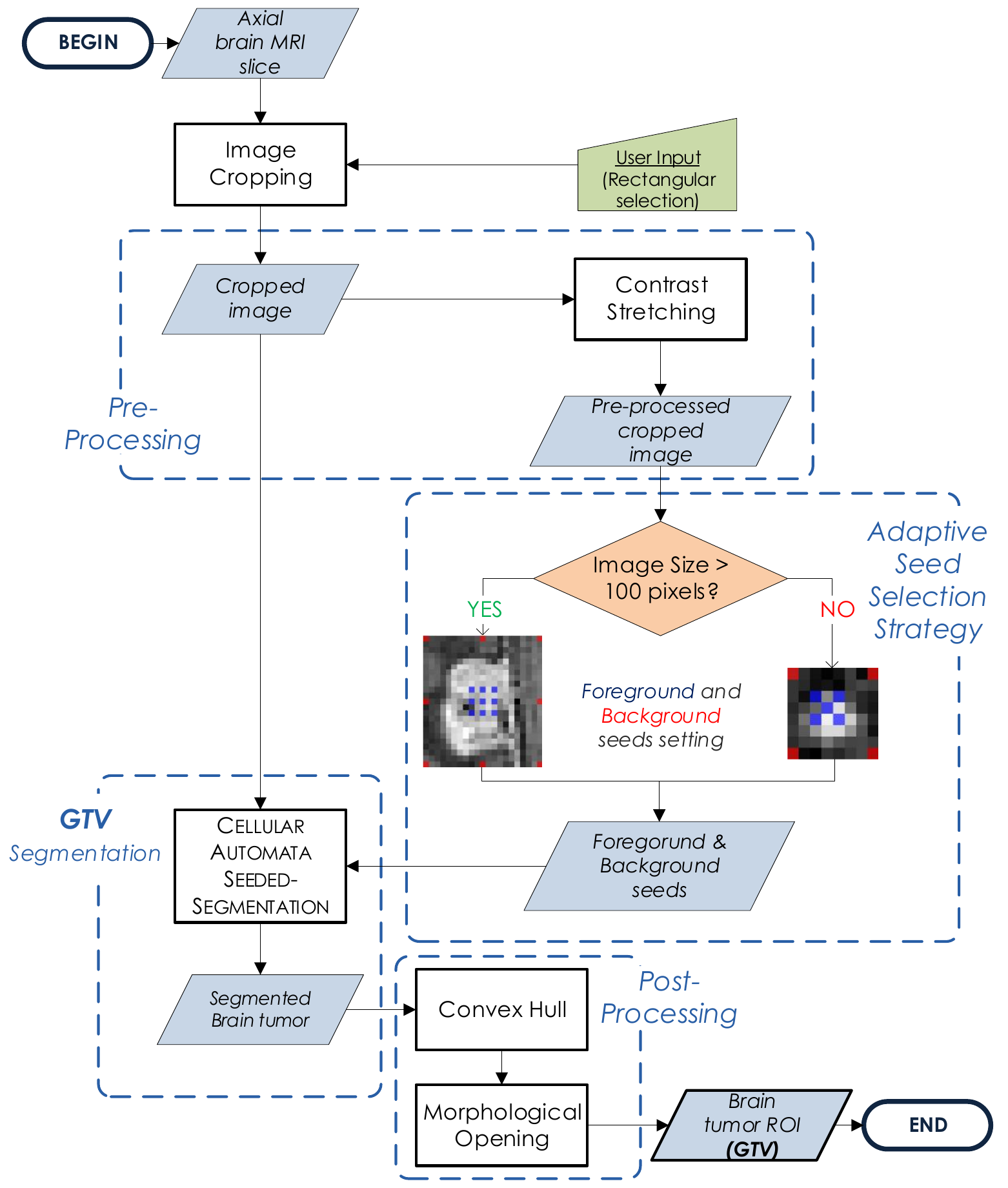}
	\caption[Flow diagram of the \textsc{GTVcut} segmentation approach]{Flow diagram of the \textsc{GTVcut} segmentation approach. The pipeline can be divided into four main stages: (\textit{i}) pre-processing to improve segmentation process; (\textit{ii}) adaptive seed selection to choose suitable foreground and background seed pixels; (\textit{iii}) GTV segmentation using a CA model; (\textit{iv}) post-processing to refine the achieved segmentation results.}
	\label{fig:GTVcut-flowDiagram}
\end{figure}

The analyzed brain metastatic cancers, wherein necrotic and enhancing tumor tissues are often mixed, exemplify the challenges encountered in radiosurgery scenarios.
Thus, interactive segmentation approaches represent a more feasible and safe solution for physician in clinical practice with respect to fully automatic approaches \cite{hamamci2012}.
Since region connectedness is ensured by the CA algorithm, segmentation results are represented by connected-regions with unbroken edges even if some necroses or cysts extend partially outside the tumor.
Dealing with connected target regions is very important in radiation therapy treatment planning, because tumor delineation is a critical step that must be performed accurately and effectively.
Moreover, since cancer imaging is usually characterized by non-uniform intensity distributions, local image statistics should be considered instead of global ones \cite{aslian2013}.
CA model is very suitable due to its own local update rules.
\textsc{GTVcut} results in a simple but effective application based on a plain cellular automata model, by initializing adaptively the foreground and background seeds in a smart fashion.
A detailed flow diagram of the proposed brain tumor segmentation approach is shown in Fig. \ref{fig:GTVcut-flowDiagram}.

According to the interaction technique employed in ‘‘GrabCut’’ \cite{rother2004}, user intervention consists just in dragging a rectangle around the target area, without
parameter settings.
Unlike ‘‘Graph cuts’’ \cite{boykov2001a} and ‘‘GrowCut’’ \cite{vezhnevets2005} algorithms, wherein the user has to define the foreground and background seeds explicitly by drawing the respective brushing pixels, our approach calculates both foreground and background seeds adaptively by using an adaptive strategy.
As a matter of fact, operator-dependency is minimized by lightening the load involving the user. This results in a very straightforward and usable ROI selection tool, which is generally provided by medical imaging clinical applications \cite{chen2009}.

\subparagraph{Pre-processing}
After the bounding region selection accomplished by the user, the cropped input images are pre-processed in order to improve the subsequent segmentation procedure.
The range of intensity values of the selected part is expanded to enhance the GTV extraction.
Thereby, a linear contrast stretching operation is applied.
This linear normalization converts input intensity values $r$ into the output values $s \in [0,1]$, by implying an expansion of the values between $r_{\text{min}}$ and $r_{\text{max}}$, and improving detail discrimination.

\begin{figure}[!t]
	\centering
	\includegraphics[width=0.9\linewidth]{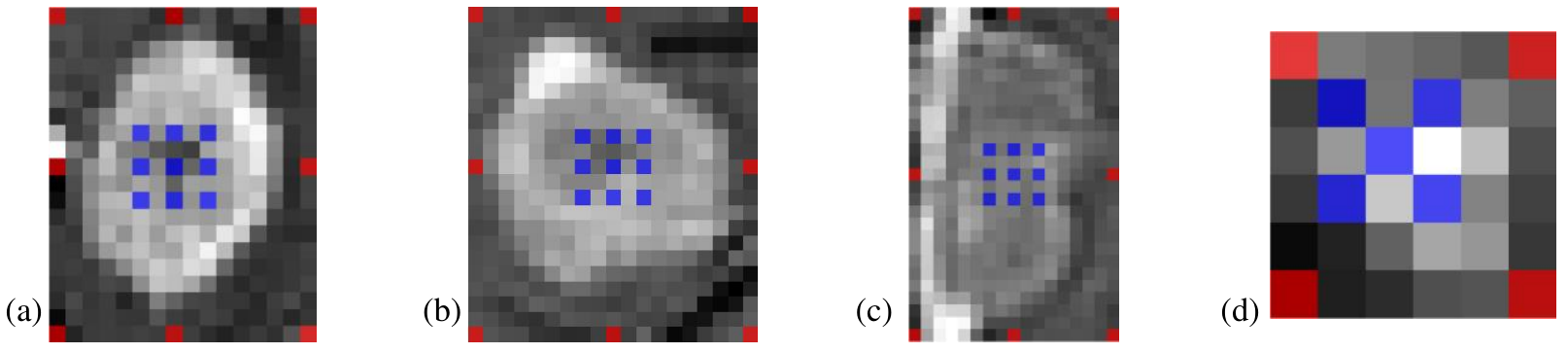}
	\caption[Adaptive seed selection according to the rectangular boundary regions (including the tumor) dragged by the user]{Adaptive seed selection according to the rectangular boundary regions (including the tumor) dragged by the user: (a, b, c) cropped image size greater than $100$ pixels (input MR slices in Fig. \ref{fig:GTVcut-inputImages}(a, b, c), respectively); (d) cropped image size less than or equals to $100$ pixels. Foreground and background seeds are shown using transparent blue and red pixels (alpha factor  $\alpha = 0.40$), respectively.}
	\label{fig:GTVcut-seedPoints}
\end{figure}

\subparagraph{Adaptive seed selection strategy}
A strategy for the choice of both foreground (tumor) and background (healthy tissue) seeds was designed \textit{ad hoc}.
Especially, according to the flow diagram in Fig. \ref{fig:GTVcut-flowDiagram}, seed selection depends on the size of the image cropped by the operator, and then it is adaptive and fully automatic.
Let $\mathcall{I}_{\text{cropped}}: \{1,2, \ldots, M\} \times \{1,2, \ldots, N\} \rightarrow [0,1]$ be the cropped MR image, whose size and center coordinates are $M \times N$ and $C \equiv (C_x, C_y) \equiv (\text{round}(M/2), \text{round}(N/2))$, respectively.
Adaptive seed selection is performed as follows:
\begin{itemize}
    \item if  $M \times N > 100$, nine foreground pixels in the central zone and eight background pixels at the border of the cropped image are chosen, such as in Fig. \ref{fig:GTVcut-seedPoints}(a, b, c), to provide significant samples for the CA algorithm initialization;
    \item otherwise, to avoid the access to pixels outside the cropped image, five foreground pixels in the central zone and four background pixels at the corners of the cropped image are chosen (see Fig. \ref{fig:GTVcut-seedPoints}d).
\end{itemize}

This discrimination is necessary because very small lesions could be present ($5$ mm diameter), such as in initial and final tumor slices.
Foreground tumor seed-points are set according to the center coordinates of the cropped image, hence the only trivial requirement for the operator is that the rectangular bounding region must be centered on the tumor.
This strategy implements a highly robust seed selection procedure.

\subparagraph{GTV segmentation}
A two-labeled CA segmentation is the most efficient scheme for classifying a hyper-intense lesion from the healthy part of the brain.
As a matter of fact, in contrast-enhanced MR brain scans, metastases have a brighter core than periphery and distinct borders differentiating them from the surrounding healthy brain tissue \cite{ambrosini2010}.
Nevertheless, the proposed method can be safely applied to benign and malignant brain tumors, since it relies on contrast medium absorption by tumor tissues.
In some instances, either central or adjacent areas of the lesions could comprise necrotic tissue \cite{chin2001,shah2012} that, for radiotherapeutic purposes, must be considered in the CTV during the treatment planning phase.
Our method is able to segment correctly heterogeneous tumors, even when foreground seeds are located in necrotic core proximity.
The pseudo-code for the proposed image segmentation \textsc{GTVcut} approach based on a CA model is given in the Algorithm \ref{pc:GTVcut} box.
Differently to “GrowCut” \cite{vezhnevets2005}, the user does not explicitly select seed-points.
In the proposed segmentation approach the user has to drag only a bounding rectangle.
Inheriting the interaction modality used in “GrabCut” \cite{rother2004}, the user is indicating the region of background and is free of any need to select foreground markers because seed selection is accomplished automatically by the adaptive strategy explained previously.
Unlike “Tumor-Cut” \cite{hamamci2012}, in which tumor segmentation is performed by a level set evolution on a tumor probability map, the proposed method is based solely on a CA model for the automated GTV extraction.
We used both the pixel similarity functions reported in Eqs. (\ref{eq:g_IFD}) and (\ref{eq:g_GM}) and an experimental comparison was performed.

\begin{algorithm}
	\caption{Pseudo-code of the \textsc{GTVcut} segmentation algorithm based on a CA model.}
	\label{pc:GTVcut}
	\textbf{Input:} Foreground (tumor) seeds $\mathcal{F}$, Background seeds $\mathcal{B}$, Cropped MRI axial slice $\mathcall{I}_\text{cropped}$\\
	\textbf{Output:} ROI segmentation result $\mathcall{S}$\\
	\textbf{Parameters:} 
	$\theta_\mathbf{p}^{(t)} \in [0,1]$ is the strength value of the cell $\mathbf{p}$ at the current time step $t$\\
	$l_\mathbf{p} \in \{0,1,2\}$ is the label of the cell $\mathbf{p}$ at the current time step $t$, according to:
	\begin{equation*}
     \begin{cases}
       l_\mathbf{p}^{(t)} = 0: \mathbf{p} \text{ is an unlabeled cell} \\
	    l_\mathbf{p}^{(t)} = 1: \mathbf{p} \text{ is assigned to the foreground class}\\
	    l_\mathbf{p}^{(t)} = 2: \mathbf{p} \text{ is assigned to the background class}
     \end{cases}
    \end{equation*}
	\begin{algorithmic}[1]
	\ForEach{pixel $\pi_i \in \mathcall{I}_\text{cropped}$}
		\Comment Initialization of the pixels in $\mathcall{I}_\text{cropped}$
		    \State $l_{\pi_i}^{(0)} \gets 0\text{; } \theta_{\pi_i}^{(0)} \gets 0$
	\EndFor
	\ForEach{foreground seed $\varphi_i \in \mathcal{F}$}
		\Comment Initialization of the foreground pixels
		    \State $l_{\varphi_i}^{(0)} \gets 1\text{; } \theta_{\varphi_i}^{(0)} \gets 1$
	\EndFor
	\ForEach{foreground seed $\beta_i \in \mathcal{B}$}
		\Comment Initialization of the background pixels
		    \State $l_{\beta_i}^{(0)} \gets 2\text{; } \theta_{\beta_i}^{(0)} \gets 1$
	\EndFor
	\While{convergence is \textbf{not} achieved}
	    \ForEach{cell $\mathbf{p} \in         \mathcall{I}_\text{cropped}$} \Comment Moore $8$-neighborhood tries to attack the cell
\LeftComment For each neighbor cell $\mathbf{q} \in \mathcall{N}(\mathbf{p})$ find $\bar{\mathbf{q}}$ with $\max{\left\{g(\mathbf{p},\mathbf{q}) \cdot \theta_\mathbf{q}^{(t)}\right\}}$
\State $\bar{\mathbf{q}} \gets \text{arg} \max\limits_{\mathbf{q} \in \mathcall{N}(\mathbf{p})} {\left\{g(\mathbf{p},\mathbf{q}) \cdot \theta_{\mathbf{q}}^{(t)}\right\}}$
\If{$g(\mathbf{p},\bar{\mathbf{q}}) \cdot \theta_{\bar{\mathbf{q}}}^{(t)} > \theta_\mathbf{p}^{(t)}$}
    \State $\theta_\mathbf{p}^{(t+1)} \gets g(\mathbf{p},\bar{\mathbf{q}}) \cdot \theta_{\bar{\mathbf{q}}}^{(t)} \text{; } l_\mathbf{p}^{(t+1)} \gets l_{\bar{\mathbf{q}}}^{(t)}$ \Comment{Update the current state}
\Else
    \State $\theta_\mathbf{p}^{(t+1)} \gets \theta_\mathbf{p}^{(t)}\text{; } l_\mathbf{p}^{(t+1)} \gets l_\mathbf{p}^{(t)}$ \Comment{Keep the previous state}
\EndIf
	    \EndFor
	\EndWhile
	\ForEach{cell $\mathbf{p} \in \left\{\mathcall{I}_\text{cropped}: l_\mathbf{p}^{\text{final}}=1\right\}$}
	    \State $\mathcall{S}_\mathbf{p} \gets 1$ \Comment{ROI segmentation output}
	\EndFor
	\end{algorithmic}
\end{algorithm}

\subparagraph{Post-processing}
A pair of simple morphological operations \cite{soille2013} is also required to refine the segmentation results achieved by the proposed CA-based approach.
Firstly, a convex hull algorithm is used to envelope the segmented lesion into the smallest convex polygon that contains the automatically segmented region as well as to remove possible holes at tumor boundaries.
In particular, this operator allows for: (\textit{i}) hole filling when necroses or cysts are fully included in the tumor, and (\textit{ii}) whole tumor detection when part of the necrotic material may extend partially outside the tumor.
Secondly, a morphological opening with a circular structuring element ($1$-pixel radius) is employed to smooth the GTV contour shape by unlinking poorly connected pixels.

\begin{figure}[!t]
	\centering
	\includegraphics[width=0.8\linewidth]{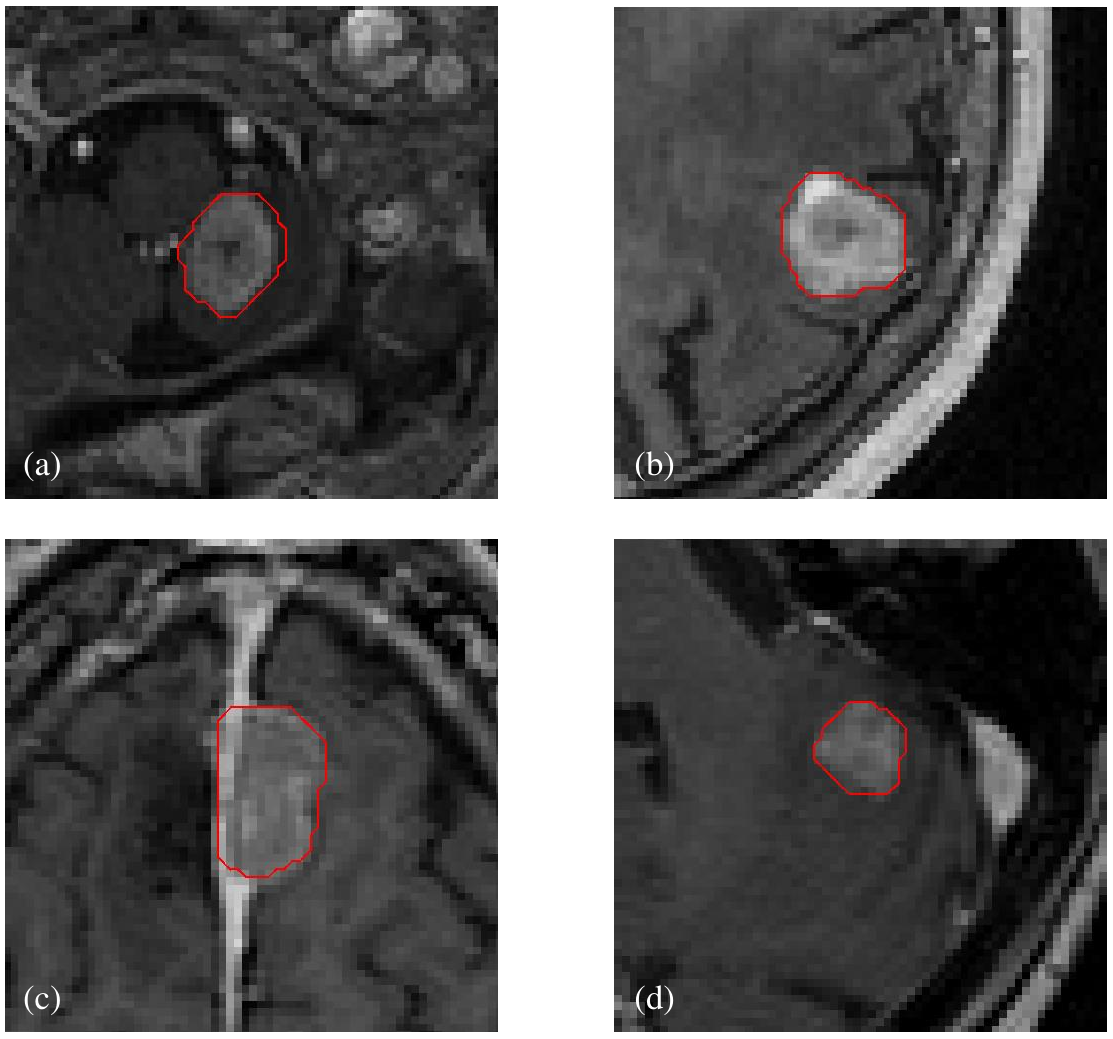}
	\caption[Brain tumor segmentation using \textsc{GTVcut} in different anatomic areas of human brain]{Brain tumor segmentation using \textsc{GTVcut} in different anatomic areas of human brain: (a) brainstem (Fig. \ref{fig:GTVcut-inputImages}a); (b) occipital lobe (Fig. \ref{fig:GTVcut-inputImages}b); (c) temporo-parietal region in proximity to the superior sagittal sinus (Fig. \ref{fig:GTVcut-inputImages}c); (d) anterior cerebellar lobe. ($4 \times$ zoom factor).}
	\label{fig:GTVcut-segRes}
\end{figure}

\paragraph{Experimental results}
Fig. \ref{fig:GTVcut-segRes} shows four GTV segmentation results achieved using the proposed \textsc{GTVcut} approach.
It is appreciable how \textsc{GTVcut} obtains correct results in different brain areas, even with inhomogeneous input MRI data.

\begin{figure}[!t]
	\centering
	\includegraphics[width=0.7\linewidth]{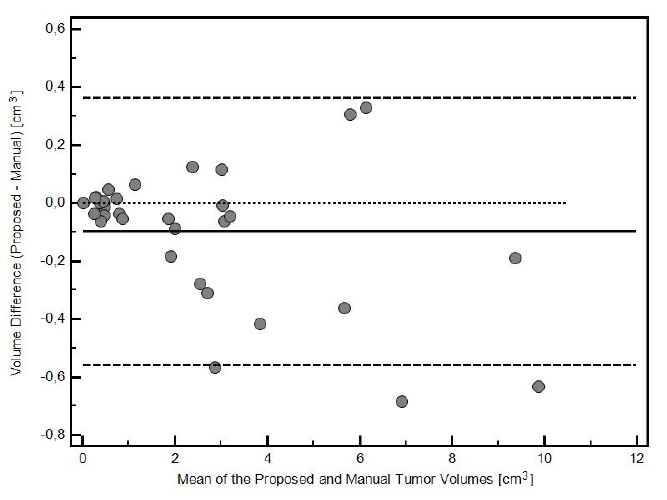}
	\caption[Bland-Altman plot of brain tumor volume measurements]{Bland-Altman plot of brain tumor volume measurements: the results yielded by the \textsc{GTVcut} method against the manual segmentations by an expert neurosurgeon (gold standard). Solid line: mean; dashed lines: $\pm 1.96$ standard deviations.}
	\label{fig:GTVcut-BlandAltman}
\end{figure}

A first clinical validation can be performed by a quantitative analysis on tumor volume measurements.
Accordingly, the proposed and manual GTVs for the brain cancers (statistical samples), considered in the MRI dataset, were compared. The agreement between the automatic and manual volume measurement can be graphically represented by the Bland-Altman plot in Fig. \ref{fig:GTVcut-BlandAltman}, which shows that more than $90\%$ of the volume differences lie within $\pm 1.96$ standard deviations around the mean difference ($0.1$ cm$^3$).
Therefore, the \textsc{GTVcut} approach achieves reproducible results also considering the different brain tumor sizes and shapes included in the analyzed real MRI dataset.

We developed and tested two different version of the proposed method:
\begin{itemize}
    \item \textsc{GTVcut}\textsubscript{IFD} implements the pixel similarity function given in Eq. (\ref{eq:g_IFD}), i.e., the absolute image feature difference;
    \item \textsc{GTVcut}\textsubscript{GM} utilizes the pixel similarity function given in Eq. (\ref{eq:g_GM}), i.e., the gradient magnitude (absolute intensity difference measure).
\end{itemize}
Moreover, all the evaluation metrics were also calculated on the same MRI datasets using our previous brain tumor segmentation method based on the FCM clustering algorithm \cite{militelloCBM2015,rundo2016WIRN}, which outperformed the most common literature segmentation methods, as it has been shown in \cite{militelloCBM2015}.
This approach was implemented in MatLab\textsuperscript{\textregistered} R2014a by exploiting the Image Processing Toolbox\textsuperscript{TM} and the Fuzzy Logic Toolbox\textsuperscript{TM}.

\begin{figure}[!t]
	\centering
	\includegraphics[width=0.5\linewidth]{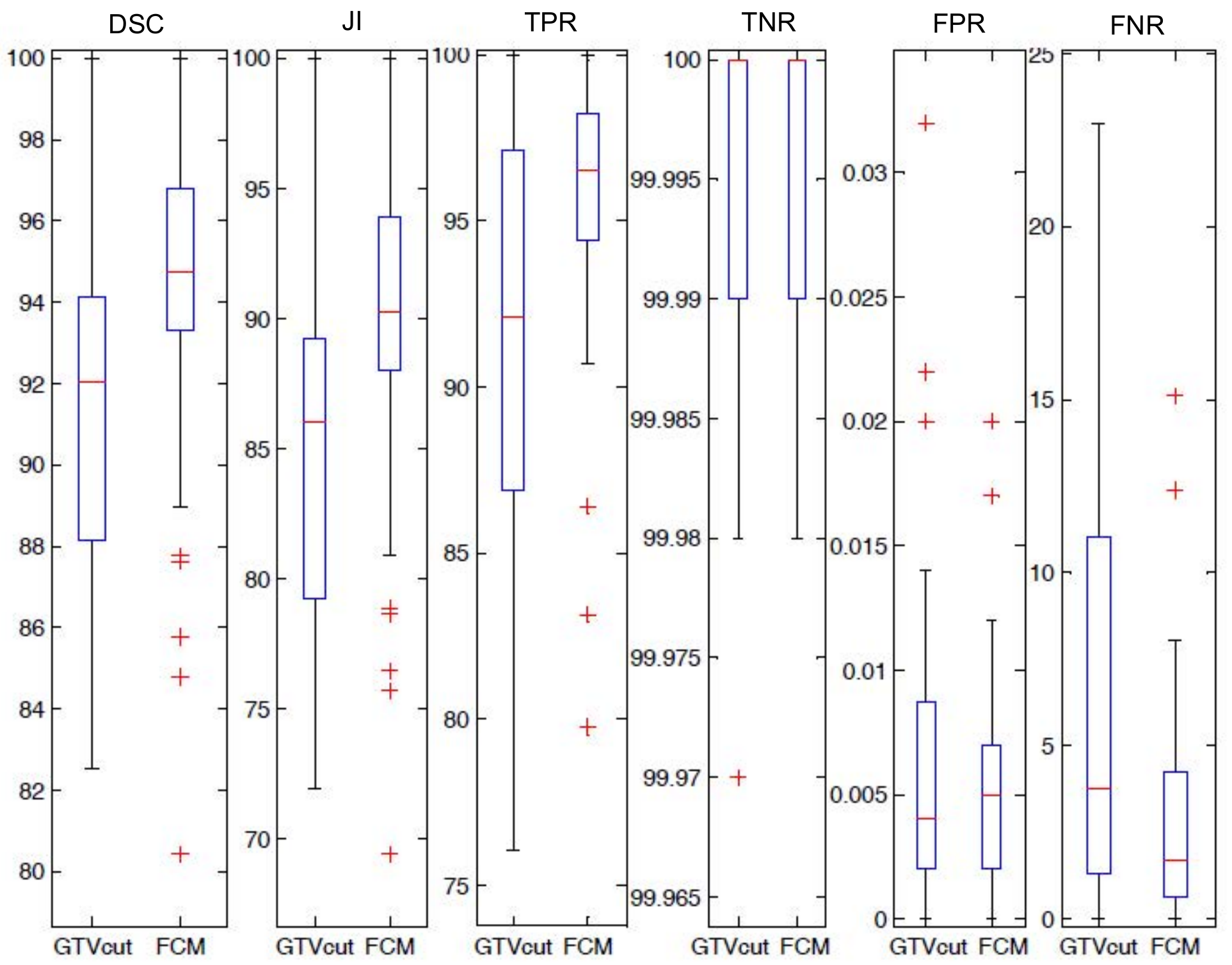}
	\caption[Boxplots of the spatial overlap-based metrics values achieved by \textsc{GTVcut} and the FCM-based method]{Boxplots of the spatial overlap-based metrics values achieved by the \textsc{GTVcut} approach and the FCM-based method in \cite{militelloIJIST2015}. The median value (i.e., the second quartile) is represented by a red line dividing the box. Whisker value is $1.5$ in all cases.}
	\label{fig:GTVcut-BoxplotsOM}
\end{figure}

Table \ref{table:GTVcut-ResOM} reports the mean and standard deviation values of spatial overlap-based metrics obtained in the experimental segmentation tests on the analyzed $32$ brain tumors (concerning the $25$ patients undergone neuro-radiosurgery) using the \textsc{GTVcut}\textsubscript{IFD}, \textsc{GTVcut}\textsubscript{GM} and FCM-based method.
High \emph{DSC} and \emph{JI} values prove the segmentation accuracy and reliability.
In addition, \emph{SEN} and \emph{TNR} average values involve the correct detection of the “true” pathological areas as well as the ability of not detecting wrong parts within the segmented tumors.
The \textsc{GTVcut}\textsubscript{GM} segmentation approach achieved slightly better results than \textsc{GTVcut}\textsubscript{IFD}.
Therefore, we chose the pixel similarity function in Eq. (\ref{eq:g_GM}) to be employed in the local transition rule of the \textsc{GTVcut} computational model.
To provide a graphical representation of the statistical distribution of the results, the corresponding boxplots of spatial overlap-based evaluation metrics achieved by \textsc{GTVcut} and the FCM-based approach are also reported in Fig. \ref{fig:GTVcut-BoxplotsOM}.
The short width of the interquartile range represented in boxplots implies that values are very concentrated.
All index distributions present a small number of outliers, thus demonstrating extremely low variability (whisker value is $1.5$ in all cases).

\begin{table}[!t]
\centering
	\caption[Values of the spatial overlap-based metrics regarding the achieved GTV segmentation results]{Values of the spatial overlap-based metrics regarding the achieved GTV segmentation results. The results are expressed as average value $\pm$ standard deviation.}
	\label{table:GTVcut-ResOM}
	\begin{scriptsize}
		\begin{tabular}{ccccccc}
			\hline\hline
			Method	& DSC	& JI	& SEN	& TNR	& FPR	& FNR \\
			\hline
			\textsc{GTVcut}\textsubscript{IFD} &	$90.83 \pm 4.12$ & $84.01 \pm 6.64$ &	$91.30 \pm 6.96$ &	$99.99 \pm 0.01$ & $0.008 \pm 0.009$ &	$6.218 \pm 6.455$ \\
			\textsc{GTVcut}\textsubscript{GM} & $90.88 \pm 4.19$ &	$84.11 \pm 6.74$ &	$91.20 \pm 7.00$ &	$99.99 \pm 0.01$ &	$0.007 \pm 0.008$ &	$6.353 \pm 6.482$	 \\
			FCM \cite{militelloIJIST2015} &	$90.88 \pm 4.19$ &	$84.11 \pm 6.74$ &	$91.20 \pm 7.00$ &	$99.99 \pm 0.01$ &	$0.007 \pm 0.008$ &	$6.353 \pm 6.482$ \\
			\hline\hline
		\end{tabular}
	\end{scriptsize}
\end{table}

\begin{figure}[!t]
	\centering
	\includegraphics[width=0.35\linewidth]{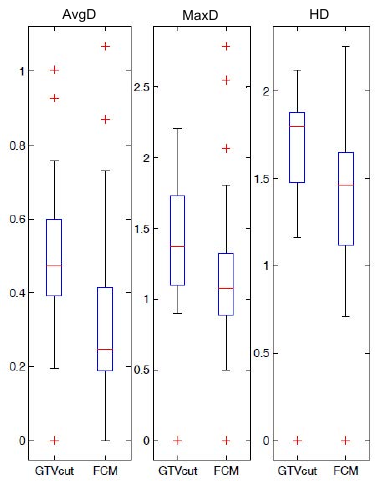}
	\caption[Boxplots of the spatial distance-based metrics values achieved by \textsc{GTVcut} and the FCM-based method]{Boxplots of the spatial distance-based metrics values achieved by the \textsc{GTVcut} approach and the FCM-based method in \cite{militelloIJIST2015}. The median value (i.e., the second quartile) is represented by a red line dividing the box. Whisker value is $1.5$ in all cases.}
	\label{fig:GTVcut-BoxplotsDM}
\end{figure}

Table \ref{table:GTVcut-ResDM} shows the spatial distance-based metrics mean and standard deviation values achieved by the \textsc{GTVcut}\textsubscript{IFD}, \textsc{GTVcut}\textsubscript{GM}, and FCM-based segmentation methods.
The achieved spatial distance-based indices are consistent with overlap-based metrics.
Hence, good performances were obtained also with heterogeneous cancers, since the CA algorithm delineates accurately irregular and complex shapes.
The achieved segmentation results are comparable with the FCM-based approach in \cite{militelloIJIST2015}.
Moreover, the distance-based metrics values calculated on the \textsc{GTVcut}\textsubscript{GM} segmentations are slightly lower than \textsc{GTVcut}\textsubscript{IFD} results, confirming the trend observed in the overlap-based metrics.
Fig. \ref{fig:GTVcut-BoxplotsDM} depicts the boxplots of the spatial distance-based evaluation metrics achieved by \textsc{GTVcut} and our previous FCM-based method in \cite{militelloIJIST2015}.

\begin{table}[!t]
\centering
	\caption[Values of the spatial distance-based metrics regarding the achieved GTV segmentation results]{Values of the spatial distance-based metrics regarding the achieved GTV segmentation results. The results are expressed as average value $\pm$ standard deviation.}
	\label{table:GTVcut-ResDM}
	\begin{scriptsize}
		\begin{tabular}{cccc}
			\hline\hline
			Method	& AvgD	& MaxD	& HD \\
			\hline
			\textsc{GTVcut}\textsubscript{IFD} & $0.492 \pm 0.223$ &	$1.426 \pm 0.495$ &	$1.650 \pm 0.453$	\\
			\textsc{GTVcut}\textsubscript{GM} & $0.487 \pm 0.223$ &	$1.414 \pm 0.483$ &	$1.638 \pm 0.443$	 \\
			FCM \cite{militelloIJIST2015} &	$0.329 \pm 0.246$ &	$1.172 \pm 0.576$	& $1.389 \pm 0.448$ \\
			\hline\hline
		\end{tabular}
	\end{scriptsize}
\end{table}

Even though there are some outliers in \emph{AvgD} and \emph{MaxD} boxplots (the whisker is set to $1.5$ in all cases), just a small deviation between the segmentations of the proposed method and those of the experienced neurosurgeon is denoted.
Considering the surgery treatment planning purpose of the application, referring to MRI spatial resolution (see characteristics in Table \ref{table:GK-MRIcharacteristics}), all distance-based metrics mean values are in the range of $1$ mm.
This confirms the great validity and accuracy of the proposed segmentation method.

\paragraph{Conclusions}
In this section, an original application of CA for semi-automatic MRI brain GTV segmentation (\textsc{GTVcut}) was presented.
The proposed technique can be used for brain tumors segmentation in neuro-radiosurgery treatments.
The developed methodology can be integrated in the current clinical practice, by supporting physicians in the target volume delineation task during Gamma Knife treatment planning. 
The proposed \textsc{GTVcut} segmentation method was tested on an MRI dataset composed of $32$ metastatic brain concerning real patients, who underwent Leksell Gamma Knife treatment, by calculating both spatial overlap-based and distance-based metrics.
The achieved experimental results are very encouraging and show the effectiveness of the proposed approach.

Considering the used CA-based models, the main experimental finding is that \textsc{GTVcut}\textsubscript{GM}, based on the gradient magnitude defined in Eq. (\ref{eq:g_GM}), obtained slightly better segmentation results than \textsc{GTVcut}\textsubscript{IFD}, which implements the image feature difference in Eq. (\ref{eq:g_IFD}).
In addition, the experimental results achieved by \textsc{GTVcut} and by our previous FCM-based approach in \cite{militelloIJIST2015,rundo2016WIRN} are comparable.
The \textsc{GTVcut} segmentation approach works better than FCM-based segmentation when heterogeneous tumors with either diffused inner necrotic material or cysts are processed.
Indeed, the employed adaptive seed selection strategy is more robust in these very critical cases with respect to the FCM clustering algorithm that could yield also disjoint output regions.
Moreover, \textsc{GTVcut} does not require any smoothing operations to reduce MRI acquisition noise and no imaging details are missed.
These small differences between the \textsc{GTVcut} and the FCM-based segmentation results, revealed in the experimental trials, can be also due to the different interactive selection tools utilized by the two approaches; the method in \cite{militelloIJIST2015} employs a free-hand “lasso” tool and \textsc{GTVcut} uses a draggable bounding rectangle.
However, the second ROI selection tool is more usable and less operator-dependent than the first one, by minimizing the user's degrees of freedom \cite{rother2004}.
Furthermore, the overall segmentation approach based on the FCM clustering algorithm in \cite{militelloIJIST2015} exploited also a sophisticated strategy for necrotic material inclusion, which has to be considered in the resulting GTV.


\subsection{Random Walker}
\label{sec:randomWalker}

The RW algorithm \cite{grady2006} encodes an image as a graph $\mathcall{G} = (\mathcall{V}, \mathcall{E})$ with nodes (vertices) $v \in \mathcall{V}$ corresponding to the voxels and edges (arcs) $e \in \mathcall{E}$ are associated to a Gaussian cost function that maps a change in image intensity to the edge weights $w_{ij}$ defined by means of the following Gaussian function:
\begin{equation}
    \label{eq:weightsRW}
	w_{ij} = e^{-\beta\norm{g_i - g_j}^2},
\end{equation}
where: $g_i$ and $g_j$ are the image intensity values at voxels $v_i$ and $v_j$, respectively; $\beta$ is a free parameter.
It is also assumed that the graph $\mathcall{G}$ is connected and undirected (i.e., $w_{ij} = w_{ji}$).

The image is then converted into a lattice where some pixels' classes are known while other pixels are unassigned.
In the original method in \cite{grady2006}, the known nodes are marked by user input.
The segmentation problem is solved by assigning a label to unknown nodes, by finding the minimum energy among all possible graph scenarios to achieve an optimal segmentation.
The RW method classifies the nodes into foreground and background classes, considering the probability that a “random walker” starting at a source node, first reaches a node with a pre-assigned label by visiting every voxel.
Every unlabeled node is assigned to the subset which is most likely a RW would reach first starting from that pixel.
The RW problem has the same solution as the combinatorial Dirichlet
problem, defined by the Dirichlet integral:
\begin{equation}
    \label{eq:Dirichlet}
	D[u] = \frac{1}{2} \int_{\Omega} |\nabla u|^2 d\Omega,
\end{equation}
for a field $u$ and a region $\Omega$.
This problem can be solved by a harmonic function that satisfies the Laplace equation:
\begin{equation}
    \label{eq:Laplace}
	\nabla^2 u = 0.
\end{equation}

More specifically, the harmonic function that satisfies the boundary conditions minimizes the Dirichlet integral, since the Laplace equation is the Euler-Lagrange equation for the Dirichlet integral \cite{courant1989}.

The combinatorial Laplacian matrix $\mathbf{L}$ \cite{dodziuk1984} can be defined as:
\begin{equation}
    \label{eq:LaplacianMatrix}
	L_{ij}=
	\begin{cases}
	\text{deg}(v_i), & \text{if } i = j\\
	-w_{ij}, & \text{if } v_i \text{ and } v_j \text{ are adjacent nodes}\\
	0, & \text{otherwise}
	\end{cases}.
\end{equation}

The matrix $\mathbf{L}$ can be decomposed into four sub-matrices:
\begin{equation}
    \label{eq:LaplacianDecompose}
	\mathbf{L}=
	\begin{bmatrix}
    \mathbf{L}_M & \mathbf{B} \\
    \mathbf{B^\top} & \mathbf{L}_U
    \end{bmatrix},
\end{equation}
where $\mathbf{L}_U$ is a sub-matrix of edge weights for the unlabeled voxels and $\mathbf{L}_M$ is a sub-matrix of edge weights for the marked nodes.
$\mathbf{B}$ is a sub-matrix of the probabilities corresponding to the labeled voxels.

Accordingly, the harmonic function can be found by solving the system of linear equations in Eq. (\ref{eq:harmFunc}):
\begin{equation}
    \label{eq:harmFunc}
	\mathbf{L}_U \mathbf{X}= -\mathbf{B}^\top \mathbf{M},
\end{equation}
where: $\mathbf{M}$ is a Boolean matrix representing the boundary conditions of the seeds; $\mathbf{X}$ is the probability for each node being a member of the labels.
A threshold of $50\%$ is chosen to discriminate the foreground from the background yielding a voxel binary mask $\mathcall{S}$, so that:
\begin{itemize}
    \item $\mathcall{S}_{\mathbf{p}_\text{fg}} \gets 1$, for foreground nodes $\mathbf{p}_\text{fg}$ with probability higher than or equal to $0.5$;
    \item $\mathcall{S}_{\mathbf{p}_\text{fg}} \gets 0$, for background nodes $\mathbf{p}_\text{bg}$ with probability lower than $0.5$.
\end{itemize}

Briefly, a probability map is  produced, and a threshold of $50\%$ is chosen to discriminate between foreground and background voxels.

\subsubsection{Brain lesion segmentation on PET images}
In medical imaging, especially when dealing with PET, the RW method is able to localize weak boundaries as part of consistent boundaries.
To obtain the BTV delineation, the RW parameters were modulated to incorporate PET information: $g_i$ and $g_j$  have been replaced with the Standardized Uptake Value (SUV) \cite{stefano2016,stefano2015,comelli2018} in the voxels $i$ and $j$:
\begin{equation}
    \label{eq:weightsSUV}
	w_{ij} = e^{-\beta(\text{SUV}_i - \text{SUV}_j)^2},
\end{equation}
where:
\begin{equation}
    \label{eq:SUV}
	\text{SUV} = \frac{c_\mathcall{I}}{i_\text{dose}/W_\text{pat}},
\end{equation}
$c_\mathcall{I}$ is the radioactivity activity concentration [kBq/ml]  measured from an image $\mathcall{I}$, $i_\text{dose}$ is
the decay-corrected amount of injected radiotracer [kBq], and
$W_\text{pat}$ is the patient’s weight [g].

The SUV is the most common semi-quantitative parameter used to estimate radiotracer accumulation within a lesion in clinical practice \cite{lucignani2004}.
It normalizes the voxel activity considering acquisition time, administered activity, and patient’s body weight \cite{zasadny1993}.
Hence, the PET image is converted into a lattice where the SUV of each voxel is assigned to the corresponding graph node $v \in \mathcall{V}$ and the edge weights $w_{ij}$ are computed accordingly.

The RW method is very sensitive to the choice of $\beta$ factor in the Gaussian-like weighting function in Eq. (\ref{eq:weightsSUV}).
$\beta$ influences how quickly the probability decreases with increasing intensity differences \cite{stefano2017}: a high $\beta$ value reduces the weight of the walker, which weakens the connection between the adjacent voxels and under-estimates the foreground volume, whilst a low $\beta$ value increases the weight, which over-estimates the target volume.

In addition, to obtain a fully automatic and user-independent method, two crucial improvements have been implemented to overcome two critical issues in the RW methodology:
\begin{enumerate}
    \item an automatic method to localize starting target and background seeds: the RW algorithm requires a set of pre-labeled seeds, which may be generated by the user, making the RW delineation sensitive to the location of the pre-labeled voxels;
    \item a strategy to adaptively determine the appropriate probability threshold rather than fixed one of $50\%$ on the probabilistic output of the RW in order to overcome the operator-dependence of the weighting factor $\beta$ in Eq. (\ref{eq:weightsRW}).
\end{enumerate}

A more detailed explanation of these improvements is reported in \cite{stefano2015,stefano2017}.

\subsubsection{Multimodal PET/MRI brain metastasis segmentation}
\label{sec:multimodalBrainSeg}

The study presented in \cite{rundoCMPB2017} investigated the impact of BTV segmentation, using MET-PET imaging, and the subsequent co-registration with MR images, utilized to delineate the GTV, in stereotactic neuro-radiosurgery therapy.
The main goal was to present a novel multimodal PET/MRI automatic segmentation method, combining complementary information, and encourage its use in future Gamma Knife treatments.

To the best of our knowledge, this approach is the first work that presents a co-segmentation method to integrate MET-PET metabolic information with anatomic MRI in stereotactic neuro-radiosurgery treatments.
The proposed method improves and combines two computer-assisted and operator-independent methods to segment BTV and GTV from PET \cite{stefano2016,stefano2015,stefano2017} and MR \cite{militelloIJIST2015,rundo2016WIRN} images, respectively.
Each single method has been already validated \cite{militelloIJIST2015,stefano2016}, comparing each of them against the most common literature segmentation methods.
The former uses the graph-based RW algorithm \cite{grady2006} and the latter is based on unsupervised FCM clustering \cite{bezdek1981,bezdek1984}.
Both segmentation methods are valid operator-independent approaches to identify the BTV and the GTV, respectively, in order to delineate a comprehensive CTV that includes metabolic and morphologic information, useful for treatment planning and patient follow-up.
Our method addresses the co-segmentation problem only on multimodal PET/MRI in the brain anatomic district imaging, by exploiting the great sensitivity and specificity of brain MET-PET in distinguishing healthy and pathological tissues.
As result, a fully automatic approach was obtained.

The co-segmentation tests on $19$ metastatic brain tumors, treated with Leksell Gamma Knife radiosurgery, were retrospectively performed to evaluate the effectiveness of the proposed approach.
Overlap-based and spatial distance-based metrics were calculated between PET and MRI segmentation results.
Statistics was also considered to quantify correlation.
Finally, a qualitative evaluation, based on a Likert score scale, was carried out by three experienced clinicians.

It is worth to note that clinical evaluation is not just a mathematical procedure, based just on tumor area difference or on GTV and BTV union, but a deep decision-making process involving several anatomic and metabolic insights, the specific patient's pathological scenario as well as the physician experience.
The CTV identification is actually a critical task accomplished by expert physicians according to tumor volumes (i.e., BTV and GTV), and deciding about the regions to be included or excluded in the planned target volume.
This may pave the way for personalized therapy, by integrating molecular imaging and allowing for a diagnosis and therapy that is specialized to the individual metabolism and disease \cite{brady2016}.

\paragraph{Dataset description}
PET brain acquisitions without head frame were performed using a time-of-flight PET/CT Discovery 690 by General Electric Medical Systems (Milwaukee, WI, USA).
Patients fasted for $4$ hours before the PET exam and were intravenous injected with MET.
The PET protocol began $10$ minutes after the injection. PET images consist of a $256 \times 256$ pixel matrix of $1.1719 \times 1.1719 \times 3.27$ mm$^3$ voxel size.

\begin{figure}[!t]
	\centering
	\subfloat[]{\includegraphics[width=0.4\textwidth]{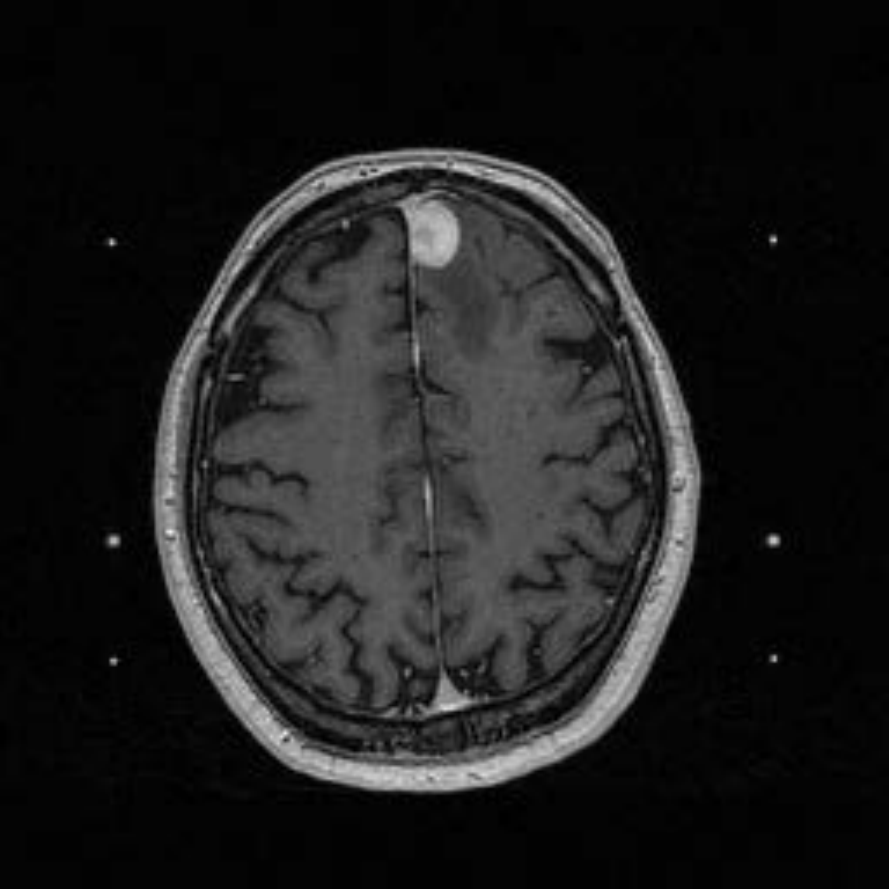}}\qquad
	\subfloat[]{\includegraphics[width=0.4\textwidth]{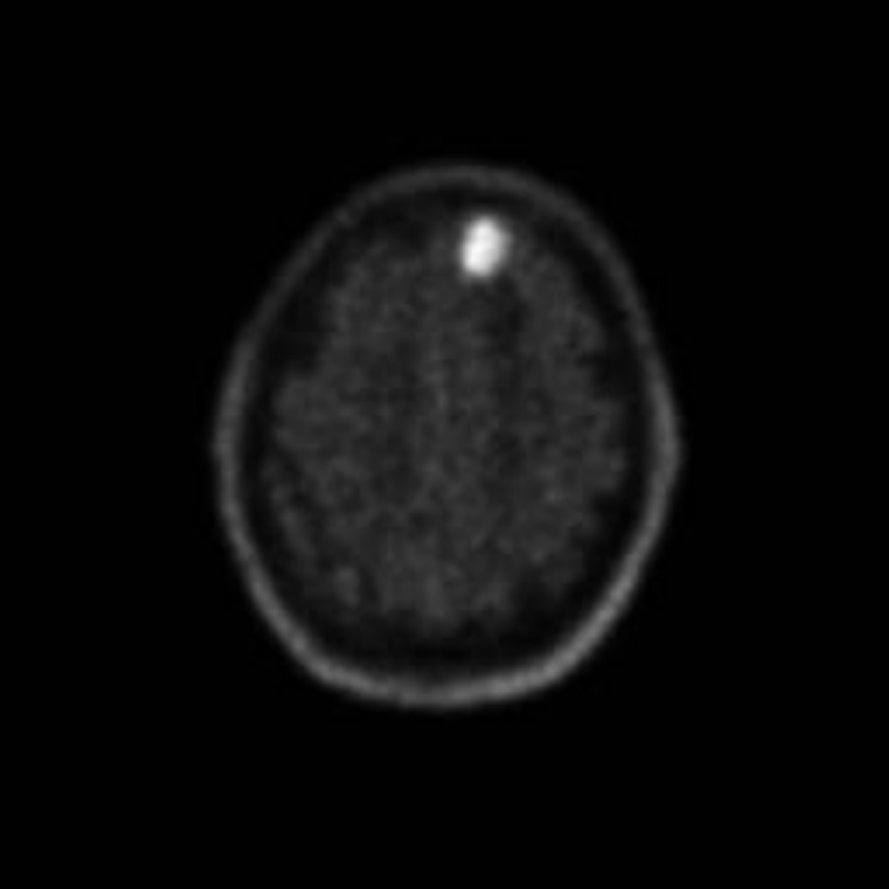}}\\
	\subfloat[]{\includegraphics[width=0.4\textwidth]{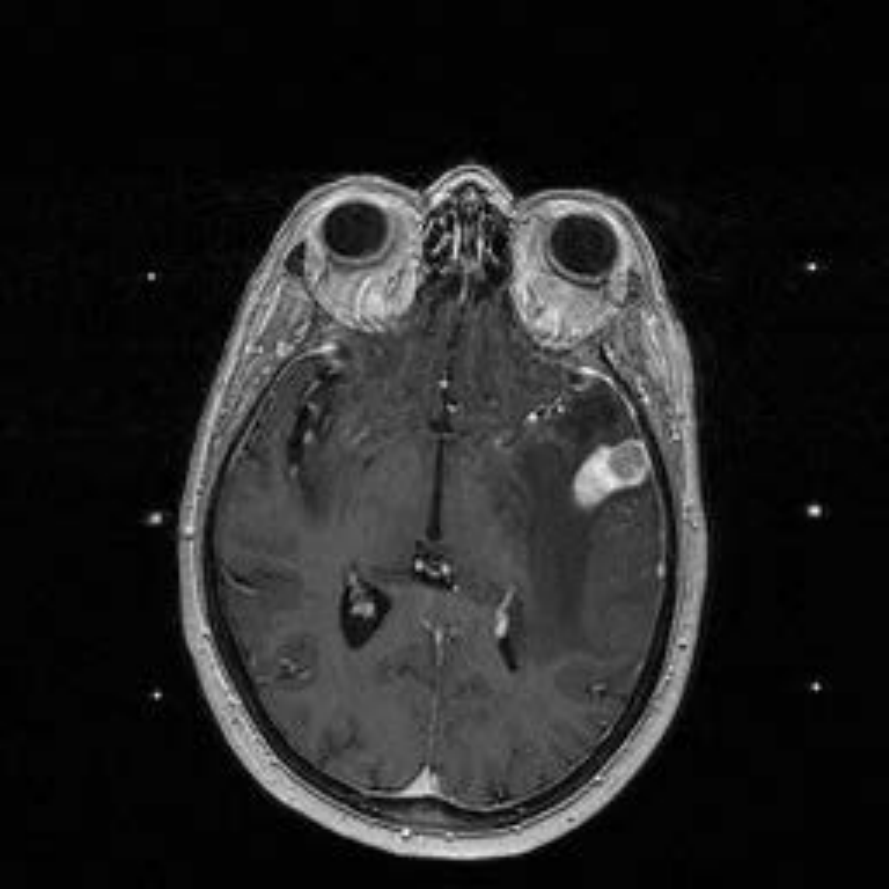}}\qquad
	\subfloat[]{\includegraphics[width=0.4\textwidth]{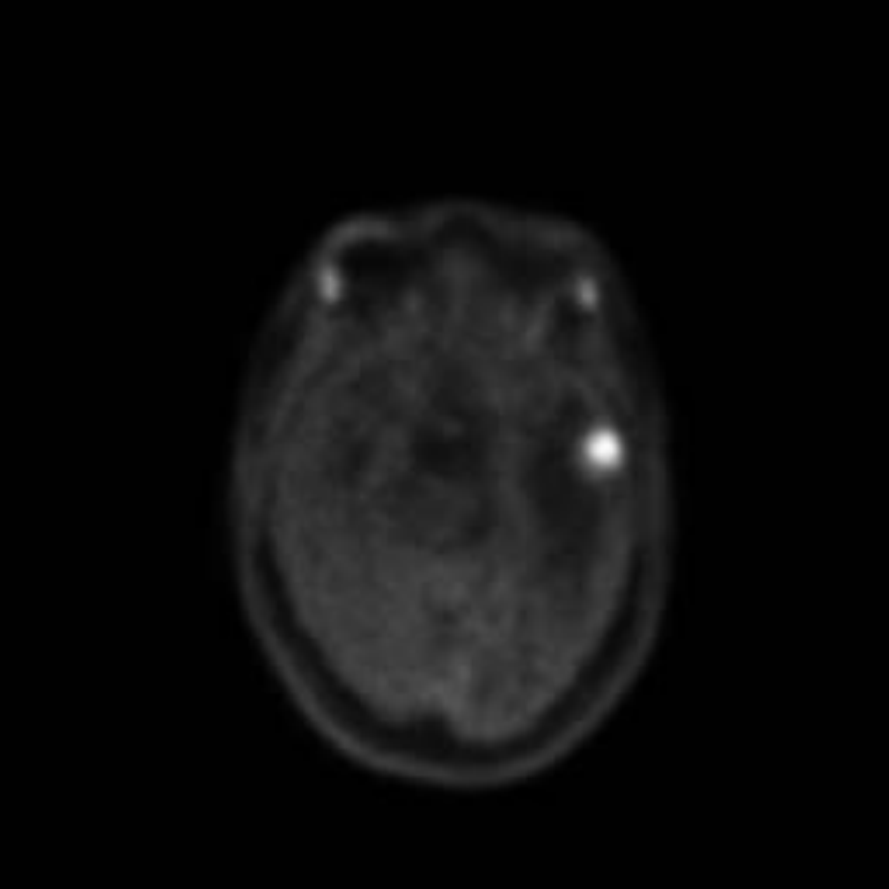}}\\
	\caption[Examples of original input brain MR and PET image pairs concerning patients with brain tumors who underwent Gamma Knife]{Examples of original input brain MR and PET image pairs concerning patients with brain tumors who underwent Gamma Knife: (a, c) MR slices and (b, d) the nearly respective corresponding PET slices (tumors $\#6$ and $\#14$). Note the different dimensions and the substantial tridimensional misalignment between corresponding MR and PET images. The images are displayed in gray-scale.}
	\label{fig:PET-MRoriginal}	
\end{figure}

All available MRI datasets were acquired on a Philips Gyroscan Intera $1.5$ T MRI scanner, before treatment, for the planning phase.
MR images are T1w FFE CE sequences, using the following MRI acquisition parameters are: TR: $25$ ms, TE: $1.808$-$3.688$ ms, matrix size: $256 \times 256$ pixels, slice thickness: $1.5$ mm, slice spacing: $1.5$ mm, pixel spacing: $1.0$ mm.
Therefore, the size of each voxel was $1.0 \times 1.0 \times 1.50$ mm$^3$.
Imaging data are encoded in the $16$-bit DICOM format.
Representative instances of input brain MR and PET image pairs are shown in Fig. \ref{fig:PET-MRoriginal}.

\begin{figure}[!t]
	\centering
	\includegraphics[width=\textwidth]{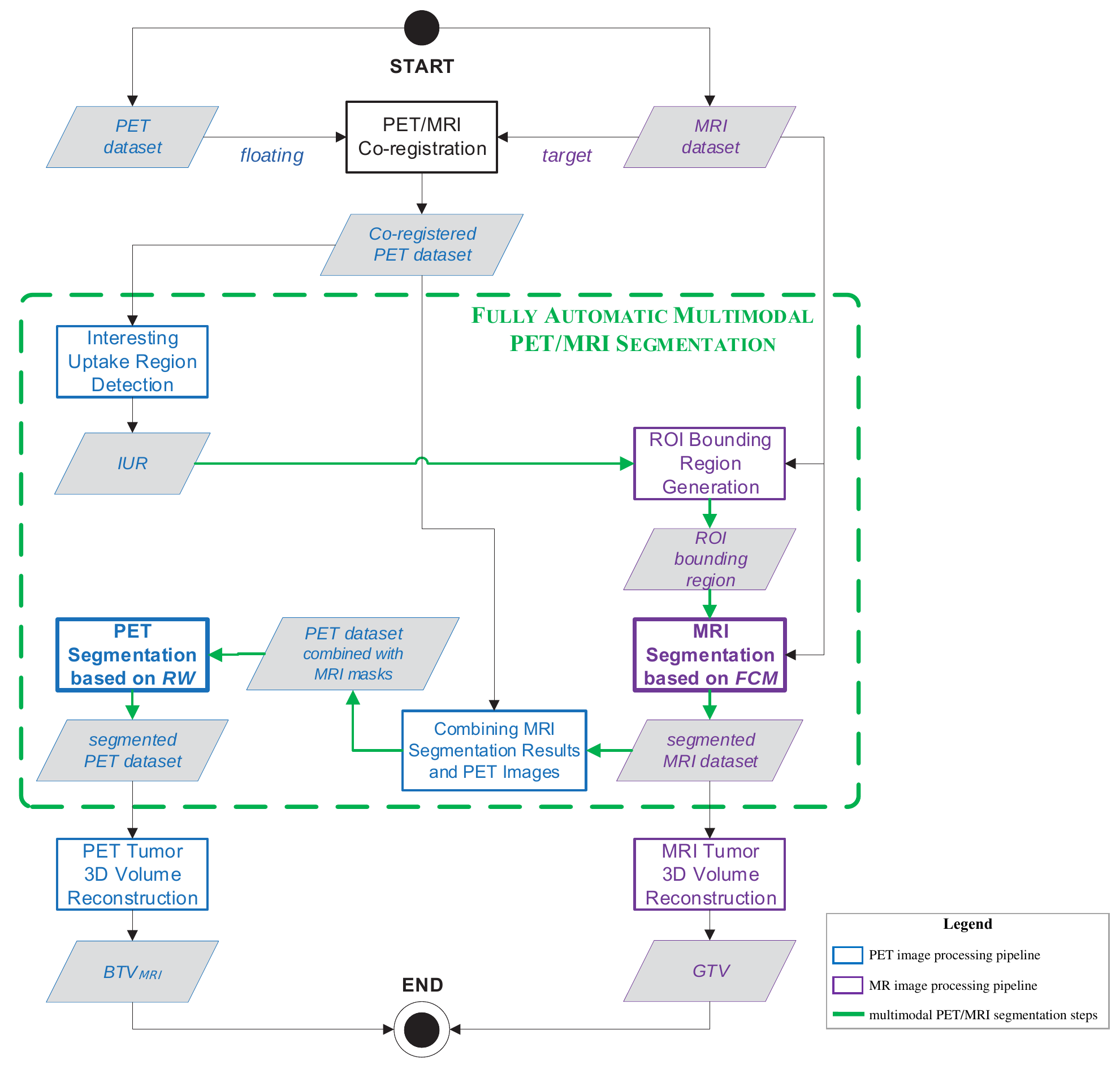}
	\caption[Flow diagram of the proposed fully automatic multimodal PET/MRI segmentation method]{Flow diagram of the proposed fully automatic multimodal PET/MRI segmentation method. The adopted graphical and color notations are explained in the legend box.}
	\label{fig:PET-MRflowdiagram}
\end{figure}

\paragraph{The proposed multimodal PET/MRI segmentation approach}
Fig. \ref{fig:PET-MRflowdiagram} outlines the overall flow diagram of the proposed multimodal PET/MRI segmentation method.
For a better understanding, PET and MR image processing pipelines are represented with blue and purple blocks, respectively.
It is worth noting the smart combination of the two single modality pipelines (green connections in Fig. \ref{fig:PET-MRflowdiagram}).
In particular, PET and MRI segmentation results mutually exploit each other.
Firstly, the IUR, obtained on co-registered PET images by the graph-based RW tumor segmentation method, is used for the generation of an area including the tumor ROI on MR brain images.
These ROI bounding regions are calculated adaptively on MR images using an LSF method \cite{li2010DRLSE} and then utilized by the MRI brain lesion segmentation method based on the FCM clustering technique.
Lastly, the MRI GTV masks are combined with PET images, which are processed by the RW algorithm, to influence and refine BTV segmentation on PET images.

All the employed parameters were fixed in the proposed computational model for all the analyzed datasets.
These settings are completely hidden to the end-users (clinicians), who have not to select any parameters.
Therefore, this does not affect the fully automation of the proposed method.
First of all, the IUR, obtained on co-registered PET images using the graph-based BTV segmentation method, is exploited for the automatic generation of a suitable region including the tumor on MR brain images.
These bounding regions, computed adaptively on MR images using an LSF method, are then utilized by the GTV segmentation method.
Lastly, GTV masks are combined with PET images to influence the BTV segmentation.
The presented multimodal segmentation approach yields the tumor boundaries in both PET and MRI modalities, conveying different information, which are not always complementary, considering that enhancement, edema, and necrosis regions are imaged differently by the modalities \cite{evanko2008}; so the tumor volumes defined on PET and on CT or MRI could be highly different \cite{song2013}.

The main key novelties of the proposed multimodal segmentation approach are:
\begin{itemize}
    \item presenting a fully automatic method for GTV segmentation, calculated automatically by exploiting the IURs detected on PET images to define a suitable bounding region for the tumor zone with a DRLSE approach \cite{li2010};
    \item computing the BTV\textsubscript{MRI}, a new BTV that can reduce radioactivity spill-in and spill-out effects, between tumor and surrounding tissues, affecting the segmented BTV (according to the morphologic GTV information) and the subsequent BTV\textsubscript{MRI} integration for a comprehensive treatment planning.
\end{itemize}
Thereby, a complete knowledge about the clinical scenario is achieved from both anatomic and metabolic imaging perspectives.
In conclusion, the tackled problem and the aim of the proposed multimodal approach are considerably different with respect to joint segmentation approaches.
Employing a decision-level fusion, the assumption of a “ground truth” joint volume, defined on fused multimodal imaging data, and the consequent comparison of the proposed method with related works is not significant, and sometimes misleading.
Under these hypotheses, a reference gold standard for the evaluation is not consistent with the purpose of our work.
This retrospective study aims to evaluate the feasibility and the clinical value of BTV integration in Gamma Knife treatment planning.

\subparagraph{PET/MRI modalities co-registration}
An image co-registration stage is mandatory to bring the different MRI and MET-PET datasets, concerning the same patient, into the same reference space.
In this way, it is possible to make quantitative and meaningful comparisons between the brain lesion segmentation results achieved by both MRI and PET image segmentation methods (see Section \ref{sec:medImageReg} for more details on medical image registration).
In our multimodality PET/MRI registration, MRI is used as reference image (target) while the PET is the source image (floating), because MRI conveys more anatomic information than PET.
Indeed, PET imaging is characterized by weak boundaries and has a lower spatial resolution than MRI.
Realignment and reslicing operations are thus required to get a one-to-one mapping between PET and MRI slices.
From an algorithmic perspective, image co-registration involves finding parameters (i.e., geometric transformation matrix) that either maximize or minimize some objective function. However an accurate interpolation method is clearly required, by which the floating and target images are sampled when being represented in different spaces \cite{maes1997,pluim2003}.
PET/MRI inter-modal $3$D registration was performed using SPM 12 (Statistical Parametric Mapping, Wellcome Trust Centre for Neuroimaging, University College, London, UK), a software package designed for the analysis of brain imaging data sequences \cite{penny2011}.
We relied on the SPM tool that is widely used in the neuroimaging community and in the clinical routine, also for Voxel-Based Morphometry (VBM).
The registration method used by SPM is based on the work by Collignon \textit{et al.} \cite{collignon1995}, where the original interpolation method has been changed to give a smoother cost function. The images are also smoothed slightly, by means of their histogram.
This makes the cost function as smooth as possible, to give faster convergence and less chance of local minima.
A $3$D rigid-body model, parameterized by $3$ translations and $3$ rotations about the different axes, is used by means of voxel-to-voxel affine transformations.
This approach is very efficient for brain anatomic district.
We chose Normalized Mutual Information (NMI) as the cost function to be optimized. The Mutual Information (MI) registration criterion states that the mutual information of the image intensity values of corresponding voxel pairs is maximal if the images are geometrically aligned. MI is the most intensively investigated criterion for registration of intra-individual human brain images \cite{cizek2004,pluim2000}.
In addition, PET/MRI registration misalignment can be large with respect to the FOVs, then a criterion invariant to image overlap statistics is very important, such as the NMI \cite{studholme1999}.
Since MET-PET and MRI multimodal images are written in a different space, an efficient resampling and interpolation method must be utilized. Although Nearest-Neighbor and Trilinear interpolations are faster, it may be better to use a higher degree approach to achieve more accurate results (i.e., reducing the deviation from an ideal low-pass filter, by avoiding artifacts mostly near the edges) \cite{thevenaz2000}.
In particular, we used the $4$th Degree B-Spline interpolation.

\begin{figure}[!t]
	\centering
	\includegraphics[width=\textwidth]{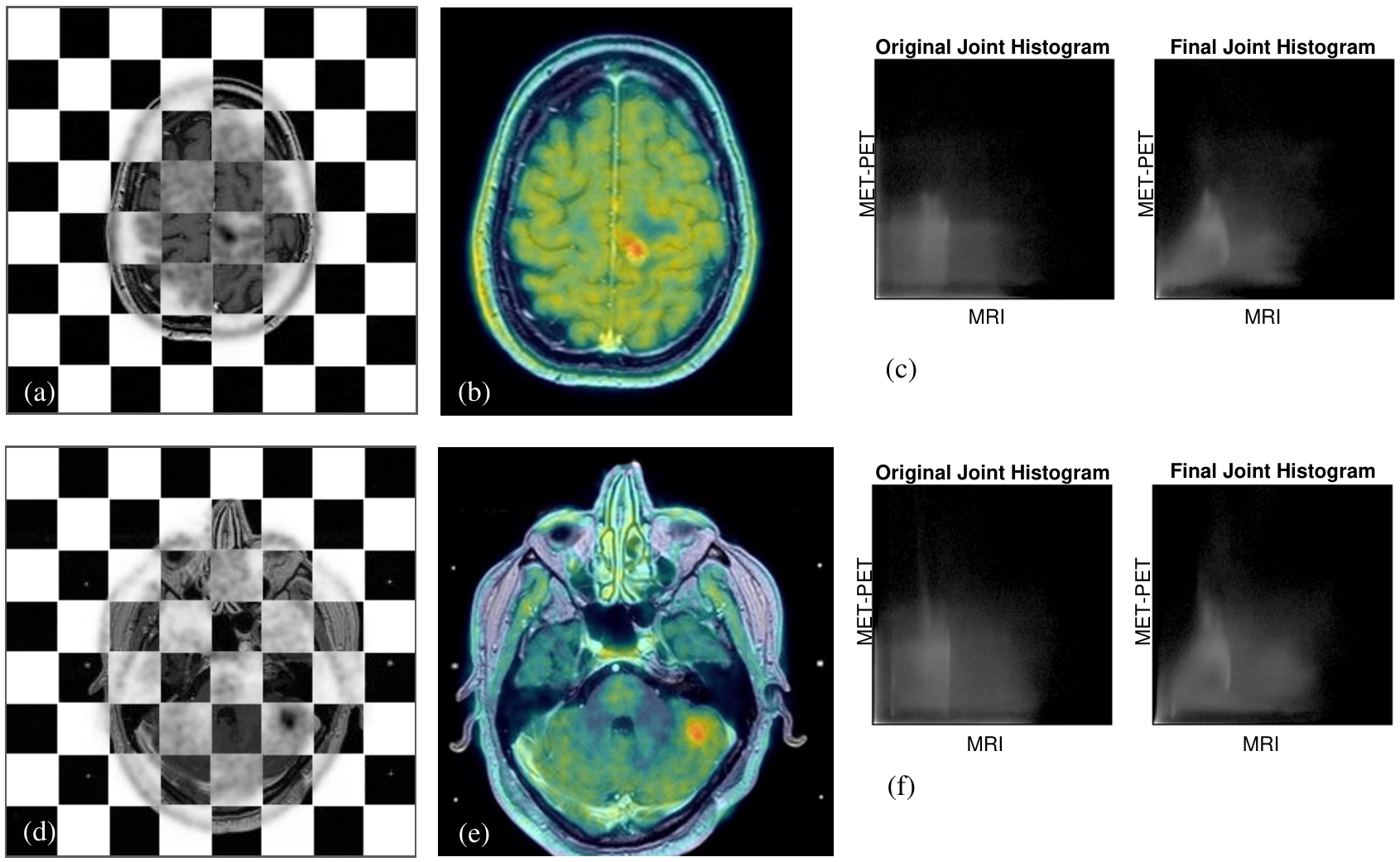}
	\caption[Instances of PET/MRI co-registration using SPM \textit{via} maximization of NMI]{Instances of PET/MRI co-registration using SPM \textit{via} maximization of NMI (tumors $\#7$ and $\#10$): (a, d) checkerboard images of co-registered PET (negative gray-scale) and MRI (normal gray-scale); (b, e) fused images (MRI in gray-scale and PET in Jet colormap); (c, f) corresponding original and final joint histograms (feature space constructed by counting the number of times a combination of gray values occurs contextually on MRI and MET-PET images) for (b) and (e), respectively.}
	\label{fig:PET-MRregistration}
\end{figure}

Two instances of PET/MRI co-registration are shown in Fig. \ref{fig:PET-MRregistration}, where the quality of the achieved registration can be qualitatively appreciated with the checkerboard images (Fig. \ref{fig:PET-MRregistration}(a, d)) as well as fused PET/MR images (Fig. \ref{fig:PET-MRregistration}(b, e)).
In addition, the joint histogram, which is a feature space constructed by counting the number of times a combination of gray values occurs contextually on source (MET-PET) and reference (MRI) images, was also computed and plotted (Fig. \ref{fig:PET-MRregistration}(c, f)).
Along the axes of each joint histogram the gray values of the two images are represented: from left to right for source (MET-PET) and from bottom to top for reference (MRI).
Original and final joint histograms prove the quality of the NMI registration process, because pixel values are mainly redistributed on the histogram diagonal.

\subparagraph{Interesting uptake region detection}
The fully automatic multimodal PET/MRI segmentation starts with the automatic identification of the PET slices with the maximum SUV (SUV\textsubscript{max}), relative to each lesion detected by means of the iterative procedure reported in Algorithm \ref{pc:IURdetection}, on PET dataset previously co-registered against the corresponding MRI.
The voxels with a SUV greater than $95\%$ of the SUV\textsubscript{max} concerning each tumor are marked as target seeds. The neighborhood of the node with SUV\textsubscript{max}, through searching in all $8$ neighboring directions, is explored to identify the voxels with a SUV less than $30\%$ of the average of target seed SUVs \cite{stefano2016}.
In this way, $8$ background nodes are identified.
Once the foreground and background seeds are automatically localized, the PET lesion is segmented using our enhanced RW method: the probability threshold to discriminate between target and background voxels is obtained slice by slice to follow the whole lesion volume.
In this way, the method takes into account the intensity gradient and contrast changes of the metabolic lesion over the entire range of PET slices.
Once the BTV has been extracted, the ROI of the PET slice with the SUV\textsubscript{max} is propagated to the corresponding MRI dataset.
In addition, the range of segmented PET slices with high uptake regions is also provided to the next processing phase, to generate efficiently a ROI bounding region for the tumor on MR images.
Such information defines the so-called IUR.

By taking advantage of the great sensitivity and specificity of $^{11}$C-labeled MET radiotracers in the discrimination among malign versus benign tissues, each brain tumor is independently processed by our algorithm.
For each patient study, the highest uptake regions are automatically analyzed.
The $\text{SUV}_\text{max}$ value is not global, but is relative to each single brain tumor present in the PET study under examination.
Therefore, in the case of $L_\text{tot}$ brain tumors a different $\text{SUV}_\text{max}(l_i)$ is identified for each lesion $l_i$, with $i \in \{1, 2, \ldots, L_\text{tot} \}$.

\begin{algorithm}
	\caption{Iterative procedure for the IUR detection and BTV delineation on PET images.}
	\label{pc:IURdetection}
	\textbf{Input:} Analyzed PET study\\
	\textbf{Output:} List $\mathcall{L}_\text{BTV}$ that stores the coordinates of the voxels included in the segmented $L_\text{tot}$ brain lesions on the processed PET study \\
	\begin{algorithmic}[1]
	    \State $\mathcall{L}_\text{BTV} \gets \{\}$ \Comment{Initialize $\mathcall{L}_\text{BTV}$ as an empty list}
		\State $i \gets 1$ \Comment{Initialize the iteration counter for the detected tumors}
		\Do
        \State  Find the slice with the highest $\text{SUV}^{(i)}_{\text{max}}$ in the PET study, except the voxels in $\mathcall{L}_\text{BTV}$
        \State Let $l_i$ be the current lesion with $\text{SUV}^{(i)}_{\text{max}}$
        \State $\text{SUV}_{\text{max}}(l_i) \gets \text{SUV}^{(i)}_{\text{max}}$ \Comment Set the $\text{SUV}_{\text{max}}(l_i)$ regarding $l_i$ to $\text{SUV}^{(i)}_{\text{max}}$
        \State $\text{BTV}_i \gets \Call{BTVsegmentation}{l_i}$ \Comment{Segment the current $l_i$ using the RW algorithm}
        \LeftComment Store the segmented $\text{BTV}_i$ voxel coordinates, by updating the list $\mathcall{L}_\text{BTV}$
        \State $\mathcall{L}_\text{BTV} \gets \mathcall{L}_\text{BTV} \cup \Call{GetVoxelCoordinates}{\text{BTV}_i}$
        \State $i \gets i + 1$ \Comment{Increment the iteration counter}
        \LeftComment Check if the $\text{SUV}^{(i)}_{\text{max}}$ in the $i$-th iteration is at least $2.0$ (minimum threshold value for considering a candidate region to be a tumor), otherwise stop the loop 
        \DoWhile{$\text{SUV}^{(i)}_{\text{max}} \ge 2.0$}
	\end{algorithmic}
\end{algorithm}

\subparagraph{ROI bounding region generation}
A coarse lesion delineation on MR images is required in order to define automatically a bounding region that encloses the actual GTV on MR slices. Hence user interaction, needed by the methods in \cite{militelloIJIST2015,rundo2016WIRN} through a ROI selection tool, will be completely prevented and avoided.
Because of the variability of brain lesions in terms of location and intensity values, the definition of a valid ROI bounding region is not a trivial problem.
A bad choice of the ROI bounding region on MR image might affect the whole FCM clustering process.
Therefore, this task must be carried out not in a static way (e.g., using a simple morphological dilation or closing with a fixed structuring element \cite{breen2000}), but rather dynamically by processing MRI input data.
This adaptive procedure starts on the lesion ROI found on the SUV\textsubscript{max} slice.
This IUR is first dilated (using a structuring element represented by a disk of $3$-pixel radius) and then utilized as a binary step function to initialize an LSF-based method.
A rough and over-estimated MRI brain lesion segmentation, based on the DRLSE formulation proposed in \cite{li2010DRLSE}, is used to generate dynamically a bounding region that includes the brain lesion in the MR slice corresponding with the SUVmax PET slice.
Unlike region growing algorithms \cite{adams1994,rundoMBEC2016}, the DRLSE method allows a controlled and regular LSF evolution, by avoiding leaking in the brain tissue area.
In short, Level Set methods represent an active contour as the zero level set of a higher dimensional function, i.e., a time-dependent LSF  $\phi(x,y,t): \Omega \rightarrow \mathbb{R}$ (where $\Omega$ is the domain), and determine the evolution of the contour \cite{osher1988}.

\begin{figure}[!t]
	\centering
	\includegraphics[width=\textwidth]{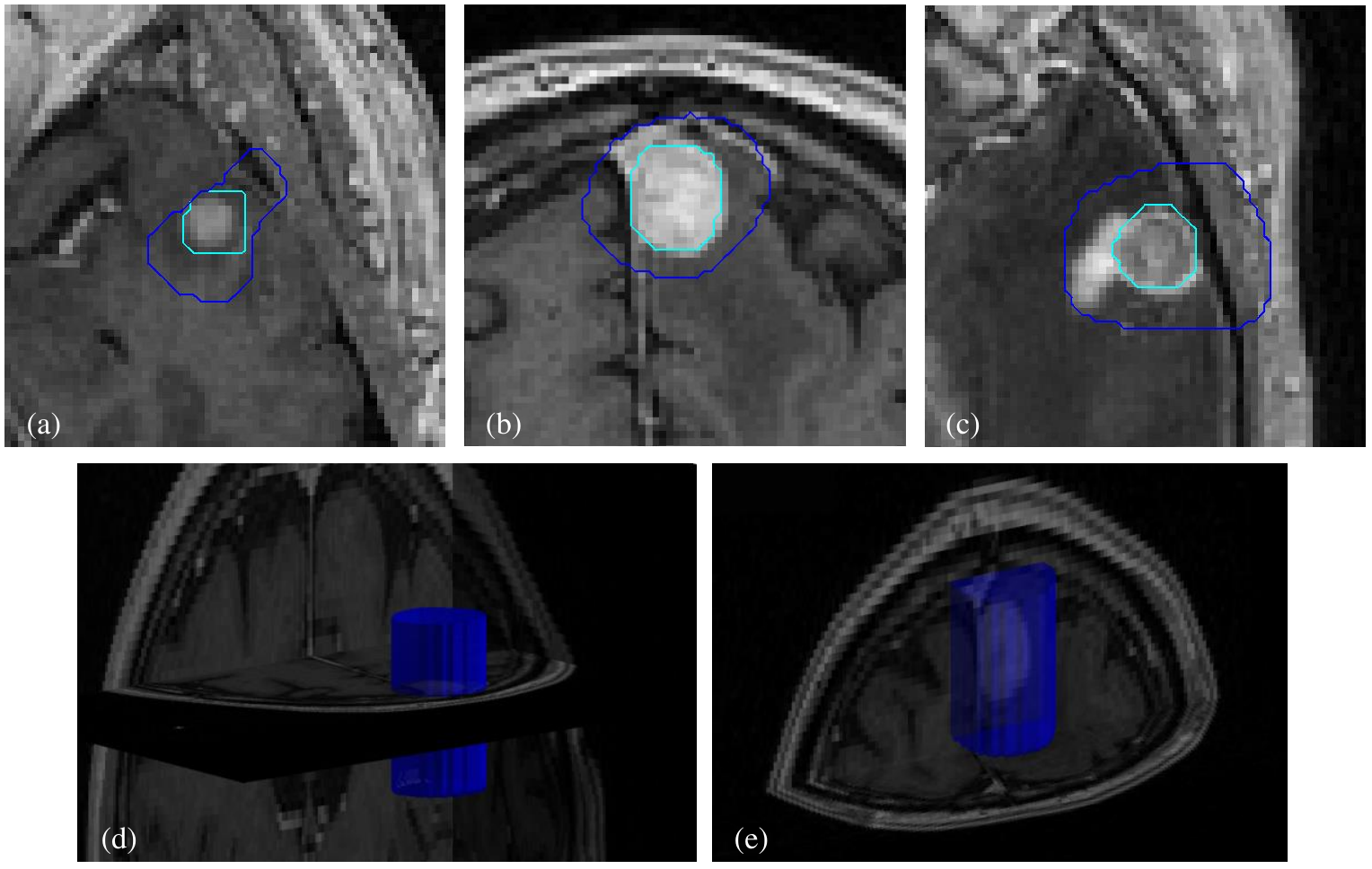}
	\caption[ROI bounding region generation on MR images]{ROI bounding region generation on MR images: (a, b, c) three examples (tumors $\#5$, $\#6$, and $\#10$) of bounding region (blue contour) determined \textit{via} a DRLSE-based method, starting from IUR (cyan contour) imported from the corresponding SUV\textsubscript{max} PET slice. Cylindroid extrusion on MRI dataset according to active PET slice range: (d) tridimensional bounding cylindroid for the tumor in the temporal brain lobe, reported in (b), rendered with transparent blue surface (Alpha blending with $\alpha = 0.70$) on $3$D orthogonal tri-planar view (middle slice is shown in the axial plane), that includes the whole GTV, as visible in (e).}
	\label{fig:boundCylind}
\end{figure}

Three examples of ROI bounding regions delineated, on original MR images, by the DRLSE-based method are shown in Fig. \ref{fig:boundCylind}(a-c).
These computations are performed just once for each dataset.
In fact, only one bounding region is determined \textit{via} DRLSE-based segmentation and it is extruded along the $z$ axis onto the other candidate MR slices.
For this purpose, the range of PET slices with high uptake regions is also provided by the IUR detection steps, to construct a cylindroid (i.e., a cylinder with irregular-shaped bases) including the whole GTV brain lesion.
Since PET and MRI tumor volumes may be distributed along different axial slices, the range of PET active slices is augmented by three slices on upper and lower sides, in order to ensure the total enclosing of the tumor into the cylindroid defined on the MRI dataset.
Fig. \ref{fig:boundCylind}d shows a $3$D volume rendering of the cylindroid extruded on the ROI bounding region shown in Fig. \ref{fig:boundCylind}a.
The whole GTV is included into the cylinder volume (see Fig. \ref{fig:boundCylind}e).

\subparagraph{MRI segmentation based on the FCM algorithm}
Typical MRI protocols, employed in stereotactic radio-surgery treatment planning, include only T1w CE-MRI of the head volume.
Therefore, no multiparametric MRI structural data (i.e., T1w, T2w, Proton Density, and FLAIR images) are generally available and brain tumor segmentation approaches based on multispectral MR images cannot be applied to Leksell Gamma Knife treatment planning.
These approaches rely on these multispectral MRI for segmenting and distinguishing cancer enhancement region, necrosis or edema \cite{dou2007}.
As a matter of fact, tumor detection and segmentation on T1w CE-MRI datasets are based on the hyper-intense enhancement regions.
Since sometimes either edema or necrotic material (hypo-intense regions in T1w CE-MR images) could be present in the analyzed tumors, necrosis inclusion in the planned target volume is required for radiosurgery purposes \cite{rundo2018next,shah2012}.
Please refer to Section \ref{sec:brainTumorSeg} for an extensive description of the method \cite{militelloIJIST2015,rundo2016WIRN}.

Unfortunately, due to the automation of the segmentation multi-modal process, the direct control of the human operator is lost during ROI bounding region manual delineation.
Some MR images included in the extruded cylindroid, especially the most distant slices from the IUR, might include no brain tumor as well as anatomic parts with pixel values similar to enhancement regions (such as bone or epidermal tissue).
For the former case, our MRI brain lesion segmentation method yields an empty ROI.
For the latter case, a shape-based control on connected-components belonging to the cluster with the highest intensity pixels is also defined, by taking advantage of the pseudospherical appearance of metastatic brain tumors \cite{ambrosini2010}.
When either skull bone or skin are included in the ROI bounding region, since the IUR is located near brain boundary, other high-valued pixel areas may be erroneously segmented by the unsupervised FCM clustering algorithm.
These areas are often characterized by a lengthened shape and may be removed by checking eccentricity (that is, the ratio between the foci distance related to an ellipse and its own major axis length) and extent (that is, the ratio between the pixels belonging to the region and the bounding box pixels) of the connected-components.
Theoretically suitable and experimentally validated feature values for this refinement step are: $\text{extent} \geq 0.6$ and $0.0 \leq \text{eccentricity} < 0.8$.

\begin{figure}[!t]
	\centering
	\includegraphics[width=\textwidth]{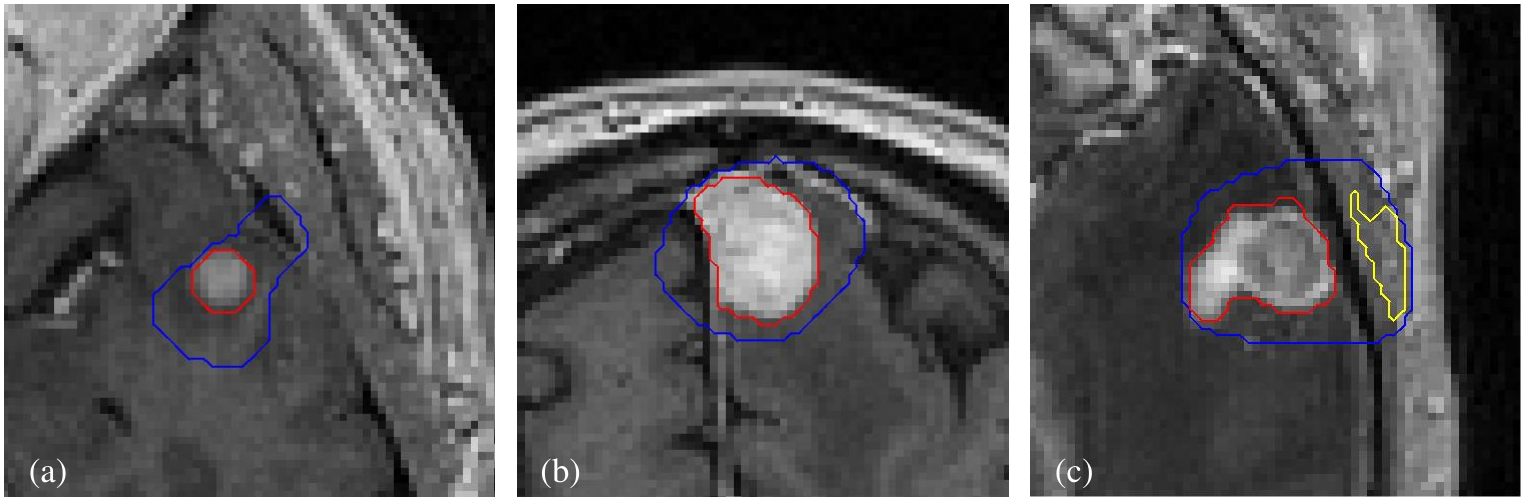}
	\caption[Examples of brain GTV segmentation on MR images, starting from the ROI bounding region]{Examples (tumors $\#5$, $\#6$, and $\#10$) of brain GTV segmentation on MR images (red contour), starting from the ROI bounding region (blue contour). A difficult case is shown in (c), where a lengthened hyper-intense region of the head skin is detected by the FCM clustering (yellow contour), but it is removed using the shape-based control.}
	\label{fig:GTVrefinement}
\end{figure}

Fig. \ref{fig:GTVrefinement} shows segmentation results on the images reported in Fig. \ref{fig:boundCylind}.
In particular, refinement steps can be appreciated in Fig. \ref{fig:GTVrefinement}c, where the region composed of the skin highest pixels included in the ROI bounding region is correctly removed by the shape-based controls (yellow boundary).

Although the literature offers $3$D approaches for brain lesion segmentation \cite{kanas2015}, we chose a $2$D methodology because $3$D FCM should be not well suitable on clusters with more general shapes \cite{huang2012}.
The analyzed brain tumors are composed of inhomogeneous cancerous tissue and necrotic material.
As a matter of fact, tumor detection and segmentation on T1w CE-MRI datasets are based on the hyper-intense enhancement regions.
Sometimes a dark area might be present due to either cystic or necrotic tissue, which must be included into the GTV for radiotherapeutic purposes.
These hypo-intense areas may negatively affect the FCM clustering process when $3$D MRI data are processed. In these cases, clustering could yield several disconnected volumes and successful post-processing steps are very difficult.
On the other hand, performing FCM cluster analysis on the single MRI slices is definitely more reliable and post-processing steps proposed in \cite{militelloIJIST2015} are able to segment the whole tumor.

\subparagraph{PET segmentation based on the RW algorithm}
The RW-based segmentation process on the input PET dataset is updated and repeated, in the same way, on the same PET dataset combined with the corresponding MRI GTV binary masks in order to achieve a new BTV, called as BTV\textsubscript{MRI}.
In PET imaging, the adaptive probability threshold is computed separately slice-by-slice to follow the whole lesion volume, by taking into account the intensity gradient and contrast changes of the metabolic lesion over the entire range of PET slices.
This key feature could be lost using a fully 3D method: the proposed method is then performed in parallel for the slices adjacent to the starting PET slice with the SUV\textsubscript{max} \cite{stefano2016}.

\begin{figure}[!t]
	\centering
	\includegraphics[width=0.8\textwidth]{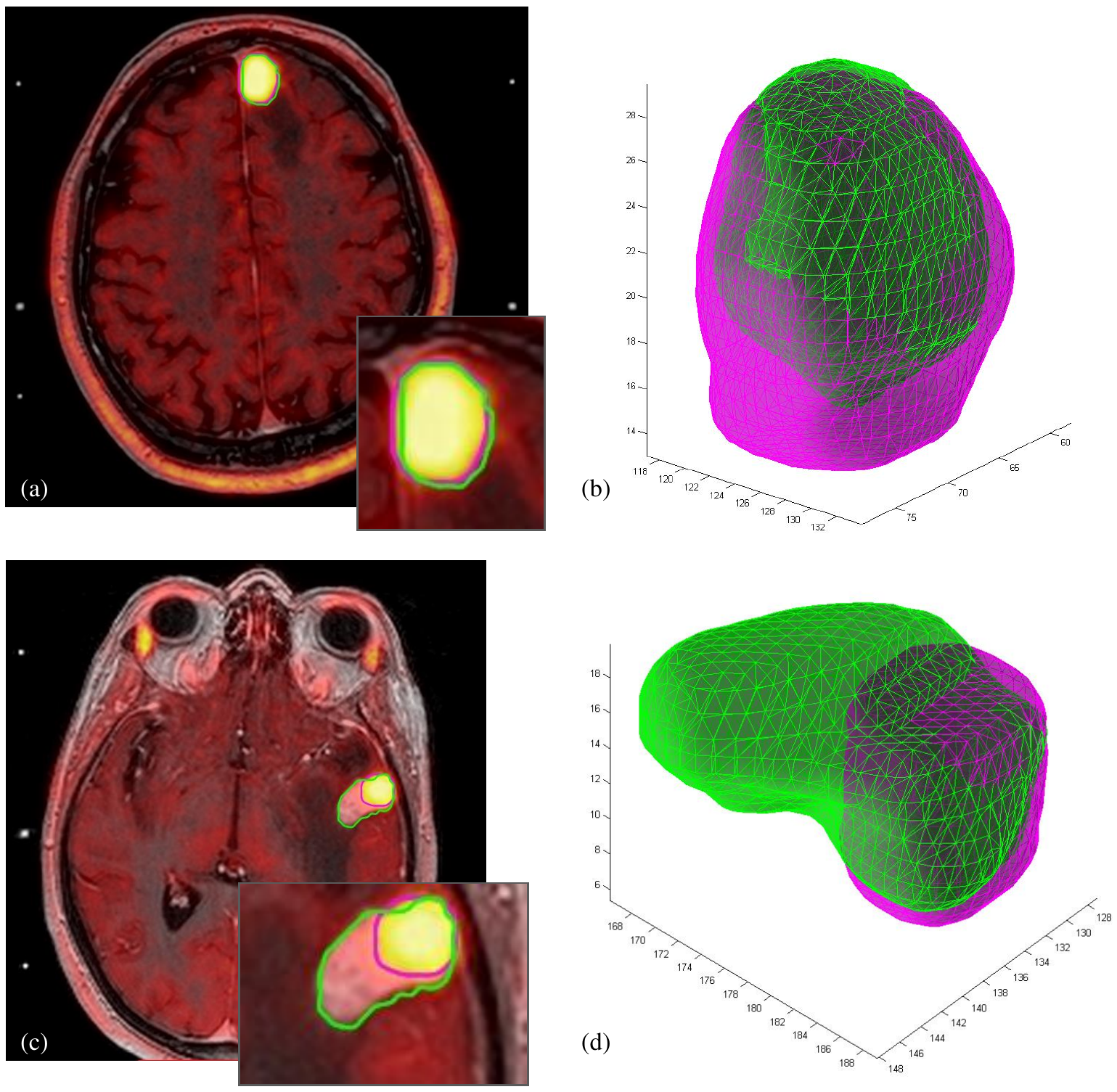}
	\caption[Results achieved by the proposed multimodal PET/MRI segmentation method]{Results achieved by the proposed multimodal PET/MRI segmentation method on the input image pairs (tumors $\#6$ and $\#14$) reported in Fig. \ref{fig:PET-MRoriginal}(a, b) and Fig. \ref{fig:PET-MRoriginal}(c, d), respectively. The BTV\textsubscript{MRI} (magenta contour) and GTV (green contour) are superimposed: (a) boundaries computed on PET and MR images are nearly overlaid; (c) the two ROIs are very different in this case. At the bottom right, the lesion regions are zoomed in both cases. Tridimensional volume reconstructions of brain tumors in (a) and (c) are shown in (b) and (d), respectively. Transparent surfaces are rendered with Alpha blending ($\alpha = 0.50$) to emphasize volume intersections.}
	\label{fig:volRenderingsBTV}
\end{figure}

\begin{figure}[!t]
	\centering
	\includegraphics[width=0.8\textwidth]{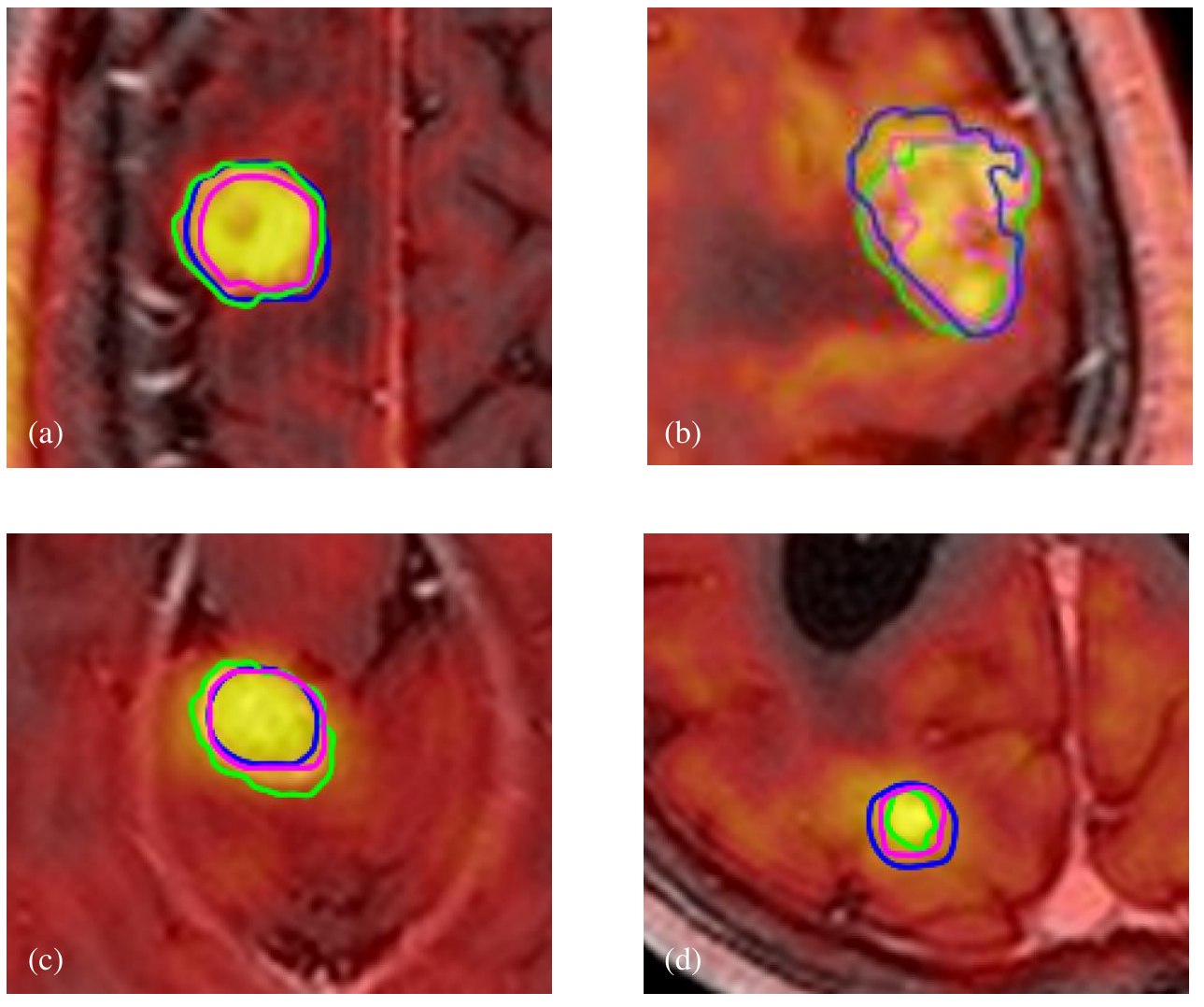}
	\caption[Graphical examples of co-segmentation results achieved by the proposed multimodal PET/MRI approach]{Graphical examples of co-segmentation results achieved by the proposed multimodal PET/MRI approach, where GTV (green contour), BTV (blue contour), and BTV\textsubscript{MRI} (magenta contour) ROIs are overlaid on the corresponding fused PET/MRI slices (displayed in Hot Look-Up Table). Four metastatic tumors in different anatomic regions of human brain are considered: (a) left temporo-parietal (tumor $\#3$), (b) right temporo-parietal (tumor $\#11$), (c) limbic (tumor $\#15$), and (d) occipital (tumor $\#13$).}
	\label{fig:cosegRes}
\end{figure}

\begin{figure}[!t]
	\centering
	\includegraphics[width=0.8\textwidth]{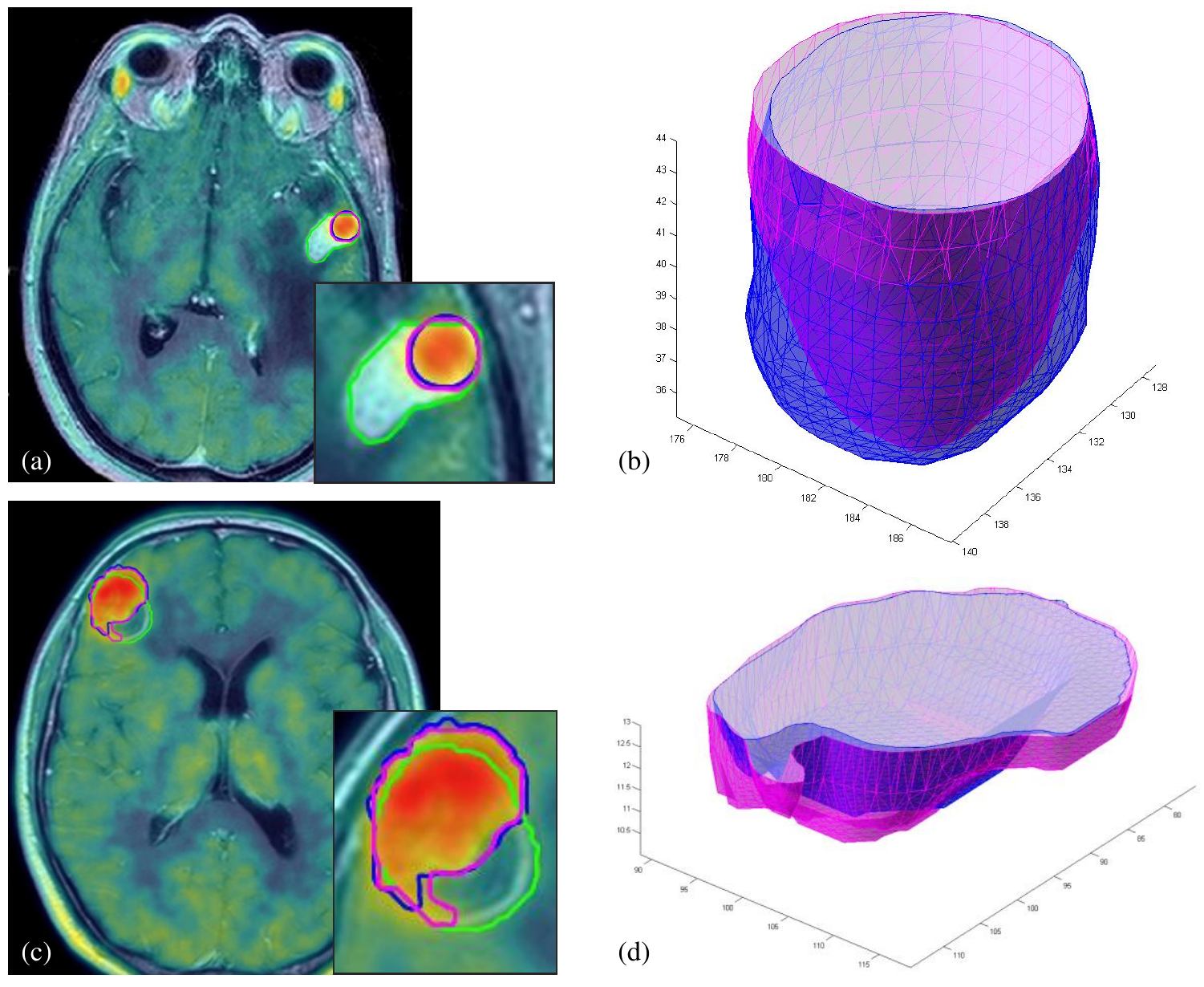}
	\caption[Segmentation results obtained by the single modality PET method (BTV) versus those obtained by the multimodal PET/MRI approach after the PET data updating based on the results of the MRI brain lesion segmentation (BTV\textsubscript{MRI})]{Segmentation results, on brain tumors $\#14$ and $\#16$, obtained by the single modality PET method (BTV) versus those obtained by the multimodal PET/MRI approach after the PET data updating based on the results of the MRI brain lesion segmentation (BTV\textsubscript{MRI}): (a, c) GTV (green contour), BTV (blue contour), and BTV\textsubscript{MRI} (magenta contour) ROIs overlaid on the corresponding fused PET/MRI slice (PET and MRI are displayed using Jet and gray-scale LUTs, respectively); (b, d) 3D representation of BTV and BTV\textsubscript{MRI} rendered with transparent surfaces (Alpha blending with $\alpha = 0.40$), respectively, cut at the corresponding slice shown in (a) and (c).}
	\label{fig:slicedVol}
\end{figure}

\subparagraph{Combining MRI segmentation results and PET images}
Lesion segmentation results on the morphologic brain MRI data are combined with the metabolic information of the co-registered PET images.
This PET dataset represents the input of the RW algorithm.
In this way, MRI GTV is utilized to combine the superior contrast of PET images with the higher spatial resolution of MR images.
Each PET slice is weighted pixel-by-pixel according to the corresponding MRI ROIs.
Nevertheless, an assumption that there is a one-to-one correspondence between BTV and GTV is unrealistic.
Lesions may have smaller uptake regions compared to GTV, such as shown in Figs. \ref{fig:volRenderingsBTV}c and \ref{fig:cosegRes}c.
In the same way, the PET lesion can show additional area compared to lesion boundaries in MR images (see Figs. \ref{fig:cosegRes}d \ref{fig:slicedVol}c).
For this reason, we incorporate, with extreme caution, the MRI GTV binary mask in the RW-based segmentation procedure.
In particular, PET pixels inside the corresponding MR target are multiplied by a gain factor to slightly emphasize probable lesion area; PET pixels outside the corresponding MRI ROIs are weakly underestimated in order to reduce the radioactivity spill-in effect---from background into the lesion---and spill-out effect---from the lesion into background---by taking advantage of the higher spatial resolution of MR images.

\paragraph{Results}

The evaluation of the data presented in this study was performed retrospectively on $19$ metastatic brain tumors, treated with Leksell Gamma Knife.
Nowadays, the clinical protocol does not usually consider co-registered PET images for Gamma Knife treatment planning.
In the proposed multimodal segmentation approach, two different PET segmentation results were computed: (\textit{i})  BTV, by considering PET images alone, and (\textit{ii})  BTV\textsubscript{MRI}, by considering PET/MRI co-segmentation.

Firstly, volumes measured on PET and MRI data are reported and, for a quantitative assessment, PET/MRI volume difference and centroid distance were calculated for each lesion.
Figs. \ref{fig:volRenderingsBTV}a and \ref{fig:volRenderingsBTV}c show experimental results achieved by the proposed multimodal PET/MRI segmentation method on the input image pairs reported in Figs. \ref{fig:PET-MRoriginal}(a, b), and Figs. \ref{fig:PET-MRoriginal}(c, d), respectively.
In addition, the corresponding GTV and BTV\textsubscript{MRI} volume rendering is reported in Figs. \ref{fig:volRenderingsBTV}b and \ref{fig:volRenderingsBTV}d, respectively.
Three other interesting examples obtained by the proposed multimodal segmentation method can be visually and qualitatively evaluated in Fig. \ref{fig:cosegRes}.

It is worth to observe that the margin of enhancement of tumors in MRI datasets is not always strongly correlated with high uptake regions in PET images.
Segmentation results of PET images combined with GTV MRI masks are very stable (magenta boundaries in Fig. \ref{fig:cosegRes}), even if GTV is quite different.
In Fig. \ref{fig:cosegRes}d, PVE on PET image \cite{stefano2014} is reduced by using MRI ROIs.

\begin{figure}[!t]
	\centering
	\includegraphics[width=0.6\textwidth]{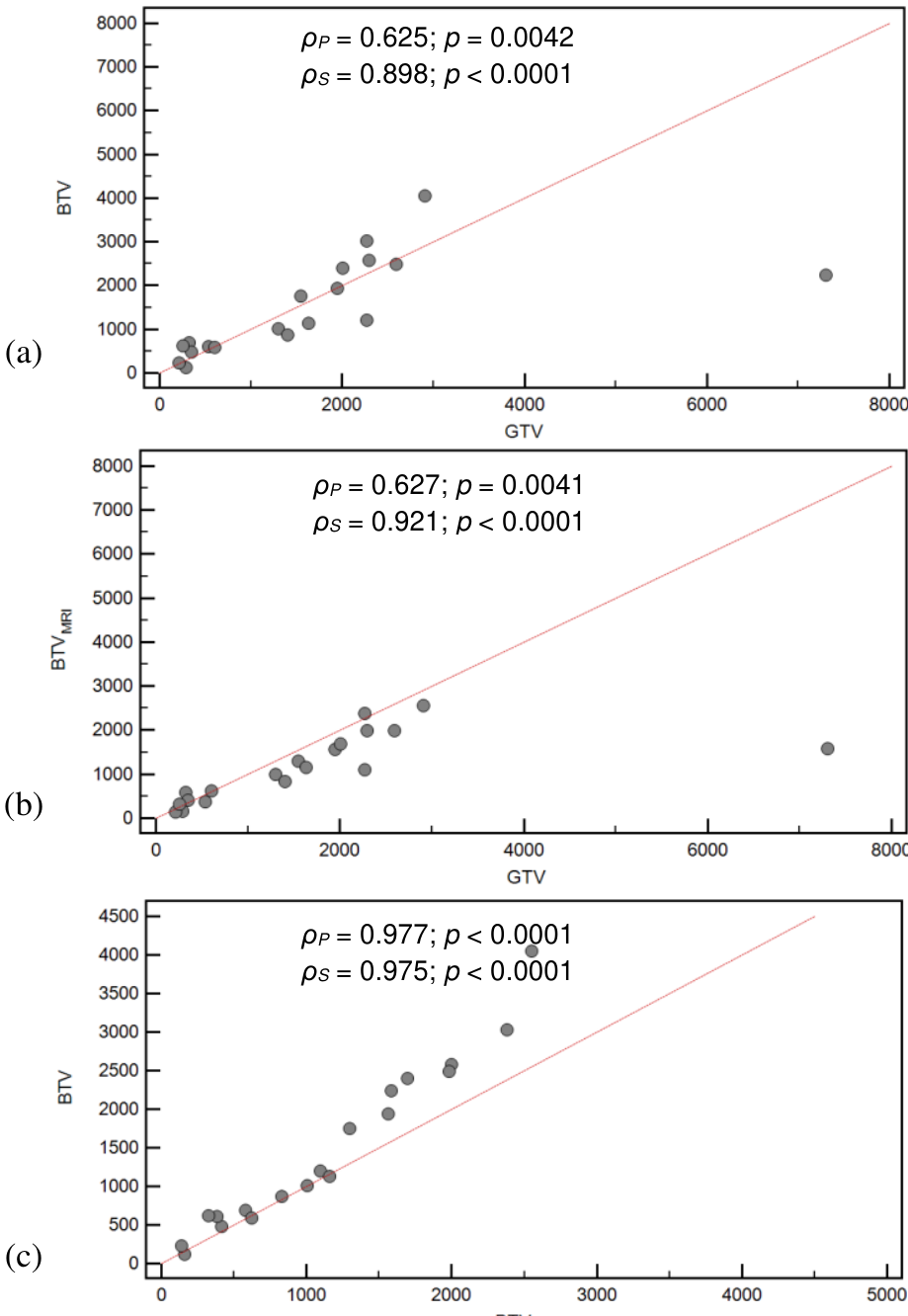}
	\caption[Scatter diagrams of tumor volume measurements on multimodal PET/MRI data]{Scatter diagrams of tumor volume measurements on multimodal PET/MRI data: (a) BTV versus GTV; (b) BTV\textsubscript{MRI} versus GTV; (c) BTV versus BTV\textsubscript{MRI}. Red dashed line is the line of equality. In the upper part of each sub-figure, Pearson and Spearman correlation coefficients with the corresponding significance level ($p$-value) are also reported.}
	\label{fig:multi-scatterDiag}
\end{figure}

\subparagraph{PET and MRI tumor volume measurements}
The volumes, calculated for each brain tumor imaged on MRI and PET modalities, are shown in Table \ref{table:multi-tumorVolumes}. Absolute average volume difference (defined as  $\text{abs}(V_1 - V_2)/V_2$) between BTV, BTV\textsubscript{MRI}, and GTV were also calculated---by generalizing the formulation in Eq. (\ref{eq:AVD})---in order to evaluate either overestimation or underestimation of PET functional imaging with respect to anatomic MRI data, before and after GTV binary mask combination into PET images.
In addition, Euclidean distances between BTV, BTV\textsubscript{MRI}, and GTV centroids (centers of mass) were reported.

\begin{table}[t]
\scriptsize
\centering
  \caption[Brain tumor volume measurements for each metastatic tumor using MR and PET imaging]{Brain tumor volume measurements for each metastatic tumor using MR and PET imaging. Absolute volume difference and centroid distance between segmented GTV, BTV, and BTV\textsubscript{MRI} are also reported. Mean value and standard deviation are reported in the last two rows.}
\label{table:multi-tumorVolumes}
\begin{tabular}{l|ccc|ccc}
\hline \hline
\multirow{2}{*}{Tumor} & \multicolumn{3}{c}{Absolute Volume Difference [voxels]}              & \multicolumn{3}{c}{Volume Centroid Distance [pixels]}              \\
                       & BTV vs GTV & BTV\textsubscript{MRI} vs GTV & BTV vs BTV\textsubscript{MRI} & BTV vs GTV & BTV\textsubscript{MRI} vs GTV & BTV vs BTV\textsubscript{MRI} \\
    \hline
    \#1	& 33.61	& 4.98	& 27.27	& 1.805	& 1.067	& 1.457 \\
    \#2	& 13.45	& 16.11	& 35.24	& 1.523	& 0.740	& 0.839 \\
    \#3	& 0.36	& 19.74	& 24.15	& 1.250	& 0.949	& 0.398 \\
    \#4	& 112.69	& 79.26	& 18.65	& 2.591	& 1.275	& 0.607 \\
    \#5	& 39.88	& 20.52	& 16.07	& 1.446	& 0.845	& 0.246 \\
    \#6	& 12.60	& 12.95	& 29.36	& 0.974	& 0.971	& 0.098 \\
    \#7	& 13.41	& 30.98	& 59.84	& 2.248	& 1.042	& 1.229 \\
    \#8	& 22.49	& 23.10	& 0.80	& 1.830	& 0.802	& 1.454 \\
    \#9	& 57.34	& 44.41	& 23.27	& 2.238	& 1.681	& 0.399 \\
    \#10	& 1.17	& 3.50	& 4.51	& 0.697	& 1.080	& 0.730 \\
    \#11	& 39.44	& 12.18	& 58.78	& 0.647	& 0.772	& 0.801 \\
    \#12	& 7.62	& 34.29	& 63.77	& 3.118	& 2.720	& 0.447 \\
    \#13	& 143.14	& 27.45	& 90.77	& 5.660	& 0.840	& 0.648 \\
    \#14	& 47.07	& 51.83	& 9.87	& 5.753	& 5.551	& 0.718 \\
    \#15	& 30.91	& 29.13	& 2.50	& 1.162	& 0.768	& 0.518 \\
    \#16	& 3.79	& 23.37	& 25.55	& 5.941	& 5.614	& 0.431 \\
    \#17	& 38.11	& 41.03	& 4.95	& 0.972	& 1.099	& 0.238 \\
    \#18	& 19.42	& 15.49	& 41.31	& 1.414	& 1.211	& 0.224 \\
    \#19	& 69.37	& 78.36	& 41.56	& 7.141	& 6.769	& 0.885 \\
    \hline
    Avg.	& 37.15	& 29.93	& 30.43	& 2.548	& 1.884	& 0.651 \\
    Std. dev.	& 37.59	& 21.37	& 24.22	& 2.018	& 1.890	& 0.395 \\
    \hline \hline
\end{tabular}
\end{table}

First of all, mean BTV and GTV values and their average absolute volume difference attest that tumor regions are imaged and represented differently on PET and MRI datasets.
Furthermore, high standard deviation in BTV and GTV measurements implies that we are dealing with metastatic lesions characterized by various types and dimensions.
On the other hand, BTV\textsubscript{MRI} measurements follow a different trend because BTV\textsubscript{MRI} segmentation is influenced by the combination of GTV masks and PET images.
Although we slightly integrate MRI ROIs with PET datasets during BTV\textsubscript{MRI} computation, this conservative choice frequently could reduce radioactivity spill-in and spill-out effects, by exploiting higher MRI spatial resolution.
This is evident in the examples shown in Fig. \ref{fig:cosegRes}.
Moreover, the BTV\textsubscript{MRI} measurement standard deviation is lower than BTV and GTV ones.
This means that multimodal PET/MRI segmentation yields more stable results with respect to single modality segmentations, since both metabolic and morphological imaging are considered.
These experimental evidences are corroborated by the achieved absolute volume difference values, where BTV\textsubscript{MRI} has about the same difference with GTV and BTV.
Volume centroid distances reflect the aforementioned situation, too.

The above observations, based on numerical experimental results, are qualitatively supported by the graphical multimodal PET/MRI segmentation results in Fig. \ref{fig:slicedVol}.
To measure the degree of dependence between GTV, BTV, and BTV\textsubscript{MRI}, both Pearson’s correlation coefficient $\rho_P$ and Spearman’s rank correlation coefficient $\rho_S$ were calculated for each comparison.
Spearman’s correlation yields different information with respect to Pearson’s correlation, since $\rho_S$ is less sensitive to strong outliers that are in the tails of both samples.
In fact, outlier importance is limited to the value of its rank.
As expected, GTV revealed greater correlation to BTV\textsubscript{MRI}, especially according to Spearman’s $\rho_S$.
In addition, BTV and BTV\textsubscript{MRI} are significantly correlated, using both Pearson and Spearman correlation coefficients.
Accordingly, scatter diagrams in Fig. \ref{fig:multi-scatterDiag} represent graphically these numerical results.

\subparagraph{Evaluation of the overall multimodal PET/MRI segmentation method}

In this section, an overall evaluation of the proposed multimodal PET/MRI segmentation method is presented. Spatial overlap-based, and distance-based metrics were calculated, to measure the similarity and correlation concerning PET and MRI segmentation approaches.
Considering these results, it was possible to argue the complementarity between PET and MR imaging modalities, combined by the proposed multimodal segmentation method. 

Table \ref{table:multi-overlapMetrics} shows tumor spatial overlap-based metrics for PET and MRI segmentation methods.
The achieved results demonstrated that GTV was, on average, more similar to BTV\textsubscript{MRI} than to BTV.
The highest mean values and the lowest standard deviation of the evaluation measures were obtained in BTV and BTV\textsubscript{MRI} comparisons.

\begin{table}[t]
\scriptsize
\centering
  \caption[Achieved tumor spatial overlap-based metrics for GTV, BTV, and BTV\textsubscript{MRI} segmentations]{Achieved tumor spatial overlap-based metrics for GTV, BTV, and BTV\textsubscript{MRI} segmentations. Total average and standard deviation calculated on $19$ brain lesions are also reported.}
\label{table:multi-overlapMetrics}
\begin{tabular}{l|ccc}
\hline \hline
\multirow{2}{*}{Tumor} & \multicolumn{3}{c|}{DSC [voxels]} \\
                       & BTV vs GTV & BTV\textsubscript{MRI} vs GTV & BTV vs BTV\textsubscript{MRI} \\
    \hline
    \#1	& 63.97	& 78.89	& 76.17\\
    \#2	& 80.55	& 81.11	& 80.24\\
    \#3	& 81.33	& 79.81	& 83.91\\
    \#4	& 40.79	& 56.32	& 79.30\\
    \#5	& 55.42	& 59.24	& 73.70\\
    \#6	& 76.43	& 73.26	& 86.76\\
    \#7	& 58.29	& 78.46	& 62.22\\
    \#8	& 68.31	& 83.90	& 80.82\\
    \#9	& 39.22	& 51.24	& 83.99\\
    \#10	& 72.59	& 74.20	& 84.35\\
    \#11	& 65.74	& 78.71	& 68.14\\
    \#12	& 46.79	& 51.72	& 69.23\\
    \#13	& 52.57	& 76.90	& 68.36\\
    \#14	& 53.38	& 54.62	& 87.37\\
    \#15	& 76.66	& 79.08	& 89.81\\
    \#16	& 67.80	& 70.59	& 86.97\\
    \#17	& 65.55	& 67.03	& 85.80\\
    \#18	& 77.30	& 79.51	& 82.39\\
    \#19	& 32.85	& 31.66	& 78.24\\
    \hline
     Avg.	& 61.87	& 68.75	& 79.36\\
     Std. dev.	& 14.64	& 14.07	& 7.79\\
    \hline \hline
\end{tabular}
\end{table}

Table \ref{table:multi-distanceMetrics} reports the values of spatial distance-based metrics achieved by the proposed multimodal PET/MRI segmentation approach in the experimental trials.
Distance-based metrics values agreed with achieved volume-based and overlap-based measurements, showing the same trend.

\begin{table}[t]
\tiny
  \caption[Achieved tumor spatial distance-based metrics for GTV, BTV, and BTV\textsubscript{MRI} segmentations]{Achieved tumor spatial distance-based metrics for GTV, BTV, and BTV\textsubscript{MRI} segmentations. Total average and standard deviation calculated on $19$ brain lesions are also reported.}
\label{table:multi-distanceMetrics}
\hspace{-1.0cm}
\begin{tabular}{l|ccc|ccc|ccc}
\hline \hline
\multirow{2}{*}{Tumor} & \multicolumn{3}{c}{HD [pixels]}              & \multicolumn{3}{c}{AvgD [pixels]}    & \multicolumn{3}{c}{MHD [pixels]}           \\
                       & BTV vs GTV & BTV\textsubscript{MRI} vs GTV & BTV vs BTV\textsubscript{MRI} & BTV vs GTV & BTV\textsubscript{MRI} vs GTV & BTV vs BTV\textsubscript{MRI} & BTV vs GTV & BTV\textsubscript{MRI} vs GTV & BTV vs BTV\textsubscript{MRI} \\
    \hline
    \#1	& 8.544	& 6.403	& 7.616	& 2.669	& 1.377	& 1.756	& 0.412	& 0.264	& 0.339\\
    \#2	& 5.831	& 4.000	& 5.831	& 1.343	& 1.163	& 1.259	& 0.340	& 0.178	& 0.195\\
    \#3	& 8.485	& 7.810	& 4.472	& 1.400	& 1.355	& 1.145	& 0.436	& 0.312	& 0.138\\
    \#4	& 5.831	& 3.606	& 5.000	& 1.090	& 0.667	& 0.778	& 3.497	& 3.257	& 0.277\\
    \#5	& 5.385	& 3.000	& 3.606	& 1.295	& 0.798	& 0.742	& 1.857	& 1.814	& 0.107\\
    \#6	& 5.099	& 4.123	& 3.000	& 1.121	& 1.243	& 0.940	& 1.208	& 1.247	& 0.029\\
    \#7	& 9.220	& 4.000	& 5.831	& 2.233	& 0.804	& 1.332	& 0.786	& 0.441	& 0.421\\
    \#8	& 10.630	& 3.162	& 10.630	& 1.312	& 0.943	& 0.815	& 0.610	& 0.239	& 0.524\\
    \#9	& 4.472	& 4.472	& 1.414	& 0.727	& 0.969	& 0.204	& 1.392	& 0.994	& 0.377\\
    \#10	& 7.810	& 6.403	& 3.162	& 1.368	& 1.369	& 0.689	& 0.285	& 0.379	& 0.286\\
    \#11	& 12.000	& 6.083	& 11.662	& 2.630	& 1.535	& 2.262	& 0.734	& 0.180	& 0.214\\
    \#12	& 4.472	& 5.388	& 2.236	& 1.898	& 1.556	& 1.167	& 1.496	& 1.389	& 0.249\\
    \#13	& 7.280	& 2.828	& 7.280	& 2.345	& 0.809	& 1.711	& 0.530	& 0.404	& 0.296\\
    \#14	& 15.621	& 14.213	& 3.606	& 3.618	& 3.281	& 0.810	& 1.115	& 1.045	& 0.233\\
    \#15	& 4.243	& 3.606	& 2.236	& 1.250	& 1.107	& 0.643	& 0.286	& 0.219	& 0.147\\
    \#16	& 11.705	& 11.402	& 7.616	& 3.503	& 3.175	& 1.906	& 0.788	& 0.754	& 0.094\\
    \#17	& 5.385	& 4.243	& 2.236	& 2.136	& 1.939	& 0.769	& 0.267	& 0.319	& 0.087\\
    \#18	& 7.071	& 5.385	& 4.472	& 1.782	& 1.420	& 1.356	& 0.374	& 0.336	& 0.062\\
    \#19	& 15.811	& 16.125	& 10.000	& 5.820	& 5.811	& 1.486	& 1.446	& 1.360	& 0.195\\
    \hline
    Avg.	& 8.15	& 6.12	& 5.36	& 2.081	& 1.648	& 1.146	& 0.940	& 0.796	& 0.225\\
    Std. dev.	& 3.57	& 3.79	& 3.04	& 1.211	& 1.235	& 0.515	& 0.787	& 0.778	& 0.131\\
    \hline \hline
\end{tabular}
\end{table}

Table \ref{table:multi-Likert} shows clinical evaluation results regarding both BTV and BTV\textsubscript{MRI} integration with GTV to determine CTV, according to the explanation in Appendix \ref{sec:LikertEval}.
In the current clinical practice the CTV is always based on the GTV (defined on MR images). As a consequence, the BTV contribution (defined on PET images) cannot get worse the CTV definition from a clinical perspective, i.e., the Likert score is always greater than $2$.
This qualitative evaluation performed by three physicians using a five-point Likert scale, defined as: 1) strong worsening in CTV definition; 2) moderate worsening in CTV definition; 3) indifferent, neither enhancement nor worsening in CTV definition; 4) moderate enhancement in CTV definition; 5) strong enhancement in CTV definition.

\begin{table}[t]
\scriptsize
\centering
  \caption[Clinical value of BTV and BTV\textsubscript{MRI}, for each brain metastasis, in CTV definition]{Clinical value of BTV and BTV\textsubscript{MRI}, for each brain metastasis, in CTV definition.}
\label{table:multi-Likert}
\begin{tabular}{l|cc}
\hline \hline
\multirow{2}{*}{Tumor} & \multicolumn{2}{c}{Clinical value} \\
                       & BTV integration & BTV\textsubscript{MRI} integration \\
    \hline
    \# 1	& 4	& 5\\
    \# 2	& 4	& 4\\
    \# 3	& 3	& 3\\
    \# 4	& 4	& 4\\
    \# 5	& 3	& 3\\
    \# 6	& 4	& 5\\
    \# 7	& 4	& 4\\
    \# 8	& 3	& 3\\
    \# 9	& 3	& 3\\
    \# 10	& 3	& 3\\
    \# 11	& 5	& 5\\
    \# 12	& 3	& 3\\
    \# 13	& 3	& 4\\
    \# 14	& 4	& 4\\
    \# 15	& 3	& 3\\
    \# 16	& 4	& 4\\
    \# 17	& 3	& 3\\
    \# 18	& 4	& 5\\
    \# 19	& 4	& 5\\
    \hline \hline
\end{tabular}
\end{table}

In $\sim26\%$ of cases, the CTV definition was strongly influenced by PET imaging.
In more than $50\%$ of cases, the CTV was strongly or moderately conditioned by metabolic imaging: only in $8/19$ cases PET volume did not modify the CTV definition.
In more than $25\%$ of cases, BTV\textsubscript{MRI} enhanced the CTV more accurately than BTV, because radioactivity spill-in and spill-out effects between tumor and surrounding tissues were reduced by the contribution of MRI ROIs in PET segmentation, shrinking false positive uptake areas.

\paragraph{Discussion and conclusions}
The proposed fully automated multimodal segmentation method introduced several novelties and advantages.
Firstly, we used two computer-assisted single modality segmentation approaches for PET and MR images, which have been already tested using evaluation metrics and validated by clinicians.
Although image co-registration is needed to bring multimodal PET/MR image information into the same reference system, the co-segmentation results are not just derived \textit{via} image registration since anatomic and metabolic are carefully integrated, by relying also on the physicians’ expertise, in the proposed multimodal PET/MRI segmentation algorithm.
These accurate and operator-independent single modality segmentation methods are certainly more effective and reliable on PET and MRI datasets, with respect to approaches that unify the multimodal information into a single graph.

State-of-the-art methods based on a hyper-graph \cite{bagci2013MedIA,han2011} or subgraphs with inter-subgraph arcs \cite{song2013} are more sensitive to registration errors, especially when multimodal images are not acquired contextually using a hybrid scanner.
In these cases, no significant anatomic and functional changes have to be assumed between the images acquired with different modalities at the same clinical phase to directly construct the hyper-graph or the interconnected subgraphs.
The proposed approach was not tailored specifically for the latest generation hybrid PET/MRI scanners.
Obviously, in the case of multimodal PET/MRI scanner \cite{catana2008,judenhofer2008}, the scenario is simplified considering that all the initial registration stages are not necessary, whereas the subsequent steps in the processing pipeline are still valid.
Methods based on hyper-graphs generally yield a single target volume on the fused multimodal images.
This may compromise the segmentation quality since PET and MRI could convey different information, resulting in a disagreement between BTV and GTV, because enhancement, edema, and necrosis regions are imaged differently by the PET and MRI modalities; so the tumor volumes defined on metabolic PET and on anatomic CT or MRI could be highly different \cite{song2013}.
The proposed multimodal method, keeping all anatomic and metabolic information and enabling a decision-level fusion \cite{conti2010,serra2018}, is more reliable on PET and MRI datasets with respect to approaches that unify the multimodal information into a single graph structure.
Result repeatability is ensured because it is able to adapt in different pathological scenarios.
This choice allows for a greater awareness on the decision-making process, carried out by the clinical staff that is going to plan the neuro-radiosurgery treatment, that brings to the CTV definition.
The CTV is determined by the clinical staff, by taking into account the specific patient’s pathological scenario.
According to the computed BTV and GTV volumes, the CTV identification is a deep decision-making process involving numerous anatomic and metabolic insights.
Therefore, the appropriateness of the multimodal segmentation is justified by the addressed scenario, by allowing the clinicians to carefully consider the possibility to include BTV information into the planned CTV and to determine a personalized therapy for each single cancer metabolism \cite{brady2016}.
Secondly, IURs detected on PET images are not blindly propagated to GTV segmentation algorithm, but properly modified according to MRI data using the developed ROI bounding region generation method.
This tailored MRI-driven method, based on LSFs, finds adaptively a valid bounding region in order to eliminate user intervention, required by the MRI brain tumor segmentation method in \cite{militelloIJIST2015}.
On the other hand, in \cite{bagci2013MedIA}, the identified seeds on PET images are propagated to the corresponding anatomic images before segmentation process beginning.
Hence, two existing and efficient single modality processing pipelines are combined in a smart fashion.
In particular, PET and MRI segmentation results are mutually exploited.
Lastly, this fully automated approach is very reproducible and reliable, also due to the brain anatomic district imaging.
We focused on the application domain regarding brain tumors that underwent Leksell Gamma Knife stereotactic radiosurgery.
In fact, the fully automatic identification of target IURs is feasible because MET-PET datasets for stereotactic neuro-radiosurgery treatment planning include only the brain area avoiding the possible presence of false positives in other anatomic regions.
Differently, in total body FDG-PET examinations user interaction to manually identify the target lesion is always needed (normal structures such as brain, heart, bladder, kidneys, and ureters normally have high FDG uptake) \cite{stefano2017}.
Accordingly, these co-segmentation software implementation platforms must provide manual seeding facilities in addition to automated multimodal segmentation algorithm.
Although a one-to-one relationship between two different structural images of the same abnormal region could be reasonable, the assumption of identical lesion contours in both functional and morphologic images is often infeasible in many anatomic regions \cite{bagci2013MedIA}.
For instance, depending on the metabolic characteristics of the cells within a lung tumor \cite{ballangan2013}, it may not take up a radiotracer in all its volume, or uptake region may be larger than the anatomic boundary of the tumor due to cellular activation in nearby tissues.
Further issues are introduced by combining more than two structural modalities, i.e., MRI and CT, with PET metabolic imaging.
Briefly, different imaging modalities (PET/MRI vs PET/CT), anatomic regions, and pathologies require a customization/adaptation of the method according to the specific clinical scenario.
In our study, PET and MRI datasets regarding the same subject were acquired at two different times (MRI is scanned a few days after MET-PET) by two different dedicated scanners.
This represents a non-trivial problem in co-segmentation domain, because an efficient and accurate multimodal co-registration is required.
On the contrary, state-of-the-art about co-segmentation methods process mostly simultaneous PET/CT or PET/MRI acquisitions.
Of course, the potential of hybrid imaging (PET/CT and PET/MRI systems) overcomes image registration issues by giving multimodal images contextually \cite{pichler2008,weissler2015}, especially in brain imaging \cite{heiss2009}.

The achieved experimental results showed that GTV and BTV segmentations are statistically correlated (Spearman’s rank correlation coefficient: $0.898$) but have not higher degree of similarity (average \emph{DSC}: $61.87 \pm 14.64$).
In fact, GTV and BTV measurements as well as evaluation metrics values (volume-based, overlap-based, and spatial distance-based metrics) corroborated that MRI and MET-PET convey different but complementary imaging information.
Furthermore, when lesion sizes are greater than $2$-$3$ times the Full Width at Half Maximum (FWHM) of the point spread function of the PET image resolution reconstructed by the PET imaging system, the underestimation of metabolic activity due to PVE can be assumed to be negligible \cite{soret2007}.
In other cases, a recovery coefficient method could be included in our algorithm, such as the approach described in \cite{gallivanone2011}.
Nevertheless, the output of the proposed multimodal PET/MRI segmentation method, named BTV\textsubscript{MRI}, accurately combines PET and MR imaging.
BTV\textsubscript{MRI} segmentation could reduce radioactivity spill-in and spill-out effects between tumor and surrounding tissues, taking advantage of the higher spatial resolution of MRI.

In addition, since it is not possible to define a gold standard CTV according to both MRI and PET images without treatment response assessment, the feasibility and the clinical value of BTV integration in Gamma Knife treatment planning were considered.
Therefore, a visual and qualitative evaluation was performed by experienced physicians to assess the clinical value of BTV integration in CTV delineation using a five-point Likert scale.
In most cases, CTV delineation was strongly or moderately influenced by PET imaging: GTV does not match with the functional area of the tumor, and metabolic imaging must be considered to assist the radiation oncologist in treatment planning.
These results agree with \cite{grosu2005a}, where in $74\%$ the expansion of the MET-PET volume was larger than the Gadolinium enhancement area.
Table 4 shows that BTV\textsubscript{MRI} influenced the CTV more accurately than BTV (in $5/19$ of cases): this finding was attributed to the higher spatial resolution of MRI, which shrank small false positive areas in PET target segmentation.
Finally, PET imaging should be usually considered by clinicians in order to determine a CTV that takes into account the “active” part of the cancer in addition to anatomic tumor boundaries.
According to these findings, it is appropriate to include PET images in stereotactic neuro-radiosurgery treatment planning.

In FDG-PET studies, target lesions must always be specified by an operator, because healthy structures (such as brain, heart, bladder, and kidneys) normally have high radiotracer uptake \cite{stefano2017}.
We do not claim that our approach is able to segment other types of imaging data.
However, there are no general approaches that can work on all acquisition modalities and in all the clinical contexts without any limitations.
As a matter of fact, multimodal PET/CT and PET/MRI segmentation techniques in the literature are quite different and have to be designed \textit{ad hoc}.
These methods have to be adapted considering the different imaging acquisition modalities in certain clinical scenarios, as well as the involved body districts and pathologies.
As concerning state-of-the-art methods on PET/MRI segmentation, we would like to underline that methods based on unifying graphs, which assume no significant anatomic and functional changes between the images acquired using different modalities to directly construct the hyper-graph \cite{han2011,bagci2013MedIA} or the subgraphs with inter-subgraph arcs \cite{song2013}, generally yield a single target volume on the fused multimodal images.
Considering a decision-level fusion (i.e., taking into account the physicians' final decision for CTV definition), rather than a feature-level fusion (i.e., processing and fusing the input imaging data into a unifying data-structure) \cite{conti2010,serra2018}, keeps all the initial information even when PET and MR images are considerably different.

This study showed that the BTV should be used to modify the GTV, by properly considering both metabolic and morphologic information.
By using our multimodal methodology, clinicians could be assisted in CTV delineation during stereotactic radiosurgery treatment planning.
Therefore, the proposed multimodal segmentation approach can be considered clinically feasible, since it can be integrated in the current clinical practice.
After the positive opinion provided by clinicians, we have the rationale to begin clinical trials to evaluate tumor response after Leksell Gamma Knife treatment execution, even when BTV is included in CTV delineation during the planning phase.

\chapter{Computational Intelligence}

\graphicspath{{Chapter5/Figs/}}
\label{chap5}

\section{Genetic Algorithms}
\label{sec:geneticAlg}

Genetic Algorithms (GAs) represent an Evolutionary Computation technique for global optimization tasks \cite{eiben2015}.
Taking inspiration from Darwin's theory of biological evolution, GAs search optimal solutions to complex problems by evolving a population $P$ of randomly created candidate solutions \cite{holland1992}.
In the most general formulation, each solution represents an ``individual'' encoded as a fixed length string of characters taken from a finite alphabet (i.e., the genes); in alternative versions, the individuals codify real-values genes \cite{ali2018}.
It is also possible to leverage GAs in other contexts by using more aggregate objects for the individuals, such as image histogram bins (see for instance \cite{hashemi2010}).

In GAs, the population of candidate solutions evolves in competition under controlled variation.
Relying on a quality measure (i.e., the fitness value), each individual competes against the other individuals according to a \emph{selection} process.
The individuals characterized by a better fitness value will be selected with higher probability, using one of the available selection strategies, e.g., \emph{roulette wheel selection} \cite{goldberg1989}, \emph{ranking selection} \cite{baker1985}, and \emph{tournament selection} \cite{miller1995}.
Starting from these individuals, new ones are generated applying both \emph{crossover} and \emph{mutation} genetic operations.
These operators allow us to mix the parent characteristics and introduce new genetic material in the population, respectively.
Taking advantage of the information accumulation about an unknown search space, GAs are able to converge to optimal solutions, adapting the individuals during the iterations through exploration and exploitation of promising subspaces.
Assuming a sufficiently large population, the \emph{schema theorem} proves that GAs asymptotically converge to the optimal solution of any problem under investigation \cite{holland1992}.

Given an optimization problem, a GA is initialized with a randomly generated population of individuals corresponding to potential solutions.
During each \emph{generation} (i.e., iteration) the quality of each individual is evaluated using the fitness function. 
Then, individuals undergo selection, crossover, and mutation to create the next population (see pseudo-code in Algorithm \ref{GAPseudo}).

\begin{algorithm}
\caption{General pseudo-code for GAs}\label{GAPseudo}
{\small \textbf{Input:} the number of individuals $|P|$\\}
{\small \textbf{Output:} the individual $C_\text{best}$ characterized by the best fitness value}
\begin{algorithmic}[1]
  \State $t \gets 0$
  \State $P^{(0)} \sim \Call{RandomInitialization}{|P|}$
  \While{$t < T_{\text{max}}$}
    \State $ \Call{EvaluateFitness}{P^{(t)}}$
    \State $P'$ $\gets  \{\}$
    \For{$i$ in $1,2, \dots, \left\lfloor \frac{|P|}{2} \right\rfloor$}
      \State $\{$Parent$_1$, Parent$_2 \} \gets \Call{SelectionOperator}{P^{(t)}}$
      \State $\eta \sim $ Uniform(0,1)
      \If{$\eta < p_c$}
		\State $\{$offspring$_1$, offspring$_2 \}\gets \Call{CrossoverOperator}{\text{Parent}_1, \text{Parent}_2}$
	\Else
		\State $\{$offspring$_1$, offspring$_2 \}$ $\gets$ \{Parent$_1$, Parent$_2$\}
      \EndIf
	\For{\textbf{each} gene $\in$ offspring$_1$}
	\State $\eta \sim $ Uniform(0,1)
	\If{$\eta < p_m$}
	  \State $\Call{MutationOperator}{\text{gene}}$
	\EndIf
	\EndFor
	\For{\textbf{each} gene $\in$ offspring$_2$}
	\State $\eta \sim $ Uniform(0,1)
	\If{$\eta < p_m$}
	  \State $\Call{MutationOperator}{\text{gene}}$
	\EndIf
	\EndFor
	\State $P'$ $\gets$ $P'$ $\cup$ $\{$offspring$_1$, offspring$_2\}$
    \EndFor
    \State $P^{(t)} \gets$ $P'$
    \State $t \gets t + 1$
  \EndWhile
  
  \State $C_\text{best} \gets \Call{SelectBestSolution}{P^{(t)}}$
  
  \end{algorithmic}
\end{algorithm}

Different termination criteria can be used: for instance, the GA could stop when a given number of generations $T_{\text{max}}$ is reached, or when the fitness value of an individual is lower than a fixed threshold value (assuming a minimization problem).
In our approach, the algorithm stops when the number of generations $T_{\text{max}}$ is reached.

As mentioned above, several \textit{selection} mechanisms exist, such as \textit{roulette wheel selection} \cite{goldberg1989}, \textit{ranking selection} \cite{baker1985}, or \textit{tournament selection} \cite{miller1995}.
Given a population $P$, consisting in $|P|$ individuals, all selection techniques have the same goal, that is, obtaining an intermediate population $P'$ composed of copies of individuals from the previous population $P$.
Each individual can be added more than one time at $P'$, depending on its fitness value (the better the fitness, the higher the chance to be selected).

Once $P'$ has been created by means of a selection procedure, the variation operators (i.e., crossover and mutation) are applied to obtain new individuals, named offspring, for the next generation.
The \textit{crossover} operator is used to recombine the genetic information from two parent individuals (Parent$_1$ and Parent$_2$).
Such operator, which plays a key role in GAs, allows for obtaining offspring solutions with better characteristics with respect to the parents.
Crossover is applied to several selected pairs of individuals with a probability value $p_c$, named crossover rate.
There are different versions of crossover, such as \textit{single-point crossover}, \textit{two-point crossover}, and \textit{uniform crossover}.

In order to arbitrarily alter one or more genes of a selected offspring, we can apply the \textit{mutation} operator, so increasing the variability inside $P'$.
This simple operator, introducing new genetic material into the population, is able to prevent the convergence of the individuals to local optimal solutions, increasing the probability to reach every
point in the search space.
The mutation operator is applied with a probability $p_m$ (i.e., mutation rate), such that some randomly selected genes of each individual are selected and their values are perturbed.
Generally, the crossover is executed with high probability values (e.g., $p_c = 0.95$), conversely the mutation is applied with very low rates (e.g., $p_m = 0.05$).
These parameter settings aim at reflecting the natural evolution process in which the characteristics (genes) of two parents are deeply blended, while mutations has a very low probability to occur.

Finally, the \textit{elitism} strategy could also be exploited, by copying the best individual (or a subset of the best individuals) directly into the next generation without modifying it (them).
This mechanism prevents the quality of the best solution from decreasing during the iterations, since the best individual is copied into the next population without undergoing the genetic operators.

\subsection{Medical image enhancement}

Medical imaging plays a key role in the clinical workflow, thanks to its capability of representing anatomical and physiological features \cite{lambin2017,rueckert2016}, by means of several different principles to measure spatial distributions of physical attributes of the human body, allowing for a better understanding of complex or rare diseases \cite{toennies2012}.
The effectiveness of such techniques can be reduced by a plethora of phenomena, such as noise and PVE \cite{toennies2012}, which might affect the measurement processes involved in imaging and data acquisition devices.
In addition, computer-aided medical image acquisition procedures generally include reconstruction methods (producing two, three or even four dimensional imaging data), which could cause the appearance of artifacts.
Image contrast and details might also be impaired by the procedures used in medical imaging, as well as by the physiological nature of the body part under investigation.

Medical images actually convey an amount of information---mainly related to high image resolution and high pixel depth---that could overwhelm the human vision capabilities in distinguishing among dozens of gray levels \cite{ortiz2013}.
Improvements in the appearance and visual quality of medical images are therefore essential to allow physicians attaining valuable information that would not be immediately observable in the original image, thus assisting them in anomaly detection, diagnosis, and treatment.
This kind of diagnosis includes two basic processes: image observation (visual perception), and diagnostic interpretation (cognition) \cite{krupinski2010}.
Errors occurring in these diagnostic and therapeutic decision-making processes may have a significant impact on patient care, by causing possible misdiagnoses.
In this context, image enhancement techniques aim at realizing a specific improvement in the quality of a given medical image.
The enhanced image is expected to better reveal certain features, compared to their original appearance \cite{deAraujo2014}.
In particular, these methods could have a significant clinical impact in the case of not commensurable dynamic range of the actual pictorial data with respect to the displayed ones (i.e., monitor luminance response), as well as when the input image is characterized either by a high level of noise or by low contrast \cite{gonzalez2002,paranjape2009}.
This also applies to specialized computer screens for diagnostic reporting.

Although the majority of the enhancement techniques are typically applied to generate improved images for a human observer, others are exploited as a pre-processing step to provide enhanced images to further algorithms for computer-assisted analyses \cite{paranjape2009}.
The first category includes techniques devoted to remove noise, enhance contrast, and sharpen the details.
The second category, partially overlapped with the former one, includes additional techniques such as edge detection and object segmentation for automated processing \cite{rangayyan2009}.
It was shown that a high-contrast medical image could lead to a better interpretation of the different adjacent tissues in the imaged body part \cite{chen2015}.
Accordingly, the resulting enhanced image---in terms of signal intensities of different tissues---can facilitate the automated segmentation, feature extraction, and classification of these tissues.

In the clinical routine, CE-MRI is a diagnostic technique that enables a more precise assessment of the imaged tissues after the administration of a Gadolinium-based contrast medium in patients \cite{sourbron2013}.
MRI is currently the most prominent modality to obtain soft-tissue imaging \cite{brown2014}, especially in oncology, since it provides significant improvements---in terms of image contrast and resolution---between lesion and healthy tissue \cite{metcalfe2013}.
However, MRI data are affected by acquisition noise \cite{styner2000} and are also prone to imaging artifacts, related to magnetic susceptibility and large intensity inhomogeneities of the static magnetic field (i.e., streaking or shadowing artifacts \cite{bellon1986}), especially using high magnetic field strengths.
These aspects make MR image enhancement a challenging task devoted to improve the outcome of automatic segmentation.

The existing image enhancement approaches generally attempt to improve the contrast level of the whole image and do not address the issues related to overlapped gray level intensities; as a consequence, neither the region contour sharpness nor the image thresholding results can be improved.
In the case of threshold-based image segmentation with two classes (i.e., foreground and background) \cite{muangkote2017}, the input image is assumed to have a bimodal histogram \cite{xue2012}.
Thus, an appropriate image enhancement method that yields medical images with a stronger bimodal distribution is required.
However, determining the best pre-processing of an image---able to preserve the structural information of the image while enhancing the underlying bimodal distribution of the histogram bins---is a complex task on a multi-modal fitness landscape that mandates the use of global optimization approaches.

\subsubsection{Related work}

Most of the existing enhancement techniques are empirical or heuristic methods---strongly related to particular types of images---which generally aim at improving the contrast level of images degraded during the acquisition process \cite{chen2017}.
As a matter of fact, finding the best gray level mapping that adaptively enhances an input image for each different sample can be considered an optimization problem \cite{draa2014,paulinas2007}.
Unfortunately, no unifying theory employing a standardized image quality measure is currently available to define a general criterion for image enhancement \cite{munteanu2004}.
In addition, in the case of medical imaging, techniques tailored on specific tasks are necessary to achieve a significant enhancement and, in general, interactive procedures involving considerable human effort are needed to obtain satisfactory results.

In order to achieve objective and reproducible measurements conveying clinically useful information, operator-dependence should be minimized by means of automated methods.
Point-wise operations in the spatial (pixel) domain, representing the simplest form of image processing, are effective solutions since efficiency requirements have also to be met.
In the case of image enhancement, they re-map each input gray level into a certain output gray level, according to a global transformation \cite{gonzalez2002}.
Thus, such kind of techniques treat images as a whole, without considering specific features of different regions or selectively distinguishing between a collection of contrast enhancement degrees or settings \cite{munteanu2004}.
Histogram Equalization (HE) is the most common image enhancement technique \cite{gonzalez2002}, but it is not suitable for medical images due to the obtained over-brightness \cite{gandhamal2017}.
Indeed, this method uniformly spreads the input gray level values according to the cumulative density function of the image histogram.
However, HE does not preserve the input mean brightness, possibly suffering from over-enhancement and giving rise to artifacts such as washed-out effect \cite{chen2003}.
This global transformation applies contrast stretching just on gray levels with the highest frequencies, causing a significant contrast loss for gray levels characterized by lower occurrences in the input histogram \cite{kim1997}.
In order to address the issues related to input mean brightness preservation, a modification of the standard HE technique, called Bi-Histogram Equalization (Bi-HE), was introduced \cite{kim1997}.
Bi-HE attempts to improve the results achieved by HE, by first splitting the original histogram into two components according to the global mean of the original image, and then separately performing the HE method on the two sub-histograms.

Other traditional global gray level transformations generally used for contrast stretching are formalized as transformation functions of the form $s = \mathcal{T}(r)$, where $\mathcal{T}(\cdot)$ maps an input intensity value $r$ into an output intensity value $s$ \cite{gonzalez2002}.
Power-law transformation---also called Gamma Transformation (GT)---is a non-linear operation of the form $\mathcal{T}(r) = c r^\gamma$, where typically $c=1$.
For instance, when the image is predominantly dark, an expansion of the intensity levels is desirable.
In such a case, GT with $\gamma < 1$ yields a brighter image by increasing the number of hyper-intense pixels.
On the contrary, by using $\gamma > 1$, the GT converts the input gray-scale range into a darker one, by increasing the occurrences of darker pixels.
Obviously, the value of $\gamma$ strongly depends on the medical application.
Accordingly, logarithmic and anti-logarithmic transformations make an image much brighter and darker, respectively.
Unfortunately, for medical images characterized by low contrast and weak edges at adjacent tissue boundaries, GT may result in merely brighter or darker images, leading to difficulties in the visualization and interpretation of different tissues.
Therefore, to adequately enhance contrast, the two different behaviors---corresponding to values $\gamma > 1$ and $\gamma < 1$---should be combined for contextually decreasing the darker pixel gray values and increasing the brighter pixel gray values.
This results in a significant improvement of the contrast, by enhancing the edges thanks to the increased gradient magnitude of the image \cite{gandhamal2017}.
This kind of contrast stretching can be achieved by using a Sigmoid intensity Transformation (ST), which darkens a wide range of hypo-intense gray levels and brightens a wide range of hyper-intense gray levels \cite{gonzalez2002}.
Such an operation indirectly increases the difference between low and high intensity values, resulting in the overall contrast enhancement of the image \cite{gandhamal2017}.

In addition to HE, which automatically yields an image with a uniform histogram, it is possible also to explicitly specify the desired shape of the output histogram.
This method, named Histogram Specification (HS), aims at matching the histogram of the gray level intensities of the input image against a desired histogram \cite{gonzalez2002}.
Unfortunately, such approaches cannot be applied in the case of image datasets with heterogeneous gray level distributions, since the histogram to be matched should be defined either \textit{a priori} for all the images in the dataset or interactively for each processed image, by separating and shaping the two underlying sub-distributions \cite{xiao2018}.
It is therefore necessary to employ global search metaheuristics to automatically identify the best solution case-by-case.

The complexity of the enhancement criteria to be met (i.e., the effective contrast stretching combined with image detail preserving) leads to the application of global search metaheuristics that allow for coping with several constraints, which are not generally tractable by means of traditional exhaustive computational approaches \cite{munteanu2004,ortiz2013,paulinas2007}.
Evolutionary methods have been widely adopted in the image enhancement domain to find the optimal enhancement kernel \cite{munteanu2004}, sequence of filters \cite{kohmura2006}, or input-output mapping transformation \cite{carbonaro1999,saitoh1999}.
Recently, Hashemi \textit{et al.} \cite{hashemi2010} proposed a GA that efficiently encodes the histogram by means of the non-zero intensity levels, by employing genetic operators that directly process images to increase the visible details and contrast of low illumination regions, especially in the case of high dynamic ranges.
The authors argued that this method yields ``natural-looking'' images, considering the visual appearance.

In this context, Genetic Programming (GP) \cite{koza1992} was shown to be a powerful framework to select and combine existing algorithms in the most suitable way.
Differently to GAs, GP evolves a population of functions, or more generally, computer programs to solve a problem.
The solutions in the computer program space can be represented as trees, lines of code, expressions in prefix or postfix notations as well as strings of variable length \cite{castelli2014}.
For instance, the authors of \cite{bianco2017} tackled the video change detection problem (among the frames of video streams) by combining existing algorithms \textit{via} different GP solutions exploiting several fusion schemes.
The fitness function was composed of different performance measures regarding change detection evaluation.
For what concerns the application of GP in image enhancement, Poli and Cagnoni \cite{poli1997} proposed an approach to yield optimally pseudo-colored images for visualization purposes, aiming at combining multiple gray-scale images (e.g., time-varying images, multi-modal medical images and multi-band satellite images) into a single pseudo-color image.
This approach relies on user interactions to determine which individual should be the winner in tournament selection, so it does not explicitly require a fitness function.
As case studies, a pair of brain MRI sequences were fused as well as the motion of the heart on echocardiographic images was synthesized into a single pseudo-color image.
GP-based semi-supervised learning was shown to be successful even in the case of noisy labeled data in classification tasks \cite{silva2018}.
In the medical domain, GP with geometrics semantic operators \cite{vanneschi2014} was employed to predict the relative position of a slice within a CT image stack when DICOM metadata are unreliable \cite{castelli2016}.

Other works exploited Swarm Intelligence techniques.
The approach presented in \cite{shanmugavadivu2014}, called Multi-Objective HE, uses Particle Swarm Optimization (PSO) \cite{kennedy1995} to enhance the contrast and preserve the brightness at the same time.
Draa and Bouaziz \cite{draa2014} employed the same encoding of individuals and histogram mapping strategy described in \cite{hashemi2010}, within an optimization strategy based on the Artificial Bee Colony (ABC) algorithm \cite{karaboga2007}.
However, since ABC natively works in a continuous space, while a discrete representation is used for the solutions (i.e., gray-level mapping), a discretization step is mandatory in the correction operation during the search phase.
An alternative approach using the ABC algorithm for image contrast enhancement has been proposed in \cite{chen2017}, wherein the optimal values for the parameters of a parametrics image transformation, namely the Incomplete Beta Function, are estimated.
Differently to the work described in \cite{draa2014}, the optimization procedure is carried out in a continuous search space.
Finally, multi objective Bat Optimization and a neuron-based model of Dynamic Stochastic Resonance were combined in \cite{singh2017} for the enhancement of brain MR images.

Image pre-processing methods often include operations at the lowest level of abstraction---wherein both input and output are gray-scale images---employing intensity transformations.
These global transformations just deal with the gray levels, without considering the pixel position, and the corresponding relationship with its neighborhood, in the image.
The main objective of pre-processing is to improve the raw imaging data, dealing with undesired artifacts or enhancing some image features important for further processing \cite{sonka2007,suzuki2003}.

Conventional pre-processing techniques generally expand a narrow input range of gray levels into a wider range of output levels, in order to improve the image detection performance.
Pre-processing tackles the problems related to image degradation correction, whether no assumption about the nature of noise/artifacts can be made as well as when neither \textit{a priori} knowledge about the acquisition device nor the objects of interest are available \cite{sonka2007}.
Thus, Soft Computing methods could be beneficially exploited---by considering, for instance, metaheuristics \cite{suresh2017} or Fuzzy Logic \cite{chaira2014}---to find the best solution in a high variety of scenarios.
In addition, interactive procedures and trial-and-error parameter tuning are generally required to obtain satisfactory results \cite{munteanu2004}.
GAs offer a very powerful framework for image enhancement since they implement an unbiased global optimization method for huge search spaces \cite{paulinas2007}.
Considering the Computational Intelligence metaheuristics, GAs are the most suitable framework because of the discrete encoding of the candidate solutions and the combinatorial structure of the search space \cite{tangherloni2018GenHap}.
More specifically, here we exploited GAs to efficiently encode the individuals by means of their corresponding image histograms.

Biomedical images usually require methods tailored on a specific task \cite{shamir2010}.
Well-designed preliminary image pre-processing techniques can provide benefits for a better ROI feature detection to be exploited in a downstream image processing pipeline.
Especially, these pre-processed data can be fed as a suitable input for an image segmentation algorithm to improve its accuracy.

Na\"ive smoothing techniques, by using both linear (e.g., average or Gaussian filters) and non-linear (e.g., median filter) operators, may improve the shape of the histogram \cite{rogowska2000}, also also by widening the separation between the two modes in bimodal gray-level histograms \cite{sahoo1988}.
These approaches are generally exploited for noise reduction, even though sharp boundaries might be replaced with a fuzzy region of varying shades of gray.
Thus, such a kind of low-pass filtering could affect the actual pictorial content, compromising the clinical effectiveness of the subsequent ROI detection and delineation phases.

In medical imaging, pre-processing steps are valuable for further computer-assisted image analysis, by making anatomical or functional structures more easily detectable.
Therefore, pre-processing involving image enhancement operations may be needed to achieve more identifiable and sharpened boundaries for medical image segmentation \cite{toennies2012}.
A better distinction among adjacent tissues in medical images can be achieved, by means of an adequate contrast performing gray-scale transformations.

It is worth noting that all works above mentioned are focused on consumer electronics or medical applications, to obtain more ``visually pleasant'' images by mainly increasing the contrast of the whole image.
As a matterof fact, the visual appearance is a fundamental feature in quality assessment and recognition tasks, such as in the case of food products \cite{ciocca2017,pereira2008}.
On the contrary, the main key novelty of MedGA consists in better revealing the two underlying sub-distributions occurring in an image sub-region characterized by a roughly bimodal histogram, overcoming the limitations of the state-of-the-art contrast enhancement methods, which could produce false edges and consequently over-segmentation when the input images are affected by noise, as in the case of MRI data \cite{gandhamal2017}.
There exist other algorithms, like HS, whose aim is similar to MedGA and consists in matching the histogram of the gray level intensities of the input MR image with a desired output histogram \cite{gonzalez2002}.
Unfortunately, such approaches cannot be applied in the case of image datasets with high variability in gray level distributions, since the histogram to be matched should be defined either \emph{a priori} for the whole dataset or interactively for each processed image, by strengthening and shaping the two underlying sub-distributions.
It is therefore necessary to employ global search metaheuristics to automatically identify the best solution, as in the case of MedGA.

Finally, supervised---such as Artificial Neural Networks (ANNs) \cite{shi2017} and SVMs \cite{zhou2002}---as well as unsupervised learning approaches---such as Self-Organizing Maps (SOMs) \cite{ortiz2013}---require a training phase and are generally exploited in image classification and recognition tasks.

\subsubsection{Clinical MRI datasets}
MedGA was applied as a pre-processing step in two segmentation tasks concerning challenging clinical MRI dataset from patients affected by: (\textit{i}) uterine fibroids; (\textit{ii})  brain metastatic cancers.
All the analyzed MRI data are encoded in the $16$-bit DICOM format.
The MRI acquisition characteristics are reported in Table \ref{table:MRIcharacteristics}.

\begin{figure}[!ht]
	\centering
	\subfloat[]{\includegraphics[width=0.45\textwidth]{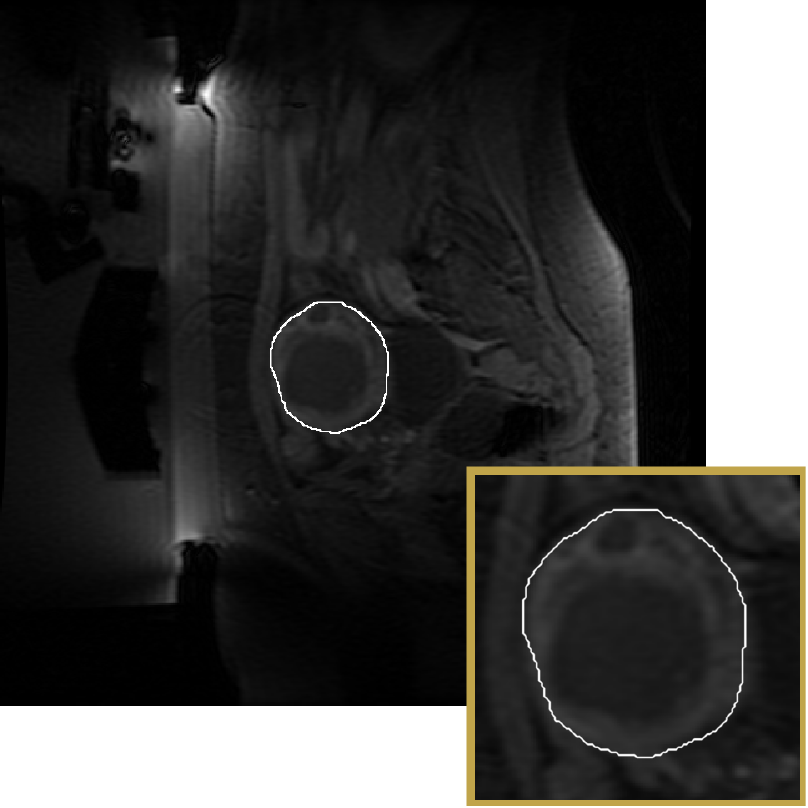}\label{InputFibA}} \quad
	\subfloat[]{\includegraphics[width=0.45\textwidth]{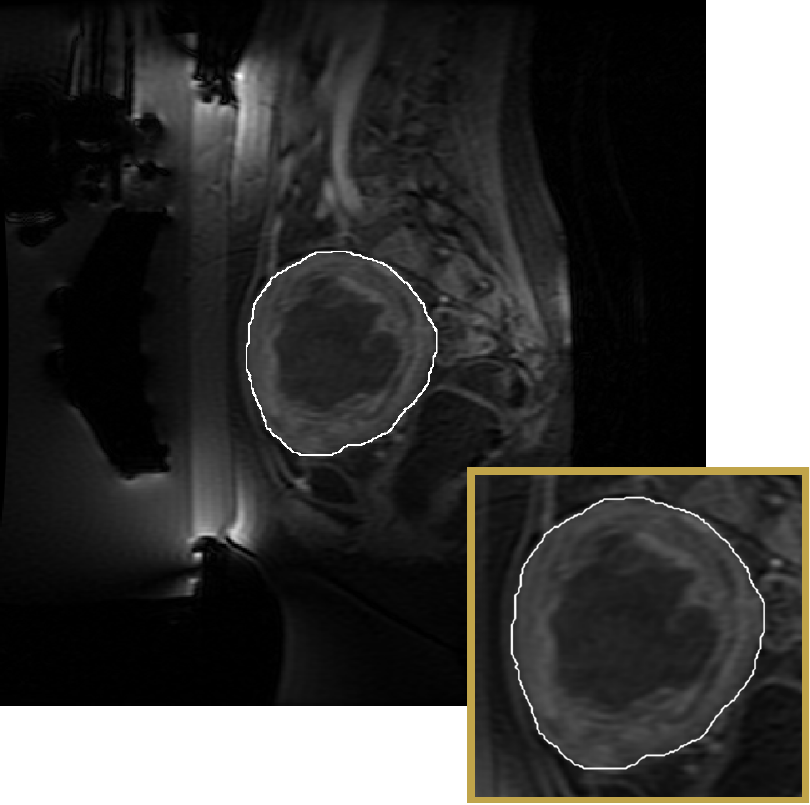}\label{InputFibB}} \\
	\subfloat[]{\includegraphics[width=0.45\textwidth]{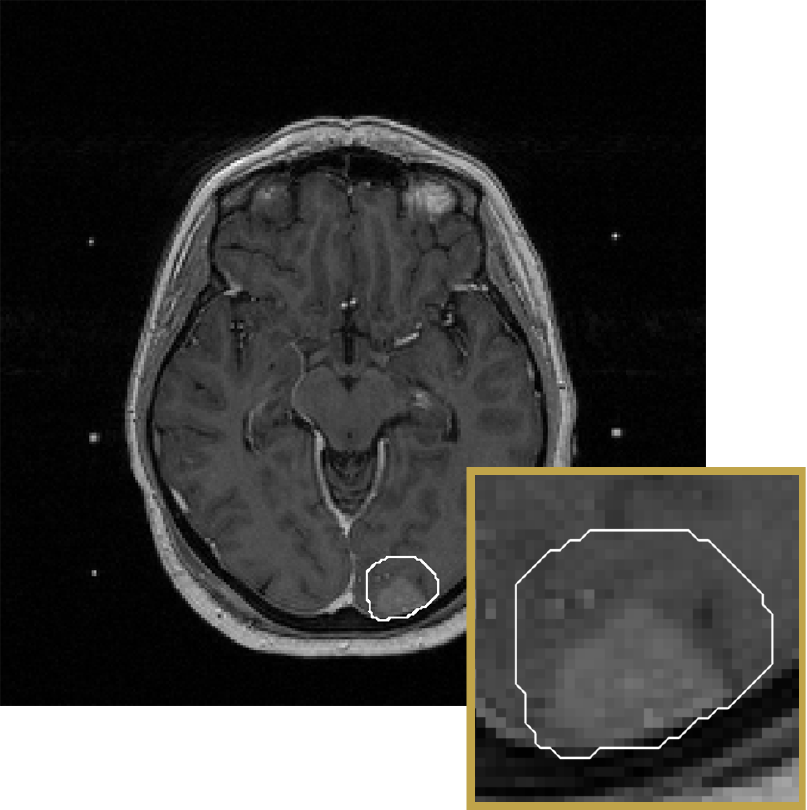}\label{InputBrainA}} \quad
	\subfloat[]{\includegraphics[width=0.45\textwidth]{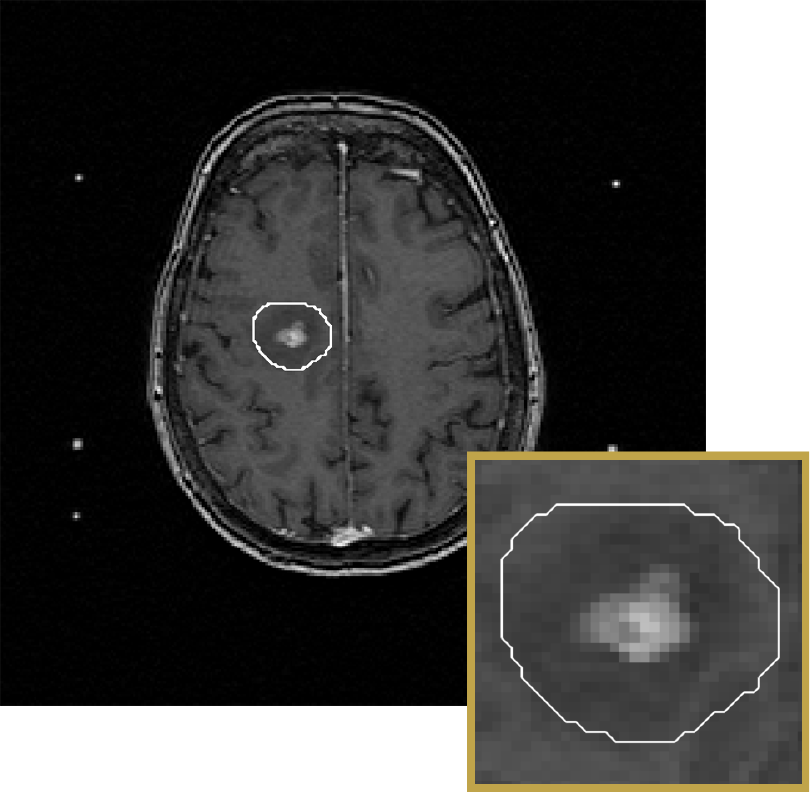}\label{InputBrainB}} \\
	\caption[Examples of input MR images characterized by nearly bimodal histograms]{Examples of input MR images: (a, b) uterine fibroid inside the uterus region; (c, d)  brain tumor inside a ROI bounding region selected by the healthcare operator.
	The image regions including the ROIs, defined by the white contour and zoomed at the bottom right of each sub-figure, are characterized by nearly bimodal histograms.}
	\label{inputImages}	
\end{figure}

\begin{table}[!h]
	\caption[MRI acquisition parameters of uterine fibroids and brain metastatic tumor dataset]{MRI acquisition parameters of uterine fibroids and brain metastatic tumor dataset.}
	\label{table:MRIcharacteristics}
	\begin{tiny}
		\centering
		\begin{tabular}{c|ccccccc}
			\hline\hline
			Dataset		& MRI sequence	& TR [ms]	& TE [ms]	& Matrix size [pixels]	& Slice spacing	[mm]	& Slice thickness [mm]	& Pixel spacing [mm] \\
			\hline
			Uterine fibroids	& T1w FSPGR+FS+C	& $150$-$260$	& $1.392$-$1.544$		& $512 \times 512$ 		& $5.0$ 	& $6.0$	  	&  $0.6641$-$0.7031$ \\
			Brain metastases	& T1w FFE			& $25$		& $1.808$-$3.688$ 		& $256 \times 256$ 	& $1.5$ 	& $1.5$ 	& $1.0$ \\
			\hline\hline
		\end{tabular}
	\end{tiny}
\end{table}

\paragraph{Uterine fibroids}
\label{sec:uterineFibroids}
Eighteen patients affected by symptomatic uterine fibroids who underwent MRgFUS therapy \cite{roberts2008} were considered.
The total number of the examined fibroids was $29$, overall represented on $163$ MR slices, since some patients presented a pathological scenario with multiple fibroids.
The analyzed images were acquired using a Signa HDxt $1.5$ T MRI scanner at two different institutions.
These MRI series were acquired after the MRgFUS treatment, executed with the ExAblate 2100 (Insightec Ltd., Carmel, Israel) HIFU equipment.
The considered MR slices were scanned using the T1w ``Fast Spoiled Gradient Echo + Fat Suppression + Contrast mean'' (FSPGR+FS+C) sequence.
This MRI protocol is usually employed for NPV assessment, since ablated fibroids appear as hypo-intense areas due to low perfusion of the contrast mean \cite{militelloTCR2014}.
Sagittal MRI sections were processed, in compliance with the current clinical routine for therapy response assessment \cite{militelloTCR2014}.
In current clinical practice, the NPV evaluation procedure is fully manual \cite{rundoMBEC2016}.
Two uterine fibroid MR slices are depicted in Figs. \ref{InputFibA} and \ref{InputFibB}.

\paragraph{Brain metastatic tumors}
\label{sec:brainTumors}
Twenty-seven brain metastases treated using a Leksell Gamma Knife stereotactic neuro-radiosurgical device \cite{leksell1951} were processed, for a total of $248$ MR slices.
All the available MRI datasets were acquired on a Philips Gyroscan Intera $1.5$ T MRI scanner, before treatment, for the planning phase.
In current radiation therapy practice, Gamma Knife treatments are planned manually by a neurosurgeon on MRI alone, by typically using T1w FFE CE-MRI sequences \cite{rundo2017NC,rundoCMPB2017}.
Thanks to the Gadolinium-based contrast agent, brain lesions appear as enhanced hyper-intense zones.
Sometimes a dark area might be present due to either edema or necrotic tissues \cite{rundo2017NC,rundo2018next}.
Two representative instances of brain tumors are shown in Figs. \ref{InputBrainA} and \ref{InputBrainB}.

\subsubsection{State-of-the-art image enhancement methods}

\begin{figure}[t]
	\centering
	\subfloat[]{\includegraphics[width=0.4\textwidth]{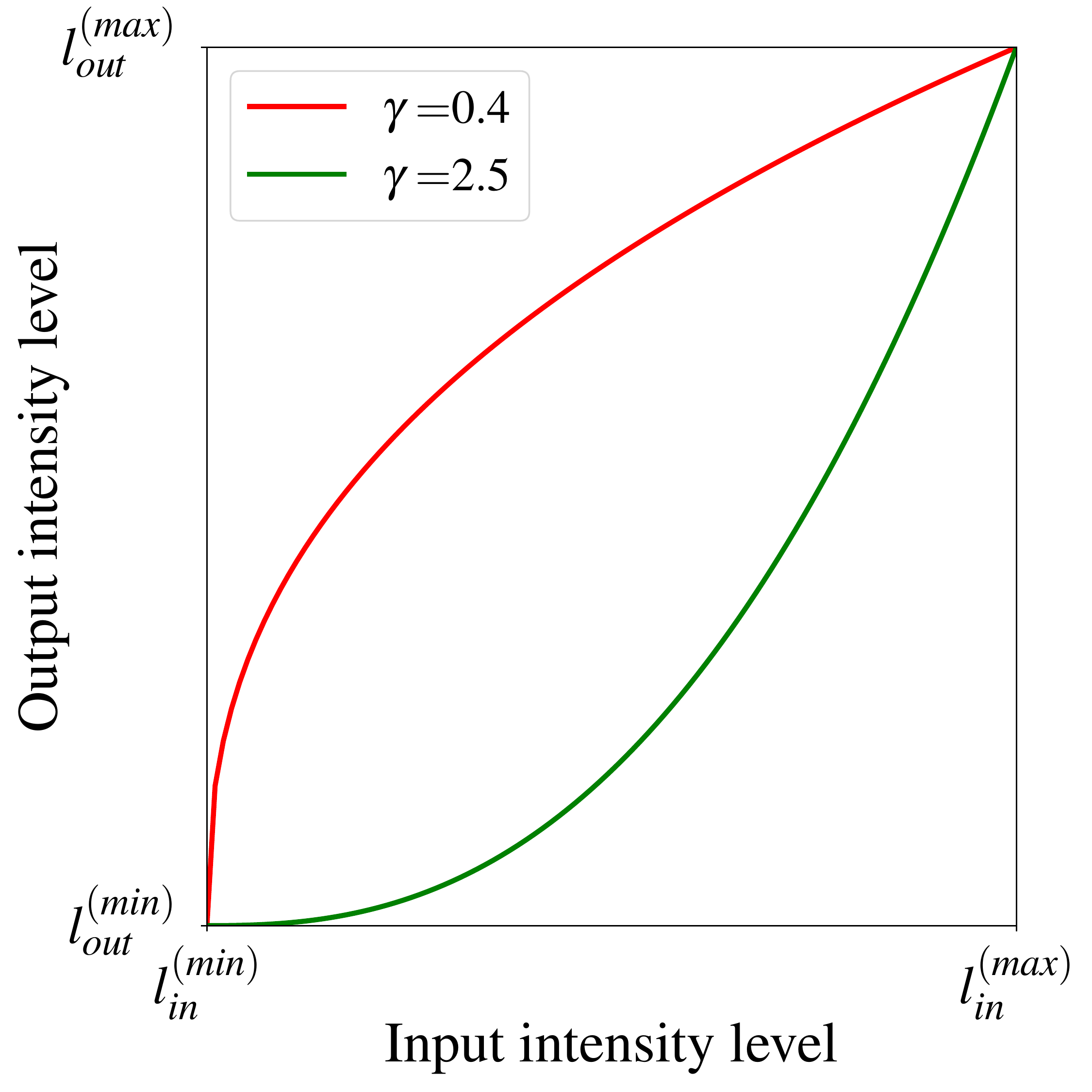}\label{gammaFunctions}}\quad
	\subfloat[]{\includegraphics[width=0.4\textwidth]{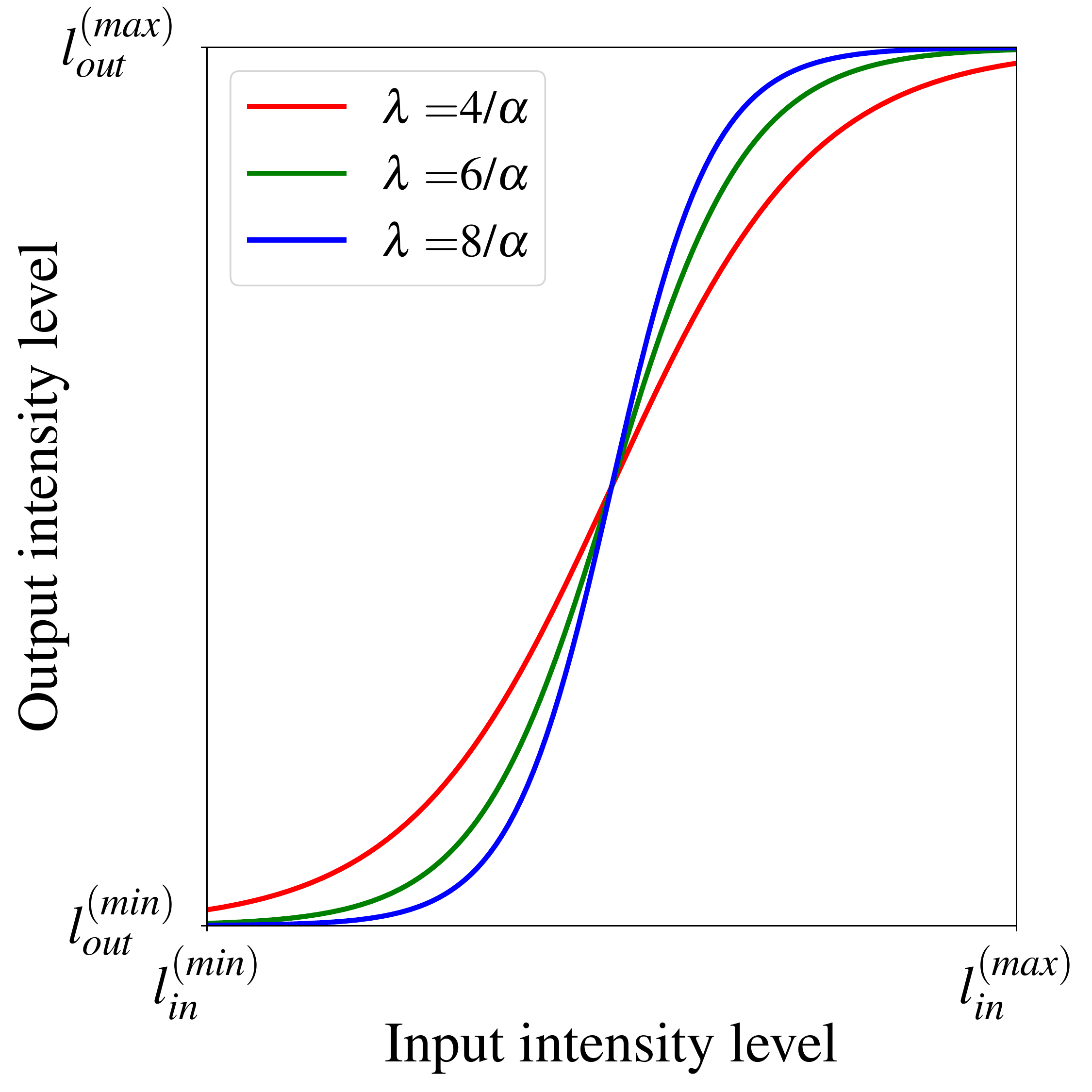}\label{sigmoidFunctions}}\\
	\caption[Plots of the global non-linear intensity transformations for image enhancement: (a) Gamma Transformation; (b) Sigmoid intensity Transformation]{Plots of the implemented global non-linear intensity transformations for image enhancement: (a) Gamma Transformation; (b) Sigmoid intensity Transformation.
	We report on the $x$-axis the input intensity range $[l_{in}^{(min)}, \dots, l_{in}^{(max)}]$, and on the $y$-axis the output intensity range $[l_{out}^{(min)}, \dots, l_{out}^{(max)}]$.}
	\label{nonLinearFunctions}
\end{figure}

To achieve a fair comparison, we considered the most common computational techniques for image enhancement and pre-processing, namely:
\begin{itemize}
	\item HE \cite{gonzalez2002}, which adjusts pixel intensities for contrast enhancement according to the normalized histogram of the original image $\mathcall{I}_\text{orig}$.
	With HE, gray levels are more uniformly distributed on the histogram, by spreading the most frequent intensity values;
	\item Bi-HE \cite{kim1997}---a modification of the traditional HE---which addresses issues concerning mean brightness preservation;
	\item GT, which is a non-linear operation using the power-law relationship $s(r) = c r^\gamma$, where $r$ and $s$ are the input and the output gray-scale values, respectively, and $c$ is a multiplication constant ($c=1$ in the following tests).
	The parameter $\gamma$ is set to values greater than $1$ (i.e., decoding gamma) to obtain a gamma expansion, or to values smaller than $1$ (i.e., encoding gamma) to realize a gamma compression (see Fig. \ref{nonLinearFunctions}a).
	In our tests we considered the values $\gamma = 0.4$ and $\gamma = 2.5$; higher (lower) values of $\gamma = 2.5$ ($\gamma = 0.4$) tend to logarithmic (anti-logarithmic) functions, resulting in an excessively bright (dark) output image, unsuitable for medical applications \cite{gandhamal2017};
	\item ST function (Fig. \ref{nonLinearFunctions}b), also called S-shaped curve, which is a global non-linear mapping defined as follows:
	\begin{equation}
	\label{Eq_sigmoid}
	s(r) = \frac{l_{in}^{\text{(max)}}}{1+\exp\left({-\lambda  (r - \alpha)}\right)},
	\end{equation}
	where $l_{in}^{\text{(max)}}=\max\{\mathcall{L}_{in}\}=\max\{\mathcall{L}'_{in}\}$ is the asymptotic maximum value of the function, $\alpha=\frac{1}{2} \left(l_{in}^{\text{(max)}}-l_{in}^{\text{(min)}}\right)$ is the midpoint value, and  $\lambda$ defines the function steepness.
	This transformation stretches the intensity around the level $\alpha$, by making the hypo-intense histogram part darker and the hyper-intense histogram part brighter.
	Thus, the difference between the minimum and maximum gray values as well as the gradient magnitude of the image are increased, obtaining strong edges \cite{gandhamal2017}.
	In our validation, we used sigmoid functions that allow to consider the entire input dynamic range, by varying the curve slope with the values $\lambda \in \left\{ \frac{4}{\alpha}, \frac{6}{\alpha}, \frac{8}{\alpha} \right\}$.
\end{itemize}

\subsubsection{MR image enhancement}
\label{sec:MedGAenhancement}
In order to address these issues, we proposed a novel image enhancement technique based on GAs \cite{holland1992}, called MedGA \cite{rundo2018MedGA1}, specifically aimed at strengthening the sub-distributions in medical images with an underlying bimodal histogram of the gray level intensities.
Among the existing soft computing methods for global optimization, GAs represent the most suitable technique because of the discrete structure of the candidate solutions and the intrinsic combinatorial structure of the problem under investigation.

Considering the possible clinical applications, MedGA improves the visual perception of a ROI in MRI data with an underlying bimodal intensity distribution \cite{rundo2018MedGA1}.
In addition, MedGA can be used as a pre-processing step, in a framework defined to realize an efficient threshold-based image segmentation with two classes (i.e., binarization), applied to MRI data \cite{rundo2018MedGA2}.
Image thresholding approaches performed on CE-MR image regions could considerably benefit from input data pre-processed by MedGA.

The main contributions of MedGA in the context of intelligent clinical systems can be briefly outlined as follows.
MedGA acts as an expert system by playing a two-fold role: (\textit{i}) image enhancement to visually assist physicians during their interactive decision-making tasks, and (\textit{ii}) improvement of the results in downstream automated processing pipelines for clinically useful measurements.
The motivations underlying MedGA rely on the design of an image enhancement technique for medical images with roughly bimodal histograms.
Indeed, our objective is to propose an intelligent model that is well-suited for effectively enhancing such a particular type of medical images.
To the best of our knowledge, MedGA is the first work that explicitly deals with the improvement of thresholding-based segmentation results \cite{xue2012}.
Therefore, our computational framework can be employed as an intelligent solution in CDSSs.
As a matter of fact, MedGA represents an interpretable computational model \cite{castelvecchi2016} that allows for the understandability of the results (i.e., the gray level histogram is readable by the user).
The compelling issues related to the interpretability of Machine Learning and Computational Intelligence methods in medicine \cite{criminisi2016} are fundamental for the adoption and the clinical feasibility of a novel CDSS \cite{cabitza2017}.
In addition, thanks to the reliable calibration step of MedGA, no user parameter setting is required.
We aim at showing how evolutionary methods can boost the state-of-the-art performance in medical image enhancement, thus fostering GAs as a new concrete support tool for the clinical practice.
Such an expert system may have a significant impact in real healthcare environments.

Even though MedGA exploits the same encoding of individuals defined in \cite{hashemi2010} and \cite{draa2014}, its purpose is very different as it is designed to explicitly strengthen the two sub-distributions of medical images characterized by an underlying bimodal histogram.
To this aim, we defined a specific fitness function that emphasizes the two Gaussian distributions composing a bimodal histogram.
This achievement plays a fundamental role for threshold-based segmentation approaches.
As a matter of fact, these approaches strongly rely on the assumption that the bimodal histogram under investigation is composed of two nearly Gaussian distributions with almost equal size and variance \cite{xue2012}.

Finally, MedGA differs from the GP-based approaches, in which the final generated solution might have large size \cite{castelli2014}, even when the GP model is implemented efficiently, representing a limitation that could significantly impair the readability and interpretability of the solutions.
Moreover, MedGA does not require any user interaction step, differently to \cite{poli1997} where the user, being directly involved in the tournament selection, controls the evolution of simple programs that enhance and integrate multiple gray-scale images into a single pseudo-color image.
Therefore, the selection is based on the output image (i.e., phenotype) rather than the structure and size of the program (i.e., genotype) \cite{poli1997}.

\begin{figure}[ht!]
	\centering
	\includegraphics[width=0.35\textwidth]{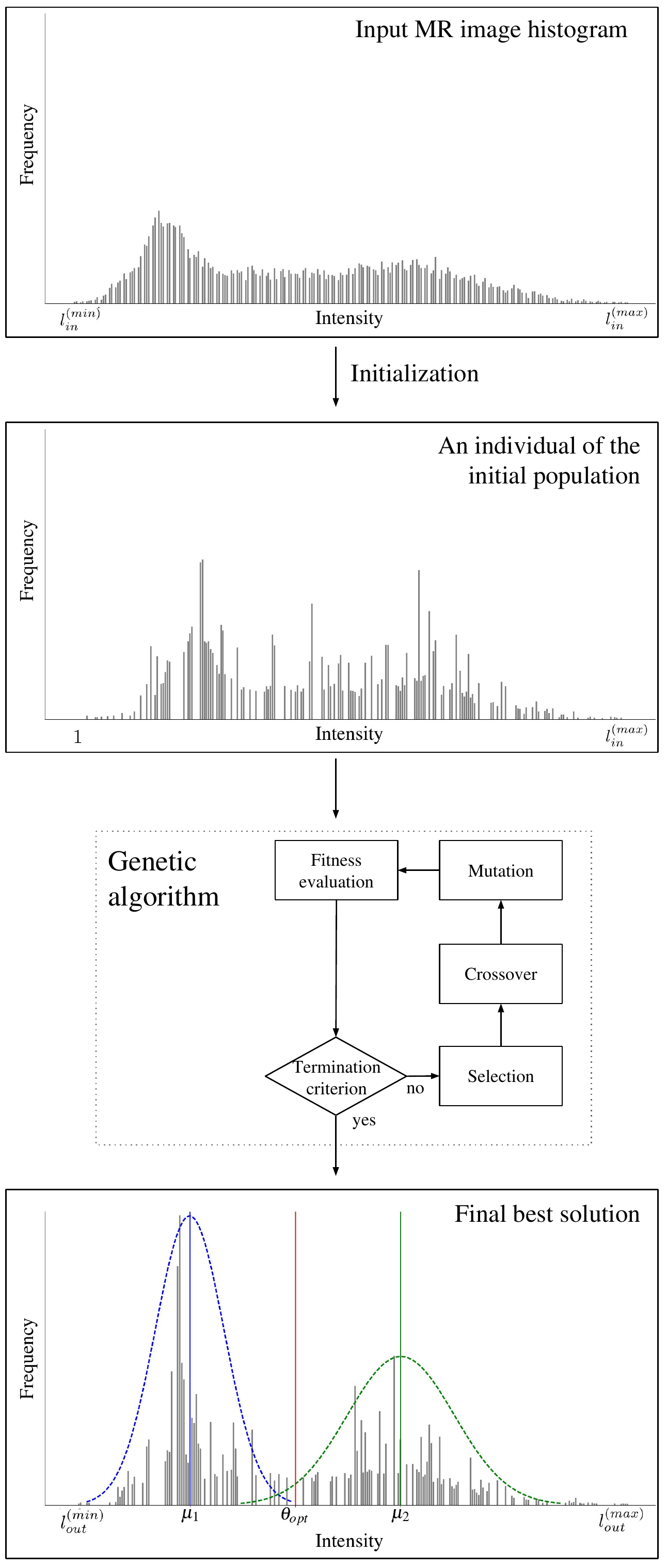}
	\caption[Workflow of MedGA]{Workflow of MedGA: the individuals are initialized according to the characteristics of the input MR image, and processed by the GA. The final best solution of MedGA strengthens the two underlying distributions in the gray levels intensity characterized by mean values $\mu_1$ and $\mu_2$, and standard deviations $\sigma_1$ and $\sigma_2$, respectively. The two distributions are highlighted in the plot with blue and green dashed lines.
	}
	\label{fullScheme}	
\end{figure}

\paragraph{The proposed GA-based image enhancement approach}
MedGA requires an input image $\mathcall{I}_\text{orig}$, with  $M$ rows and $N$ columns, depicting a ROI included in a bounding region whose pixel values are approximately characterized by an underlying bimodal histogram.
Therefore, as a first step, either a computational method or a user must detect a region including the ROI.
Then, the input MR image is masked with this bounding region and the whole image is cropped according to the smallest rectangle enclosing the bounding region.
Fig. \ref{fullScheme} outlines the overall procedure of MedGA, by presenting the initialization phase as well as the flow diagram of the proposed GA for image enhancement.
A linear normalization is applied on the initial full range of the masked MR image to balance out the pixel distribution for the following bin rearrangement by means of MedGA.
It is worth noting that no additional pre-processing operation (e.g., low-pass or high-pass filtering) is needed by MedGA.
The final best solution found by MedGA will emphasize the two underlying bimodal Gaussian distributions occurring in the bimodal MR image, for the subsequent image thresholding phase according to the optimal adaptive threshold $\theta_\text{opt}$ computed using the efficient IOTS algorithm \cite{ridler1978,trussell1979}.

Hereafter, we present MedGA in detail, showing the structure of individuals and genes and explaining the used selection, crossover and mutation operators.
Finally, we explain the fitness function that allows for obtaining two well-separated normal distributions starting from the input histogram computed on the original image, while preserving the input mean brightness.

\subparagraph{Individual's structure and gene encoding}
MedGA exploits a very efficient and comprehensive individual's structure.
The size of each individual is equal to $n$, where $n$ is the number of gray levels (different from $0$) of the input MR image.

Let $l_{in}^{\text{(min)}}$, $l_{in}^{\text{(max)}}$, $l_{out}^{\text{(min)}}$, and $l_{out}^{\text{(max)}}$ be the minimum non-zero and maximum gray levels of input and output images, respectively.
Assuming that $l_{in}^{\text{(min)}} \leq l_{out}^{\text{(min)}}$ and $l_{in}^{\text{(max)}} \geq l_{out}^{\text{(max)}}$, the linear normalization on $\mathcall{L}_{in}$ gives rise to the extended range of the non-zero gray levels, that is, the ordered set $\mathcall{L}'_{in}=[1, \dots, l_{in}^{\text{(max)}}]\subset \mathbb{N}$ (typically, $l_{in}^{\text{(min)}} > 1$).
This normalization operation, which employs only values of gray levels already representable in the initial dynamic range, does not alter the image content and allows MedGA to process additional intensity levels with respect to the initial full range $\mathcall{L}_{in}$, by considering the variability within the analyzed MRI datasets.
During each iteration $t = 1, 2, \dots, T$, each individual $C_i^t=[C_i^t(1), C_i^t(2), \ldots, C_i^t(n)]$ (with $i = 1, \ldots, |P|$) is defined as a circular array of integer numbers of size $n$, where $n=|\mathcall{L}'_{in}|$ corresponds to the number of different gray levels belonging to $\mathcall{L}'_{in}$ identified in the input MR image (i.e., the gray levels whose frequency is greater than zero in the input MR image).
The $n$ values are then sorted in ascending order so that the intensity levels $C_i^t(j)$ (with $j = 1, \ldots, n$) codified by the individual can be mapped onto the intensity levels of the input MR image (i.e., the gray level frequencies of the input MR image are assigned to the corresponding intensity levels of the individual).

In order to evaluate the fitness value of the individuals, we apply the following transformation $\mathcal{T}$ that re-maps each input gray level $r$ into $s$:
\begin{equation}
s = \mathcal{T}(C_i^t(j)) = \mathcal{T}(r),
\end{equation}
where $r \in \mathcall{L}_{in}=[l_{in}^{\text{(min)}}, \dots,  l_{in}^{\text{(max)}}] \subset \mathbb{N}$ and $s \in \mathcall{L}_{out}=[l_{out}^{\text{(min)}}, \dots,  l_{out}^{\text{(max)}}] \subset \mathbb{N}$ are intensity values in the input and output gray-scale ranges, respectively.
By so doing, a direct mapping between the gray levels of the original image $\mathcall{I}_\text{orig}$ and the output $\mathcall{I}_\text{enh}$ is defined: each gray level in the original histogram is replaced with the gray level value contained by the same position in the final best solution $C_\text{best} \in P$.

Since GAs are unbiased global optimization methods \cite{paulinas2007}, as population initialization method we used a simple uniform distribution to create the initial histogram bins.
Generally, computational approaches requiring initialization settings should be initialized randomly in order to increase the convergence speed and the probability to avoid local optima determined by fixed settings.
Thus, the following steps are used to initialize the population $P$:
\begin{enumerate}
	\item each individual $C_i^t \in P$ is randomly initialized by sampling $n$ integer values from the discrete uniform distribution in $\mathcall{L}'_{in}$;
	\item these $n$ values are then sorted in ascending order allowing for a mapping with the intensity levels of the input MR image.
\end{enumerate}
During the initialization, if an integer value is sampled more than once then the frequency values of the input MR image, corresponding to these gray level intensities, are summed-up and assigned to the same gray level of the individual.
After constructing the initial population, the fitness values are calculated for all individuals.
In every generation, the size of the population $P$ is kept constant to $|P|$.

\subparagraph{Selection operator}
For the selection of individuals, MedGA exploits a \textit{tournament} strategy for three main reasons:
(\textit{i}) the selection pressure can be controlled by setting the tournament size $k$ (with $k \ll |P|$);
(\textit{ii}) the fitness evaluations are performed only on the $k$ individuals selected for the tournaments, and not on the whole population;
(\textit{iii}) this technique could be easily implemented on parallel architectures \cite{miller1995}.

When a tournament is executed, $k \ll |P|$ individuals are randomly selected from the population $P$ and the winner, that is, the individual with the best fitness, is deterministically selected and added to $P'$.
This procedure is usually performed with reinsertion, that is, the individual can compete to multiple tournaments during a generation.
The size of the tournament $k$ balances the selection pressure: if $k$ is large, then the individuals characterized by worse fitness have a reduced probability to be selected for the intermediate population $P'$.
MedGA exploits this selection mechanism since it has three advantages: (\textit{i}) the fitness evaluations are required only for the individuals selected for the tournaments; (\textit{ii}) the selection pressure can be increased or decreased simply varying the value of $k$;  (\textit{iii}) this technique is suitable for parallel implementations.

\begin{figure}[!t]
	\centering
	\includegraphics[width=0.4\textwidth]{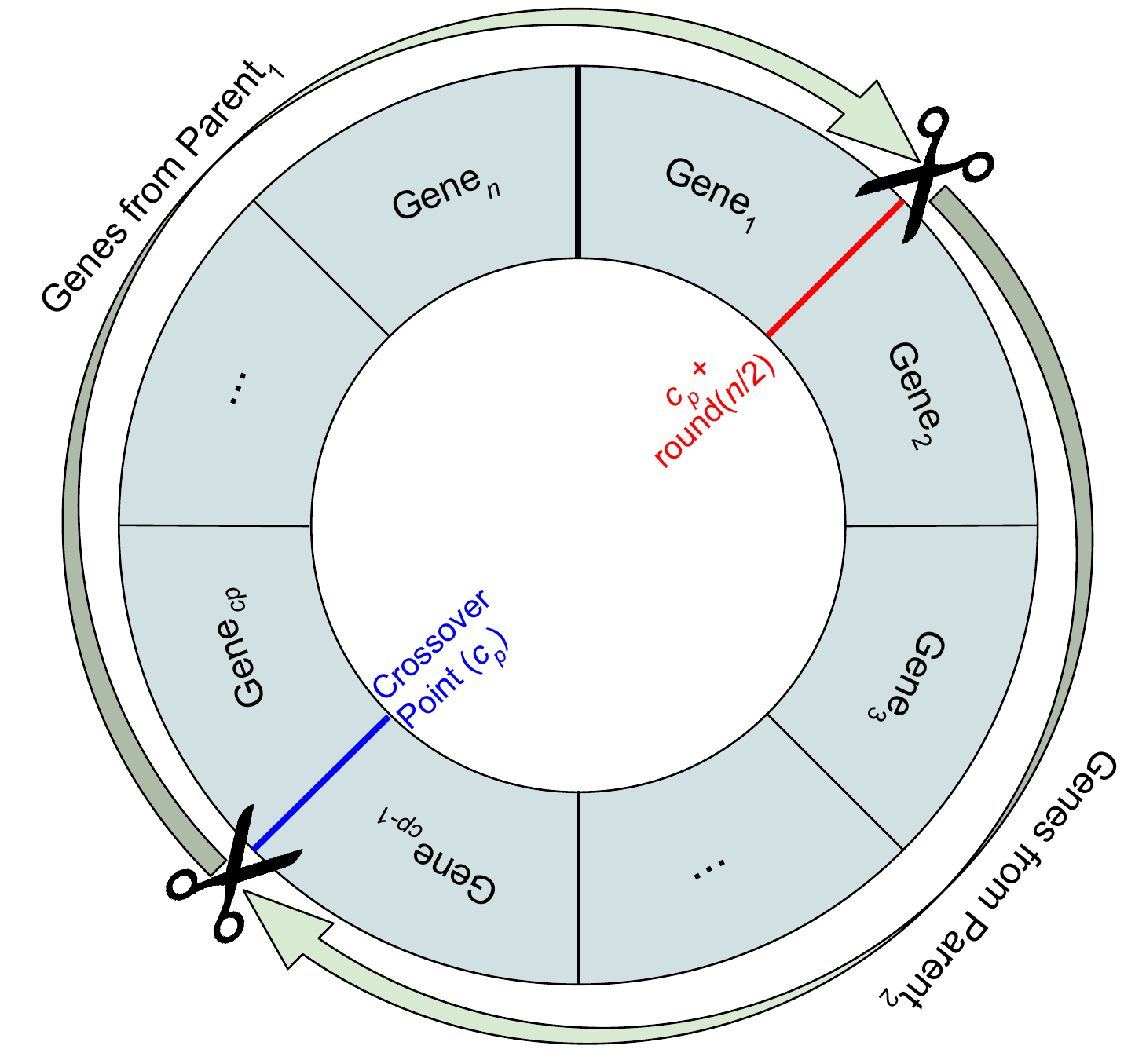}
	\caption[Graphical representation of the single-point crossover strategy]{Graphical representation of the single-point crossover strategy used for generating the first offspring from two parents (Parent$_1$ and Parent$_2$). The second offspring is generated changing the Parent$_1$'s genes with those of Parent$_2$ and vice-versa.}
	\label{crossoverImage}
\end{figure}

\subparagraph{Crossover and mutation operators}
At each iteration of MedGA, a number of individuals properly selected from
the current population are inserted into the intermediate population $P'$, and modified by means of crossover and mutation operators.
A \textit{single-point crossover} operator is applied with a given probability $p_c$ to the individuals selected by the tournament strategy and belonging to the first intermediate population $P'$.
MedGA employs the following \textit{single-point crossover} to blend the genetic information of two parents, resulting in a mixing ratio fixed to $0.5$ (see scheme in Fig. \ref{crossoverImage}):
\begin{enumerate}
		\item one crossover point $c_p$ is randomly selected from the ordered set $[1, 2, \ldots, n]$. Let $h_p$ be equal to $\mbox{round}(\frac{n}{2})$;
		\item if $c_p > \mbox{round}(\frac{n}{2})$, the first offspring is generated using the genes $1, \dots, c_p - h_p - 1$ and $c_p, \dots, n$ from Parent$_1$, and the genes $c_p - h_p, \dots, c_p - 1$ from Parent$_2$.
		Otherwise, it is generated using the genes $c_p, \dots, c_p + h_p - 1$ from Parent$_1$, and the genes $1, \dots, c_p - 1$ and $c_p + h_p, \dots, n$ from Parent$_2$;
		\item the second offspring is generated by replacing the Parent$_1$'s genes with those of Parent$_2$ and vice-versa.
\end{enumerate}

The \textit{mutation} operator is applied with probability $p_m$ to each element $C_i^t(j) \in C_i^t=[l_{out,i}^{(\text{min})}, \ldots, l_{out,i}^{(\text{max})}]$ of each offspring from the population $P'$, where $l_{out,i}^{(\text{min})}$ and $l_{out,i}^{(\text{max})}$ are the minimum and maximum non-zero gray levels encoded by $C_i^t$ during the current iteration, respectively.
In particular, if the gray level intensity encoded in $C_i^t(j)$ is smaller than the optimal threshold $\theta^t_{\text{opt},i}$ evaluated by IOTS at that iteration for the individual $C_i^t$, then an integer is randomly sampled from the uniform distribution in $[l_{out, i}^{(min)}, \ldots, \theta_{\text{opt},i}-1] \subset \mathbb{N}$ to update the value $C_i(j)$; otherwise, an integer is randomly sampled from the uniform distribution in $[\theta_{\text{opt}, i}^t, \ldots, l_{out,i}^{(\text{max})}] \subset \mathbb{N}$ to update the value $C_i(j)$.

Finally, to prevent the quality of the best solution from decreasing during the optimization, MedGA also exploits an \textit{elitism} strategy, so that the best individual from the current population is copied into the next population without undergoing the genetic operators.

\subparagraph{Fitness function}
In order to obtain a bimodal histogram separation that yields better results for further automated image processing phases, we propose a fitness function that aims at realizing two well-separated normal distributions with equal distance from the optimal threshold $\theta_\text{opt}$.
For each iteration $t$ (with $t = 1, 2, \dots, T$) and for each individual $C_i^t$ (with $i = 1, 2, \dots, |P|$), first the mean values $\mu_{1,i}^t$ and $\mu_{2,i}^t$, concerning the two sub-distributions in the histogram $\mathcall{H}_{1,i}^t$ and $\mathcall{H}_{2,i}^t$, respectively, and the corresponding  optimal threshold $\theta_{\text{opt},i}^t$, are efficiently computed using the IOTS algorithm \cite{ridler1978,trussell1979}.
Then, for each individual $C_i^t$ the following fitness function is calculated:
\begin{equation}
      \begin{array}{lcl}
	    \mathcal{F}(C_i^t) & = & \tau_1 + \tau_2 + \tau_3, \quad \mbox{where:}\\[1ex]
	    \tau_1 & = & \abs{2 \cdot \theta_\text{opt,i}^t - \mu_{1,i}^t - \mu_{2,i}^t}, \\[1ex]
	    \tau_2 & = & \abs{\omega_{1,i}^t - 3 \sigma_{1,i}^t}, \\[1ex]
	    \tau_3 & = & \abs{\omega_{2,i}^t - 3 \sigma_{2,i}^t}.
      \end{array}
\end{equation}

The terms $\omega_{1,i}^t=\frac{1}{2}(\theta_{\text{opt},i}^t - \min \limits_{j \in \{1, \ldots, n\}} \{C_i^t(j)\})$ and $\omega_{2,i}^t=\frac{1}{2}(\max \limits_{j \in \{1, \ldots, n\}} \{C_i^t(j)\} - \theta_{\text{opt},i}^t)$ correspond to the half width of $\mathcall{H}_{1,i}^t$ and $\mathcall{H}_{2,i}^t$, respectively, while $\sigma_{1,i}^t$ and $\sigma_{2,i}^t$ are the standard deviations of $\mathcall{H}_{1,i}^t$ and $\mathcall{H}_{2,i}^t$, respectively.
The three terms of the fitness function $\mathcal{F}(\cdot)$ cooperate together to achieve the desired image enhancement:
$\tau_1$ aims at maintaining the mean values $\mu_{1,i}^t$ and $\mu_{2,i}^t$ equidistant from the yielded optimal threshold $\theta_{\text{opt},i}^t$, while
$\tau_2$ and $\tau_3$ are designed to force the sub-histograms $\mathcall{H}_{1,i}^t$ and $\mathcall{H}_{2,i}^t$, respectively, to approximate normal distributions.
Especially, we exploited the empirical $3$-$\sigma$ rule, which states that approximately $99.73\%$ of the values lie within $3\sigma$ according to $\text{Pr}(\mu - 3\sigma \leq X \leq \mu + 3\sigma) \approx 0.9973$, where $\mu$, $\sigma$, and $X$ represent the mean, the standard deviation and an observation from a normally distributed random variable, respectively.

\begin{figure}[!t]
	\centering
	\subfloat[]{\includegraphics[width=0.2\textwidth]{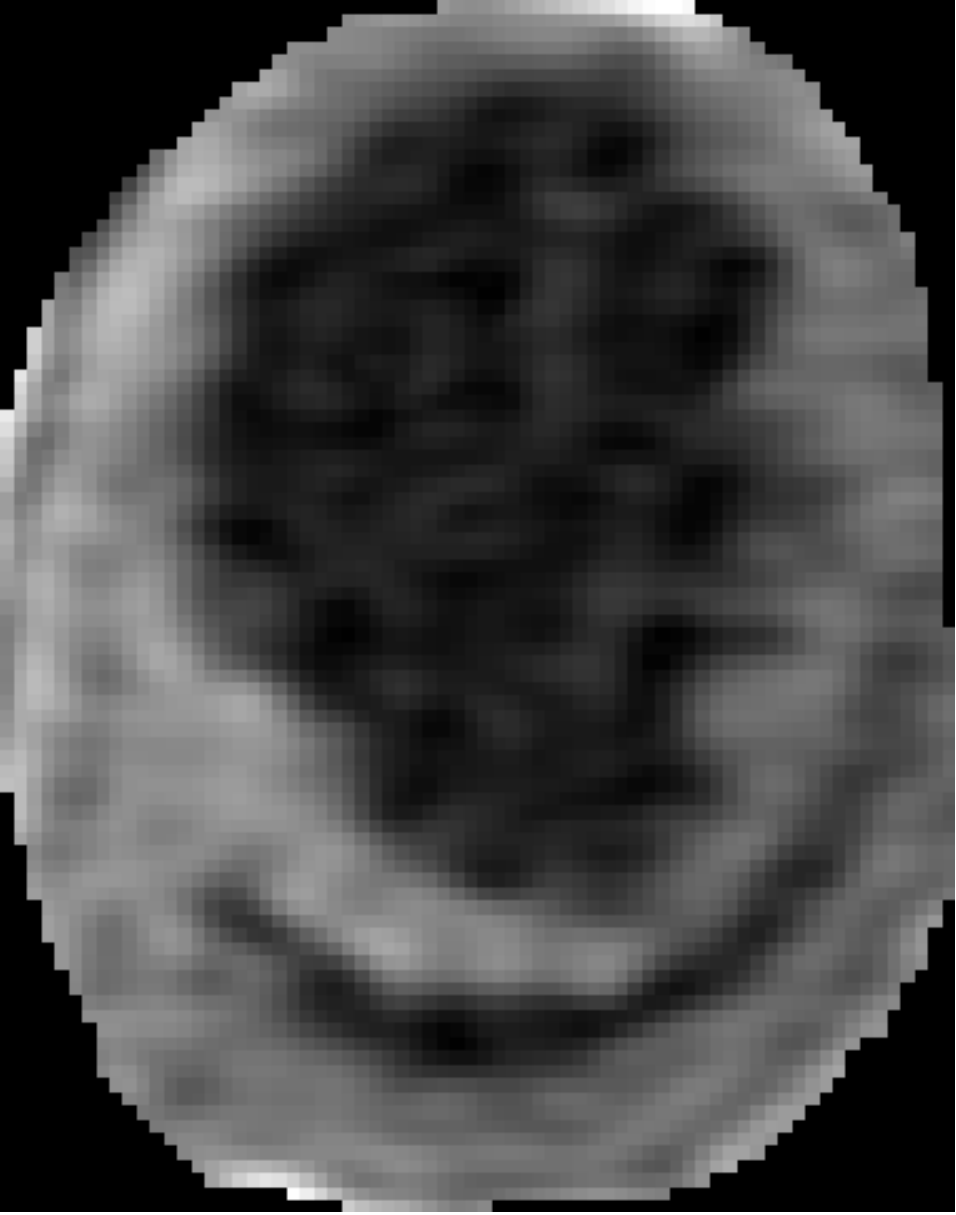}\label{OrigFibroid}} \quad
	\subfloat[]{\includegraphics[width=0.45\textwidth]{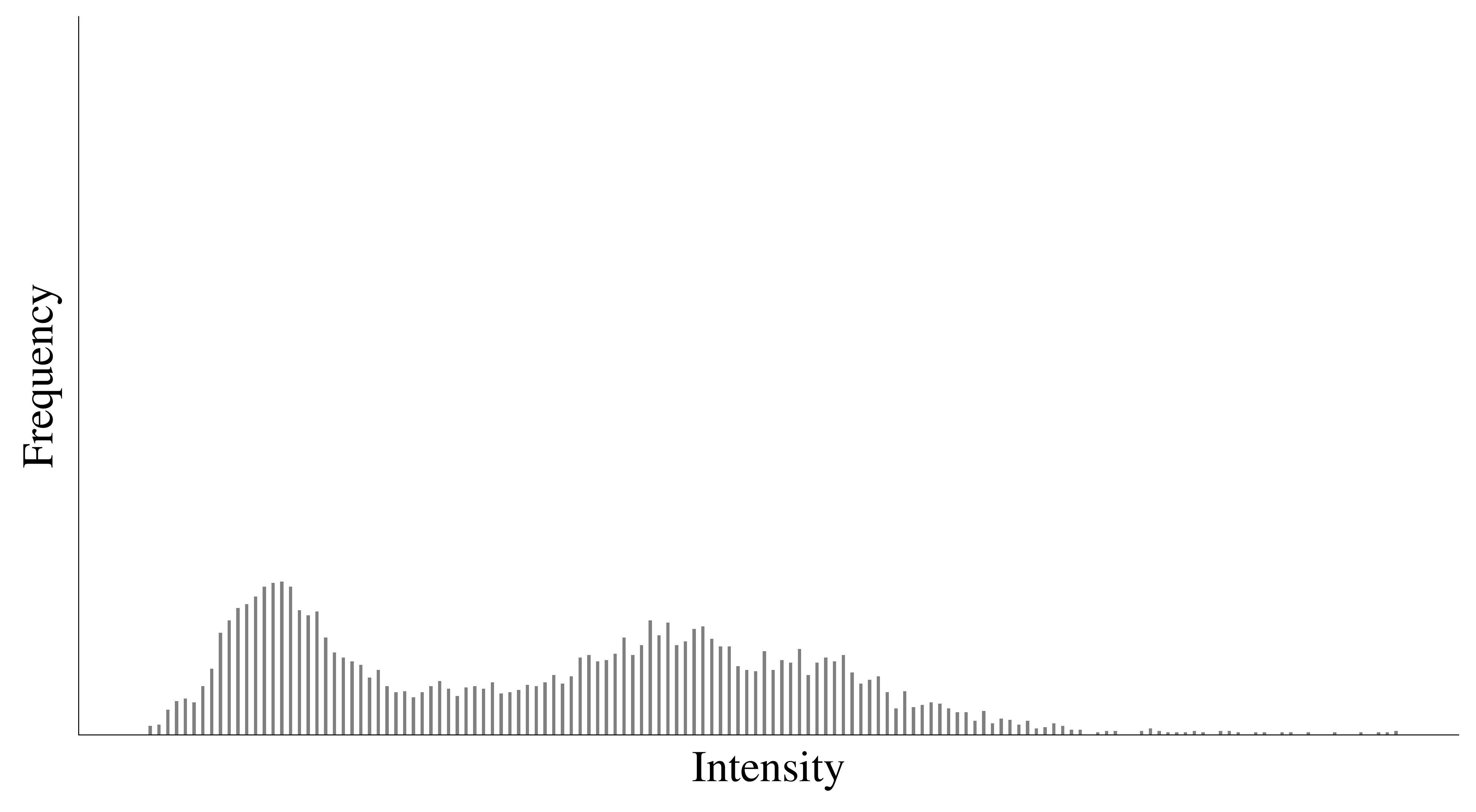}\label{OrigFibroid_Hist}} \\
	\subfloat[]{\includegraphics[width=0.2\textwidth]{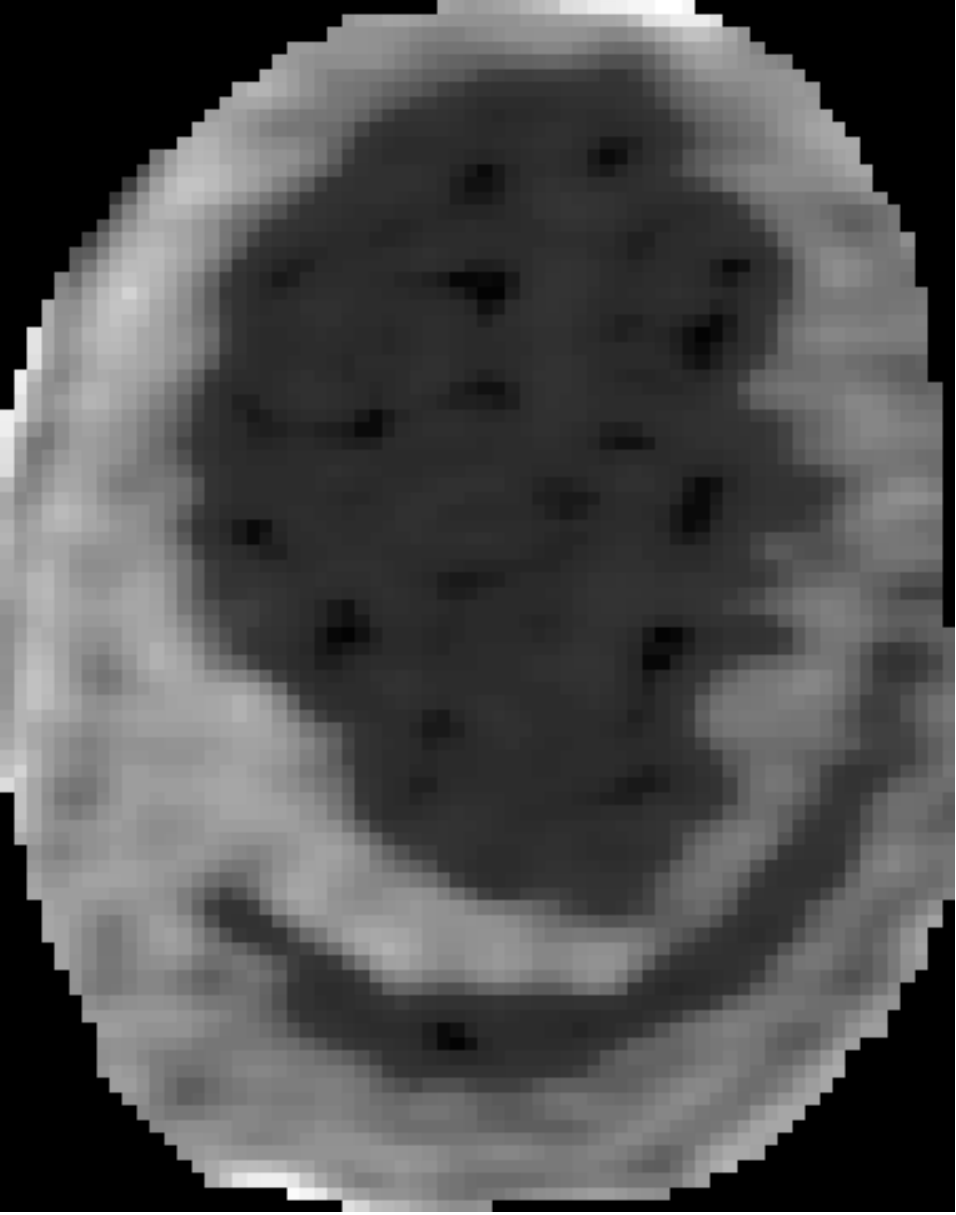}\label{EnhFibroid}} \quad
	\subfloat[]{\includegraphics[width=0.45\textwidth]{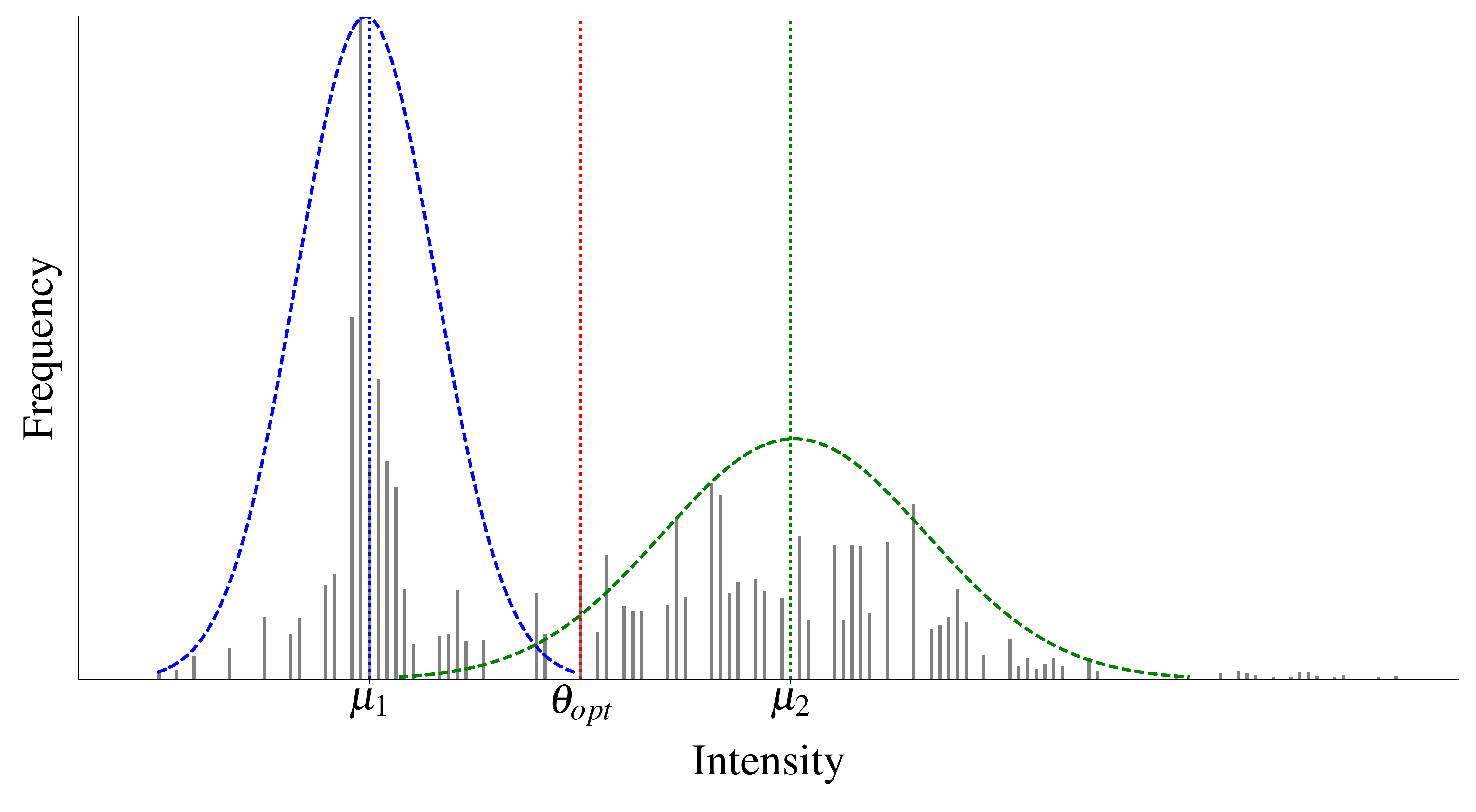}\label{EnhFibroid_Hist}} \\
	\caption[Enhanced uterine fibroid MR image and histrogram obtained by MedGA]{Enhanced image obtained by MedGA on an example of uterine fibroid (size: $89 \times 70$ pixels): (a)  normalized input image using linear contrast stretching on the initial full range of the masked MR image; (c) resulting image after the application of the pre-processing using MedGA.
		The histograms corresponding to the sub-images in (a) and (c) are shown in (b) and (d), respectively.
		The final histogram emphasizes the two underlying distributions in the gray levels intensity characterized by mean values $\mu_1$ and $\mu_2$, and standard deviations $\sigma_1$ and $\sigma_2$, respectively. 
		The two distributions are highlighted in the plot with blue and green dashed lines.}
	\label{FibroidEnhancement}	
\end{figure}

\begin{figure}[!t]
	\centering
	\subfloat[]{\includegraphics[width=0.25\textwidth]{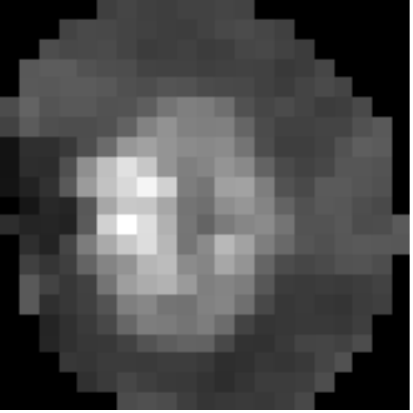}\label{OrigBrain}} \quad
	\subfloat[]{\includegraphics[width=0.45\textwidth]{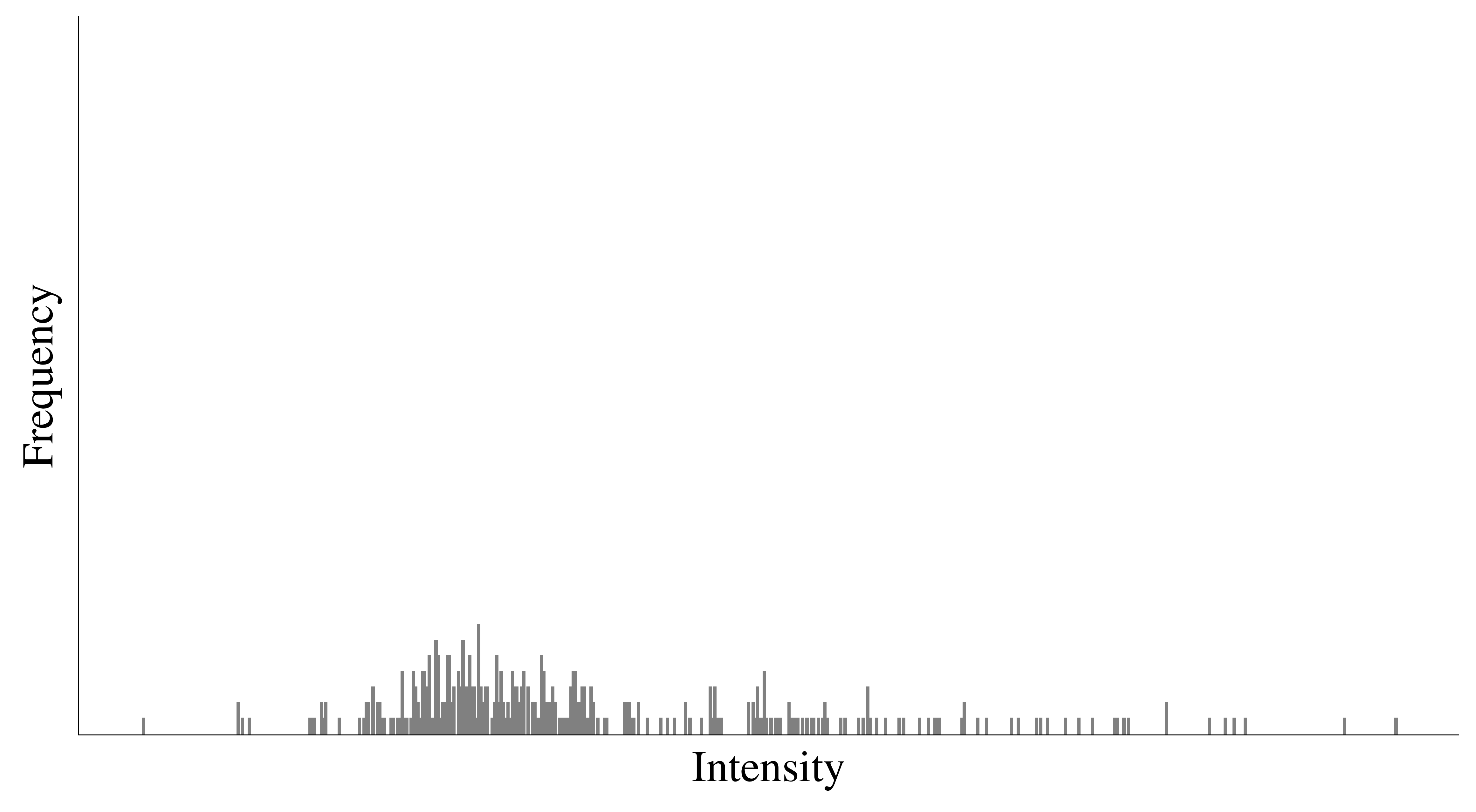}\label{OrigBrain_Hist}} \\
	\subfloat[]{\includegraphics[width=0.25\textwidth]{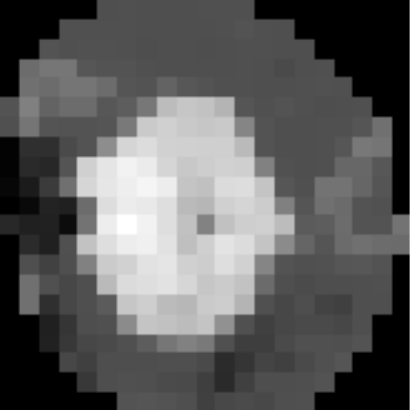}\label{EnhBrain}} \quad
	\subfloat[]{\includegraphics[width=0.45\textwidth]{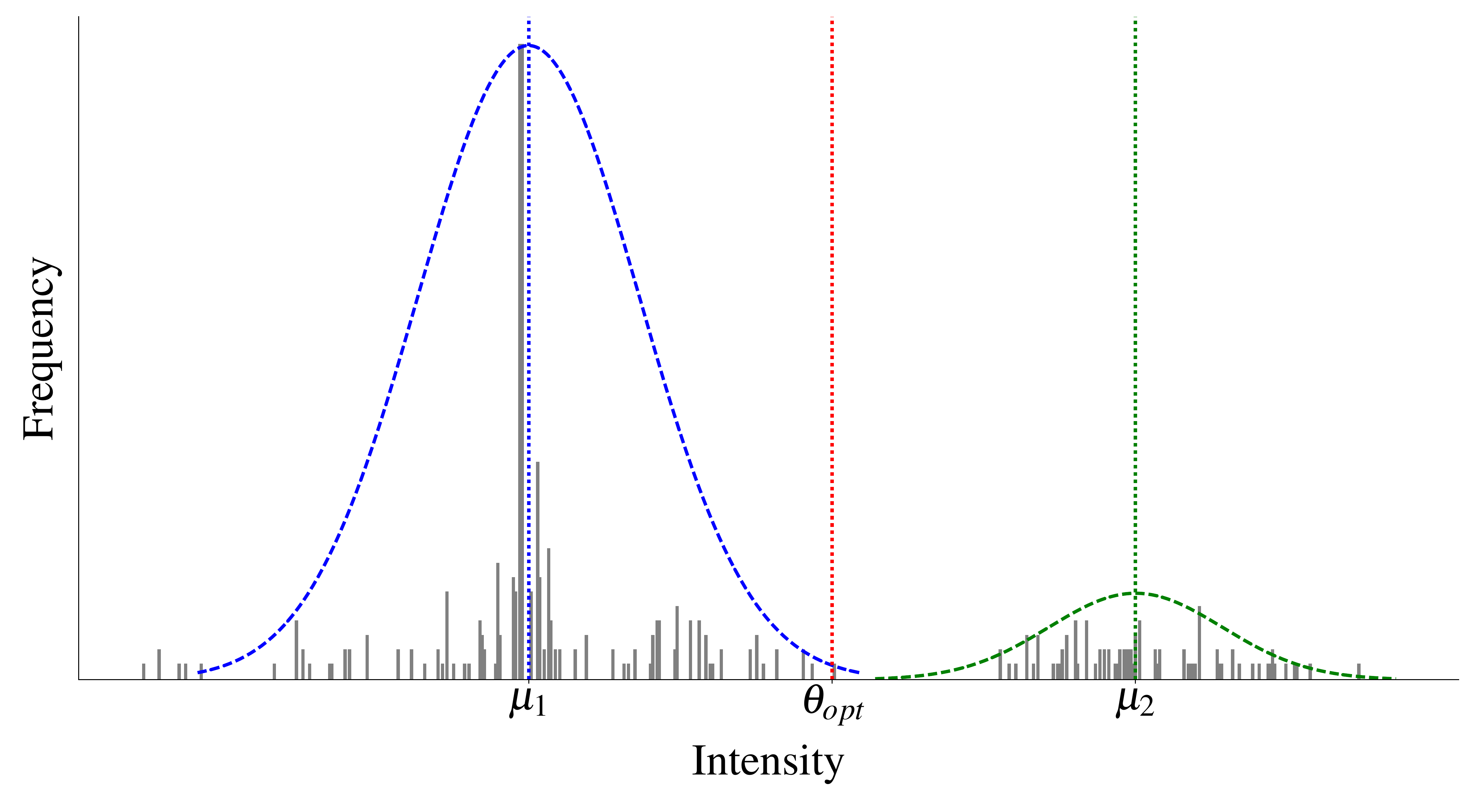}\label{EnhBrain_Hist}} \\
	\caption[Enhanced brain tumor MR image and histrogram obtained by MedGA]{Enhanced image obtained by MedGA on an example of brain tumor (size: $21 \times 21$ pixels): (a) normalized input image using linear contrast stretching on the initial full range of the masked MR image; (c) resulting image after the application of the pre-processing using MedGA.
	The histograms corresponding to the sub-images in (a) and (c) are shown in (b) and (d), respectively.
	The final histogram emphasizes the two underlying distributions in the gray levels intensity characterized by mean values $\mu_1$ and $\mu_2$ and standard deviations $\sigma_1$ and $\sigma_2$, respectively. 
	The two distributions are highlighted in the plot with blue and green dashed lines.}
	\label{BrainEnhancement}	
\end{figure}

Two examples of image enhancement results, achieved by MedGA on a uterine fibroid and on a brain tumor, are shown in Figs. \ref{FibroidEnhancement} and \ref{BrainEnhancement}, respectively.
In the case of uterine fibroids, the proposed method enhances the input MR image by making fibroid regions more uniform and with sharper edges in terms of both visual human perception and automated image segmentation.
The histogram in Fig. \ref{EnhFibroid_Hist} highlights that the output image is characterized by a more defined bimodal distribution compared to the initial image (Fig. \ref{OrigFibroid_Hist}), which presents approximately a trimodal gray level distribution.
In the  case of brain tumors, MedGA enhances the underlying bimodal distribution related to contrast-enhancing tumoral tissue and brain healthy tissues on CE-MR images.
This visual achievement is endorsed by the  histogram of the enhanced image (see Fig. \ref{EnhBrain_Hist}) that shows two more distinct peaks with respect to the initial gray level distribution  (see Fig. \ref{OrigBrain_Hist}).

Although \cite{draa2014,hashemi2010} used the same individual structures, both the fitness function and the aim of the GA are very different.
They defined two similar fitness functions that combine the number of edge pixels and the intensity of such pixels.
The goal of these studies is the image contrast enhancement for consumer electronics products.
The proposed method aims to enhance the image thresholding results on MRI data, generating two well-distinct normal distributions based on the input histogram, where the final best solution emphasizes the underlying bimodal distribution of the input histogram.

\subparagraph{MedGA parameter setting and calibration}
To analyze the performances of MedGA and identify the best settings for the image enhancement problem, we considered a calibration set consisting of $80$ medical images randomly selected from the available uterine fibroid MRI dataset, and we varied the settings of MedGA used throughout the optimization process, that is: (\textit{i}) the size of the population $|P| \in \{50, 100, 150, 200\}$; (\textit{ii}) the crossover probability $p_c \in \{ 0.8, 0.85, 0.9, 0.95, 1.0 \}$; (\textit{iii}) the mutation probability $p_m \in \{ 0.01, 0.05, 0.1, 0.2\}$; (\textit{iv}) the size of the tournament selection strategy $k \in \{ 5, 10, 15, 20\}$.
In all tests MedGA was run for $T_{\text{max}} = 100$ iterations.
Each MedGA execution was performed by varying one setting at a time, for a total of $320$ different settings tested and a total number of $320\times 80 = 25,600$ MedGA executions.

The results of these tests (data not shown) highlighted that, for each value of $|P|$, the best settings in terms of fitness values achieved are:
\begin{enumerate}
 \item $|P|=50$, $p_c=0.85$, $p_m=0.01$, $k=15$;
 \item $|P|=100$, $p_c=0.9$, $p_m=0.01$, $k=20$;
 \item $|P|=150$, $p_c=0.85$, $p_m=0.01$, $k=20$;
 \item $|P|=200$, $p_c=0.85$, $p_m=0.01$, $k=20$.
\end{enumerate}

\begin{figure}[!t]
	\includegraphics[width=0.75\textwidth]{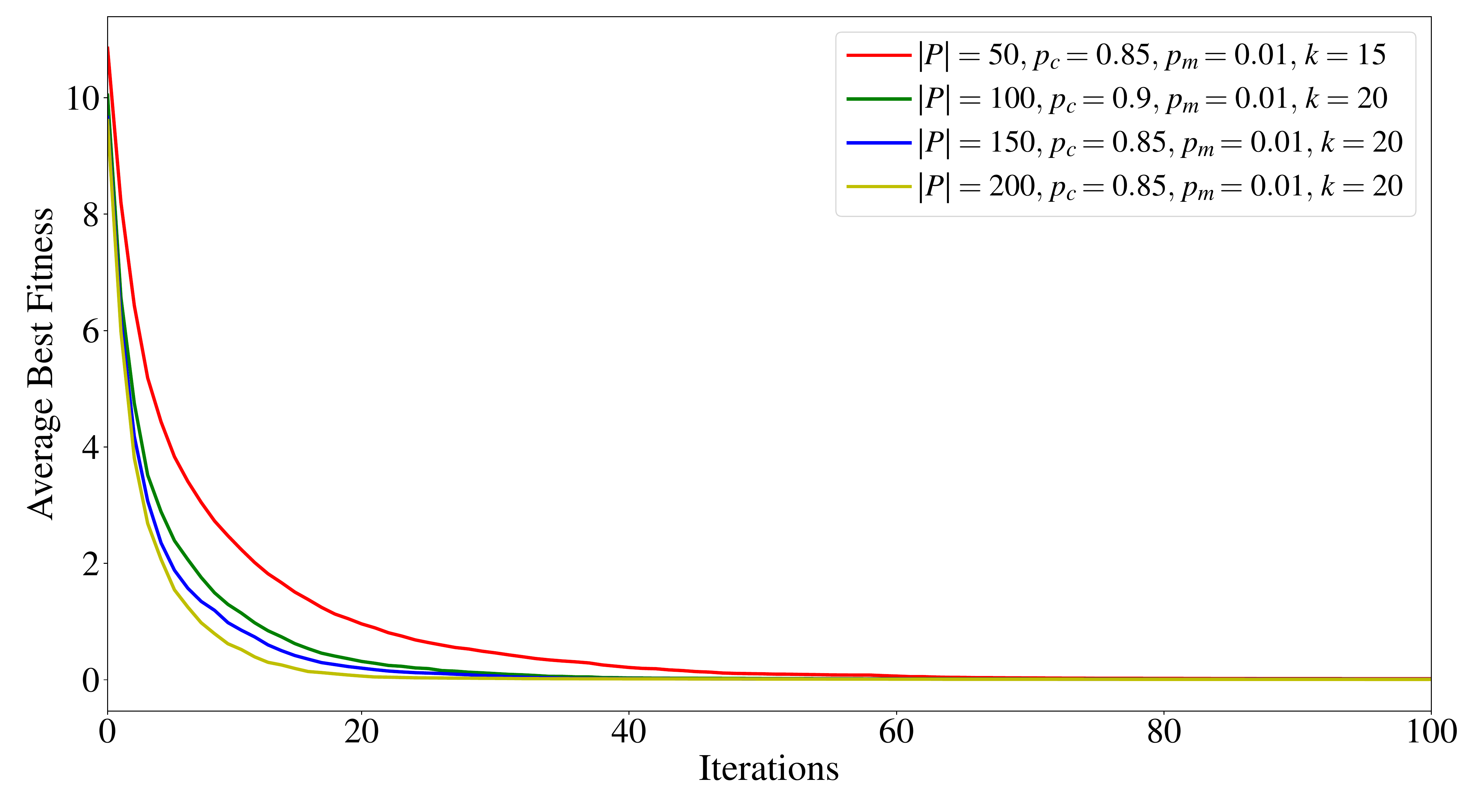}
	\centering
	\caption[Comparison of the ABF achieved by MedGA considering different parameterizations]{Comparison of the ABF achieved by MedGA with the best parameterizations found for each value of $|P|$ tested here.
	The average was computed over the results of the optimization of $80$ MR images.}
	\label{fig:settingscomparison}	
\end{figure}

Figure \ref{fig:settingscomparison} reports the comparison of the performances achieved by MedGA with these settings, where the Average Best Fitness (ABF) was computed by taking into account, at each iteration of MedGA, the fitness value of the best individuals over the $80$ optimization processes.
It is clear from the plot that, despite the final ABF values are comparable in all settings, the convergence speed increases with the size of the population, as well as the running time required by MedGA; therefore, to choose the best settings, we analyzed the computational performances concerning the $4$ tests listed above.

Considering that an MRI series related to a single patient contains on average $10$ slices with ROI fibroids, we tested the clinical feasibility of MedGA by calculating the total execution time for enhancing $10$ randomly chosen MR images.
For what concerns the tests $1$-$4$ described above, the executions lasted on average $672.12$ s, $1290.15$ s, $1987.8$ s, and $2669.74$ s, respectively, for the optimization of the same batch of $10$ images running a single core of the Intel\textsuperscript{\textregistered} Xeon\textsuperscript{\textregistered} E5-2440 CPU with $2.40$ GHz clock frequency.
On the other hand, by exploiting the $6$ cores of the same CPU to execute the parallel version of MedGA, we achieved up to $3.6\times$ speed-up with respect to the sequential version.
The results achieved using the parallel version of MedGA confirm the importance of HPC solutions in the field of real healthcare environment to obtain clinically feasible outcomes, that is, enhancing MR images in reasonable time for medical imaging practice.

By considering both the performance of MedGA in terms of ABF and the running time required to process 80 images, we selected the parameter settings $|P|=100$, $p_c=0.9$, $p_m=0.01$, $k=20$ as the best trade-off characterized by a good convergence speed and an adequate running time (for this specific application), and we exploited this configuration for all tests reported and discussed in what follows.

\paragraph{Results}

Tables \ref{table:ImageEnhancementFibroid} and \ref{table:ImageEnhancementBrain} show the image pre-processing results achieved by each method on the uterine fibroid and brain tumor MRI datasets, respectively, using the metrics reported in Appendix \ref{sec:enhanceEval}.
In both MR image analysis applications, MedGA remarkably achieves the highest Peak Signal-to-Noise Ratio (\emph{PSNR}) mean values with respect to the state-of-the-art methods, generally involving the highest signal quality.

\begin{table}[!t]
	\centering
	\begin{scriptsize}
	\caption[Image enhancement evaluation metrics achieved by MedGA and the classical image pre-processing approaches on the uterine fibroid MRI dataset]{Values of the image enhancement evaluation metrics, achieved by MedGA and the classical image pre-processing approaches, calculated on the MRI dataset with $18$ patients affected by uterine fibroids and expressed as mean and standard deviation values. Numbers in bold indicate the best values.}
	\label{table:ImageEnhancementFibroid}
	\begin{tabular}{c|cc|cc|cc|cc}
		\hline\hline
		\multirow{2}{*}{Method} & \multicolumn{2}{c|}{PSNR} & \multicolumn{2}{c|}{$\#$DE} & \multicolumn{2}{c|}{AMBE} & \multicolumn{2}{c}{SSIM}\\
		& Mean & Std. Dev. & Mean & Std. Dev. & Mean & Std. Dev.  & Mean & Std. Dev.\\ \hline
		HE				& 30.994	& 1.949		& \textbf{975.465} 		& 475.951 	& 0.085	  	& 0.029 	& 0.859	& 0.044 \\
		Bi-HE			& 31.880	& 2.046		& 907.177 		& 415.703 	& 0.038 	& 0.020 	& 0.907	& 0.032 \\
		GT $\gamma=0.4$	& 30.194	& 2.170		& 717.555	  & 401.186	& 0.212		 & 0.019		& 0.823 & 0.024   \\
		GT $\gamma=2.5$	& 29.952	& 2.127		& 965.012	  			& 380.967	& 0.261	  	& 0.012	 & 0.586 & 0.075  \\
		ST $\lambda=4/\alpha$		& 33.971	& 1.874		& 872.594	  & 396.016	& 0.040		 & 0.014	& 0.880 & 0.023    \\
		ST $\lambda=6/\alpha$		& 32.286	& 1.975		& 869.032	  & 378.277	& 0.060		 & 0.021	& 0.715 & 0.056    \\
		ST $\lambda=8/\alpha$		& 31.353	& 2.029		& 841.420	  & 348.674	& 0.073		 & 0.025	& 0.613 & 0.070    \\
		MedGA	 		& \textbf{37.366} & 2.347 & 866.604	  & 409.604	& \textbf{0.033} & 0.011	 & \textbf{0.928} & 0.025     \\
		\hline\hline
	\end{tabular}

	\vspace{1cm}
	
		\caption[Image enhancement evaluation metrics achieved by MedGA and the classical image pre-processing approaches on the brain tumor MRI dataset]{Values of the image enhancement evaluation metrics, achieved by MedGA and the classical image pre-processing approaches, calculated on the MRI dataset composed of $27$ brain tumors and expressed as mean and standard deviation values. Numbers in bold indicate the best values.}
		\label{table:ImageEnhancementBrain}
		\begin{tabular}{c|cc|cc|cc|cc}
			\hline\hline
			\multirow{2}{*}{Method} & \multicolumn{2}{c|}{PSNR} & \multicolumn{2}{c|}{$\#$DE} & \multicolumn{2}{c|}{AMBE} & \multicolumn{2}{c}{SSIM}\\
			& Mean & Std. Dev. & Mean & Std. Dev. & Mean & Std. Dev.  & Mean & Std. Dev.\\ \hline
			HE				& 34.215	& 1.447		& 38.779 		& 21.287 	& 0.124	  	& 0.049 	& 0.756	& 0.112 \\
			Bi-HE				& 36.758	& 1.718		& 44.923 		& 26.252 	& \textbf{0.042}	  	& 0.024 	& \textbf{0.932}	& 0.021 \\
			GT $\gamma=0.4$	& 33.193	& 0.841		& 21.271	  & 18.467	& 0.229		 & 0.020		& 0.713 & 0.065   \\
			GT $\gamma=2.5$	& 33.520	& 1.113		& \textbf{45.119}	  			& 27.444	& 0.229	  	& 0.028	 & 0.457 & 0.096  \\
			ST $\lambda=4/\alpha$		& 36.812	& 0.923		& 43.574	  & 27.307	& 0.055		 & 0.019	& 0.848 & 0.048    \\
			ST $\lambda=6/\alpha$		& 35.270	& 0.942		& 43.779	  & 26.492	& 0.079		 & 0.028	& 0.645 & 0.105    \\
			ST $\lambda=8/\alpha$		& 34.435	& 0.991		& 44.072	  & 26.508	& 0.090		 & 0.034	& 0.543 & 0.121    \\
			MedGA	 		& \textbf{37.751} & 1.990 & 43.534	  & 22.598	& 0.079 & 0.039	 & 0.881 & 0.053     \\
			\hline\hline
		\end{tabular}	
	
	\end{scriptsize}
\end{table}

Considering the results in Table \ref{table:ImageEnhancementFibroid}, HE over-enhances the processed uterine fibroid MR images, as denoted by the highest mean number of detected edges (\#DE) value \cite{canny1986}, while Bi-HE allows for the preservation of the mean brightness, as also indicated by the lower mean value of Absolute Mean Brightness Error (\emph{AMBE}) \cite{arriaga2014,chen2003}.
For what concerns the other techniques, on the one hand, GT with $\gamma = 0.4$ yields better results compared to GT with $\gamma = 2.5$, especially in the case of the Structural Similarity Index (\emph{SSIM}) metrics \cite{wang2004}; on the other hand, all metrics related to the tested ST functions show that their performances decrease as the value of $\lambda$ increases.
As it can be observed in Figs. \ref{FibroidResults} and \ref{BrainResults}, MedGA strengthens the ROI edges by enhancing details and features useful for image binarization; this result confirms, from a qualitative perspective, the quantitative results presented above.
From an overall view of the metrics values, we can claim that the approaches obtaining the highest values of the \#DE measure (i.e., HE and GT with $\gamma = 2.5$) could imply a considerable over-enhancement of the output image, according to the other image quality metrics.

The results on brain tumor MRI data reported in Table \ref{table:ImageEnhancementBrain} show a slightly different trend, also due to the small size of the pre-processed cropped sub-images.
As a first evidence, both GTs do not preserve the input mean brightness considering the \emph{AMBE} measure.
Interestingly, GT with $\gamma = 2.5$ and $\gamma = 0.4$ achieve the highest and the lowest \emph{\#DE} values, respectively.
Bi-HE strongly improves the enhancement metrics obtained by HE, by generally reporting the best results.
Consistently with the metrics calculated on uterine fibroid MRI data, all the results concerning ST functions get worse when the value of $\lambda$ increases.
The highest \emph{SSIM} mean value is achieved by Bi-HE, revealing the best structural information, even though MedGA obtains the best signal quality in terms of \emph{PSNR} mean values.

These findings are also corroborated by a visual inspection of Figs. \ref{FibroidResults} and \ref{BrainResults}, where the enhanced images using HE and GT with $\gamma = 2.5$ present an inadequate appearance for image observation and interpretation.
Overall, these results highlight that MedGA generally outperforms the conventional image enhancement approaches considering signal and perceived image quality, while preserving the input mean brightness.

\paragraph{Conclusion and future work}
Differently to the state-of-the-art methods that aim at improving the contrast level of the whole image to obtain ``visually pleasant'' images, MedGA is the first work that explicitly addresses the challenging issues pertaining to the enhancement of medical images characterized by a nearly bimodal gray level histogram distribution.
Our intelligent image enhancement system can visually assist physicians during their interactive decision-making tasks as well as improve the results in downstream automated processing pipelines for clinically useful measurements \cite{rueckert2016}.

MedGA deals with the practical problems regarding the interpretability of Machine Learning and Computational Intelligence methods in medicine \cite{cabitza2017}.
As a matter of fact, the final best solution (i.e., the output gray level histogram) and the enhanced image are understandable by physicians.
The efficient encoding of the individuals---taking inspiration by \cite{hashemi2010}---coupled with effective HPC solutions, allows for a clinically feasible computational framework.
By defining a specific fitness function to emphasize the two Gaussian distributions composing a bimodal histogram, while the existing approaches based on Evolutionary Computation or Swarm Intelligence techniques were conceived for a different purpose, i.e., improving the perceived visual information in terms of image contrast.
Thanks to this \textit{ad hoc} fitness function (differently to the GP-based image enhancement method in \cite{poli1997}, where no fitness function is defined because the user interactively selects the best solution) and the robustness achieved by means of the calibration step for GA's parameter setting, MedGA does not require any user interaction step.
Moreover, MedGA differs from the GP-based approaches, in which the final generated solution could have large size \cite{castelli2014}, so heavily affecting the readability and interpretability of the provided solutions.
The main impact of this contribution consists in showing how an effective method based on evolutionary computation can outperform the existing methods in medical image enhancement.

As a future extension, in the case of large size images (e.g., $1000 \times 1000$ pixels), Graphics Processing Units (GPUs) can represent an enabling technology for real-time radiology applications \cite{eklund2013}, since the running time of histograms computation can be considerably reduced by using a parallel implementation \cite{scheuermann2007}.

Considering the achieved results in terms of $\#$DE (see Tables \ref{table:ImageEnhancementFibroid} and \ref{table:ImageEnhancementBrain}), MedGA's performance could be further improved---in terms of contrast---by integrating a novel component in the fitness function, which explicitly relies on the number of detected edges.
Since this additional component would have a different purpose and a different magnitude, a multi-objective optimization method should be taken into account.
In particular, MedGA could be extended by means of an effective evolutionary approach, such as Non-dominated Sorting Genetic Algorithm (NSGA-III) \cite{deb2014}, to simultaneously optimize both conflicting objectives, which consist in maximizing the number of edges while minimizing the distance between the optimal threshold and the two normal distributions.

\subsection{Computational framework to improve MR image thresholding}
\label{sec:MedGAsegmentation}
Medical image segmentation concerns both detection and delineation of anatomical or physiological structures from the background, distinguishing among the different components included in the image \cite{bankman2009}.
This important task allows for the extraction of clinically useful information and features in medical image analysis \cite{joao2016,wang2012}.
Accordingly, computer-assisted approaches enable quantitative imaging \cite{duncan2000}, aiming at accurate and objective measurements from digital images regarding a ROI \cite{gillies2015,yankeelov2016}.
Indeed, image segmentation is still one of the most challenging research areas especially in medical image analysis \cite{duncan2000}.
Accurately delineating the ROIs is a critical task, since manual segmentation procedures are time-expensive, error-prone, and operator-dependent (i.e., not ensuring result repeatability).

In Pattern Recognition, among low-level intensity-based techniques, which are widely adopted in scenarios with real-time constraints, the most basic unsupervised image segmentation approach is global thresholding that essentially reduces to a pixel classification problem \cite{gonzalez2002} (see Section \ref{sec:imageThresholding} for the theoretical prerequisites).
In particular, image binarization classifies the input pictorial data into exactly two classes (i.e., foreground and background), given a threshold intensity value \cite{rogowska2000}.
This global threshold value is efficiently computed by operating on the image histogram alone.
Unfortunately, image binarization techniques work properly only for input images with a bimodal histogram \cite{xue2012}.
In practice, different types of regions in an image could overlap, thus affecting the bimodality conditions of the gray level histogram.
As a matter of fact, the histogram modes semantically correspond to different types of regions.
Image pre-processing can definitely improve the result accuracy achieved by computer-assisted segmentation methods \cite{rogowska2000}, by sharpening the peaks of the two sub-distributions.
The resulting histogram is supposed to be thus more strongly bimodal, dealing also with blurred region contours and the related Mach band effect pertaining to edge-detection in the human visual system \cite{daffner1980,krupinski2010}.
As a matter of fact, in radiology, this phenomenon is accentuated in the edges of adjacent regions that slightly differ in terms of gray level intensities \cite{ratliff1984}.

No existing pre-processing technique addresses the issues related to medical image enhancement for subsequent binarization using adaptive thresholding \cite{xue2012}.
Literature methods may be inadequate when dealing with low contrast images \cite{li2016}, producing false edges and under-/over-segmentation when input images are affected by noise, as in the case of MRI data \cite{gandhamal2017}.
In order to overcome these limitations and automatically determine a suitable optimal threshold, this computational framework exploits  MedGA \cite{rundo2018MedGA1}, for improving bimodal histogram separation \cite{rundo2018MedGA1}.
MedGA is designed \emph{ad hoc} to help basic thresholding techniques by improving the threshold selection between the two underlying distributions in CE-MR images with a nearly bimodal histogram.

Therefore, in \cite{rundo2018MedGA2}, we used MedGA as a pre-processing step, before performing threshold-based two-class image segmentation of MR images characterized by a nearly bimodal histogram distribution.
As case study, MedGA was applied to two processing pipelines, based on the efficient IOTS algorithm \cite{ridler1978,trussell1979}, concerning two different clinical contexts that require MR image analysis:
\begin{itemize}
	\item uterine fibroid segmentation in MRgFUS \cite{militelloTCR2014};
	\item brain metastatic cancer segmentation in neuro-radiosurgery therapy \cite{meier2016}.
\end{itemize}
In both cases, a precise and reliable and reproducible segmentation is mandatory.
Unfortunately, these tasks are generally carried out manually by experienced physicians.
The achieved image enhancement results were quantitatively evaluated using image enhancement and medical image segmentation metrics.

\paragraph{MR image segmentation using adaptive thresholding}
Image enhancement techniques can facilitate the user interpretation of an image, as well as improve the automated interpretation.
Therefore, we use MR image segmentation as an important processing goal \cite{starck2003}.
After the application of MedGA, the enhanced images are segmented using the IOTS algorithm \cite{ridler1978,trussell1979}, which is the most straightforward automated segmentation approach.

Differently to existing image enhancement techniques, we aimed at designing an appropriate pre-processing method that is able to reliably enhance a particular type of medical images for further automated image processing phase.
In our case, MedGA tackles the problems related to medical images with a roughly bimodal histogram, by strengthening the two underlying sub-distributions.
The main goal of MedGA is to yield pre-processed medical images well-suited for classic threshold-based segmentation techniques, in order to improve ROI delineation.
By so doing, the computational load---required for achieving accurate segmentation results---is transferred from the pixel classification stage to the pre-processing phase by means of an effective Soft Computing technique.

\begin{figure}[t]
	\centering
	\includegraphics[width=0.8\linewidth]{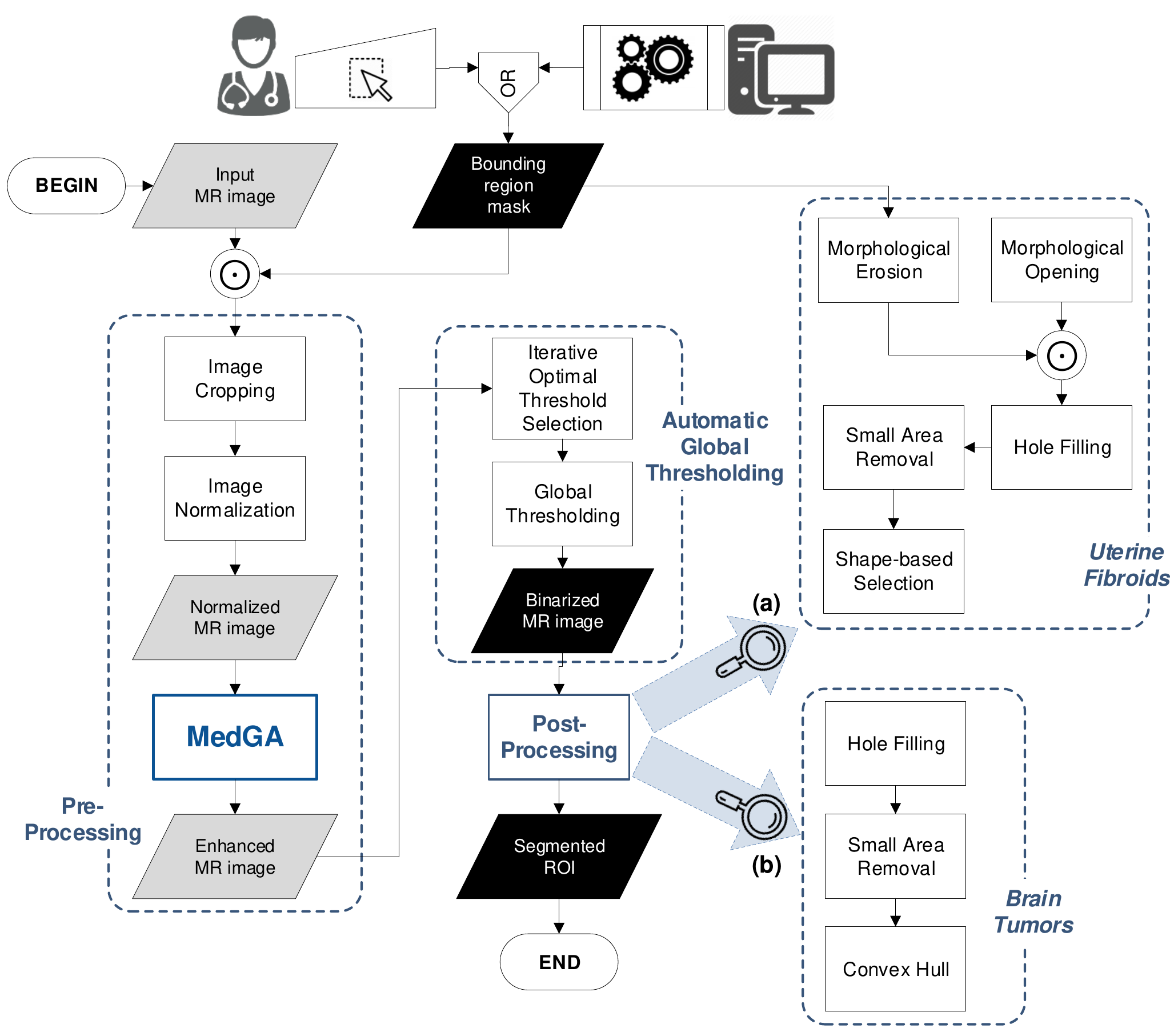}
	\caption[Flow diagram of the proposed computational framework that integrates MedGA as a pre-processing step for MR image segmentation based on the efficient IOTS algorithm]{Flow diagram of the proposed computational framework that integrates MedGA as a pre-processing step for MR image segmentation based on the efficient IOTS algorithm \cite{ridler1978,trussell1979}.
	Note that two slightly different post-processing pipelines were developed for (a) uterine fibroids and (b) brain tumors.
	Gray and black data blocks denote MRI gray-scale images and binary masks, respectively.}
	\label{flowdiagram_MedGA}	
\end{figure}

Fig. \ref{flowdiagram_MedGA} shows the overall flow diagram of the proposed computational framework that integrates MedGA for the thesholding-based segmentation of bimodal MR images.
We developed two slightly different post-processing pipelines to refine the results achieved by this efficient adaptive thresholding technique.
These post-processing steps, here applied to perform uterine fibroid and brain tumor segmentation, are described in what follows.
MedGA is able to enhance images in segmentation tasks involving both hyper- and hypo-intense ROIs in CE-MR images, also dealing with data unbalanceness (i.e., the number of foreground pixels is either much higher or lower than the number of background pixels).

\subparagraph{Uterine fibroid segmentation}
Firstly, uterus region delineation is required.
This task can be accomplished manually by the user or automatically by means of computational methods to reduce operator-dependency, as described in \cite{militelloCBM2015}.
ROIs are represented by tissues with low contrast mean absorption (i.e., NPV), thus the pixels with lower values with respect to the achieved threshold are yielded in the binarized MR image.
However, segmentation approaches have to take into account NPV inhomogeneities, due to sonication spots during the MRgFUS treatment.

The used post-processing refinement steps are the following (Fig. \ref{flowdiagram_MedGA}a):
\begin{enumerate}
	\item morphological opening with a circular structuring element ($2$-pixel radius) to separate possible loosely connected hypo-intense regions;
	\item some regions at the boundary of the uterus bounding region mask could present similar intensity values to gray levels characterizing fibroid regions, so being included in the thresholding output. To eliminate this ambiguity, it is appropriate to apply a morphological erosion (with a circular structuring element of $5$-pixel radius) to the ROI binary mask, and then the logical pixel-by-pixel product (i.e., Hadamard multiplication) with the image resulting from the previous step is performed;
	\item a hole filling algorithm is necessary to deal with possible holes in fibroid regions also due to non-uniform distribution of ablated tissue caused by sonication spots;
	\item segmentation is further improved through a connected-component labeling based operation by removing objects that are smaller than a certain area (i.e., $120$ pixels) and characterized by similar intensity with respect to the fibroids to be treated, because there may be regions or artifacts caused by very small dark areas;
	\item some lengthened connected-component with sufficiently large area could be present (i.e., due to other anatomical	structures or to acquisition artifacts). Fibroids, in fact, present a spherical or semi-spherical shape \cite{verkauf1993} that can be denoted by means of the parameters of the various connected-components. This connected-component based selection considers the eccentricity and the extent of the detected regions. Specifically, experimental reference values to discern fibroids from the rest	of connected-components are: $0.3 \leq \text{extent} < 0.8$ and $ 0.0 \leq \text{eccentricity} < 0.8$, according to \cite{militelloCBM2015}. Lastly, any connected-component, which has passed the shape-based control and whose centroid distance is more than a given upper limit (i.e., $\frac{\sqrt{M^2+N^2}}{3}$) from the MR image center, is removed.
\end{enumerate}

\subparagraph{Brain tumor segmentation}
The accurate and reproducible measurement of tumor size and its changes over time is crucial  for diagnosis, treatment planning, as well as monitoring of response to oncologic therapy for brain tumors \cite{meier2016}.
As a preliminary step, the user has to interactively select a bounding region that includes the tumor zone (by means of a free-hand ``lasso'' tool).
Since the areas to segment are enhancement regions, the pixels that have higher intensities than the threshold are selected during the image binarization phase.
Brain metastatic cancers may contain also necrotic material, so these inner necrotic cores could affect the achieved enhancement region segmentation.
Therefore, some refinement steps are useful to cope with this situation.

The used post-processing pipeline is described in the following (see Fig. \ref{flowdiagram_MedGA}b):
\begin{enumerate}
	\item hole filling algorithm to consider also necrotic areas;
	\item adaptive post-processing steps based on the size of the input image, consisting in small area removal (considering $4$-connectivity) with minimum threshold equal to $30$ pixels on images with size greater than $300$ pixels, or $10$ pixels otherwise;
	\item to allow also for large bounding regions, shape-based selection is applied in the case of at least two connected-components, according to: $\text{extent} \geq 0.6$ and $0.0 \leq \text{eccentricity} < 0.8$ (see \cite{rundoCMPB2017}). However, when a single connected-component is present, these controls are avoided;
	\item brain metastases have a pseudo-spherical appearance \cite{ambrosini2010}, therefore a convex hull algorithm is employed to envelope the segmented lesion into the smallest convex polygon containing this region \cite{zimmer1997}.
\end{enumerate}

\begin{figure}[!t]
	\centering
	\subfloat[]{\includegraphics[width=0.2\textwidth]{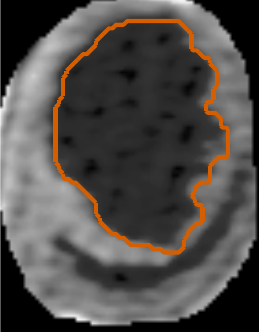}\label{SegFibroid}} \quad\quad\quad
	\subfloat[]{\includegraphics[width=0.25\textwidth]{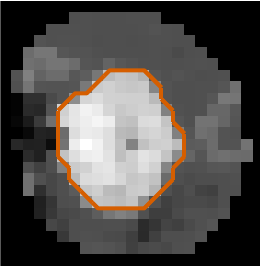}\label{SegBrain}} \\
	\caption[Segmentation results achieved by the IOTS algorithm on the MR images pre-processed by MedGA]{Segmentation results achieved by the IOTS algorithm on the MR images pre-processed by MedGA: (a) uterine fibroid delineation on the image given in Fig. \ref{EnhFibroid}; (b) brain tumor delineation on the image given in \ref{EnhBrain}.}
	\label{SegmentationMedGA}	
\end{figure}

The thresholding-based segmentation results achieved on the MR images pre-processed by MedGA in Figs. \ref{EnhFibroid} and \ref{EnhBrain} are shown in Figs. \ref{SegFibroid} and \ref{SegBrain}, respectively.

\begin{figure}[!t]
	\subfloat[]{\centering\includegraphics[width=\textwidth]{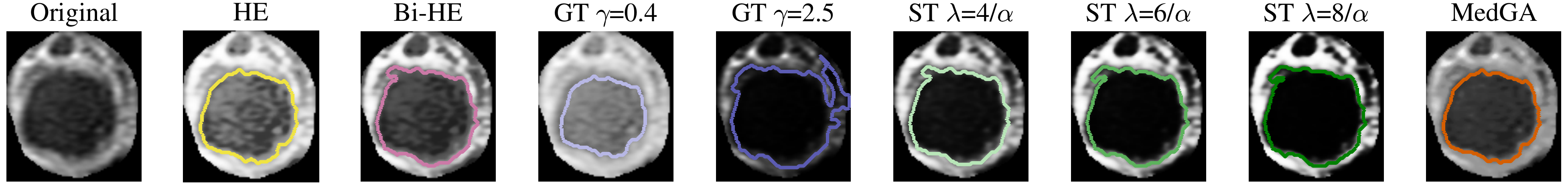}\label{fibroidResultsA}}\\
	\subfloat[]{\centering\includegraphics[width=\textwidth]{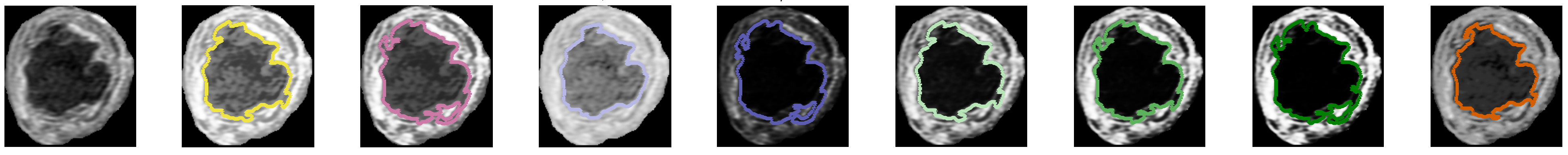}\label{fibroidResultsB}}\\
	\caption[Segmentation results achieved by MedGA and the competing methods on uterine fibroids]{Segmentation results on the uterine fibroids shown in Figs. \ref{InputFibA} and \ref{InputFibB}, achieved by the processing pipeline presented in Fig. \ref{flowdiagram_MedGA}a exploiting the state-of-the-art image pre-processing approaches (namely: HE, Bi-HE, GT $\gamma=2.5$, GT $\gamma=0.4$, ST $\lambda=4/\alpha$, ST $\lambda=8/\alpha$, ST $\lambda=6/\alpha$) and MedGA.}
	\label{FibroidResults}
\end{figure}

\begin{figure}[!t]
	\centering
	\subfloat[]{\centering\includegraphics[width=\textwidth]{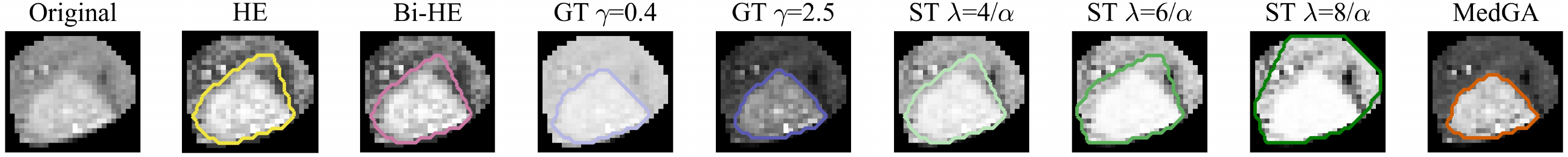}\label{brainResultsA}}\\
	\subfloat[]{\centering\includegraphics[width=\textwidth]{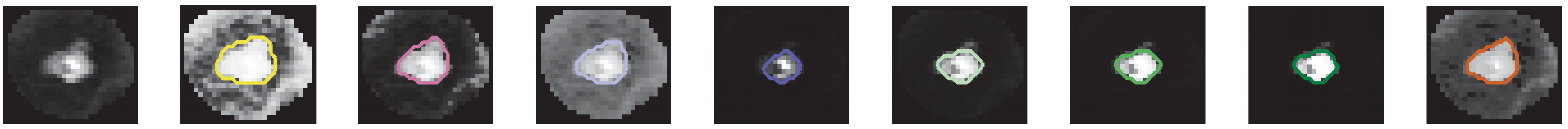}\label{brainResultsB}}\\
	\caption[Segmentation results achieved by MedGA and the competing methods on brain tumors]{Segmentation results on the the brain tumors in Figs. \ref{InputBrainA} and \ref{InputBrainB}, achieved by the processing pipeline in Fig. \ref{flowdiagram_MedGA}b by exploiting the state-of-the-art image pre-processing approaches (namely: HE, Bi-HE, GT $\gamma=2.5$, GT $\gamma=0.4$, ST $\lambda=4/\alpha$, ST $\lambda=8/\alpha$, ST $\lambda=6/\alpha$) and MedGA.}
	\label{BrainResults}	
\end{figure}

\begin{figure}[!t]
	\centering
	\subfloat[]{\includegraphics[width=.25\textwidth]{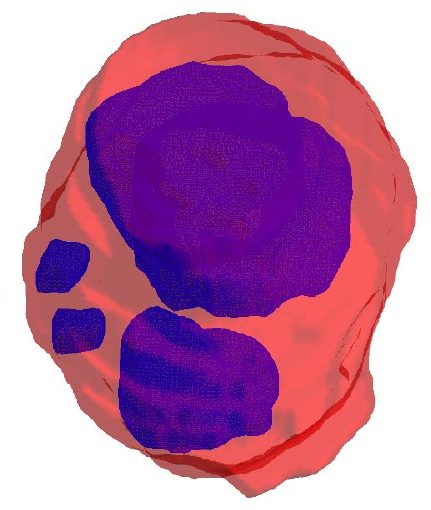}\label{renderingFibroid1}}\qquad
	\subfloat[]{\includegraphics[width=.25\textwidth]{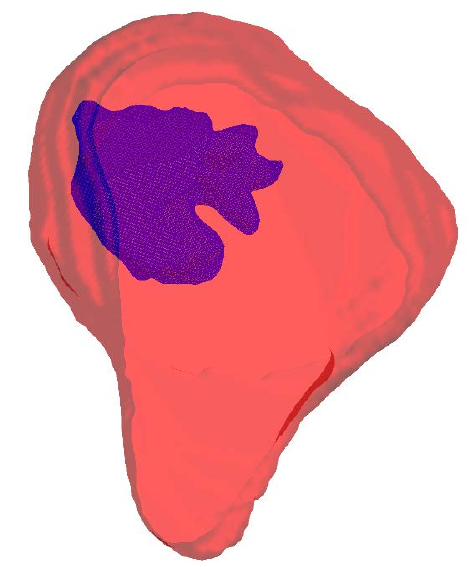}\label{renderingFibroid2}}\\
	\subfloat[]{\includegraphics[width=.25\textwidth]{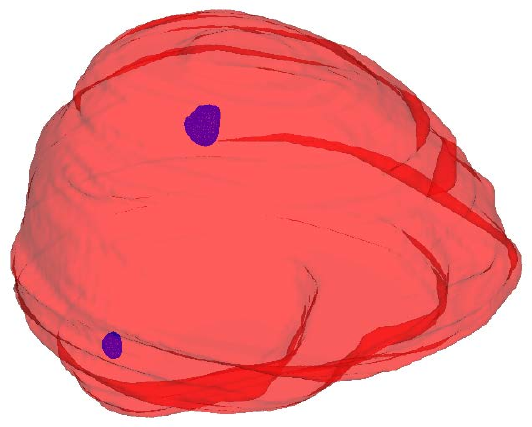}\label{renderingBrain1}}\qquad
	\subfloat[]{\includegraphics[width=.25\textwidth]{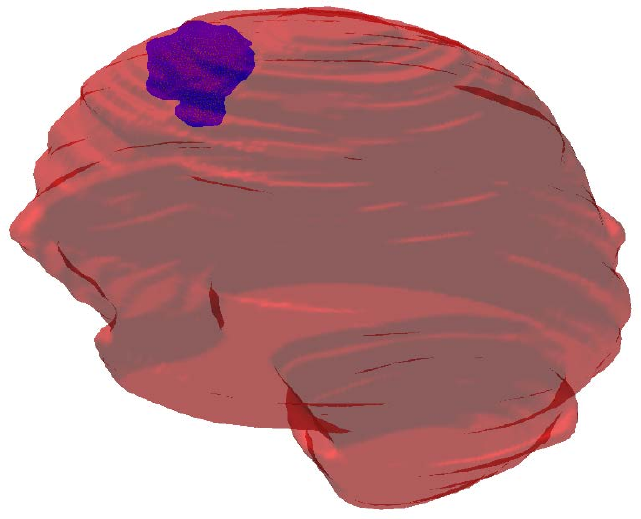}\label{renderingBrain2}}\\
	\caption[Tridimensional reconstruction of the segmented ROIs]{Tridimensional reconstruction of the segmented ROIs (blue volumetrics models) in their real location with respect to the enclosing organ (transparent red surface): (a, b) uterine fibroids within the uterus, segmented using the method in \cite{militelloCBM2015}; (c, d) brain tumors within the whole brain, achieved using a skull stripping algorithm.
	Transparent surfaces are rendered with alpha blending ($\alpha  = 0.40$).}
	\label{volumeRenderings}
\end{figure}

\paragraph{Results}
This section presents the experimental results achieved by MedGA, according to the evaluation metrics defined in Appendix \ref{sec:segEval}.
In order to achieve a comprehensive comparison between MedGA and the other pre-processing techniques listed above, we exploited the entire set of MRI data consisting in $18$ patients affected by uterine fibroids and $27$ brain metastatic cancers.
Figs. \ref{FibroidResults} and \ref{BrainResults}  show two examples of uterine fibroid and brain tumor MR images, respectively, which were pre-processed by means of the comparison methods considered in this work and segmented using the processing pipelines described above.
Fig. \ref{volumeRenderings} shows two pairs of examples concerning tridimensional reconstructions of the ROIs, i.e., uterine fibroids and brain tumors, allowing us to display their actual locations in the whole uterus and brain (represented by means of a transparent red surface), respectively.
This visualization allows for an intuitive and comprehensive representation of complex data \cite{walter2010}.

\begin{figure}[t]
	\centering\includegraphics[width=1\textwidth]{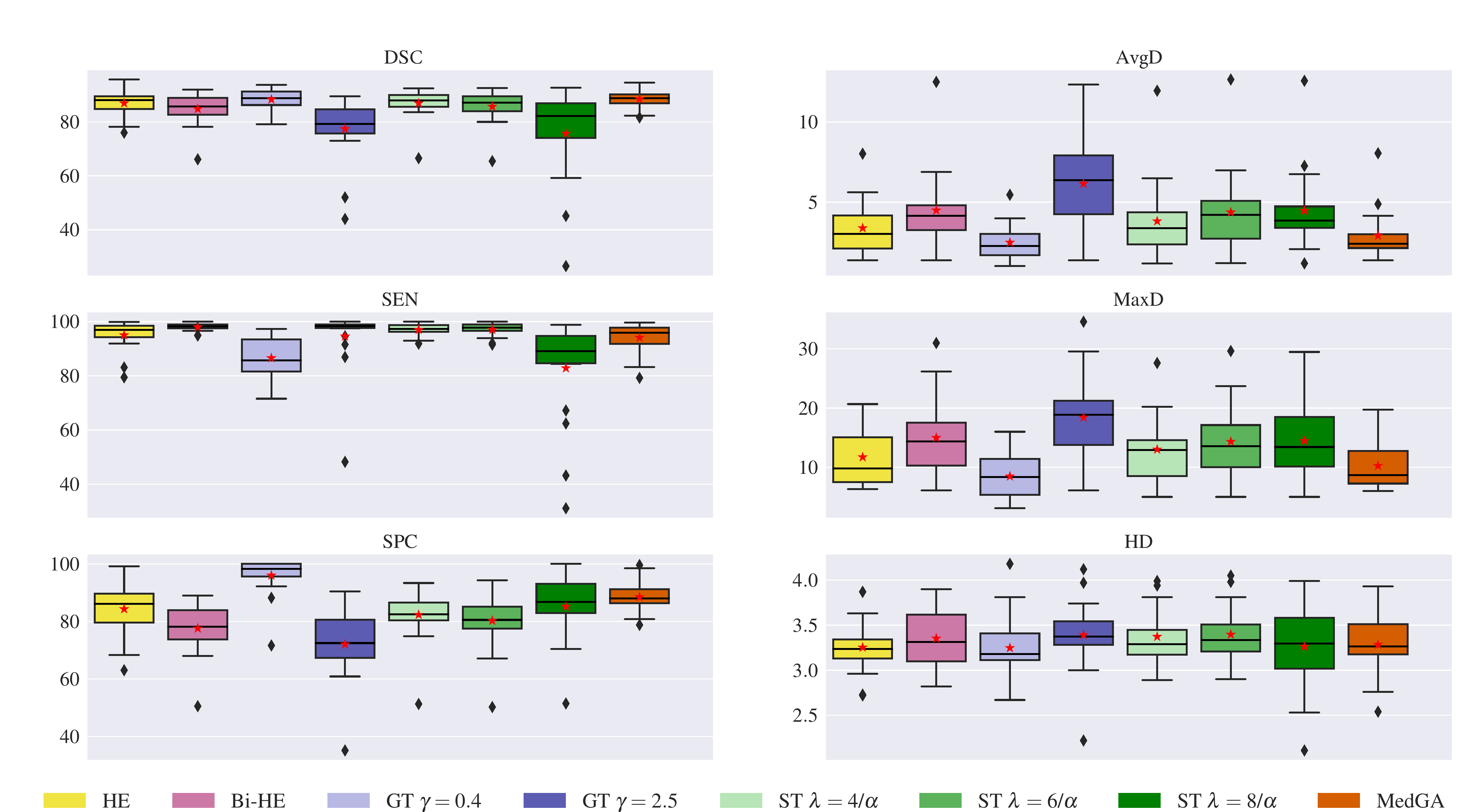}
	\caption[Boxplots of the overlap-based and distance-based metrics obtained by MedGA and the competitor methods on the uterine fibroid MRI dataset]{Boxplots of the overlap-based and distance-based metrics (left and right columns, respectively) obtained on the MRI dataset composed of $18$ patients affected by uterine fibroids who underwent MRgFUS treatment. The lower and the upper bounds of each boxplot represent the first and third quartiles of the statistical distribution, respectively. The median (i.e., the second quartile) and the mean values are represented by a black solid line and a red star, respectively. Whisker value is $1.5$ in all cases, and outliers are displayed as black diamonds.}
	\label{boxplotFibroidSeg}	
\end{figure}

\begin{figure}[t]
	\centering \includegraphics[width=\textwidth]{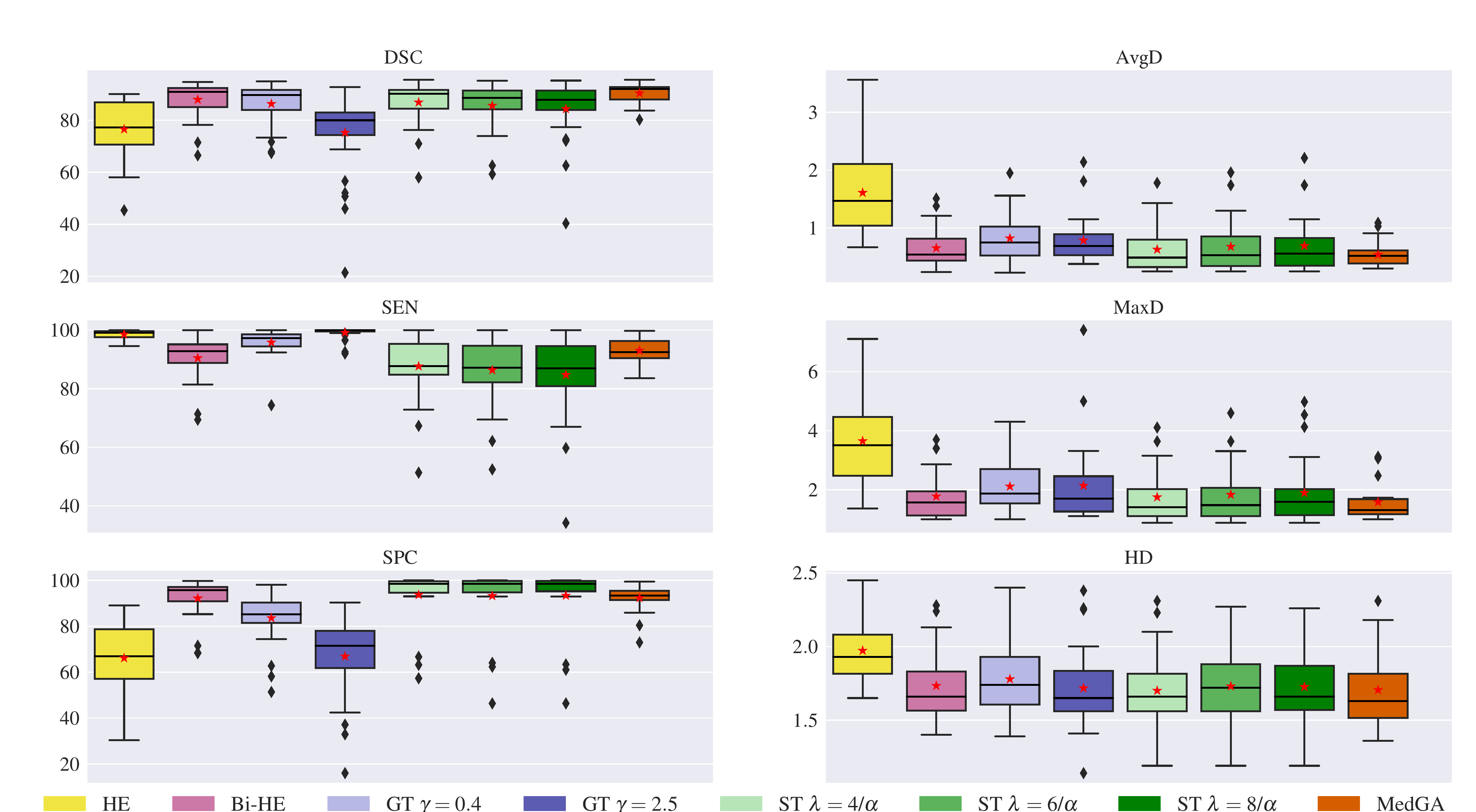}
	\caption[Boxplots of the overlap-based and distance-based metrics obtained by MedGA and the competitor methods on the brain tumor MRI dataset]{Boxplots of the overlap-based and distance-based metrics (left and right columns, respectively) obtained on the MRI dataset composed of $27$ brain metastatic tumors underwent stereotactic neuro-radiosurgery. The lower and the upper bounds of each boxplot represent the first and third quartiles of the statistical distribution, respectively. The median (i.e., the second quartile) and the mean values are represented by a black solid line and a red star, respectively. Whisker value is $1.5$ in all cases, and outliers are displayed as black diamonds.}
	\label{boxplotBrainSeg}	
\end{figure}

The quantitative segmentation results achieved by using the pipeline in Fig. \ref{flowdiagram_MedGA}a, employing the different pre-processing approaches, on the analyzed MRI dataset composed of $18$ patients affected by uterine fibroids are depicted in the boxplots in Fig. \ref{boxplotFibroidSeg}, reporting both overlap-based and distance-based metrics values.
Analogously, the boxplots concerning the segmentation results, achieved by using the pipeline in Fig. \ref{flowdiagram_MedGA}b on the analyzed MRI dataset consisting in $27$ brain metastases, are shown in Fig. \ref{boxplotBrainSeg}.

In the literature, it has been shown that a \emph{DSC} above $70\%$ is generally regarded as a satisfactory level of agreement between two segmentations (i.e., manual and automated delineations) in clinical applications \cite{zijdenbos1994}.
Since the MR image segmentation methods above desscribed obtain a \emph{DSC} appreciably higher than $70\%$ regardless the pre-processing technique, we can consider that the processing pipelines in Figs. \ref{flowdiagram_MedGA} are clinically valuable, so allowing for a fair comparison on segmentation performance among the state-of-the-art pre-processing algorithms.

In both cases, the segmentation results concerning the images pre-processed using MedGA achieved the highest mean and median \emph{DSC} values, with low standard deviation.
GT with $\gamma = 0.4$ and Bi-HE obtained the second best performances for uterine fibroid and brain tumor MR image segmentation, respectively.
So, we can claim that MedGA shows the highest accuracy and reliability in the two considered MRI analysis tasks.
This evidence is also confirmed by the boxplots, where the distributions for MedGA present significantly less than 10\% outliers in all the overlap-based metrics, thus evidencing extremely low statistical dispersion.
As a matter of fact, MedGA is the only technique that significantly supports the IOTS algorithm in both dark (i.e., uterine fibroid NPV) and bright (i.e., brain tumor enhancement region) ROI extraction.
In agreement with the image enhancement results discussed in Section \ref{sec:MedGAenhancement}, GT with $\gamma = 0.4$ considerably outperforms GT with $\gamma = 2.5$.
The decreasing trend, related to ST when the value of $\lambda$ increases, is also confirmed.
ST with $\lambda=4/\alpha$ achieved good results in both cases.
Brain tumor MR images pre-processed by means of HE achieved low \emph{DSC} values, but better results are obtained on uterine fibroid MR segmentation with respect to Bi-HE.
Overall, the achieved segmentation performance shows the great accuracy and reliability of the proposed evolutionary method.
Considering \emph{SEN} and \emph{SPC}, MedGA yielded the best trade-off between these two often conflicting  measures that should be always considered and combined together.
These metrics reveal that the other techniques could involve over-estimation or under-estimation of the segmented regions.

The achieved spatial distance-based indices are consistent with overlap-based metrics, also observing the corresponding boxplots shown in Figs. \ref{boxplotFibroidSeg} and \ref{boxplotBrainSeg}, respectively.
Hence, MedGA allows also for accurate results in terms of distance between the automated and the manual boundaries.
Note that generally the boxplots pertaining to MedGA results present the lowest statistical dispersion (in terms of box width and number of outliers), so implying a lower standard deviation with respect to the conventional techniques.
Therefore, the use of MedGA as a pre-processing step allows for considerably robust and reliable segmentation results. 
These experimental findings are endorsed by the segmentation examples shown in Figs. \ref{FibroidResults} and \ref{BrainResults}.

\paragraph{Conclusion}
We proposed MedGA \cite{rundo2018MedGA1,rundo2018MedGA2}, a novel Soft Computing method based on Evolutionary Computation, designed \emph{ad hoc} for enhancing MR image segmentation results of MRI data characterized by an underlying bimodal histogram.
This image pre-processing approach exploits a GA that aims at emphasizing bimodal histogram separation, and consequently optimize the subsequent segmentation phase based on the efficient IOTS algorithm \cite{ridler1978,trussell1979}.

In Pattern Recognition, among the low-level intensity-based techniques, the most straightforward unsupervised image segmentation technique is global thresholding \cite{xue2012}.
In order to overcome the limitations related to the assumptions underlying threshold selection methods and automatically determine a suitable optimal threshold, MedGA tackles this complex problem, introducing a fitness function tailored to better separate the two underlying sub-distributions of the gray level intensities.
Unlike the traditional image enhancement techniques that, generally, improve the contrast level of the whole image, MedGA focuses on MR image sub-regions characterized by a roughly bimodal histogram, making it valuable in clinical contexts, especially involving CE-MRI analysis.
As a matter of fact, the Soft Computing approaches presented in \cite{draa2014,hashemi2010} explicitly consider in the fitness function both the number of edge pixels and the intensity of these pixels, thus achieving high $\#$DE values that would consistently lead to over-enhanced images, possibly yielding also inaccurate ROI segmentations.

MedGA was integrated into two processing pipelines concerning different clinical contexts that require MR image analysis: (\textit{i}) uterine fibroid segmentation in MRgFUS treatments, and (\textit{ii}) brain metastatic cancer segmentation in neuro-radiosurgery therapy.
Overall, the MedGA pre-processing outperformed, in terms of image quality and segmentation accuracy, the conventional image enhancement approaches.
According to the achieved experimental results, MedGA was shown to be an appropriate and reliable solution when employed as a medical image pre-processing method.
Even considering the statistical dispersion of the segmentation evaluation metrics, MedGA achieved the most robust and repeatable segmentation results.
More generally, in addition to the clinical contexts investigated here, the application of MedGA can be extended also to real-world problems involving the analysis of images characterized by an underlying bimodal histogram, as in the case of bright-field and fluorescence microscopy imaging \cite{meijering2012}.

MedGA has been currently used only for off-line image analysis.
Our Python-coded implementation requires approximately two minutes for each image (running on a computational platform equipped with a a $6$-core Intel\textsuperscript{\textregistered} Xeon\textsuperscript{\textregistered} E5-2440 CPU at $2.40$ GHz, $16$ GB RAM, and CentOS 7 operating system), using the selected GA parameters.
In the case of tomography image stack analysis, we achieved a sublinear speed-up with respect to the number of the available cores, by developing a Master-Slave version to distribute on multiple cores the computations pertaining to different slices \cite{pinho2013}.
An additional speed-up could be definitely obtained by porting the current Python code into a faster compiled programming language (e.g., C/C++) \cite{behnel2011}.
Consequently, leveraging efficient programming languages and HPC paradigms, MedGA may become a clinically feasible solution as a pre-processing step in real-time radiology applications.

In the near future, we plan to apply MedGA as an image pre-processing phase also in other clinical contexts requiring MR image analysis and segmentation to provide useful insights for differential diagnosis and prognosis, such as in the case of breast cancer \cite{alNajdawi2015,hassanien2014} and meningiomas \cite{hale2018}, also for differentiating tumor grade.

\section{Particle Swarm Optimization}
\label{sec:PSO}

Swarm Intelligence studies the collective behavior of decentralized, self-organized natural or artificial systems \cite{eberhart2001}.
These models consist typically in a population of simple agents interacting locally each other and with their environment.
The agents follow very simple rules and, although there is no centralized control structure dictating how individual agents should behave, local interactions among such agents---often affected by a certain degree of stochasticity---lead to a complex intelligent emergent global behavior, with effects that would not have been expected by each individual.

PSO is a population-based stochastic optimization algorithm, introduced in 1995 by Kennedy and Eberhart \cite{kennedy1995}, which searches an optimal solution in the computable search space \cite{eberhart1995}.
This global optimization technique, which is inspired by the collective movement of bird flocks and fish schools, results in a metaheuristic for solving non-linear optimization problems \cite{tangherloni2018ASoC}.
Generally, such problems cannot be solved exactly by an explicit method because in practice the mathematical expression of the objective function, also called fitness function $\mathcal{F}:\Omega \rightarrow \mathbb{R}$, is not available, where $\Omega$ represents the search space.
Instead of computing the optimum's position $\textbf{x}^* \in \Omega$, a sufficiently good not necessarily optimal point  $\hat{\textbf{x}}^* \in \Omega$, called sub-optimal solution, is obtained by a metaheuristic in the given search space.

PSO can be also seen as an evolutionary technique which, in contrast to GAs \cite{holland1992} and traditional ESs \cite{back1993} that use the competitive characteristics of biological survival, exploits cooperative and social aspects.
Starting from a widely diffused population (swarm), individual components (particles) tend to move through the search space, eventually clustering in regions where minima are identified.
Briefly, PSO simulates natural movement evolution for searching a solution with higher quality.
This population-based algorithm is suitable for optimization problems with solutions encoded as real-valued vectors (i.e., $\mathbf{x}_i \in \mathbb{R}^D$, with $i=1,2, \ldots, N$) \cite{kennedy1995}.

In the traditional PSO formulation, a \textit{swarm} of $N$ individuals (\textit{particles}) moves inside a bounded $D$-dimensional search space to the aim of identifying the optimal solution, collecting and sharing information about the best position found, with respect to a user-defined fitness function, and cooperating to identify the optimal solution for a complex problem.
Typically, the particles are randomly initialized in the search space with a uniform distribution.
The movement of particles in the (bounded) search space is influenced by two different attractors: (\textit{i}) the social component that leads particles towards the (current) best position $\mathbf{g} \in \mathbb{R}^D$ identified by the whole swarm (or by a specified particle's neighborhood), and (\textit{ii}) the cognitive component that drives each particle towards its own best position $\mathbf{b}_i \in \mathbb{R}^D$ found so far.
Social and cognitive attractions are weighted by means of two constants, named the social ($c_\text{soc} \in [0,2]$) and cognitive ($c_\text{cog} \in [0,2]$) factors, respectively.
In order to prevent a chaotic exploration of the search space, the movement of particles is further weighted by using an inertia factor $w$.
Since a deterministic movement of particles would probably lead to local optima, both attractors are multiplied by two vectors $\mathbf{r}_1$ and $\mathbf{r}_2$ of random numbers, uniformly sampled from the interval $(0,1)$.
In addition, to avoid a chaotic behavior of the swarm, the change of particle velocity is attenuated by an inertia weight $w \in \mathbb{R}^+$.

Formally, the velocity of the $i$-th particle at iteration $t$ is updated as follows:
\begin{small}\begin{align}
\label{eq:update_standard}
\mathbf{v}_i(t) &= w \cdot \mathbf{v}_i(t-1) + \\
&+ c_\text{soc} \cdot \mathbf{r}_1 \odot \left( \mathbf{x}_i(t-1) - \mathbf{g}(t-1) \right) + \nonumber \\
&+ c_\text{cog} \cdot \mathbf{r}_2 \odot \left( \mathbf{x}_i(t-1) - \mathbf{b}_i(t-1) \right) \nonumber,
\end{align}\end{small}\noindent
where $\odot$ denotes the component-wise multiplication operator (i.e., Hadamard product).
Afterwards, the positions of particles are updated by calculating:
\begin{equation}
\label{eq:posUpdate}
\mathbf{x}_i(t) = \mathbf{x}_i(t-1) + \mathbf{v}_i(t)\text{, for all }i=1, \dots, N.
\end{equation}

The quality of the positions within the search space encoded by each particle is evaluated by means of a \emph{fitness function} $\mathcal{F}(\cdot)$.
The hyper-surface described by the fitness values over the set of feasible solutions is called the fitness landscape.
The movement of particles is the result of the two attractors, which are determined according to the fitness function $\mathcal{F}$ defined for the problem under investigation: during each iteration of the PSO, $\mathbf{b}_i$ and $\mathbf{g}$ are updated according to the new positions.
Summarizing, the $i$-th particle is able to memorize its own best position $\mathbf{b}_i$ from the past, thus creating a kind of “nostalgia” to return there.
When the fitness function is to be maximized, if $\mathcal{F}(\mathbf{x}_i(t))$, then $\mathbf{b}_i(t) \gets \mathbf{x}_i(t)$, otherwise the previous individual best position $\mathbf{b}_i(t-1)$ is kept. On the other hand, the particle $\mathbf{x}_i$ has also a social behavior, because it follows the swarm in its global best position $\mathbf{g}(t) \gets \text{arg} \max\limits_{\forall \mathbf{b}_i(t)}{\{\mathcal{F}(\mathbf{b}_i(t))\}}$.

This search procedure may bring some particles to regions outside the feasible space of solutions for the optimization problem.
To avoid this situation, the search space is bounded (according to the domain knowledge) and some specific boundary conditions are exploited to keep the particles inside the feasible space of solutions.
For instance, the so-called \textit{damping} boundary conditions could be considered, in which a random bounce on the limit of the search space is simulated to let the particle go back to the feasible space of solutions \cite{xu2007bound}.
The boundaries of the $d$-th dimension of the search space are denoted by $\beta_{\text{min}_d}$ and $\beta_{\text{max}_d}$, for each $d=1, \dots, D$, with $\beta_{\text{min}_d}, \beta_{\text{max}_d} \in \mathbb{R}$.
The particle velocity is usually clamped to a maximum value $v_{\text{max}_d} \in \mathbb{R}^+$ along each $d$-th dimension of the search space (with $d=1, \dots, D$), to avoid divergence during the optimization process \cite{poli2007}.
The velocity of particles may diverge during the optimization process; for this reason, its value is usually clamped to a maximum value $v_{\text{max}_d} \in \mathbb{R}^+$ along each $d$-th dimension of the search space, with $d=1, \dots, D$ \cite{poli2007pso}.

Note that the values used for the $N$, $c_\text{soc}$, $c_\text{cog}$, $w$ parameters, and the vector of the allowed maximum velocity values $\mathbf{v}_\text{max}: \mathbf{v}_i \in [-\mathbf{v}_\text{max},\mathbf{v}_\text{max}]$, typically set by the user and strongly problem-dependent, have a deep impact on the performance of PSO \cite{poli2007pso}, both in terms of convergence speed and quality of the best solution found.
PSO performance strongly depends on a proper choice of the aforementioned settings (i.e., $N$, $c_\text{soc}$, $c_\text{cog}$, $w$, the maximum number of iterations $T_\text{max}$) \cite{eberhart2000,trelea2003}, and the vector of the allowed maximum velocity values $\mathbf{v}_\text{max}$; in addition, even the initial distribution of particles can largely affect the results \cite{cazzaniga2015}.
Unfortunately, these settings are problem-dependent, so that a proper setting for PSO can be established only by having an in-depth knowledge of the shape and roughness of the fitness landscape.
These parameters are typically tuned by the user and may heavily influence the convergence speed and quality of the best solution found \cite{poli2007pso}.
In general, the identification of the best settings is a complex, lengthy and time-consuming meta-problem. 
Hence, an intense research has been devoted to self-tuning, settings-free, or adaptive variations of PSO and other evolutionary and Swarm Intelligence based algorithms.
TRIBES is a PSO version able to automatically modify the particles' behavior, as well as the topology of the swarm, resulting in a settings-free version of the algorithm \cite{clerc2006}.
Other examples are represented by the Parameter-Less Evolutionary Search that dynamically selects the settings by exploiting the statistical properties of the population \cite{papa2013}.
More recently, Nobile \emph{et al}. proposed in \cite{nobile2015ppso,nobile2018fstPSO} a novel PSO variant that exploits a set of fuzzy rules \cite{zadeh1965} to dynamically change, for each particle of the swarm, the values of $w$, $c_\text{soc}$, and $c_\text{cog}$.
In this way, PSO's \textit{reactive} individuals are turned into \textit{proactive} optimizing agents \cite{nobile2015ppso}.
The resulting version of the PSO, called Fuzzy Self-Tuning PSO (FST-PSO) \cite{nobile2018fstPSO}, achieved promising performance on benchmark functions \cite{tangherloni2017cec} as well as in the parameter estimation of biochemical systems \cite{nobile2018cec}.
Finally, in order to avoid particles to get trapped in local optima, reboot strategies for PSO that reinitialize particle positions may be useful \cite{spolaor2017}.

In addition to the basic PSO algorithm, several modifications are possible to enhance the achieved results:
\begin{itemize}
    \item the best position in a neighborhood of each particle $\mathbf{x}_i$ could be considered, named $\mathbf{b}'_i$, instead of the global best position $\mathbf{b}$ \cite{suganthan1999}. Therefore, $\mathbf{b}'_i$ replaces $\mathbf{g}$ in the updating rules in Eq. (\ref{eq:update_standard});
    \item Evolutionary Computation may also be integrated, giving rise to a hybrid PSO. Genetic operators, such as mutation and crossover, can be used to preserve exploration capability in the various iterations, especially in the later stage of the evolution process \cite{talbi2004}. Such a case could occur even at the early stage for a particle that is very close to the global best position $\mathbf{g}$, and the velocity will tend to zero. To avoid premature convergence and stagnation, after the particle positions updating, pairs of particles are selected for crossover with probability $p_c$, which is a random number generated uniformly in the interval $[0,1]$. For each pair, two child particles are generated by a crossover rule and replace their parents, maintaining the population size constant to $N$ \cite{wachowiak2004}. The authors of \cite{lovbjerg2001} proposed the following crossover rule for parents $\mathbf{x}_i$ and $\mathbf{x}_j$, with $i \neq j$:
    \begin{equation}
    \label{eq:crossPop}
    \begin{cases}
    \mathbf{x}'_i = p_c\mathbf{x}_i + (1-p_c)\mathbf{x}_j \\
    \mathbf{x}'_j = p_c\mathbf{x}_j + (1-p_c)\mathbf{x}_i
    \end{cases}.
    \end{equation}
    The velocities $\mathbf{v}_i$ and $\mathbf{v}_j$  are also updated from the velocities of the parents $\mathbf{v}'_i = \mathbf{v}'_i \cdot \nu$, $\mathbf{v}'_j = \mathbf{v}'_j \cdot \nu$, where: $\nu = (\mathbf{v}_i + \mathbf{v}_j) / || \mathbf{v}_i + \mathbf{v}_j ||$;
    \item grouping the particles into sub-populations is a further alternative. Any clustering method can be used to perform this subdivision. Another random number $p_{\text{sp}_c} \in [0,1]$ is specified to represent the probability of intra-population crossover. Crossover among different sub-populations occurs with probability  $1-p_{\text{sp}_c}$;
    \item a constriction factor $\chi$  was also introduced in Eq. (\ref{eq:update_standard}) to control the movement of $\mathbf{x}_i$, by balancing both convergence and explosive particle movements \cite{clerc2002}. The $\chi$ coefficient must hold the following conditions:
    \begin{equation}
    \label{eq:costriction}
    \chi = \frac{2\kappa}{|2 - \varphi - \sqrt{\varphi^2 - 4\varphi}|}, \varphi=c_\text{cog} + c_\text{soc}, \varphi > 4, \kappa=[0,1].
    \end{equation}
\end{itemize}

\subsection{Multimodal biomedical image registration using PSO}
\label{sec:medImageReg}
\begin{figure}[t]
	\centering \includegraphics[width=0.8\textwidth]{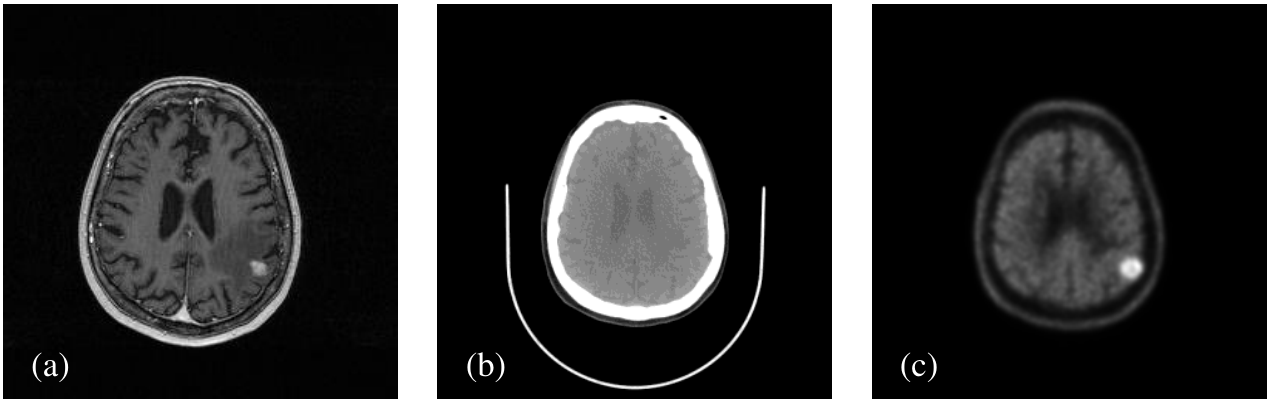}
	\caption[Different medical image modalities concerning the same subject affected by brain tumor]{Examples of different medical image modalities concerning the same subject affected by brain tumor: (a) MRI; (b) CT; (c) PET.}
	\label{fig:multiModal_medImageReg}	
\end{figure}

Image co-registration is a fundamental task in medical imaging because it enables to integrate different images into a single representation (i.e., the same reference system), allowing physicians and researchers to access at this complementary information more easily and accurately \cite{ou2014,rohr2000}.
Fused image data can improve medical diagnosis, surgery planning and simulation, as well as intra-operative navigation.
The choice of the modality depends strongly on the medical task.
In radiotherapy planning, for instance, dose calculation is based on the CT data, while tumor delineation is often better performed on the corresponding MRI scans, especially in soft-tissue imaging (e.g., brain or prostate) \cite{maes1997}.
The same anatomical district of the human body is often imaged with different modalities (Fig. \ref{fig:multiModal_medImageReg}).
In particular, morphologic images, which define the anatomy of organs or pathological tissues, can be integrated with functional ones, which describe the cellular physiology or the metabolism (i.e., PET imaging).
These images are used in a complementary fashion to gain additional insights into biological phenomena and pathologies.

Even though combining appropriately different modalities is definitely clinically useful, multimodal images even concerning the same subject generally differ by local geometrical differences.
Therefore, in order to perform quantitative and precise evaluations on biomedical imaging data, such images have to be mapped into the same coordinate system by means of the alignment process, named biomedical image co-registration. Medical image registration algorithms can be conveniently applied in: (\textit{i}) intra-modality matching, for patient follow-up (e.g., tumor response assessment over the time); (\textit{ii}) inter-modality matching, for comparisons and quantitative measurements concerning images acquired with different modalities.
The aim of this section is to provide an overview on biomedical image registration using PSO.

\subsubsection{Theoretical background on biomedical image registration}
An image co-registration stage is mandatory to integrate and quantitatively compare biomedical imaging data originating from different modalities.
As a matter of fact, image registration is able to bring the different medical datasets, concerning the same patient, into the same space.
In this way, it will be possible to make quantitative and meaningful comparisons between the ROIs, such as organs and tumors, obtained by processing several image series from different scanners and modalities \cite{pluim2003}.

Registration approaches can be principally distinguished in feature-based and intensity-based schemes \cite{rohr2000}:
\begin{itemize}
    \item feature-based techniques find correspondences between geometrical image features (e.g., points, lines, contours, surfaces) or landmarks (external or anatomical). Either image features or landmarks are first extracted from the input images, and then a transformation is established according to the correspondence between the found features. However, segmenting and finding correspondences are very difficult tasks;
    \item intensity-based techniques directly exploit image intensities to compute the transformation that maximizes a similarity metrics by searching in a certain space of transformations and comparing intensity patterns. The main advantage of these schemes is that explicit image segmentation or feature extraction is not required.
\end{itemize}

\begin{figure}[t]
	\centering \includegraphics[width=0.8\textwidth]{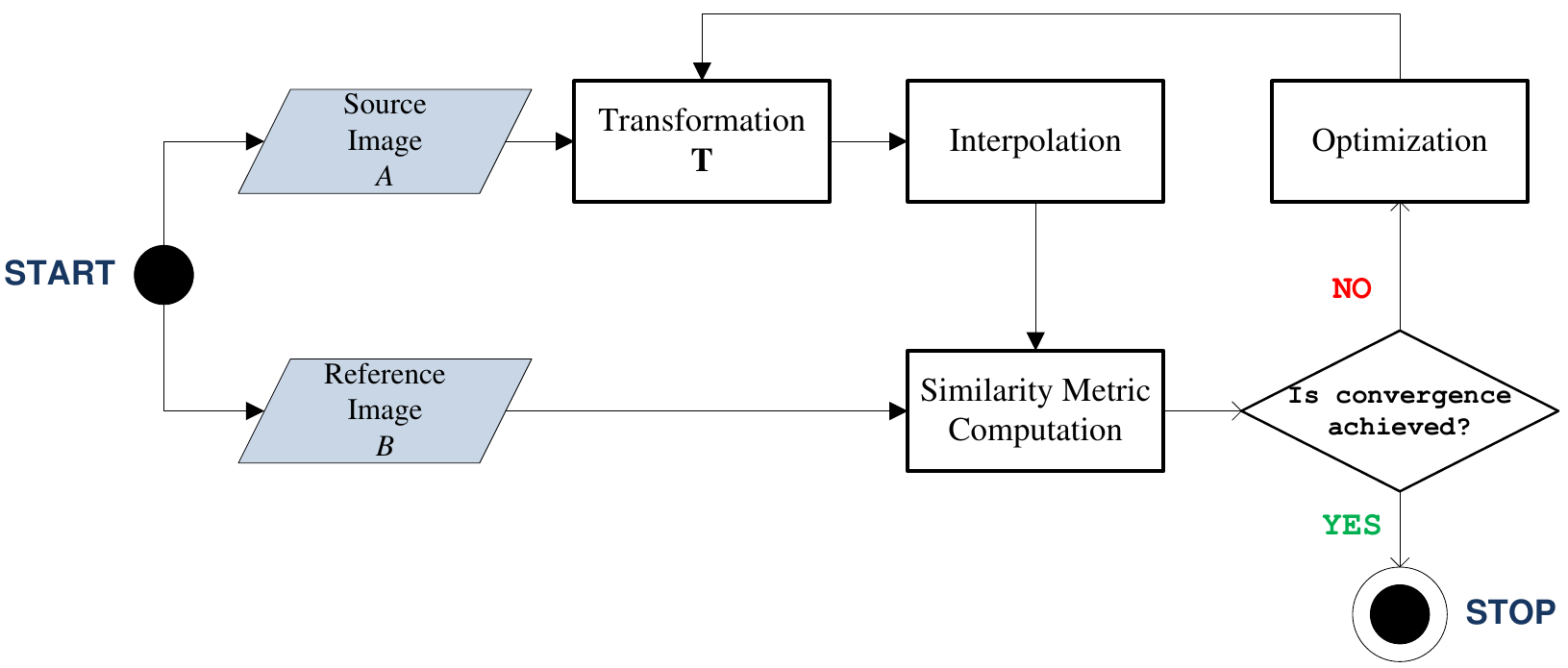}
	\caption[Flow diagram of the biomedical image registration process, by using an intensity-based technique]{Flow diagram of the biomedical image registration process, by using an intensity-based technique.}
	\label{fig:medImageReg_flowDiag}	
\end{figure}

Intensity-based registration techniques are the most suitable for applying Swarm Intelligence approaches.
Fig. \ref{fig:medImageReg_flowDiag} shows the overall flow diagram of the biomedical image registration process, by using an intensity-based technique.

From an algorithmic perspective, image co-registration involves finding the parameters, i.e., a geometric transformation matrix $\mathbf{T}$, which either maximize or minimize some objective function, also called fitness function $\mathcall{F}$.
Therefore, image co-registration can be modeled as an iterative procedure by successive refinements.
In each iteration, the current estimate of the transformation $\hat{\mathbf{T}}$ is used to calculate a similarity measure.
Afterwards, the optimization algorithm makes another (hopefully better) estimate of the transformation, evaluates the similarity measure again and continues to iterate until the convergence condition is achieved (i.e., no transformation can improve the value of the similarity metrics above a preset tolerance threshold $\varepsilon$ or the number of possible iterations $T_\text{max}$ is achieved).
The multimodal registration process starts from two images $A$ and $B$ probably characterized by different FOVs: $A$ is used as source image (floating) while $B$ is the reference image (target).
The reference imaging modality is usually selected according to the higher spatial resolution and image content. For instance, in PET/MRI registration, MR images are chosen as target images because they convey more anatomical information than PET images.

The geometric transformation $\mathbf{T}$ is defined only in the region of overlap of the image FOVs, and must take into account image sampling and resolution.
It is also important to note that the images are discrete. The discretization is determined by the sampling grids, which are different when FOVs are not the same \cite{hill2001}.
Even if the images $A$ and $B$ have exactly the same sampling grid, the grid points do not normally coincide in the volume of overlap and an interpolation step is therefore necessary.
Realignment and reslicing operations are mandatory to get a one-to-one mapping between different modality slices.
Especially, in iterative registration algorithms, an accurate interpolation method is required \cite{maes1997,pluim2003} (e.g., B-Spline interpolation \cite{thevenaz2000}). 
In intensity-based registration techniques, the three core components are: (\textit{i}) the search space (i.e., parameter space); (\textit{ii}) the similarity metrics (i.e., objective function); (\textit{iii}) the search strategy (i.e., optimization algorithm).

\paragraph{Search space}
The search space $\Omega$ is the set of potential transformations used to align the images.
Each point in the parameter space corresponds to a different estimate of the transformation.
Accordingly, the parameter space can be thought as a high-dimensionality function in which the value of each location corresponds to the value of the similarity measure for that transformation estimate. Geometric transformations may be rigid, affine, and elastic.

Three dimensional (3D) rigid-body registration has six degrees of freedom, regarding:
\begin{itemize}
    \item translations along the three axes of the reference system  $x$, $y$, and $z$, denoted by the displacements  $t_x$, $t_y$, and $t_z$, respectively;
    \item rotations around the  $x$, $y$, and $z$   axes, denoted by the angles  $\alpha$, $\beta$, and $\gamma$, respectively.
\end{itemize}

In 3D rigid-body registration, the mapping of the coordinates $\mathbf{p}_A = (x, y, z)^\top$ into the transformed coordinates $\mathbf{p}'_A = (x', y', z')^\top$ can be suitably formulated as a matrix multiplication in homogeneous coordinates with the geometric transformation matrix $\mathbf{T}$.
Some registration algorithms increase the number of degrees of freedom by allowing for anisotropic scaling (giving nine degrees of freedom) and skews (giving twelve degrees of freedom).
A transformation that includes scaling and shearing as well as the rigid-body parameters is referred to as affine, and has the important characteristics that it can be described in matrix form and that all points, straight lines and planes are preserved.
Rigid-body registration is widely used in medical applications where the structures of interest are either bony or are enclosed in bones (e.g., head, neck, pelvis, leg or spine), but the errors are likely to be larger.
The use of an affine transformation rather than a rigid-body one does not significantly increase the applicability of image registration, as there are not many organs that only stretch or shear.
Tissues usually deform in more complicated ways.
However, errors introduced by the scanner may occur, resulting in scaling or skew terms, and affine transformations are sometimes used to overcome these problems \cite{hill2001}.
For most organs in the body, many more degrees of freedom are necessary to describe the tissue deformation with adequate accuracy, thus elastic or non-rigid methods are required to cope with local differences between the images.
However, for global elastic transformation, the number of parameters to be optimized is generally too large (often many thousands) to be feasible in practice.
Therefore, two-step intensity-based registration approaches are typically used.
In the first step, the global affine medical image registration is used to establish a one-to-one mapping between the two images to be registered.
Afterwards, the images are registered up to small local elastic deformation.

\paragraph{Similarity metrics}
The similarity metrics is an indicator that quantifies the degree of closeness between features or intensity values of two images. The Sum of Squared intensity Differences (SSD), correlation coefficient, ratio image uniformity are often utilized in intra-modality registration \cite{hill2001}.
Because of the similarity of the intensities in the images being registered, these subtraction, correlation and ratio techniques are pretty intuitive. With inter-modality registration, the situation is quite different: there is, in general, no simple relationship between the intensities in the images $A$ and $B$ \cite{pluim2003}.

MI, denoted by $I(A,B)$, is an information theoretic concept for estimating the degree of dependence of the random variables $A$ and $B$, with marginal probability distributions $p_A(a)$ and $p_B(b)$, by measuring the distance between the joint distribution $p_{AB}(a,b)$ and the distribution associated to the case of complete statistical independence $p_A(a) \cdot p_B(b)$ \cite{maes1997,viola1997}:
\begin{equation}
\label{eq:mutInf}
I(A,B) = \sum\limits_{a,b}p_{AB}(a,b) \cdot \log{\frac{p_{AB}(a,b)}{p_A(a) \cdot p_B(b)}}.
\end{equation}

Let the random variables $A$ and $B$ represent the image intensity values $a$ and $b$ concerning the pairs of voxels in the two images to be registered, respectively.
Estimations for the joint distribution and the marginal distributions can be simply obtained by normalizing the joint and marginal histograms of the overlapping parts of both images.
In general, the input images are also smoothed slightly, by means of their histogram.
This makes the cost function (i.e., similarity metrics) as smooth as possible to give faster convergence and less chance of being trapped in local minima.
The intensities $a$ and $b$ are related through the geometric transformation $\mathbf{T}$.
The MI registration criterion states that the images $A$ and $B$ are geometrically aligned by the transformation $\mathbf{T}$ for which $I(A,B)$ is maximal.
Therefore, the objective of intensity-based registration is to find an estimation of the transformation $\mathbf{T}$ that best aligns the source image $A$ against the reference image $B$:  $\hat{\mathbf{T}} = \text{arg} \max\limits_{\mathbf{T} \in \Omega}{\{I(\mathbf{T}(A),B)\}}$.

The results presented by Maes \textit{et al.} \cite{maes1997} proved that sub-voxel registration differences with respect to the stereotactic reference solution can be obtained for CT/MRI and PET/MRI matching without using any prior knowledge about the gray-value content of both images and the correspondence between them.
As a matter of fact, MI is the most intensively investigated criterion for registration of intra-individual human brain images \cite{ou2014}.
NMI is also frequently used as the cost function to be optimized \cite{studholme1999}.
Especially, when misalignment can be large with respect to the imaged FOVs, a criterion invariant to image overlap statistics should be used:
\begin{equation}
\label{eq:normMutInf}
I_N(A,B) = \frac{H(A) + H(B)}{H(A,B)},
\end{equation}
where $H(\cdot)$ and $H(\cdot,\cdot)$ are the marginal and the joint entropies, respectively.

\paragraph{Optimization of the similarity metrics}
Intensity-based registration techniques determine the registration transformation $\mathbf{T}$ by optimizing a certain voxel similarity measure.
Unfortunately, parameter spaces for image registration are frequently not so simple.
There are often multiple optima within the parameter space, and the registration can fail if the optimization algorithm converges to the wrong optimum.
Some of these optima may be very small, caused either by interpolation artifacts or a local good match between features or intensities \cite{hill2001}.
As explained previously, these small optima can often be removed from the parameter space by smoothing the images before the registration.

Local methods, such as Powell's direction set method \cite{powell1964}, Nelder-Mead simplex algorithm, conjugate gradient, Levenberg-Marquardt algorithm \cite{bernon2001}, are usually employed in image registration \cite{maes1999}.
Because many similarity metrics (i.e., functions of transformation parameters) are generally irregular and rough, especially in multimodal image registration, local methods are more accurate when the initial orientation is very close to the transformation that yields the best registration \cite{wachowiak2004}.
These strategies are also susceptible to premature convergence to local optima, especially for non-smooth functions.
An approach to address this issue is to apply multiresolution techniques, whereby images are aligned at increasing resolutions with initial orientations from the previous (lower) resolution registration result.
However, these hierarchical methods frequently become trapped in local optima, as the global optimum may not be present in lower resolutions \cite{hill2001}.
Global optimization is often required for dealing with the most general and tricky situations.
Such global approaches include simulated annealing, tabu search, GAs, and ESs.
Efficiency is the primary reason that local techniques are preferred for registration. Efficient global optimization may gain acceptance if a significant improvement in accuracy can be demonstrated.

For intensity-based registration, the problem is even more complicated.
The desired optimum when registering images using voxel similarity measures is not often the global optimum, but it is one among the local optima \cite{hill2001}.
A solution to this problem is to start the algorithm in the proximity of the correct optimum, which is within the portion of the parameter space in which the algorithm is more likely to converge to the correct optimum than the incorrect global one.
In practical terms, this requires that the starting estimate of the registration transformation is reasonably close to the correct solution.

\subsubsection{Multimodal medical image registration approaches based on PSO}
In this section, multimodal medical image registration approaches that use PSO as searching strategy are described and critically reviewed according to our work in \cite{rundo2016SSCI}.
First, the introduction of an initial orientation term in the PSO formulation is explained.
Then, the different PSO modifications are described.

\paragraph{Introduction of an initial position in the standard PSO formulation}
In a large amount of the optimization techniques, such as PSO, standard test functions are employed for benchmark testing and continuous optimization algorithms assessment.
In these cases, the initial set of parameters has little importance.
However, in many practical applications, \textit{a priori} knowledge of the characteristics of $\mathbf{x}^*$ could be available for which $\mathcal{F}(\mathbf{x}^*)$ is the global optimum \cite{wachowiak2004}.
This situation is certainly true in biomedical image registration, since the users of clinical imaging systems are generally skilled physicians.
These clinicians can supply a trustworthy indication of the correct orientation, by choosing an accurate initial transformation.
Although co-registration is required because of both medical image complexity and human subjectivity or error, registration algorithms can definitely benefit from an accurate initial guess.

This may also occur with PSO, as $\mathbf{v}_\text{max}$ and constriction coefficient $\chi$ only prevent particle straying from the region of feasible solutions.
However, if the particles were arranged according the user's initial orientation, while swarming, they may have a higher probability of discovering a region that contains $\mathbf{x}^*$.
Briefly, in addition to the swarming effect around the current global best $\mathbf{g}$ and each particle recollection of its personal best $\mathbf{b}_i$, the initial orientation $\mathbf{x}_\text{init}$ can be also introduced into the velocity $\mathbf{v}_i$ of each particle \cite{wachowiak2004}.
The formulation in Eq. (\ref{eq:update_standard}) is modified into Eq. (\ref{eq:update_initPos}):
\begin{small}\begin{align}
\label{eq:update_initPos}
\mathbf{v}_i(t) &= w(t) \cdot \mathbf{v}_i(t-1) + c_\text{soc} \cdot \mathbf{r}_1 \odot \left( \mathbf{x}_i(t-1) - \mathbf{g}(t-1) \right) +  \\
&+ c_\text{cog} \cdot \mathbf{r}_2 \odot \left( \mathbf{x}_i(t-1) - \mathbf{b}_i(t-1) \right) + c_\text{ret} \cdot \mathbf{r}_3 \odot \left( \mathbf{x}_\text{init} - \mathbf{x}_i(t-1) \right)\nonumber,
\end{align}\end{small}\noindent
where $c_\text{ret}$ is the acceleration constant for returning to the initial orientation and $\mathbf{r}_3$ is a uniformly distributed random number in $[0,1]$.
Stochastic and evolutionary global optimization techniques, including PSO, can generally discover the promising region or the “basin of attraction” in the search landscape.
However, they typically exhibit slow convergence to the global optimum $\mathbf{x}^*$, even though the complexity of similarity metrics computation needs fast convergence.
For these reasons, a local method is applied to the best point in the promising region found by the PSO \cite{wachowiak2004}.
Powell’s direction set algorithm \cite{powell1964} is well-suited for this purpose, because it does not require any derivative computation \cite{maes1999}.

\paragraph{Versions of medical image registration approaches using PSO}
Several literature works addressed the issues related to biomedical image registration using PSO.
In addition to the “traditional” plain PSO, other versions, including initial orientation knowledge and/or ESs, have been proposed in the literature.
These more advanced registration approaches aim at balancing out exploration and exploitation, avoiding premature convergence.

In \cite{wachowiak2004}, three main variants of biomedical image registration based on PSO are described and compared:
\begin{enumerate}
    \item[(\textit{i})] hybrid PSO with crossover operators for positions and velocities updating \cite{lovbjerg2001}. However, for normal velocity updates, Eq. (\ref{eq:update_standard}) with the constriction factor $\chi$ is exploited. The convergence criteria are $T_\text{NoImprove} = 20$ iterations in which there is no improvement in $\mathcal{F}(\textbf{x})$, or reaching the maximum number of iterations $T_\text{max}$. After the PSO convergence, the Powell’s local optimization method \cite{powell1964} is applied to the best point in the swarm;
    \item[(\textit{ii})] hybrid PSO with crossover and sub-populations, analogous to the previous one except that five sub-populations were initially determined with the $K$-Means clustering algorithm \cite{lloyd1982}. Additionally, after the convergence, the Powell’s method is applied to the best points in each sub-population, resulting in the final registration transformation;
    \item[(\textit{iii})] PSO with constriction coefficient and relaxed convergence criteria is the most “controlled” of the three techniques. A “loose” local optimization through thw Powell’s method is applied to $\mathbf{x}_\text{init}$, resulting in $\mathbf{x}'_\text{init}$ around which particles are generated. PSO with the constriction factor $\chi$ is applied, the velocities are updated according to Eq. (\ref{eq:update_initPos}) and the convergence criteria are more relaxed. If during some iteration $\mathcal{F}(\mathbf{x}_i) < \mathcal{F}(\mathbf{g})$, then $\mathbf{g}$ is set to $\mathbf{x}_i$, but if $||\mathbf{x}_i-\mathbf{g}|| < \varepsilon$ the iteration is still considered to be a non-improving iteration. Convergence is faster, as the iteration counter is not reset to zero for improving points very close to $\mathbf{g}$. The Powell’s local optimization is applied to $\mathbf{g}$ after the convergence (a function value change less than $0.005$).
\end{enumerate}

Although the constriction coefficient prevents the particles from straying out of the space of feasible solutions, the particles have a greater probability of being drawn out of local optima by the additional term.
The authors of \cite{wachowiak2004} argued that the modifications were designed \textit{ad hoc} for image registration and this term improved registration accuracy.
In other applications, however, there may be no prior knowledge of the location of the global optimum.
In these cases, the last version may prevent particles from moving towards the global optimum, and Eq. (\ref{eq:update_standard}) should be used for the velocity update.
If a feasible region, wherein the correct transformation likely lies, cannot be identified, then the other PSO hybridizations (i.e., the use of crossover and sub-populations) are recommended.
Some knowledge of the correct orientation might greatly improve the search.
Both the initial orientation term and the constriction coefficient prevent the search from straying too far from the global optimum.
Hybridization with ES operators appears to improve accuracy by diversifying particle locations.
As shown in the last version, convergence criteria during the global search can be relaxed, as local optimization can find the global optimum if a particle is sufficiently close to it.

The authors of \cite{schwab2015} investigated four different plain PSO versions for the registration of the images, without integrating other optimization approaches: (\textit{i}) standard PSO corresponds to the standard PSO introduced in 2007 by Bratton and Kennedy \cite{bratton2007};
(\textit{ii}) standard PSO with variable inertia weight represents a modification of the earlier described standard PSO based on \cite{wachowiak2004}, where the inertia weight $w(t)$ monotonically decreases during the iterations;
(\textit{iii}) standard PSO with initial orientation, relaxed convergence criteria and constriction coefficient is another alteration presented by Wachowiak \textit{et al.} \cite{wachowiak2004}.
This version includes the initial orientation of the volumes to one another, according to Eq. (\ref{eq:update_initPos});
(\textit{iv}) PSO with constriction coefficient, relaxed convergence criteria and variable influence of the initial orientation is a modification of the previous version.
Since the involvement of the initial orientation may prevent the convergence of the swarm, the influence of the component $\mathbf{x}_\text{init}$ in Eq. (\ref{eq:update_initPos}) should decrease in each iteration:
\begin{equation}
    \label{eq:decInitPos}
    \begin{cases}
    c_\text{soc}(t) = c_\text{soc}(t-1) - \Delta \varphi \\
    c_\text{ret}(t) = c_\text{ret}(t-1) - \Delta \varphi \\
    \varphi = c_\text{cog} + c_\text{soc} + c_\text{ret}\\
    \Delta \varphi = \frac{c_{\text{ret}_\text{min}}-c_{\text{ret}_\text{max}}}{T_\text{max}} \xrightarrow{c_{\text{ret}_\text{min}} = 0} \Delta \varphi = \frac{c_{\text{ret}_\text{max}}}{T_\text{max}}
    \end{cases}.
    \end{equation}

Chen \textit{et al.} \cite{chen2009reg} investigated an extension of the PSO to a Hybrid Particle Swarm Optimizer (HPSO), by integrating two methods from the GAs into the standard PSO. The authors argued that classical PSO is convenient for 2D-2D registration, but is less efficient for 3D-3D alignment. They solved this problem by using a hybrid algorithm. The chosen optimization metrics is again MI.
An alternative non-linear 2D-2D affine registration technique for MR and CT modality images of human brain slices was presented in \cite{das2011}, using a correlation function as objective function.
Both GA and PSO schemes were considered in a multiresolution domain (using the Haar wavelet transform) to decrease the sensitivity of the registration procedure to local maxima and achieve an idea of the initial orientation of the images to be registered.

Finally, in \cite{abdel2017}, a hybrid approach is proposed, which uses a modified MI by including the spatial image information into the computation of the similarity metrics by means of a linear combination of image pixel intensity and image Gradient Vector Flow (GVF) \cite{xu1998} intensity.
The values of the weighing factors are determined using numerical experimentation.
By so doing, statistical and spatial information can be integrated.

\subsubsection{Discussion}
Regarding possibile experimental findings that can be drawn from the state-of-the-art works, an accurate and fair comparison among the different literature approaches is not so straightforward.
Indeed, different medical datasets were used in experimental trials, due to the unavailability of public medical benchmarks for intensity-based image registration.
In addition, all the approaches used a small number of images for 2D or 3D registration tests, making the experimental findings much less significant.

Swarm Intelligence techniques have been shown to be effective and powerful in a wide variety of computer science areas.
Depending on the problem nature, continuous or discrete search spaces may be properly defined.
Accordingly, the different Swarm Intelligence approaches, such as PSO, may provide more efficient solution encoding either in continuous or discrete optimization problems.
However, although each optimization technique was first designed for a particular purpose, the majority of evolutionary algorithms were adapted from continuous to discrete search space, and vice-versa.
Other Swarm Intelligence techniques, such as the Bat Algorithm (BA) \cite{yang2010} and the ABC algorithm \cite{karaboga2007}, could be useful to improve the PSO performance in biomedical image registration.
In this context, the elastic registration \cite{fornefett2001} is still a challenging problem because of the thousands of parameters to be optimized \cite{klein2010elastix}.
However, to the best of my knowledge, no literature work has addressed yet this challenging issue using Swarm Intelligence techniques.

In conclusion, although more accurate and robust comparisons must be performed with other global and local optimization paradigms, PSO achieves encouraging results in biomedical image registration.
This approach deserves certainly further study and represents a promising open research topic for multimodal medical image registration.

\makeatletter
\def\Hline{%
\noalign{\ifnum0=`}\fi\hrule \@height 1pt \futurelet
\reserved@a\@xhline}
\makeatother

\def\x{{\mathbf x}}
\def\L{{\cal L}}

\chapter{Deep Neural Networks}
\label{chap6}
\graphicspath{{Chapter6/Figs/}}

\section{Convolutional Neural Networks}
\label{sec:CNNs}

In the latest years, DNNs have been exploited to learn a hierarchy of increasingly complex features from the processed data, enabling multiple levels of abstraction \cite{lecun2015}.
This kind of models allows for raw data processing differently to conventional Machine Learning classifiers the are built on top of hand-engineered features, such as Local Binary Patterns (LBP) \cite{ojala2002}, Scale-Invariant Feature Transform (SIFT) \cite{lowe2004}, and Histogram of Oriented Gradients (HOG) \cite{dalal2005}.

In traditional multi-layer networks, each neuron is densely or fully connected to every neuron of the subsequent layer.
However, in Pattern Recognition exploiting only local sub-structural information could be useful.
Inspired by the idea of the neurocognitiron \cite{fukushima1981,fukushima1983}, it was shown that pixels that are close together in the image (e.g., adjacent pixels) tend to be strongly correlated and can represent meaningful features such as edges, while pixels that are far apart in the image tend to be weakly correlated or uncorrelated.

CNNs are deep feed-forward neural network architectures that are suitable to work on data structured with grid-like topology, such as images, videos, and time-series. 
CNNs are effective in training and generalization much better than common neural networks.
As a matter of fact, efficient neural architectures are a hot topic in computer science engineering, also considering hardware devices with limited resources \cite{conti2017}.

Many standard feature representations used in Computer Vision problems are based upon local features within the image.
In order to reflect this powerful property, a CNN architecture takes advantage of sparse connectivity: the local sub-structure within a image is captured by constraining each neuron to depend only on a spatially local subset of the variables of the previous layer.
The set of neurons in the input layer that influences the activation of a neuron is known as the neuron's receptive field.
This property needs for less parameters to learn, so allowing for a more efficient training.
In such a practical scenario, leveraging GPUs \cite{eklund2013,smistad2015} has been representing an important technological enabling factor in effective training \cite{litjens2017,shen2017}.

The main components of a CNN \cite{krizhevsky2012,lecun1990} are:
\begin{itemize}
    \item \textit{convolutional layers}, defining linear convolution operations organized in feature maps. Each unit is connected to local patches in the feature maps of the previous layer by means of a set of weights (i.e., filter bank) allowing for weight-sharing;
    \item \textit{non-linear activation layers}, applied after each convolutional layer or fully-connected layer, allowing for the learning of multimodal functions. Considering the universal approximation theorem---firstly proved by Cybenko \cite{cybenko1989} for sigmoidal activation function and then extended by Hornik \cite{hornik1991} to an arbitrary bounded and non-constant activation functions---feed-forward neural networks with a single hidden layer of finite size can learn any continuous function on a compact subset of $\mathbb{R}^n$. In addition to the traditional non-linear activation layer (e.g., the sigmoid and the hyperbolic tangent), which could give rise to the vanishing gradient phenomenon \cite{bengio1994}.
    Thereby, relying on a biological basis \cite{hahnloser2000}, piecewise linear activation functions---such as the Rectified Linear Unit (ReLU)  \cite{nair2010}---have been introduced in DNNs. ReLU efficiently allows for scale-invariance and sparse activation. ReLU variants are ReLU the Leaky ReLU \cite{maas2013}, and the Parametric ReLU (PReLU) \cite{he2015};
    \item \textit{pooling layers}, denoting an operation that replaces the output of the net at a certain location with a summary statistics of the neighborhood's outputs \cite{boureau2010}. Examples are max pooling and average pooling.
\end{itemize}

The convolutional and pooling layers in CNNs are directly inspired by the classic notions of simple cells and complex cells in visual neuroscience \cite{hubel1962}.
Deep neural networks exploit the property that many natural signals are compositional hierarchies (i.e., higher-level features are obtained by composing lower-level ones).

In supervised learning, the training phase relies on the minimization of an objective function, also called loss function $\mathcal{L}(\cdot)$.
As long as the modules are relatively smooth functions of their inputs
and of their internal weights, the gradients can be computed using the backpropagation algorithm \cite{rumelhart1986}.
In this context, the backpropagation procedure aims at computing the gradient of an objective function with respect to the weights of a multi-layer architecture.
This algorithm applies the chain rule for derivatives.
Regarding the difficulties encountered during DNN training, Haeffele and Vidal \cite{haeffele2017} provided the conditions to guarantee that local minima are globally optimal and that a local descent strategy can reach a global minima from any initialization.
In particular, these conditions require both the network output and the regularization to be positively homogeneous functions of the network parameters, with the regularization being designed to control the network size.
Multi-layer architectures can be trained by the Stochastic Gradient Descent (SGD) \cite{bottou2010}, especially when the optimization problem is convex or pseudo-convex \cite{hovhannisyan2016}.
Alternatives are adaptive methods, such as AdaGrad \cite{duchi2011}, ADADELTA \cite{zeiler2012}, and Adam \cite{kingma2014}.

Another novelty in DNN training is the dropout technique, which randomly drop neurons from the neural network during training and, thus, prevents the co-adaptation of features \cite{srivastava2014}.
Initially, dropout was proposed as an implicit regularization method, in \cite{baldi2013,cavazza2017} it was demonstrated the equivalence between dropout and a fully deterministic model for Matrix Factorization in which the factors are regularized by the sum of the product of squared Euclidean norms of the columns.

However, in medical imaging it is also important to achieve high classification rates on potentially perturbed test data \cite{litjens2017}.
The analysis of state-of-the-art deep classifiers' robustness to perturbation at test time is therefore an important step for validating the models’ reliability to unexpected (possibly adversarial) nuisances that might occur when deployed in uncontrolled environments \cite{fawzi2017}, such as clinical scenarios.

Considering the most recent computational methods in medical image segmentation, along with traditional Pattern Recognition techniques~\cite{rundo2018next}, significant advances have been proposed in CNN-based architectures.
For instance, to overcome the limitations related to accurate image annotations, DeepCut~\cite{rajchl2017} relies on weak bounding box labeling~\cite{rother2004,rundo2017NC}.
Moreover, Wang \textit{et al.}~\cite{wang2018} proposed an interactive  segmentation framework by incorporating CNNs into a bounding box and scribble-based segmentation pipeline.
This method aims at learning features for a CNN-based classifier from bounding box annotations.

Among the architectures devised for biomedical image segmentation~\cite{havaei2017,kamnitas2017}, U-Net~\cite{ronneberger2015} showed to be a noticeably successful solution, thanks to the combination of a contracting path, for coarse-grained context detection, and a symmetric expanding path, for fine-grained localization.
This fully CNN is capable of stable training with reduced samples.
The authors of V-Net~\cite{milletari2016} extended U-Net for volumetric medical image segmentation, by introducing also a different loss function based on the \emph{DSC}.
Oktay \textit{et al.}~\cite{oktay2018} presented an Attention Gate (AG) model for medical imaging, which aims at focusing on target structures or organs.
AGs were introduced into the standard U-Net, so defining Attention U-Net, which achieved high performance in multi-class image segmentation without relying on multi-stage cascaded CNNs.
Recently MS-D Net~\cite{pelt2017} was shown to yield better segmentation results in biomedical images than U-Net~\cite{ronneberger2015} and SegNet~\cite{badrinarayanan2017}, by creating dense connections among features at different scales obtained by means of dilated convolutions.
By so doing, features at different scales can be contextually extracted using fewer parameters than full CNNs.
Finally, also image-to-image translation approaches---e.g., pix2pix~\cite{isola2016}, leveraging conditional adversarial neural networks---were exploited for image segmentation.

\subsection{Cross-dataset generalization abilities of CNNs: Prostate zonal segmentation of multi-centric MRI datasets}

Even though CNNs achieved state-of-the-art performance for automatic medical image segmentation, they have not yet shown sufficiently accurate and reliable results for clinical use \cite{wang2018}.
As a matter of fact, CNN-based architectures are limited by the lack of image-specific adaptation and the lack of generalizability to previously unseen object classes.

In the specific case of prostate MR image analysis, MRI plays a decisive role in PCa diagnosis and disease monitoring (even in an advanced status~\cite{padhani2017}), revealing the internal prostatic anatomy, prostatic margins, and PCa extent~\cite{villeirs2007}.
According to the zonal compartment system proposed by McNeal, the prostate WG can be partitioned into the CG and PZ \cite{selman2011}.
In prostate imaging, T2w MRI serves as the principal sequence~\cite{scheenen2015}, thanks to its high resolution that enables to differentiate the hyper-intense PZ and hypo-intense CG in young male subjects~\cite{hoeks2011}.

Considering prostate zonal segmentation, an improved PCa diagnosis requires a reliable and automatic zonal segmentation method, since  manual delineation is time-consuming and operator-dependent~\cite{muller2015,rundo2017Inf}.
Moreover, in clinical practice, the generalization ability among multi-institutional prostate MRI datasets is an unresolved issue due to large anatomical inter-subject variability and the lack of a standardized pixel intensity representation for MRI (such as in the case of CT-based radiodensity measurements expressed in HUs)~\cite{klein2008}.
Hence, we aim at automatically segmenting the prostate zones on three multi-institutional T2w MRI datasets to evaluate the generalization ability of CNN-based architectures.
This task is challenging because images coming from multi-institutional datasets are characterized by different contrasts, visual consistencies, and heterogeneous characteristics~\cite{vanOpbroek2015}.

\subsubsection{Multi-institutional MRI datasets}
\label{sec:multiDatasetsProstate}
We segmented the CG and PZ from the WG on three completely different multi-parametric prostate MRI datasets, namely:
\begin{itemize}
\item[$\#1$] dataset ($21$ patients/$193$ MR slices with prostate), acquired with a whole body Philips Achieva $3.0$ T MRI scanner at the Cannizzaro Hospital (Catania, Italy) \cite{rundo2017Inf}.
MRI parameters: matrix size $= 288 \times 288$ pixels; slice thickness $= 3.0$ mm; inter-slice spacing $=4$ mm; pixel spacing $= 0.625$ mm; number of slices per image series (including slices without prostate) $= 18$.
Average patient age: $65.57 \pm 6.42$ years;
\item[$\#2$] Initiative for Collaborative Computer Vision Benchmarking (I2CVB) dataset ($19$ patients/$503$ MR slices with prostate), acquired with a whole body Siemens TIM $3.0$ T MRI scanner at the Hospital Center Regional University of Dijon-Bourgogne (Dijon, France) \cite{lemaitre2015}.
MRI parameters: matrix size $\in \{308 \times 384, 336 \times 448, 360 \times 448, 368 \times 448 \}$ pixels; slice thickness $= 1.25$ mm; inter-slice spacing $= 1.0$ mm; pixel spacing $\in \{0.676, 0.721, 0.881, 0.789 \}$ mm; number of slices per image series $= 64$.
Average patient age: $64.36 \pm 9.69$ years;
\item[$\#3$] National Cancer Institute -- International Symposium on Biomedical Imaging (NCI-ISBI) 2013 Automated Segmentation of Prostate Structures Challenge dataset ($40$ patients/$555$ MR slices with prostate) \textit{via} The Cancer Imaging Archive (TCIA) \cite{prior2017}, acquired with a whole body Siemens TIM $3.0$ T MRI scanner at Radboud University Medical Center (Nijmegen, The Netherlands) \cite{TCIA}.
The prostate structures were manually delineated by five experts.
MRI parameters: matrix size $\in \{256 \times 256, 320 \times 320, 384 \times 384\}$ pixels; slice thickness $\in \{3.0, 4.0\}$ mm; inter-slice spacing $\in \{3.6, 4.0\}$ mm; pixel spacing $\in \{0.500, 0.600, 0.625\}$ mm; number of slices per image series ranging from $15$ to $24$.
Average patient age: $63.90 \pm 7.17$ years.
\end{itemize}

All the analyzed MR images are encoded in the $16$-bit DICOM format.
It is worth noting that even MR images from the same dataset have intra-dataset variations (such as the matrix size, slice thickness, and number of slices).
Furthermore, inter-rater variability for the CG and PZ annotations exists, as different physicians delineated them.
For clinical feasibility~\cite{hoeks2011}, we analyzed only axial T2w MR slices---the most commonly used sequence for prostate zonal segmentation---among the available sequences.
In our multi-centric study, we conducted the following seven experiments resulting from all  possible training/testing conditions:
\begin{itemize}
\item Individual dataset $\#1$, $\#2$, $\#3$: training and testing on dataset $\#1$ ($\#2$, $\#3$, respectively) alone in $4$-fold cross-validation, and testing also on whole datasets $\#2$ and $\#3$ ($\#1$ and $\#3$, $\#1$ and $\#2$, respectively) separately for each round;
\item Mixed dataset $\#1/\#2$, $\#2/\#3$, $\#1/\#3$: training and testing on both datasets $\#1$ and $\#2$ ($\#2$ and $\#3$, $\#1$ and $\#3$, respectively) in $4$-fold cross-validation, and testing also on whole dataset $\#3$ ($\#1$, $\#2$, respectively) separately for each round;
\item Mixed dataset $\#1/\#2/\#3$: training and testing on whole datasets $\#1$, $\#2$, and $\#3$ in $4$-fold cross-validation.
\end{itemize}
For the $4$-fold cross-validation, we partitioned the datasets $\#1$, $\#2$, and $\#3$ into $4$ folds by using the following patient indices: $\{ [1, \ldots, 5]$, $[6, \ldots, 10]$, $[11, \ldots, 15]$, $[16, \ldots, 21] \}$, $\{ [1, \ldots, 5]$, $[6, \ldots, 10]$, $[11, \ldots, 15]$, $[16, \ldots, 19] \}$, and $\{ [1, \ldots, 10]$, $[11, \ldots, 20]$, $[21, \ldots, 30]$, $[31, \ldots, 40] \}$, respectively. Finally, the results from the different cross-\sloppy validation rounds were averaged to obtain a final descriptive value.
We chose a $4$-fold cross-validation scheme to conduct reliable and fair training/testing phases, according to the number of patients provided by each dataset, as well as to calculate the evaluation metrics on a statistically significant test set.
Our experimental settings were designed to evaluate the CNN's generalization abilities  among different MRI acquisition options, such as different devices and functioning  parameters.
For instance, both intra-/inter-scanner evaluations can be carried out, as dataset $\#1$'s scanner is different from those of datasets $\#2$ and $\#3$.

Our study adopts a selective delineation approach to focus on internal prostatic anatomy: the CG and PZ, denoted by $\mathcall{R}_{CG}$ and $\mathcall{R}_{PZ}$, respectively.
Let the entire image and the WG region be $\mathcall{I}_\Omega$ and $\mathcall{R}_{WG}$, respectively, the following relationships can be defined:
\begin{equation}
	\label{eq:globalConstraints}
	\mathcall{I}_\Omega = \mathcall{R}_{WG} \cup \mathcall{R}_\text{bg} \mbox{ and } \mathcall{R}_{WG} \cap \mathcall{R}_\text{bg} = \varnothing,
\end{equation}
where $\mathcall{R}_\text{bg}$ represents background pixels.
Relying on \cite{qiu2014,villeirs2007}, $\mathcall{R}_{PZ}$ was obtained by subtracting $\mathcall{R}_{CG}$ from $\mathcall{R}_{WG}$ meeting the constraints:
\begin{equation}
	\label{eq:segConstraints}
	\mathcall{R}_{WG} = \mathcall{R}_{CG} \cup \mathcall{R}_{PZ} \mbox{ and } \mathcall{R}_{CG} \cap \mathcall{R}_{PZ} = \varnothing.
\end{equation}

\begin{figure}[!t]
 \centering
  \includegraphics[width=0.9\textwidth]{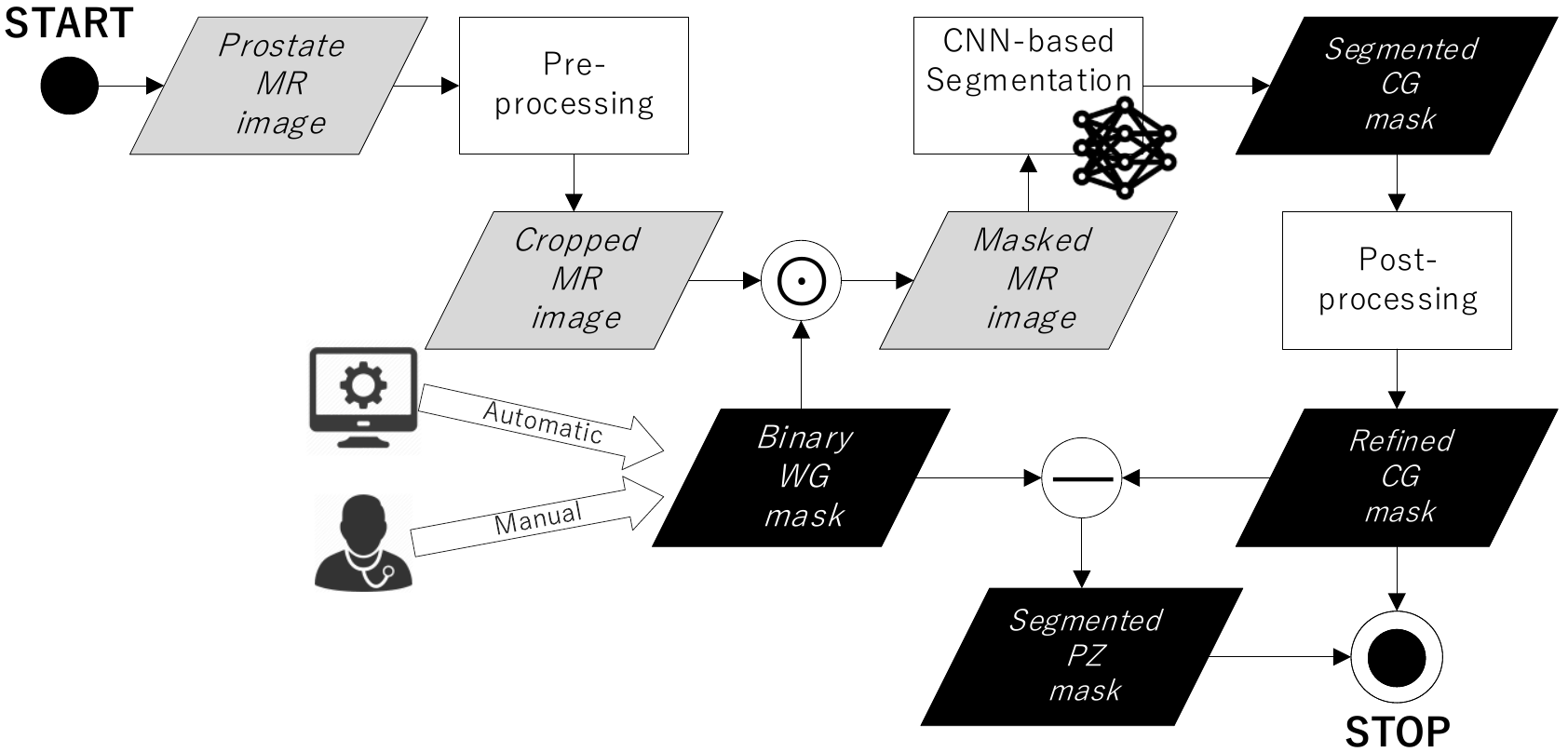}
  \caption[Workflow of the overall CNN-based prostate zonal segmentation approach]{Workflow of the overall CNN-based prostate zonal segmentation approach. The gray and black data blocks denote gray-scale images and binary masks, respectively.}
  \label{fig:WorkFlow}
\end{figure}

The overall prostate zonal segmentation method is outlined in Fig. \ref{fig:WorkFlow}.

\subsubsection{Quantitative comparison of the state-of-the-art CNN-based architectures}
\label{sec:ProstZonCompSotA}
In the preliminary study in \cite{rundoWIRN2018}, we automatically segmented the CG and PZ using Deep Learning to evaluate the generalization ability of CNNs on two different MRI prostate datasets---namely, the datasets $\#1$ and $\#2$---by means of a $4$-fold cross-validation scheme.
However, this is challenging since multi-centric datasets are generally characterized by different contrast, visual consistencies, and image characteristics.
Therefore, prostate zones on T2w MR images were manually annotated for supervised learning, and then automatically segmented using a mixed scheme by (\textit{i}) training on either each individual dataset or both datasets and (\textit{ii}) testing on both datasets, using CNN-based architectures: SegNet~\cite{badrinarayanan2017}, U-Net~\cite{ronneberger2015}, and pix2pix~\cite{isola2016}.
In such a context, we compared the segmentation performances achieved with/without pre-training~\cite{tajbakhsh2016}.

\paragraph{Pre-processing}
To fit the image resolution of the dataset $\#1$, we center-cropped the images of the dataset $\#2$ and resized them to $288 \times 288$ pixels.
Furthermore, the images of these datasets were masked using the corresponding prostate binary masks to omit the background and only focus on extracting the CG and PZ from the WG.
This operation can be performed either by an automated method~\cite{rundo2017Inf} or previously provided manual WG segmentation~\cite{lemaitre2015}. For better training, we randomly cropped the input images from $288 \times 288$ to $256 \times 256$ pixels and horizontally flipped them.

\paragraph{Investigated CNN-based Architectures}
During the training of all architectures, the $\mathcal{L}_{DSC}$ loss function (i.e., a continuous version of \emph{DSC}) was used  \cite{milletari2016}:
\begin{equation}
	\label{eq:DSCloss}
	\mathcal{L}_{DSC} = - \frac{2\sum_{i=1}^{N} s_i \cdot r_i}{\sum_{i=1}^{N} s_i + \sum_{i=1}^{N} r_i},
\end{equation}
where $s_i$ and $r_i$ represent the continuous values of the prediction map (i.e., the result of the final layer of the CNN) and the ground truth at the $i$-th pixel ($N$ is the total number of pixels to be classified).

\subparagraph{SegNet} is a CNN architecture for semantic pixel-wise segmentation~\cite{badrinarayanan2017}.
It consists of an encoder/decoder network followed by a pixel-wise classification layer.
Since our classification task involves only one class, the soft-max operation and ReLU activation function at the final layer were removed for stable training.

We implemented SegNet using PyTorch.
During training, we used the SGD \cite{bottou2010} with a learning rate of $0.01$, momentum of $0.9$, weight decay of $5\times10^{-4}$, and batch size of $8$.
It was trained for $50$ epochs and the learning rate was multiplied by $0.2$ at the $20$-th and $40$-th epochs.

\subparagraph{U-Net} is a fully CNN capable of stable training with a reduced number of samples~\cite{ronneberger2015}, combining pooling operators with up-sampling operations.
The general architecture is an encoder-decoder with skip connections between mirrored layers in the encoder/decoder stacks. By so doing, high resolution features from the contracting path are combined with the up-sampled output for better localization.
We utilized four scaling operations.

U-Net was implemented using Keras on top of TensorFlow.
We used SGD with a learning rate of $0.01$, momentum of $0.9$, weight decay of $5 \times 10^{-4}$, and batch size of $4$.
Training was executed for $50$ epochs, multiplying the learning rate by $0.2$ at the $20$-th, and $40$-th epochs.

\subparagraph{pix2pix} is an image-to-image translation method with conditional adversarial networks \cite{isola2016}. As a generator, U-Net is used to translate the original image into the segmented one \cite{ronneberger2015}, preserving the highest level of abstraction.
The generator and discriminator include $8$ and $5$ scaling operations, respectively.

We implemented pix2pix on PyTorch.
Adam was used as an optimizer with a learning rate of $0.0002$ and $0.01$ for the discriminator and generator, respectively.
The learning rate for generator was multiplied by $0.1$ every $20$ epochs.
It was trained for $50$ epochs with a batch size of $12$.\\

\paragraph{Post-processing}
Two simple morphological steps were applied on the obtained CG binary masks to smooth boundaries and avoid disconnected regions:
\begin{itemize}
	\item a hole filling algorithm on the segmented $\mathcall{R}_{CG}$ to remove possible holes in the predicted map;
    \item  a small area removal operation to delete connected components with area less than $\lfloor |\mathcall{R}_{WG}|/8 \rfloor$ pixels, where $|\mathcall{R}_{WG}|$ denotes the number of the pixels contained in the WG segmentation.
This criterion effectively adapts according to the different dimensions of the $\mathcall{R}_{WG}$ (ranging from the apical to the basal prostate slices).
\end{itemize}


\paragraph{Influence of pre-training}
In medical imaging, due to the lack of training data, ensuring CNN's proper training convergence is difficult from scratch.
Therefore, pre-training the models on a different application and then fine-tuning them is common \cite{tajbakhsh2016}.

To evaluate cross-dataset generalization ability \textit{via} pre-training, we compared the performances of the three CNN-based architectures with/without pre-training on a similar application. We used a relatively large dataset of $50$ manually segmented examples from the PROMISE12 challenge \cite{litjens2014}.
Since this competition focuses only on WG segmentation without providing prostate zonal labeling, we pre-trained the architectures on WG segmentation.
To adjust this dataset to our experimental setup, the images of this dataset were resized from $512 \times 512$ to $288 \times 288$ pixels and randomly cropped to $256 \times 256$ pixels; because our task only focuses on slices with prostate, we also omitted initial/final slices without prostate, so the number of slices for each sample was fixed to $25$.

\paragraph{Results}
This section explains how the three CNN-based architectures segmented the prostate zones, evaluating their cross-dataset generalization ability.

Table \ref{table:PrelResults} shows the $4$-fold cross-validation results obtained in the different experimental conditions.
When training and testing are both performed on the dataset $\#1$, U-Net outperforms the other architectures on both CG and PZ segmentation; however, it experiences problems with testing on the dataset $\#2$ due to the limited number of training images in the dataset $\#1$.
In such a case, pix2pix generalizes better thanks to its internal generative model.
When trained on the dataset $\#2$ alone, U-Net yields the most accurate results both in intra- and cross-dataset testing.
This probably derives from the dataset $\#2$'s relatively larger training data as well as U-Net's good generalization ability when sufficient data are available.
Moreover, SegNet reveals rather unstable results, especially when trained on a limited amount of data.
Finally, when trained on the mixed dataset, all three architectures---especially U-Net---achieve good results on both datasets without losing accuracy compared to training on the same dataset alone.
Therefore, using mixed MRI datasets during training can considerably improve the performance in cross-dataset generalization towards other clinical applications.
Comparing the CG and PZ segmentation, when tested on the dataset $\#1$, the results on the PZ are generally more accurate, except when trained on the dataset $\#2$ alone; however, for the dataset $\#2$, segmentations on the CG are generally more accurate.

\begin{figure}[!t]
	\centering
	\subfloat[]{\includegraphics[height=2.5cm]{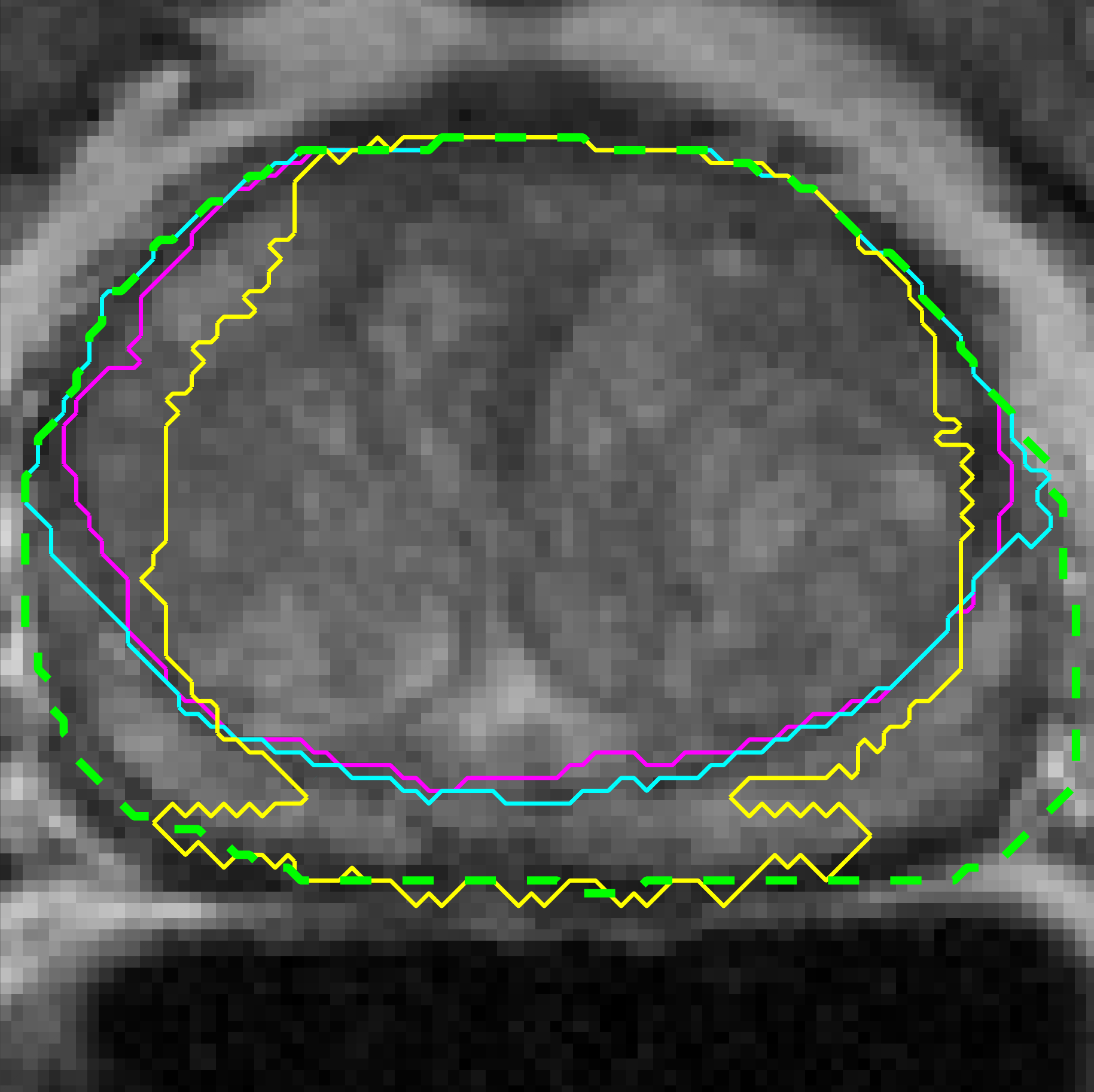}\label{fig:ResultImagesA}}\qquad
\subfloat[]{\includegraphics[height=2.5cm]{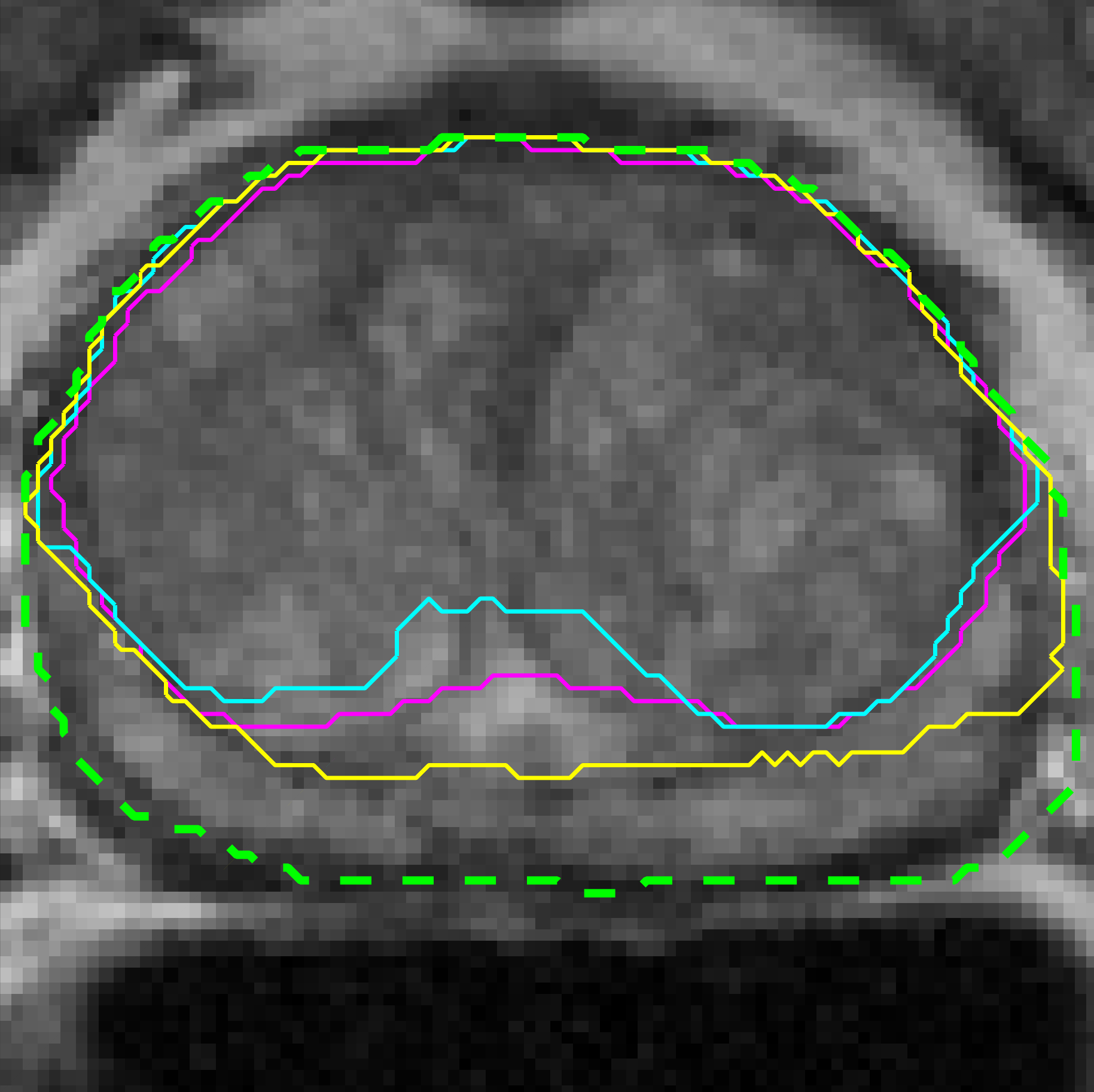}\label{fig:ResultImagesB}}\qquad
\subfloat[]{\includegraphics[height=2.5cm]{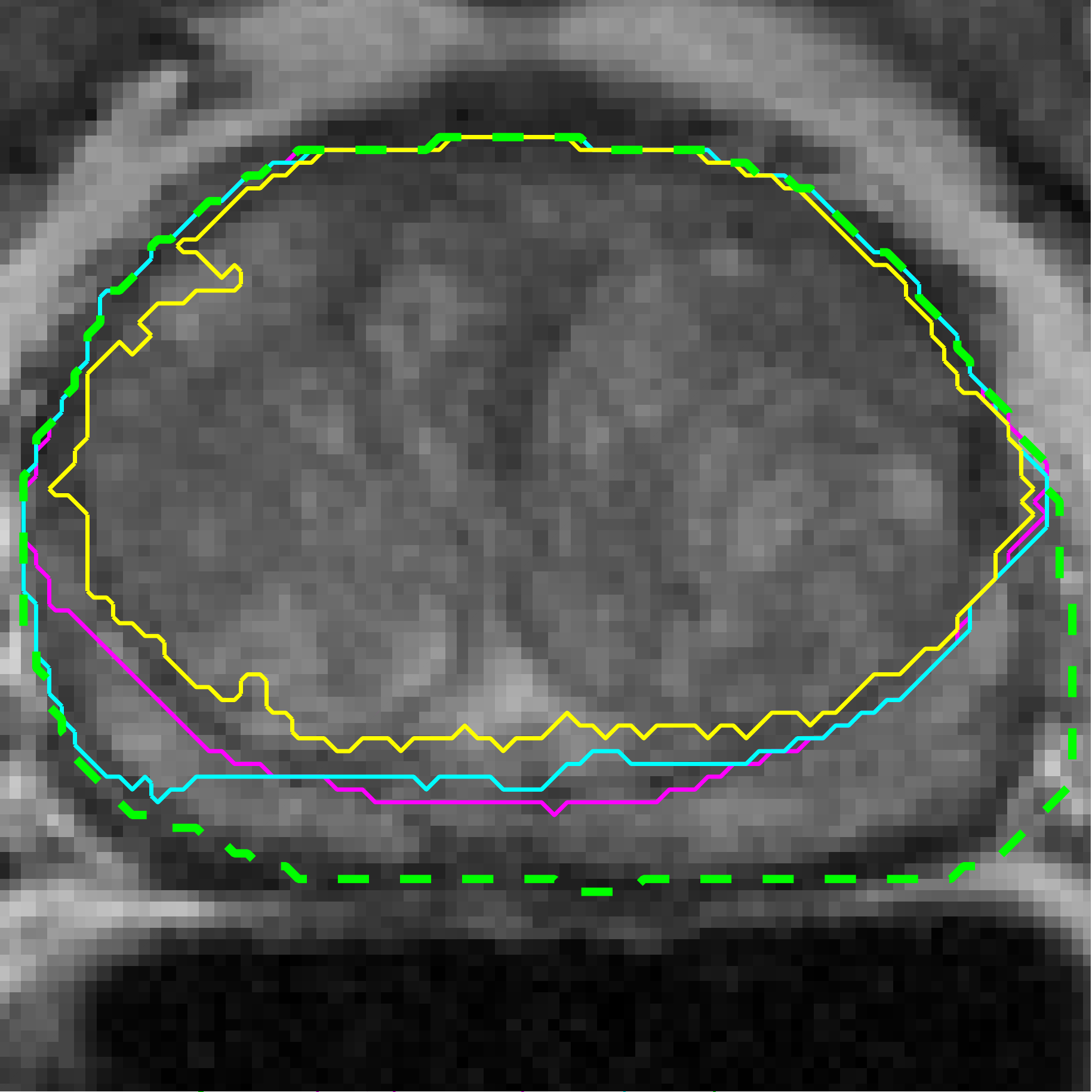}\label{fig:ResultImagesC}}\qquad
\subfloat[]{\includegraphics[height=2.5cm]{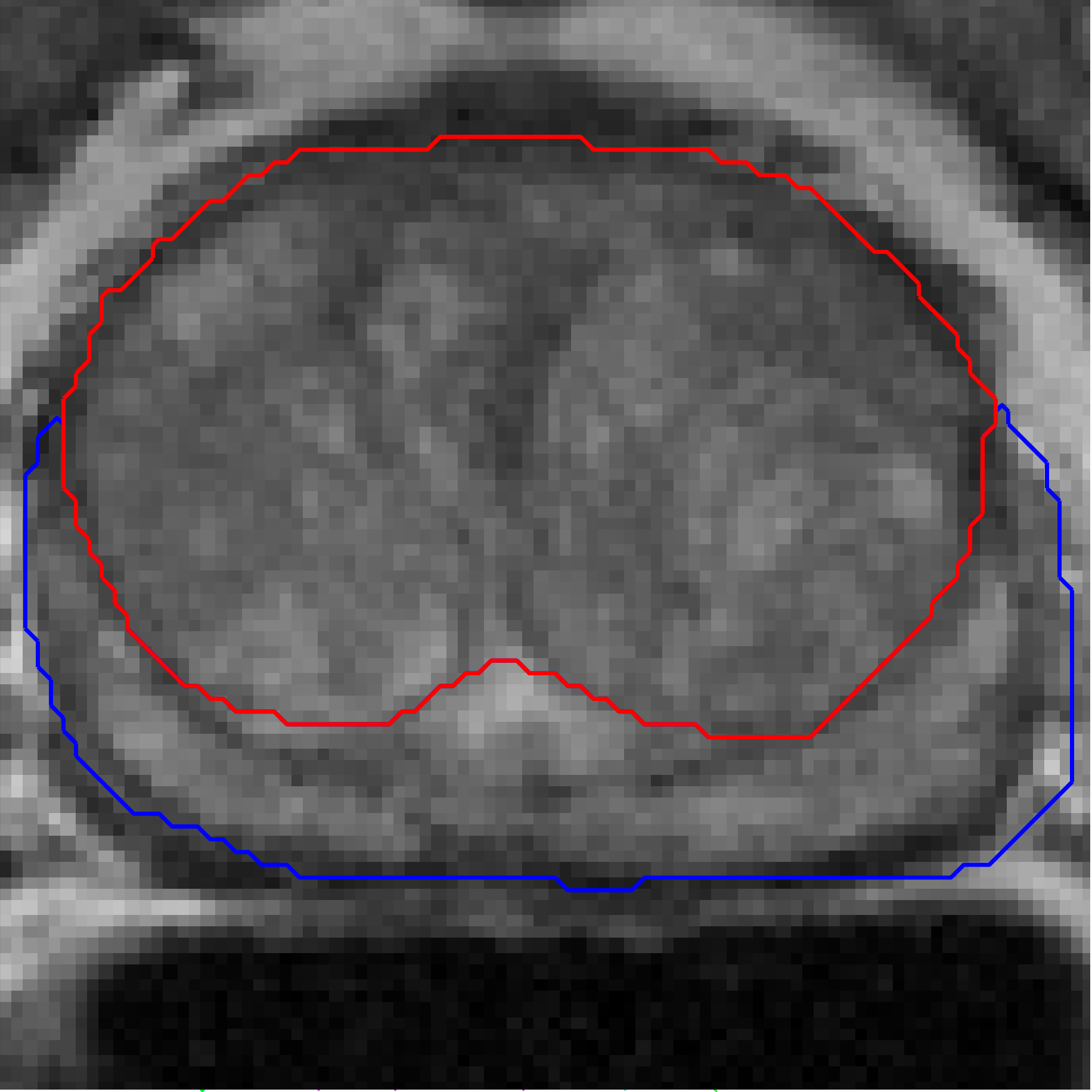}\label{fig:ResultImagesD}} \\
\subfloat[]{\includegraphics[height=2.5cm]{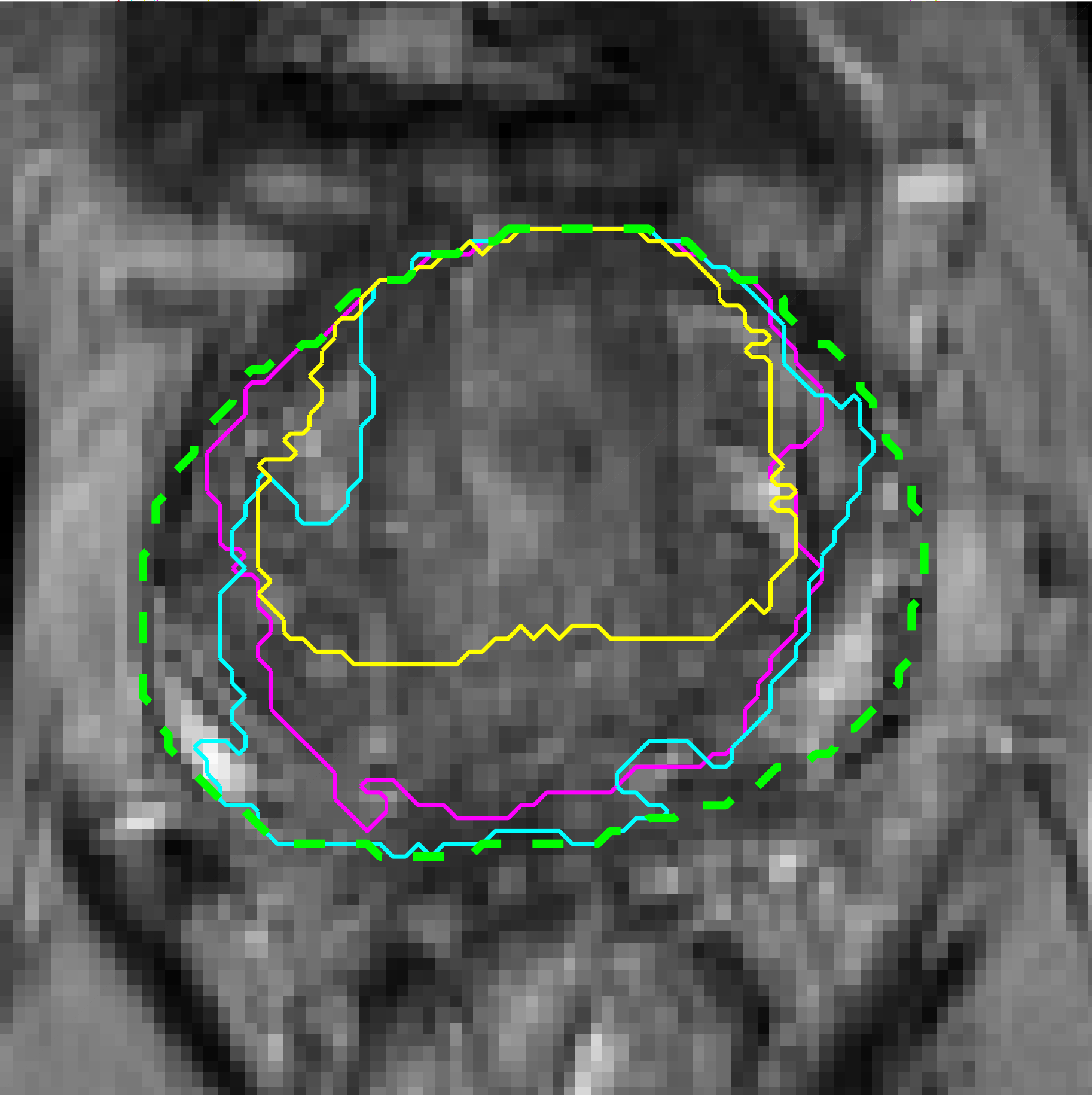}\label{fig:ResultImagesE}}\qquad
\subfloat[]{\includegraphics[height=2.5cm]{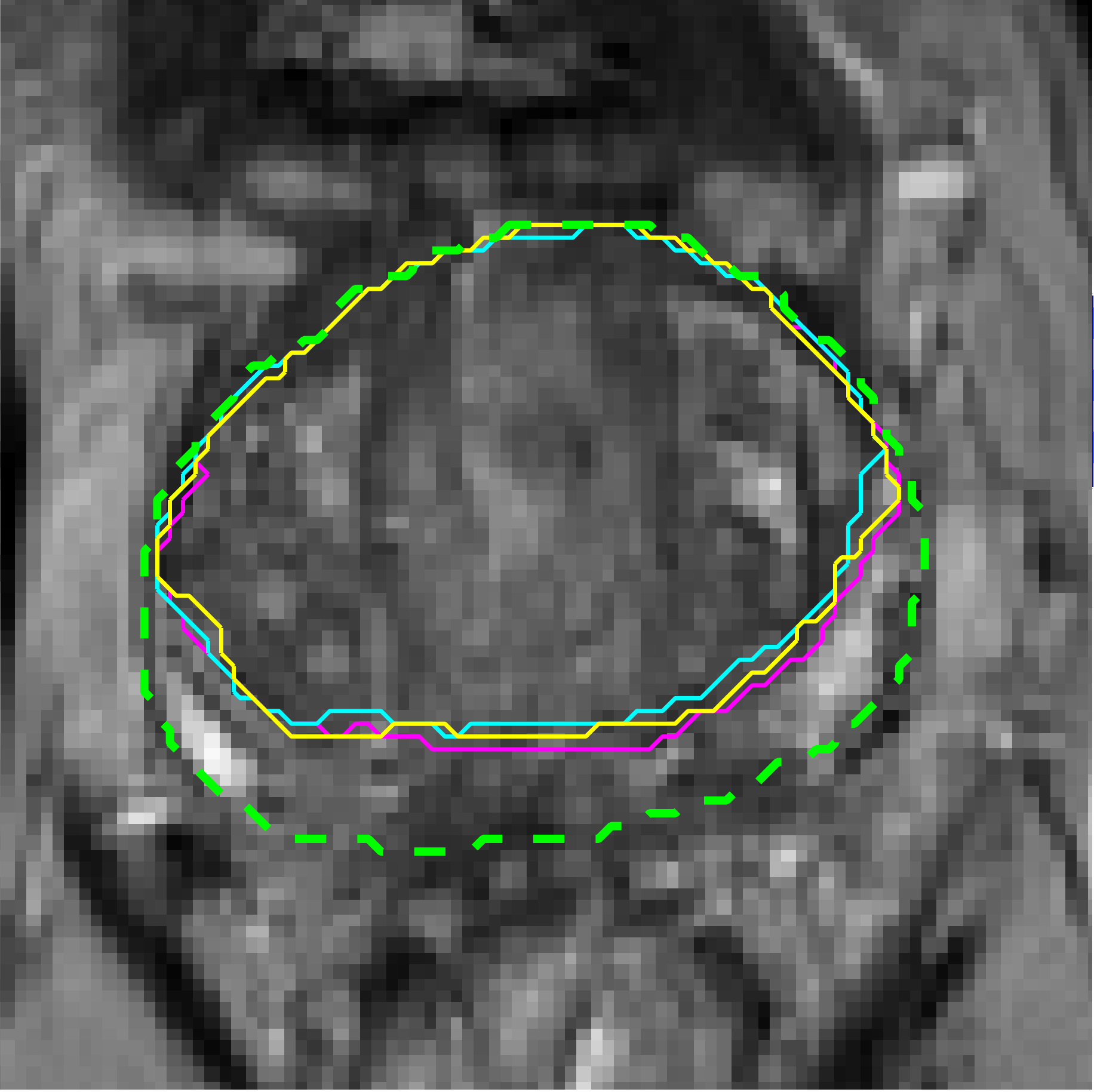}\label{fig:ResultImagesF}}\qquad
\subfloat[]{\includegraphics[height=2.5cm]{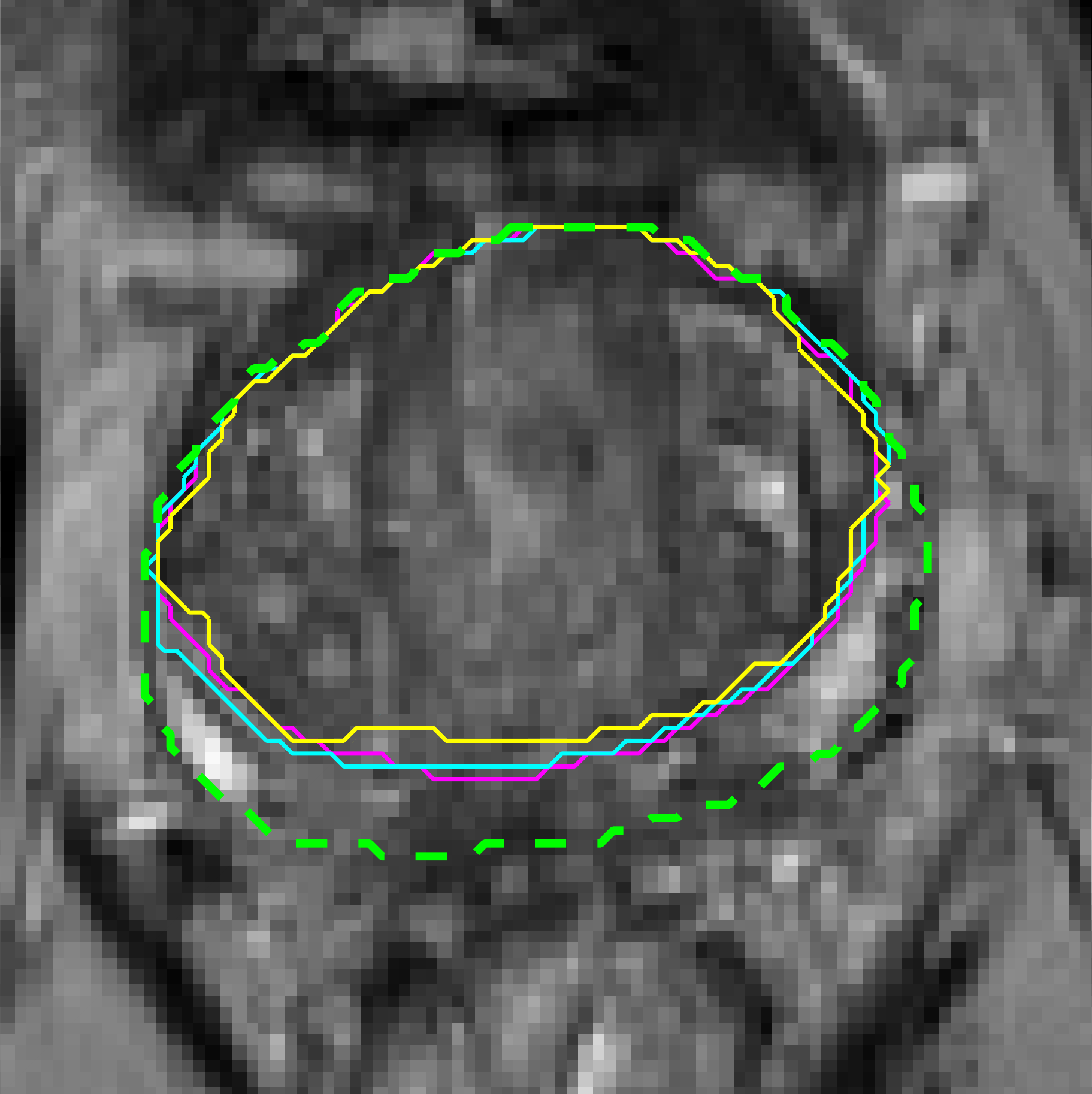}\label{fig:ResultImagesG}}\qquad
\subfloat[]{\includegraphics[height=2.5cm]{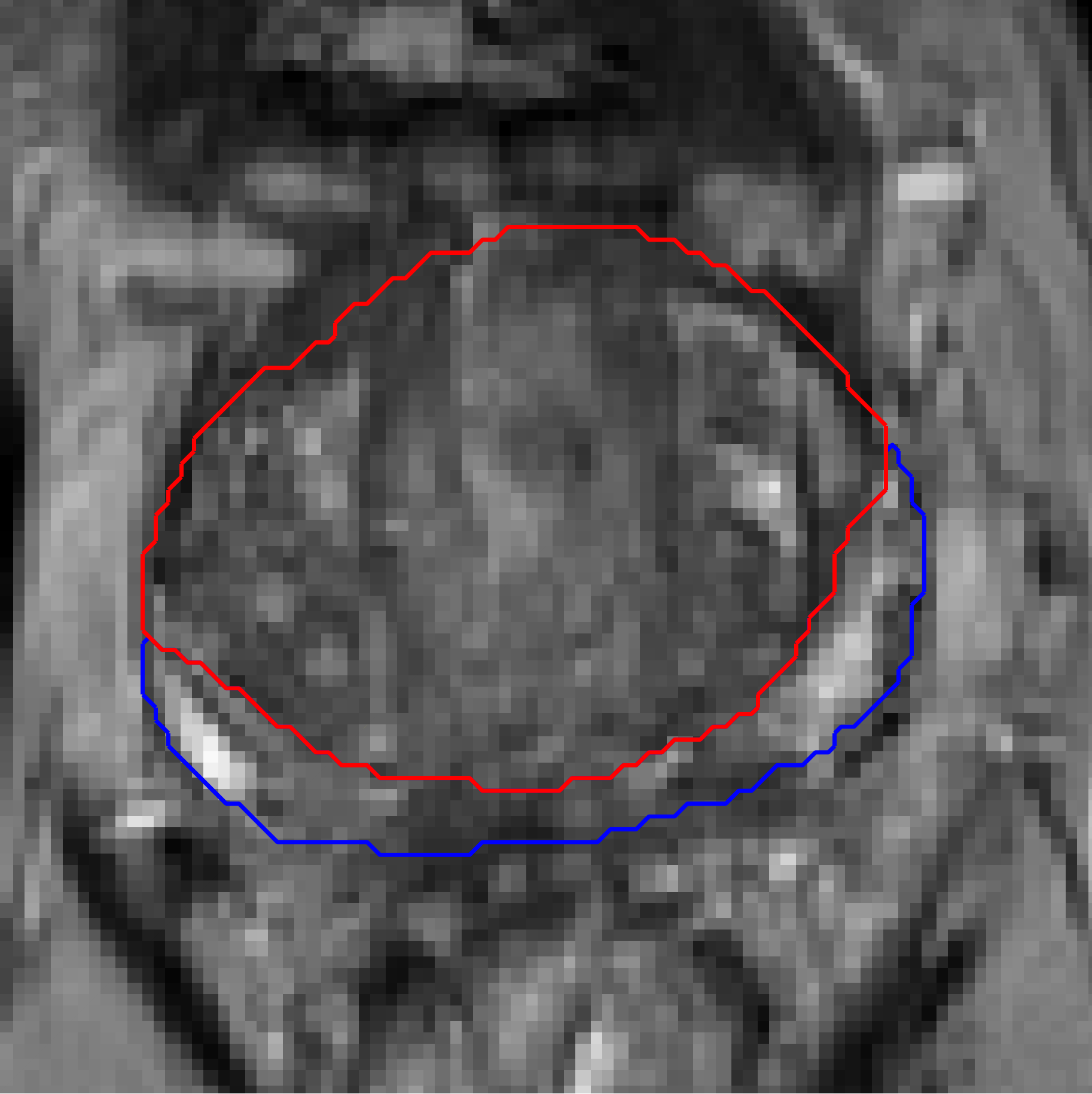}\label{fig:ResultImagesH}} \\
	\caption[Examples of prostate zonal segmentation in pre-training/fine-tuning]{Examples of prostate zonal segmentation in pre-training/fine-tuning. The first row concerns testing on dataset $\#1$, trained on: (a) dataset $\#1$; (b) dataset $\#2$; (c) mixed dataset. The second row concerns testing on dataset $\#2$, trained on: (e) dataset $\#1$; (f) dataset $\#2$; (g) mixed dataset.
    The $\mathcall{R}_{CG}$ segmentation results are represented with magenta, cyan, and yellow solid contours for SegNet, U-Net, and pix2pix, respectively.
The dashed green line denotes the $\mathcall{R}_{WG}$ boundary.
The last column (sub-figures (d, h)) shows the gold standard for $\mathcall{R}_{CG}$ and $\mathcall{R}_{PZ}$ with red and blue lines, respectively.
The images are zoomed with a $4\times$ factor.}
	\label{fig:ResultImages}	
\end{figure}

Fine-tuning after pre-training sometimes leads to slightly better results than training from scratch, when trained only on a single dataset.
However, its influence is generally negligible or rather negative, when trained on the mixed dataset. This modest impact is probably due to the ineffective data size for pre-training.
For a visual assessment, two examples (one for each dataset) are shown in Fig. \ref{fig:ResultImages}.
Relying on the gold standards in Figs. \ref{fig:ResultImages}(d, h), it can be seen that U-Net generally achieves more accurate results compared with SegNet and pix2pix.
This finding confirms the trend revealed by the \emph{DSC} values in Table \ref{table:PrelResults}.

\begin{table*}[ht!]
\centering
\begin{footnotesize}
  \caption[Prostate zonal segmentation results of the three CNN-based architectures in $4$-fold cross-validation assessed by the \emph{DSC} metrics]{Prostate zonal segmentation results of the three CNN-based architectures in $4$-fold cross-validation assessed by the \emph{DSC} metrics (presented as the average and standard deviation). The experimental results are calculated on the different setups of (\textit{i}) training on either each individual dataset or both datasets and (\textit{ii}) testing on both datasets. Numbers in bold indicate the highest \emph{DSC} values for each prostate region (i.e., CG and PZ) among all architectures with/without pre-training (PT).}

\label{table:PrelResults}
\begin{tabular}{c|l|c|cc|cc}
\hline\hline
\multirow{2}{*}{\hspace{30pt}}   & \multicolumn{1}{c|}{\multirow{2}{*}{\textbf{Architecture}}} & \multirow{2}{*}{\textbf{Zone}} & \multicolumn{2}{c|}{\textbf{Testing on Dataset $\#1$}} & \multicolumn{2}{c}{\textbf{Testing on Dataset $\#2$}} \\
                            & \multicolumn{1}{c|}{}                                      &        & \textit{Average}            & \textit{Std. Dev.}            & \textit{Average}          & \textit{Std. Dev.}        \\ \hline

\multirow{12}{*}{\begin{turn}{-90}\textbf{\shortstack{Training on\\Dataset $\#1$}}\end{turn}} & \multirow{2}{*}{SegNet (w/o PT)}                                   & CG     &  80.20                &       3.28              &       74.48           &     5.82             \\
                            &                                                           & PZ     &            80.66        &            11.51          &       59.57           &          12.68         \\

 & \multirow{2}{*}{SegNet (w/ PT)}                                   & CG     &  83.38                &       3.22               &       72.75           &     2.80              \\
                            &                                                           & PZ     &            87.39        &            3.90          &       66.20           &          5.64         \\
\cline{2-7}

                            & \multirow{2}{*}{U-Net (w/o PT)}                                    & CG     &       84.33             &         2.37             &        74.18          &       3.77            \\
                            &                                                           & PZ     &        88.98           &         2.98             &       66.63           &         1.93          \\

 & \multirow{2}{*}{U-Net (w/ PT)}                                   & CG     &  \textbf{86.88}                &       1.60               &       70.11           &     5.31              \\
                            &                                                           & PZ     &            \textbf{90.38}        &            3.38          &       58.89           &          7.06         \\
\cline{2-7}
                            & \multirow{2}{*}{pix2pix (w/o PT)}                                  & CG     &	82.35	&	2.09	&	\textbf{76.61}	&	2.17                 \\
                            &                                                           & PZ     &            87.09	&	2.72	&	73.20	&	2.62                  \\

 & \multirow{2}{*}{pix2pix (w/ PT)}                                   & CG     &  80.38                &       2.81               &       76.19           &     5.77              \\
                            &                                                           & PZ     &            83.53        &            5.65          &       \textbf{73.73}           &          2.40         \\
\hline
\multirow{12}{*}{\begin{turn}{-90}\textbf{\shortstack{Training on\\Dataset $\#2$}}\end{turn}} & \multirow{2}{*}{SegNet (w/o PT)}                                   & CG     &         76.04	&	2.05	&	87.07	&	2.41               \\
                            &                                                           & PZ     &                   77.25	&	3.09	&	82.45	&	1.77                  \\

 & \multirow{2}{*}{SegNet (w/ PT)}                                   & CG     &  77.99                &       2.15               &       87.75           &     2.83              \\

                            &                                                           & PZ     &        76.51	&	2.70	&	82.26	&	2.09                   \\
\cline{2-7}

                            & \multirow{2}{*}{U-Net (w/o PT)}                                    & CG     &      78.88	&	0.88	&	88.21		&	2.10                   \\
                            &                                                           & PZ     &         74.52	&	1.85	&	\textbf{83.03}		&	2.46                  \\
 & \multirow{2}{*}{U-Net (w/ PT)}                                   & CG     &  \textbf{79.82}                &       1.11               &       \textbf{88.66} &     2.28              \\
                            &                                                           & PZ     &        \textbf{74.56}	&	5.12	&	82.48	&	2.47                   \\
\cline{2-7}
                            & \multirow{2}{*}{pix2pix (w/o PT)}                                  & CG     &       77.90	&	0.73	&	86.95	&	2.93                   \\
                            &                                                           & PZ     &                   66.09	&	3.07	&	81.33	&	0.90                   \\
 & \multirow{2}{*}{pix2pix (w/ PT)}                                   & CG     &  77.21                &       1.02               &       85.94           &     4.31              \\
                            &                                                           & PZ     &        67.39	&	5.04	&	80.07	&	0.84                   \\
\hline

\multirow{12}{*}{\begin{turn}{-90}\textbf{\shortstack{Training on\\Mixed Dataset}}\end{turn}} & \multirow{2}{*}{SegNet (w/o PT)}                                   & CG     &          84.28	&	3.12	&	87.92	&	2.80                   \\

                            &                                                           & PZ     &        87.74	&	1.66	&	82.21	&	0.79                   \\

 & \multirow{2}{*}{SegNet (w/ PT)}                                   & CG     &  86.08                &       1.92               &       87.78           &     2.75              \\

                            &                                                           & PZ     &        89.53	&	3.28	&	82.39	&	1.50                   \\
\cline{2-7}
                            & \multirow{2}{*}{U-Net (w/o PT)}                                    & CG     &          \textbf{86.34}	&	2.10	&	\textbf{88.12}		&	2.34                   \\

                            &                                                           & PZ     &           90.74	&	2.40	&	\textbf{83.04}	&	2.30                  \\

 & \multirow{2}{*}{U-Net (w/ PT)}                                   & CG     &  85.82 	               &       1.98               &       87.42           &     1.89              \\

                            &                                                           & PZ     &        \textbf{91.44}	&	2.15	&	82.17	&	2.11                   \\
\cline{2-7}
                            & \multirow{2}{*}{pix2pix (w/o PT)}                                  & CG     &   83.07	&	3.39	&	86.39	&	3.16                  \\

                            &                                                           & PZ     &        83.53	&	2.36	&	80.40	&	1.80                   \\

 & \multirow{2}{*}{pix2pix (w/ PT)}                                   & CG     &  82.08                &       4.37               &       85.96           &     5.40              \\
                            &                                                           & PZ     &        83.04	&	4.20	&	80.60	&	1.49                   \\
\hline\hline       
\end{tabular}
\end{footnotesize}
\end{table*}

\clearpage

\paragraph{Discussion and conclusions}

Our preliminary results show that CNN-based architectures can segment prostate zones on two different MRI datasets to some extent, leading to valuable clinical insights; CNNs suffer when training and testing are performed on different MRI datasets acquired by different devices and protocols, but this can be mitigated by training the CNNs on multiple datasets, even without pre-training.
Generally, considering different experimental training and testing conditions, U-Net outperforms SegNet and pix2pix thanks to its good generalization ability.
Furthermore, this study suggests that significant performance improvement \textit{via} fine-tuning may require a remarkably large dataset for pre-training.

As future developments, we plan to improve the results by refining the predicted binary masks for better smoothness and continuity, avoiding disconnected segments; furthermore, we should enhance the output delineations considering the three-dimensional spatial information among slices.

\subsubsection{USE-Net: Incorporating SE blocks into U-Net}
\label{sec:USE-Net}

In this research context, we proposed a novel CNN, called USE-Net \cite{rundo2018Neurocomp}, which incorporates Squeeze-and-Excitation (SE) blocks~\cite{hu2017} into U-Net after every Encoder (Enc USE-Net) or Encoder-Decoder block (Enc-Dec USE-Net).
The rationale behind the design of USE-Net is to exploit adaptive channel-wise feature recalibration to boost the generalization performance.
Considering the three datasets above mentioned, this study adopted a mixed scheme for cross- and intra-dataset generalization: (\textit{i}) training on either each individual dataset or pair of datasets, and (\textit{ii}) testing on all three datasets with all possible training/testing combinations.
To the best of our knowledge, we tackle for the first time CNN-based prostate zonal segmentation on T2w MRI alone.
We compared USE-Net against three CNN-based architectures using both spatial overlap-/distance-based metrics: U-Net, pix2pix, and Mixed-Scale Dense Network (MS-D Net)~\cite{pelt2017}, along with a semi-automatic continuous max-flow model~\cite{qiu2014}.

\begin{figure}[!t]
	\centering
	\includegraphics[width=0.9\textwidth]{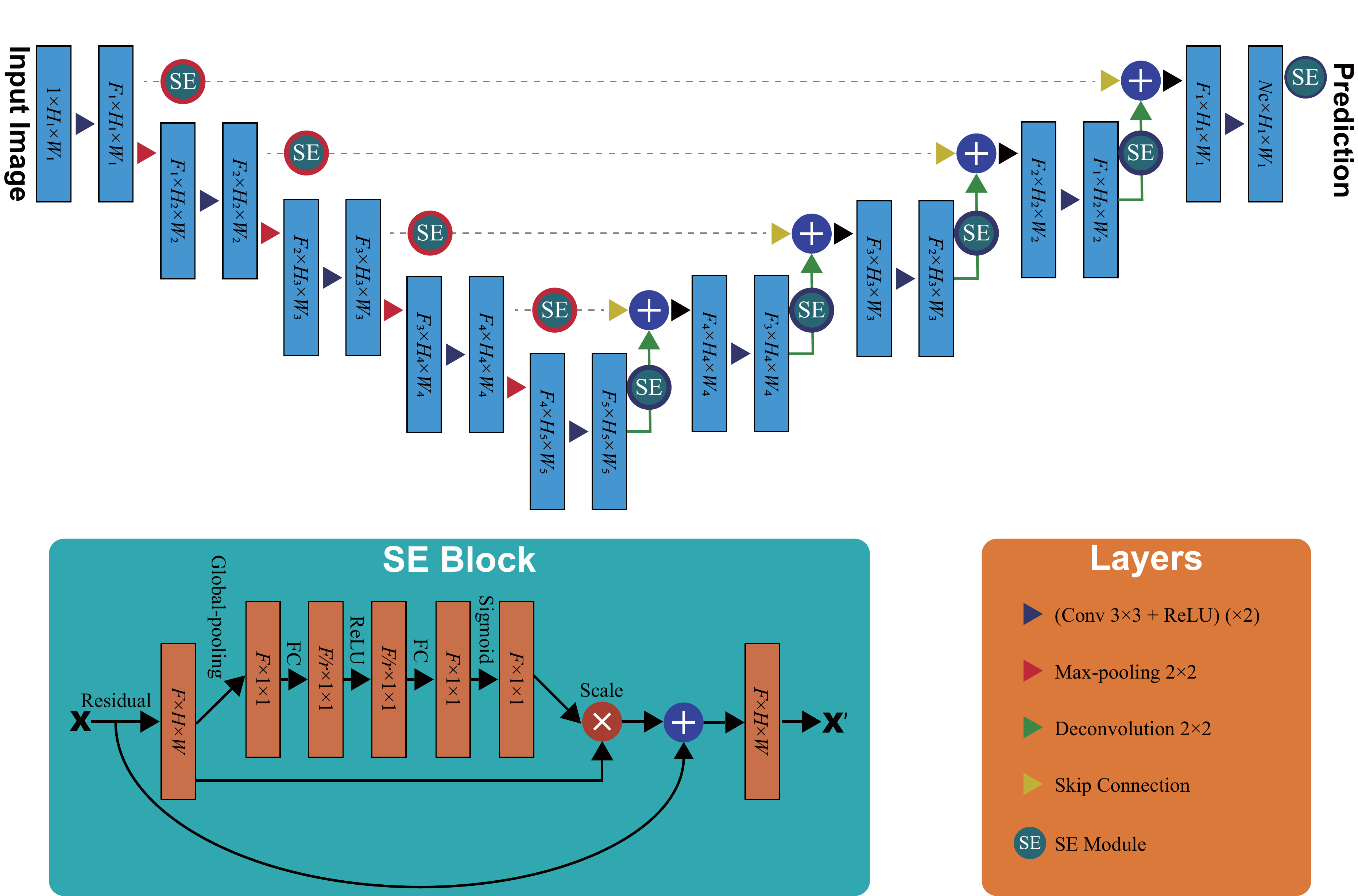}
	\caption[Scheme of the proposed USE-Net architecture]{Scheme of the proposed USE-Net architecture: Enc USE-Net has only $4$ (red-contoured) SE blocks after every encoder, whilst Enc-Dec USE-Net has $9$ SE blocks integrated after every encoder/decoder (represented with red/blue contours, respectively).}
	\label{fig:usenet}
\end{figure}

\paragraph{Contributions}
Our main contributions are:

\begin{itemize}
\item \textbf{Prostate zonal segmentation:}
our novel Enc-Dec USE-Net achieves accurate CG and PZ segmentation results on T2w MR images, remarkably outperforming the other competitor methods when trained on all datasets used for testing in multi-institutional scenarios.
\item \textbf{Cross-dataset generalization:} this first cross-dataset study among three medical datasets shows that training on the union of multi-institutional data\-sets generally outperforms training on each dataset during testing, realizing both intra-/cross-dataset generalization---thus, we may train CNNs by feeding samples from multiple different datasets for improving the performance.
\item \textbf{Deep Learning for medical imaging:} this research reveals that SE blocks provide excellent intra-dataset generalization in multi-insti\-tutional scenarios, when the testing is performed on samples from the datasets used during the training.
Therefore, adaptive mechanisms (e.g., feature recalibration in CNNs) may be a valuable solution in medical imaging applications involving multi-institutional settings.
\end{itemize}

We propose to introduce SE blocks~\cite{hu2017} following every Encoder (Enc USE-Net) or Encoder-Decoder (Enc-Dec USE-Net) of U-Net~\cite{ronneberger2015}, as shown in Fig. \ref{fig:usenet}.
As pointed out before, U-Net allows for a multi-resolution decomposition/composition technique~\cite{suzuki2006}, by combining encoders/decoders with skip connections between them ~\cite{yao2018};
in our implementation, encoders and decoders consist of four pooling operators that capture the context and up-sampling operators that conduct precise localization, respectively.

We introduce SE blocks to enhance image segmentation, expecting an increased representational power from modeling the channel-wise dependencies of convolutional features~\cite{hu2017}.
These blocks were originally envisioned for image classification using adaptive feature recalibration to boost informative features and suppress weak ones at minimal computational burden.
Enc USE-Net and Enc-Dec USE-Net are investigated to evaluate the effect of strengthened feature recalibration.

SE blocks can be formally described as follows:

\subparagraph{Squeeze}
Let $\UU = [\uu_1, \uu_2, \dots, \uu_{F}]$ be an input feature map, where $\uu_f \in \mathbb{R}^{H \times W}$ is a single channel with size $H \times W$.
Through spatial dimensions $H \times W$, a global average pooling layer generates channel-wise statistics $\zz \in \RR^{F}$, whose $f$-th element is given by:
\begin{equation}\label{squeeze}
z_f = \frac{1}{H \times W}\sum_{h=1}^{H} \sum_{w=1}^{W} [\uu_f]_{i,j}.
\end{equation}

\subparagraph{Excitation}
To limit model complexity and boost generalization, two fully-connected layers and the ReLU~\cite{nair2010} function $\delta$ transform $\zz$ with a sigmoid activation function $\sigma(\cdot)$:
\begin{equation}\label{excitation}
\mathbf{s} = \sigma(g(\mathbf{z}, \mathbf{W})) = \sigma(\WW_2\delta(\WW_1\mathbf{z})),
\end{equation}
where $\WW_1 \in \RR^{\frac{F}{r} \times F}$, $\WW_2 \in \RR^{F \times \frac{F}{r}}$, and $r$ is the reduction ratio.
In order to obtain an adaptive recalibration that ignores less important channels and emphasizes important ones (allowing for non-mutual exclusivity among multiple channels, differently from one-hot encoding), $\UU$ is rescaled into $\widetilde{\XX} = [\widetilde{\xx}_1, \widetilde{\xx}_2, \dots, \widetilde{\xx}_{F}]$ by applying Eq.~(\ref{eq:Fscale}):
\begin{equation}\label{eq:Fscale}
\widetilde{\mathbf{x}}_f = \mathbf{F}_\text{scale}(\mathbf{u}_f, s_f) = s_f \cdot \mathbf{u}_f, \mbox{ for } f = 1,2, \ldots, F,
\end{equation}
where $\FF_\text{scale}(\uu_f, s_f)$ represents the channel-wise multiplication between the feature map $\uu_f \in \RR^{H \times W}$ and the scalar $s_f \in [0,1]$.

\paragraph{Pre-processing}
\label{sec:PreProc}
To fit the image resolution of dataset $\#1$, we either center-cropped or zero-padded the images of datasets $\#2$ and $\#3$ to resize them to $288 \times 288$ pixels.
Afterwards, all images in the three datasets were masked using the corresponding prostate binary masks to omit the background and only focus on extracting the CG and PZ from the WG.
This operation can be performed either by an automated method~\cite{rundo2017Inf} or previously provided manual WG segmentation~\cite{lemaitre2015}.
As a simple form of Data Augmentation (DA), we randomly cropped the input images from $288 \times 288$ to $256 \times 256$ pixels and horizontally flipped them.

\paragraph{Post-processing}

Two efficient morphological operations were applied on the obtained $\mathcall{R}_{CG}$ binary masks to smooth boundaries and deal with disconnected regions:
\begin{itemize}
	\item a hole filling algorithm on the segmented $\mathcall{R}_{CG}$ to remove possible holes in a predicted map;
    \item  a small area removal operation dealing with connected components smaller than $\lfloor |\mathcall{R}_{WG}|/8 \rfloor$ pixels, where $|\mathcall{R}_{WG}|$ denotes the number of pixels contained in WG segmentation.
This adaptive criterion takes into account the different sizes of $\mathcall{R}_{WG}$ (ranging from the apical to the basal prostate slices).
\end{itemize}

\paragraph{Comparison against the state-of-the-art methods}

After our preliminary study in \cite{rundoWIRN2018}, we discarded SegNet \cite{badrinarayanan2017} due to the poor performance in this particular task.
Therefore, we compared USE-Net against three supervised CNN-based architectures (i.e., U-Net \cite{ronneberger2015}, pix2pix \cite{isola2016}, and MS-D Net \cite{pelt2017}) and the unsupervised continuous max-flow model proposed in~\cite{qiu2014}.
Therefore, in this comparative analysis we introduced also a very recent CNN (i.e., MS-D Net \cite{pelt2017}) and a very powerful state-of-the-art method tailored for prostate zonal segmentation \cite{qiu2014}.
All the investigated CNN-based architectures were trained using the $\mathcal{L}_{DSC}$ loss function (i.e., a continuous version of the \emph{DSC})~\cite{milletari2016} in Eq. (\ref{eq:DSCloss}).

\subparagraph{USE-Net and U-Net}
Using four scaling operations, U-Net and USE-Net were implemented on Keras with TensorFlow backend.
We used the SGD method~\cite{bottou2010} with a learning rate of $0.01$, momentum of $0.9$, weight decay of $5 \times 10^{-4}$, and batch size of $4$.
Training was executed for $50$ epochs, multiplying the learning rate by $0.2$ at the $20$-th and $40$-th epochs.

\subparagraph{pix2pix}
  This image-to-image translation method with conditional adversarial networks was used to translate the original image into the segmented one~\cite{isola2016}.
The generator and discriminator (both U-Nets in our implementation) include eight and five scaling operations, respectively.
We developed pix2pix on PyTorch. Adam~\cite{kingma2014} was used as an optimizer with a learning rate of $0.01$ for the generator---which was multiplied by $0.1$ every $20$ epochs---and $2 \times 10^{-4}$ for the discriminator. 
Training was executed for $50$ epochs with a batch size of $12$.

\subparagraph{MS-D Net}
This dilated convolution-based method, characterized by densely connected feature maps, is designed to capture features at various image scales \cite{pelt2017}.
It was implemented on PyTorch with a depth of $100$ and width of $1$.
We used Adam~\cite{kingma2014} with a learning rate of $1 \times 10^{-3}$ and trained it for $100$ epochs with a batch size of $12$.

\subparagraph{Continuous max-flow model}
This model~\cite{qiu2014} exploits duality-based convex relaxed optimization~\cite{yuan2010} to achieve better numerical stability (i.e., convergence) than classic graph cut-based methods~\cite{freedman2005}.
This semi-automatic approach simultaneously segments both $\mathcall{R}_{WG}$ and $\mathcall{R}_{CG}$ under the constraints given in Eq.~(\ref{eq:segConstraints}), relying on user intervention.
The initialization procedure consists in two closed surfaces defined by a thin-plate spline interpolating $10$-$12$ control points interactively selected by the user (considering both the axial and sagittal views).
These 3D partitions estimate the intensity probability density functions associated with three sub-regions of background, CG, and PZ.
This allows for defining the region appearance models for global optimization-based multi-region segmentation~\cite{yuan2010}.

Since the supervised CNN-based architectures rely on the gold standard $\mathcall{R}_{WG}$ for zonal segmentation, we apply the continuous max-flow method on CG for single-region segmentation for fair comparison.
Moreover, in our tests, a very accurate slice-by-slice $\mathcall{R}_{CG}$ initialization is provided by eroding the gold standard CG with a circular structuring element (radius: $6$ pixels).

\paragraph{Experimental results}
This section shows how the CNN-based architectures and the continuous max-flow model segmented the prostate zones, through the evaluation of their cross-dataset generalization ability.

\begin{figure}[!t]
\centering
	\subfloat[][]
    {\includegraphics[width=\textwidth]{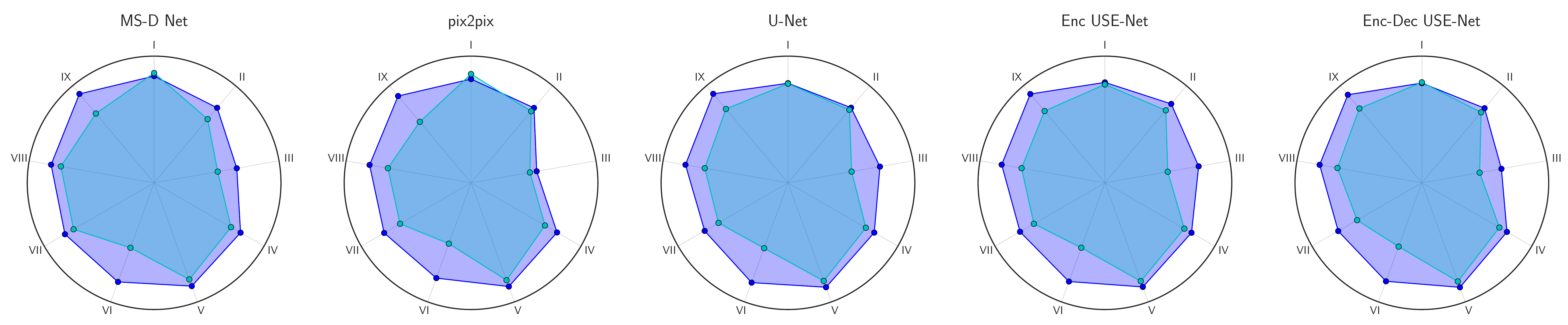}} \\
    \subfloat[][]
    {\includegraphics[width=\textwidth]{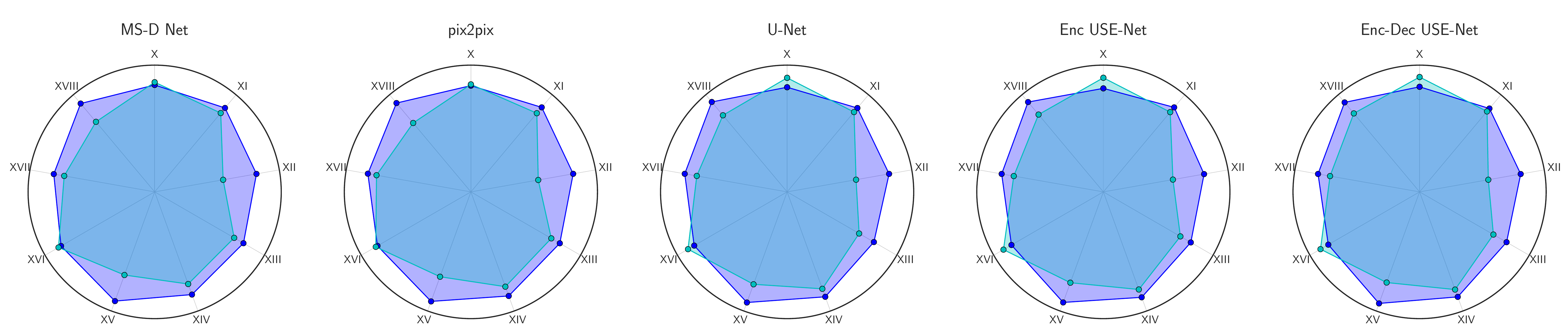}} \\
     \subfloat[][]
    {\includegraphics[width=\textwidth]{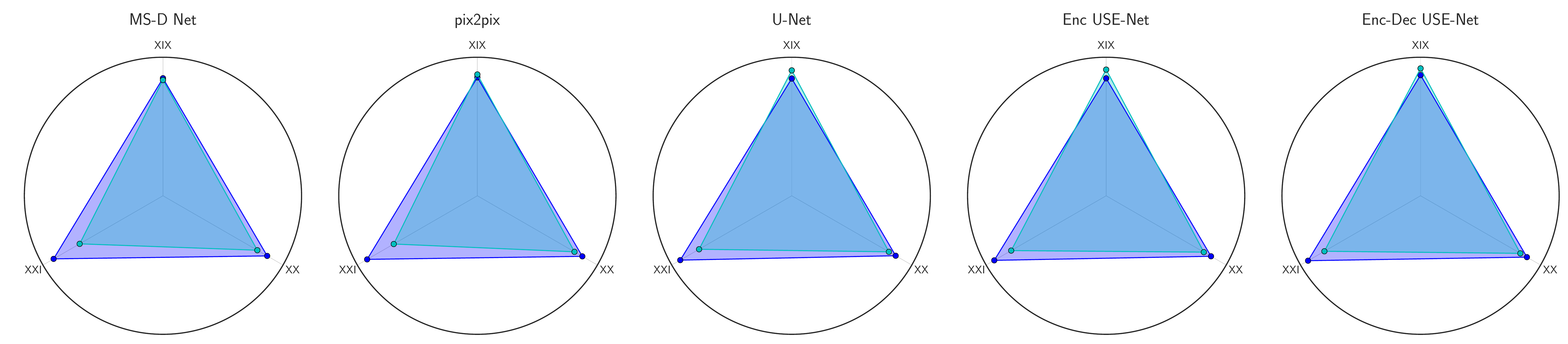}} \\
  \caption[Kiviat diagrams showing the \emph{DSC} values achieved by each method under different conditions]{Kiviat diagrams showing the \emph{DSC} values achieved by each method under different conditions.
  $\mathcall{R}_{CG}$ and $\mathcall{R}_{PZ}$ results are denoted by blue and cyan colors, respectively. Each variable represents a ``training-set $\rightarrow$ test-set'' condition as follows:
(a) one-dataset training: I) $\#1\rightarrow\#1$; II) $\#1\rightarrow\#2$; III) $\#1\rightarrow\#3$; IV) $\#2\rightarrow\#1$; V) $\#2\rightarrow\#2$; VI) $\#2\rightarrow\#3$; VII) $\#3\rightarrow\#1$; VIII) $\#3\rightarrow\#2$; IX) $\#3\rightarrow\#3$.
(b) two-dataset training: X) $\#1/\#2\rightarrow\#1$; XI) $\#1/\#2\rightarrow\#2$; XII) $\#1/\#2\rightarrow\#3$; XIII) $\#1/\#3\rightarrow\#1$; XIV) $\#1/\#3\rightarrow\#2$; XV) $\#1/\#3\rightarrow\#3$; XVI) $\#2/\#3\rightarrow\#1$; XVII) $\#2/\#3\rightarrow\#2$; XVIII) $\#2/\#3\rightarrow\#3$.
(c) three-dataset training: XIX) $\#1/\#2/\#3\rightarrow\#1$; XX) $\#1/\#2/\#3\rightarrow\#2$; XXI) $\#1/\#2/\#3\rightarrow\#3$.}
\label{fig:Kiviat}
\end{figure}

Table \ref{table:USENetresults} shows the $4$-fold cross-validation results, as assessed by the \emph{DSC} metrics, obtained under different training/testing conditions.
For visual comparison, the Kiviat diagrams~\cite{demvsar2006} for each CNN-based architecture are also displayed in Fig. \ref{fig:Kiviat}.
Here, we can observe the impact of leaving dataset $\#3$ out of the training set and, at the same time, using it as test set: the corresponding spokes III, VI, and XII generally show lower performance, probably due to the peculiar image characteristics of dataset $\#3$ (comprising the highest number of patients) that are not learned during the training phase on datasets $\#1/\#2$.
In general, Enc USE-Net performs similarly to U-Net, which stably yields satisfactory results.
More interestingly, Enc USE-Net obtains considerably better results when trained/tested on multiple datasets.
Enc-Dec USE-Net (characterized by a higher number of SE blocks with respect to Enc USE-Net) consistently and remarkably outperforms the other methods on both CG and PZ segmentation when trained on all the investigated datasets, also performing well when trained and tested on the same datasets.

\begin{table*}[!t]
\scriptsize
\centering
  \caption[Prostate zonal segmentation results of the CNN-based architectures and the unsupervised continuous max-flow model]{Prostate zonal segmentation results of the CNN-based architectures and the unsupervised continuous max-flow model (proposed by Qiu~\textit{et al.}~\cite{qiu2014}) in $4$-fold cross-validation assessed by \emph{DSC} (presented as the mean value $\pm$ standard deviation). The supervised experimental results are calculated under the different seven conditions described in Section \ref{sec:multiDatasetsProstate}.
  Numbers in bold indicate the best \emph{DSC} values (the higher the better) for each prostate region (i.e., $\mathcall{R}_{CG}$ and $\mathcall{R}_{PZ}$) among all architectures.\vspace{0.2cm}}
  
\label{table:USENetresults}
\begin{tabular}{p{3.3em}|l|ll|ll|ll}
\Hline
\multirow{2}{*}{\hspace{30pt}}   & \multicolumn{1}{c|}{\multirow{2}{*}{\textbf{Method}}} & \multicolumn{2}{c|}{\textbf{Testing on dataset $\#1$}} & \multicolumn{2}{c|}{\textbf{Testing on dataset $\#2$}} & \multicolumn{2}{c}{\textbf{Testing on dataset $\#3$}}\\
                            & \multicolumn{1}{c|}{}                                      & \multicolumn{1}{c}{\textit{CG}}            & \multicolumn{1}{c|}{\textit{PZ}}            & \multicolumn{1}{c}{\textit{CG}}          & \multicolumn{1}{c|}{\textit{PZ}} & \multicolumn{1}{c}{\textit{CG}}          & \multicolumn{1}{c}{\textit{PZ}}       \\ \hline

\parbox[t]{2mm}{\multirow{5}{*}{\rotatebox[origin=c]{270}{\textbf{\shortstack{\\Training on\\dataset\\ $\#1$\vspace{0.4mm}}}}}} & MS-D Net
                                        &  $\textbf{84.3}\pm1.6$                &       $\textbf{86.7}\pm1.6$               &       $77.3\pm4.1$           &     $65.6\pm11.2$ &       $66.2\pm3.7$           &     $\textbf{50.9}\pm1.2$             \\

                            & pix2pix                                        &  $81.9\pm2.2$                &       $85.9\pm5.0$               &       $77.2\pm2.9$           &     $73.7\pm4.3$ &       $52.5\pm3.2$           &     $47.1\pm1.3$             \\
                            & U-Net                                        &  $78.6\pm4.1$                &       $78.3\pm7.6$               &       $77.4\pm5.4$           &     $\textbf{75.3}\pm1.4$ &       $73.6\pm6.2$           &     $50.9\pm1.5$             \\

& Enc USE-Net                                        &  $79.3\pm3.5$                &       $77.7\pm2.7$               &       $\textbf{81.3}\pm1.5$           &     $74.7\pm1.8$ &       $\textbf{75.0}\pm4.2$           &     $50.3\pm1.2$             \\
                            & Enc-Dec USE-Net                                        &  $78.8\pm2.9$                &       $79.4\pm7.9$               &       $76.9\pm5.5$           &     $72.7\pm1.7$ &       $63.7\pm14.6$           &     $46.3\pm1.8$             \\
\hline
\parbox[t]{2mm}{\multirow{5}{*}{\rotatebox[origin=c]{270}{\textbf{\shortstack{\\Training on\\dataset\\ $\#2$\vspace{0.4mm}}}}}} & MS-D Net
                                        &  $78.7\pm1.1$                &       $70.0\pm4.4$               &       $86.8\pm3.7$           &     $81.1\pm0.5$ &       $83.2\pm1.0$           &     $54.6\pm0.8$             \\

                            & pix2pix                                        &  $78.3\pm0.9$                &       $67.3\pm3.2$               &       $87.1\pm2.9$           &     $81.8\pm1.0$ &       $80.0\pm2.5$           &     $51.1\pm1.5$             \\
                            & U-Net                                        &  $78.6\pm1.0$                &       $70.9\pm3.2$               &       $87.7\pm2.0$           &     $82.4\pm2.4$ &       $\textbf{83.8}\pm1.8$           &     $\textbf{54.9}\pm1.8$             \\

& Enc USE-Net                                        &  $\textbf{78.8}\pm1.4$                &       $\textbf{72.3}\pm5.6$               &       $87.4\pm2.5$           &     $82.6\pm2.1$ &       $82.9\pm2.5$           &     $54.5\pm2.0$             \\

                            & Enc-Dec USE-Net                                        &  $77.5\pm2.1$                &       $70.6\pm5.5$               &       $\textbf{87.8}\pm2.7$           &     $\textbf{82.8}\pm1.9$ &       $82.7\pm1.5$           &     $53.6\pm1.0$             \\
\hline
\parbox[t]{2mm}{\multirow{5}{*}{\rotatebox[origin=c]{270}{\textbf{\shortstack{\\Training on\\dataset\\$\#3$\vspace{0.4mm}}}}}} & MS-D Net
                                        &  $\textbf{81.2}\pm1.3$                &       $\textbf{73.3}\pm3.7$               &       $82.5\pm1.9$           &     $\textbf{74.7}\pm2.0$ &       $91.6\pm1.1$           &     $71.4\pm5.6$             \\

                            & pix2pix                                        &  $79.1\pm5.6$                &       $64.6\pm22.1$               &       $81.2\pm4.2$           &     $66.6\pm19.1$ &       $89.4\pm4.8$           &     $62.8\pm10.0$             \\
                            & U-Net                                        &  $75.9\pm3.4$                &       $63.3\pm5.0$               &       $82.1\pm2.9$           &     $66.6\pm8.4$ &       $\textbf{91.7}\pm2.4$           &     $76.1\pm4.1$             \\

& Enc USE-Net                                        &  $77.3\pm3.6$                &       $64.7\pm6.4$               &       $\textbf{82.7}\pm4.3$           &     $66.7\pm15.9$ &       $91.5\pm3.2$           &     $74.0\pm7.8$             \\

                            & Enc-Dec USE-Net                                        &  $76.1\pm4.2$                &       $58.9\pm13.7$               &       $81.8\pm4.8$           &     $67.6\pm13.2$ &       $90.7\pm3.1$           &     $\textbf{76.6}\pm7.8$             \\
\hline \hline
\parbox[t]{2mm}{\multirow{5}{*}{\rotatebox[origin=c]{270}{\textbf{\shortstack{\\Training on\\datasets\\$\#1/\#2$\vspace{0.4mm}}}}}} & MS-D Net
                                        &  $84.4\pm3.1$                &       $86.5\pm2.7$               &       $86.4\pm2.8$           &     $81.2\pm1.3$ &       $81.7\pm2.3$           &     $54.9\pm2.5$             \\

                            & pix2pix                                        &  $\textbf{83.8}\pm2.6$                &       $84.8\pm3.1$               &       $\textbf{87.1}\pm2.7$           &     $81.0\pm0.4$ &       $\textbf{82.1}\pm2.5$           &     $54.0\pm1.8$             \\
                            & U-Net                                        &  $82.6\pm3.3$                &       $90.0\pm2.7$               &       $86.4\pm2.0$           &     $82.2\pm2.7$ &       $81.8\pm2.1$           &     $55.3\pm2.5$             \\

& Enc USE-Net                                        &  $81.7\pm5.4$                &       $90.0\pm2.1$               &       $87.0\pm2.1$           &     $82.2\pm1.8$ &       $80.8\pm2.7$           &     $\textbf{55.8}\pm1.7$             \\

                            & Enc-Dec USE-Net                                        &  $85.4\pm1.8$                &       $\textbf{87.7}\pm2.5$               &       $85.9\pm2.1$           &     $\textbf{82.9}\pm1.4$ &       $81.1\pm2.7$           &     $55.1\pm2.1$             \\
\hline
\parbox[t]{2mm}{\multirow{5}{*}{\rotatebox[origin=c]{270}{\textbf{\shortstack{\\Training on\\datasets\\$\#1/\#3$\vspace{0.4mm}}}}}} & MS-D Net
                                        &  $\textbf{85.4}\pm1.8$                &       $87.7\pm2.5$               &       $80.9\pm2.7$           &     $72.6\pm3.7$ &       $91.0\pm2.9$           &     $72.2\pm1.9$             \\

                            & pix2pix                                        &  $85.2\pm1.6$                &       $86.8\pm2.4$               &       $\textbf{82.7}\pm1.9$           &     $\textbf{75.7}\pm3.6$ &       $91.5\pm1.9$           &     $71.0\pm3.6$             \\
                            & U-Net                                        &  $84.8\pm0.4$                &       $90.4\pm2.8$               &       $82.1\pm2.9$           &     $72.5\pm4.5$ &       $\textbf{92.6}\pm1.5$           &     $78.9\pm4.0$             \\

& Enc USE-Net                                        &  $83.8\pm1.4$                &       $\textbf{91.1}\pm1.4$               &       $81.6\pm3.7$           &     $71.9\pm8.1$ &       $92.5\pm1.9$           &     $79.6\pm2.1$             \\

                            & Enc-Dec USE-Net                                        &  $83.3\pm3.2$                &       $90.4\pm2.1$               &       $81.5\pm4.6$           &     $71.8\pm6.7$ &       $92.2\pm2.4$           &     $\textbf{80.8}\pm1.8$             \\
\hline
\parbox[t]{2mm}{\multirow{5}{*}{\rotatebox[origin=c]{270}{\textbf{\shortstack{\\Training on\\datasets\\$\#2/\#3$\vspace{0.4mm}}}}}} & MS-D Net
                                        &  $81.0\pm1.3$                &       $72.5\pm5.6$               &       $86.2\pm2.5$           &     $77.4\pm4.7$ &       $91.7\pm0.9$           &     $69.8\pm3.6$             \\

                            & pix2pix                                        &  $\textbf{81.1}\pm1.1$                &       $\textbf{73.4}\pm3.3$               &       $87.4\pm2.2$           &     $79.6\pm5.7$ &       $92.0\pm1.3$           &     $71.3\pm3.4$             \\
                            & U-Net                                        &  $79.2\pm2.0$                &       $65.7\pm6.3$               &       $88.1\pm2.9$           &     $81.4\pm2.6$ &       $92.9\pm1.1$           &     $\textbf{77.6}\pm3.0$             \\

& Enc USE-Net                                        &  $79.8\pm1.8$                &       $70.3\pm7.6$               &       $\textbf{88.5}\pm2.4$           &     $\textbf{82.0}\pm3.2$ &       $92.8\pm1.0$           &     $76.3\pm2.7$             \\

                            & Enc-Dec USE-Net                                        &  $79.4\pm2.5$                &       $67.4\pm8.9$               &       $88.2\pm2.9$           &     $82.0\pm4.1$ &       $\textbf{93.7}\pm0.6$           &     $76.1\pm3.4$             \\
\hline \hline
\parbox[t]{2mm}{\multirow{5}{*}{\rotatebox[origin=c]{270}{\textbf{\shortstack{\\Training on\\datasets\\$\#1/\#2/\#3$\vspace{0.4mm}}}}}} & MS-D Net
                                        &  $84.8\pm4.5$                &       $83.6\pm6.9$               &       $86.8\pm2.6$           &     $78.6\pm5.1$ &       $91.1\pm1.0$           &     $69.4\pm4.5$             \\

                            & pix2pix                                        &  $85.5\pm2.6$                &       $87.6\pm3.5$               &       $87.5\pm2.0$           &     $80.9\pm5.3$ &       $91.8\pm1.3$           &     $69.7\pm4.8$             \\
                            & U-Net                                        &  $84.6\pm1.9$                &       $90.5\pm3.0$               &       $86.6\pm2.0$           &     $80.9\pm3.3$ &       $92.9\pm1.1$           &     $77.2\pm2.0$             \\

& Enc USE-Net                                        &  $84.8\pm2.3$                &       $91.1\pm2.5$               &       $87.4\pm1.8$           &     $81.4\pm4.4$ &       $93.2\pm0.7$           &     $79.1\pm3.5$             \\

                            & Enc-Dec USE-Net                                        &  $\textbf{87.1}\pm3.6$                &       $\textbf{91.9}\pm2.1$               &       $\textbf{88.6}\pm1.5$           &     $\textbf{83.1}\pm2.9$ &       $\textbf{93.7}\pm1.0$           &    $\textbf{80.1}\pm5.5$             \\
\hline \hline
\textbf{\,\,None} & Qiu \textit{et al.} \cite{qiu2014}
                                        &  $78.0\pm4.9$                &       $75.3\pm6.4$               &       $71.0\pm7.0$           &     $77.3\pm2.6$ &       $82.1\pm1.5$           &     $61.9\pm4.6$             \\
\Hline
\end{tabular}
\end{table*}


\begin{figure}[!t]
\captionsetup[subfigure]{labelformat=empty}
\centering
	\subfloat[]{\includegraphics[width=0.85\textwidth]{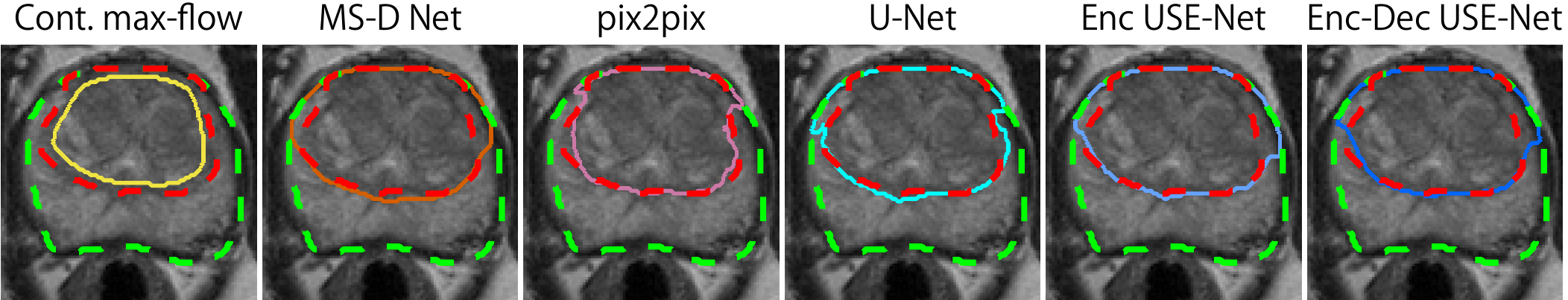}\label{sfig:segResD1a}}\\
    \addtocounter{subfigure}{-1}
    \captionsetup[subfigure]{labelformat=parens}
    \vspace{-1.1cm}
	\subfloat[]{\includegraphics[width=0.85\textwidth]{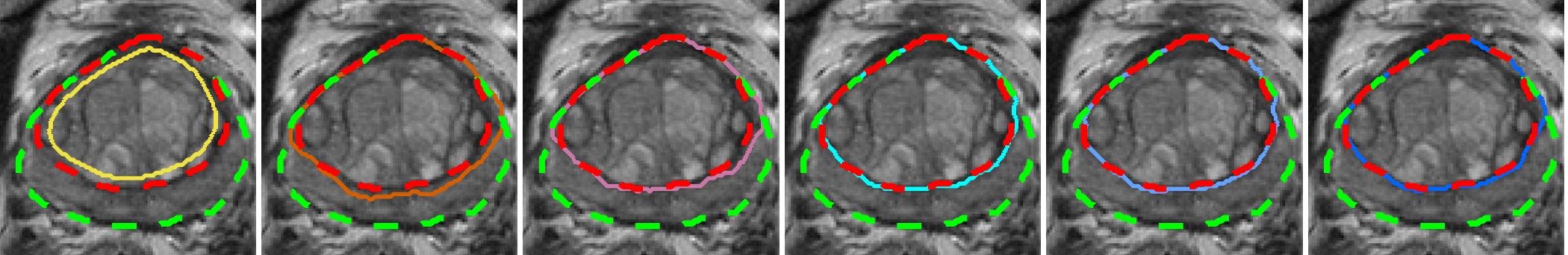}\label{sfig:segResD1b}}\\
    \captionsetup[subfigure]{labelformat=empty}
    \vspace{-0.4cm}
    	\subfloat[]{\includegraphics[width=0.85\textwidth]{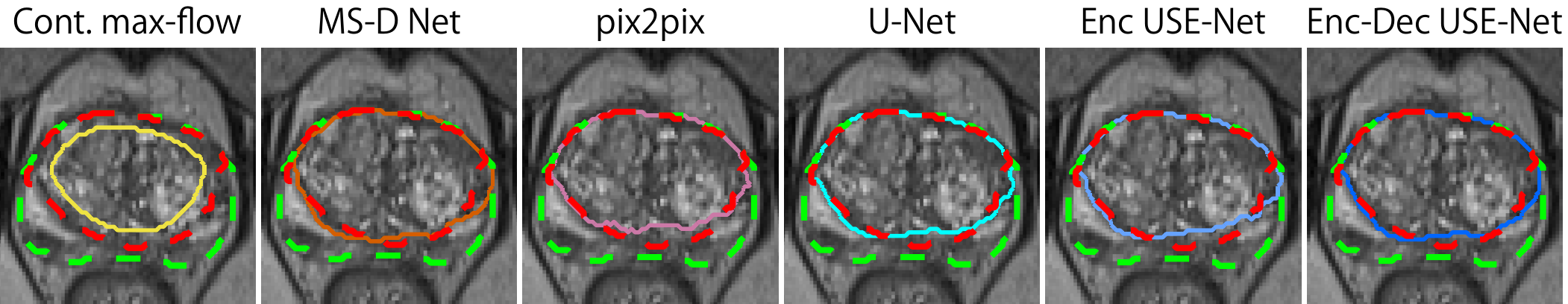}\label{sfig:segResD2a}}\\
        \addtocounter{subfigure}{-1}
        \captionsetup[subfigure]{labelformat=parens}
    \vspace{-1.1cm}
	\subfloat[]{\includegraphics[width=0.85\textwidth]{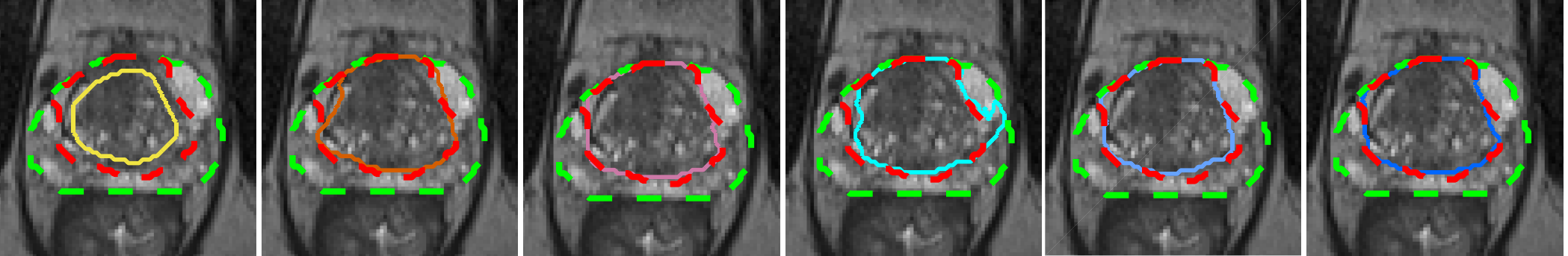}\label{sfig:segResD2b}}\\
    \captionsetup[subfigure]{labelformat=empty}
    \vspace{-0.4cm}
    	\subfloat[]{\includegraphics[width=0.85\textwidth]{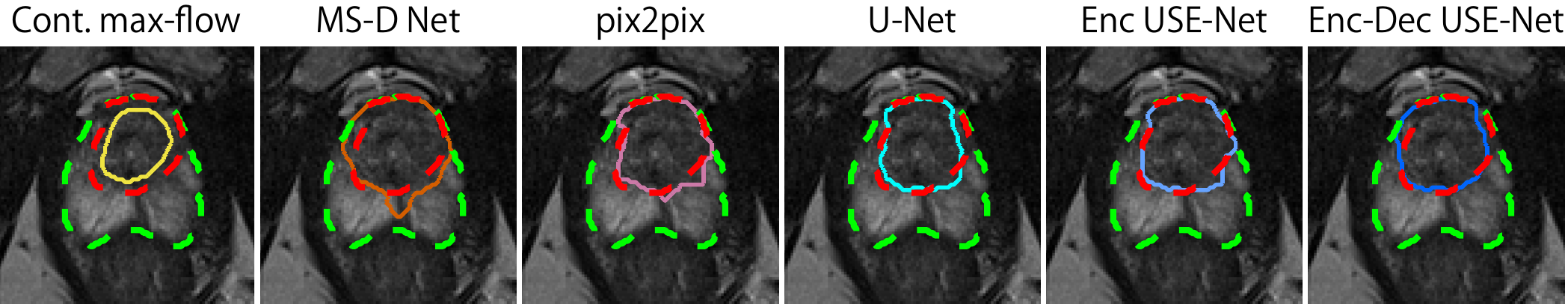}\label{sfig:segResD3a}}\\
        \addtocounter{subfigure}{-1}
        \captionsetup[subfigure]{labelformat=parens}
    \vspace{-1.1cm}
	\subfloat[]{\includegraphics[width=0.85\textwidth]{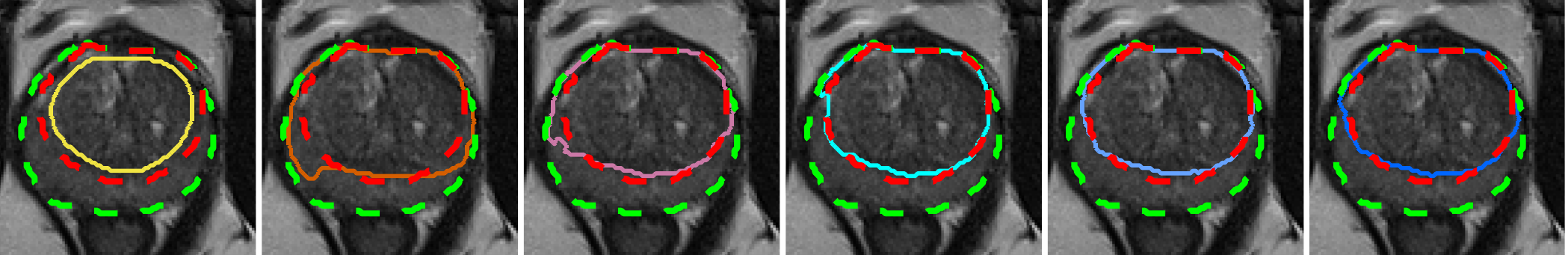}\label{sfig:segResD3b}}\\
	\caption[Segmentation results obtained by the six investigated methods (under the three-dataset training condition)]{Segmentation results obtained by the six investigated methods (under the three-dataset training condition) on two different images for each dataset: (a) $\#1$; (b) $\#2$; (c) $\#3$. Automatic $\mathcall{R}_{CG}$ segmentations (solid lines) are compared against the corresponding gold standards (dashed red line). $\mathcall{R}_{PZ}$ segmentations can obtained from $\mathcall{R}_{CG}$ and $\mathcall{R}_{WG}$ (dashed green line) the constraints according to Eq.~(\ref{eq:segConstraints}).}
	\label{fig:segResD}
\end{figure}

\begin{figure}[t!]
\centering
	\subfloat[][Central Gland]
    {\includegraphics[width=\textwidth]{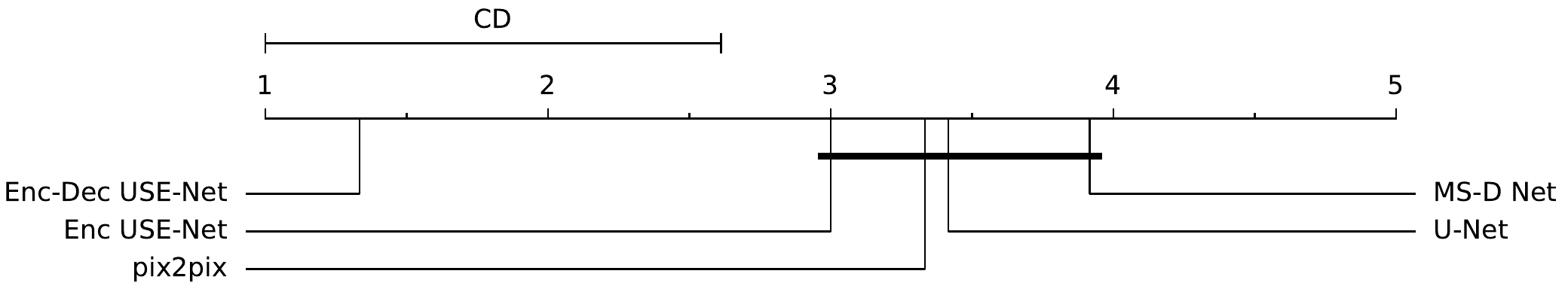}\label{sfig:BonferroniCG_triple}} \\
    \subfloat[][Peripheral Zone]
    {\includegraphics[width=\textwidth]{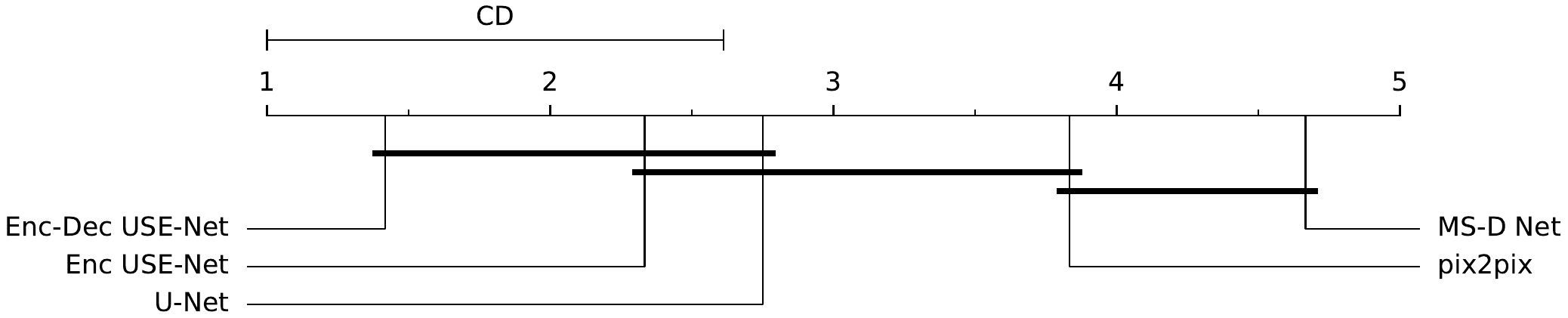}\label{sfig:BonferroniPZ_triple}} \\ 
  \caption[Critical Difference (CD) diagram comparing the \emph{DSC} values achieved by all the investigated CNN-based architectures using the Bonferroni-Dunn's \textit{post hoc} test]{Critical Difference (CD) diagram comparing the \emph{DSC} values achieved by all the investigated CNN-based architectures using the Bonferroni-Dunn's \textit{post hoc} test with $95\%$ confidence level for the three-dataset training conditions. Bold lines indicate groups of methods whose performance difference was not statistically significant.}
  \label{fig:Bonferroni_three}
\end{figure}

In order to visualize the achieved results, example images segmented by each method are compared in Fig. \ref{fig:segResD} under the three-dataset training condition.
The critical difference diagram (Fig. \ref{fig:Bonferroni_three}) using the Bonferroni-Dunn's \textit{post hoc} test also confirms this trend, considering \emph{DSC} values for every round of the $4$-fold cross-validation.

\begin{figure}[ht!]
\centering
	\subfloat[][Central Gland]
    {\includegraphics[width=\textwidth]{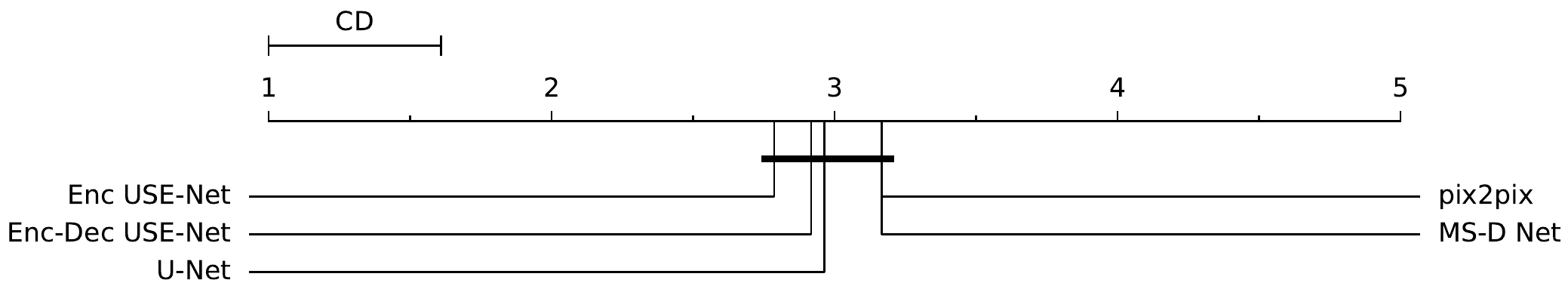}\label{sfig:BonferroniCG_all}} \\
    \subfloat[][Peripheral Zone]
    {\includegraphics[width=\textwidth]{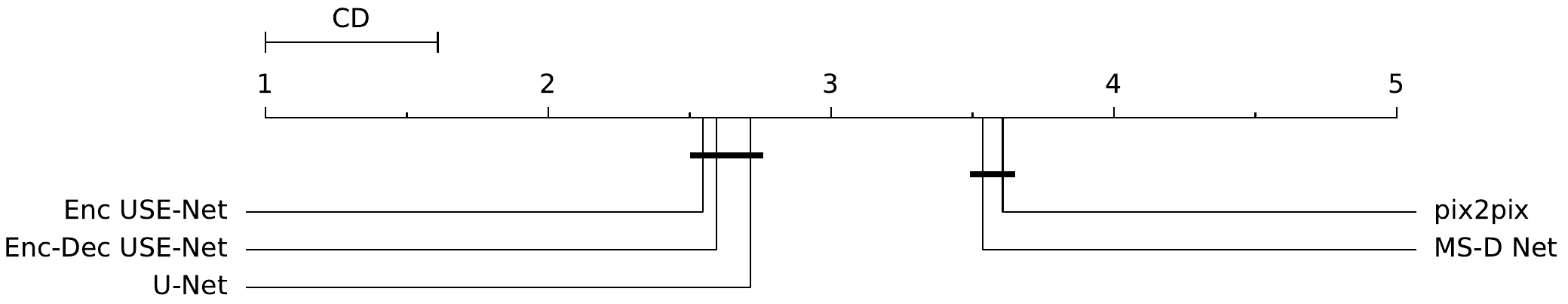}\label{sfig:BonferroniPZ_all}} \\ 
  \caption[Critical Difference (CD) diagram comparing the \emph{DSC} values achieved by all the investigated CNN-based architectures using the Bonferroni-Dunn's \textit{post hoc} test]{Critical Difference (CD) diagram comparing the \emph{DSC} values achieved by all the investigated CNN-based architectures using the Bonferroni-Dunn's \textit{post hoc} test with $95\%$ confidence level considering all training/testing combinations. Bold lines indicate groups of methods whose performance difference was not statistically significant.}
  \label{fig:Bonferroni_all}
\end{figure}

However, as shown in Fig. \ref{fig:Bonferroni_all}, Enc-Dec USE-Net shows less powerful cross-dataset generalization when trained and tested on different datasets, achieving slightly lower average performance than Enc USE-Net (considering all training/testing combinations).
This implies that the SE blocks' adaptive feature recalibration---boosting informative features and suppressing weak ones---provides excellent intra-dataset generalization when testing is performed on multiple datasets used during training (i.e., when training samples from every testing dataset are fed to the model).

On the contrary, pix2pix achieves good generalization when trained and tested on different datasets, especially under mixed-dataset training conditions, thanks to its internal generative model.
MS-D Net generally works better in single dataset scenarios, using a limited amount of training samples, according to \cite{pelt2017}.
The unsupervised continuous max-flow model achieves comparable results to the supervised ones only when trained and tested on different datasets.
However, this semi-automatic approach is outperformed by the supervised methods when trained and tested on the same datasets, as it underestimates $\mathcall{R}_{CG}$.

The results also reveal that training on multi-institutional datasets generally outperforms training on each dataset during testing on any dataset/zone, realizing both intra-/cross-dataset generalization.
For instance, training on datasets $\#1$ and $\#2$ generally outperforms training on dataset $\#1$ during testing on all datasets $\#1$, $\#2$, and $\#3$, without losing accuracy. 

Therefore, training schemes with mixed MRI datasets can achieve reliable and excellent performance, potentially useful for other clinical applications.
Comparing the CG and PZ segmentation, the results on the CG are generally more accurate, except when trained and tested on dataset $\#1$; this could be due to intra- and inter-scanner generalization, since dataset $\#1$'s scanner is different from those of datasets $\#2$ and $\#3$.

Hence, Enc-Dec USE-Net obtained high performance also in terms of difference between the automated and the manual boundaries.

\paragraph{Discussion and conclusions}

The novel CNN architecture introduced in this work, Enc-Dec USE-Net, achieved accurate prostate zonal segmentation results when trained on the union of the available datasets in the case of multi-institutional studies---significantly outperforming the competitor CNN-based architectures, thanks to the integration of SE blocks~\cite{hu2017} into U-Net~\cite{ronneberger2015}.
This also derives from the presented cross-dataset generalization approach among three prostate MRI datasets, collected by three different institutions, aiming at segmenting $\mathcall{R}_{CG}$ and $\mathcall{R}_{PZ}$; Enc-Dec USE-Net's segmentation performance considerably improved when trained on multiple datasets with respect to individual training conditions.
Since the training on multi-institutional datasets analyzed in this work achieved good intra-/cross-dataset generalization, CNNs could be trained on multiple datasets with different devices/protocols to obtain better outcomes in clinically feasible applications. 
Moreover, our research also implies that state-of-the-art CNN architectures properly combined with innovative concepts, such as feature recalibration provided by the SE blocks~\cite{hu2017}, allow for excellent intra-dataset generalization when tested on samples coming from the datasets used for the training phase.
Therefore, such adaptive mechanisms may be a valuable solution in medical imaging applications involving multi-institutional settings.

As future developments, we will refine the output images considering the 3D spatial information among the prostate MR slices.
Finally, for better cross-dataset generalization, we plan to use domain adaptation \textit{via} transfer learning by maximizing the distribution similarity \cite{vanOpbroek2015}.
In this context, Generative Adversarial Networks (GANs)~\cite{goodfellow2014,han2018} and Variational Auto-Encoders (VAEs)~\cite{kingma2013} represent useful solutions.

\section{Generative Adversarial Networks}
\label{sec:GANs}
\begin{figure}[t]
  \centering
  \centerline{\includegraphics[width=8.5cm]{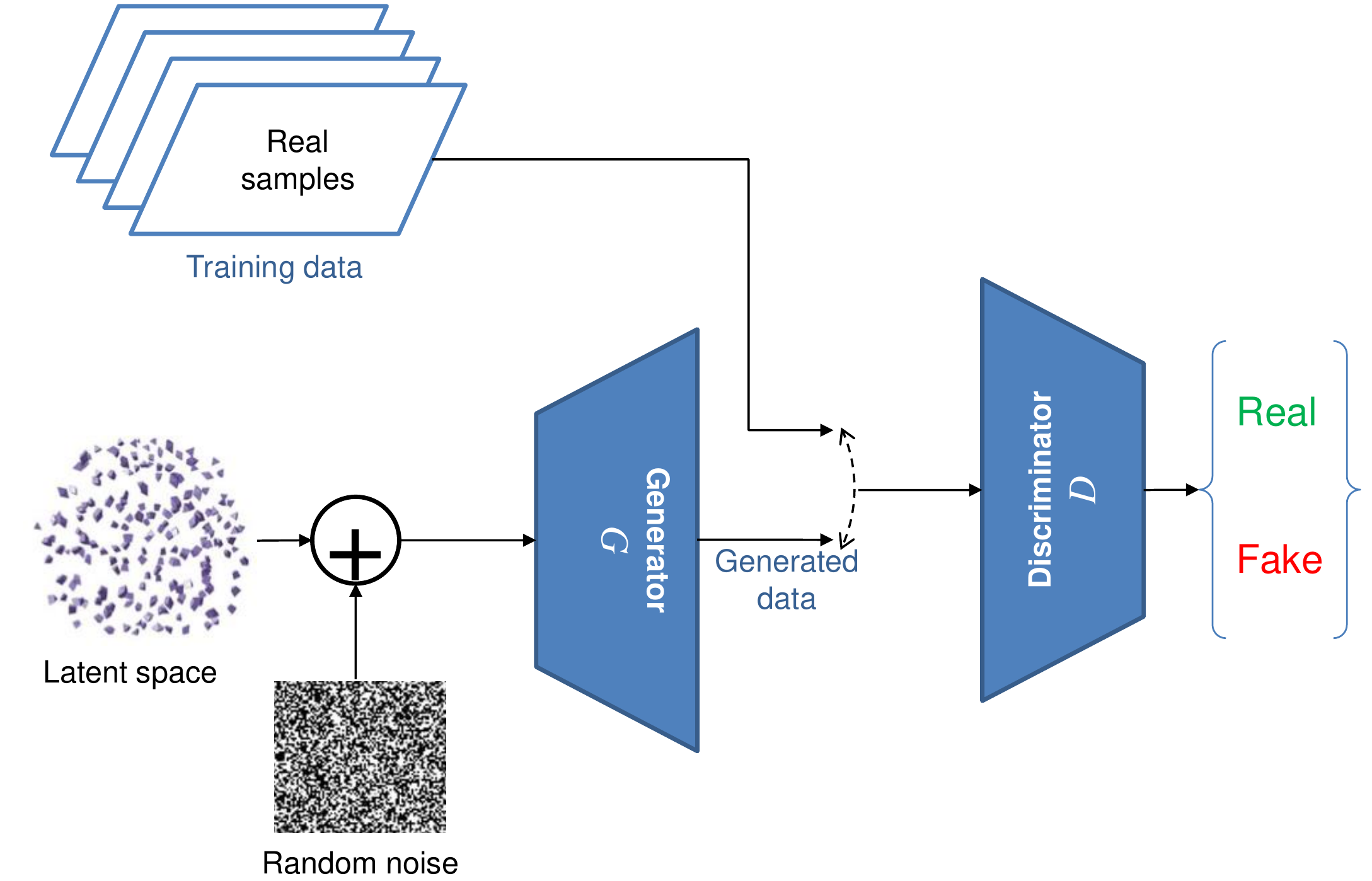}}
\caption[Conceptual scheme of GANs]{Conceptual scheme of GANs.}
\label{fig:GANscheme}
\end{figure}

GANs introduce a framework for unsupervised learning aiming at estimating generative models \textit{via} an adversarial process by simultaneously training two models~\cite{goodfellow2014}.
Originally proposed by Goodfellow \textit{et al.} in 2014~\cite{goodfellow2014}, GANs have shown remarkable results in image generation~\cite{zhu2017}.
Let $p_{\rm data}$ be a generating distribution over data $\mathbf{x}$.
The generator $G(\mathbf{z}; \theta_{g})$ is a mapping to data space that takes a prior on input noise variables $p_{\mathbf{z}}(\mathbf{z})$, where $G$ is a neural network with parameters $\theta_{g}$.
Similarly, the discriminator $D(\mathbf{x}; \theta_{d})$ is a neural network with parameters $\theta_{d}$ that takes either real data or synthetic data and outputs a single scalar probability that $\mathbf{x}$ came from the real data.
The discriminator $D$ maximizes the probability of classifying both training examples and samples from $G$ correctly while the generator $G$ minimizes the likelihood (Fig. \ref{fig:GANscheme}); it is formulated as a minimax two-player game with value function $\mathcal{V}(G,D)$:
\begin{equation}
\min_{G} \max_{D} \mathcal{V}(D,G) = \mathbb{E}_{\mathbf{x} \sim p_{{\text{data}}} (\mathbf{x})} [\log D(\mathbf{x})] + \mathbb{E}_{\mathbf{z} \sim p_{\mathbf{z}} (\mathbf{z})} [\log (1 - D(G(\mathbf{z})))].
\end{equation}
This can be reformulated as the minimization of the Jensen-Shannon (JS) divergence between the distribution $p_{\rm data}$ and another distribution $p_{g}$ derived from $p_{\mathbf{z}}$ and $G$.

Along with classic Image Processing methods, CNNs have recently revolutionized medical image analysis~\cite{shen2017}, including MRI segmentation~\cite{havaei2017,kamnitas2017}.
However, CNN training demands large-scale annotated training data that are laborious to obtain in the medical domain~\cite{ravi2017}.
Unfortunately, obtaining such massive medical data is challenging; consequently, better training requires intensive DA techniques, such as geometric/intensity transformations of original images~\cite{milletari2016,ronneberger2015}.

However, those reconstructed images intrinsically resemble the original ones, leading to limited performance improvement in terms of generalization abilities; thus, generating realistic (similar to the real image distribution) but completely new images is essential.
In this context, GAN-based DA has excellently performed in general computer vision tasks.
It attributes to GAN's good generalization ability from matching the generated distribution from noise variables to the real one with a sharp value function.
Especially, Shrivastava \emph{et al.} (SimGAN) outperformed the state-of-the-art with a relative $21\%$ improvement in eye-gaze estimation~\cite{shrivastava2017}.

\subsection{GAN-based synthetic brain MR image generation}
\begin{figure}[t]
  \centering
  \centerline{\includegraphics[width=8.5cm]{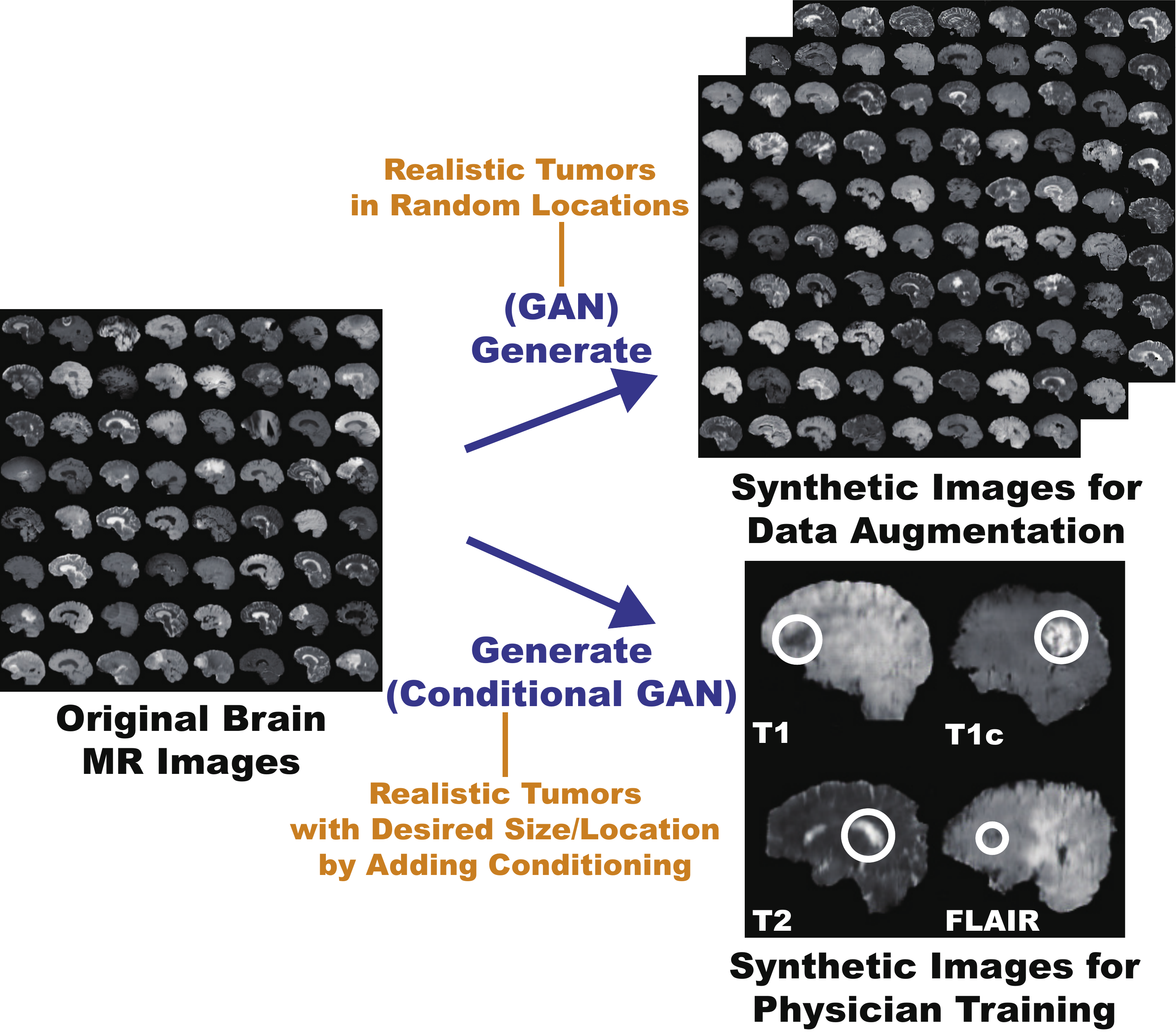}}
\caption[Potential applications of the proposed GAN-based synthetic brain MR image generation]{Potential applications of the proposed GAN-based synthetic brain MR image generation: (1) DA for better diagnostic accuracy by generating random realistic images giving insights in classification; (2) physician training for better understanding various diseases to prevent misdiagnosis by generating desired realistic pathological images.}
\label{fig:GANappl}
\end{figure}

The generation of realistic medical images completely different from the original samples is worth to investigate.
Our work in \cite{han2018} aims to generate synthetic multi-sequence brain MR images using GANs, which is essential in medical imaging to increase diagnostic reliability, such as \textit{via} DA in computer-assisted diagnosis as well as physician training and teaching (Fig.~\ref{fig:GANappl})~\cite{prastawa2009}.
However, this is extremely challenging---MR images are characterized by low contrast, strong visual consistency in brain anatomy, and intra-sequence variability.
Our novel GAN-based approach for medical DA adopts Deep Convolutional GAN (DCGAN)~\cite{radford2016} and Wasserstein GAN (WGAN)~\cite{arjovsky2017} to generate realistic images, and an expert physician validates them \textit{via} the visual Turing test (described in Appendix \ref{sec:TuringEval})~\cite{salimans2016}.

We mainly addressed two research questions:
\begin{itemize}
\item \textbf{GAN selection:} Which GAN architecture is well-suited for realistic medical image generation?
\item \textbf{Medical DA:} How can we handle MR images with specific intra-sequence variability?\\
\end{itemize}

Our main contributions are as follows:
\begin{itemize}
\item \textbf{MR image generation:} This research shows that WGAN can generate realistic multi-sequence brain MR images, possibly leading to valuable clinical applications: DA and physician training.
\item \textbf{Medical image generation:} This research provides how to exploit medical images with intrinsic intra-sequence variability towards GAN-based DA for medical imaging.
\end{itemize}

GANs generate highly realistic images, without a well-defined objective function associated with difficult training accompanying oscillations and mode collapse---i.e., a common failure case where the generator learns with extremely low variety. Whereas VAEs~\cite{kingma2014}, the other most used deep generative models, have an objective likelihood function to optimize, and could so generate blurred samples because of the injected noise and imperfect reconstruction~\cite{mescheder2017}.

Therefore, many medical imaging researchers have begun to use GANs recently, such as in image super-resolution~\cite{mahapatra2017}, anomaly detection~\cite{schlegl2017}, and estimating CT images from the corresponding MR images~\cite{nie2017}.
As GANs allow for adding conditioning on the class labels and images, they often use such conditional GANs~\cite{mirza2014} to produce desired images, while it makes learning robust latent spaces difficult.

Differently from a very recent work of GANs for biological image synthesis (fluorescence microscopy)~\cite{osokin2017}, to the best of our knowledge, this is the first GAN-based realistic brain tumor MR image generation approach aimed at DA and physician training.
Instead of reconstructing real brain MR images themselves with respect to geometry/intensity, a completely different approach---generating novel realistic images using GANs---may become a clinical breakthrough.

Towards clinical applications utilizing realistic brain MR images, we generate synthetic brain MR images from the original samples using GANs.
Here, we compare the two most used GANs, namely DCGAN and WGAN, to find a well-suited GAN between them for medical image generation---it must avoid mode collapse and generate realistic MR images with high resolution.

\begin{figure}[t]
  \centering
  \centerline{\includegraphics[width=0.9\textwidth]{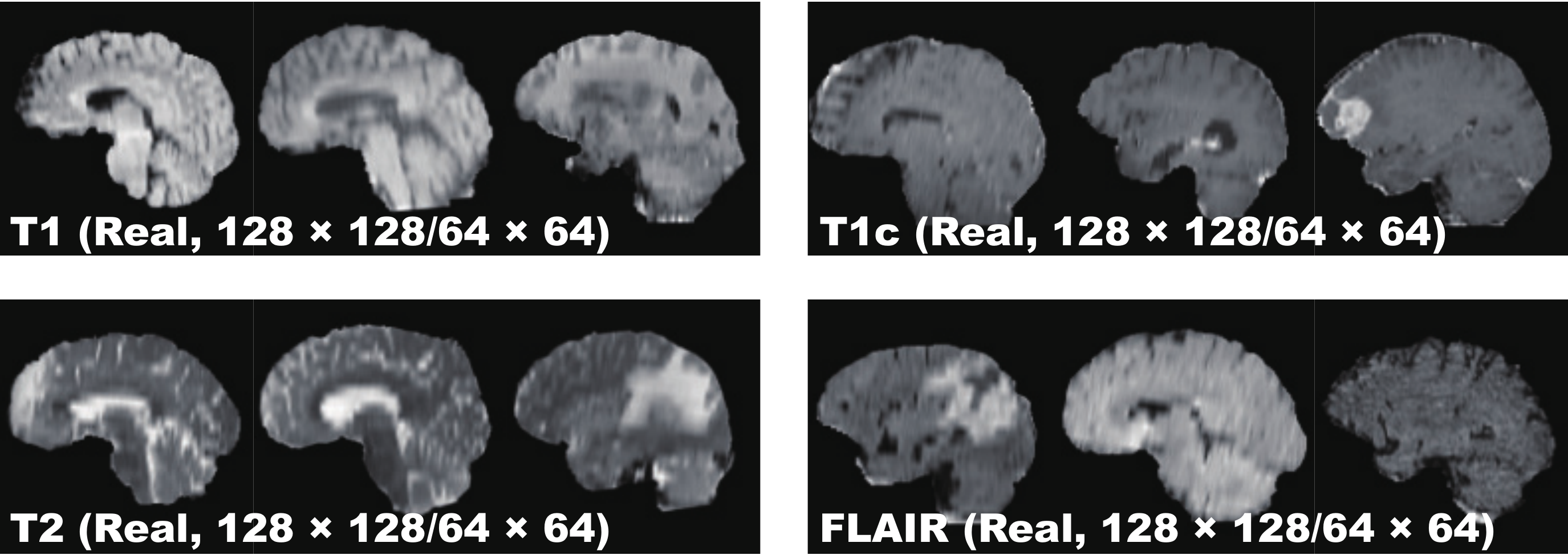}}
\caption[Examples of real MR images from the BRATS 2016 training dataset used for GAN training]{Examples of real MR images used for training the GANs: the resized sagittal multi-sequence brain MRI scans of patients with HGG on the BRATS 2016 training dataset~\cite{menze2015}.}
\label{fig:realBRATS}
\end{figure}

\subsubsection{The BRATS 2016 dataset}
This study exploits a dataset of multi-sequence brain MR images to train GANs with sufficient data and resolution, which was originally produced for the Multimodal BRATS challenge~\cite{menze2015}.
In particular, the BRATS 2016 training dataset contains $220$ High-Grade Glioma (HGG) and $54$ Low-Grade Glioma (LGG) cases, with T1w (T1), T1w-CE (T1c), T2w (T2), and FLAIR sequences---they were skull stripped and resampled to isotropic $1\mbox{mm} \times 1\mbox{mm} \times 1\mbox{mm}$ resolution with image dimension $240 \times 240 \times 155$; among the different sectional planes, we use the sagittal multi-sequence scans of patients with HGG to show that our GANs can generate a complete view of the whole brain anatomy (allowing for visual consistency among the different brain lobes), including also severe tumors for clinical purpose.

\subsubsection{Proposed GAN-based image generation approach}
This section describes the details of the GAN-based MR image generation.

\paragraph{Pre-processing}
We select the slices from $\#80$ to $\#149$ among the whole $240$ slices to omit initial/final slices, since they convey a negligible amount of useful information and could affect the training.
The images are resized to both $64 \times 64$ and $128 \times 128$ from $240 \times 155$ for better GAN training (DCGAN architecture results in stable training on 64 $\times$ 64~\cite{radford2016}, and so $128 \times 128$ is reasonably a high-resolution).
Fig.~\ref{fig:realBRATS} shows some real MR images used for training; each sequence contains $15,400$ images with $220$ patients $\times$ $70$ slices ($61,600$ in total).

\paragraph{GAN-based MR image generation}
DCGAN and WGAN generate six types of images as follows:
\begin{itemize}
\item T1 sequence ($128 \times 128$) from the real T1;
\vspace{-0.11in}
\item T1c sequence ($128 \times 128$) from the real T1c;
\vspace{-0.11in}
\item T2 sequence ($128 \times 128$) from the real T2;
\vspace{-0.11in}
\item FLAIR sequence ($128 \times 128$) from the real FLAIR;
\vspace{-0.11in}
\item Concat sequence ($128 \times 128$) from concatenating the real T1, T1c, T2, and FLAIR (i.e., feeding the model with samples from all the MRI sequences);
\vspace{-0.11in}
\item Concat sequence ($64 \times 64$) from concatenating the real T1, T1c, T2, and FLAIR.
\end{itemize}
Concat sequence refers to a new ensemble sequence for an alternative DA, containing the features of all four sequences. We also generate $64 \times 64$ Concat images to compare the generation performance in terms of image size.\\

\noindent \textbf{DCGAN.} DCGAN~\cite{radford2016} is a standard GAN~\cite{goodfellow2014} with a convolutional architecture for unsupervised learning; this generative model uses up-convolutions interleaved with ReLu non-linearity and batch-normalization.\\

\noindent \textbf{DCGAN implementation details.} We use the same DCGAN architecture~\cite{radford2016} with no $\tanh$ in the generator, ELU as the discriminator, all filters of size $4 \times 4$, and a half channel size for DCGAN training. A batch size of $64$ and Adam optimizer with $2.0 \times 10^{-4}$ learning rate were implemented.\\

\noindent \textbf{WGAN.} WGAN~\cite{arjovsky2017} is an alternative to traditional GAN training, as the JS divergence is limited, such as when it is discontinuous; this novel GAN achieves stable learning with less mode collapse by replacing it to the Earth Mover (EM) distance (a.k.a. the Wasserstein-1 metrics):
\begin{equation}
W(p_{g},p_{r}) = \inf_{p \in \prod (p_{g},p_{r})} \mathbb{E}_{(\mathbf{x},\mathbf{x}') \sim p} \| \mathbf{x} - \mathbf{x}' \|,
\end{equation}
where $\prod (p_{g},p_{r})$ is the set of all joint distributions $p$ whose marginals are $p_{g}$ and $p_{r}$, respectively. In other words, $p$ implies how much mass must be transported from one distribution to another. This distance intuitively indicates the cost of the optimal transport plan.\\

\noindent \textbf{WGAN implementation details.} We used the same DCGAN architecture~\cite{radford2016} for WGAN training. A batch size of 64 and Root Mean Square Propagation (RMSprop) optimizer with $5.0 \times 10^{-5}$ learning rate were implemented. 

\begin{figure}[t]
  \centering
  \centerline{\includegraphics[width=0.9\textwidth]{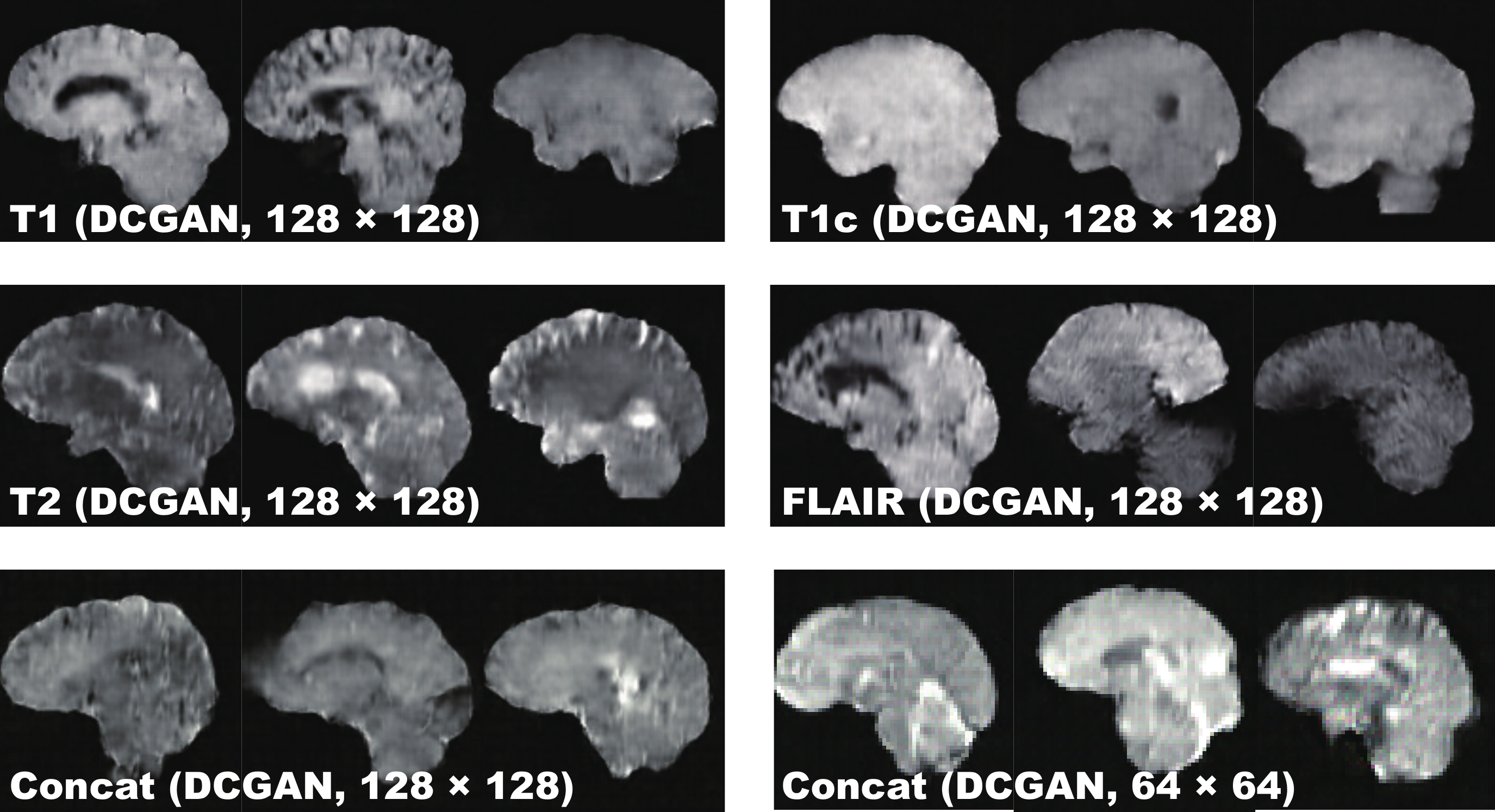}}
\caption[Examples of synthetic MR images yielded by DCGAN]{Examples of synthetic MR images yielded by DCGAN.}
\label{fig:synthDCGAN}
\end{figure}

\paragraph{Clinical validation using the visual Turing test}
In the visual Turing test (see Appendix \ref{sec:TuringEval} for more details), an expert physician was asked to constantly classify a random selection of $50$ real/$50$ synthetic MR images as real or synthetic shown in a random order for each GAN/sequence, without previous training stages revealing which is real/synthetic; Concat images were classified together with real T1, T1c, T2, and FLAIR images in equal proportion. 

\subsubsection{Results}
This section shows how DCGAN and WGAN generate synthetic brain MR images. The results include instances of synthetic images and their quantitative evaluation of the realism by an expert physician. The training took about $2$ ($1$) hours to train each $128 \times 128$ ($64 \times 64$) sequence on an NVIDIA (Santa Clara, CA, USA) GeForce GTX 980 GPU, increasingly learning realistic features.

\paragraph{MR images generated by DCGAN} Fig.~\ref{fig:synthDCGAN} illustrates examples of synthetic images by DCGAN. The images look similar to the real samples.
Concat images combine appearances and patterns from all the four sequences used in training. Since DCGAN's value function could be unstable, it often generates hyper-intense T1-like images analogous to mode collapse for $64 \times 64$ Concat images, while sharing the same hyper-parameters with $128 \times 128$.

\begin{figure}[t]
  \centering
  \centerline{\includegraphics[width=0.9\textwidth]{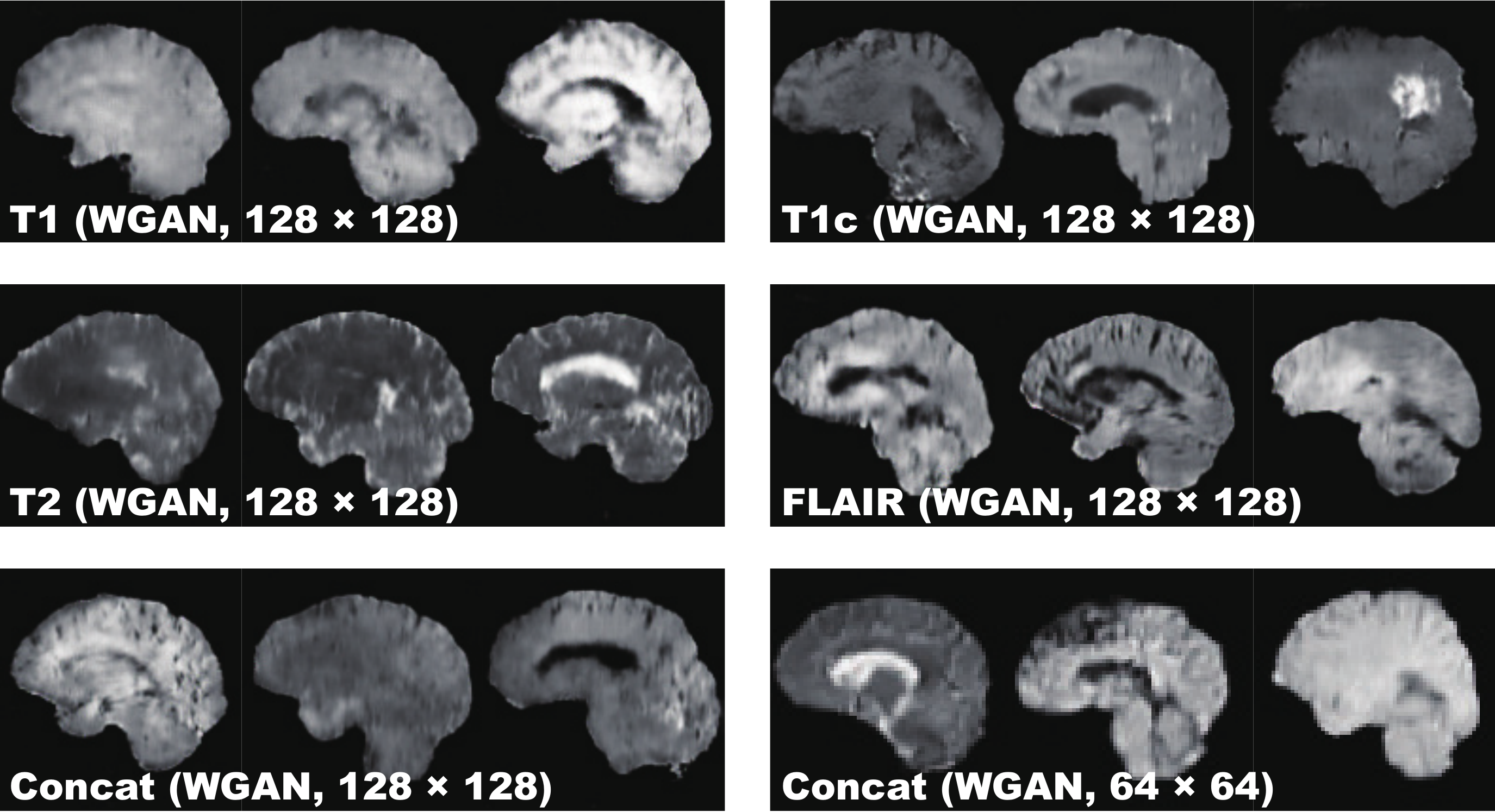}}
\caption[Examples of synthetic MR images yielded by WGAN]{Examples of synthetic MR images yielded by WGAN.}
\label{fig:synthWGAN}
\end{figure}

\paragraph{MR images generated by WGAN} Fig.~\ref{fig:synthWGAN} shows the example output of WGAN in each sequence. Outperforming remarkably DCGAN, WGAN successfully captures the sequence-specific texture and the appearance of the tumors while maintaining the realism of the original brain MR images. As expected, $128 \times 128$ Concat images tend to have more messy and unrealistic artifacts than $64 \times 64$ Concat ones, especially around the boundaries of the brain, due to the introduction of unexpected intensity patterns.

\paragraph{Visual Turing test results}

Table~\ref{tab:VTT1} shows the confusion matrix concerning the visual Turing test. Even the expert physician found classifying real and synthetic images challenging, especially in lower resolution due to their less detailed appearances unfamiliar in clinical routine, even for highly hyper-intense $64 \times 64$ Concat images by DCGAN; distinguishing Concat images was easier compared to the case of T1, T1c, T2, and FLAIR images because the physician often felt odd from the artificial sequence. WGAN succeeded to deceive the physician significantly better than DCGAN for all the MRI sequences except FLAIR images ($62\%$ to $54\%$).

\begin{table*}[!t]
\caption[Visual Turing test results by a physician for classifying real \textit{vs} synthetic images]{Visual Turing test results by a physician for classifying real \textit{vs} synthetic images. It should be noted that proximity to 50\% of accuracy indicates superior performance (chance = $50\%$).}
\label{tab:VTT1}
\centering
\begin{scriptsize}
\begin{tabular}{lccccc}
\hline\hline\noalign{\smallskip}
\bfseries  & \multicolumn{1}{c}{\bfseries Accuracy} \hspace{-0.1in}  (\%) & \bfseries Real Selected as Real & \bfseries Real as Synt & \bfseries Synt as Real & \bfseries Synt as Synt \\\noalign{\smallskip}\hline\noalign{\smallskip}
T1 (DCGAN, $128 \times 128$) & 70 & 26 & 24 & 6 & 44\\
T1c (DCGAN, $128 \times 128$) & 71 & 24 & 26 & 3 & 47\\
T2 (DCGAN, $128 \times 128$) & 64 & 22 & 28 & 8 & 42\\
FLAIR (DCGAN, $128 \times 128$) & 54 & 12 & 38 & 8 & 42\\
Concat (DCGAN, $128 \times 128$) & 77 & 34 & 16 & 7 & 43\\
Concat (DCGAN, $64 \times 64$) & 54 & 13& 37& 9&41\\
\noalign{\smallskip}\hline\noalign{\smallskip}
T1 (WGAN, $128 \times 128$) & 64 & 20 & 30 & 6 & 44\\
T1c (WGAN, $128 \times 128$) & 55 & 13 & 37 & 8 & 42\\
T2 (WGAN, $128 \times 128$) & 58 & 19 & 31 & 11 & 39\\
FLAIR (WGAN, $128 \times 128$) & 62 & 16 & 34 & 4 & 46\\
Concat (WGAN, $128 \times 128$) & 66 & 31 & 19 & 15 & 35\\
Concat (WGAN, $64 \times 64$) & 53 & 18 & 32 & 15 & 35\\
\noalign{\smallskip}\hline\hline\noalign{\smallskip}
\end{tabular}
\end{scriptsize}
\end{table*}

\subsubsection{Conclusion}
Our preliminary results show that GANs, especially WGAN, can generate $128 \times 128$ realistic multi-sequence brain MR images that even an expert physician is unable to accurately distinguish from the real, leading to valuable clinical applications, such as DA and physician training. This attributes to WGAN's good generalization ability with a sharp value function. In this context, DCGAN might be unsuitable due to both the inferior realism and mode collapse in terms of intensity. We only use the slices of interest in training to obtain desired MR images and generate both original/Concat sequence images for DA in medical imaging.

This study confirms the synthetic image quality by the human expert evaluation, but a more objective computational evaluation for GANs should also follow, such as Classifier Two-Sample Tests (C2ST)~\cite{lopezPaz2017}, which assesses whether two samples are drawn from the same distribution. Currently this work uses sagittal MR images alone, so we will generate coronal and transverse images in the near future. As this research uniformly selects middle slices in pre-processing, better data generation demands developing a classifier to only select brain MRI slices with/without tumors.

Towards DA, while realistic images give more insights on geometry/intensity transformations in classification, more realistic images do not always assure better DA, so we have to find suitable image resolutions and sequences; that is why we generate both high-resolution images and Concat images, yet they looked more unrealistic for the physician. For physician training, generating desired realistic tumors by adding conditioning requires exploring extensively the latent spaces of GANs.

Overall, our novel GAN-based realistic brain MR image generation approach sheds light on diagnostic and prognostic medical applications; future studies on these applications are needed to confirm our encouraging results.

\subsection{PGGAN-based DA for MRI brain tumor detection}
As reported in the previous section, the classical images transformed  geometric/intensity have intrinsically a similar distribution with respect to the original ones, leading to limited performance improvement; thus, generating realistic (i.e., similar to the real image distribution) but completely new samples is essential to fill the real image distribution uncovered by the original dataset \cite{han2018}.
In this context, GAN-based DA is promising, as it has shown excellent performance in computer vision, revealing good generalization ability.
Especially, SimGAN outperformed the state-of-the-art with $21\%$ improvement in eye-gaze estimation~\cite{shrivastava2017}.

Also in medical imaging, realistic retinal image and CT image generation have been tackled using adversarial learning~\cite{chuquicusma2017,costa2018}; a very recent study reported performance improvement with synthetic training data in CNN-based liver lesion classification, using a small number of $64 \times 64$ CT images for GAN training~\cite{fridAdar2018}. However, GAN-based image generation using MRI, the most effective modality for soft-tissue acquisition, has not yet been reported due to the difficulties from low-contrast MR images, strong anatomical consistency, and intra-sequence variability; in our previous work~\cite{han2018}, we generated $64 \times 64$/$128 \times 128$ MR images using conventional GANs and even an expert physician failed to accurately distinguish between the real/synthetic images.

According to the results achieved in \cite{han2018}, we aimed at generating highly-realistic and original-sized $256\times256$ images, while maintaining clear tumor/non-tumor features using GANs \cite{hanWIRN2018}.
Therefore, our goal is to generate GAN-based synthetic T1w-CE brain MR images---the most commonly used sequence in tumor detection thanks to its high-contrast~\cite{militelloIJIST2015}---for CNN-based tumor detection.
This computer-assisted brain tumor MRI analysis task is clinically valuable for better diagnosis, prognosis, and treatment~\cite{meier2016}.

\begin{figure}[t]
  \centering
  \centerline{\includegraphics[width=12cm]{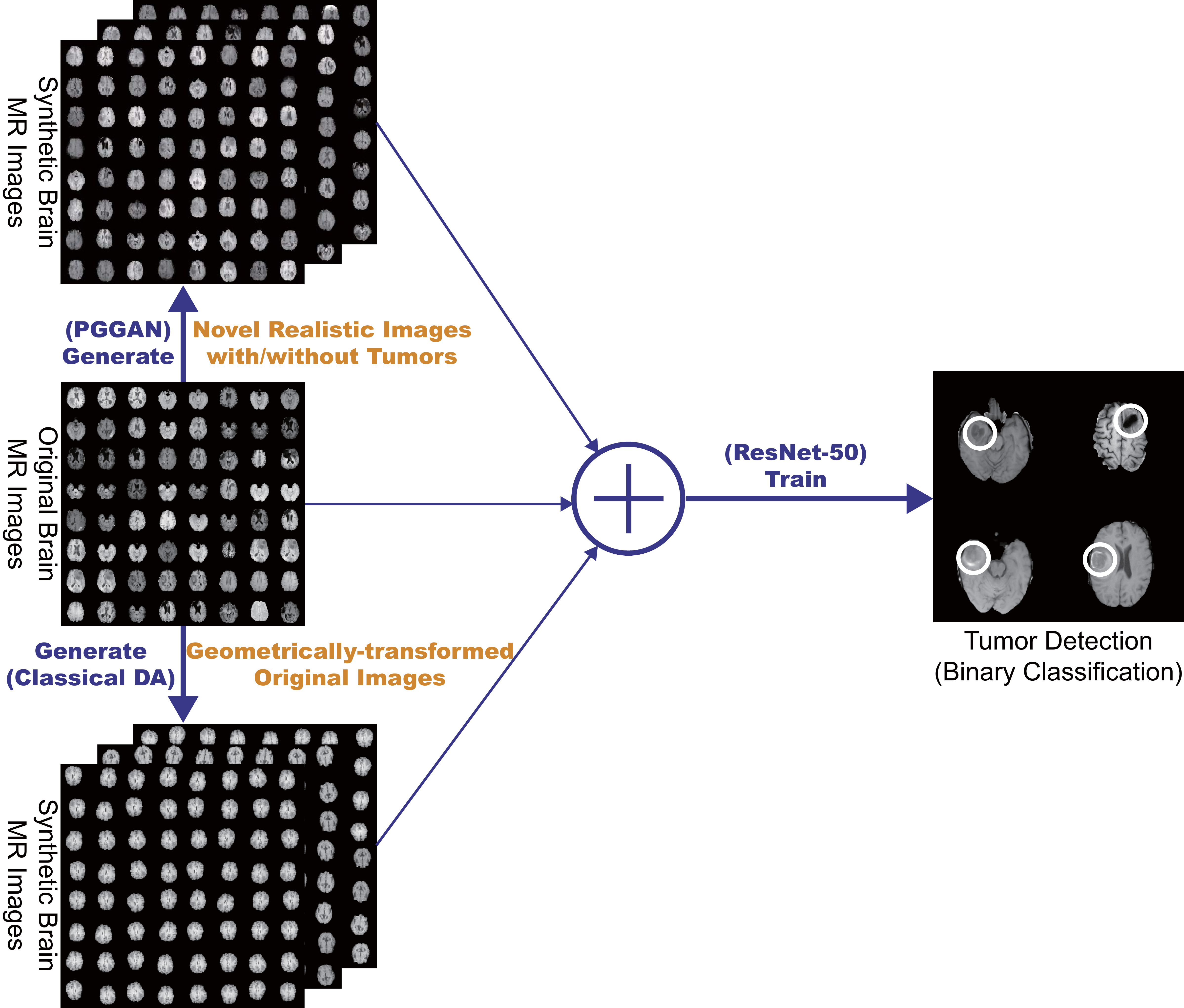}}
\caption[PGGAN-based DA for improved tumor detection]{PGGAN-based DA for improved tumor detection: the PGGANs method generates a number of realistic brain tumor/non-tumor MR images and the binary classifier uses them as additional training data.}
\label{fig:PGGAN-DA}
\end{figure}

Generating $256\times256$ images is extremely challenging: (\textit{i}) GAN training is unstable with high-resolution inputs and severe artifacts appear due to strong consistency in brain anatomy; (\textit{ii}) brain tumors vary in size, location, shape, and contrast.
However, it is beneficial, because most CNN architectures adopt around $256\times256$ input sizes (e.g., InceptionResNetV2~\cite{szegedy2017}: $299\times299$, ResNet-50~\cite{he2016}: $224\times224$) and we can achieve better results with original-sized image augmentation---towards this, we use Progressive Growing of GANs (PGGANs), a multi-stage generative training method~\cite{karras2018}.
Moreover, an expert physician evaluates the generated images' realism and tumor/non-tumor features \textit{via} the visual Turing test~\cite{salimans2016}.
Using the synthetic images, our novel PGGAN-based DA approach achieves better performance in CNN-based tumor detection, when combined with classical DA (Fig.~\ref{fig:PGGAN-DA}).

Recently, researchers applied GANs to medical imaging, mainly for image-to-image translation, such as segmentation~\cite{xue2017}, super-resolution~\cite{mahapatra2017}, and cross-modality translation~\cite{nie2017}.
Since GANs allow for adding conditional dependency on the input information (e.g., category, image, and text), they used such conditional GANs~\cite{mirza2014} to produce the desired corresponding images.
However, GAN-based research on generating large-scale synthetic training images is limited, while the biggest challenge in this field is handling small datasets.

Differently to a very recent DA work for $64 \times 64$ CT liver lesion ROI classification~\cite{fridAdar2018}, to the best of our knowledge, this is the first GAN-based whole MR image augmentation approach.
This work also firstly uses PGGANs to generate $256 \times 256$ medical images. Along with classical transformations of real images, a completely different approach---generating novel realistic images using PGGANs---may become a clinical breakthrough.

Our main contributions are as follows: 
\begin{itemize}
\item \textbf{MR image generation:} This research explains how to exploit MRI data to generate realistic and original-sized $256\times256$ whole brain MR images using PGGANs, while maintaining clear tumor/non-tumor features.
\item \textbf{MR image augmentation:} This study shows encouraging results on PGGAN-based DA, when combined with classical DA, for better tumor detection and other medical imaging tasks.
\end{itemize}

In the context of the high-resolution images generation, several multi-stage generative training methods have been proposed, such as the StackGAN architecture~\cite{zhang2017}.
Composite GAN exploits multiple generators to separately generate different parts of an image~\cite{kwak2016}.
The PGGANs method adopts multiple training procedures from low resolution to high to incrementally generate a realistic image \cite{karras2018}.

\subsubsection{Dataset preparation}
We exploited again the BRATS 2016 dataset \cite{menze2015}.
We selected the slices from $\#30$ to $\#130$ among the whole $155$ slices to omit initial/final slices, since they convey a negligible amount of useful information and negatively affect the training of both PGGANs \cite{karras2018} and ResNet-50 \cite{he2016}.
For tumor detection, our whole dataset ($220$ patients) was divided into: (\textit{i}) a training set ($154$ patients); (\textit{ii}) a validation set ($44$ patients); (\textit{iii}) a test set ($22$ patients). Only the training set is used for the PGGAN training to be fair.
Since tumor/non-tumor annotations were based on $3$D volumes, these labels are often incorrect/ambiguous on $2$D slices; so, we discard (\textit{i}) tumor images tagged as non-tumor, (\textit{ii}) non-tumor images tagged as tumor, (\textit{iii}) unclear boundary images, and (\textit{iv}) too small/big images; after all, our datasets consist of:

\begin{itemize}
\item training set ($5,036$ tumor/$3,853$ non-tumor images);
\item validation set ($793$ tumor/$640$ non-tumor images);
\item test set ($1,575$ tumor/$1,082$ non-tumor images).
\end{itemize}

\begin{figure}[t]
  \centering
  \centerline{\includegraphics[width=12cm]{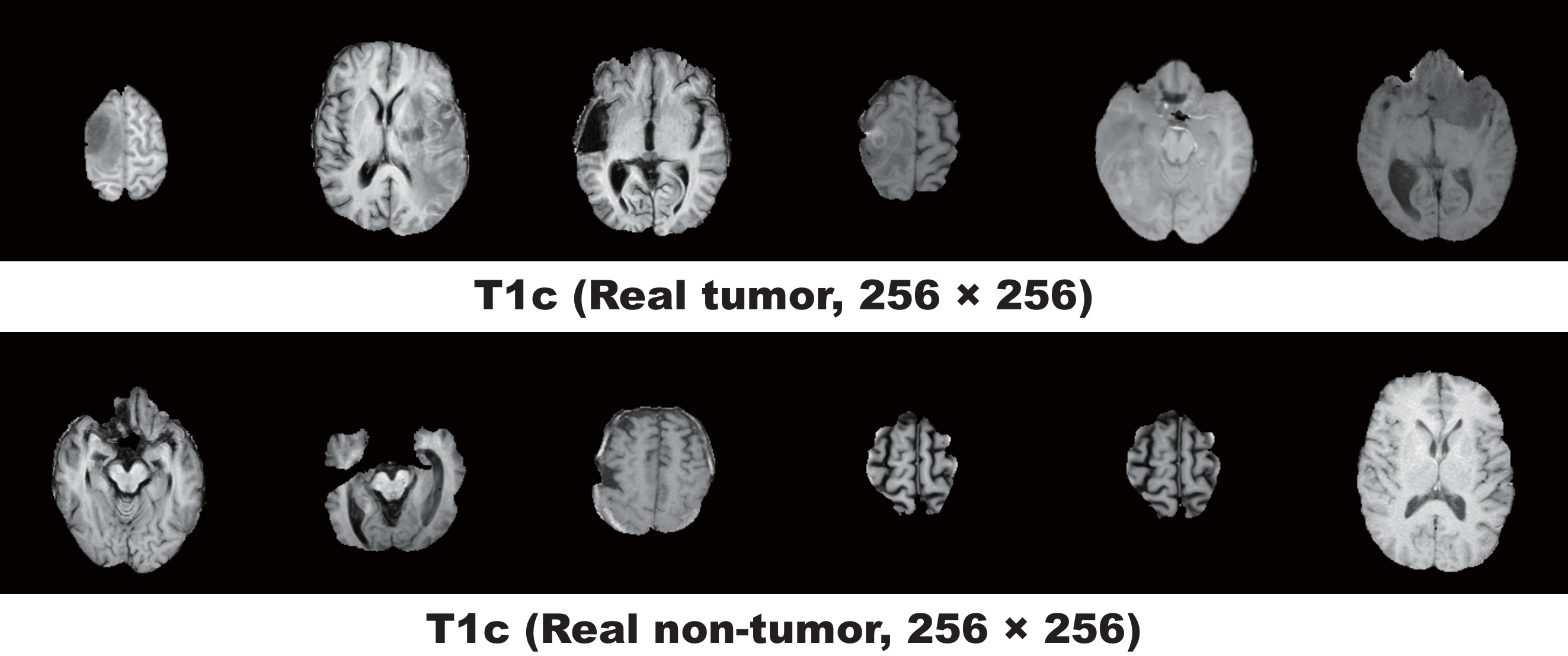}}
\caption[Examples of real $256 \times 256$ MR images used for PGGAN training]{Examples of real $256 \times 256$ MR images used for PGGAN training.}
\label{fig:realTum}
\end{figure}

The images from the training set are zero-padded to reach a power of $2$, $256 \times 256$ from $240 \times 240$ pixels for better PGGAN training.
Fig.~\ref{fig:realTum} shows examples of real MR images.

\begin{figure}[H]
  \centering
  \centerline{\includegraphics[width=12cm]{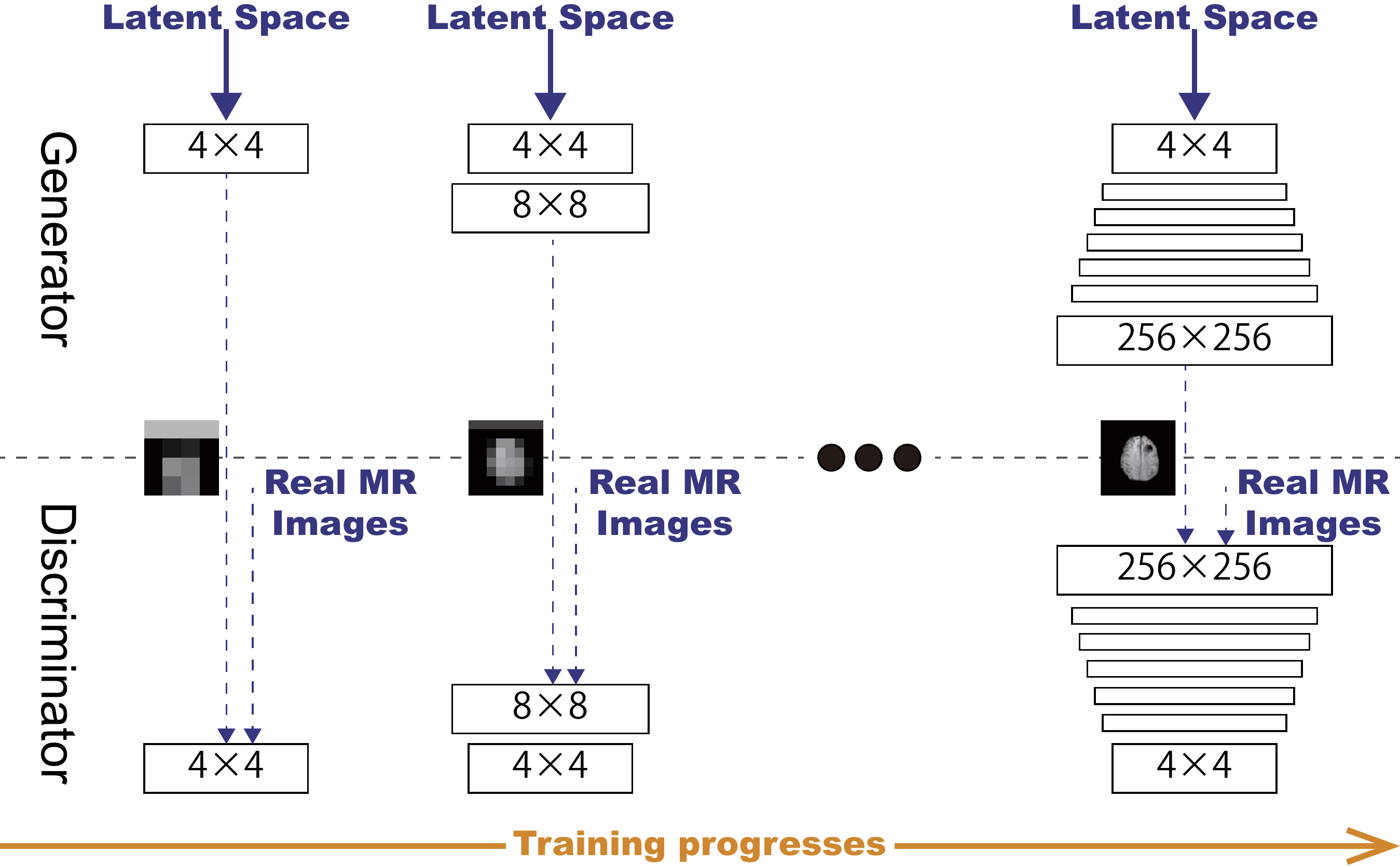}}
\caption[PGGANs architecture for synthetic $256 \times 256$ MR image generation]{PGGANs architecture for synthetic $256 \times 256$ MR image generation.}
\label{fig:PGGANscheme}
\end{figure}

\subsubsection{PGGAN-based image generation}

\paragraph{PGGANs} is a novel training method for GANs with progressively growing generator and discriminator~\cite{karras2018}: starting from low resolution, newly added layers model fine-grained details as training progresses.
As Fig.~\ref{fig:PGGANscheme} shows, we adopt PGGANs to generate highly-realistic and original-sized $256 \times 256$ brain MR images; tumor/non-tumor images are separately trained and generated.

\paragraph{PGGAN implementation details} We used the PGGAN architecture with the Wasserstein loss using gradient penalty~\cite{gulrajani2017}. Training lasts for $100$ epochs with a batch size of 16 and $1.0 \times 10^{-3}$ learning rate for Adam optimizer.

\subsubsection{Tumor detection using ResNet-50}
\paragraph{Pre-processing} To fit ResNet-50's input size, we center-crop the whole images from $240 \times 240$ to $224 \times 224$ pixels.
\paragraph{ResNet-50} is a residual learning-based CNN with $50$ layers~\cite{he2016}: unlike conventional learning unreferenced functions, it reformulates the layers as learning residual functions for sustainable and easy training. We adopt ResNet-50 to detect tumors in brain MR images, i.e., the binary classification of images with/without tumors.

\begin{figure}[t]
  \centering
  \centerline{\includegraphics[width=10cm]{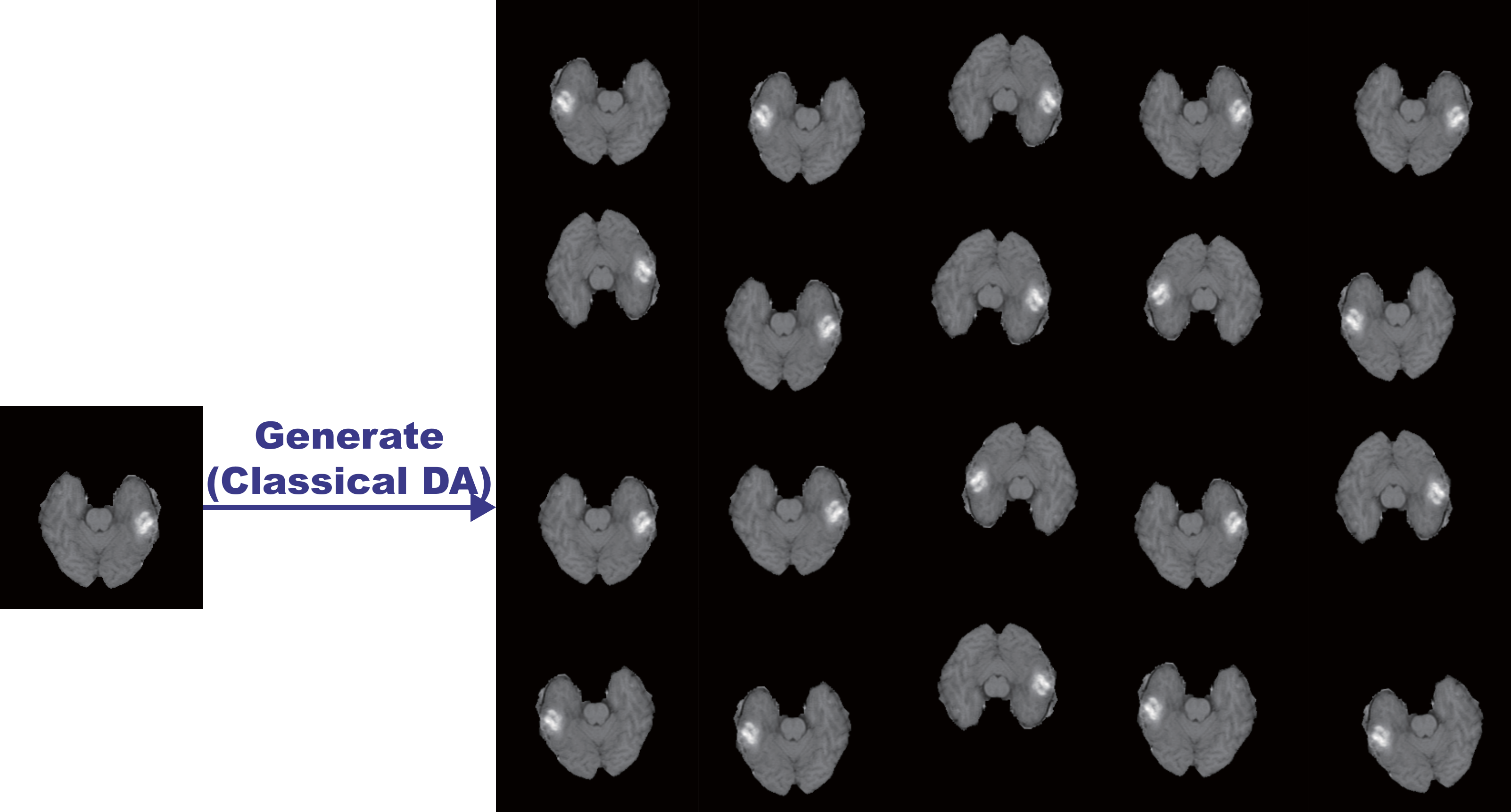}}
\caption[Examples of real MR image and its geometrically-transformed synthetic images]{Examples of real MR image and its geometrically-transformed synthetic images.}
\label{fig:ClassicDA}
\end{figure}

To confirm the effect of PGGAN-based DA, the following classification results are compared: (\textit{i}) without DA, (\textit{ii}) with $200,000$ classical DA ($100,000$ for each class), (\textit{iii}) with $200,000$ PGGAN-based DA, and (\textit{iv}) with both $200,000$ classical DA and $200,000$ PGGAN-based DA; the classical DA adopts a random combination of horizontal/vertical flipping, rotation up to $10$ degrees, width/height shift up to $8\%$, shearing up to $8\%$, zooming up to $8\%$, and constant filling of points outside the input boundaries (Fig.~\ref{fig:ClassicDA}).
For better DA, highly-unrealistic PGGAN-generated images are manually discarded.

\paragraph{ResNet-50 implementation details}
We use the ResNet-50 architecture pre-trained on ImageNet with a dropout of $0.5$ before the final softmax layer, along with a batch size of $192$, $1.0 \times 10^{-3}$ learning rate for Adam optimizer, and early stopping of $10$ epochs.

\subsubsection{Clinical validation using the visual Turing test}
To quantitatively evaluate (\textit{i}) how realistic the PGGAN-based synthetic images are, (\textit{ii}) how obvious the synthetic images' tumor/non-tumor features are, we supply, in a random order, to an expert physician a random selection of:
\begin{itemize}
\item $50$ real tumor images;
\item $50$ real non-tumor images;
\item $50$ synthetic tumor images;
\item $50$ synthetic non-tumor images.
\end{itemize}

Then, the physician is asked to constantly classify them as both (\textit{i}) real/synthetic and (\textit{ii}) tumor/non-tumor, without previous training stages revealing which is real/synthetic and tumor/non-tumor; here, we only show successful cases of synthetic images, as we can discard failed cases for better DA.
The so-called visual Turing test~\cite{salimans2016} is used to probe the human ability to identify attributes and relationships in images, also in evaluating the visual quality of GAN-generated images~\cite{shrivastava2017}.
Similarly, this applies to medical images in clinical environments~\cite{chuquicusma2017,fridAdar2018}, wherein physicians' expertise is critical.

\paragraph{Visualization using t-SNE}
To visually analyze the distribution of both (\textit{i}) real/synthetic and (\textit{ii}) tumor/non-tumor images, we use t-Distributed Stochastic Neighbor Embedding (t-SNE)~\cite{maaten2008} on a random selection of:
\begin{itemize}
\item $300$ real non-tumor images;
\item $300$ geometrically-transformed non-tumor images;
\item $300$ PGGAN-generated non-tumor images;
\item $300$ real tumor images;
\item $300$ geometrically-transformed tumor images;
\item $300$ PGGAN-generated tumor images.
\end{itemize}

Only $300$ images per each category are selected for better visualization. t-SNE is a Machine Learning algorithm for dimensionality reduction to represent high-dimensional data into a lower-dimensional (2D/3D) space. It non-linearly adapts to input data using perplexity, which balances between the data's local and global aspects.

\subparagraph{t-SNE implementation details}
We used t-SNE with a perplexity of $100$ for $1,000$ iterations to obtain a $2$D visual representation.

\paragraph{Results}
This section shows how PGGANs generates synthetic brain MR images. The results include instances of synthetic images, their quantitative evaluation by an expert physician, and their influence on tumor detection.

\subparagraph{MR images generated by PGGANs}

\begin{figure}[t]
  \centering
  \centerline{\includegraphics[width=12cm]{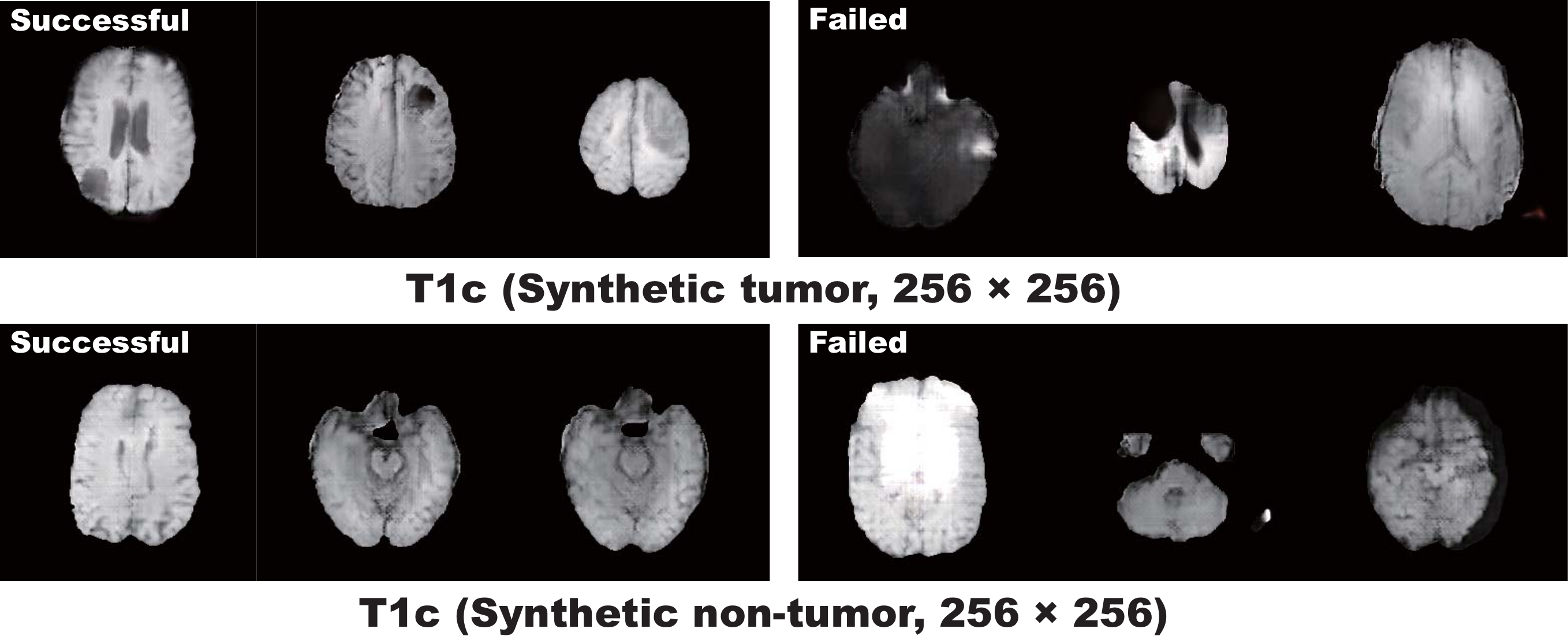}}
\caption[Examples of synthetic MR images yielded by PGGANs]{Examples of synthetic MR images yielded by PGGANs: (a) successful cases; (b) failed cases.}
\label{fig:synthTum}
\end{figure}

Fig.~\ref{fig:synthTum} illustrates examples of synthetic tumor/non-tumor images by PGGANs. In our visual confirmation, for about $75\%$ of cases, PGGANs successfully captures the T1c-specific texture and tumor appearance while maintaining the realism of the original brain MR images; however, for about $25\%$ of cases, the generated images lack clear tumor/non-tumor features or contain unrealistic features, such as hyper-intensity, gray contours, and odd artifacts. 

\begin{table*}[!t]
\caption[Binary classification results for detecting brain tumors with/without DA]{Binary classification results for detecting brain tumors with/without DA.}
\label{tab:PGGAN1}
\centering
\begin{scriptsize}
\begin{tabular}{lccccc}
\hline\hline\noalign{\smallskip}
\bfseries Experimental condition \ & \multicolumn{1}{c}{ Accuracy}  \ & Sensitivity \ & Specificity \\\noalign{\smallskip}\hline\noalign{\smallskip}
ResNet-50 (w/o DA) \ & $90.06\%$ \ & $85.27\%$ \ & $97.04\%$ \\
ResNet-50 (w/ 200k classical DA) \ & $90.70\%$ \ & $88.70\%$ \ & $93.62\%$ \\
ResNet-50 (w/ 200k PGGAN-based DA) \ & $62.02\%$ \ & $\mathbf{99.94}\%$ \ & $6.84\%$ \\
ResNet-50 (w/ 200k classical DA + 200k PGGAN-based DA) \ & $\mathbf{91.08}\%$ \ & $86.60\%$ \ & $\mathbf{97.60}\%$ \\
\noalign{\smallskip}\hline\hline\noalign{\smallskip}
\end{tabular}
\end{scriptsize}
\end{table*}

\subparagraph{Tumor detection results}

Table~\ref{tab:PGGAN1} shows the classification results for detecting brain tumors with/without DA techniques. As expected, the test accuracy improves by $0.64\%$ with the additional $200,000$ geometrically-transformed images for training. When only the PGGAN-based DA is applied, the test accuracy decreases drastically with almost 100\% of sensitivity and $6.84\%$ of specificity, because the classifier recognizes the synthetic images' prevailed unrealistic features as tumors, similarly to anomaly detection.

However, surprisingly, when it is combined with the classical DA, the accuracy increases by $1.02\%$ with higher sensitivity and specificity; this could occur because the PGGAN-based DA fills the real image distribution uncovered by the original dataset, while the classical DA provides the robustness on training for most cases.

\begin{table*}[!t]
\caption[Visual Turing test results by a physician for classifying real \textit{vs} synthetic images and tumor \textit{vs} non-tumor images]{Visual Turing test results by a physician for classifying Real ($\mathsf{R}$) \textit{vs} Synthetic ($\mathsf{S}$) images and Tumor ($\mathsf{T}$) \textit{vs} Non-tumor ($\mathsf{N}$) images.} 
\label{tab:PGGAN2}
\centering
\begin{footnotesize}
\begin{tabular}{ccccc}
\hline\hline\noalign{\smallskip}
\multicolumn{1}{c}{Real/Synthetic Classification}  \ \ \ &  $\mathsf{R}$ as $\mathsf{R}$ \ \ \ &  $\mathsf{R}$ as $\mathsf{S}$ \ \ \ &  $\mathsf{S}$ as $\mathsf{R}$ \ \ \ &  $\mathsf{S}$ as $\mathsf{S}$ \\\noalign{\smallskip}\hline\noalign{\smallskip}
$78.5\%$ \ \ \ & $58$ \ \ \ & $42$ \ \ \ & $1$ \ \ \ & $99$ \\
\noalign{\smallskip}\Hline\noalign{\smallskip}
\multicolumn{1}{c}{Tumor/Non-tumor Classification} \ \ \ &  $\mathsf{T}$ as $\mathsf{T}$ \ \ \ &  $\mathsf{T}$ as $\mathsf{N}$ \ \ \ &  $\mathsf{N}$ as $\mathsf{T}$ \ \ \ &  $\mathsf{N}$ as $\mathsf{N}$\\
\noalign{\smallskip}\hline\noalign{\smallskip}
$90.5\%$ \ \ \ & $82$ \ \ \ & $18$ ($\mathsf{R}:5$, $\mathsf{S}:13$) \ \ \ & $1$ ($\mathsf{S}:1$) \ \ \ & $99$\\
\noalign{\smallskip}\hline\hline\noalign{\smallskip}
\end{tabular}
\end{footnotesize}
\end{table*}

\subparagraph{Visual Turing test results}
Table~\ref{tab:PGGAN2} shows the confusion matrix for the visual Turing test. Differently from our previous work on GAN-based $64 \times 64$/$128 \times 128$ MR image generation, the expert physician easily recognizes $256 \times 256$ synthetic images \cite{han2018}, while tending also to classify real images as synthetic; this can be attributed to high resolution associated with more difficult training and detailed appearance, making artifacts stand out, which is coherent to the ResNet-50's low tumor detection accuracy with only the PGGAN-based DA. Generally, the physician's tumor/non-tumor classification accuracy is high and the synthetic images successfully capture tumor/non-tumor features. However, unlike non-tumor images, the expert recognizes a considerable number of tumor images as non-tumor, especially on the synthetic images; this results from the remaining real images' ambiguous annotation, which is amplified in the synthetic images trained on them.

\begin{figure}[t]
  \centering
  \centerline{\includegraphics[width=12cm]{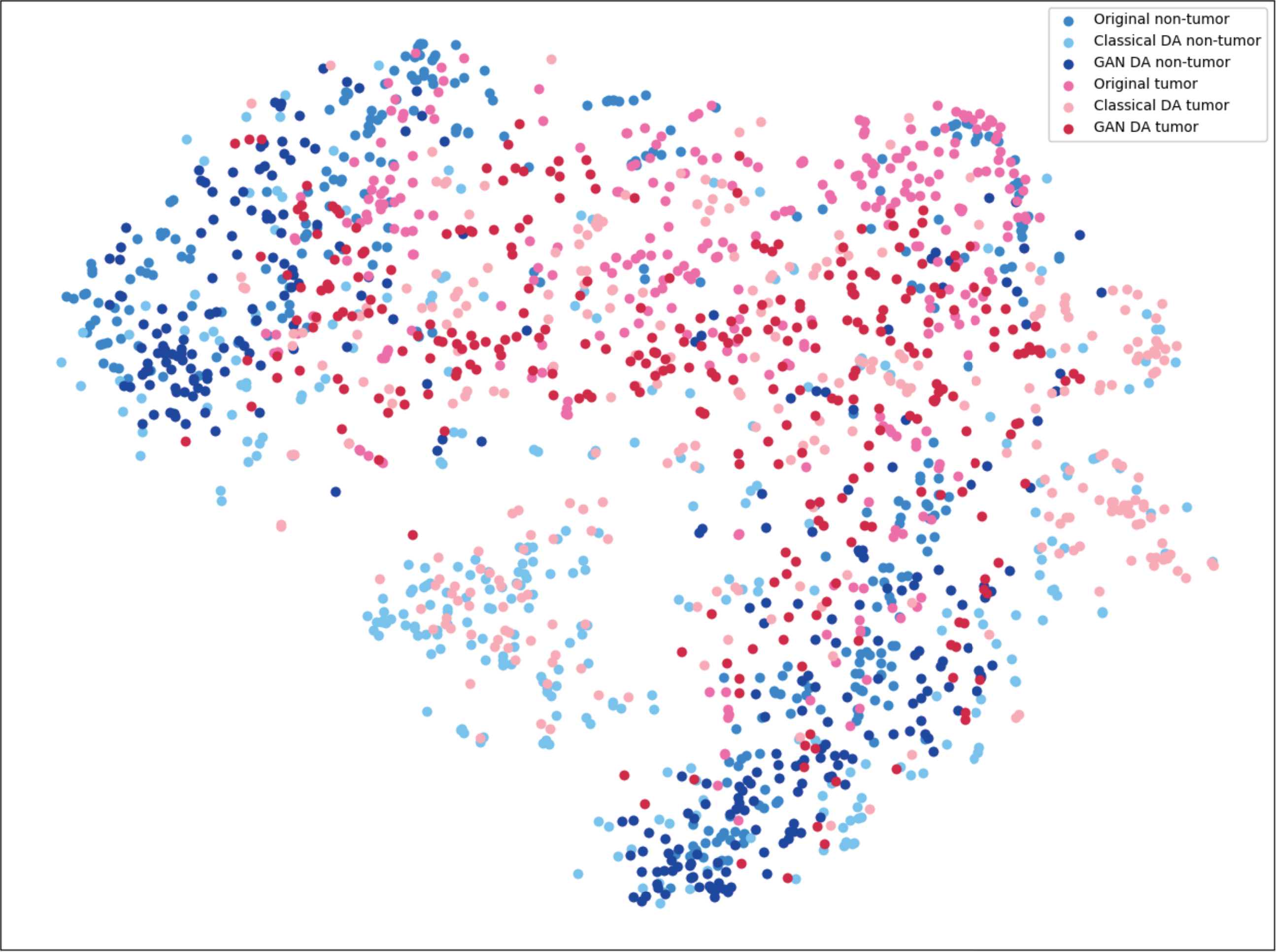}}
\caption[t-SNE result for tumor/non-tumor images]{t-SNE result on six categories, with $300$ images per each category: (a) real tumor/non-tumor images; (b) geometrically-transformed tumor/non-tumor images; (c) PGGAN-generated tumor/non-tumor images.}
\label{fig:tSNE}
\end{figure}

\subparagraph{t-SNE results}
As presented in Fig.~\ref{fig:tSNE}, tumor/non-tumor images' distribution shows a tendency that non-tumor images locate from top left to bottom right and tumor images locate from top right to center, while the distinction is unclear with partial overlaps. Classical DA covers a wide range, including zones without any real/GAN-generated images, but tumor/non-tumor images often overlap there.
Meanwhile, PGGAN-generated images concentrate differently from real images, while showing more frequent overlaps than the real ones; this probably derives from those synthetic images with unsatisfactory realism and tumor/non-tumor features.

\paragraph{Conclusion}
\label{sec:conclusion}
Our preliminary results show that PGGANs can generate original-sized $256 \times 256$ realistic brain MR images and achieve higher performance in tumor detection, when combined with classical DA.
This occurs because PGGANs' multi-stage image generation obtains good generalization and synthesizes images with the real image distribution unfilled by the original dataset. However, considering the visual Turing test and t-SNE results, yet unsatisfactory realism with high resolution strongly limits DA performance, so we plan to (\textit{i}) discard unrealistic images or (\textit{ii}) generate only realistic images, and then (\textit{iii}) refine synthetic images more similar to the real image distribution.

For (\textit{i}), classifier two-sample tests~\cite{lopezPaz2017}, assessing whether two samples are drawn from the same distribution, can help discard images not from the real image distribution, as manual removal is demanding. Regarding (\textit{ii}), we can map an input random vector onto each training image~\cite{schlegl2017} and generate images with suitable vectors, to control the divergence of the generated images; virtual adversarial training can be also integrated to control the output distribution.
Lastly, (\textit{iii}) can be achieved by GAN-based image-to-image translation, such as CycleGAN~\cite{zhu2017} or the UNsupervised Image-to-image Translation (UNIT) framework~\cite{liu2017}, considering SimGAN's remarkable performance improvement after refinement~\cite{shrivastava2017}.
At the time of writing, taking inspiration form the SimGAN's rationale~\cite{shrivastava2017}, we are refining the MR images synthesized by the PGGAN-based simulator by exploiting a refiner Neural Network.
Moreover, we should further avoid real images with ambiguous/inaccurate annotation for better tumor detection.
Finally, conditional GANs~\cite{mirza2014} could be exploited to generate labeled synthetic images from arbitrary segmentation masks~\cite{rezaei2017}, thus allowing us to synthesize brain MR images representing tumors in specific positions.
By so doing, the generated samples may be employed in DA for medical image segmentation tasks.

Overall, our novel PGGAN-based DA approach sheds light on diagnostic and prognostic medical applications, not limited to tumor detection; future studies are needed to extend these early encouraging results.
\chapter{Conclusion}
\label{chap7}

\graphicspath{{Chapter7/Figs/}}

\section{Final remarks}
\label{sec:finRemarks}
The research activities presented in this thesis have been motivated by the need to develop novel and effective methods for the analysis of biomedical images, considering the unique challenges imposed by this kind of data \cite{gillies2015,meijering2016}.
Recalling the conceptual scheme in Fig. \ref{fig:graphAbstract} (\textbf{Chapter \ref{chap1}}), the appropriate combination of different computational techniques can shed light on new discoveries in healthcare and biology (\textbf{Chapter \ref{chap2}}).
The key contributions of the computational methods proposed in this thesis are presented and discussed in the following.

Classical Image Processing methods (\textbf{Chapter \ref{chap3}}) are efficient and still effective, especially in specific situations.
Despite their simplicity, region-based and morphological approaches (Sections \ref{sec:regionBased} and \ref{sec:morphApproaches}, respectively) worked properly when the input images hold the assumptions underlying the algorithms.
Global thresholding \cite{militelloCBM2015,rundo2016WIRN} as well as the combination of region splitting-and-merging with region growing \cite{rundoMBEC2016} yielded accurate uterine fibroid MR image segmentation results for MRgFUS treatment response evaluation \cite{rundo2018SIST}. 
In clonogenic assays, CHT was suitable for fully automatic well detection in multi-well plates and local thresholding was used for calculating the area covered by the cell colony \cite{militelloCBM2017}.
The watershed transform was valuable for cell nuclei detection and counting, when dealing with overlapping cells in high-throughput microscopy imaging experiments \cite{rundo2018ACDC}.

Pattern Recognition techniques (\textbf{Chapter \ref{chap4}}), with particular focus on unsupervised fuzzy clustering (Section \ref{sec:fuzzyClustering}) and graph-based methods (Section \ref{sec:graphMethods}), were exploited for multispectral and multimodal medical imaging data processing.
Especially, the FCM clustering algorithm was applied to several clinical contexts, including brain tumor MR image segmentation \cite{militelloIJIST2015} and necrosis extraction \cite{rundo2018next} as well as prostate gland segmentation on multispectral MR images \cite{rundo2017Inf}.
\textsc{GTVcut}, a CA-based GTV delineation approach, was proposed by exploiting a transition function (based on gradient magnitude) that solves the shortest path problem \cite{rundo2016ACRI,rundo2017NC}.
The combination of fuzzy clustering with the RW algorithm can provide valuable solutions for multimodal image segmentation, such as in the case of brain metastasis delination on co-registered PET/MR images for neuro-radiosurgery treatment planning \cite{rundoCMPB2017}.

Computational Intelligence (\textbf{Chapter \ref{chap5}}) was shown to be able to considerably increase the performance of traditional Image Processing techniques.
A novel Evolutionary Computation method (Section \ref{sec:geneticAlg}) for medical image enhancement, called MedGA \cite{rundo2018MedGA1}, was proposed to improve the threshold-based segmentation results \cite{rundo2018MedGA2}.
As a matter of fact, MedGA aimed at overcoming the limitations of thresholding approaches that strongly rely on the assumption that the histogram under investigation is bimodal.
Moreover, multimodal image co-registration approaches using PSO (Section \ref{sec:PSO}) were critically reviewed and analyzed \cite{rundo2016SSCI}.
This global optimization metaheuristic may considerably improve the reliability and robustness of the multimodal medical image co-registration process with respect to local-based search methods.

In the latest years, DNNs (\textbf{Chapter \ref{chap6}}) have been exploited in biomedical image analysis achieving state-of-the-art performance.
In this context, the generalization abilities of CNNs (Section \ref{sec:CNNs}) in medical image segmentation for multi-institutional studies was investigated \cite{rundoWIRN2018}.
The generalization abilities of CNNs in medical image segmentation for multi-institutional studies are addressed by conceiving a novel architecture, called USE-Net \cite{rundo2018Neurocomp}, which integrates adaptive feature recalibration by incorporating Squeeze-and-Excitation blocks into U-Net.
Regarding the generation of realistic medical images, the GANs (Section \ref{sec:GANs}) were exploited to synthesize completely new brain tumor MR slices \cite{han2018}.
Then, in order to yield real-sized MR images, the PGGANs method achieved promising results in data augmentation for brain tumor MRI detection \cite{hanWIRN2018}.

Therefore, the synergies emerging from the combination of diverse computer science areas are valuable to develop accurate and comprehensive approaches in clinical and laboratory routine.
These computer-assisted biomedical image analysis methods could be beneficial for the definition of imaging biomarkers as well as for quantitative medicine and biology.

Finally, the community of computer scientists in biomedical imaging cannot ignore the interpretability of the results yielded by advanced computational models \cite{castelvecchi2016}.
As a matter of fact, these compelling issues must be concretely addressed for the adoption of CDSSs in real environments, allowing for the physicians' awareness during complex clinical decision-making tasks \cite{cabitza2017}.

\section{Future work}
\label{sec:futWork}

Quantitative cancer studies represent the immediate future in clinical practice \cite{gatenby2013,yankeelov2016}.
The proper combination of different imaging modalities provides an in-depth understanding of molecular processes that can guide to personalized medicine \cite{evanko2008}.

The achievements related to quantitative imaging lead to valuable applications in radiomics \cite{aerts2014,gillies2015,lambin2017,lambin2012} and radiogenomics \cite{pinker2018,rosenstein2014,rutman2009,sala2017} research, so conveying clinically useful information.
Especially, radiogenomics studies can provide important information about tumor heterogeneity as well as treatment response \cite{brindle2008}, by integrating genomic information \cite{yuan2012}.
In these scenarios, haplotype assembly can be valuable for personalized therapy by relying on Single Nucleotide Polymorphims (SNPs) \cite{tangherloni2019MedHPC,tangherloni2018GenHap}.
Therefore, in the near future the main objective of biomedical research is moving towards the acquisition and the integration of heterogeneous datasets \cite{lahat2015,serra2018} (i.e., imaging modalities and high-throughput technologies) to gain insights allowing for precision medicine \cite{brady2016}.

From the technological perspective, HPC can be an enabling factor for feasible computational solutions in clinical and laboratory practice.
GPUs are used today in a wide range of applications in computational biology \cite{nobile2016} and medical image analysis \cite{eklund2013,smistad2015}.
As a matter of fact, GPUs are pervasive, energy-efficient and can dramatically accelerate parallel computing.
In the field of biomedical imaging, GPUs are in some cases crucial for enabling practical use of computationally demanding algorithms.
For instance, GPUs enabled the training of DNNs in reasonable time \cite{shen2017,tajbakhsh2016}.
In this context, our research group developed a first GPU-powered version of the online training algorithm for SOMs \cite{kohonen1982,kohonen1990}.
We aim at exploiting this efficient method in medical image classification based on textural and structural features \cite{ortiz2013}.
To this end, efficient and reproducible feature computation is a hot-topic in radiomics research \cite{apte2018}.
Leveraging GPUs for efficient imaging feature extraction is effective to speed-up the feature extraction \cite{gipp2012,tsai2017}.
We are currently improving the computational performance of Haralick's feature extraction based on the GLCM \cite{haralick1979,haralick1973}, by assessing the accuracy relying on state-of-the-art tools for radiation therapy \cite{deasy2003} and radiographic phenotyping \cite{vanGriethuysen2017}.
With reference to real-time medical diagnosis tools implemented on reconfigurable logic devices, FPGAs can provide higher throughputs with respect to the corresponding software applications \cite{vitabile2011}.
Therefore, FPGA-based solutions could be exploited to accelerate the developed algorithms on specialized processing nodes \cite{franchini2013}, implementing Systems-on-a-Chip (SoCs) for specific processing tasks \cite{cong2011}.
For instance, geometric operations (i.e., reflections, rotations, translations, and uniform scaling) could be parallelized by means of co-processing hardware architectures natively supporting Clifford Algebra \cite{franchini2015}. 

Finally, to address the issues regarding the curse of dimensionality, effective Dimensionality Reduction (DR) techniques, which allow also for online performance, could be required \cite{cirrincione2015}.
The Growing Curvilinear Component Analysis (GCCA) \cite{cirrincione2018} enables a non-linear distance preserving reduction technique \cite{cirrincione2018SIST}, by means of a self-organized incremental neural network architecture, leading to applications in life sciences \cite{barbiero2019}.
In the case of competing objectives, multi-objective optimization techniques, such as those based on GAs \cite{deb2014,konak2006}, can be exploited for feature selection in complex problems \cite{ortiz2013}.


\begin{spacing}{0.9}


\bibliographystyle{apalike}
\cleardoublepage
\renewcommand\bibname{Bibliography}
\bibliography{References/references} 



\end{spacing}


\begin{appendices} 

\chapter{Validation methodology} 
\label{appendixA}

\section{Medical image segmentation evaluation metrics}
\label{sec:segEval}
We evaluate the segmentation methods by comparing the segmented images ($\mathcall{S}$) with respect to the corresponding gold standard manual segmentation ($\mathcall{G}$) \cite{taha2015,fenster2005,zhang2001}.
In addition to evaluate the automated segmentation $\mathcall{S}$ against a gold standard $\mathcall{G}$, these metrics can be extended to compare any pair of segmentation instances $(\mathcall{A}_1$, $\mathcall{A}_2$).

\subsubsection{Volume-based metrics}
\label{sec:volumeMetrics}
Accurate prostate volume measurement is crucial in diagnostic phases.
Therefore, the volume $V_\mathcall{S}$ computed automatically by the proposed method, should be compared with the actual volume $V_\mathcall{G}$, contoured manually by an expert radiologist.

First, to evaluate either over-estimation or under-estimation of the automated volume measurements with respect to the reference measure, the absolute average volume difference ($AVD$) between the $V_\mathcall{S}$ and $V_\mathcall{G}$ values is calculated as:
\begin{equation}
    \label{eq:AVD}
    AVD = \frac{\text{abs}(V_\mathcall{S} - V_\mathcall{G})}{V_\mathcall{G}}.
\end{equation}

However, the quantitative measure that best points out the similarity between the volumes $V_\mathcall{S}$ and $V_\mathcall{G}$ is the volumetric similarity ($VS$), namely the absolute difference between volumes divided by the volume sum:
\begin{equation}
    \label{eq:VS}
    VS = \frac{\text{abs}(|V_\mathcall{S}| - |V_\mathcall{G}|)}{|V_\mathcall{S}| + |V_\mathcall{G}|} = 1 - VD,
\end{equation}
where $VD$ is the volumetric distance and $|\cdot|$ indicates the volume cardinality (in terms of the segmented voxels).

\subsubsection{Spatial overlap-based metrics}
\label{sec:overlapMetrics}
These metrics quantify the spatially-overlapping segmented ROI.
Let true positives be $TP = \mathcall{S} \cap \mathcall{G}$, false negatives be $FN = \mathcall{G} - \mathcall{S}$, false positives be $FP = \mathcall{S} - \mathcall{G}$, and true negatives be $TN = \mathcall{I}_\Omega - \mathcall{G} - \mathcall{S}$.
In what follows, we denote the cardinality of the pixels belonging to a region $\mathcall{A}$ as $|\mathcall{A}|$.

\begin{itemize}
	\setlength\itemsep{1em}
	\item \textit{Dice similarity coefficient}~\cite{dice1945} is the most used measure in medical image segmentation to compare the overlap of two regions \cite{zou2004}:
	\begin{equation}
	\label{eq:DSC}
	DSC = \frac{2 \cdot \abs{TP}}{\abs{\mathcall{S}} + \abs{\mathcall{G}}} \cdot 100.
	\end{equation}
	In the literature, it has been shown that a \emph{DSC} above $70\%$ are generally regarded as a satisfactory level of agreement between two segmentations (i.e., manual and automated delineations) in clinical applications 
	\cite{zijdenbos1994, xue2007, qiu2014}.
	\item \textit{Jaccard index} \cite{jaccard1901}, also known as \textit{Intersection over Union (IoU)}, is another similarity measure, which is the ratio between the cardinality of the intersection and the cardinality of the union calculated on the two segmentation results:
	\begin{equation}
	\label{eq:JI}
	JI = \frac{\abs{\mathcall{S} \cap \mathcall{G}}}{\abs{\mathcall{S} \cup \mathcall{G}}} \cdot 100 = \frac{DSC}{2-DSC}.
	\end{equation}
	As it can be seen in Eq. (\ref{eq:JI}), \textit{JI} is strongly related to \textit{DSC}. Moreover \textit{DSC} is always larger than \textit{JI} except at extrema $\{0\%,100\%\}$, where they take equal value.
	\item \textit{Sensitivity} measures measures the portion of positive (foreground) pixels correctly detected by the proposed segmentation method with respect to the gold standard segmentation:
	\begin{equation}
	\label{eq:SEN}
	SEN = TPR =\frac{\abs{TP}} {\abs{TP} + \abs{FN}} \cdot 100.
	\end{equation}
	\item \textit{Specificity} indicates the portion of negative (background) pixels correctly identified by the automatic segmentation against the gold standard segmentation:
	\begin{equation}
	\label{eq:TNR}
	TNR = \frac{\abs{TN}}{\abs{TN} + \abs{FP}} \cdot 100.
	\end{equation}
	However, this formulation is ineffective when data are unbalanced (i.e., the ROI is much smaller than the whole image). Consequently, we use the following definition:
	\begin{equation}
	\label{eq:SPC}
	SPC = \left(1 - \frac{\abs{FP}}{\abs{\mathcall{S}}} \right)  \cdot 100.
	\end{equation}
	\item \textit{False positive ratio} denotes the presence of false positives compared to the reference region:
	\begin{equation}
	\label{eq:FPR}
	FPR = \frac{\abs{FP}}{\abs{FP + TN}} \cdot 100.
	\end{equation}
	\item \textit{False negative ratio} is dually defined as:
	\begin{equation}
	\label{eq:FNR}
	FNR = \frac{\abs{FN}}{\abs{FN + TP}} \cdot 100.
	\end{equation}
\end{itemize}

\subsubsection{Spatial distance-based metrics}
\label{sec:distanceMetrics}
As precise boundary tracing plays an important role in clinical practice, overlap-based metrics have limitations in evaluating segmented images.
In order to measure the distance between the two ROI boundaries, distance-based metrics can be considered.
Let the manual contour $\mathcal{G}$ consist in a set of vertices $\{ \mathbf{g}_a: a = 1, 2, \dots, A \}$ and the automatically-generated contour $\mathcal{S}$ consist in a set of vertices $\{ \mathbf{s}_b: b = 1, 2, \dots, B \}$.
We calculate the absolute distance between an arbitrary element $\mathbf{s}_b \in \mathcal{S}$ and all the vertices in $\mathcal{G}$ as follows:
\begin{equation}
\label{eq:contDist}
d(\mathbf{s}_b, \mathcal{G}) = \min_{a \in \{ 1,2, \dots, A \}} \| \mathbf{s}_b - \mathbf{g}_a \|.
\end{equation}

\begin{itemize}
	\setlength\itemsep{1em}
	\item \textit{Average absolute distance} measures the average difference between the ROI boundaries of $\mathcall{S}$ and $\mathcall{G}$:
	\begin{equation}
	\label{eq:AvgD}
	AvgD = \frac{1}{B}\sum \limits_{b=1}^B d(\mathbf{s}_b, \mathcal{G}).
	\end{equation}
	\item \textit{Maximum absolute distance} represents the maximum difference between the ROI boundaries of $\mathcall{S}$ and $\mathcall{G}$:
	\begin{equation}
	\label{eq:MaxD}
	MaxD = \max_{b \in \{ 1, 2, \dots, B \}}  d(\mathbf{s}_b, \mathcal{G}).
	\end{equation}
	\item \textit{Hausdorff distance} \cite{cardenes2009} measures the extent between the set of vertices $\mathcal{S}$ and $\mathcal{G}$, corresponding to the boundaries of regions $\mathcall{S}$ and $\mathcall{G}$, respectively.
	It is defined as:
	\begin{equation}
	\label{eq:HD}
	HD = \max \{ h(\mathcal{G}, \mathcal{S}), h(\mathcal{S}, \mathcal{G}) \},
	\end{equation}
	
	where $h(\mathcal{G}, \mathcal{S}) = \max_{\mathbf{g} \in \mathcal{G}} \{ \min_{\mathbf{s} \in \mathcal{S}} \{ d(\mathbf{g}, \mathbf{s}) \} \}$, $h(\mathcal{S},\mathcal{G}) = \max_{\mathbf{s} \in \mathcal{S}} \{ \min_{\mathbf{g} \in \mathcal{G}} \{ d(\mathbf{s}, \mathbf{g}) \} \}$, and $ d(\mathbf{g}, \mathbf{s})$ is the Euclidean distance in an $n$-dimensional Euclidean space $\mathbb{R}^n$.
	\item Lastly, in order to take into account the correlation of all samples belonging to two different points clouds, a variant of the Mahalanobis distance (\textit{MHD}) is used.
	Again, let $V_{\mathcall{S}_1}$ and $V_{\mathcall{S}_2}$ be the sets of voxels segmented by two different procedures.
	The \textit{MHD} between $V_{\mathcall{S}_1}$ and $V_{\mathcall{S}_2}$ is:
	\begin{equation}
	\label{eq:MHD}
	MHD = \sqrt{\left(\mu_{V_{\mathcall{S}_1}} - \mu_{V_{\mathcall{S}_2}}\right)^\top \Sigma^{-1} \left(\mu_{V_{\mathcall{S}_1}} - \mu_{V_{\mathcall{S}_2}}\right)},
	\end{equation}
where $\mu_{V_{\mathcall{S}_1}}$ and $\mu_{V_{\mathcall{S}_2}}$ are the means of the two $3$D segmented volumes.
Furthermore, $\Sigma$ is the common covariance of the two sets and is given by $\Sigma = \frac{\abs{V_{\mathcall{S}_1}} \Sigma_{V_{\mathcall{S}_1}} + \abs{V_{\mathcall{S}_2}}\Sigma_{V_{\mathcall{S}_2}}}{\abs{V_{\mathcall{S}_1}} + \abs{V_{\mathcall{S}_2}}}$.

\end{itemize}

\section{Medical image enhancement evaluation metrics}
\label{sec:enhanceEval}
In this section, the definition of the metrics typically used to evaluate image enhancement approaches are recalled, which will be exploited to assess the performance of MedGA.
These metrics are essential to quantitatively evaluate the effects of image enhancement techniques, since measuring the ``quality'' of an image might be strongly subjective.
In particular, we benefit here of the metrics considered in \cite{hashemi2010} to assess the capability of the image enhancement approaches in improving contrast, details and human visual perception.

Let $\mathcall{I}_\text{orig}$ and $\mathcall{I}_\text{enh}$ be the original input image and the enhanced image, respectively, consisting in $M$ rows and $N$ columns.
Considering that the range of gray levels of the output image $\mathcall{L}_{out}=[l_{out}^\text{(min)}, \dots, l_{out}^\text{(max)}]$ could be different from the original range $\mathcall{L}_{in}=[l_{in}^\text{(min)}, \dots, l_{in}^\text{(max)}]$, as a first step before computing the metrics we re-map the output pixel intensities (i.e., the gray levels) onto the original range, as follows:
\begin{equation}
 \label{remapFunz}
 \tilde{\mathcall{I}}_\text{enh}(a,b) = \frac{ \left(\mathcall{I}_\text{enh}(a,b) - l_{out}^\text{(min)}\right) \cdot \left(l_{in}^\text{(max)} - l_{in}^\text{(min)}\right)} {\left(l_{out}^\text{(max)} - l_{out}^\text{(min)}\right)} + l_{in}^\text{(min)}\text{,}
\end{equation}
where $a=1, \dots, M$ and $b=1, \dots, N$.
Note that we actually consider the extended range $\mathcall{L}'_{in}$ for $I_\text{orig}$.

The Peak Signal-to-Noise Ratio (\emph{PSNR}) denotes the ratio between the maximum possible intensity value of a signal and the distortion between the input and output images:
\begin{equation}
\label{psnr}
\begin{split}
PSNR & = 10 \cdot \log_{10} \left( \frac{(l_{in}^\text{(max)})^2}{MSE} \right)\\
& = 20 \cdot \log_{10} \left( \frac{l_{in}^\text{(max)}}{\sqrt{MSE}} \right),
\end{split}
\end{equation}
where $MSE=\frac{1}{M \times N} \sum_{a=1}^M \sum_{b=1}^N \norm{\mathcall{I}_\text{orig}(a,b) - \tilde{\mathcall{I}}_\text{enh}(a,b)}^2 $ is the Mean Squared Error, which allows to compare the pixel values of $\mathcall{I}_\text{orig}$ to those of $\tilde{\mathcall{I}}_\text{enh}$.

Furthermore, the \emph{PSNR} is usually expressed in terms of the logarithmic Decibel scale.
With regard to our application, we employ only a limited portion of the full dynamic range of 16-bit images, such as in the case of DICOM madical images; we thus use as the largest possible value the maximum intensity value present in the original image (i.e., $l_{in}^\text{(max)} = \max\{\mathcall{L}_{in}\} = \max\{\mathcall{L}'_{in}\}$) instead of the maximum representable value in a 16-bit image (i.e., $2^{16}-1 = 65,535$).

\cite{munteanu2004} stated that good contrast and enhanced images are characterized by high numbers of \textit{edgels} (i.e., pixels belonging to an edge), and that an enhanced image should have a higher intensity of the edges, compared to its non-enhanced counterpart \cite{saitoh1999}.
Therefore, a good enhancement technique should yield satisfactory results in the case of standard vision processing tasks, such as segmentation or edge detection \cite{starck2003}.
Here, to evaluate image enhancement of MRI data, we employ the \cite{canny1986}'s method, which is a highly reliable and mathematically well-defined edge detector.
This approach deals with weak edges and accurately determines edgels, by applying a double threshold (to identify potential edges) and a hysteresis-based edge tracking.
Let $\mathcall{M}_\text{Canny}$ be the edge map yielded by the Canny's edge detector, which is a binary image wherein only edgels are set to $1$.
The number of detected edges ($\#\mbox{DE}$) in $\mathcall{M}_\text{Canny}$ is computed as:
\begin{equation}
	\label{numEdges}
	\#\mbox{DE} = \sum_{a=1}^M \sum_{b=1}^N \mathcall{M}_\text{Canny}(a,b).
\end{equation}

An additional metrics, called Absolute Mean Brightness Error (\emph{AMBE}) \cite{chen2003, arriaga2014}, can be employed to measure the brightness preservation of the enhanced image:
\begin{equation}
	AMBE = \frac{\abs{\mathbb{E}[\mathcall{I}_\text{orig}] - \mathbb{E}[\tilde{\mathcall{I}}_\text{enh}]}}{L} ,
\end{equation}
where $\mathbb{E}[\cdot]$ denotes the expected (mean) value of a gray level distribution.
\emph{AMBE} is normalized in $[0,1]$, dividing by $L = l_{in}^\text{(max)} - l_{in}^\text{(min)}$, which is the dynamic range of the input gray-scale (in our case, $\mathcall{L}'_{in}$).
Note that low values of \emph{AMBE} denote that the mean brightness of the original image is preserved.

Finally, we consider an alternative quality metrics called Structural Similarity Index (\emph{SSIM}) \cite{wang2004}, used to assess the image degradation perceived as variations in structural information \cite{bhandari2016}.
The structural information defines the attributes that represent the structure of objects in the image, independently of the average luminance and contrast.
In particular, local luminance and contrast are taken into account since overall values of luminance and contrast can remarkably vary across the whole image.
\emph{SSIM} is based on the degradation of structural information---assuming that human visual perception is highly adapted for extracting structural information from a scene---and compares local patterns of pixel intensities.
As a matter of fact, natural image signals are highly structured, since pixels are strongly dependent on each other, especially those close by.
These dependencies convey important information about the structure of the objects in the viewing field.
Let $\mathbf{X}$ and $\mathbf{Y}$ be the $\mathcall{I}_\text{orig}$ and $\mathcall{I}_\text{enh}$ image signals, respectively; \emph{SSIM} combines three relatively independent terms:
\begin{itemize}
	\item the luminance comparison  $l(\mathbf{X}, \mathbf{Y}) = \frac{2 \mu_{\mathbf{X}} \mu_{\mathbf{Y}} + \kappa_1}{\mu_{\mathbf{X}}^2 + \mu_{\mathbf{Y}}^2 + \kappa_1}$;
	\item the contrast comparison $c(\mathbf{X}, \mathbf{Y}) = \frac{2 \sigma_{\mathbf{X}} \sigma_{\mathbf{Y}} + \kappa_2}{\sigma_{\mathbf{X}}^2 + \sigma_{\mathbf{Y}}^2 + \kappa_2}$;
	\item the structural comparison $s(\mathbf{X}, \mathbf{Y}) = \frac{\sigma_{\mathbf{X} \mathbf{Y}} + \kappa_3}{\sigma_{\mathbf{X}} \sigma_{\mathbf{Y}} + \kappa_3}$;
\end{itemize}
where $\mu_{\mathbf{X}}$, $\mu_{\mathbf{Y}}$, $\sigma_{\mathbf{X}}$, $\sigma_{\mathbf{Y}}$, and $\sigma_{\mathbf{X}\mathbf{Y}}$ are the local means, standard deviations, and cross-covariance for the images $\mathbf{X}$ and $\mathbf{Y}$, while $\kappa_1, \kappa_2, \kappa_3 \in \mathbb{R}^+$ are regularization constants for luminance, contrast, and structural terms, respectively, exploited to avoid instability in the case of image regions characterized by local mean or standard deviation close to zero.
Typically, small non-zero values are employed for these constants; according to \cite{wang2004}, an appropriate setting is $\kappa_1 = (0.01 \cdot L)^2$, $\kappa_2 = (0.03 \cdot L)^2$, $\kappa_3 = \kappa_2/2$, where $L$ is the dynamic range of the pixel values in $\mathcall{I}_\text{orig}$  (represented in $\mathcall{L}'_{in}$).
\emph{SSIM} is then computed by combining the components described above:
\begin{equation}
\label{eq:SSIM}
	SSIM = l(\mathbf{X}, \mathbf{Y})^\alpha \cdot c(\mathbf{X}, \mathbf{Y})^\beta \cdot s(\mathbf{X}, \mathbf{Y})^\gamma,
\end{equation}
where $\alpha$, $\beta$, $\gamma > 0$ are weighting exponents.
As reported in \cite{wang2004}, if $\alpha = \beta = \gamma = 1$ and $\kappa_3 = \kappa_2/2$, the \emph{SSIM} becomes:
\begin{equation}
	SSIM = \frac{\left( 2 \mu_{\mathbf{X}} \mu_{\mathbf{Y}}  + \kappa_1 \right) \left( 2 \sigma_{\mathbf{X} \mathbf{Y}} + \kappa_2 \right) } {\left( \mu_{\mathbf{X}}^2 + \mu_{\mathbf{Y}}^2  + \kappa_1 \right) \left( \sigma_{\mathbf{X}}^2 + \sigma_{\mathbf{Y}}^2  + \kappa_2 \right)}.
\end{equation}
\emph{SSIM} generalizes the Universal Quality Index (\emph{UQI}), defined in \cite{wang2002}, which corresponds to the specific settings having $\kappa_1 = \kappa_2 = 0$ and yields unstable results when either $\left( \mu_{\mathbf{X}}^2 + \mu_{\mathbf{Y}}^2 \right)$ or $\left( \sigma_{\mathbf{X}}^2 + \sigma_{\mathbf{Y}}^2 \right)$ tends to zero.
Notice that the higher the \emph{SSIM} value, the higher the structural similarity, implying that the enhanced image $\mathcall{I}_\text{enh}$ and the original image $\mathcall{I}_\text{orig}$ are quantitatively similar.

\section{Quantitative validation in clonogenic assays}
\label{sec:validClonAssay}
This Section describes the procedure to validate our method based on ACC alone \cite{militelloCBM2017} against the conventional cell colony counting.

Generally, for higher robustness, several wells bearing the same experimental configuration are considered.
Therefore, a total of different $r$ repetitions (i.e., number of wells with the number of cells and treatment conditions) is performed.

Moreover, the obtained values are averaged over three counting measurements for each well, performed by a staff composed of three different expert biologists, represent our ground-truth used as a reference for the fully automated approach.

\subsection{Conventional evaluation}
According to \cite{franken2006}, the conventional cell colony counting takes into account the Plating Efficiency (PE) and the Surviving Fraction (SF) calculated using Eqs. (\ref{eq:plateEff}) and (\ref{eq:survFrac}), respectively:

\begin{equation}
    \label{eq:plateEff}
       \text{PE} = \frac{\text{\# grown colonies}}{\text{\# plated cells}} \times 100,
\end{equation}

\begin{equation}
    \label{eq:survFrac}
       \text{SF} = \frac{\text{\# colonies grown after treatment}}{\text{\# plated cells}} \times \text{PE}.
\end{equation}

The PE represents the fraction of colonies formed from cells that were not exposed to the treatment, considering the number of grown colonies and the number of plated cells, and must be determined for each experiment.
The SF represents the surviving fraction of cells after any treatment (i.e., irradiation or cytotoxic agent), calculated taking into account the PE of untreated cells (i.e., control well), the number of colonies grown after treatment and the number of plated cells.

\subsection{Area-based evaluation}
As for the conventional method in laboratory practice, in our approach the SF is measured considering that the ACC of untreated cells is $100$, expressed as a percentage, by normalizing as in Eq. (\ref{eq:survFracNorm}):

\begin{equation}
    \label{eq:survFracNorm}
       \text{SF}_\text{Treated cells} = \frac{\text{ACC}_\text{Treated cells}}{\text{ACC}_\text{Untreated cells}} \times 100.
\end{equation}

\section{Clinical evaluation based on a Likert scale}
\label{sec:LikertEval}
The multimodal segmentation method proposed in \cite{rundoCMPB2017} computes the tumor volumes defined in MRI as well as in MET-PET, simultaneously.
Two boundaries are separately generated: one on MRI (i.e., GTV) and one on PET (i.e., BTV or BTV\textsubscript{MRI}).
Ideally, both GTV and BTV should be validated against a manual delineation by the experts, considered as our “ground-truth”, using the same PET/MRI datasets.
However, due to the current clinical protocol in Leksell Gamma Knife treatment planning, the manual reference contours are available either on MR or PET images only, but not on the fused PET/MR images.
Currently, it is not possible to define a real “gold standard” CTV according to both morphologic MRI and metabolic PET images without treatment response evaluation, since the treatment is planned on MRI datasets alone in the clinical practice.
As a matter of fact, ground truth refers to having exact knowledge of the tumor size.
In PET imaging, and consequently in multimodal PET/MR imaging, the metabolic ground truth is impossible to be obtained for images concerning living humans
Thus, the validation of PET methods with real patient images lacks of an actual gold standard.
In addition, manual delineation by expert physicians is subject to intra- and inter-operator variability, especially in tumor segmentation on PET images.
Histological images provide the only valid ground truth for quantitative segmentation evaluation in human studies.
Obviously, the histology of brain metastases treated using neuro-radiosurgery is unavailable.
For this reason, the proposed monomodal PET image segmentation method was assessed using phantom experiments \cite{stefano2016}.
The same algorithm has been already applied in a real clinical environment to assess the clinical feasibility of the proposed automated approach \cite{stefano2016}.

So, to assess the clinical utility and feasibility of the proposed multimodal method and to evaluate retrospectively the clinical impact of BTV integration in Gamma Knife treatment planning, a qualitative evaluation, using a five-point Likert scale ranging from $1$ (worst) to $5$ (best), was carried out by a team of experienced physicians. 
The clinical staff, composed of a neurosurgeon, a nuclear medicine physician, and a radiation oncologist, jointly analyzed the brain tumors without any information of the used segmentation method (BTV or BTV\textsubscript{MRI}).
By comparing their perspectives by different clinical backgrounds, the physicians were able to gain insights about the tumors and provide a more careful evaluation for each case.
As a matter of fact, a Likert scale, rather than providing fixed criteria, allows clinicians to give a score based on overall impression, focusing mainly on the clinical value of BTV in an accurate treatment planning.
Although possibly prone to greater inter-observer variability, this approach may be more familiar to some physicians and may perhaps be more straightforward to apply for observers with previous experience in brain PET imaging.

The used Likert scale, according to retrospective clinical value for treatment planning, was defined as:
\begin{enumerate}
    \item strong worsening in CTV definition;
    \item moderate worsening in CTV definition;
    \item indifferent, neither enhancement nor worsening in CTV definition;
    \item moderate enhancement in CTV definition;
    \item strong enhancement in CTV definition.
\end{enumerate}

\section{Visual Turing test}
\label{sec:TuringEval}
To quantitatively evaluate how realistic the synthetic images are, 
a visual Turing test performed by an expert could be useful \cite{geman2015}.
The so-called Visual Turing Test~\cite{salimans2016} uses binary questions to probe a human ability to identify attributes and relationships in images \cite{geman2015}.
For these motivations, it is commonly used to evaluate GAN-generated images, such as for SimGAN~\cite{shrivastava2017}.
This applies also to medical images in clinical environments~\cite{chuquicusma2017}, wherein physicians' expertise is critical.

\end{appendices}

\printthesisindex 

\end{document}